Gran Sasso Science Institute

ASTROPARTICLE PHYSICS
DOCTORAL PROGRAMME

Cycle XXXVI - AP 2020/2025

# 3D Tracking with the CYGNO/ INITIUM experiment

PHD CANDIDATE
**David José Gaspar Marques**

ADVISORS

**Elisabetta Baracchini**
Gran Sasso
Science Institute

GRAN SASSO
SCIENCE INSTITUTE
SCHOOL OF ADVANCED STUDIES
Scuola Universitaria Superiore

GRAN SASSO SCIENCE INSTITUTE

SCHOOL OF ADVANCED STUDIES
Scuola Universitaria Superiore

Gran sasso science institute

Doctoral thesis research proposal

PhD program in particle and astroparticle physics
XXXVI Cycle

# 3D Tracking with the CYGNO/INITIUM experiment

Author:

David José Gaspar Marques

Thesis advisor:

Prof. Elisabetta Baracchini
(Gran sasso science institute)

June of 2025

The research reported in this thesis was performed with financial support of the following institutions/programs:

- European Union's Horizon 2020 research and innovation programme from the European Research Council (ERC) grant agreement No 818744.

Funded by the European Union

European Research Council
Established by the European Commission

# Acknowledgements

[ENG]

First and foremost, I thank Professor Elisabetta Baracchini, my Ph.D. and thesis advisor, for all the knowledge shared over the past four years. It has been a long journey, filled with challenges, discoveries, and a final outcome of which I am truly proud. I also express my gratitude to the CYGNO group and in particular Davide Pinci, for welcoming me as a new researcher in the collaboration and for their invaluable guidance and support throughout my work. My thesis has greatly benefited from being developed within such a knowledgeable and collaborative research environment.

A cheerful thanks to the *young Cygners* – Giorgio, Atul, Samuele, Flaminia, and more recently Stefano, Melba, and Little Flowers (Davide) – with whom I spent countless hours in the office, the labs, and the car. The sense of camaraderie and friendship we shared made my Ph.D. journey much more enjoyable. A special thanks to Samuele and Giorgio for the many Campari moments after the long days in the lab.

Within this adventure in Italy, which came with its fair share of challenges, I thank the class of the 36° cycle: Stefano, Cecilia, Michele, Simone, Ulyana, and (also flatmates) Alessio and Alessandro. As the only outsider, I felt genuinely welcomed and included. I'm grateful to all of you who, in your own unique ways, brought joy to my daily life and taught me a bit more about the Mamma Mia land. *Grazie*.

Far away in the Peninsula, a big thanks to my friends – Afonso, Jordão, Gabriel, Brito, and, deeper in the countryside, Miguel and Catarina – who made every trip back to Portugal a bit more relaxed and memorable. Knowing you're always there has been a real source of strength and joy and a reminder that I've got good friends, no matter where I am.

Finally, the biggest thanks go to my parents, José and Isabel, and my sister, Rita, who always gave me the warmest welcome and goodbye each time I arrived from or left for Italy. It's truly comforting to know there's always a home to come back to. Without your support, none of this would have been possible. *Obrigado*.

# Agradecimentos

[PT]

Em primeiro lugar, agradeço à Professora Elisabetta Baracchini, a minha orientadora de doutoramento e tese, por todo o conhecimento partilhado ao longo destes últimos quatro anos. Tem sido uma longa jornada, repleta de desafios e descobertas que culminaram num resultado final do qual me orgulho vivamente. Agradeço também ao grupo CYGNO, e em particular ao Davide Pinci, por me terem acolhido como um novo investigador na colaboração e pelo apoio e orientação que me deram ao longo do meu trabalho. Sem dúvida que a minha tese beneficiou por ter sido desenvolvida neste ambiente de investigação tão competente e colaborativo.

Um alegre agradecimento aos *young Cygners* – Giorgio, Atul, Samuele, Flaminia e mais recentemente, Stefano, Melba e Little Flowers (Davide) – com quem passei inúmeras horas no escritório, nos laboratórios e no carro. O espírito de camaradagem e amizade que partilhámos tornou a minha jornada de doutoramento muito mais divertida. Um agradecimento especial ao Samuele e ao Giorgio pelos muitos momentos Campari após os longos dias de laboratório.

Nesta aventura em Itália, que também veio com os seus desafios, agradeço à turma do 36° ciclo: Stefano, Cecilia, Michele, Simone, Ulyana, e (também colegas de casa) Alessio e Alessandro. Mesmo sendo o único de fora, senti-me genuinamente acolhido e incluído. Fico grato a todos vós que, de maneira diferente e peculiar cada um, trouxeram alegria ao meu dia-a-dia e me ensinaram um pouco mais sobre a terra do Mamma Mia. *Grazie*.

Lá longe, na Península, um grande agradecimento aos meus amigos – Afonso, Jordão, Gabriel, Brito e, mais no interior, Miguel e Catarina – que tornaram cada viagem a Portugal um pouco mais relaxante e memorável. Saber que estão lá sempre tem sido uma verdadeira fonte de força e alegria, e um lembrete de que tenho bons amigos, independentemente de onde esteja.

Finalmente, o maior agradecimento vai para os meus pais, José e Isabel, e à minha irmã, Rita, que sempre me deram uma calorosa receção e despedida quando chegava ou partia para Itália. É verdadeiramente reconfortante saber que há sempre uma casa para onde voltar. Sem o vosso apoio, nada disto teria sido possível. *Obrigado*.

# Abstract

The nature of dark matter (DM) remains one of the most fundamental open questions in modern physics. Among the main DM candidates are Weakly Interacting Massive Particles (WIMPs), which are expected to interact with ordinary matter through scattering. Detecting such interactions requires highly sensitive and low-background detectors. Directional detection offers a promising way to overcome the limits of conventional WIMP search techniques – especially the neutrino floor – by identifying the direction of WIMP-induced nuclear recoils (NR) and thereby inferring the incoming WIMP direction. The motion of Earth around the Sun and within the Galaxy creates an apparent WIMP wind, which results in an anisotropic distribution of recoil tracks in detectors on Earth, an effect that no other background can mimic. This thesis contributes to this effort through the development and application of 3D reconstruction techniques for the CYGNO/INITIUM experiment.

CYGNO employs a $\mathrm{He{:}CF_4}$ gas mixture at atmospheric pressure and a triple-GEM amplification system, producing scintillation light that is read by CMOS sensors and photomultiplier tubes (PMTs). This work presents the development of a 3D reconstruction strategy by combining CMOS (XY) and PMT (Z) data. A dedicated framework was developed to extract information from PMT waveforms, synchronize it with CMOS data, and reconstruct the full 3D topology and direction of ionizing tracks. This reconstruction was applied to LIME data, CYGNO's largest prototype, operating underground at LNGS.

The 3D analysis was initially focused on alpha particles, whose ionization profiles closely mimic WIMP-induced NRs. By combining CMOS and PMT data, alpha tracks were reconstructed with accurate direction and head-tail determination. This method was used to characterize the alpha background in LIME, including spatial localization within different sectors of the detector. The results confirmed the presence of $^{222}\mathrm{Rn}$ diffused in the sensitive gas volume and the likelihood that its decay daughters are produced with a positive charge. The spatial and angular distributions of alphas also revealed contamination from $^{238}\mathrm{U}$ and $^{232}\mathrm{Th}$ in the detector materials, namely field cage rings and GEMs, showcasing the robustness of the developed analysis. These findings helped explain discrepancies between Monte Carlo simulations and measured data and led to design improvements in future CYGNO detectors, namely CYGNO-04.

Finally, the thesis reports the first steps toward implementing Negative Ion Drift (NID) operation with MANGO prototype. A novel analysis strategy is presented for the literature first-ever reported NID PMT waveforms. These results highlight the potential of the CYGNO/INITIUM approach for directional, low-background dark matter detection and support its development into future large-scale detectors.

# Contents

# Introduction

Over the past several decades, a wide range of observations from astrophysics, cosmology, and particle physics have converged into a coherent framework suggesting that most of the Universe is composed of an elusive form of matter known as Dark Matter (DM). Despite its dominant presence, the fundamental nature of DM remains unknown, as no interactions beyond gravity have been conclusively observed. Nevertheless, it is widely accepted that DM accounts for the vast majority of matter in the cosmos. To incorporate DM into a unified theoretical model, numerous hypotheses have been proposed. Among the most compelling and extensively studied candidates are Weakly Interacting Massive Particles (WIMPs) – hypothetical elementary particles that could constitute DM. WIMPs are particularly appealing because they could elastically scatter off atomic nuclei, potentially producing observable nuclear recoils (NRs) in detectors on Earth. This possibility has driven a wide range of direct detection efforts.

Although no definitive evidence for WIMPs has yet emerged, the search continues using increasingly sophisticated technologies aimed at achieving extreme sensitivity and minimizing background noise. Direct detection experiments are typically conducted in underground laboratories to shield them from cosmic-ray-induced backgrounds. Additional strategies for background suppression include the use of passive and active shielding, selecting construction materials with low intrinsic radioactivity, and limiting the analysis to a central fiducial volume within the detector. To enhance sensitivity to high-mass WIMPs, experiments with heavy targets aim to maximize exposure by constructing large detectors with ample sensitive volumes – such as dual-phase liquid noble gas Time Projection Chambers (TPCs). Conversely, probing low-mass WIMPs requires detectors with extremely low energy thresholds. This can typically achieved using cryogenic bolometers, which detect the heat generated by particle interactions, often in conjunction with ionization or scintillation signals for background rejection. However, maintaining effective electron/nuclear recoil (ER/NR) discrimination at such low thresholds remains a significant technical challenge. Ultimately, all direct detection strategies face an intrinsic limitation from coherent elastic neutrino-nucleus scattering (CE$\nu$NS). This so-called neutrino fog generates nuclear recoils from solar, atmospheric, and diffuse supernova neutrinos, potentially mimicking or obscuring a genuine WIMP signal.

Directional DM searches offer a promising and innovative approach that may over-come several limitations of conventional detection methods. Due to the motion of the Sun and Earth through the galaxy, an apparent WIMP wind is expected from a fixed direction in galactic coordinates. As a result, WIMP-induced NRs should exhibit a dipole-like angular distribution – an unmistakable signature that no terrestrial background can replicate, as the Earth's rotation randomizes any local anisotropies. The only directional background arises from solar neutrinos, whose recoil distribution points toward the Sun and overlaps only minimally with the WIMP signal. Thus, directionality provides a powerful means to bypass the neutrino fog and allow the unambiguous discovery of DM. This approach typically employs gaseous TPCs with high-resolution readouts capable of reconstructing the 3D shape and orientation of recoil tracks. Such detectors provide both directional sensitivity and (ER/NR) discrimination. By selecting appropriate target gases, sensitivity to both spin-independent (SI) and spin-dependent (SD) WIMP interactions can be achieved.

The CYGNO/INITIUM project aims to gradually improve its techniques and detector performances through the deployment of increasingly larger prototypes, with the final goal of building a large-volume TPC – on the order of 30 $m^3$ – optimized for rare event detection, including DM searches and solar neutrino studies. It operates with a helium and carbon tetrafluoride ($He:CF_4$) gas mixture at atmospheric pressure, making it sensitive to both SI and SD interactions. The primary ionization is amplified using a triple-GEM (Gas Electron Multiplier) stack. This amplification produces secondary scintillation light from $CF_4$ de-excitation, which is read out by a combination of CMOS cameras and photomultiplier tubes (PMTs). This hybrid optical readout enables 3D reconstruction of track topology and energy loss per unit length ($dE/dx$), providing both ER/NR discrimination and directional detection capabilities. In synergy with the CYGNO experiment, the ERC Consolidator Grant INITIUM project – from which this thesis was supported – aims to integrate Negative Ion Drift (NID) technology into CYGNO's 3D optical readout TPC. By adding an electronegative gas such as $SF_6$, the primary ionization electrons are captured, resulting in the drift of negative ions rather than electrons. This technique greatly reduces diffusion during drift, enhancing track reconstruction and significantly improving directional sensitivity.

This thesis is largely dedicated to the reconstruction of the 3D geometry, topology, and direction of alpha particles. This is achieved by merging the data acquired from the two different sensors that compose the CYGNO optical readout – CMOS, for the XY plane, and PMTs, for the Z coordinate. The 3D analysis builds upon and specializes the reconstruction techniques also developed in this thesis for extracting the basic information from the PMT waveforms, within the general framework of CYGNO data reconstruction. While applicable to all CYGNO detectors, the 3D analysis was first implemented for the reconstruction and identification of alpha particles in Run 4 of the LIME underground campaign. LIME is currently the largest CYGNO detector and has been operating underground at the Laboratori Nazionali del Gran Sasso (LNGS) since 2022, having collected the most extensive dataset among all CYGNO detectors and pro-totypes. The primary goal was to evaluate the performance of the analysis on real data – in particular, its efficiency and capabilities – and to contribute to the preliminary studies of the LIME background spectrum, where inconsistencies between data and MC simulations were observed. The study of alpha particles was also motivated by the similarity of their tracks to those of nuclear recoils, especially in terms of charge deposition density and track topology. As will be shown, the measurement of the 3D length of alpha tracks proved to be a particularly valuable variable for characterizing the alpha spectrum in LIME. The reconstruction of ionization tracks in 3D marks a significant milestone for the CYGNO/INITIUM project.

In *Chapter 1*, a brief review of the theoretical and experimental arguments supporting the current DM paradigm is provided. The main detection strategies are discussed, with particular emphasis on those exploiting the directional signature of the WIMP wind to distinguish DM from backgrounds and confirm its galactic origin.

In *Chapter 2*, a detailed description of the CYGNO/INITIUM approach is presented, highlighting its distinguishing features – such as the combined use of CMOS-PMT sensors and GEMs. The various stages of the CYGNO timeline detectors are discussed, with particular emphasis on LIME, to provide the necessary context for the development of techniques used to analyze and associate PMT signals with those from the CMOS.

In *Chapter 3*, an introduction to the CMOS data reconstruction tools used in CYGNO is provided. The chapter then focuses on the reconstruction of the basic properties of PMT signals, which is one of the main focuses of this thesis. It also explains how CYGNO combines the information from multiple PMTs to reconstruct not only the track's Z coordinate, but also its XY position. Finally, preliminary tools for associating PMT waveforms with CMOS images are discussed.

In *Chapter 4*, the LIME overground studies carried out within the scope of this thesis are presented. These include the development of PMT trigger strategies within the CYGNO framework. Using LIME overground data and the newly developed PMT analysis, a study was conducted to measure the cosmic muon flux at ground level, near sea level. This served as an initial benchmark for assessing the capabilities and relevance of the PMT analysis.

In *Chapter 5*, a full description of the 3D analysis of alpha particles is provided. A brief discussion of particle interactions with matter is included to contextualize the techniques developed. The CMOS analysis, partly adapted from previous students' work, is specialized for alpha particle events. From this, the XY projection and the inclination relative to the X axis are extracted, along with the absolute Z position of the event. The PMT data is then used to determine the longitudinal path $\Delta Z$ and the track's inclination relative to the XY plane, completing the full 3D geometry. The track's sense is inferred from the position of the Bragg peak in the waveforms, resulting in a full 3D vectorial description of the ionization track.

In *Chapter 6*, the 3D analysis is applied to LIME underground data from Run 4 to characterize the alpha background. This is done by studying the emission positions and directions of alpha particles in various regions of LIME. Special attention is given to the study of the $^{222}$Rn decay chain, which was initially considered the likely source of discrepancies observed between data and MC.

In *Chapter 7*, an overview of the progress made in implementing NID operation is reported, with a focus on the algorithm developed for analyzing the first-ever observed NID PMT waveforms.

Finally, a commented summary of the results obtained throughout this thesis is presented in the Conclusions.

Chapter 1

# Overview of the Dark Matter paradigm

In recent decades, numerous astrophysical and cosmological observations have revealed that our understanding of the Universe is incomplete. These observations point to the existence of an unseen mass which cannot be explained with known particles and forces. Due to its lack of electromagnetic interaction, this mass is often referred to as *dark matter*. Despite decades of experiments providing clues about its behavior – mainly by setting limits on its properties such as mass and interaction cross-section – the exact nature of dark matter (DM) remains elusive. Strong evidence suggests that dark matter constitutes approximately 84% of the total mass of the Universe and plays a crucial role in the formation of galaxies and large cosmic structures.

The mysterious nature of dark matter has brought together researchers from astronomy, cosmology, and particle physics in the quest to explore new theories beyond the Standard Model. Advanced underground detectors on Earth are being developed to detect interactions between Standard Model (SM) particles and dark matter, employing techniques that push the limits of signal identification and background rejection. This field has also driven several R&D efforts, particularly in the development of micro-pattern gas detectors (MPGDs), which are designed to achieve high gain and high granularity, granting detectors and experiments the low energy thresholds and precise tracking required for dark matter searches. Despite these efforts, direct evidence of dark matter has yet to be found, and this ongoing enigma continues to drive research and experimentation.

This chapter provides an overview of the various pieces of evidence for dark matter (Section 1.1), potential particle candidates (Section 1.2), and detection methodologies (Section 1.3). Finally, the chapter concludes with a discussion of the directional approach in DM searches (Section 1.4). The discussion is performed envisioning the forthcoming chapters, focusing on the directionality signature of dark matter particles and its exploitation in gaseous TPCs. Throughout the chapter, several references to the studies conducted in this thesis are made, particularly those involving 3D reconstruction with alpha particles, head-to-tail asymmetry, and the identification of Rn-induced backgrounds.

## 1.1 Evidences of dark matter

Numerous experimental and observational findings support the existence of dark matter in the universe. These include a broad range of astronomical observations, from within individual galaxies to large-scale cosmic structures. This section provides a brief summary of some of the most robust evidence across different cosmological scales. For a more in-depth review on this topic, the reader is referred to [1].

### 1.1.1 Galaxy rotation curves

One of the most well-known and compelling pieces of evidence for the existence of dark matter comes from the study of rotation curves in spiral galaxies, such as the Milky Way. These studies, which analyze the circular velocities of stars and gas within galaxies, reveal that the velocities do not decrease as expected with increasing distance from the center. Instead, they remain nearly constant [2,3].

According to Newton's laws of gravitation, and assuming that the galaxy's mass is concentrated near its center, one would expect the orbital velocity, $v(r)$, of a star or gas cloud to be inversely proportional to its distance from the galaxy's center, $r$, following the equation:

$$v(r) = \sqrt{\frac{GM(r)}{r}} \tag{1.1}$$

where $G$ is the gravitational constant and $M(r)$ is the total mass enclosed within radius $r$. This implies that the farther a star is from the center, the slower its velocity should be.

Contrary to expectations, large sets of rotation curves reveal that the measured velocity remains constant at large radii, as exemplified in Figure 1.1. This contradicts the expected decline and suggests the presence of additional, unseen mass forming a halo around spiral galaxies. According to Equation 1.1, for the velocity to stay constant as $r$ increases, this additional halo must have a mass distribution that increases proportionally with $r$, i.e., $M(r) \propto r$. Assuming a spherical distribution for this mass, its density is often modeled as $\rho(r) \propto 1/r^2$, which would provide enough gravitational influence to maintain a constant total mass, $M(r)$, in the outer regions of the galaxy.

### 1.1.2 Cosmic microwave background anisotropies

Indicators of DM existence can also be inferred from measurements of cosmological-scale quantities, such as the Cosmic Microwave Background (CMB). The CMB is the faint radiation remnant of the Big Bang that permeates the entire universe. It represents the oldest light visible to Earth, essentially serving as a snapshot of the universe during its early phases. At the time of the Big Bang, the universe was extremely hot and dense, filled with a plasma of charged particles. As the universe expanded and cooled, electrons and protons combined to form neutral atoms in a process called recombination, allowing light to travel freely. This event, known as *decoupling*, is what is now imprinted in the CMB.

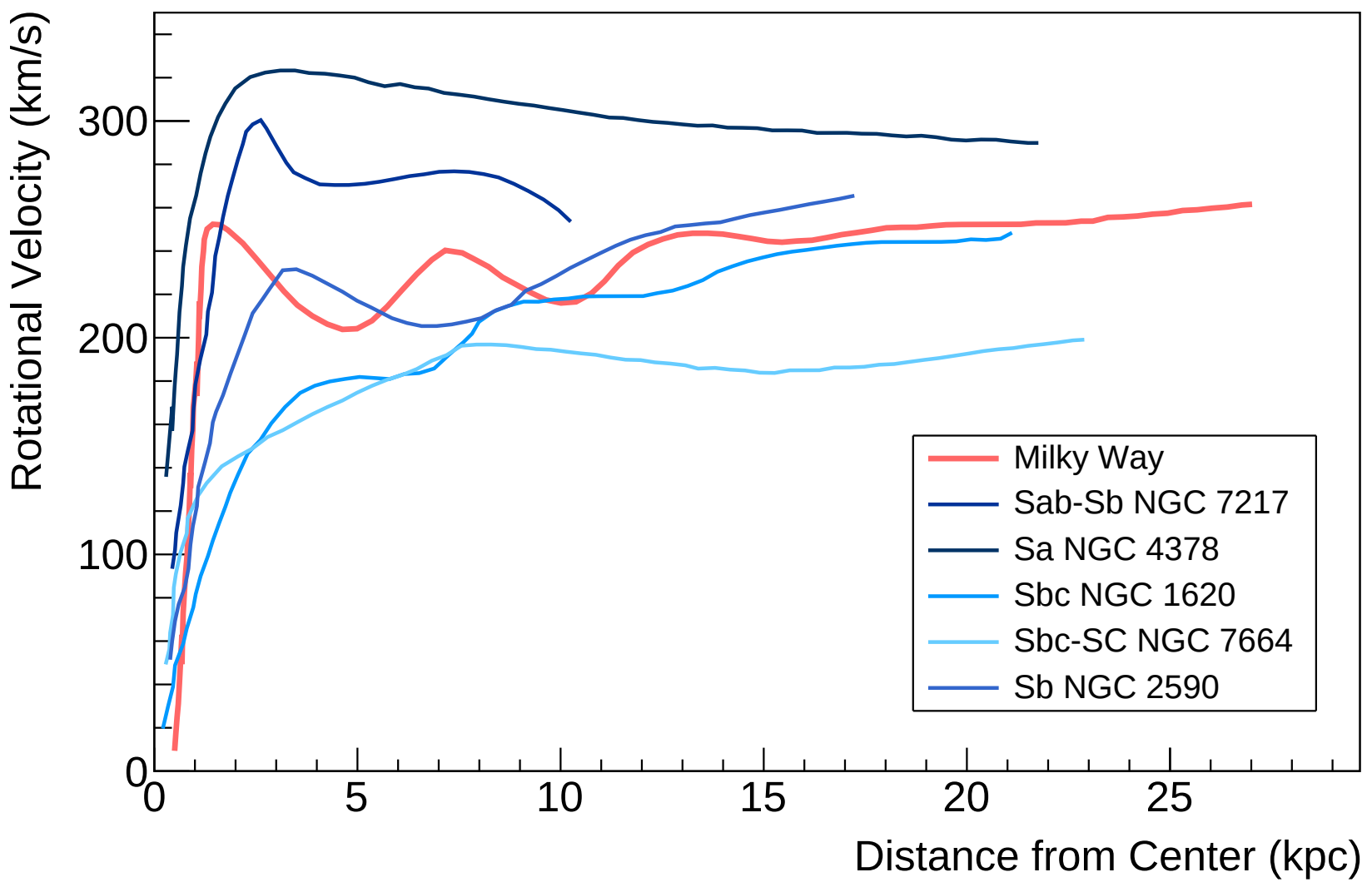

**Figure 1.1:** Rotational velocities for different galaxies, as a function of the distance from the center of the galaxy. Adapted from [4].

Although the CMB was originally emitted as visible and infrared light, the expansion of the universe has stretched the wavelengths of the radiation (a process called *redshift*), causing it to be observed in the microwave band today. The temperature of the CMB is remarkably uniform, averaging around 2.725 K [5], but tiny fluctuations are present, representing small variations in the density of the early universe, on the order of $\Delta T/T \sim 10^{-5}$. These fluctuations eventually led to the formation of galaxies and other cosmic structures. The temperature anisotropies can be decomposed into spherical harmonic functions, as follows:

$$T(\theta, \phi) = \sum_{lm} a_{lm} Y_{lm}(\theta, \phi) \tag{1.2}$$

The features of the CMB can also be studied through the $l$ power spectrum, given by:

$$D_l = \frac{l(l+1)}{2\pi} \langle |a_{lm}|^2 \rangle \tag{1.3}$$

Focusing on the temperature power spectrum ($D_l^{TT}$), $l = 0$ corresponds to the component related to the black body radiation, while $l = 1$ represents the dipole anisotropy component introduced by the motion of the Local Group relative to the CMB. Higher multipoles ($l \geqslant 2$), on the other hand, provide information related to density

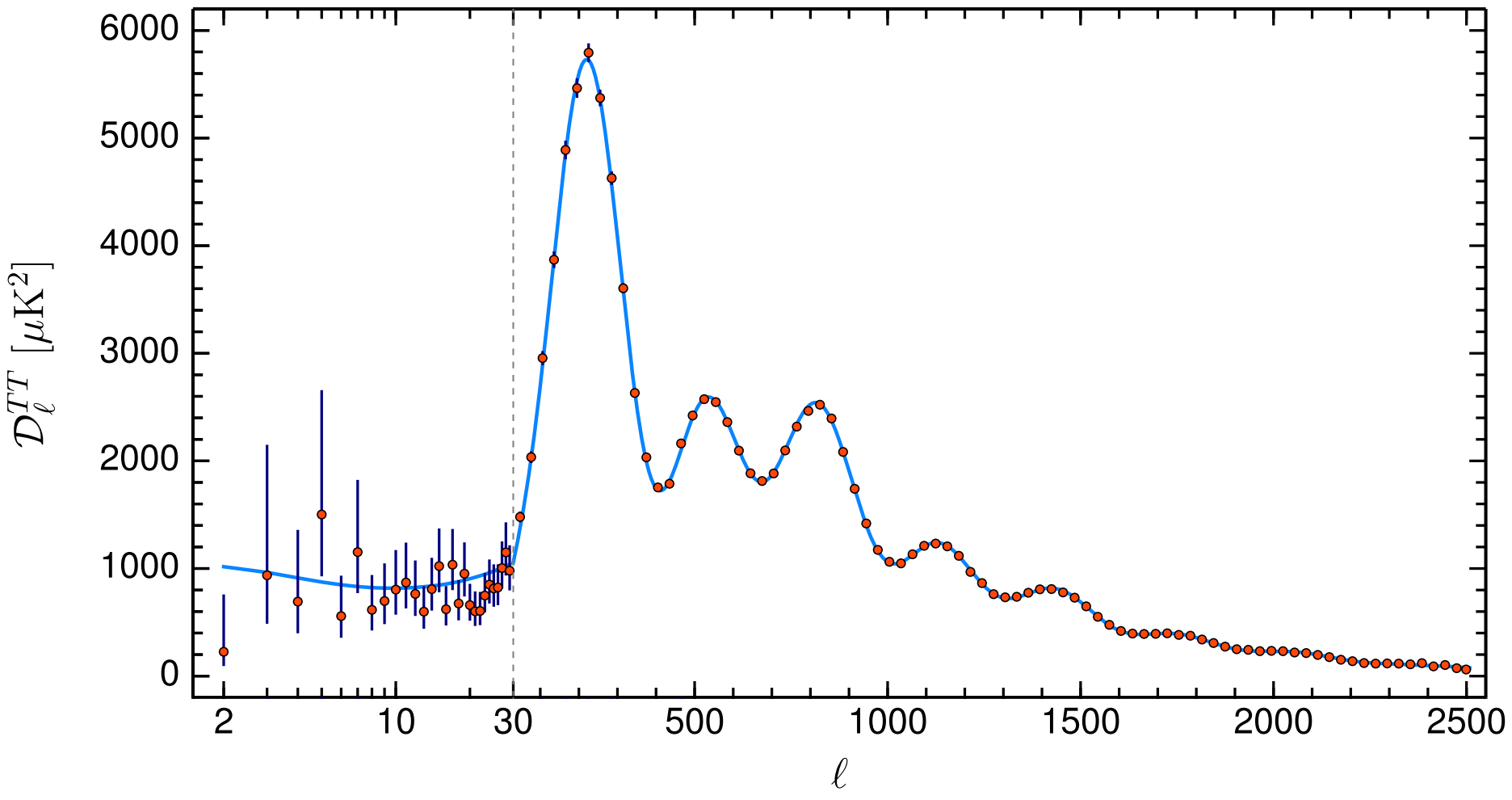

**Figure 1.2:** Temperature power spectrum measured by the Planck collaboration. The ΛCDM theoretical spectrum best fit is plotted in light blue. Note that the horizontal axis changes at $l = 30$, transitioning from logarithmic to linear. Adapted from [5].

perturbations in the early universe.

The Planck collaboration has measured this quantity as a function of the spherical harmonics, with the results shown in Figure 1.2 [5]. The blue line represents the minimal ΛCDM model, which involves only 6 parameters, including dark matter and dark energy. The best-fit values are listed in Table 1.1, assuming a total $\Omega = 1$, where $\Omega_i$ is defined as:

$$\Omega_i = \frac{\rho_i}{\rho_c} \tag{1.4}$$

Here, $\Omega_i$ expresses the ratio of the energy density ($\rho_i$) of each component to the critical density, $\rho_c$, which is required for a flat universe. This framework, the ΛCDM model, is the most widely supported theory to describe the composition and evolution of the universe, including the Big Bang [6]. In this model, the total energy density ($\Omega$) is composed of different components, as also illustrated in Table 1.1. Specifically, there is radiation, matter (including both baryonic and non-baryonic matter like DM), and the cosmological constant, which accounts for dark energy, the “force” responsible for the expansion of the universe.

The results obtained by the Planck collaboration, combined with the high precision of the ΛCDM model in describing the observed data, further support the hypothesis that dark matter is essential for explaining the development of the universe in its early stages, as evidenced by the CMB.

**Table 1.1:** Summary of the Planck collaboration's combined results on the $\Omega$ parameters of the $\Lambda$CDM model, fitted to the measured temperature anisotropies shown in Figure 1.2. Retrieved from [5].

| Component | Density parameter | Plack results at $1\sigma$ CL |
|---|---|---|
| Radiation | $\Omega_r$ | $\sim 9 \cdot 10^{-5}$ |
| Baryonic matter | $\Omega_b$ | $0.0489 \pm 0.003$ |
| Dark Matter | $\Omega_{DM}$ | $0.2607 \pm 0.0019$ |
| Dark Energy | $\Omega_\Lambda$ | $0.6889 \pm 0.0056$ |

### 1.1.3 Gravitational lensing

According to Einstein's theory of General Relativity, any massive object causes a distortion of the space-time continuum, proportional to its mass. As particles and light travel through space, their paths are distorted towards the direction of large, concentrated masses they encounter.

Following this, an observer can see an object (source) behind a massive object due to the deflection of the electromagnetic radiation emitted by the source, even when the source is geometrically blocked (as in "shadowed") by the massive object. Additionally, since the source emits radiation in all directions and the gravitational bending of space-time occurs in three dimensions, the massive object can create multiple images of the same source.

In this context, the massive object in the middle of the optical system is called a *lens*, leading to the process being referred to as *gravitational lensing*. This process is illustrated in Figure 1.3a. The change in angle, $\theta$, from the straight path of a light ray passing at a distance $r$ from a lens of mass $M$ can be approximated as [7]:

$$\theta = \sqrt{\frac{4GM}{rc^2}} \tag{1.5}$$

where $c$ is the speed of light in vacuum and $G$ is the gravitational constant.

From this formula, the mass of the lens can be inferred by measuring and studying the lensing intensity. A special term is used for cases where the central mass is so small that it does not produce a noticeable deformation of the image but only reduces the light intensity: *microlensing* [8].

Using this phenomenon, the effects of missing mass in the universe have been observed, particularly by studying the merging of galaxy clusters. An example worth mentioning is the Bullet Cluster, formed by the collision of two smaller clusters. Scientists compared the gravitational potential measured through the lensing effect, with the region of strongest X-ray emission. A displacement between these two was observed, suggesting the presence of dark matter in the two pre-existing galaxies. This dark matter seemingly passed through the collision point with minimal interaction and formed two separate regions. Meanwhile, the central part of the new cluster is populated by visible intergalactic gases that slowed down due to friction during the collision [9]. This effect is visible in Figure 1.3b. The displacement between the locations of baryonic mass emitting X-rays and the gravitational lensing supports the hypothesis of unaccounted-for mass in the universe and within galaxies, which is not electromagnetically visible but detectable through gravity.

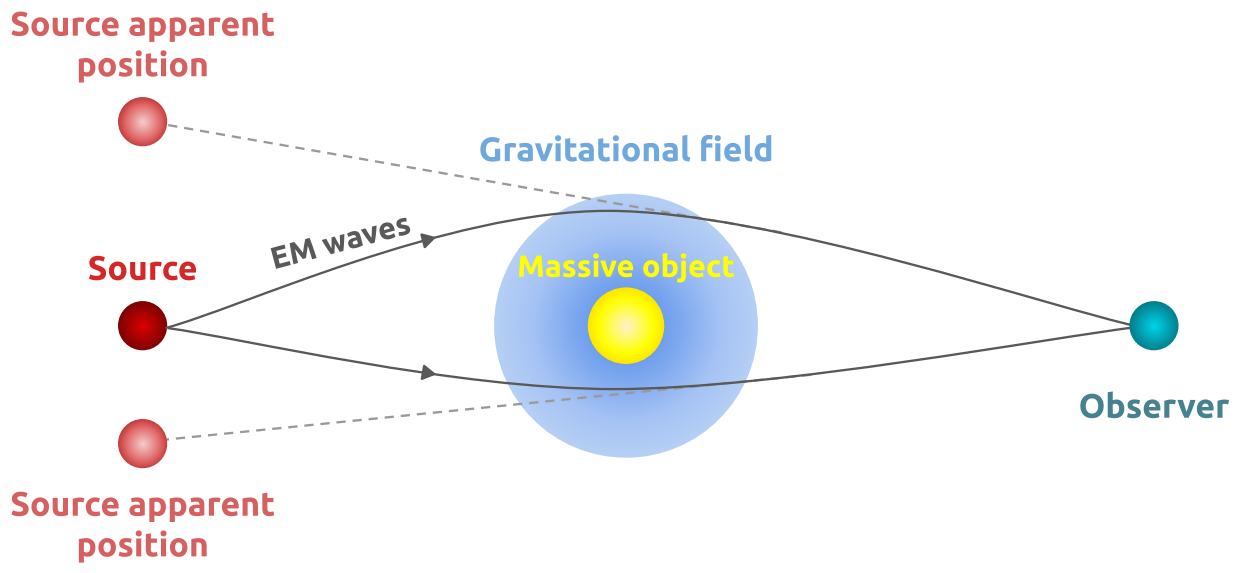

(a)

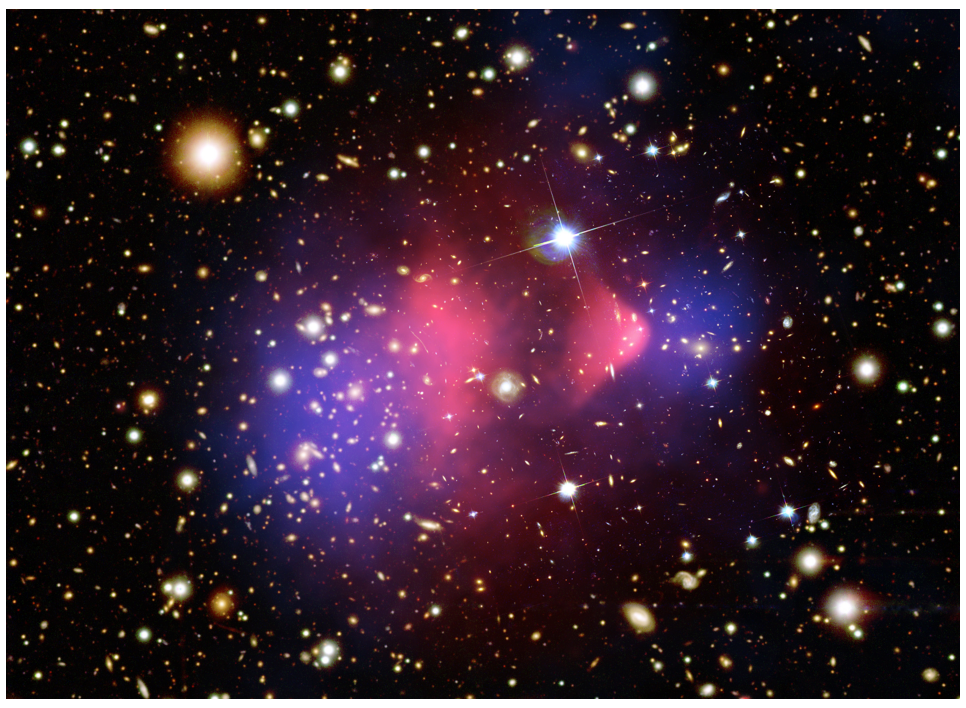

(b)

**Figure 1.3:** (a) Schematic of the gravitational lensing effect, adapted from [10]. (b) Galaxy cluster known as the *Bullet Cluster*, formed after the collision of two large clusters. The pink clumps, detected by Chandra, represent the X-rays emitted by baryonic ("normal") matter. The blue clumps represent the locations where most of the mass is found, as inferred through measurements of gravitational lensing. These are superimposed on an optical image obtained by the Magellan and Hubble Space Telescopes, showing galaxies in orange and white. In summary, the separation between the majority of the mass (blue) and normal matter (pink) indicates that most of the matter in this galaxy cluster is is not visible/dark. Retrieved from [11].

## 1.2 Dark Matter as a particle

Following the evidence for a missing massive component in the universe, observed at astronomical and cosmological scales, several theories have been developed to explain the inconsistencies arising from gravitational probes. Among the various theories attempting to explain dark matter, the most well-motivated contender is the Weakly Interacting Massive Particle (WIMP). The work presented in this thesis is based on the existence of dark matter as WIMPs, and the studies carried out focus on optimizing the detection and characterization of these particles. Therefore, in this section, a brief introduction to WIMPs, their characteristics, and detection techniques is provided. Other possible dark matter theories are briefly mentioned in Section 1.2.2 to offer the reader a broader perspective on the dark matter problem and potentially encourage further exploration of this topic and the cited works.

From the dark matter evidence presented above, several characteristics can be de-duced for any model that considers dark matter as a particle:

- **Dark & Neutral:** Dark matter gets its name due to the lack or very weak electromagnetic interactions with standard model particles, which would have otherwise been detected by now.

- **Non-baryonic & Abundant:** Based on the Planck CMB measurements, dark matter is implied to be non-baryonic with an abundance of $\Omega_{dm} \sim 0.26$, a significant and much larger density than that found for regular (baryonic) matter, as shown in Table 1.1.

- **Weakly-interacting:** While up to now only gravitational-affecting evidences of DM exist, weak-interactions (not necessarily referring to the electroweak force) between DM and SM particles are hoped in order to allow its detection.

- **Stable:** Assuming dark matter consists of weakly interacting particles, and considering their presence since the early universe and current measurements of gravitational influence, these particles must be stable to persist over such long timescales.

- **Cold:** The *hotness* of dark matter is determined by its energy at the decoupling moment. *Hot dark matter* models require relativistic particles with smaller masses, while *cold dark matter* allows for non-relativistic, heavier particles. Although still debated, most large-scale structure simulations suggest a cold dark matter scenario [12–14].

There is currently no SM particle that fits this profile, meaning an extension of the SM is required to explain dark matter. As will be shown, some SM extensions naturally predict the existence of particles that satisfy these characteristics, such as WIMPs.

### 1.2.1 WIMP candidate

Weakly Interacting Massive Particles are among the most studied and supported candidates to explain the dark matter problem, based on the premise that they are neutral and interact only at the weak scale with standard model particles.

In the hot early universe, WIMPs ($\chi$) and SM particles ($\phi$) were in thermal equilibrium, so the production of WIMPs through SM self-annihilation was balanced by their own self-annihilation, i.e.:

$$\chi + \bar{\chi} \Longleftrightarrow \phi + \phi^{\star} \tag{1.6}$$

At some point, as the universe expanded and its temperature and density decreased, dark matter decoupled from the SM particles, fixing the abundance of DM particles from that moment until today. This process is referred to as *freeze-out*. The DM number density after freeze-out ($n_\chi$) is found by solving the Boltzmann equation, assuming WIMPs are in kinetic equilibrium with the thermal bath. This results in:

$$\frac{dn_\chi}{dt} = -Hn_\chi - \langle \sigma_A v \rangle \left( n_\chi^2 - n_{\chi,equil}^2 \right) \tag{1.7}$$

where H is the Hubble parameter (i.e., the expansion rate of the Universe) [15], $n_{\chi,equil}$ is the number density in equilibrium, and $\langle \sigma_A v \rangle$ is the thermally averaged product of the annihilation cross-section and the velocity flux for the process $\chi + \bar{\chi} \implies \phi + \phi^\star$ [16]. This equation shows how the WIMP number density decreases due to the expansion of the universe (described by H) and the self-annihilation of DM particles (the second term). This number density is often related to the total entropy density of the universe, $s$, resulting in the comoving number density of DM, $Y_\chi$, which is given by:

$$Y_\chi = \frac{n_\chi}{s} \tag{1.8}$$

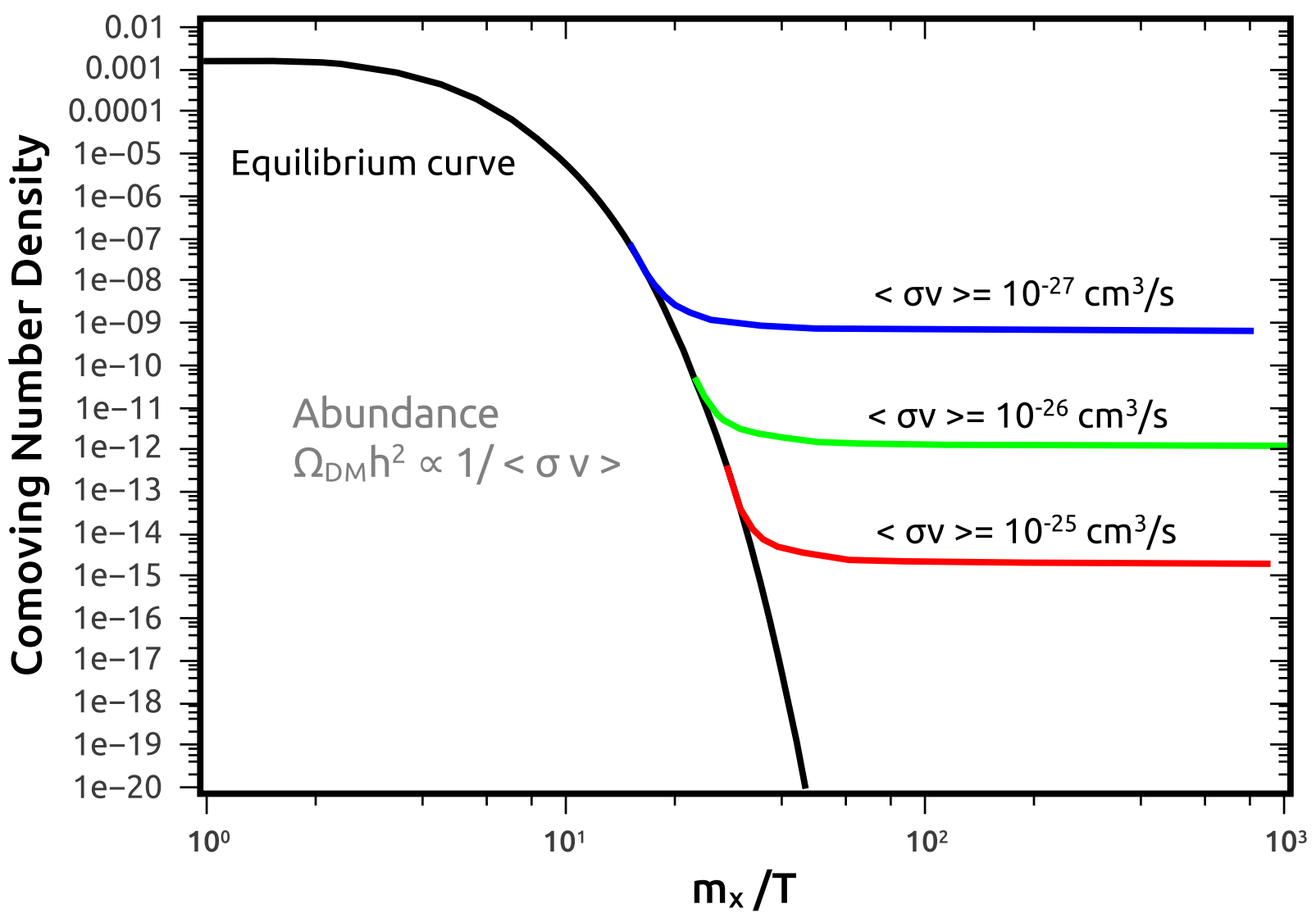

**Figure 1.4:** Evolution of the comoving number density as a function of the ratio $m_\chi/T$ in the context of DM thermal freeze-out. The remaining DM abundance in the universe ($\Omega_{DM}h^2$ in the figure) is determined by the DM self-annihilation cross-section $\langle \sigma_A v \rangle$. Retrieved from [17].

Figure 1.4 shows the evolution of $Y_\chi$ as a function of the universe's evolution, highlighting the freeze-out mechanism, i.e., the point at which DM stops annihilating and its number density $Y_\chi$ becomes constant. The figure also illustrates that as the annihilation cross-section ($\langle \sigma_A v \rangle$) increases, the freeze-out abundance decreases, while the time at which decoupling occurs shifts to a later point.

The final relic density, computed using the Boltzmann equation (Equation 1.7), can then be related to the annihilation cross-section in such a way that, numerically, at the freeze-out, one obtains [18,19]:

$$\langle \sigma_A v \rangle \approx 3 \times 10^{-26} \, \mathrm{cm}^3/\mathrm{s} \tag{1.9}$$

For typical velocities of $v \sim 0.1c$, the resulting cross-section is on the scale of electroweak interactions ($\sim 10^{-36}$ cm$^2$). This noteworthy coincidence makes the WIMP an ideal candidate for dark matter, closing the gap between particle physics and cosmology, and is thus often referred to as the *WIMP miracle*. This has driven scientists to focus on WIMPs as a potential solution to the dark matter problem through beyond the Standard Model (BSM) theories.

The most well-known BSM theory is perhaps the Supersymmetry (SUSY) model [20], which predicts the existence of a particle that fits the WIMP scenario. SUSY was introduced to address several challenges in physics, such as the hierarchy problem, which concerns the large disparity between the Higgs boson's mass and the Planck mass [21]. In SUSY theories, every SM particle has a corresponding supersymmetric partner. Supersymmetric particles are assigned an R-parity of -1, essential for "preventing" the decay of protons. This also implies that the lightest SUSY particle (LSP) – one of several particles introduced by these models as BSM candidates – is stable because it cannot decay into SM particles. As a result, if the LSP is also electrically neutral, it becomes a strong dark matter candidate. Within supersymmetric theories, the properties of the LSP can vary, leading to several sub-candidates, such as the neutralino [22] or the heavy gluino [23]. Further discussions on the LSP from SUSY theories can be found in references such as [24].

While SUSY theories have historically strengthen the WIMP dark matter theory, the lack of significant results from both the LHC and direct WIMP searches has raised questions about their validity [25]. Nevertheless, the WIMP dark matter theory remains one of the most supported models, and further discussions on its detection are presented later in this chapter.

### 1.2.2 Alternative dark matter models

Besides WIMPs, there are other less well-established ideas that attempt to explain dark matter in different ways, including the Modified Newtonian Dynamics (MOND) hypothesis and the recently revived idea that dark matter is composed of Primordial Black Holes. Other theories consider dark matter as a particle but with much lower mass ranges, such as the sterile neutrino, with masses on the keV scale, and axions, with masses ranging from $\mu$eV to meV [26]. A brief overview of these models is provided in this section.

**MOND:** The MOND theories [27] are based on modifications to the Newtonian gravitational force itself. At its core, MOND introduces a new acceleration constant, $a_0$, which modifies the gravitational force *only* at large scales, such as in galaxy clusters, and successfully predicts the anomalies observed at these scales [28]. However, when these modifications are applied to other measurements, such as the Cosmic Microwave Background or the Bullet Cluster, the theory struggles to account for the observations [29]. As a result, while MOND is an interesting approach, it remains an incomplete theory.

**Primordial Black holes:** Primordial Black Holes (PBHs) are hypothesized to have formed in the early universe, shortly after Big Bang nucleosynthesis (BBN), from the collapse of extremely dense regions. Unlike black holes formed from stellar collapse, PBHs could have a broader range of masses. Being produced before the BBN, they are decoupled from the rest of the baryonic matter and thus are not included in the evaluation of its abundance [30]. If a significant population of PBHs exists, their gravitational influence could explain the observed effects of dark matter. As non-luminous objects, PBHs would interact with regular matter primarily through gravity, making them a viable dark matter candidate. However, no positive discovery of PBHs has been made to date.

**Axions:** Axions are hypothetical particles introduced by Peccei and Quinn to resolve the strong charge parity (CP) problem in Quantum Chromodynamics (QCD) [31]. In the QCD Lagrangian, a term violates CP symmetry, with the parameter $\theta$ determining the strength of this violation. However, there is no theoretical limit on $\theta$, and if it were too large, it would produce a neutron electric dipole moment, which has never been observed. This requires $\theta$ to be extremely small, less than $10^{-11}$ [32]. To address this, Peccei and Quinn proposed a $U(1)$ chiral symmetry that prevents $\theta$ from becoming large. This symmetry spontaneously breaks at a scale $f_a$, generating a pseudo-Nambu-Goldstone boson, the *axion* [33]. When the QCD Lagrangian is rewritten in terms of the axion, it cancels the CP-violating component, resolving the fine-tuning issue. The axion also couples to gluons and photons, with its mass tied to its interaction strength with gluons. If $f_a$ is sufficiently large, the axion becomes stable, weakly interacting, and could serve as a dark matter candidate [34]. Its coupling with photons allows for the correct relic density to form in the early universe, and it may decay into photon pairs, offering a potential detection method. Axion-like particles (ALPs), although not directly addressing the original Standard Model issues, are also considered potential dark matter candidates, and there are ongoing experimental efforts to detect them [35–37].

**Sterile neutrinos:** This family of neutrinos offers another theory in which dark matter is explained as a particle. The Standard Model predicts three types of neutrinos, which are massless and have left-handed chirality. However, the observation of neutrino flavor oscillations indicates that neutrinos must have mass [38]. A possible explanation for this discrepancy involves the introduction of a sterile neutrino, a singlet under the standard gauge group that interacts only weakly with SM particles via a mixing mass matrix. The seesaw mechanism explains how the sterile neutrino mass is linked to the SM neutrinos, with a large term in the mixing matrix causing the masses of the neutrinos to be inversely correlated – when one mass is large, the other is small. In this scenario, if the sterile neutrino mass exceeds 1 $\mathrm{keV/c^2}$ and the mixing angles are small, this new particle can become non-relativistic early in the universe, contributing to the formation of a thermal relic density consistent with cold dark matter [39]. The extremely weak interaction ensures the stability of the sterile neutrino, which gradually increases its abundance through interactions with SM particles via the freeze-in mechanism rather than freeze-out [40]. This process allows dark matter to decouple from SM particles at early times, with its density increasing slowly until it is suppressed by the expansion of the universe. A key way to detect sterile neutrinos is by observing their decay into monochromatic photons [41, 42].

A more in-depth discussion of the mentioned and other less common dark matter candidates, their motivations, and potential detection methodologies can be found in sources such as [26, 43, 44].

## 1.3 WIMP direct detection

In the WIMP dark matter framework, there are three main strategies to detect these particles: (1) using accelerators to detect the products of induced collisions; (2) through direct detection of interactions between WIMPs and SM particles; and (3) via their self-annihilation or decay in the cosmos. The different types of interactions used to search for WIMPs are illustrated in Figure 1.5 and are summarized below. For a more detailed discussion, see [26].

**Indirect detection:** In the framework of indirect searches, as illustrated in Figure 1.5, scientists search for SM particles resulting from the decay or *self-annihilation* of dark matter particles, using both space- and ground-based cosmic ray detectors. Dark matter self-annihilation plays a key role in the freeze-out mechanism that determines the present dark matter density, as mentioned earlier. This annihilation rate will be higher in regions of the universe with a greater dark matter density. Therefore, an excess in the flux of SM particles – primarily in the form of gamma rays, neutrinos, and particle-antiparticle pairs – coming from these regions could be interpreted as a signature of dark matter self-annihilation [45].

**Collider detection:** In this approach, the goal is to eventually *produce* dark matter particles through the collisions of SM particles accelerated to very high energies. A notable example is proton-proton collisions at the Large Hadron Collider (LHC). However, since dark matter particles are not expected to produce a directly visible signal in LHC detectors due to their very low interaction cross-section, the expected signal is instead the missing transverse momentum. This refers to the imbalance in the transverse plane of the beam, which would result from processes like $pp \to \chi + \bar{\chi} + \Delta$, where $\Delta$ could be a photon or a hadronic jet [46].

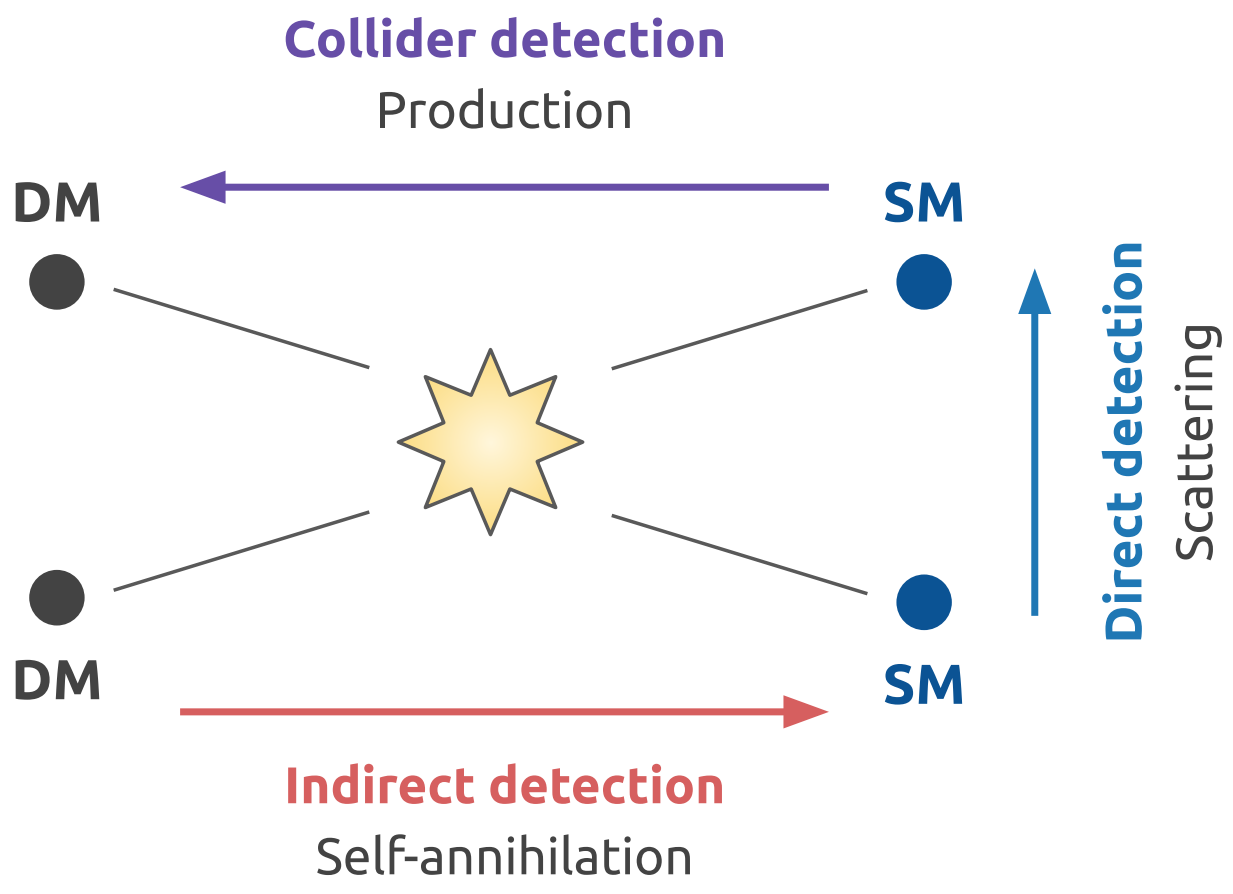

**Figure 1.5:** Different possible interaction channels between DM and SM particles, associated with their respective detection methods: direct detection via DM-SM scattering, indirect detection via DM-DM annihilation, and collider detection via SM-SM production. Inspired by [47].

**Direct detection:** The direct detection of WIMPs aims to detect the *scattering* of a SM particle by a dark matter one. Such events are typically searched for in underground, low-background detectors, where a dark matter signal would potentially appear as an excess over the well-understood background. This excess can manifest through various signatures, such as energy, temporal, or directional dependencies, which will be further discussed in Section 1.3.2.

Since the CYGNO experiment, in which this thesis is based, is focused on the directional direct search of WIMP-nuclei scatter interactions, the remaining discussion of this chapter will concentrate on this topic.

### 1.3.1 WIMP signal rate

In the WIMP DM theory, these particles form a halo around the Milky Way, producing an apparent wind of dark matter particles at Earth as it rotates along with the Sun in the galaxy. If WIMPs interact with SM particles, the movement of the Earth can be exploited to detect this wind by placing a large volume of material on Earth and measuring the rate of recoils in standard matter caused by a WIMP when it passes and collides with it. It is important to note that dark matter can potentially interact with both nuclei and electrons, although the preferred channel is nuclear scattering. This preference is particularly due to the expected WIMP mass, typically of the order $\mathcal{O}(1 - 100)$ GeV/c$^2$, which enhances the coupling to nuclei compared to electrons.

Assuming a WIMP DM model and considering WIMP-nuclei scattering, the rate of these events can be calculated for a given experiment, taking into account some general considerations. Following [48,49], the expected differential rate of WIMP-induced nuclear recoils is described by:

$$\frac{d^2R}{dE_r d\Omega_r} = \frac{N_A \rho_0}{A m_\chi} \cdot M \cdot \int_{v_{min}(E_r)} \frac{d^2\sigma_{\chi-A}}{dE_r d\Omega_r} \cdot v \cdot f(v)\ d^3v \tag{1.10}$$

where:

- $\mathbf{d^2R/dE_r d\Omega_r}$: The double differential rate of WIMP-nucleus interactions per unit mass of the detector, as a function of the recoil energy $E_r$ and solid angle $\Omega_r$.
- $\mathbf{N_A}$: Avogadro's number, representing the number of atoms per mole.
- $\boldsymbol{\rho_0}$: Local dark matter density.
- $\mathbf{A}$: Atomic mass of the target nucleus.
- $\mathbf{m_\chi}$: The mass of the WIMP particle.
- $\mathbf{M}$: Total mass of the detector.
- $\mathbf{v}$: Velocity of the WIMP with respect to the target nucleus.
- $\mathbf{f(v)}$: Velocity distribution of WIMPs in the detector's rest frame.
- $\mathbf{d^2\sigma_{\chi-A}/dE_r d\Omega_r}$: Double differential WIMP-nucleus cross-section as a function of recoil energy $E_r$ and solid angle $\Omega_r$.
- $\mathbf{v_{min}(E_r)}$: The minimum velocity of the WIMP required to produce a recoil with energy $E_r$, dependent on scattering kinematics.

The different components of Equation 1.10, namely the minimum velocity $v_{min}$, the velocity distribution $f(v)$, and the interaction cross-section $\sigma_{\chi-A}$, can be calculated independently. The following sections present the derivations of these components.

#### 1.3.1.1 Minimum velocity ($v_{min}$)

The minimum velocity appearing in the rate of dark matter-induced nuclear recoils can be deduced from the WIMP-nuclei kinematics. Under the WIMP DM model, the average velocity of WIMPs is expected to be non-relativistic, of the order of $\sim 10^{-3}c$, while a nucleus at rest has an energy of $\sim 10^{-5}c$. The collision between these two particles results in an elastic collision, where the nucleus recoils with energy $E_r$. In this scenario, the momentum transferred during such a collision, $q$, can be expressed as:

$$|q| = 2\mu v \cos(\theta_r) \tag{1.11}$$

where $\mu = \frac{m_\chi m_A}{m_\chi + m_A}$ is the reduced mass of the WIMP-nucleus system, with $m_\chi$ being the WIMP mass and $m_A$ the target nucleus mass. The WIMP velocity is denoted as $v$, and $\theta_r$ is the angle between the WIMP's initial direction and the nucleus' recoil direction in the center-of-mass frame. The maximum recoil energy $E_{r_{max}}$ of the nucleus is obtained for $\theta_r = 0$, and is given by:

$$E_{r_{max}} = \frac{|q|^2}{2m_A} = \frac{2v^2\mu^2}{m_A} \tag{1.12}$$

Given the WIMP's initial energy of $E_i = \frac{1}{2} m_\chi v^2$, the recoil energy can be expressed as:

$$E_r = r \cdot E_i \tag{1.13}$$

where $r = \frac{4 m_\chi m_A}{(m_\chi + m_A)^2}$ is a dimensionless kinematic factor, which ranges from 0 to 1 and represents the efficiency of momentum transfer. This factor reaches its maximum value, $r = 1$, when the masses of the two colliding particles are equal, i.e., $m_\chi = m_A$. This feature illustrates how an experiment can optimize its sensitivity to specific WIMP masses. In the case of CYGNO, the use of a light target such as $He$ enhances the sensitivity to low WIMP masses, approximately around 4 $GeV/c^2$.

From Equation 1.12, it is also possible to deduce the minimum WIMP velocity required to produce a recoil energy $E_r$, which is given by:

$$v_{min} = \sqrt{\frac{m_A E_r}{2\mu^2}} \tag{1.14}$$

This minimum velocity sets the lower limit for the integration in the differential WIMP scattering rate, as shown in Equation 1.10.

#### 1.3.1.2 WIMP-nuclei interaction cross-section

While the exact mechanism of WIMP coupling remains unclear, it is commonly assumed that the WIMP interacts coherently with the entire nucleus through elastic scattering, based on the relative velocity between the WIMP and the nucleus, as well as the WIMP's mass. To describe this interaction in its simplest form, both spin-independent (SI) and spin-dependent (SD) couplings are considered. The differential cross section is obtained by coherently summing the contributions from SI and SD interactions, which, in the non-relativistic limit, can be written as:

$$\frac{d\sigma_{\chi-A}}{dE_r} = \frac{m_A}{2\mu^2 v^2}\left(\sigma_{SI} F_{SI}^2(E_r) + \sigma_{SD} F_{SD}^2(E_r)\right) \tag{1.15}$$

Here, $\sigma_{SI/SD}$ refers to the SI/SD WIMP-nucleus cross section at zero momentum, and the factor F represents the form factor for both cases. The form factor accounts for and corrects the loss of coherence in the interaction between the two particles at higher momentum transfer. At high values of q, the de Broglie wavelength becomes comparable to or smaller than the nucleus, leading to a non-coherent response. This effect is more pronounced for heavy or large WIMPs and nuclei [50], as will be demonstrated later in the resulting WIMP rates for different detector materials (Figure 1.6).

The SI cross section ($\sigma_{SI}$) can be expressed as the sum of the contributions from both protons (Z) and neutrons in the nucleus, as follows:

$$\sigma_{SI} = \frac{4\mu^2}{\pi} \left[Zf_p + (A - Z)f_n\right]^2 \tag{1.16}$$

where $f_p$ and $f_n$ are the coupling constants of WIMPs to protons and neutrons, respectively. In many WIMP models, the reasonable assumption is that $f_p = f_n = f$, which simplifies the above equation to:

$$\sigma_{SI} = \frac{4\mu^2}{\pi} f^2 A^2 = \sigma_{SI}^n \frac{\mu^2}{\mu_n^2} A^2 \tag{1.17}$$

with $\mu_n$ representing the reduced mass of the WIMP-nucleon (proton or neutron) system, and $\sigma_{SI}^n$ being the WIMP-nucleon interaction cross-section. With this modification, the interaction of a WIMP with a generic nucleus can be generalized and expressed as a function of the single cross sections with the nucleons constituting the nucleus. Another important aspect of this equation is that the cross section is enhanced by a factor of $A^2$, which increases the sensitivity of experiments using heavier elements.

For the spin-dependent counterpart of the interaction cross section, it can be written as [51]:

$$\sigma_{SD} = \frac{32}{\pi} G_F^2 \mu^2 \frac{J + 1}{J} \left[a_p \langle S_p \rangle + a_n \langle S_n \rangle\right]^2 \tag{1.18}$$

where $G_F$ is the Fermi constant, J is the nuclear spin, and $a_{p/n} \langle S_{p/n} \rangle$ represents the product of the SD coupling constants of WIMPs with protons and neutrons, and their respective total spin operators' expectation values. As can be seen from this equation, to be sensitive to SD interactions, an odd atomic number for the atom is required to increase the total expectation value of the spin. This is because, in atoms with an even number of protons and neutrons, the contributions from these nucleons typically (or nearly) cancel out [52].

Considering the total cross-section ($\sigma_{\chi-A}$) of interaction as the sum of the two SI and SD components, respectively represented by Equation 1.17 and 1.18, $\sigma_{\chi-A}$ for a nucleus

**Table 1.2:** Summary of the nuclear spin characteristics for typical nuclei used in the detection of WIMP SD interactions with SM particles. The last two columns indicate the affinity of each element to SD interactions, based on the expression in Equation 1.19. Adapted from [52].

| Nucleus | Z | Odd nucleon | J | $\langle S_p \rangle$ | $\langle S_n \rangle$ | $\frac{4\langle S_p \rangle (J+1)}{3J}$ | $\frac{4\langle S_n \rangle (J+1)}{3J}$ |
|---|---|---|---|---|---|---|---|
| $^{1}$H | 1 | p | 1/2 | 0.500 | 0.0 | 1.0 | 0 |
| $^{19}$F | 9 | p | 1/2 | 0.441 | -0.109 | 0.91 | $6.4\times10^{-5}$ |
| $^{73}$Ge | 32 | n | 9/2 | 0.030 | 0.378 | 0.0015 | 0.23 |
| $^{129}$Xe | 54 | n | 1/2 | 0.028 | 0.359 | 0.0031 | 0.52 |

containing an odd number of protons can be written as:

$$\sigma_{\chi-A} = \sigma_{SI} + \sigma_{SD} = \sigma_{SI}^{n} \frac{\mu^2}{\mu_n^2} A^2 + \sigma_{SD}^{p} \frac{\mu^2}{\mu_p^2} \frac{4\langle S_n \rangle (J+1)}{3J} \tag{1.19}$$

From this equation, it is clear that the term $\frac{4\langle S_n \rangle (J+1)}{3J}$ dictates the sensitivity of a particular element to SD interactions. In Table 1.2, some typical nuclei used in SD-sensitive experiments are shown. For each element, as observed, the sensitivity primarily arises from the interaction of the odd nucleon in the nucleus. When represented graphically, the SD results are typically reported assuming WIMPs couple only with neutrons ($a_p = 0$) for atoms with an odd number of neutrons, and vice versa for atoms with an odd number of protons.

An interesting conclusion that can be taken from Table 1.2 is that fluorine shows one of the largest sensitivity to SD interactions, only beat by hydrogen, making it one of the greatest contender for SD WIMP searches. This, as discussed later, represents another advantage of the CYGNO gas mixture, He:$CF_4$, which due to this effect becomes specially sensitive to this type of coupling.

#### 1.3.1.3 DM halo

**Velocity profile:** The Standard Halo Model (SHM) is often used to describe the dark matter halo of the Milky Way. It assumes a spherical, isotropic distribution of DM particles with a density profile proportional to $\rho(r) \propto r^{-2}$, representing an isothermal system of collisionless particles. The velocity distribution of these DM particles follows a Maxwell-Boltzmann distribution:

$$f(\mathbf{v}) = \begin{cases} \alpha \cdot \exp(-v^2/v_p^2) & \text{if } |\mathbf{v}| < v_{esc} \\ 0 & \text{if } |\mathbf{v}| > v_{esc} \end{cases} \tag{1.20}$$

Here, $\alpha$ is a normalization factor, and $v_p$ is the most probable speed of WIMPs near Earth, estimated at approximately 230 $km/s$ [53,54]. More detailed studies suggest that $v_p$ ranges between 200 and 279 $km/s$ due to uncertainties in the Milky Way's flatness. In the SHM, the Boltzmann distribution of velocity is truncated at the local escape speed $v_{esc}$ to account for particles that are so fast they escape the galaxy, effectively limiting the maximum available velocities for WIMPs. This escape velocity depends on the distance to the center of the galaxy and has been measured to be around 544 $km/s$ [13,55] at the Solar position of 8.2 $kpc$ from the center.

**Local DM density:** Another key parameter is the local DM density around us, i.e., at the Solar distance from the center, denoted as $\rho_0$. The most widely accepted value is 0.3 $GeV/cm^3$ [56], although more recent estimates suggest values closer to 0.4 $GeV/cm^3$, depending on the assumptions of different density profiles, such as the Frenk Navarro White profile [13].

Regarding the overall validity of the SHM, it should be noted that the description of the DM halo surrounding the Milky Way as an isotropic and isothermal sphere is a rather oversimplification. In the cold DM cosmology framework, structures form hierarchically, leading to triaxial and anisotropic DM halos [57]. In these works, it has also been shown that the DM halo velocity profile may differ considerably from the Maxwellian distribution typically assumed. Nonetheless, the SHM remains a useful benchmark for calculating the expected rate of DM events in detectors worldwide.

#### 1.3.1.4 Typical numbers

To summarize the discussion so far and give the reader an idea of the event rates being discussed, an example of the total rate of events ($R$) is provided, obtained with an approximation [58] of the integral shown in Equation 1.10, using typical numbers in this field:

$$R \approx \frac{100}{A} \cdot \frac{500}{m_\chi[GeV]} \cdot \frac{\sigma}{1\ pb} \cdot \frac{\rho_0}{0.3\ \left[\frac{GeV}{cm^3}\right]} \cdot \frac{v}{220\ \left[\frac{km}{s}\right]} \cdot 10^{-2}\ \frac{events}{kg\ \times\ day} \tag{1.21}$$

where the typical cross section of weak strength for masses at the electroweak scale was considered. Given the low numbers involved, and since no experiment has yet provided decisive proof of the existence of dark matter, scientists must continually increase the scope and size of their experiments to improve the likelihood of detecting an interaction, or at least rule out another section of the cross-section-mass exclusion plot, as shown later in Section 1.3.4.

The following section discusses the different types of WIMP signatures that can be explored and their effects on the event rate required to identify a dark matter signal. Additionally, common background sources in dark matter experiments are explored, along with the techniques employed to mitigate them.

### 1.3.2 Experimental signatures

The WIMP scattering rate in Equation 1.10 shows several dependencies that can be explored by experiments. In particular, it is possible to study the energy dependence of the recoils, the rate's time variation due to Earth's movement within the galaxy, and the directional distribution of the recoil rate, which, as will be shown, exhibits a distinct pattern when observed in galactic coordinates. Each of these signatures presents different challenges in measurement, leading to varying levels of reliability in the results. These dependencies are reviewed in the following sections.

#### 1.3.2.1 Energy dependence

The distribution of WIMP-induced recoil energies depends on several factors, including the masses of both colliding particles, the WIMP velocities, and their cross-section of interaction. Assuming no direction information in the recoil (isotropic scattering direction), and given the WIMP velocity distribution, the differential recoil rate shown in Equation 1.10 is found to be described by a steeply falling function [59]:

$$\frac{dR}{dE_{nr}} \propto \exp\left(-\frac{E_{nr}}{E_0} \cdot r\right) \tag{1.22}$$

where $E_0$ represents the most probable kinetic energy of an incoming WIMP, $E_0 = \frac{1}{2} \cdot m_\chi v_0^2$, and $r$ is the kinematic factor mentioned in Section 1.3.1.1, i.e., $r = \frac{4m_\chi m_A}{(m_\chi+m_A)^2}$. Figure 1.6 shows the rate of WIMP-induced nuclear recoils by a 100 GeV WIMP, as a function of the recoiling energy for different elements, highlighting the exponentially decaying nature of the event rate. The recoil energies are shown in terms of nuclear recoil equivalents ($keV_{nr}$), which differ from their electron recoil counterparts ($keV_{ee}$) due to quenching effects - how the particle's energy loss is partitioned among different energy loss channels.

The outcome of Equation 1.22, also illustrated in Figure 1.6, shows that for the same WIMP mass, the interaction rate increases with the atomic number, reflecting how the spin-independent cross-section scales with $A^2$. This dependency, often referred to as the "materials signal", would strongly support the existence of WIMPs if observed across different experiments using various materials [59]. However, as the atomic mass increases, the form factor suppression in the cross-section becomes stronger (as mentioned earlier in Section 1.3.1.2), leading to a more rapid reduction in the rate's slope, which limits the sensitivity enhancement, particularly at higher recoil energies.

While not visible in this particular plot, another effect expected for lower-mass WIMPs is the abrupt cutoff in the interaction rate due to the escape velocity, as shown in [58]. This escape velocity sets a limit on the maximum energy that WIMPs can possess, and thus, it also restricts the maximum nuclear recoil energy that can be transferred during interactions. This phenomenon further reduces the interaction rate for low-mass WIMPs at higher recoil energies.

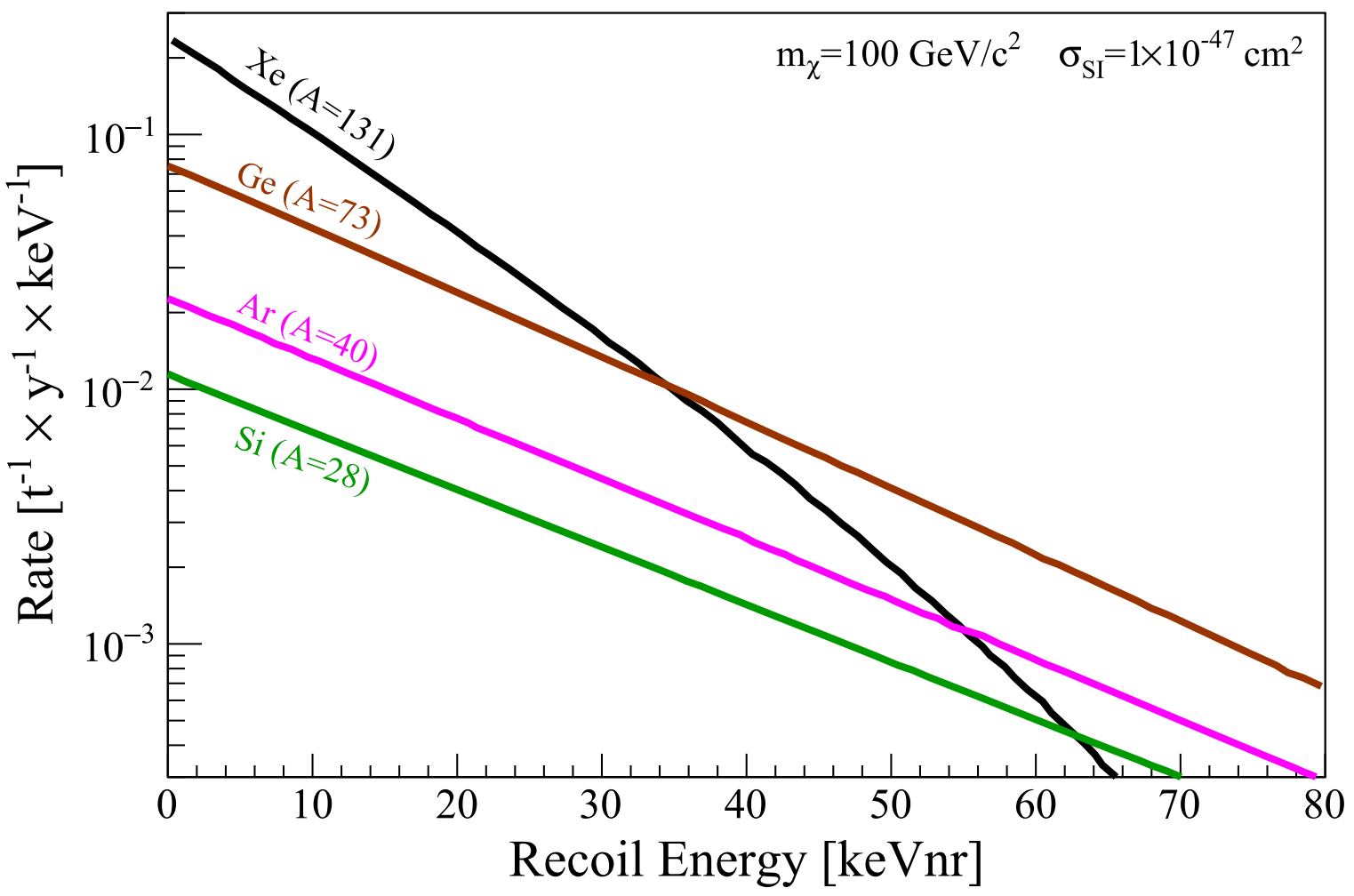

**Figure 1.6:** Nuclear recoil spectra induced by a 100 GeV/c$^2$ mass WIMP in different common target elements, assuming a SI interaction cross-section of $1 \times 10^{-47}$ cm$^2$. Retrieved from [60].

The main limitation of energy-only measurement experiments in dark matter searches is that a significant number of events need to be observed to identify a DM signal and extract meaningful information about the WIMP mass. This is due to the overlap of different $m_\chi$-dependent rate curves, making it challenging to distinguish between various WIMP masses. Ideally, such experiments should be conducted in parallel with other experiments using different detector materials to enhance the analysis. Additionally, uncertainties in the local WIMP halo density and velocity distribution can introduce further systematic errors in determining the WIMP mass, especially in low event rate scenarios. The feasibility and precision of deducing the WIMP mass from a single event, considering different materials, masses, and detection efficiencies, are discussed in more detail in [61].

#### 1.3.2.2 Time dependence

The velocity relevant for calculating and detecting WIMP-induced interactions is affected by Earth's orbital motion around the Sun. During half of the year, Earth's velocity adds to the Sun's motion as it orbits the galactic center, increasing the apparent velocity of the WIMP wind that Earth experiences while crossing the dark matter halo. In the other half of the year, Earth's velocity is in the opposite direction, reducing the apparent WIMP wind velocity. This seasonal variation in velocity, illustrated in Figure 1.7, creates a time-dependent modulation in the WIMP scattering rate.

This phenomenon effectively alters the distribution of recoil energies observed throughout the year. Given the Sun's rotation velocity of approximately 220 km/s and Earth's yearly-averaged orbital velocity of roughly 30 $km/s$, this results in an annual modulation in the interaction rate on the order of $\mathcal{O}(5)\%$ [50]. Assuming such modulation, the differential event rate can be expressed as [63]:

$$\frac{dR}{dE} \simeq R_0 + R_m \cdot \cos(w(t - t_0)) \tag{1.23}$$

where $R_0$ represents the time-averaged signal rate and $R_m$ is the modulation amplitude, modeled as a sinusoidal function with frequency $w = 2\pi/year$ and phase $t_0$. This characteristic signature of WIMP-induced recoils could provide strong evidence for the existence of dark matter. However, such detection requires an experiment with long-term exposure, spanning at least one year but preferably more, and exceptional stability in the detector's response over this extended period. Moreover, many environmental factors also exhibit seasonal variations, thus requiring careful control of these external influences to avoid false modulations.

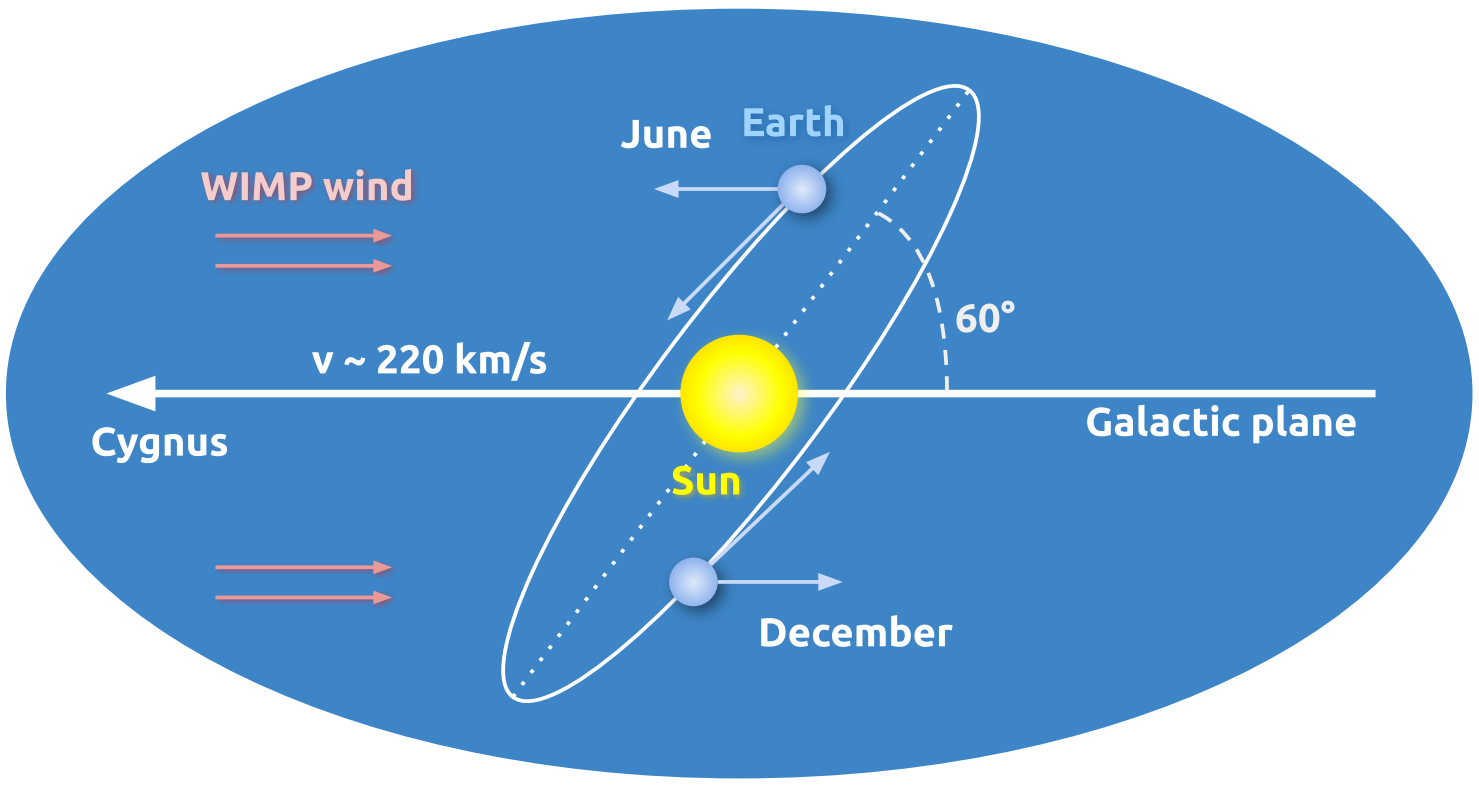

**Figure 1.7:** Schematic of the solar system's movement in the galactic plane, highlighting Earth's velocity components, which affect the velocity at which WIMPs collide with Earth detectors. In galactic coordinates, the solar system is moving towards the Cygnus constellation, creating an apparent *WIMP wind* from that direction. Illustration inspired by [62].

The DAMA experiment [64] is one of the most prominent direct detection experiments to explore this DM signature, using sodium iodide crystals. DAMA has been operational for several decades and has reported an annual modulation in its signal, which could be indicative of WIMP interactions with the detector [65]. However, the results from DAMA have been controversial, as they have not been universally accepted by the scientific community. The primary reason for the skepticism is the lack of independent confirmation, as several other experiments have excluded the dark matter parameter space where DAMA claim to have "seen" dark matter. Despite this, DAMA continues to be a critical piece in the dark matter search, raising important questions not only about detection approaches and the properties of dark matter, but also about the openness of data and analysis methods by large experiments to the public.

#### 1.3.2.3 Direction dependence

Following up on the previous argument, an even stronger signature of the WIMP halo felt on Earth is its directional component. As the Earth orbits the center of the galaxy, it moves with the Sun towards the Cygnus constellation. This motion gives Earth the appearance of traveling through a "wind" of WIMPs, expected to originate from the direction in which the Sun is moving.

If the detector's angular resolution is sufficient, it may be possible to identify an anisotropy in the distribution of recoil directions, induced by this precise motion of Earth. Assuming $\phi$ is the angle between the direction of the incoming WIMP and the nuclear recoil originated by it in the galaxy frame, the differential recoil rate can be written as [60]:

$$\frac{d^2R}{dE_{nr}d\phi} \propto \exp\left(\frac{2(v_{lab}\cos(\phi) - v_{min})^2}{3v_p^2}\right) \tag{1.24}$$

In this expression, $v_{lab}$ is the velocity of Earth with respect to the galactic rest frame, $v_{min}$ is the minimum WIMP velocity needed to generate a signal (as defined in Section 1.3.1.1), and $v_p$ is the characteristic WIMP velocity in the halo.

The dependence of the rate on the exponential of $\cos(\phi)$ emphasizes the anisotropic nature of the WIMP-induced recoil distribution, which becomes highly peaked toward the Cygnus constellation. The Earth's rotation on its axis introduces a time-dependent (daily) modulation in the observed recoil direction, but in galactic coordinates, the recoil distribution always remains concentrated toward Cygnus.

As illustrated in Figure 1.8, the angular flux of dark matter arriving at Earth in the SHM is directed from the Cygnus constellation in galactic coordinates. This implies that the WIMP-induced recoils observed in Earth-based detectors by this flux will also be concentrated in that direction. Nevertheless, the distribution of recoils will be smeared due to factors such as collision kinematics and the detector's angular resolution, as discussed further in Section 1.4. The result of this is a clear dipole signature in the distribution of WIMP-induced recoils when observed in galactic coordinates.

In Figures 1.9a and 1.9b, it is shown instead a schematic of Earth's rotation on its axis, along with the expected recoil rate modulation during the day as a function of the angle $\phi$, for both DM and solar neutrinos induced recoils. Solar neutrinos are a well-known background originating from the Sun that pose a significant challenge for many experiments that do not have directionality measurement. Without directionality and with few total number of events, it is difficult (or virtually impossible) to distinguish between recoils caused by WIMPs and those caused by solar neutrinos, thus limiting the WIMP parameter space that can be explored, as discussed in Section 1.3.4.

Interestingly, the recoils induced by solar neutrinos and the dipole effect from DM coming from the Cygnus constellation have different behaviors. Both signals oscillate throughout the day due to Earth's rotation, but they do not overlap in detector coor-dinates, as shown in Figure 1.9b. This characteristic makes it possible to distinguish them from one another. Additionally, any other local background will not exhibit this daily oscillation and will instead peak at a specific angle in the detector frame, further differentiating itself from the WIMP and Solar signals.

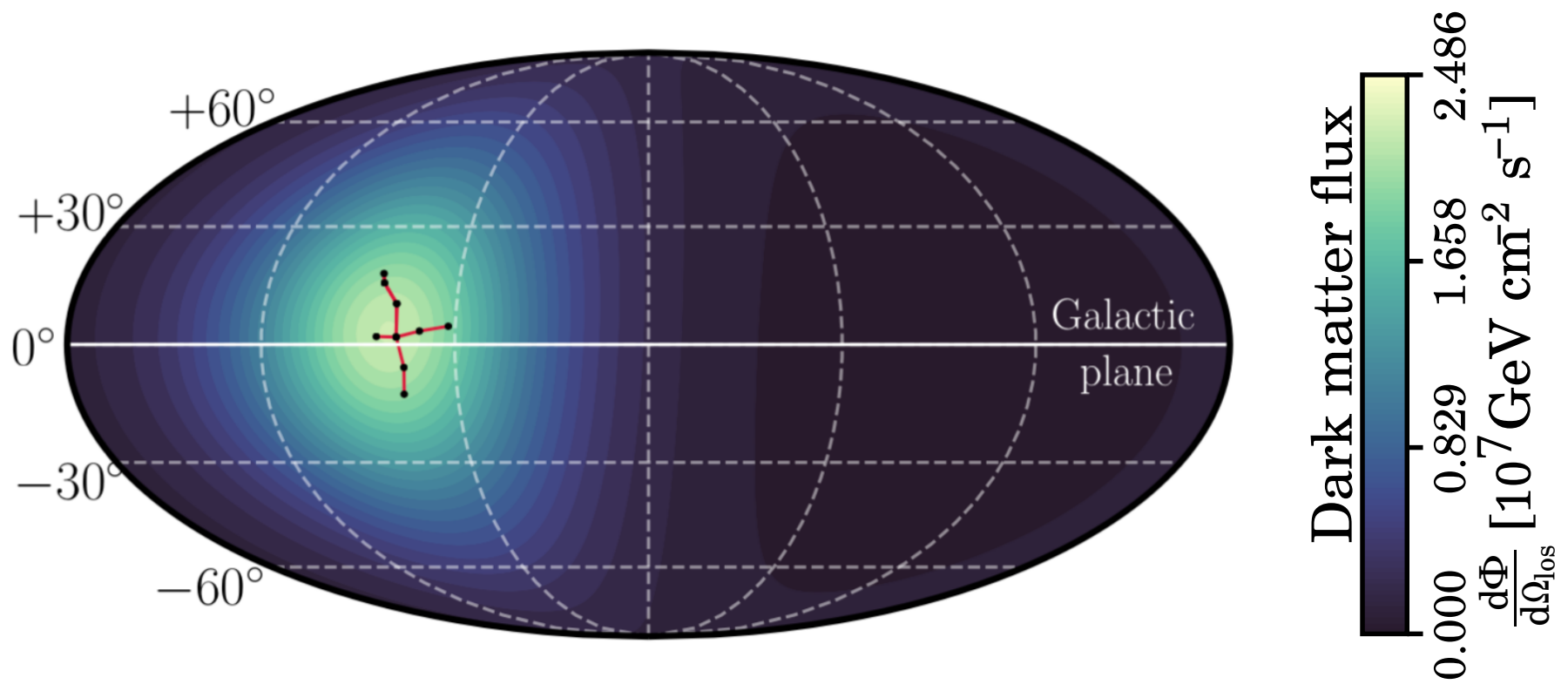

**Figure 1.8:** Angular dependence of the DM flux arriving at Earth under the Standard Halo Model, shown in galactic coordinates with the galactic plane crossing the horizontal axis. Earth and the solar system are moving towards the Cygnus constellation, highlighted in red in the figure. The recoils caused by these WIMPs will follow a similar shape to the DM flux, smeared due to detector effects, but maintaining the characteristic dipole pattern. Retrieved from [66].

This distinctive dipole directional signature, both in galactic coordinates and as a daily modulation in detector coordinates, creates the possibility to discriminate between signal and background events. No other source can mimic the specific direction signature of WIMP-induced recoils, including solar neutrinos, which allows experiments to more effectively and efficiently explore the DM WIMP parameter space. Additionally, beyond the clear identification of a dark matter signal, the directional information would allow the study and characterization of the cosmological properties of the dark matter halo within the SHM theory, as well as of the solar neutrino flux itself.

While being a clear signature of dark matter, measuring the directionality of an event can present experimental challenges: it involves balancing the detector's medium density to ensure that nuclear recoils are long enough to reconstruct their initial direction and sense, while maintaining an overall high detector exposure to dark matter. The benefits and challenges of measuring the directional signature of dark matter, as well as the current approaches being employed, are further explored in Section 1.4.

### 1.3.3 Detection techniques

When a WIMP collides with an atom in a given detector, it causes a recoil, and the energy from this interaction can be deposited in the detector through three distinct channels: *excitation*, leading to molecular de-excitation followed by the emission of light; *ionization*, where electron-ion pairs are generated, and the $e^-$ are typically drifted and amplified afterward; and *heat*, which induces vibrations and results in a temperature increase in the medium.

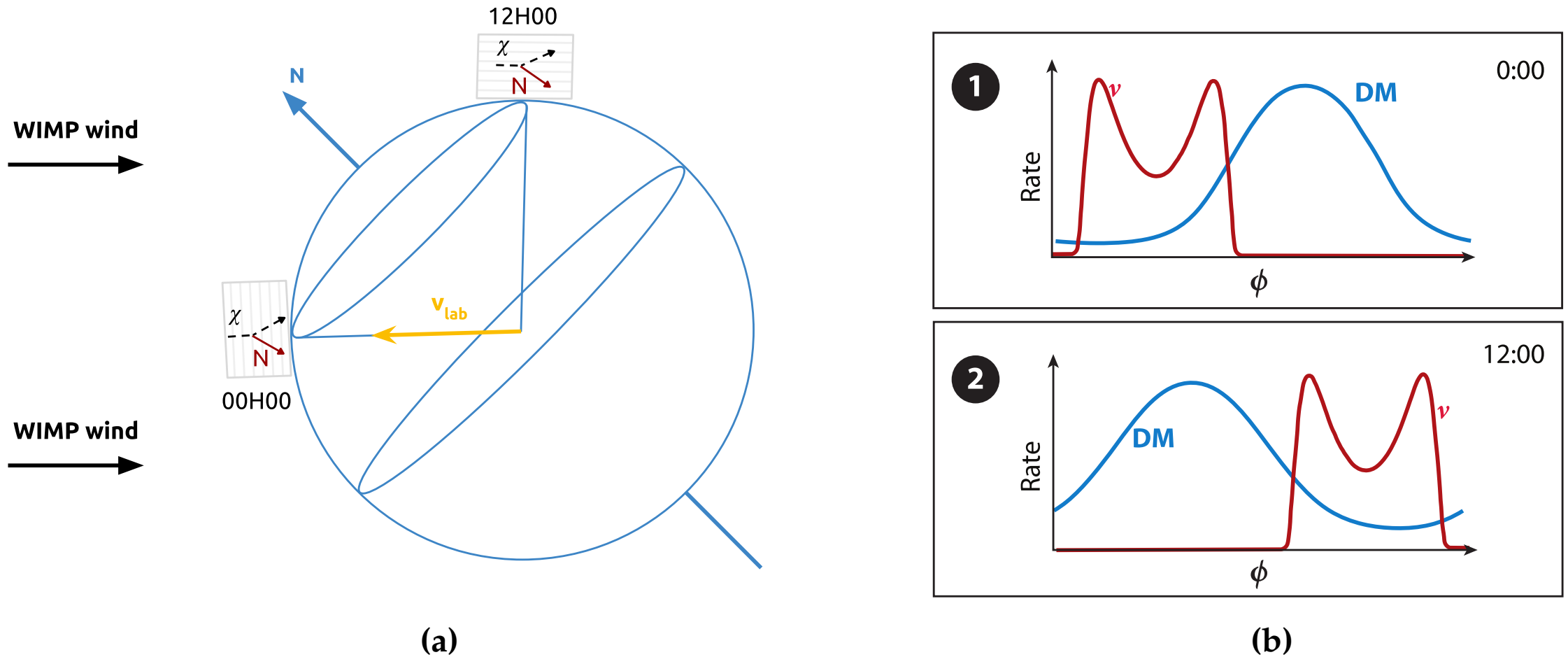

**Figure 1.9:** Illustration of the daily oscillation signature generated by Earth's rotation on its axis, showing (a) a schematic of Earth under the influence of the WIMP wind at different hours of the day and (b) the resulting diphase in the nuclear recoil rate originating from the WIMP wind and solar neutrinos. Retrieved from [67].

For nuclear recoils, the expected signal induced by WIMPs, the energy loss partitioning must be taken into account to correlate the measured energy with the actual kinetic energy of the WIMP. Nuclear recoils lose a significant portion of their energy through the heat channel, which is often invisible to experiments. This leads to the concept of the *ionization quenching factor (QF)*, defined as the fraction of the recoil's energy deposited through ionization (the more typical visible channel).

Since the partition of energy deposition across different channels depends on the particle, its mass, and energy, the quenching effect can be used to identify the type of interacting particle and discriminate signal from background events. Different experiments employ various combinations of measurements to infer the nature of the events and differentiate between WIMP and background signals. These methods include analyzing the event's energy partition, the topology and direction of the ionization cloud, the exploitation of differing lifetimes of excited state de-excitations, or combinations of these and other approaches. Depending on the technique and materials used, experiments are sensitive to different parameter spaces (mass and cross-section) within the WIMP search region. This section provides an overview to the most common techniques, following [50] and Figure 1.10, and is followed by a summary of the experiments' sensitivities to different WIMP masses, as detailed in Section 1.3.4.

### Scintillation crystals

High-purity scintillating crystals, such as NaI(Tl) and CsI(Tl) (Figure 1.10a), are used in dark matter searches due to their high scintillation light yield which grants the ability of reaching low energy thresholds. These crystals can be configured in arrays to achieve large detector masses and can operate stably over extended periods. When charged particles interact with the crystal, part of the deposited energy excites the crystal lattice, producing visible or UV light upon de-excitation. The addition of dopants like thallium enhances scintillation efficiency by introducing intermediate energy levels within the crystal lattice. The emitted light is typically detected by PMTs, either directly or via light guides. Contaminants introduced during the crystal growth process can increase background levels, and the technique often misses the ability to define a specific fiducial volume. Despite these challenges, scintillating crystals are widely used in experiments searching for the annual modulation of the WIMP signal, such as DAMA/LIBRA [65], which developed highly radiopure NaI(Tl) crystals or experiments such as COSINE-100 [68] and SABRE [69] that have been developed to replicate DAMA's approach.

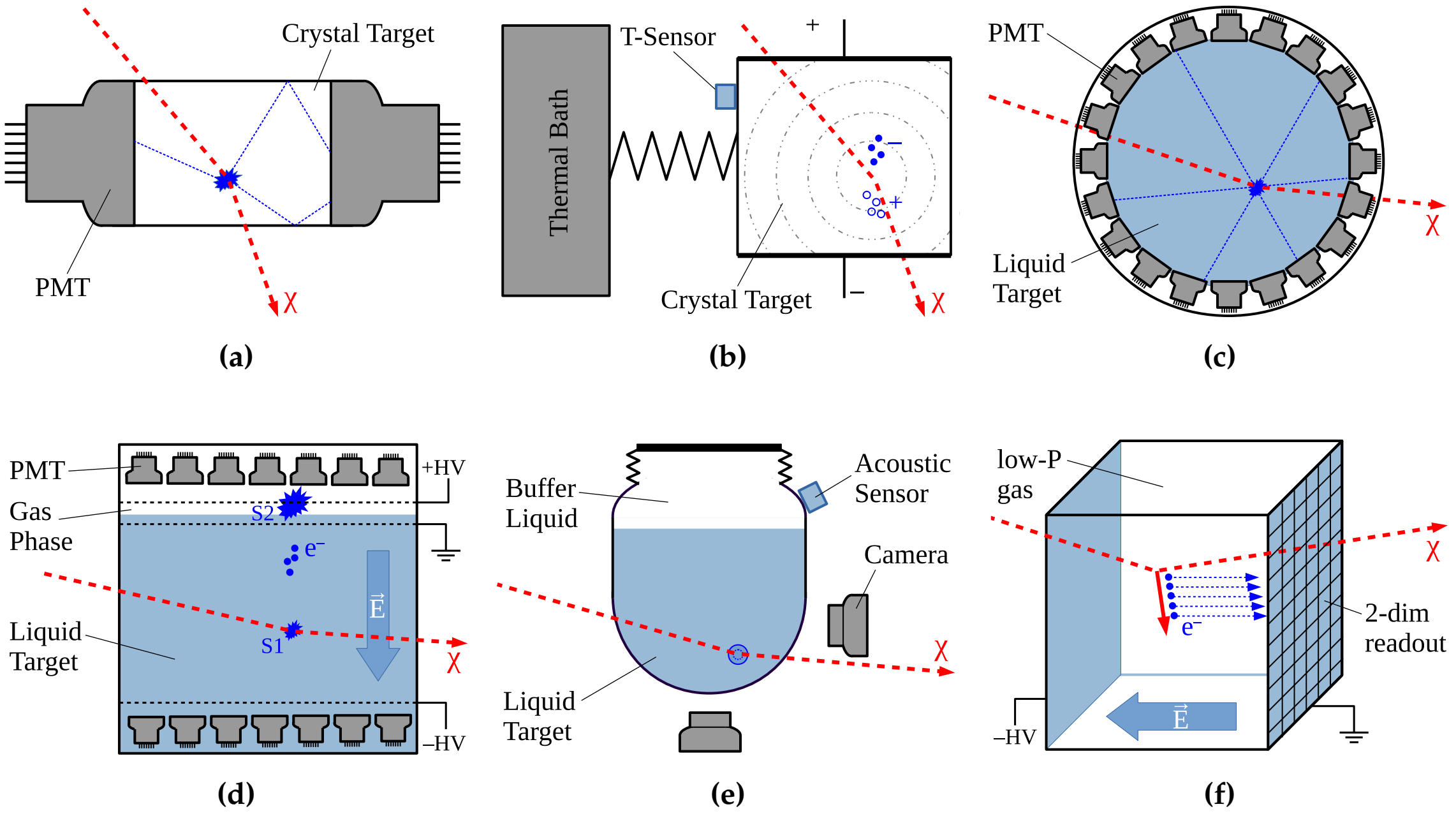

**Figure 1.10:** Summary of the common techniques and detectors used for direct WIMP searches: (a) scintillation crystals, (b) bolometers, (c) single-phase and (d) dual-phase noble liquid/gas detectors, (e) bubble chamber, and (f) directional (gas) detectors. Retrieved from [60].

### Bolometers

Crystals cooled to cryogenic temperatures (below 50 mK), as in Figure 1.10b, are highly sensitive detectors for small energy deposits, as lowering the temperature reduces the crystal's heat capacity, making it more responsive to small energy depositions. When particles interact with the crystal, they lose some of their energy as phonons (vibrations), causing a small temperature increase in the crystal. This temperature change, $\Delta T$, depends on the deposited energy, the crystal's mass, and its heat capacity (which decreases with $T^3$). Both ERs and NRs primarily lose energy through heat, but they differ in the amount transferred to ionization or scintillation, with NRs typically transferring less energy to ionization than ERs. This difference allows for ER/NR discrimination, as mentioned earlier, by simultaneously measuring the temperature change and either the ionization charge or scintillation light.

Transition edge sensors and neutron transmutation doped germanium thermistors are commonly used to detect these small temperature changes, as their resistance is highly temperature-dependent at cryogenic levels. Experiments like EDELWEISS and SuperCDMS use bolometers with germanium and silicon crystals. SuperCDMS employs the iZIP detector technology [70], which allows for ER/NR discrimination by measuring ionization and phonon signals down to 10 $keV$. Other experiments, such as CRESST [71] and COSINUS [72], use $CaWO_3$ and NaI(Tl) crystals, respectively, to simultaneously detect scintillation and phonon signals. Given the low energy required to produce phonons in these materials, this family of experiments often focuses on the lower WIMP mass range due to their very low energy thresholds, reaching down to 1 $keV_{nr}$ and below.

**Noble liquid**

Noble liquids like argon and xenon are highly effective for particle detection due to their ability to be easily ionized and their strong scintillation properties. When an ionizing particle passes through these materials, it excites the atoms, generating excited states in both triplet and singlet forms. The de-excitation of these states leads to the emission of UV photons with different decay times. The type of ionizing particle affects the distribution between triplet and singlet states, enabling particle identification by analyzing this property in the scintillation signal. Large volumes of noble liquids (Figure 1.10c) can be instrumented with light detectors to capture this primary scintillation light, and because they remain in liquid form at low temperatures, detectors with masses in the tonne range can be built. Examples of such single-phase detectors include DEAP-3600 [73] and XMASS [74].

In addition to single-phase detectors, noble liquids are also used in dual-phase TPCs (Figure 1.10d), which allow for the measurement of both scintillation light and ionization charge. In these setups, the primary scintillation light is collected by photosensors, while ionization electrons are drifted by an electric field toward a region that separates the liquid and gaseous phases. In the gas phase, these electrons are accelerated and produce photons through the electroluminescence process, amplifying the original charge signal. By combining the information from the primary and secondary scintillation signals, the deposited energy, particle type, and interaction position can be determined. Experiments like XENONnT [75] and DarkSide [76] employ this dual-phase TPC technology. The scalability of these detectors makes them leading technologies in the DM search paradigm, with the largest experiments reaching the neutrino floor, an ultimate limit to their sensitivity (see Section 1.3.4).

In the context of WIMP dark matter searches, the use of xenon has the added benefit of being a large atom ($A = 131$), which enhances sensitivity to SI coupling due to the increase in cross-section (see Section 1.3.1.2). This also makes the experiments more focused and sensitive to larger WIMP masses due to kinematic affinity. Additionally, xenon's two naturally occurring isotopes, which have an odd number of neutrons, make it sensitive to SD coupling as well.

#### Bubble chamber

Bubble chambers (Figure 1.10e) utilize superheated liquids kept at a temperature just above their boiling point, where bubbles form when energy exceeding a fixed threshold is deposited due to a local phase transition. For bubbles to nucleate at a specific temperature, a critical amount of energy must be deposited within a critical length, making bubble formation dependent on the stopping power of the particle. Since ERs and NRs have different stopping powers, the chamber's operational parameters can be adjusted so that *only* NRs create bubbles, making the detector largely insensitive to the ER background. Discrimination between NRs and alpha particles is possible by recording the acoustic signal generated during bubble growth with a piezoelectric sensor, as alpha particles tend to produce multiple bubbles over longer distances ($\sim 10\ \mu m$), while NRs generate bubbles over much shorter distances ($\sim 10\ nm$). The number and position of the bubbles can be recorded providing millimeter precision imaging.

Certain liquids with high fluorine content (like $C_3F_8$ or $C_4F_{10}$) are used to enhance sensitivity to spin-dependent interactions, while liquids combined with heavier elements like iodine (e.g., $CF_3I$) improve sensitivity to spin-independent couplings. A limitation of this technique is that bubble chambers function as threshold detectors, meaning they only detect the presence of events without measuring the precise energy deposited. This prevents direct constraints on mass-cross section parameters in the eventuality of a WIMP signal. The PICO experiment [77] is a notable example and often referred to as the leader of this detection method.

#### Gas (directional) detectors

Many gaseous detectors are being developed for the detection of low-mass WIMPs. The NEWS-G experiment [78] uses a Spherical Proportional Counter (SPC) to search for sub-GeV WIMPs. It operates at a pressure of $3\ bar$, making use of various gas mixtures, including noble gases and hydrogen-rich targets like $CH_4$. The SPC consists of a grounded copper sphere with a central anode. By applying a voltage to the anode, an electric field is created that pulls the electrons toward the center, where they then multiply when near the anode. This configuration enables the detection of single ionization electrons, allowing for a very low energy threshold. Another example is the TREX-DM experiment [79], which uses a high-pressure gaseous TPC with neon or argon as the target. It is equipped with Micromegas for amplification and readout, providing low intrinsic radioactivity and very high spatial granularity.

Low-pressure TPCs can also be used, typically in detectors focused on reconstructing the angular distribution of events, i.e., directional detectors (Figure 1.10f). The lower density of the medium results in longer particle tracks, which simplifies directional reconstruction and enhances ER/NR discrimination due to the different energy deposition patterns along the track path. The CYGNO detectors are also part of the gas directional detector family, although operating at an intermediate pressure (1 $\mathrm{bar}$). This allows for a balance between the ability to reconstruct the direction of low-energy events and the effective exposure to dark matter. A more detailed discussion of directional detectors is presented later in Section 1.4.3, and of CYGNO detectors in Chapter 2.

### 1.3.4 Current limits

As mentioned earlier, the rate of WIMP-induced NRs mainly depends on the interaction cross-section between the colliding particles and the WIMP mass. These two are correlated, and thus, if a DM signal or signature is found, this results in an *allowed region* of the WIMP cross-section-mass parameter space. The higher the number of detected events and/or the certainty in the results, the more stringent this region will be. On the other hand, if no events are detected, given a certain detector exposure and background knowledge, an upper limit on the parameter space can be set. For this reason, and since no DM has been effectively detected, this plot of the limits on the WIMP interaction is often referred to as the *Dark Matter exclusion plot*.

Figure 1.11 shows a summary of the published SI exclusion limits for several experiments, many of which were referred to above as examples of different detection techniques. Most of the experiments in this exclusion plot only measure the energy dependency of the rate of expected events, and so far, no definitive discovery has been made. The plot should be read such that all the area above the lines has been excluded from the possibility of constituting a WIMP particle. DAMA, which also measures the annual modulation signature, has identified two allowed regions around $\sigma \sim 10^{-40}\ \mathrm{cm}^2$, at $m_\chi \sim 10$ and $70\ \mathrm{GeV/c^2}$. These DAMA regions, although present, have been covered and excluded by an overwhelming number of other experiments. The year attached to the name of each experiment reflects the year of the published result. This is an important detail, as large experiments continuously improve their data analysis and increase their exposure over time, leading to more stringent limits.

As mentioned before, given that the spin-dependent contribution to the cross-section is only relevant for atoms with unpaired protons or neutrons, for most atoms, this contribution can be considered negligible. Therefore, the $\sigma_n^{SI}$ (nucleon-independent, as discussed) is often presented separately from the limits for $\sigma_n^{SD} - m_\chi$ and $\sigma_p^{SD} - m_\chi$. Although not shown, the SD sensitivities are about $10^6$ times weaker than the SI ones, and the PICO experiment is the leading one in this regard, thanks to the large exposure provided by the dense liquid and the high concentration of fluorine, the element with the highest coupling factor for SD interactions (see Table 1.2). Such results can be seen, for instance, in [77].

As can be seen in the figure, the limits show a minimum (higher sensitivity) where the WIMP mass is closest to the atomic mass of the detector target material, highlighting the kinematic affinity. As the WIMP mass increases, the number of particles to scatter off decreases with $1/m_\chi$, since the local DM density is constant. As mentioned, the right side of the plot is predominantly populated by dual-phase detectors such as XENONnT, LUX, and PandaX, which are the leading technologies in this region.

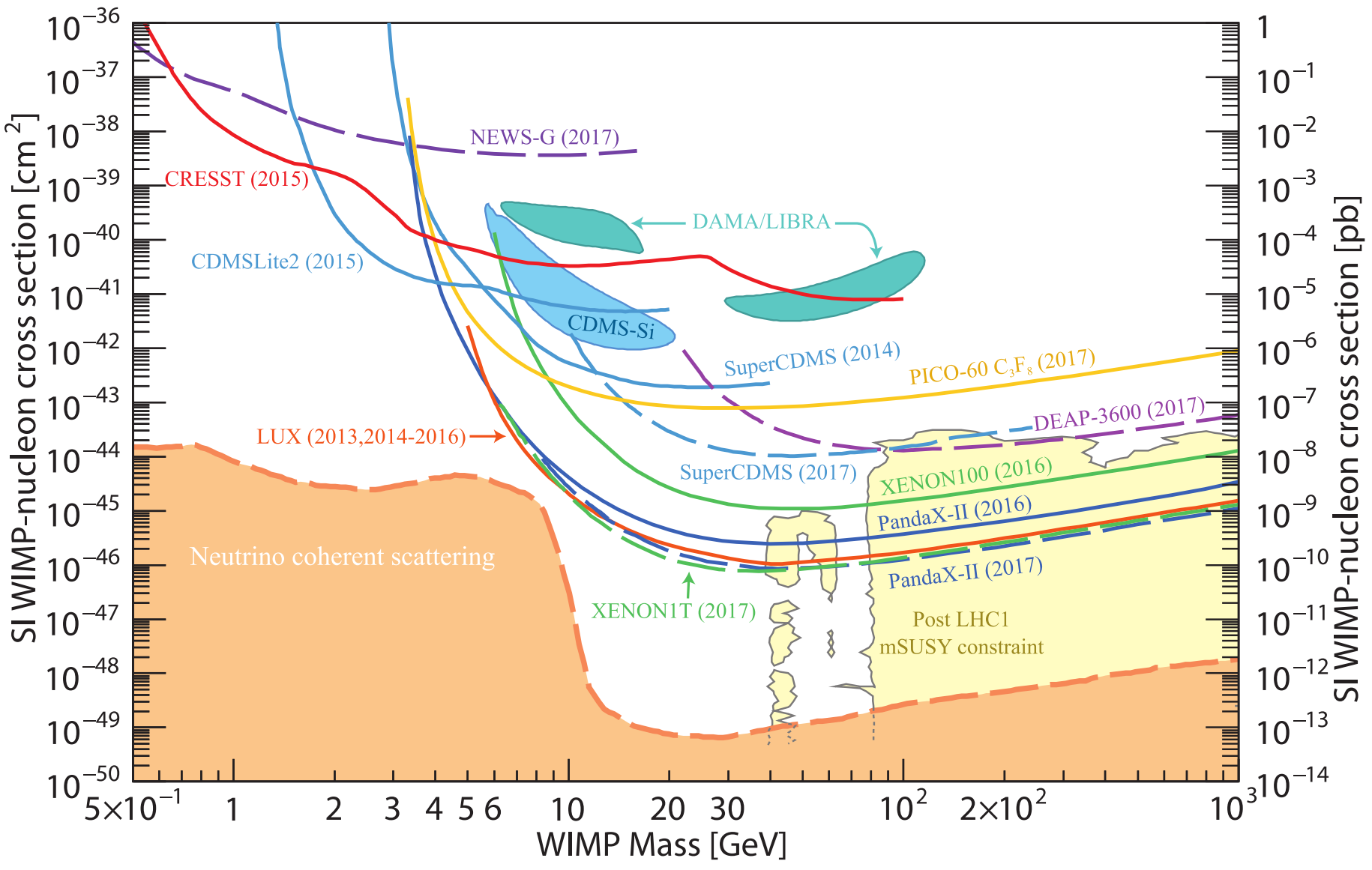

**Figure 1.11:** Current status of the limits on the cross-section for SI WIMP-nucleon interactions, as a function of the WIMP mass, based on the SHM model. Each line represents the most stringent lower limit set by each experiment in different subsequent years. The orange region at the lower part indicates the irreducible background from CEνNS interactions, which can mimic WIMP signals (see text for more details). This plot was selected and retrieved from [80] due to its detailed illustration of the various limits. A more up-to-date (but less detailed) version can be found in [81].

As WIMP masses decrease towards the left side, the limit is driven by the detector's energy threshold, which directly constrains the minimum detectable NR energy and, consequently, the WIMP mass. This results in a sharp rise in sensitivity for most experiments. In this low WIMP mass region, the greatest contenders are experiments capable of reaching very low energy thresholds, such as CRESST or SuperCDMS. This is also the region that the CYGNO experiment expects to explore and cover, given the light mass elements and low threshold achieved, as further discussed in Chapter 2.

To reach such high sensitivities, most of these experiments are placed in deep underground facilities where they are shielded from cosmic radiation. Many of these experiments employ purification systems and advanced particle identification techniques to discriminate against background signals. The detectors are also generally built with very high purity materials to minimize internal backgrounds as much as possible. A deeper overview of the background sources for general DM searches is provided in Section 1.3.5.

The "neutrino floor/fog", identified in the image as "neutrino coherent scattering", represents a major challenge in dark matter detection, as it sets a strong limit on the ability of experiments to identify potential DM signals [82]. This phenomenon occurs due to the irreducible background originating from interactions between neutrinos and nuclei, specifically through coherent elastic neutrino-nucleus scattering ($CE\nu NS$). In this process, nuclear recoils from atmospheric, diffuse supernovae background, and solar neutrinos produce signals indistinguishable from those induced by dark matter WIMPs. As a result, experiments cannot explore this region in a nearly background-free environment.

The neutrino floor in lower WIMP mass band ($< 10\ GeV/c^2$) is populated only by solar neutrinos, especially those originating from the $^8B$ chain, which generate signals in the detector compatible with WIMPs of masses $\mathcal{O}(1)\ GeV$. On the other side of the spectrum, at high WIMP masses, atmospheric neutrinos, produced by cosmic ray interactions, dominate this region, restricting sensitivity to WIMPs above $100\ GeV$. Meanwhile, neutrinos from core-collapse supernovae limit searches for WIMPs around $20\ GeV$.

Recent experiments like XENONnT [83] and PandaX-4T [84] have already detected signals consistent with $CE\nu NS$ from $^8B$ solar neutrinos, further confirming this limitation. Recent discussions in the dark matter community have reframed this limit as "neutrino fog" [85], which refines the concept, suggesting that DM signals aren't completely hidden but "obscured" by background fluctuations. This makes detection below the neutrino fog harder but not impossible. To better overcome this limitation, a potential solution could be the exploration of the third WIMP halo signature – directionality. As outlined later in Section 1.4, detecting the directional signature of WIMP-induced events could offer a way to undoubtedly detect DM signals and navigate through the neutrino fog without the need for high statistics, especially at low WIMP masses, where solar neutrinos can be discriminated from WIMPs coming from the Cygnus constellation.

### 1.3.5 Background sources

Given the very low expected interaction rate between DM WIMPs and materials in a detector, as shown in the general calculation performed in Section 1.3.1.4, the background induced by the environment where the experiment is placed – *external background* – as well as the background generated by the detector materials themselves – *internal background* – need to be precisely monitored, controlled, studied, and overall described through an accurate background model. In this section, an overview of the different types of background sources and their origins is provided as a background for some of the studies presented later in this thesis, namely in Chapter 6.

#### 1.3.5.1 Cosmic radiation

Cosmic rays are particles that originate from deep space and interact with Earth's atmosphere, producing hadronic and electromagnetic showers. These showers generate a wide variety of particles, including electrons, positrons, neutrinos, neutrons, pions, and others. The largest component of particles detected at ground level are muons, which are generated by the decay of charged pions and kaons. These muons, arriving at the ground with energies between $MeV$ and $GeV$ at a rate of approximately 100 $m^{-2}s^{-1}$, are deeply penetrating and can interact with particle detectors, representing a significant background to be suppressed for most rare-event search experiments. In Section 4.2, a study of the cosmic muon flux at ground level, performed with the LIME detector (more details on LIME in Chapter 4) is presented, along with a more detailed explanation of the cosmic muon flux's origin, shape, and overall characteristics.

To shield experiments from cosmic ray background, large physics experiments are typically placed deep underground in specialized laboratories designed for this purpose, such as the Laboratori Nazionali del Gran Sasso (LNGS) in Italy, or in repurposed (or even simultaneously) used old mines, like the SNOLAB laboratories in Canada. By placing experiments underground, the cosmic muon flux is drastically reduced, by factors of around $\mathcal{O}(10^5 - 10^8)$, depending on the depth of the laboratory and the type of rock providing the "shielding". To better normalize and compare the depth of different laboratories, their depth is often expressed in meter water equivalent (m.w.e.), which represents the amount of water required to reduce the cosmic muon flux by the same amount. LNGS, where the CYGNO experiment is located and where much of the data used for studies in this thesis were based, has a total m.w.e. of approximately 3600 $m$, with a muon flux reduction factor of around $10^6$.

Muons can still impact underground experiments in two primary ways: they can either directly interact with the detectors, or they can produce neutrons through inelastic interactions with the surrounding rock. These neutrons, which are highly penetrating, can escape the rock and reach the experimental setup, where they scatter in the detector materials and increase the levels of nuclear recoil background. Muons can also traverse the rock and directly produce neutrons in interactions with the experimental setup itself, leading to the term *cosmogenic neutrons*.

In general, the deeper the laboratory (in m.w.e.), the better the experiment is shielded from cosmic muon and neutron backgrounds. However, factors such as accessibility and ease of construction also play a significant role in determining where an experiment is located. For instance, LNGS is one of the easiest laboratories to access due to its construction directly from a highway tunnel, which has made it a preferred choice for many experiments, even when compared to potentially deeper laboratories.

#### 1.3.5.2 External background

In addition to the cosmic background, which arrives in the form of muons or cosmogenic neutrons, other background sources exist within the local environment surrounding the experiment. These include the atmosphere, construction materials, and the surrounding rock, the latter particularly in the case of underground experiments. The primary sources of natural radioactivity, referred to as *primordial isotopes*, are the $^{232}\mathrm{Th}$, $^{238}\mathrm{U}$, $^{235}\mathrm{U}$, and $^{40}\mathrm{K}$ isotopes. These are named for their extremely long half-lives (greater than 1 million years), which result in their presence in virtually all materials on Earth. The radioactive decay chains of these isotopes include alpha and beta emitters, leading to the emission of these particles at various energies. Gamma particles can also be produced from the de-excitation of a nucleus following decays in these chains.

In addition to these isotopes, other contaminants, such as $^{85}\mathrm{Kr}$ and $^{137}\mathrm{Cs}$, from anthropogenic or cosmogenic origins, can be found in everyday materials within laboratories. Radon ($^{222}\mathrm{Rn}$) is another common contaminant often present in the air, which, as it will be shown, can have a significant impact on the overall background of experiments.

**Electrons & Gammas**

From the decay of these isotopes, various particles are emitted, and depending on their type and energy, different interactions can occur. Starting with beta electrons, which can be emitted in these decay chains with energies up to $\mathcal{O}(1)$ MeV, they are typically absorbed by the materials from which they originate, particularly if emitted within the material bulk. In contrast, gamma rays can easily escape the materials and reach the experiment. In both cases, when these particles interact with the detector, they produce electron recoil-like events. Since the contaminants are randomly distributed and their decay directions are isotropic, the resulting diffuse background also exhibit an approximately isotropic distribution.

### Alphas & Neutrons

Alpha particles originated from decays near the exterior of the detector typically do not directly impact it, as alphas high stopping power causes them to be fully absorbed within a few centimeters of air, preventing them from entering the detector. However, alpha particles can induce neutron emission when absorbed by a nucleus through $(\alpha, n)$ reactions. These neutrons, known as *radiogenic*, have energies on the order of $\mathcal{O}(10)$ MeV and behave similarly to cosmogenic neutrons, penetrating the detector and producing nuclear recoil background. Additionally, the spontaneous fission of $^{238}U$ and $^{235}U$ can also result in the emission of radiogenic neutrons.

Both cosmogenic and radiogenic neutrons contribute to a diffuse environmental neutron flux in underground laboratories. The flux direction is roughly isotropic, due to the production mechanisms mentioned and the random scattering of neutrons in surrounding materials. The specific properties of the underground neutron flux are unique to each laboratory, and understanding this flux is crucial for all experiments conducted in these facilities. For this reason, as mentioned later in Section 2.2.2.4.3, one of the objectives of the CYGNO experiment, during its runs with LIME underground, was also to measure the neutron flux at LNGS, improving upon the existing data [86].

### Radon Progeny Recoils

In addition to gamma and neutron backgrounds, a specific contaminant present in the air that affects many experiments is radon. This inert noble gas has two primary isotopes: $^{222}Rn$, produced in the $^{238}U$ decay chain, and $^{220}Rn$, produced in the $^{232}Th$ decay chain. When considering radon contamination, the contribution of $^{220}Rn$ is usually disregarded due to its short half-life (approximately 1 minute) [87, 88], compared to the 3.8-day half-life of $^{222}Rn$. Since the parent elements of radon are found in nearly all materials on Earth, and because radon is a gas, it can easily escape from rocks and other materials, leading to its accumulation in the atmosphere. Due to its relatively long half-life, radon can diffuse over large distances, depending on the laboratory's ventilation system, and reach many experiments in closed environments like underground laboratories.

Given its origin and gaseous nature, Rn can negatively impact an experiment in various ways. Using CYGNO's gas TPC technology as an example (see Chapter 2 for more details), Rn can naturally diffuse into the detection medium from the detector walls and materials in contact with it, behaving as an internal-like background. Additionally, Rn can enter the detector from the outside atmosphere through leaks in the piping or by diffusing through plastic barriers [88]. Furthermore, it has been observed that Rn and its decay daughters can attach to water molecules [89]. This becomes particularly relevant for gas TPCs, which are often sensitive to changes in humidity, thus potentially exacerbating radon contamination.

The $^{222}Rn$ decay chain, shown in Figure 6.27, includes several high-energy alpha particles (on the order of $\mathcal{O}(1)$ MeV) and two beta decays with energies of 1.2 MeV and 3.3 MeV. These decays can result in both alpha and electron recoil background if they occur directly inside the detectors.

Another feature of the $^{222}$Rn isotope is that, upon decay, its immediate daughter, $^{218}$Po, is produced with a positive electric charge ($^{218}$Po$^{+}$) approximately 90% of the time [89]. The two subsequent daughters, $^{214}$Pb$^{+}$ and $^{214}$Bi$^{+}$, are expected to behave similarly. As charged particles, these daughters drift under an electric field, and in the case of a TPC, they will drift toward the cathode, where they will eventually decay[1]. When these daughters decay at the cathode, two possible outcomes can occur at the point of interaction: either an alpha particle is emitted toward the cathode, and a nuclear recoil is directed toward the sensitive volume, or vice versa. Alphas can generally be distinguished from other particles due to their high energy and range, but the NR originating from these events, such as from $^{214}$Pb, can mimic a WIMP-like interaction [92]. Given the properties of Rn and its decay products, the alphas and NR recoils originating from the decays of (mainly) $^{218}$Po and $^{214}$Po in the sensitive volume are often referred to as *Radon Progeny Recoils (RPRs)* [93]. RPRs are an important and significant source of background in DM searches due to the overwhelming presence of radon everywhere, difficulty in predicting and modelling its presence in an experiment, and the ultimate fact that RPRs can induce DM-like signals.

Once the Rn daughters reach the cathode, besides directly leading to an RPR event, they can undergo electrodeposition, forming a thin layer of isotopes on the cathode. This process, also known as cathode *plate-out*, results in a recurrent stable background determined by the initial amount of Rn in the sensitive volume. This issue increases the amount of RPR events and their persistence with time, presenting a significant challenge for underground rare-event searches [88, 94]. Furthermore, Rn daughters have been observed to potentially undergo incrustation and diffusion into solid materials, as seen in glass [95], further worsening the effects of Rn plate-out and contributing to a long-lasting decaying background.

### Shielding

To limit the impact of external backgrounds on the physics studies, experiments use what is called *shielding*, which involves placing specific materials around the detector to block incoming external particles. This shielding can be passive, consisting of a static material surrounding the detector, or active, if additional analysis are performed to tag external events.

To passively block gamma rays, a high Z material is used, as the photon interaction cross-section scales with the power of Z. Typical materials for blocking these particles include copper ($_{29}$Cu) and lead ($_{82}$Pb) due to their abundance, cost-effectiveness, and ease of purification and handling. Experiments often also use these materials simulta-neously, with copper serving as the internal shielding. This approach makes use the high Z of lead to block the majority of external gamma rays, while copper is used to block any residual background from the lead shielding itself, which is harder to produce with low radioactivity compared to copper.

[1] As a side note, the creation of polarized Rn decay products is a well-known issue, as it can pose a health hazard. These polarized isotopes can attach to aerosol particles and be inhaled by individuals [90, 91].

For neutrons, the opposite is required, i.e., a low $A$ material such as water or polyethylene. Neutrons primarily interact through elastic scattering, losing energy until they thermalize. The kinematic affinity, similar to that for WIMPs, is highest when the two particles have the same mass. Therefore, neutrons lose the most energy when colliding with hydrogen. The average energy loss of a neutron in an elastic collision ($\Delta E/E$) is given by:

$$\frac{\Delta E}{E} = \frac{1}{2}\left[1 - \left(\frac{A-1}{A+1}\right)^2\right] \tag{1.25}$$

This relation highlights that the maximum energy transfer for neutrons occurs with atoms having $A = 1$, i.e., hydrogen. Since water ($H_2O$) is also a cheap, safe, and readily available material, it is the most commonly used material for creating large neutron shieldings.

As mentioned, active shielding can also be used to mitigate backgrounds. A common method is the use of a water Cherenkov tank around the main detector to detect any residual muons that have traversed the rock above the laboratory and passed through the water, producing Cherenkov radiation[2]. By identifying the timing of these muons, events occurring within a subsequent time window (such as NRs from cosmogenic neutrons) can be tagged and discarded. This technique is employed in large dark matter experiments such as XENONnT [96] and DarkSide-20k [97]. Other systems involve surrounding the detector with a liquid scintillator enriched with elements prone to neutron radiative capture, such as boron or gadolinium, which emit gamma rays when neutrons are captured [98]. This serves as both passive shielding by stopping external neutrons and active shielding by allowing these events and subsequent ones to be tagged.

To mitigate the background induced by radon and its progeny (RPRs), the focus also includes controlling and monitoring the atmosphere around the detector. For example, air circulation systems are often installed throughout the laboratories to prevent the accumulation and subsequent infiltration of gases, including radon. At the detector level, experiments frequently flush the surroundings with nitrogen or argon, as these gases are pure, clean, and inert. By applying positive pressure with these gases, the accumulation and infiltration of other gases like radon is minimized, while also reducing the presence of humidity, which is also linked to the presence of radon. This technique is used in certain portions of the CYGNO detectors, with the additional objective of preventing the infiltration of helium into the PMTs through the glass photocathodes, a known issue that can reduce PMT efficiency [99]. To further mitigate the presence of radon, CYGNO has implemented dedicated gas traps and filters in the gas pipelines, as discussed later in Section 2.2.2.4.2.

[2] The Cherenkov effect occurs when a charged particle moving through a dielectric medium with refractive index $n$ travels faster than the speed of light in that medium ($v > c/n$). The constructive interference of radiation emitted from the de-excitation of the locally polarized medium leads to the emission of a detectable light cone.

#### 1.3.5.3 Internal background

The internal background refers to all backgrounds induced by the experiment itself, including contributions from the sensitive medium, detector-related materials, and external background shielding. Unlike external backgrounds, internal backgrounds are not easily shielded, and particles with short ranges, such as alpha particles, can have a non-negligible contribution.

The origin of these backgrounds, similar to some external ones, lies in primordial long-lived isotopes found in virtually all materials, such as $^{232}\mathrm{Th}$, $^{238}\mathrm{U}$, $^{235}\mathrm{U}$, and $^{40}\mathrm{K}$. Since these isotopes cannot be simply removed from the materials, it is crucial for dark matter experiments to properly account for and model their background shapes. With comprehensive energy and particle background models, it becomes possible to mathematically determine the likelihood that an event is a signal of interest rather than a background. This methodology is applied not only to DM searches but also to other rare-event searches, such as neutrinoless double-beta decay [100].

One of the most important steps in the background modeling process is measuring the actual radioactivity contamination of each detector material. This task can be achieved through various methods and tools, including the following:

- **Neutron Activation Analysis (NAA):** A non-destructive technique in which samples are irradiated with neutrons, and gamma emissions are measured to estimate isotope concentrations (e.g., $^{238}\mathrm{U}$, $^{232}\mathrm{Th}$, $^{40}\mathrm{K}$) with sensitivities down to $\mathcal{O}(0.01)$ $\mu\mathrm{Bq/kg}$. It requires tens of grams of material and only a few weeks of measurement time. The sample becomes activated and cannot be reused afterward.

- **Inductively Coupled Plasma Mass Spectrometry (ICP-MS):** A destructive method in which a liquid or dissolved solid is ionized in argon plasma, and ions are separated based on their mass-to-charge ratio. Small amounts of material (less than 1 g) are required, with short measurement times of a few days. It achieves sensitivities of $\mathcal{O}(1)$ $\mu\mathrm{Bq/kg}$, with the main limitation being its high cost.

- **Alpha Spectroscopy:** This technique measures the energy of alpha particles emitted by radionuclides for surface-only analysis, due to the short range of alpha particles. It requires small amounts of material (less than 1 g), with measurement times on the order of months, and offers sensitivities of $\mathcal{O}(1)$ $\mathrm{mBq/kg}$.

- **Gamma Spectroscopy:** A non-destructive method to identify radioactive isotopes through the detection of gamma rays emitted during the isotope's decay. It requires large samples and measurement times ranging from weeks to months, with sensitivities of $\mathcal{O}(10 - 100)$ $\mu$Bq/kg. High-purity germanium detectors provide precise energy measurements and are the most commonly used technique in low-background setups. At LNGS, a specialized facility, STELLA, is available and has been used by CYGNO to measure the radioactivity of some of their detector materials.

In addition to measuring the materials' radioactivity, several precautionary steps should be taken to minimize radioactive impurities. For instance, materials should be stored in low-radioactivity environments to prevent further activation. This includes minimizing exposure to the surface and avoiding flight transportation, as both increase contact with cosmic rays, especially the latter.

Moreover, several radiopurification techniques are typically employed. These range from surface cleaning with strong acids to remove the external layer of material, which is often more contaminated due to machining and handling processes. Electropolishing can also be used for the same purpose. This involves immersing the material in a bath where an electric field is applied, causing surface ions to dissolve and be removed. To further improve the cleanliness of the bulk (internal) of the detector material, techniques like electroplating can be used. This process is essentially the reverse of electropolishing: the material is immersed in a bath where ions from the surface of the detector receive electrons, slowly forming a layer of pure material, one layer at a time. In DM searches, this technique has been famously used by NEWS-G [101]. Other large DM experiments, such as XENON [102] and DarkSide [103], which use noble gases, use methods like cryogenic distillation. In this, the different vapor pressures and volatilities of the elements are exploited to separate different gases and purify the main one.

Other post-data taking analysis tools can also be used to reduce background. These may include detector-specific variables of interest that help distinguish between ERs and NRs (a crucial step to determine if an event is signal or background), often related to the energy deposition patterns of these events. Another common data selection method used in experiments is detector fiducialization. Fiducialization involves removing certain geometric regions of the detector from the analysis to exclude events occurring in specific areas, often near the borders where the detector materials are in contact with the sensitive volume. This is especially relevant for the detection of alpha particles, as mentioned previously in Section 1.3.5.2.

In the case of CYGNO detectors, through the combined use of the CMOS camera and PMT sensors, both the position and the 3D direction (sense and orientation) of alpha particles can be retrieved. This further helps in pinpointing their material and decay origin, allowing for the their identification as possibly RPRs/background. This is the topic of discussion of Chapter 6.

## 1.4 Directional dark matter searches

The directional dependence of WIMP-induced nuclear recoils is one of the strongest, yet most difficult to detect signatures of dark matter and the WIMP halo. As mentioned in Section 1.3.2.3, the direction of NRs induced in the detector should point opposite to the direction of the laboratory's motion in the galactic rest frame, often approximated as the direction of the Cygnus constellation. In galactic coordinates, as illustrated in Figure 1.8, a dipole-like structure appears in the angular distribution of WIMP-induced NRs, centered approximately at $(90°, 0°)$, coinciding with the Cygnus direction. In detector coordinates instead, a daily oscillation appears due to the Earth's rotation, as shown in Figure 1.9.

The differential rate in energy and angle for WIMPs in an Earth-based detector is shown in Equation 1.24, which highlights several important parameters. For instance, the term $v_{lab} \cdot \cos(\phi)$ indicates that WIMPs arriving from a specific direction will have higher velocities, making them more easily detectable, and therefore amplifying the strong directional anisotropy. In contrast, any non-WIMP-induced signal is expected to be isotropic in galactic coordinates, making it highly unlikely (or impossible) for a background source to replicate this distinct anisotropic feature in the recoil spectrum.

This anisotropy can be visualized in different ways. Based on previous considerations, the recoil angle can be evaluated as:

$$\cos(\phi) = \frac{v_{min}}{v_{lab}} = \sqrt{\frac{m_A \cdot E_{nr}}{2 \cdot \mu_A \cdot v_{lab}^2}} \qquad (1.26)$$

where $\mu_A$ is the reduced mass between the detector element and the WIMP, and $v_{min}$ has been previously defined in Section 1.3.1.1. In this equation, $\phi$ can vary from 0° to 90°, and as $\phi$ approaches 0°, the angular distribution exhibits a dipole structure as previously mentioned, with the ratio in the rate between the two poles reaching up to 100× [104]. The extent and separation of this dipole will largely depend on the experimental performance, particularly the angular resolution.

On the other hand, when $v_{min} < v_{lab}$, the angle $\phi$ peaks at different values, resulting in a ring-like structure in the angular distribution of NRs, centered around the same dipole direction. This ring-like feature is considerably less prominent than the simple dipole and requires improved angular resolution and an higher amount of events to be clearly observed [105]. For this reason, experiments often focus on the dipole shape, which is sufficient to positively identify a WIMP signal, given a reasonably large exposure and low energy threshold.

### 1.4.1 Head-tail asymmetry

The sensitivity to the 3D direction of NRs, combined with energy measurements, provides a powerful tool for identifying dark matter signals. When discussing directionality, two detector parameters are often highlighted. The first, and most direct, is the angular resolution of the detector, which determines the precision with which the nuclear recoil angle can be measured. Depending on this resolution, the visibility of the dipole pattern mentioned earlier can become more or less distinct (more smeared), potentially appearing flat if the resolution is insufficient. In this context, to reliably distinguish between a flat and a dipole distribution, an angular resolution of around $30 \times 30\ deg^2$ – a reasonable target for most directional detectors – is considered sufficient [106].

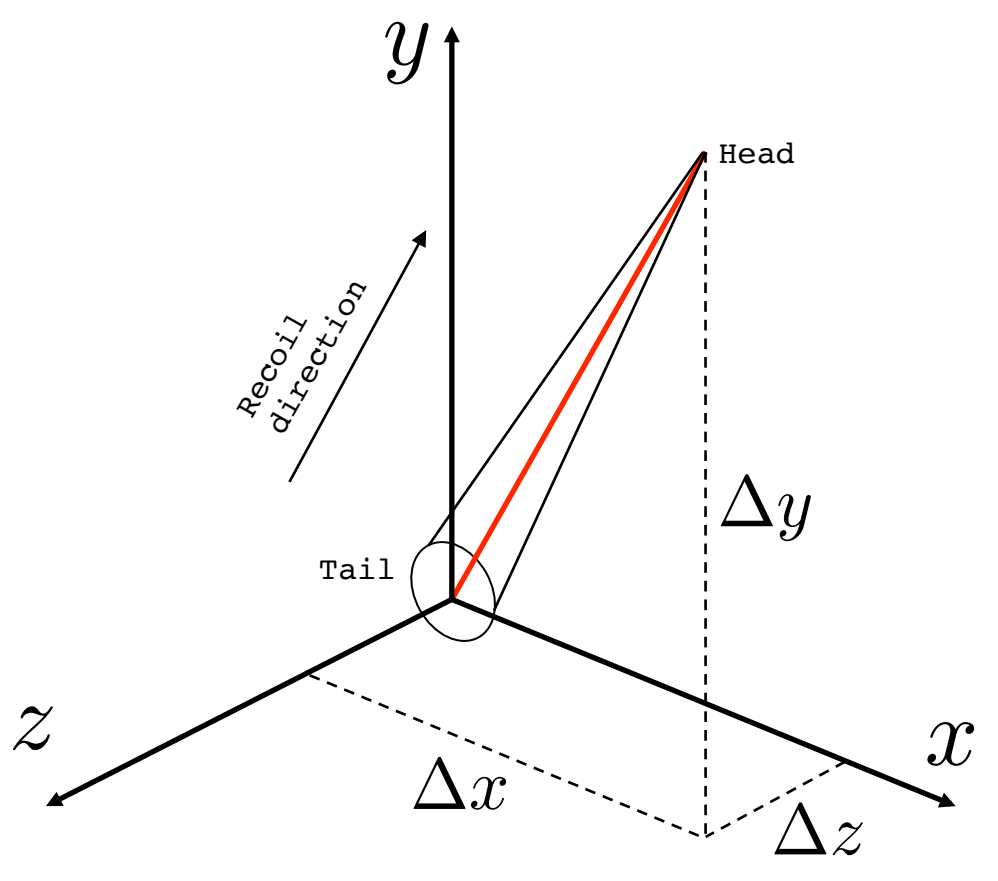

**Figure 1.12:** Schematic of the head-tail (HT) asymmetry in nuclear recoils. The cone surrounding the red line (main track) illustrates the ionization density along the nuclear recoil path. Image retrieved from [109].

The other key directional component is the ability to determine the ionization track's *head-tail (HT)* asymmetry. In nuclear recoils, the ionization track forms a cone, with a higher ionization density near the start (tail) and decreasing towards the end (head), as depicted in Figure 1.12. This HT asymmetry provides the sense (or verse) of the nuclear recoil. When combined with the angular measurement of the recoil, it allows for the determination of the recoil's 3D or vector direction [107, 108].

The ability to recognize the HT asymmetry (ranging from 0% to 100%) directly affects the ability to discern the WIMP dipole. In the thesis of G. Dho [104], examples of the angular distribution of WIMP-induced NRs in the SHM assumption were calculated for various HT recognition abilities. As shown in Figure 1.13, for a 100% recognition ability, a clear dipole is visible in the direction of the Cygnus constellation. However, with only 50% recognition – meaning the correct HT sense is determined correctly only half of the time – two parallel "dipoles" appear in galactic coordinates. In this scenario, the ability to identify a dark matter signal is significantly reduced. Moreover, if the angular resolution used in the simulation were poorer, a nearly flat distribution would appear, making it effectively indistinguishable from a flat background.

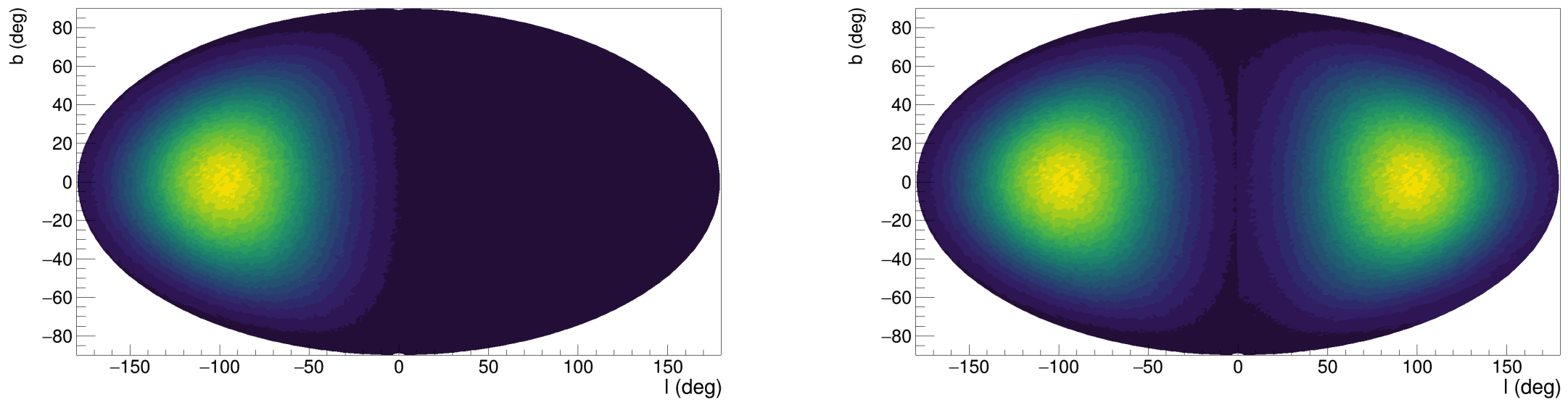

**(a)** HT = 100%

**(b)** HT = 50%

**Figure 1.13:** Examples of the angular distribution of nuclear recoils induced by WIMPs in the SHM model. Two scenarios are shown: assuming a head-tail recognition of (a) 100%, where the dipole effect is clearly visible, and (b) 50%, which corresponds to no capability of track sense reconstruction, as the sense is only correctly identified half of the time. These simulations were performed by G. Dho in his thesis [104].

### 1.4.2 Directionality advantages

Once the directional information is extracted from the NR angular distribution, several properties become accessible in the context of DM searches. For example, the presence or absence of NR angular anisotropy can be used to directly support or refute the existence of DM. If DM is observed by other experiments, directional detectors can positively identify the same DM with higher significance, given the expected direction of the flux. Finally, with enough events, the directional topology of the DM flux allows scientists to perform "WIMP astronomy", which refers to investigating the properties of the WIMP flux and Standard Halo Model, and its connections to other fields of astrophysics and the study of the cosmos. The inclusion of directional information also enhances the ability to identify background signals, whether they are internally emitted by the detector, externally from the laboratory environment, or cosmological such as neutrinos from the Sun. In this section, a brief overview of these advantages and their interplay with the current status of dark matter physics is provided.

#### 1.4.2.1 Exclusion limits

The striking dipole feature in the angular distribution of nuclear recoils induced by WIMPs is one of the strongest advantages of the directionality approach, as no other background can mimic this directional dependence. This is further enhanced by the Earth's rotation around its axis and its tilt relative to the galactic plane, as discussed in Section 1.3.2.3. With such features, directional detectors can, with a similar exposure, establish much stronger limits in the WIMP cross-section-mass exclusion plot due to their high ability to reject backgrounds.

As mentioned in Section 1.3.4, one of the largest directional background sources (and possibly the only one) in dark matter searches that generates a signal similar to the WIMP directional signal is the events produced by solar neutrinos interacting with nuclei (CE$\nu$NS). These create nuclear recoils that are indistinguishable from WIMP-induced ones based solely on energy information. Using energy alone, solar neutrinos could only be rejected in the presence of a large number of events and a better understanding of the solar neutrino fluxes. However, with directional information, solar neutrinos can be distinguished from WIMPs. This is possible because the directions of the Sun and the Cygnus constellation never overlap throughout the year, assuming sufficient angular resolution [106].

This phenomenon, already referred to in Section 1.3.2.3, is shown again in a different perspective in Figure 1.14a. In this figure, the typical NR event rates induced by a 9 $GeV$ DM WIMP on fluorine and solar neutrinos are shown in blue and red, respectively, in galactic coordinates on the $6^{th}$ of September, when the separation between the two is the greatest. In the galactic plane, the Sun follows the Ecliptic line (represented by the red line in the plot), with its neutrinos tracing a path around it. Earth instead, is always moving towards the Cygnus direction. The difference between these two signals is strongest around the date mentioned and weakest on February $26^{th}$. This effect is illustrated in Figures 1.14b and 1.14c, where the recoils induced in a detector on Earth by these two sources are shown on these two dates. While in February the two signals are closer to each other in galactic coordinates, it is still possible to distinguish them, and the overlap can be considered negligible given a reasonable angular resolution of $\mathcal{O}(30)$ $deg^2$ [106]. Additionally, the rotation of the Earth around its axis also dislocates the two signals, as already shown in Figure 1.9, further increasing the distinguishability of these two sources. Aside from solar neutrinos, any other local background will not oscillate during the day or year (in galactic coordinates), and thus does not constitute a directional background that could be confused with the WIMP-induced one.

These images highlight the power of the directionality signature in identifying dark matter WIMPs, both in galactic coordinates and detector coordinates. As a result, it has been found that the current limits on the WIMP cross-section/mass detection can be improved by around a factor of 5 when the recoil angular distribution is also taken into account [105, 110, 111]. As an added benefit, the same technique can be used to identify and study solar neutrinos, as they also represent an interesting astrophysical component of particles arriving at Earth.

#### Impact of *3D* and *Head-Tail*

As mentioned, the performance of the detector plays an important role in the ability to detect a dark matter signal, particularly regarding the accuracy of detecting its angular (anisotropic) distribution. Several studies, including [105, 110–112], have examined the physics reach given different detector readout and performance scenarios. The results are summarized in Figure 1.15, which shows the discovery limit in the WIMP search parameter space as a function of the WIMP mass and detector total exposure for various readout scenarios. These scenarios include performing different vectorial reconstructions of the ionization track (1D, 2D, and 3D) and whether or not the nuclear recoil

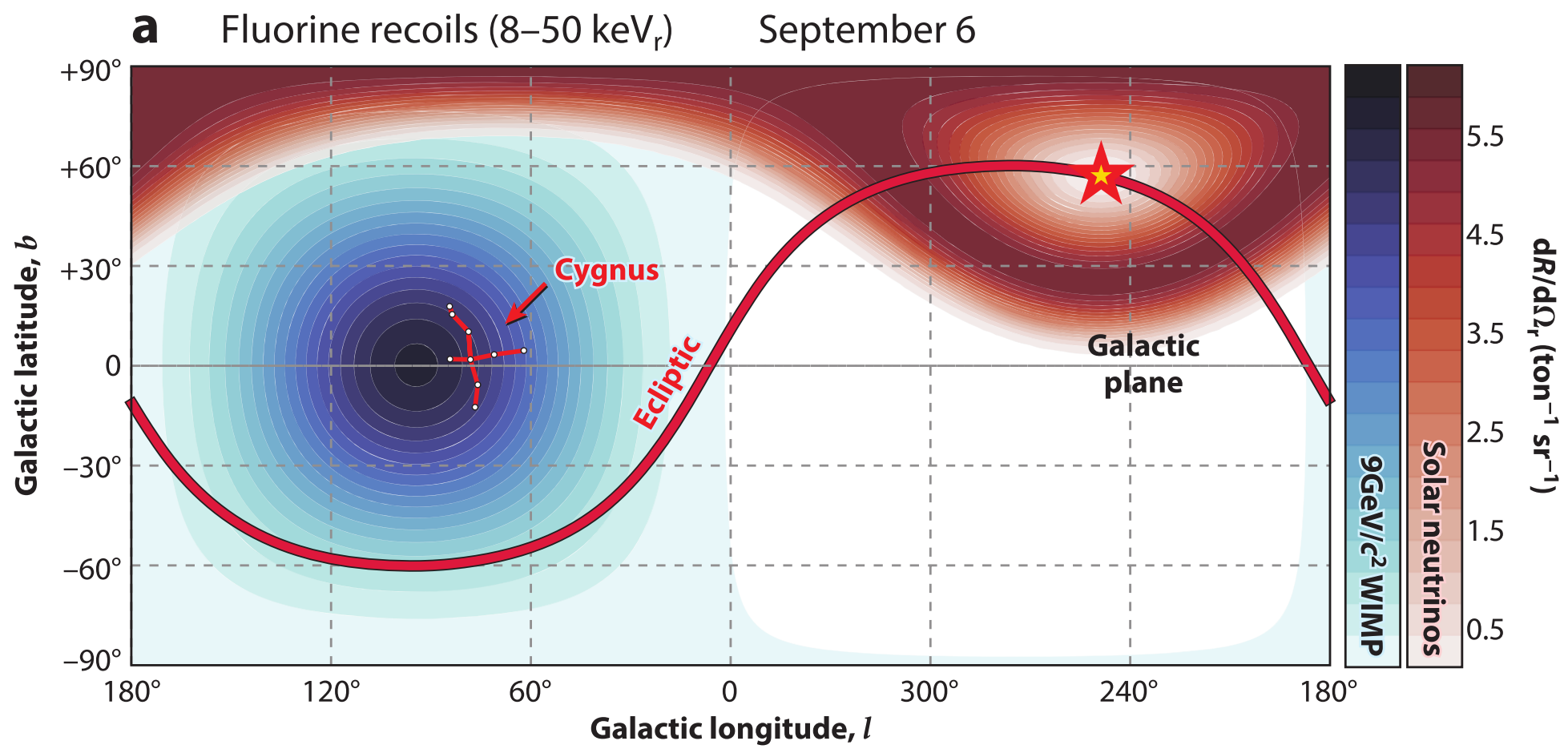

**(a)**

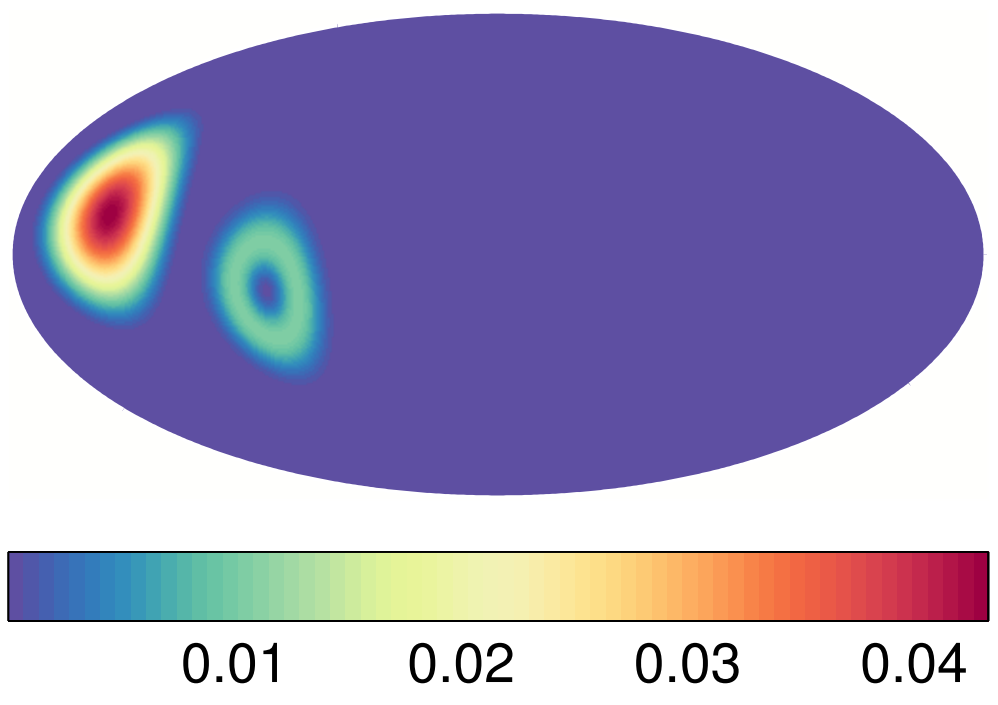

**(b)**

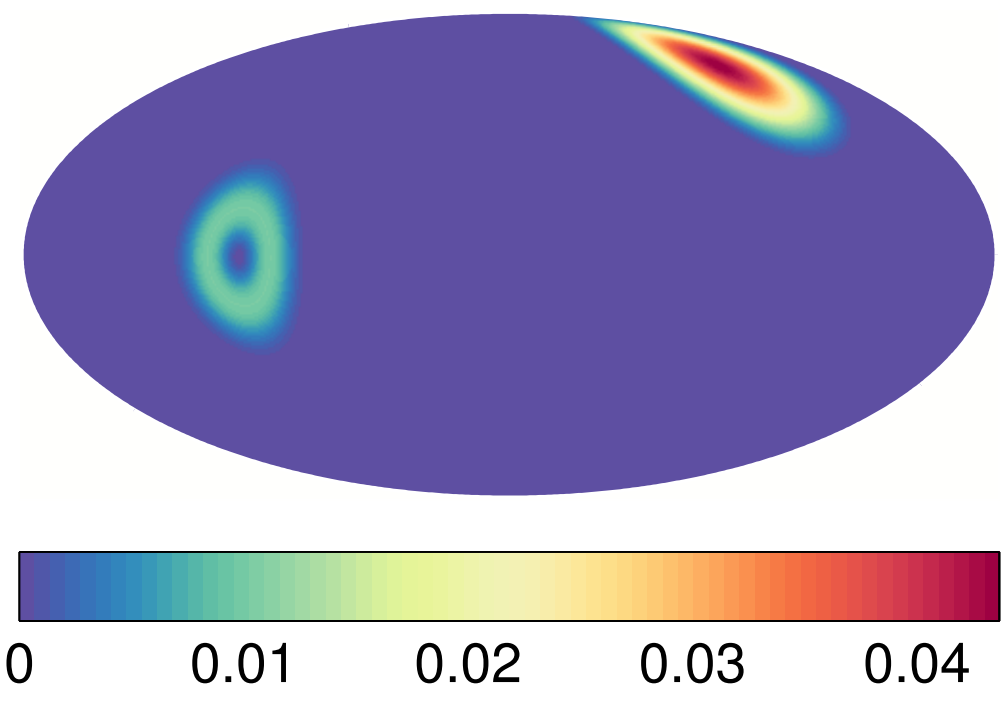

**(c)**

**Figure 1.14:** (a) Event rates from a 9 GeV WIMP (*blue*) and solar neutrinos (*red*) displayed in galactic coordinates on September 6$^{th}$, with the plane of the galaxy running horizontally and the Sun following the *Ecliptic* line. The star marks the current position of the Sun for this time of the year. The panels below show similar information, highlighting the difference between the expected positions of the angular distributions of recoils induced by WIMPs and solar $\nu$ at different times of the year: (b) February 26$^{th}$ and (c) September 6$^{th}$, the dates corresponding to the highest and lowest differences between the two recoil sources. Further details on these two original works can be found in [67] and [106], respectively.

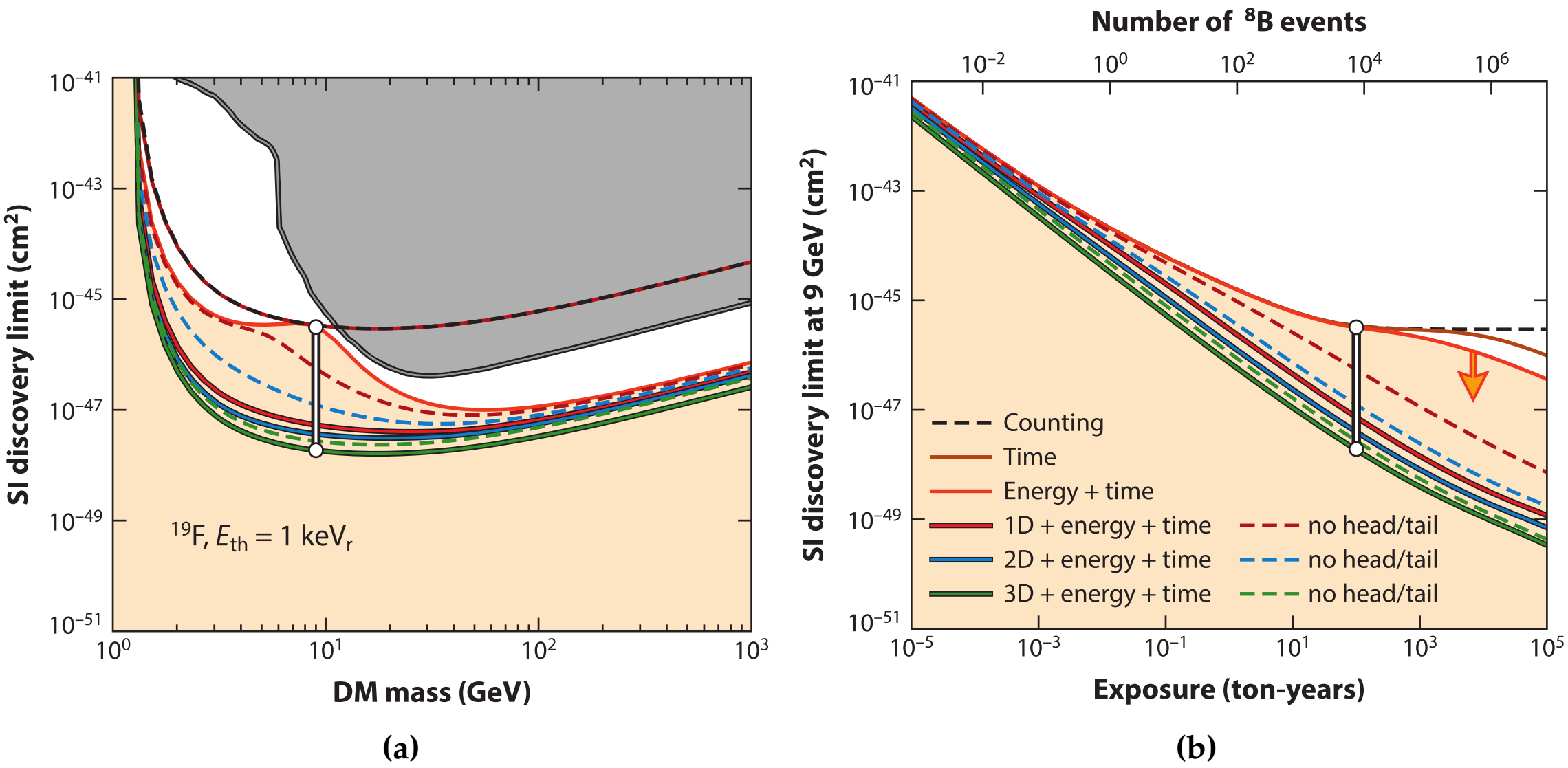

**Figure 1.15:** Effects of the readout and directionality capabilities of the detector – including vectorial reconstruction (1D, 2D, 3D) of the ionization track, as well as the inclusion or exclusion of HT recognition – when setting limits in the DM-SM interaction parameter space, in the presence of a neutrino background. In (a), the discovery limit for a fixed detector exposure is shown, while in (b), the same is shown for a fixed DM mass and varying exposures. The white lines in both plots represent the intersection of the two plots. For more details, see the original work in [67].

head-tail asymmetry is detectable.

Figure 1.15 shows how directional detectors can be compared to the current excluded limits (grey region) given enough exposure. Directional detectors can venture into the neutrino fog (yellow region), effectively mitigating the problem caused by solar neutrino-induced recoils in the detector through directional discrimination. To achieve this, the readout requirements include: an energy resolution better than 20%, HT recognition above 75%, and angular resolutions better than 30 $\mathrm{deg}^2$. Non-directional detectors could eventually enter the neutrino fog, although the exposure required to discriminate between solar and WIMP-induced signals is likely too high and not justifiable. Thus, directional information remains the best solution for exploring this region of the WIMP parameter space.

It is also visible in the left panel that when no head-tail recognition is assumed, the limits that can be imposed are considerably reduced, as it becomes much harder to identify the dark matter characteristic dipole effect, as shown in Figure 1.13. This highlights the importance of HT recognition for claiming discovery, particularly when only a small number of events are available. Indeed, in [113], it was shown that the number of events required to identify dark matter can decrease by a factor of 100 when HT recognition is achieved, compared to when no HT information is available, assuming a reasonable 20 keV energy threshold, perfect background discrimination, and a sulfur target.

The best-case scenario, as expected, involves a detector that can measure the energy, time, direction, 3D geometry, and head-tail asymmetry of the nuclear recoil. This would provide a full description of the ionization track, leading to a nearly complete understanding of the particle responsible for the interaction and potentially its origin. The ability to reconstruct the vectorial direction and 3D geometry of NRs is also crucial for detecting the daily oscillation in the NR event rate caused by Earth's rotation since, depending on the time of day, the recoil will shift axes in the detector's reference frame, as discussed in [109] and referred to in Section 1.3.2.3 and Figure 1.9.

For 3D, directional, and head-tail reconstruction, gas TPCs are one of the most promising techniques, as further discussed in Section 1.4.3. While the exposure of gas TPCs may potentially limit the technique, their ability to optimize the interaction medium to fulfill all necessary requirements is a significant advantage. By designing a gas mixture that is light enough to allow long tracks, where the 3D geometry can be reconstructed, and ensuring high granularity to infer both direction and HT asymmetry, these parameters and others can be effectively measured. As demonstrated, reconstructing these characteristics of the ionization track greatly enhances the potential for identifying a dark matter signal, particularly below the neutrino floor, where energy-only experiments are limited by the lack of directionality discrimination.

#### 1.4.2.2 Positive discovery of DM and halo characterization

The anisotropy in the angular distribution of nuclear recoils in the detector is one of the strongest features to be observed from the WIMP Dark Matter halo model, as it can be distinguished from other types of signals even with as few as tens of events [113]. As expected, this number will nonetheless depend on the detector's capabilities, especially regarding head-tail recognition, angular resolution, and energy threshold, as demonstrated earlier.

The power of directionality is shown not only in the reduced number of signals needed to distinguish a WIMP source from other background sources but also in its ability to serve as a unique observable for positively identifying and claiming the discovery of a dark matter signal. The ability to establish the galactic origin of the observed signals is a distinctive feature that provides a definitive and unequivocal identification of a DM signal, something that cannot be accomplished through the sole analysis of the DM energy spectrum.

Additionally, once (or *if*) a DM signal is positively identified, directional detectors will be further valuable in investigating the other properties of the DM halo. Specifically, these detectors could provide constraints on the physical properties of the DM halo, such as the WIMP velocity distribution and the matter profile shape, as these factors influence the rate of WIMP-induced nuclear recoils as a function of both energy and direction, as shown in Section 1.3.2. Moreover, a deeper understanding of the DM halo velocity profile could offer new insights into the history of the Milky Way's formation and improve the understanding of the remaining physical parameters of dark matter, such as WIMP mass and interaction cross-section [105]. Finally, other phenomena such as halo mergers and dark matter streams (localized DM structures) could also be explored once directional information is obtained [114, 115].

### 1.4.3 Current developments in Gas TPCs for directional searches

To identify the direction of a nuclear recoil and subsequently determine the original direction of the particle that induced it, a medium that allows the recoil track to considerably extend in space is necessary. In broad terms, a gaseous target at a pressure close to atmospheric can produce ionization tracks of $\mathcal{O}(1)$ $mm$ for nuclear recoil energies on the order of $keV_{nr}$. Gas TPCs are, therefore, a strong candidate for this purpose, as they can also be optimized and fine-tuned to detect specific types of events and energies, balancing crucial parameters such as track length, energy threshold, and detector exposure.

Besides gas TPCs, which generally perform recoil imaging, other techniques make use of the directional dependence of the detector response to perform DM directional searches. One such example involves *anisotropic crystals*, as in the ADAMO project [116], where a $ZnWO_4$ scintillating crystal is used. In this case, one of the crystal's symmetry axes has a different quenching factor, leading to a different signal depending on the recoil direction (detector directional dependence). Another approach uses aligned *carbon nanotubes* (CNTs), as seen in experiments like ANDROMEDA [117]. Here, CNTs are aligned parallel to each other, allowing recoils induced from WIMPs to either hit the nanotube walls or be channeled through the tubes if they come from a specific direction. In both techniques, the detector structure itself is sensitive to the recoil direction, and consequently, to the direction of the particle causing the recoil. *Nuclear emulsions* are another technique used for DM directional searches, in experiments such as NEWSdm [118]. In this approach, an emulsion of silver crystals is embedded within a gelatin binder. When a charged particle passes through, it leaves a 3D track in the silver grains, which becomes visible after a chemical development process. Similar to gas detectors, this track can be used to determine the original direction of the recoil, allowing for a directional analysis of WIMP-like events. These and other readout technologies for directional detection are discussed in more detail in [119].

Compared to these alternatives, gas TPCs (as detailed in Section 2.1) do not have a preferential directional axis for detection and are generally considered among the most suitable detector types for directional DM searches. This is due to their architecture and operating principles: a gas TPC is a three-dimensional detector with a sensitive volume (gas) enclosed between a cathode and anode, which provide an electric field capable of drifting ionized electrons, and maintained uniform by a field cage. The electric field in gas TPCs prevents the recombination of freed electrons from ionizing interactions, thus preserving the original (primary) ionization signal. This signal is then often amplified using one or more micro-pattern amplification systems, which increase the initial charge and create a measurable signal across a wide range of energies and particles. These am-plification systems are typically designed to offer high amplification gains and excellent spatial resolution, both of which are crucial for accurately reconstructing the direction and other parameters of low-energy nuclear recoils.

TPCs are intrinsically 3D detectors, making them an optimal choice for directional DM searches. Low energy thresholds are achievable due to the typically low W-value[3] of the gas mixtures used, which is on the order of tens of $eV$, and the amplification gains which can reach factors of up to $10^6$. The low density of gases (relative to liquids and solids) allows for longer tracks, which enhances the ability to identify the direction of the recoil and the intrinsic topology of energy release, i.e., the track's $dE/dx$. The $dE/dx$ profile can be used to discriminate between different particles – such as electron and nuclear recoils, alphas, and muons – while possibly also providing a measure of the head-tail asymmetry in nuclear recoils. While gas TPCs may face some exposure limitations due to the lower density of the detection medium compared to liquid or solid targets, these can be overcome by scaling the detector to large sizes. This has been supported by the approval for the construction of a gas TPC with up to 20.000 $m^3$ for neutrino physics [120], illustrating the feasibility of large gas TPCs for DM searches as well.

In the context of DM searches, the TPC technology has a particular downside related to the lack of a trigger due to the nature of DM signals. This makes it challenging to measure the position of the event in the longitudinal (or drift field) direction. In liquid and dual-phase TPCs, this is often overcome by recording the primary scintillation light, as it is typically large enough. However, many gas TPCs do have enough scintillation yield at low energies to detect the primary scintillation. To address this, experiments use the good 3D reconstruction of ionization tracks achievable in gases to *fit the diffusion* of the track: by measuring the diffusion of the ionized electrons that occurs during their drift, it is possible to determine the distance traveled by those electrons along the longitudinal axis, effectively retrieving the position of the event in the longitudinal plane, also called *absolute Z* [121]. This process is only possible when the granularity of the amplification and readout system is much smaller than the track itself, allowing for an accurate characterization of the electron diffusion.

The measurement of the absolute $Z$ position of an event is necessary to allow the fiducialization of the detector along the longitudinal direction, which provides a powerful tool for identifying and excluding potential background events, such as alpha particles and NRs emitted by the detector materials (especially the cathode). In addition to fitting the diffusion, several groups have successfully tested and implemented negative ion drift (NID) operation, which involves doping the gas mixture with an electronegative gas component. By doing so, the electronegative gas molecules absorb the ionized electrons, becoming negative ions. In this process, different species of negative ions can be created, and by measuring the time difference it takes these anions to reach the anode, it is possible to retrieve the event's absolute $Z$ position [122]. Moreover, negative ions are heavier than electrons, which reduces the observed diffusion during drift. This reduction in diffusion can further improve the 3D tracking capabilities and enhance the head-tail recognition of the ionization track. A more comprehensive discussion of this topic can be found in Chapter 7.

[3]The W-value of a gas (or any medium) refers to the average energy required to produce an electron-ion pair through ionization.

In the remainder of this section, a brief overview of some of the current gas TPC approaches for directional DM searches is provided as a point of reference for the studies and discussions presented in this thesis.

**DRIFT**

The Directional Recoil Identification From Tracks (DRIFT) experiment [123] is a leader in directional dark matter detection. It employs a 1 $m^3$ gaseous dual back-to-back TPC system, coupled with Multi-Wire Proportional Counters (MWPCs) for amplification and readout. The two TPCs share a 0.9 $\mu m$ aluminized-Mylar cathode, forming two separate sensitive volumes, each equipped with a MWPC anode. The readout plane consists of stainless steel wires placed orthogonally to another set of wires, enabling the registration of both $X$ and $Y$ coordinates. The detector's sensitivity to the directional head-tail asymmetry signature of nuclear recoil tracks was demonstrated with a $^{252}Cf$ neutron source down to 38 $keV_{nr}$ [107].

The TPC operates with a gas mixture of $CS_2$:$CF_4$:$O_2$ at a ratio of 30:10:1 Torr. This mixture operates in Negative Ion Drift mode, where $CS_2^-$ ions, rather than electrons, act as primary ionization charge, and drift towards the anode, minimizing diffusion and improving the tracking capabilities. The presence of trace amounts of $O_2$ leads to the formation of "minority charge carriers" (different species of anions) that drift at different velocities than the standard $CS_2^-$ ions, allowing the determination of the event's absolute $Z$ position by measuring the time each anion takes to reach the anode, as referred above. This method also enhances the overall spatial resolution of the detector, allowing for more precise directional reconstruction of nuclear recoil events.

An alpha-tagging system was also employed in the DRIFT-IId detector to identify and tag surface backgrounds from radon progeny recoils (RPRs), which can be mistaken for WIMP-induced NRs, as described in Section 1.3.5.2. This technique involves using a very thin central cathode that allows the alpha particle emitted during the RPR to enter the sensitive volume. The associated NR can then be identified and excluded from WIMP analysis. With this method, alongside their high ER rejection capabilities, DRIFT-IId achieved one of the best limits for spin-dependent WIMP-proton interactions among directional detectors, setting a limit of 0.16 pb for an 80 GeV WIMP mass and providing additional constraints on the interaction cross-section down to 9 GeV WIMP masses [124].

The results and techniques employed by the DRIFT collaboration have been used as motivation and scientific basis for many of the studies conducted in this thesis, especially regarding the identification of alpha backgrounds through their track length, and position and angle of emission.

### MIMAC

The MIMAC (MIcro-TPC MAtrix of Chambers) experiment operates a similar dual TPC approach, at the Modane Underground Laboratory in France. It uses a common 6 $\mu m$ aluminized Mylar cathode and pixelated Micromegas for amplification and readout. The target gas is a mixture of $CF_4$ with 28% $CHF_3$ and 2% $C_4H_{10}$, maintained at a low pressure of 50 $mbar$. The pixelated structure features 200 $\mu m$ wide pixels, connected by 256×256 readout strips, with an effective pitch of 424 $\mu m$. These strips are read at 50 $MHz$, providing the XY coordinates of events as a function of time. The third spatial coordinate, Z, can then be determined using the time profile of the track and the known drift velocity of electrons in the gas. MIMAC achieves a competitive energy threshold of 2 $keV_{ee}$, with a dynamic range that extends up to 62 $keV_{ee}$, limited by the saturation of its ADC channels [93].

MIMAC uses its 3D track reconstruction capabilities to discriminate between ERs and NRs. With a Boosted Decision Tree algorithm, the experiment achieved a rejection power of $10^5$ for ERs, while maintaining an NR efficiency of $86.49 \pm 0.17\%$ across the entire energy range. The experiment also successfully measured radon progeny recoils (RPRs, as discussed in Section 1.3.5.2), demonstrating its ability to identify and reconstruct these low-energy events in 3D [93]. The results obtained in this thesis in Chapter 6 within the topic of 3D reconstruction of RPRs, have also been compared to those obtained by the MIMAC experiment.

MIMAC has also demonstrated the ability to detect the motion of electrons (towards the anode) in the cathode, which allows for the determination of the absolute Z position of the ionization event by measuring the time difference between the cathode and anode signals, thereby allowing to fiducialize the sensitive volume. Furthermore, the MIMAC collaboration developed the COMIMAC facility [125], a tabletop ion and electron beam line coupled with a gas chamber featuring a Micromegas for charge amplification and strips for XY readout. This facility is used for measuring the quenching factor of various gas mixtures, a critical parameter in dark matter searches that quantifies the fraction of energy lost in nuclear recoils through ionization relative to the total energy, as discussed in Section 1.3.3. Results from quenching factor measurements conducted with this setup are available in [126, 127].

### DMTPC

The Dark Matter Time-Projection Chamber (DMTPC) [128] employs a TPC coupled with external optical sensors (CCDs and PMTs) together with charge readout. The detector operates using $CF_4$ gas at pressures ranging from 30 to 100 $Torr$. The electron signal amplification is carried out using mesh planes (ground and anode) spaced approximately 1 $mm$ apart, with a strong electric field applied between them. As the primary electrons undergo amplification, secondary scintillation light is generated, which is then captured by the CCD cameras and PMTs positioned outside the chamber. This external positioning of optical detectors helps prevent contamination of the target gas, while still allowing the collection of scintillation data for event reconstruction.

The total amount of light measured by the sensors in DMTPC provides an estimate of the event's energy, while the CCDs capture the 2D track pattern. Timing data from the PMTs, combined with the charge readout, are used to construct a preliminary 3D representation of the tracks. Additionally, the integrated charge collected on the ground and anode planes provides an independent energy measurement, which can improve the overall energy estimation. This technique resembles the one used in CYGNO detectors, although CYGNO employs CMOS sensors instead of CCDs, and operates with a gas mixture of $He:CF_4$ at atmospheric pressure.

From a 10-liter detector, the DMTPC experiment obtained initial constraints on the SD WIMP–proton cross-section [129] and demonstrated sensitivity to directionality, achieving a head-tail efficiency of approximately 75% at 200 keV [130]. While this experiment is quite similar to CYGNO and could serve as a comparison and validation for the techniques employed, the DMTPC collaboration has ceased its operations.

**CYGNUS proto-collaboration**

The CYGNUS proto-collaboration brings together many of the leading efforts in directional dark matter research. The goal is to create a multi-site directional DM observatory at the tonne-scale, which would also serve as a neutrino observatory [110]. By integrating different detectors from around the world, the directional signature becomes even stronger, as it could potentially be measured simultaneously in different lab reference frames, all pointing in the same direction in galactic coordinates, as discussed earlier in Section 1.3.2.3. The use of various detectors and different readout techniques, managed by different groups, further reduces systematics and bias related to data analysis, strengthening the sensitivity of the approach.

The ongoing research among CYGNUS collaborators focuses on lowering the energy threshold to $\mathcal{O}(1)$ $keV_{ee}$, while maintaining good directionality, head-tail recognition, and background rejection. The different groups are also testing various amplification and readout stages, making use of MPGDs, charge, and optical readouts, including MWPCs, PMTs, CCDs, and CMOS sensors. Additionally, different gas mixtures are being explored, along with the possibility of employing the NID operation mode. The CYGNO experiment plays an important role within this proto-collaboration, making significant contributions, particularly in the development and implementation of optical readout through the combined use of CMOS and PMTs, as well as optimization and testing of NID operation at atmospheric pressure.

A thorough feasibility study was published in [110], showing simulations of CYGNUS performances using different gas mixtures and readouts. The current plan is to use a gas mixture that includes $He$ and $SF_6$ in a TPC that operates at or close to atmospheric pressure. The first goal would be to build a detector of volume around $\mathcal{O}(10^2)$ $m^3$ which, given the simulations, would already be competitive for SD WIMP-proton cross-section measurements. The long-term vision is to develop a much larger detector, possibly $\mathcal{O}(10^5)$ $m^3$, which could be used to perform dark matter astronomy.

**CYGNO project**

The CYGNO (CYGNus module with Optical readout) project is developing a TPC that uses a $He{:}CF_4$ gas mixture in a 60:40 ratio, operated at atmospheric pressure. A triple GEM stack is employed for charge amplification, while the secondary scintillation light emitted during amplification is detected by the external optical readout, composed of CMOS and PMT sensors. By combining the XY information retrieved from the CMOS with the $\Delta Z$ from the PMTs, ionization events can be reconstructed in 3D, along with the topological information of the charge deposition. A detailed description of the CYGNO experiment is provided in Chapter 2.

The work presented in this thesis was developed within the CYGNO project, and one of the main topics of discussion and results is the capabilities of CYGNO to perform 3D reconstruction of ionization events using the combination of CMOS and PMT signals. With the information from these sensors, it is also possible to retrieve both the direction and sense of particles in CYGNO detectors, through the track's $dE/dx$ and HT asymmetry. As discussed throughout this chapter, the ability to reconstruct NRs in three dimensions, detect their initial direction, and recognize their head-tail asymmetry, greatly contributes to positively identifying a dark matter signal, as well as studying the properties of the DM WIMP halo.

Chapter 2

# The CYGNO/INITIUM project

The CYGNO/INITIUM project [131] represents a novel approach in the field of directional Dark Matter searches, aimed at developing a high-precision, optically read out, 3D tracking gaseous Time Projection Chamber. The experiment targets the detection of low-mass WIMPs ($0.5 - 50$ GeV) and, potentially, solar neutrino spectroscopy [132]. The detection medium is a gas mixture of $He:CF_4$ at atmospheric pressure, providing sensitivity to both spin-dependent (SD) and spin-independent (SI) interactions. The ultimate goal of the CYGNO/INITIUM project is to install an $O(1)$ $m^3$ demonstrator (Section 2.2.3) at the Laboratori Nazionali del Gran Sasso (LNGS) by 2025, serving as a proof of concept for the technology, performance, and scalability of the experimental approach, towards the realization of a $\mathcal{O}(30)$ $m^3$ experiment with competitive DM sensitivity.

In the CYGNO approach, the charge amplification stage consists of a stack of three Gas Electron Multipliers (GEMs) (Section 2.1.2), while the readout is achieved using both a scientific CMOS camera (Section 2.1.3.2) and Photomultiplier Tubes (PMTs) (Section 2.1.3.1), which record the light produced during the electron avalanche. This is made possible by the scintillation properties of the gas (Section 2.1.1), particularly $CF_4$, which emits scintillation light in the visible range. By combining the two-dimensional (XY) projection from the CMOS camera with the longitudinal track tilt ($\Delta Z$) reconstructed from the PMT signal, a 3D reconstruction of ionizing events can be achieved. Additionally, the high granularity of the camera allows for a detailed reconstruction of energy deposition along the particle's path, enabling the identification of topology (hence particle identification), direction, and head-to-tail discrimination.

The synergic INITIUM project (Section 2.3), an ERC Consolidator initiative, aims to develop a negative ion drift (NID) approach within the CYGNO optical readout framework. The NID approach enhances the spatial resolution of the technique, as ions experience lower diffusion during drift compared to electrons. Furthermore, by identifying different anion species, it is possible to determine the absolute position of events in the drift region, providing valuable information to reject background signals originating from the detector materials.

The CYGNO/INITIUM project consists of approximately 60 members, primarily based in Italy, Portugal, Brazil, and the United Kingdom. CYGNO is part of a larger proto-collaboration called CYGNUS [110], which aims to establish a multi-site directional Dark Matter observatory. This observatory seeks to test the DM-WIMP hypothesis beyond the neutrino floor by utilizing directional discrimination to detect neutrinos from known sources such as the Sun and supernovae.

This chapter provides an introduction to the CYGNO project, beginning with a description of the experimental technique (Section 2.1). It then outlines the detectors developed within the CYGNO timeline (Section 2.2), with particular emphasis on LIME (Section 2.2.2), as most of the studies presented in this thesis were conducted using it. The chapter then concludes with an overview of the INITIUM project (Section 2.3).

## 2.1 Experimental approach

The CYGNO experimental technique is based on a gas Time Projection Chamber (TPC) approach, using $He:CF_4$ gas in a 60:40 ratio, at atmospheric pressure and room temperature. A triple Gas Electron Multipliers (GEM) stack is employed simultaneously for charge amplification and light production. The optical readout system consists of both CMOS cameras and PMTs.

A TPC is a type of detector commonly used in particle physics, both in low- and high-energy experiments, designed to track the trajectories of charged particles with high precision. The basic operating principle of a TPC can be illustrated through Figure 2.1, which shows a typical CYGNO-like detector. A TPC generally consists of a volume filled with a fluid (gas or liquid), in which a trail of electron-ion pairs is created when an *incident particle* passes through. In some TPCs, such as the XENON experiment [75], the light produced during the primary ionization is also recorded, although this is not attempted in CYGNO due to the lower $CF_4$ light yield with respect to $Xe$, which prevents sensitivity to primary scintillation light at keV energies.

The charges created during ionization, depending on their polarity, will drift toward the cathode or anode under the influence of an electric field generated by the voltage difference between these two electrodes. As the electrons start reaching the anode, they are first accelerated and multiplied through an electron avalanche process to enhance the intensity of the primary signal. This amplification can be achieved using various methods, many of which rely on micro-pattern gas detectors (MPGDs)[1]. In CYGNO, this amplification is achieved through a triple GEM stack.

Depending on the readout, both charge and light from this multiplication process can be captured and measured. In CYGNO, the $He:CF_4$ gas scintillates, emitting photons that are then recorded by the CMOS cameras and PMTs. The topological information of the ionization track retrieved from these two sensors is then used to perform background rejection, 3D reconstruction, and determine the particle's initial direction and sense.

[1] MPGDs are a group of gaseous ionization detectors consisting of micro-structures with sub-millimeter distances. Examples of MPGDs include GEMs, micro-strips, and MicroMegas. For a detailed review of MPGDs, see [133].

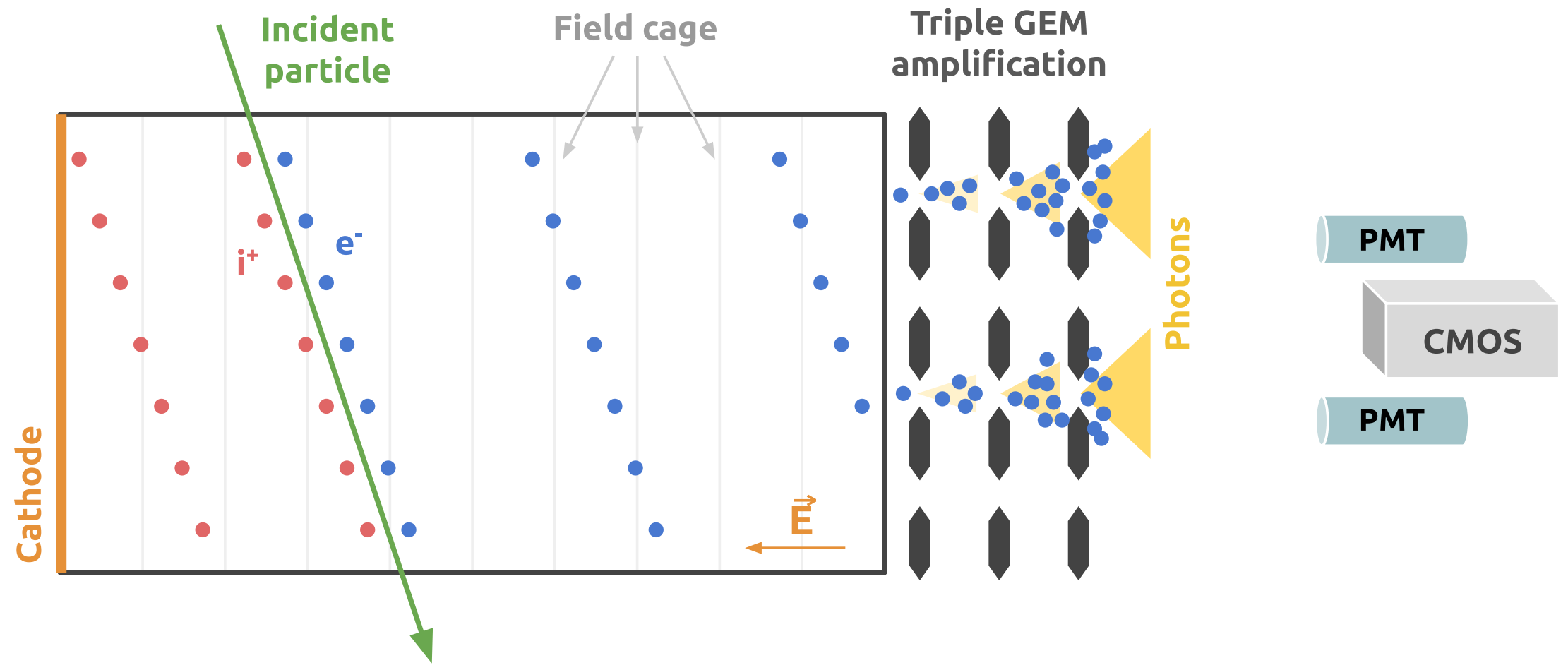

**Figure 2.1:** Working principle of a gaseous Time Projection Chamber. After an ionizing event, the freed electrons are guided towards the amplification planes through an electric field, where they are amplified and then readout. In the CYGNO case, the amplification is carried out through GEMs, and the readout performed with CMOS cameras and PMTs.

The CYGNO/INITIUM technique incorporates many features of a typical gaseous TPC used for DM searches (see also Section 1.4.3), while introducing several innovative advancements:

1. **$He$:$CF_4$($+SF_6$) gas mixture:** This multi-target gas mixture provides sensitivity to both SI and SD interactions, with the latter being possible by the high fluorine content and the presence of an unpaired proton in the fluorine nucleus. The presence of helium enhances sensitivity to low $\mathcal{O}(\text{GeV})$ WIMP masses due to its kinetic affinity, while the $CF_4$ scintillation in the visible range enables the use of very high-granularity CMOS sensors (see Section 2.1.3.2). The potential inclusion of $SF_6$ allows the operation of the TPC in negative ion mode (see Section 2.3).

   - **Atmospheric pressure:** Operating the TPC with this gas mixture at atmospheric pressure improves the total detector volume-to-target mass ratio with respect to typical directional DM search TPCs (see Section 1.4.3), increasing the sensitivity to dark matter interactions. The chosen pressure and components ratio provide a density similar to that of air, which, compared to liquid TPCs, allows particles to travel much longer paths, granting tracking and directional capabilities down to keV scale.

2. **GEM amplification:** The amplification is carried out using a stack of three 50 $\mu m$ GEMs, which are part of the MPGDs family of detectors. GEMs provide large amplification gains that, when stacked, can reach values of $\mathcal{O}(10^6)$. These gains, given the typical energy required to produce an electron-ion pair in a TPC (30 - 40 eV), help achieve the low energy thresholds of order $\mathcal{O}(\text{keV}_{nr})$ required for dark matter searches. High electron gains are particularly crucial for CYGNO, as they compensate for the reduction of light due to the small solid angle coverage of the CYGNO optical system (see Section 2.1.3.3).

Additionally, the high granularity of GEMs – on the order of $\mathcal{O}(10^2)$ $\mu m$ (see Section 2.1.2) – pairs well with the very high resolution of the CMOS sensors used in CYGNO. GEMs are also preferred for this application because they can be manufactured in large sizes, allowing a relatively straightforward scalability. They also do not require a dedicated electron "collection" plane, allowing the photons emitted at the GEM plane to travel freely toward the optical sensors.

3. **Optical readout:** The optical readout in CYGNO is carried out using both CMOS sensors and PMTs. CMOS sensors provide high-granularity readouts when paired with optimized optical systems. In CYGNO, the CMOS optical readout achieves a pixel resolution of $\mathcal{O}(150)$ $\mu m$, which allows to extract the topological and directional features from the 2D projection of ionization tracks obtained through this sensor. The fast response of the PMTs, on the other hand, allows for the accurate time-resolved detection of the scintillation light, i.e., the description of the particle's path in the drift direction (see Section 3.2).

   The combination of these two sensors provides the 3D reconstruction of ionizing events by correlating the XY information from the CMOS with the light longitudinal profile ($\Delta Z$) obtained from the PMTs [134] (see Chapter 3). Since the two sensors are decoupled, it is also possible to perform two independent measurements of the energy released, theoretically improving the energy resolution.

4. **Track topology:** By coupling the high-granularity CMOS camera with an appropriate optical system and readout, it is possible to study the track topology and its energy deposition profile with high precision. Different particles release energy at varying rates and patterns: for instance, alpha particles and nuclear recoils produce straight, short, and highly ionizing tracks, while electron recoils exhibit less dense tracks with variable lengths – ranging from a few millimeters for energies of $\mathcal{O}(keV)$ to tens of centimeters for energies above 100 $keV$ (see Figure 2.4). Therefore, understanding the ionization track's *topology* enables particle identification and, consequently, discrimination between electron and nuclear recoils.

   - **Directionality:** The charge profile along the ionization track, as seen by both the CMOS and PMT sensors, also encodes the track's direction and sense – that is, the full 3D vector. Alpha particles and electron recoils exhibit higher energy deposition at the end of their tracks, corresponding to the Bragg peak in the case of alphas, and to hard scatters or spiraling motion of electrons as they approach thermal energy. In contrast, nuclear recoils typically display a higher charge deposition at the beginning of their tracks, giving rise to the so-called head-tail asymmetry (see Section 1.4.1). These different characteristics can be used to infer the particle's initial direction.

- **Fiducialization:** With very high spatial resolution, it is also possible to study the track topology, specifically the width of the ionization cloud, to infer the position of the ionizing event along the drift direction. This is achieved through a fit to diffusion, in which the track width is analyzed and compared to the diffusion induced at the amplification stage and the transverse diffusion in the gas. This analysis is further explored with alpha particles in Section 5.2.3. The resulting information allows to effectively fiducialize the TPC and perform rejection of background events originating near the detector's boundaries.

The 3D topology of ionization tracks obtained with the two sensors can thus be used to discriminate between signal and background, while also providing insights into the initial direction of ionizing events and their position within the detector. These characteristics are all crucial for directional dark matter searches (for more details, see Section 1.4.2).

5. **Practical advantages:** The operation of a TPC filled with a gas at room temperature and atmospheric pressure gas also offers several practical advantages. Operating at near-atmospheric pressure reduces the need for vacuum or high-pressure sealed systems, while working at room temperature eliminates the requirement for cooling systems. These two factors simplify the implementation of the technology and facilitate the development and testing of various CYGNO prototypes and detectors. Moreover, under these conditions, the scaling of the experiment is more straightforward and cost-effective, particularly when compared to other cryogenic solutions currently used in DM searches (see Section 1.3.3). Additionally, these operating conditions provide flexibility, allowing for the adjustment and optimization of the gas components and the proportions of the mixture.

Given this general overview of the CYGNO experimental approach, the remainder of this section will focus on a more detailed discussion of several key aspects, specifically the gas mixture (Section 2.1.1), the triple GEM amplification (Section 2.1.2), and the CYGNO optical readout (Section 2.1.3), consisting of CMOS sensors and PMTs.

### 2.1.1 Gas mixture

The choice of gas mixture used in a TPC is critical to the experiment, as it directly influences the detector's response to ionizing radiation and impacts the overall physics reach. In CYGNO, the gas mixture used is $He:CF_4$ in a 60:40 ratio, operating at atmospheric pressure and room temperature. In this section, some of the most important gas properties – such as scintillation, diffusion, and range – will be discussed.

### Scintillation

Gas luminescence is a well-established process. When charged particles pass through a medium, they can deposit some of their energy in the form of atomic excitation, which often results in light emission (not necessarily in the visible range) as the molecules de-excite. The spectrum and intensity of the emitted light depend strongly on the properties of the medium.

Tetrafluoromethane, $CF_4$, is a well-known scintillation gas [135] that exhibits two primary peaks in its wavelength spectrum: one in the ultraviolet (UV) region, around 290 $nm$, and another in the visible region, around 620 $nm$. The emission in the visible region originates from the de-excitation of a Rydberg state of the neutral $CF_3^*$ molecule, which is produced through the fragmentation of $CF_4$ [136]. In contrast, the UV emission results from other dissociative ionizations of $CF_4$ [137]. The other component of the CYGNO gas mixture, helium, is a noble gas, making it chemically stable and inert, which is particularly useful for particle detectors. Helium is also known for generating significantly high electron gains when coupled with other gases [138].

When $He$ is mixed with $CF_4$, given its low density, it allows CYGNO detectors to operate at atmospheric pressure with a gas density close to that of air, providing a good balance between particle range in the gas and sensitivity (exposure) to WIMP interactions. The exact ratio between $He$ and $CF_4$ has been studied, particularly comparing the 60:40 and 70:30 $He:CF_4$ mixtures [139]. The 60:40 ratio was found to maximize the light yield – measured as photons detected per $keV$ of energy release – and improve the stability of the detector over long periods, while maintaining similar energy resolution in both cases.

For gases containing $CF_4$, the photon-to-electron ratio during the avalanche process is typically in the range of $\mathcal{O}(10^{-2} - 10^{-1})$ $ph/e^-$ [140,141], with the specific value for the $He:CF_4$ 60:40 gas mixture being 0.07 $ph/e^-$. The W-value for this mixture is estimated to be approximately 35 $eV$, derived from the W-value of $CF_4$ (34.3 $eV$ [142]) using the method outlined in [143]. The complete emission spectrum of the $He:CF_4$ 60:40 mixture employed in CYGNO detectors is shown in Figure 2.2 [136,144].

### Diffusion

Gas mixtures containing $CF_4$ are known to exhibit low electron diffusion due to the large scattering cross-section between electrons and $CF_4$ molecules [138], making this gas one of the best choices for imaging detectors, as it better preserves the track's topological characteristics. The properties of the CYGNO gas mixture at atmospheric pressure in terms of diffusion and drift velocity have been calculated using Garfield++ [2] [147,148], with the results shown in Figure 2.3.

Specifically, this gas mixture shows a drift velocity of approximately 5.6 $cm/\mu s$ at 800 $V/cm$, which is the typical operating drift field for CYGNO detectors. For this field, the diffusion coefficients are on the order of 125 $\mu m/\sqrt{cm}$ for transverse diffusion and 105 $\mu m/\sqrt{cm}$ for longitudinal diffusion. In comparison, the Ar:$CO_2$ gas mixture, used in similar approaches, has a diffusion coefficient of 250 $\mu m/\sqrt{cm}$ [149]. The possible addition of $SF_6$ to the gas mixture is expected to significantly reduce the diffusion, due to the much higher mass of negative ions compared to electrons, as discussed in more detail in Chapter 7.

[2] Garfield++ is a toolkit for detailed simulation of gas and semiconductor particle detectors. Its primary application area is in MPGDs. Notable works involving Garfield can be found in [145,146].

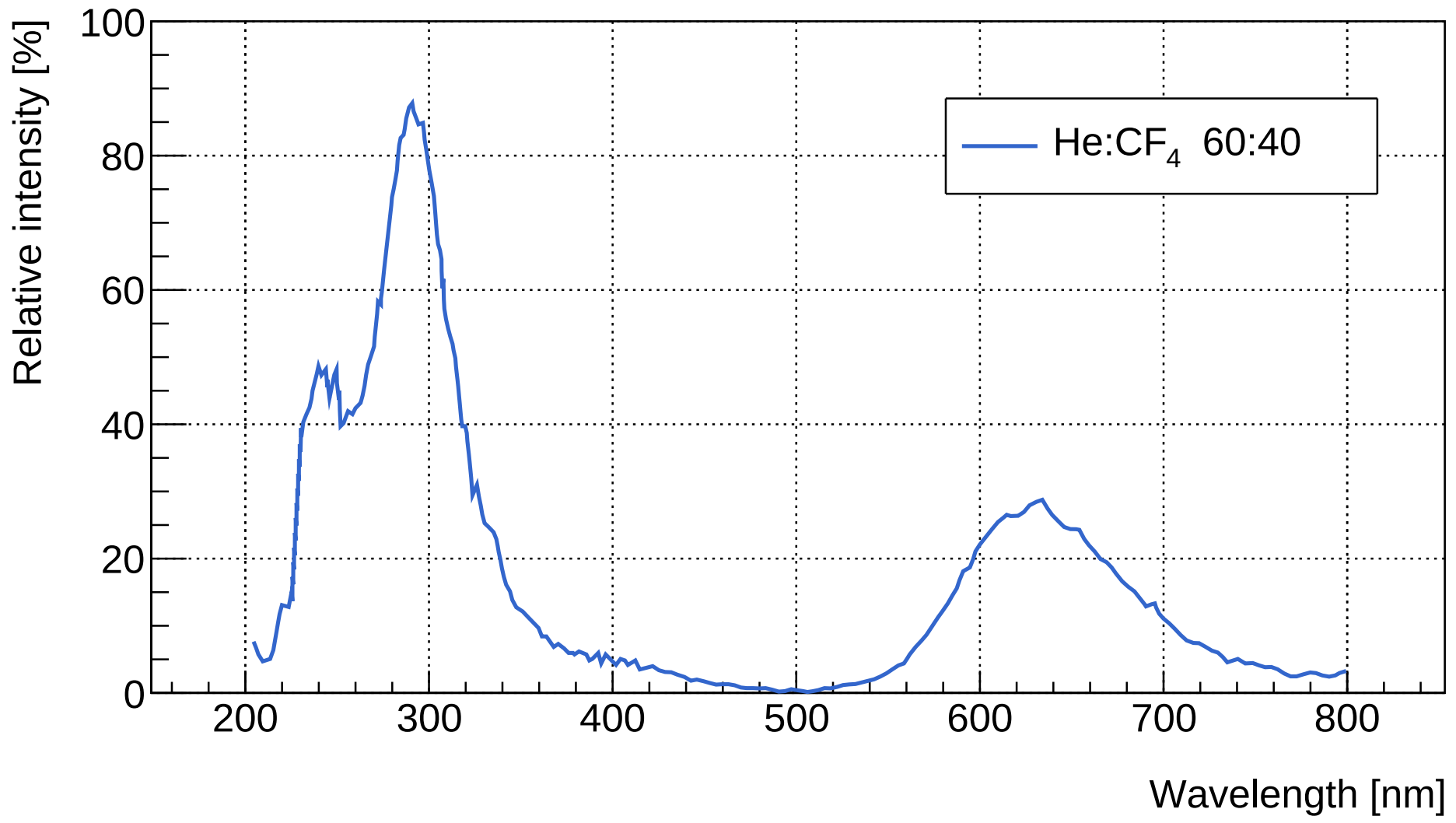

**Figure 2.2:** Emission spectrum of the CYGNO gas mixture, He:$CF_4$ 60:40, at atmospheric pressure and room temperature. Retrieved from [136].

### Range

As mentioned, using a detection medium with a density close to that of air allows for long ionization tracks, even at energies as low as a few keV. In this context, the 3D ranges – defined as distance between the production and absorption points – of electron and helium recoils were simulated for this gas mixture using both GEANT4 and SRIM software, as a function of their kinetic energy. The results of these simulations are shown in Figure 2.4. From these results, for the nuclear recoil energies expected in CYGNO detectors (up to around 100 $keV_{nr}$), helium recoils show ranges below 1 mm, reflecting their larger energy deposition ($dE/dx$) compared to electrons. On the other hand, electron recoils exhibit a sparse energy deposition, leading to 3D ranges that extend up to nearly 10 cm for the same range of energies. These differences in energy deposition, track length, and other topological properties are the basis of the particle identification strategies used to separate and analyze different types of particle interactions in CYGNO detectors, as discussed in more detail throughout Chapter 3.

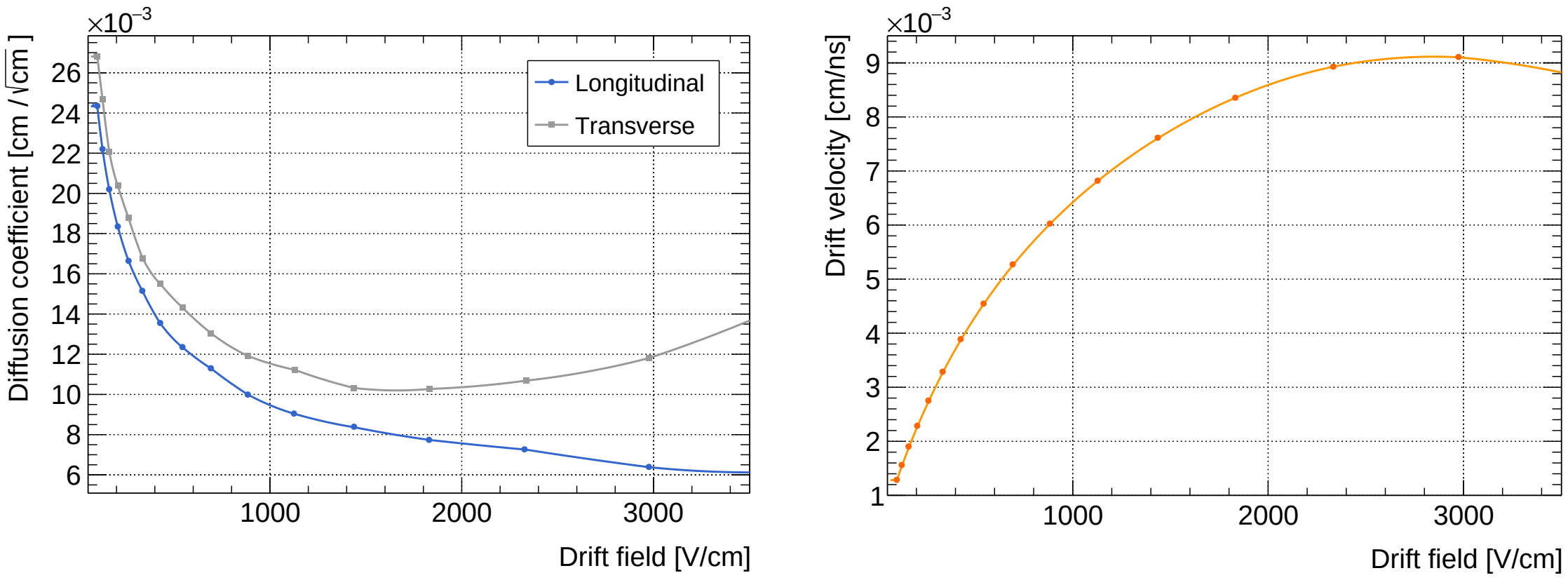

**(a)** **(b)**

**Figure 2.3:** Properties of the CYGNO gas mixture, He:$CF_4$ 60:40, highlighting the (a) transverse and longitudinal diffusion coefficients, and (b) drift velocity in the typical range of drift fields. Values obtained from Garfield++ simulations.

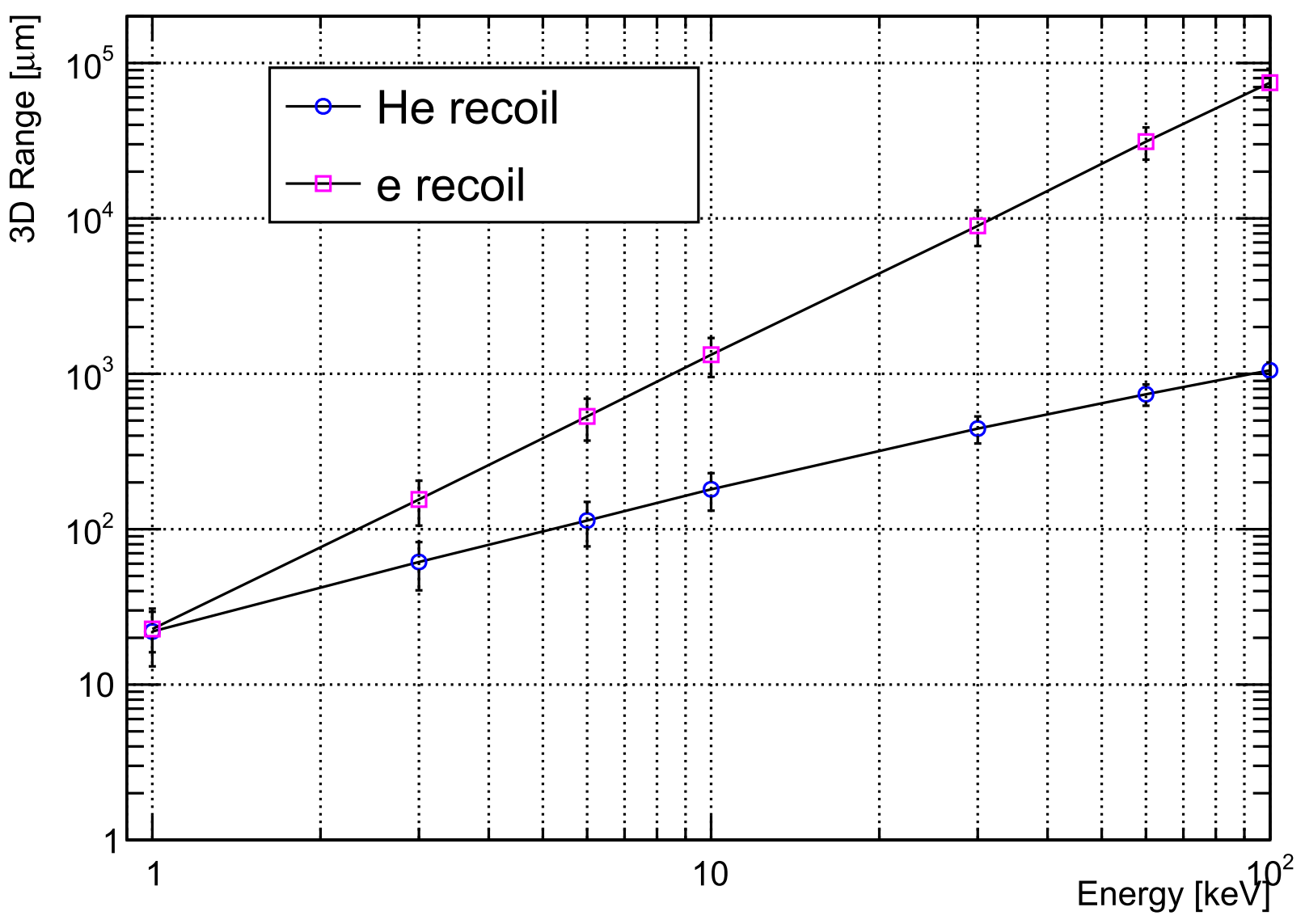

**Figure 2.4:** Simulated average 3D range of He and electron recoils in He:$CF_4$ 60:40, as a function of the kinetic energy of the particle. Retrieved from [131].

#### In the context of DM searches

In the context of Dark Matter searches, the choice of a light target with an atomic number smaller than 10, combined with a low energy threshold of around 1 keV, enables CYGNO to explore the region of DM masses below 10 $\mathrm{GeV/c^2}$ in the DM parameter space, an area that remains largely unexplored to this day, as shown in Figure 1.11. This is possible because the energy transfer in elastic scattering is maximized when the two particles involved have the same atomic number. Thus, Helium (A = 4) optimizes sensitivity to WIMP masses in the few $\mathrm{GeV/c^2}$ mass range. Furthermore, since He is a small atom, it produces longer recoils compared to heavier atoms, making it easier to identify the direction and sense of the track.

The addition of fluorine to the mixture grants sensitivity to spin-dependent coupling due to its asymmetry in the number of protons and neutrons. Furthermore, fluorine has one of the highest expectation values for proton spin within its nucleus, which increases the interaction cross-section (see Section 1.3.1.2).

With this choice of gas mixture, CYGNO becomes one of the few experiments in the field sensitive to both SI and SD couplings at WIMP masses below 10 $\mathrm{GeV/c^2}$. This is illustrated later in Figure 2.29, where the dark matter sensitivity of CYGNO-30 – a 30 $\mathrm{m^3}$ CYGNO-like detector (Section 2.2.4) – is shown alongside the current and projected limits of other similar experiments.

### 2.1.2 Triple GEM amplification

The amplification of primary ionization charges in CYGNO detectors is achieved using a stack of three 50 $\mu$m thick gas electron multipliers. These structures provide very high electron gains, compensating for the signal loss caused by the limited solid angle coverage of the CYGNO optical readout (approximately $\mathcal{O}(10^{-4})$, as discussed in Section 2.1.3.3), while maintaining relatively high spatial granularity.

GEMs were introduced to the particle physics community by Fabio Sauli in 1997 [150] as a detection and amplification device, offering advantages over the micro-strip detectors by enabling more stable operation over longer periods of time [138]. GEMs consist of a 50 − 125 $\mu$m thick Kapton foil, copper-clad on both sides. More recent developments have demonstrated that GEMs with thicknesses extending up to the millimeter scale can also be created and used in particle physics detectors [151]. Through photolithography and acid etching processes (for 50 $\mu$m GEMs [152]), or drilling (for thicker GEMs [153]), holes are created and geometrically distributed across the surface of the GEM and through the Kapton foil. Figure 2.5b shows a microscopic view of a thin (50 $\mu$m thick) GEM, as typically used in CYGNO detectors. For these, the hole diameter is typically 70 $\mu$m, with a pitch of 140 $\mu$m. The etching process results in a double-conical shape, with a narrower diameter at the center, as shown in Figure 2.5a.

When an electrical potential difference is applied between the two copper layers, a strong electric field (ranging from 30 to 100 kV/cm) is generated within the GEM holes, as shown in Figure 2.5a. This field makes the electrons crossing the holes initiate an

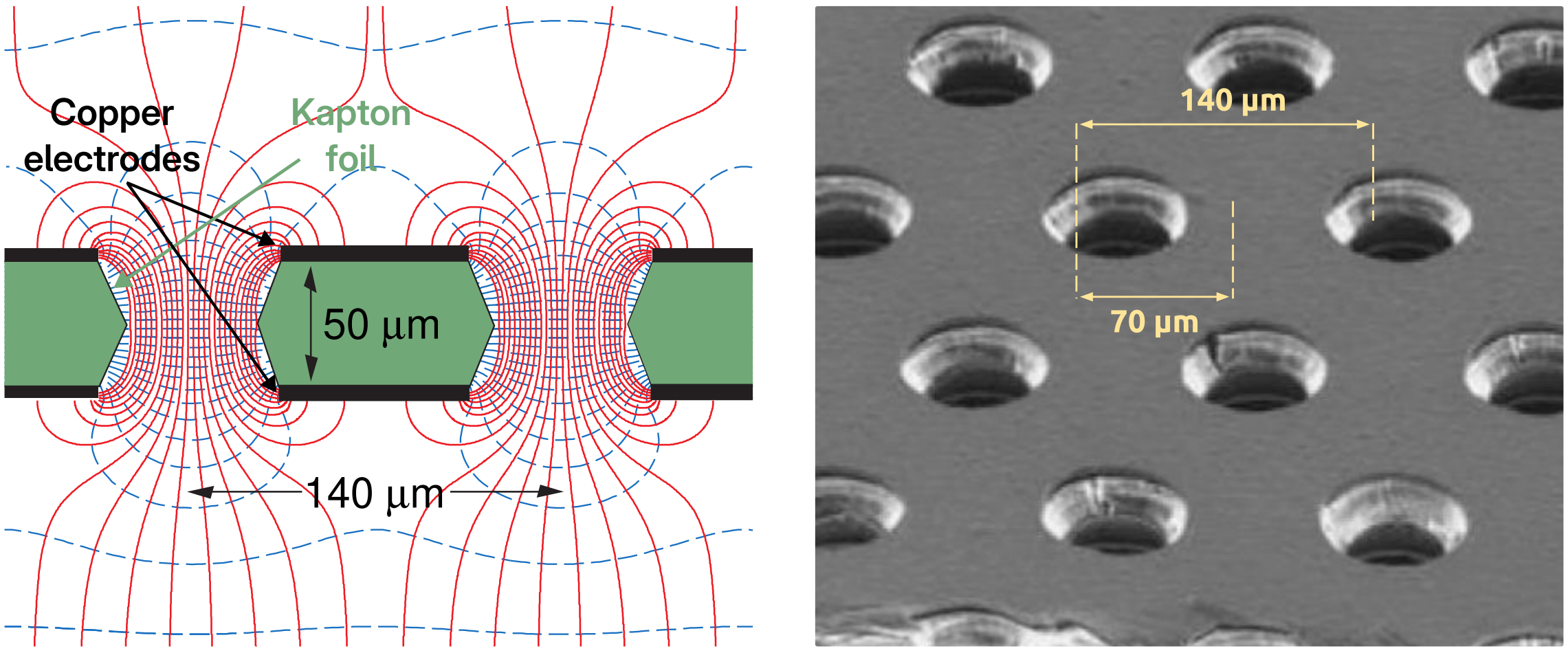

**(a)** **(b)**

**Figure 2.5:** Example of the structure of a thin GEM, illustrating a (a) a GEM cross-section and the respective field lines obtained when a voltage difference is applied between the top and bottom electrodes [149], and (b) an electron microscope picture of a GEM surface, showing the different holes and pitches involved [138].

avalanche, amplifying the initial electron signal with gains on the order of $\mathcal{O}(10^1 - 10^2)$ per GEM. By stacking multiple GEMs together, a net gain of approximately $\mathcal{O}(10^4 - 10^6)$ can be achieved [154].

GEMs were selected for the CYGNO amplification stage due to their scalability, transparency, high granularity, and high achievable gains. The ongoing interest in optimizing these structures by CYGNO researchers has driven several R&D studies, with some of the latest being reported in G. Dho's PhD thesis [104], and published in [155].

### 2.1.3 Optical Readout

The concept of optical readout involves detecting the light produced in particle detectors as interactions occur within the medium. Light is generated due to the de-excitation of medium molecules, which are initially excited by the passage and/or interaction of particles. This light is referred to as *primary scintillation* if it results from the initial interaction, or *secondary scintillation* if produced during the charge amplification process.

In CYGNO detectors, only the secondary scintillation produced at the GEMs during electron multiplication is recorded. This light is proportional to the amount of primary ionization charges, with the proportionality being determined by the gain of the amplification stage and the photon-to-electron conversion ratio of the medium. The primary scintillation light is not recorded due to its very low intensity and the geometrical constraints of the detector design. In [154], an alternative detector layout was tested in which the PMT was placed behind a transparent cathode to observe the primary scintillation. In this configuration, the primary light was indeed detectable, but only for energy deposits on the order of $\mathcal{O}(\text{MeV})$ and above, making it unfeasible for dark mat-ter searches, though potentially useful for other physics cases. Additionally, this layout did not allow the construction of the projected back-to-back TPC design envisioned for the CYGNO-04 detector (see Section 2.2.3). For these two reasons, the focus was placed only on recording of the secondary light emission.

Optical readout offers an alternative to traditional charge readout, providing several advantages, such as significantly higher signal-to-noise ratios, thanks to the very low noise levels of modern optical sensors. Charge-sensitive readouts, on the other hand, are more susceptible to electronic noise caused by external electromagnetic interference, or poor grounding. This often leads to the need of the development of more complex electronic circuits and/or advanced analysis techniques to address such issues [156,157]. In contrast, optical sensors are generally easier to integrate and less affected by these types of noise. In the context of a TPC, optical sensors can typically be placed outside the sensitive volume, minimizing the risk of contaminating the detection medium. This, however, comes at the cost of reducing the solid angle coverage as the distance between the sensors and the light emission region increases (see Section 2.1.3.3). This limitation, in the case of CYGNO, is compensated by the use of high-gain GEM structures (see Section 2.1.2).

As a directional gas detector, CYGNO's objective is to measure the direction, sense, and describe the energy deposition and 3D geometry of ionization tracks. This is achieved through its high-granularity and fast optical readout, using a scientific CMOS camera and photomultipliers. The following subsections provide a brief technical description of these two sensors to contextualize the operating characteristics of CYGNO's optical readout.

#### 2.1.3.1 PMT

Photomultiplier tubes (PMTs) are photon detectors that were introduced in the first half of the 20th century and have since been widely used across various scientific fields, including physics. One of their most notable applications was in the Kamiokande neutrino detector, which led to the Nobel Prize in Physics for its advancements in neutrino research [158].

PMTs are highly sensitive devices designed to detect very low levels of light. They operate by converting incoming photons into photoelectrons, which then undergo a multiplication process, resulting in a detectable signal at the end of the process. Using Figure 2.6 as a reference, the components and operating principle of a PMT can be described as follows:

1. **Photon detection**: When photons strike the *photocathode* deposited on the PMT entry window, they cause electrons to be ejected via the photoelectric effect. These photoelectrons are then directed towards the next stage, the electron multiplier, by the *focusing electrode*.

   The materials used for the entrance window and photocathode determine the quantum efficiency (QE) of the PMT at different wavelengths. Common window materials include borosilicate glass, which is suitable for general applications and offers good transmittance for visible light, but has limited sensitivity to UV light. Quartz windows, on the other hand, are used when UV light detection is necessary, as they provide better transparency in this wavelength range. For typical PMTs, the quantum efficiency can range from around 20% to 40% at peak sensitivity.

The PMTs used in CYGNO detectors are the Hamamatsu R7378 PMTs [159], which feature a bialkali borosilicate photocathode and 22 $mm$ diameter window (*faceplate*). The QE of these PMTs is later shown in Figure 2.9, along with the CMOS sensitivity and gas emission spectrum.

2. **Electron multiplication**: The electron multiplier consists of multiple electrodes, known as *dynodes*, which are maintained at progressively higher potentials relative to the photocathode. The primary electrons, guided by the focusing electrode, arrive at the first dynode with an approximate energy of 100 $eV$. Upon impinging on it, the primary electron causes the emission of several *secondary electrons*. These are then accelerated towards the next dynode, where they generate additional secondary electrons. This process continues through each dynode in the chain, creating a cascade effect that exponentially increases the number of electrons at each stage.

   The gain in this multiplication process depends on the PMT and can generally be parameterized as [160]:

$$G \propto (\Delta V)^{\alpha \cdot N} \tag{2.1}$$

   where $\Delta V$ is the voltage difference applied between each consecutive dynode, $\alpha$ is a multiplication coefficient dependent on the dynode material, and $N$ is the number of dynodes. Typically, the gains achieved are of the order of $\mathcal{O}(10^5 - 10^8)$ per electron, with the number of dynodes typically ranging between 9 and 12.

3. **Signal generation**: When the electron avalanche reaches the anode, a current pulse is generated. This pulse can then be converted into a voltage (drop) by passing it through a resistor, and subsequently read out and recorded by an oscilloscope or a digitizer. PMTs offer excellent time resolutions, with rise and fall times on the order of $< 5$ $ns$, allowing precise timing measurements for photon detection.

There are two main configurations of PMTs: head-on PMTs, where the photosensitive surface (*faceplate* in Figure 2.6) is located at the front, and side-on PMTs, where the photosensitive surface is positioned on the side. In the case of CYGNO detectors, since the light is produced at the GEM plane and always emitted from the same direction, entering almost parallel to the PMT surface, the PMTs used are of the head-on type and are positioned facing the GEMs. A schematic of this configuration can be seen, for instance, for the LIME detector in Figure 2.17a.

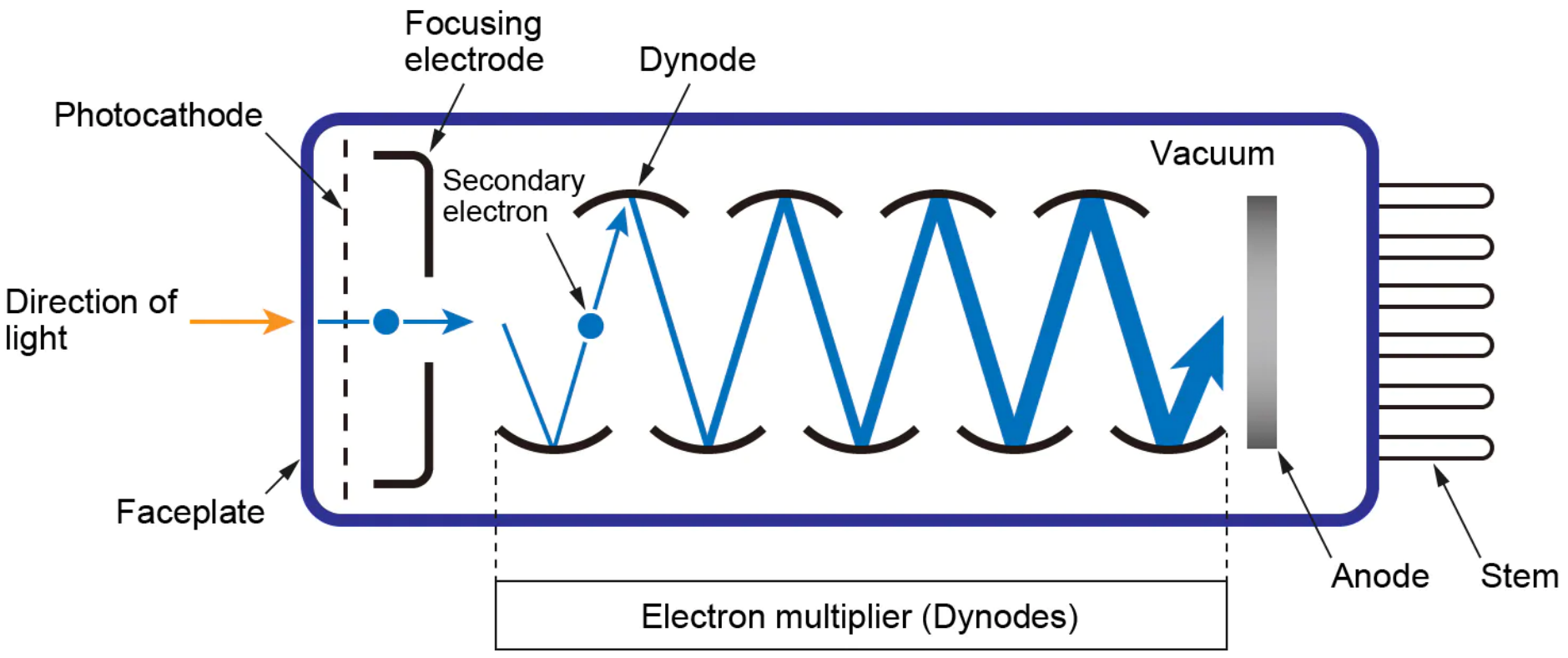

**Figure 2.6:** Schematic of the working principle of a PMT. The light impinges on the photocathode, which releases electrons inside the PMT. These are then accelerated towards the dynodes, and multiply in each one, leading to a total gain of the order of $\mathcal{O}(10^5 - 10^8)$. Figure taken from [161].

In the context of TPCs, the total charge collected by the PMTs is proportional to the light emitted, and thus to the energy deposited by the particle. This charge is also dependent on the solid angle between the emitted light and the PMT position. In fact, when multiple PMTs are used to image a single area, the relationship between the light observed by each PMT can be used to reconstruct the position of the light emission in that same area, as further illustrated in Section 3.2.3. In CYGNO detectors, this information is used to compare the position of an ionizing event as seen by the CMOS and by the PMTs, which in turn is used to match the information retrieved from both sensors, as explained in more detail in Section 3.3.

In TPCs, the PMT signal reflects the development of the ionization track along the drift direction by encoding the time of arrival of electrons from the track at the amplification stage. As particles traverse the gas and ionize it, the generated electrons drift towards the anode, each arriving at a distinct time based on their distance from it. The light produced by these electrons at the amplification region follows this arrival time pattern, leading to multiple light peaks in the PMT waveform, therefore representing the particle's path in the drift direction. In this context, the proximity of these peaks indicates, instead, the electron density in the ionization track, which corresponds to the particle's $dE/dx$.

The analysis of PMT signals in CYGNO detectors is a core topic of this thesis. A more detailed discussion of the PMT signal characteristics is provided in Section 3.2, where examples of these signals and a description of the variables developed for analyzing PMT waveforms are also presented. The information extracted from these signals is subsequently used in Chapter 5 to determine the traveled path and direction of alpha particles along the drift plane.

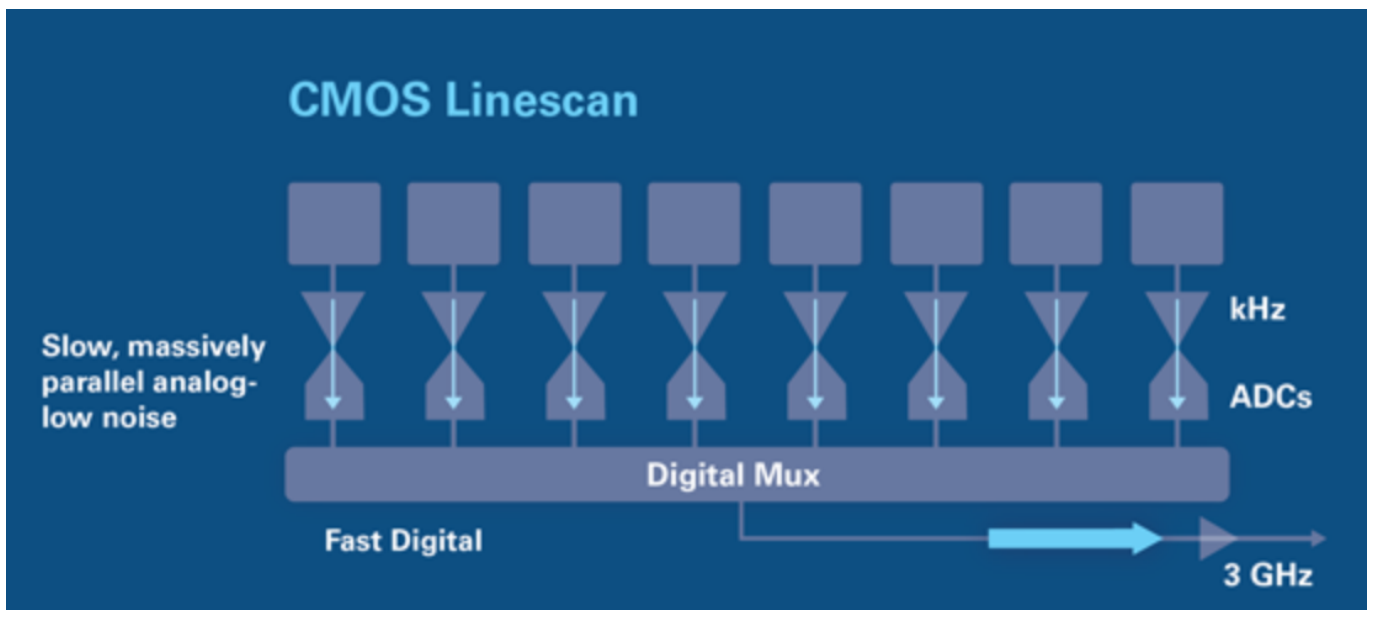

**Figure 2.7:** Operating principle of a CMOS sensor, illustrating how the charge from each individual pixel is independently amplified and read out in parallel. Adapted from [164].

#### 2.1.3.2 CMOS

Scientific CMOS (sCMOS) cameras are advanced imaging devices designed to convert light into electronic signals. They are widely utilized in scientific research due to their high sensitivity, low noise, and broad dynamic range.

CMOS sensors function by converting incoming photons into electrical signals at the pixel level. Each pixel contains a photodiode that generates a charge when exposed to light through the photoelectric effect. This charge is then stored and converted into a voltage by an individual amplifier integrated within each pixel, as shown in Figure 2.7. CMOS sensors also often integrate noise-correction and digitization circuits, further enhancing performance. This integrated circuitry design minimizes electronic noise and allows for a massively parallel readout from each pixel, resulting in high total bandwidth and frame rates. The quantum efficiency of CMOS-based cameras typically ranges from 50% to 95%, depending on the model and the wavelength of the incident light. Furthermore, these sensors offer a wide dynamic range, allowing for the detection of both very dim and bright light sources without saturating the sensor. These characteristics make CMOS cameras particularly well-suited for scientific imaging applications, such as fluorescence microscopy and astronomy [162, 163], where high sensitivity and minimal noise are critical.

In CYGNO detectors, various Hamamatsu sCMOS camera models have been tested and characterized. As new and more optimized sensors are released by the company, CYGNO often evaluates their suitability in the framework of optical readout and DM searches. Specifically, CYGNO has tested three prominent Hamamatsu CMOS camera models: ORCA-Flash, ORCA-Fusion, and ORCA-Quest. A summary of their key characteristics relevant to CYGNO is presented in Table 2.1.

The models, organized by release date – ORCA-Flash being the oldest and ORCA-Quest the most recent – demonstrate progressive improvements in several aspects with each generation. In particular, the combination of larger sensor dimensions, a higher number of pixels, and smaller pixel sizes enhances the effective imaging resolution achievable in each of these cameras.

The progression in camera performance regarding readout noise is also notable. Fig-

**Table 2.1:** Summary of the key features of the three CMOS cameras tested in CYGNO detectors, developed by Hamamatsu. With each successive generation (top to bottom), the cameras show improvements across various parameters, including pixel size, sensor dimensions, and RMS noise [165].

| **Model** | **Resolution [# pixels]** | **Pixel size [μm]** | **Sensor size [$mm^2$]** | **Noise RMS [$e^-$]** | **Frame rate [Hz]** |
|---|---|---|---|---|---|
| ORCA Flash | 2048 × 2048 | 6.5 × 6.5 | 13.312 × 13.312 | 1.4 | 30 |
| ORCA Fusion | 2304 × 2304 | 6.5 × 6.5 | 14.976 × 14.976 | 0.7 | 5.4 |
| ORCA Quest | 4096 × 2304 | 4.6 × 4.6 | 18.841 × 10.598 | 0.27 | 5 |

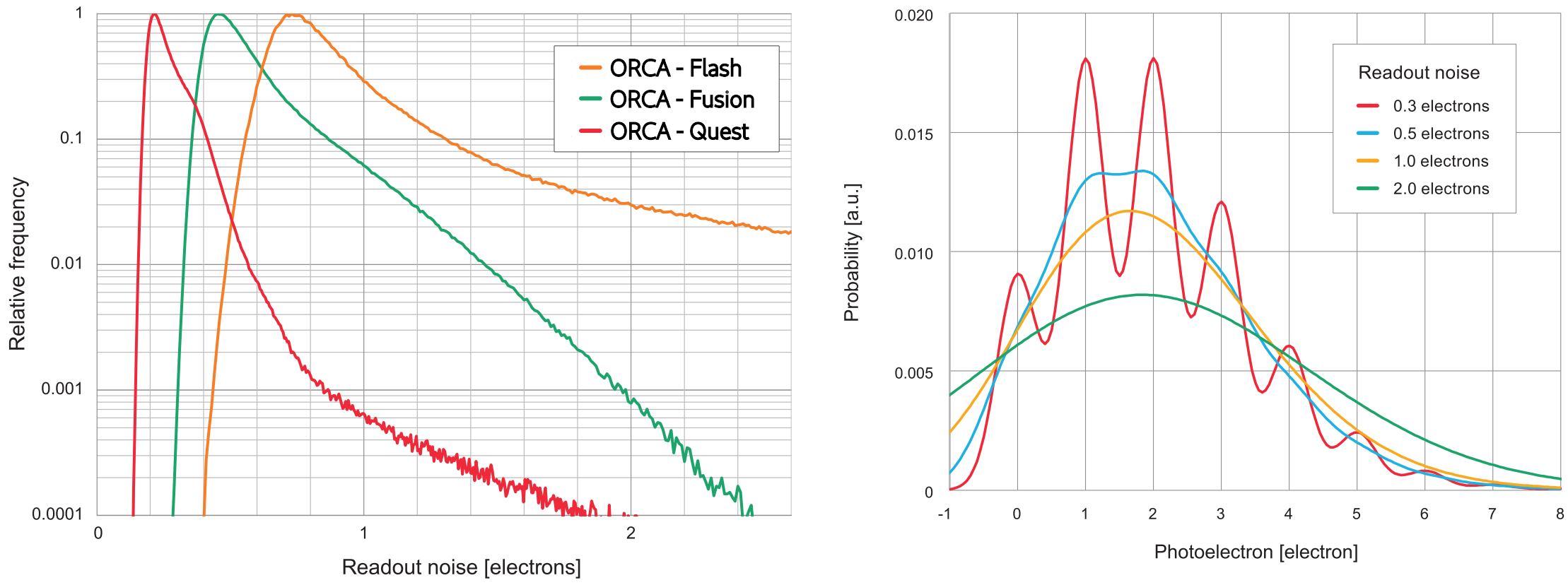

**(a)** **(b)**

**Figure 2.8:** Hamamatsu camera noise characteristics: In (a), the readout noise for different camera models is shown. In (b), the performance for single-photon readout at various noise levels is illustrated. At 0.3 electrons RMS, single photon counting becomes possible. Adapted from [166].

ure 2.8a illustrates the readout noise for each of the mentioned camera models. As shown, for each subsequent model, the noise amplitude decreases, with the peak value lowering in each iteration, reaching approximately 0.3 electrons RMS for the highest-performing model (Quest). With such a low noise level, single photoelectron counting becomes possible. This is demonstrated in Figure 2.8b, which shows a simulation of the photoelectron probability distribution at different noise levels, assuming an average of 2 photoelectrons generated per pixel. As seen, higher noise levels result in a blurrier distribution, and only at noise levels of 0.3 electrons RMS is it possible to reliably identify the number of photoelectrons generated, thus allowing for single photoelectron resolving [166, 167].

The frame rates listed in Table 2.1 correspond to the cameras' *slow scan modes*, which achieve minimal noise levels at the expense of reduced frame rates. This slower speed is not a significant drawback for CYGNO, as the event rates expected in underground dark matter searches are several orders of magnitude lower than these frame rates.

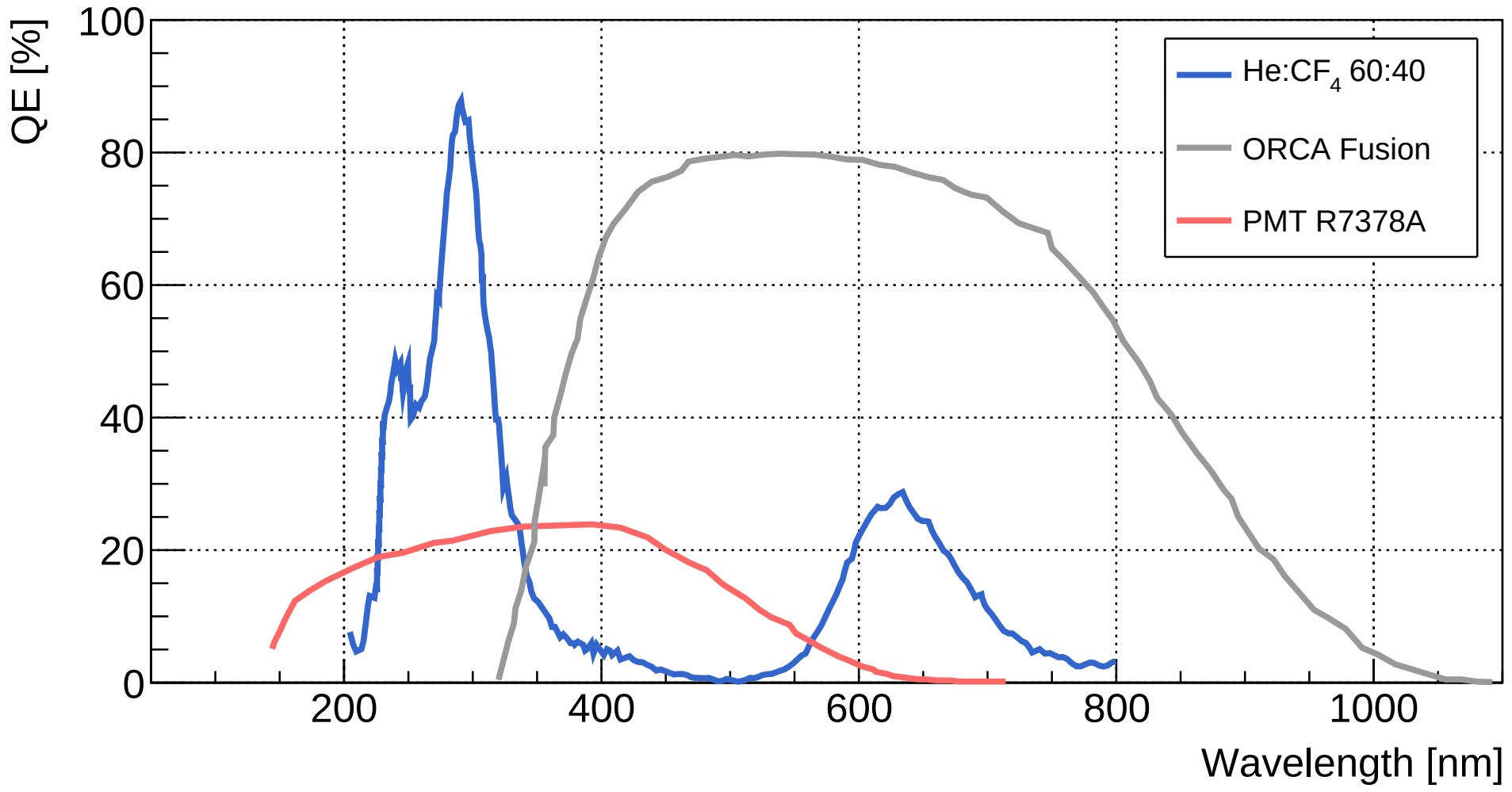

**Figure 2.9:** Normalized emission spectrum of He:$CF_4$ 60:40 (adapted from [136]) and the quantum efficiencies of the optical sensors used in CYGNO/INITIUM detectors: the ORCA-Fusion CMOS and R7378 PMTs, both from Hamamatsu.

The CYGNO/INITIUM results discussed in this thesis were obtained using the Hamamatsu ORCA-Fusion, the second-generation camera, as the most recent model had not yet been released. The Fusion (and the other models) offers a quantum efficiency which aligns well with the emission spectrum of the CYGNO/INITIUM gas mixture (see Section 2.1.1). The quantum efficiencies of the sensors used in CYGNO detectors – the ORCA-Fusion CMOS and the R7378 PMTs (both from Hamamatsu) – are shown in Figure 2.9, along with the gas emission spectrum. The excellent match between the quantum efficiency of these sensors, combined with their low noise levels and high granularity, made them an ideal choice for implementation in CYGNO detectors.

#### 2.1.3.3 Optical system

When coupled with an optical system, the centimeter-scale CMOS sensors can image very large areas. In fact, the size of the sampled area and the image granularity result from the interplay between the sensor's physical dimensions, pixel size, and the characteristics of the optical system. These parameters can be optimized to meet specific performance goals. In this section, the optical system used in CYGNO/INITIUM detectors and its main characteristics are described.

In CYGNO/INITIUM detectors, the light source is the scintillation light emitted at the GEM plane during the electron amplification process, and the receiver is the CMOS sensor, with the lens serving as the intermediary between the two. The optical system in CYGNO detectors can be generally described using the diagram shown in Figure 2.10.

This sketch presents the optical system along an axis connecting the object/source (*OP*), the sensor/receiver (*SP*), and the lens (region between *EP* and *XP*) [168]. It illustrates the operation of a generic thick lens, showing how photons in the yellow region emitted by the source are focused onto the sensor by the lens. The different components of this optical system and their corresponding elements in CYGNO detectors can be summarized as follows:

- *OP* is the object plane from which the light is emitted, i.e., the GEMs.
- *SP* is the sensor plane, where the CMOS photosensor is located.
- *EP* and *XP* are, respectively, the entrance and exit planes of the lens attached to the camera.
- *H* and *H′* represent the hyperfocal planes, corresponding to the position of a thin lens with the same behavior as the thick one being considered here.
- *D* represents the radius of the lens aperture, meaning the area through which photons can enter.
- *dA* is the infinitesimal area from where photons are emitted, and *u* is the tangent of the opening angle.
- *s* is the distance between *dA* and the hyperfocal plane *H*, while *s′* is the distance between the receiver (sensor) and *H′*.
- $f_F$ is the focal distance, with *F* being the focal point. In the general case, where the refractive indexes of the media before and after the lens are equal, then $F' = F$ and $f'_R = f_F = f$.

Using this schematic, and assuming that the planes formed by the GEMs (*OP*), the lens (*EP* and *XP*), and the CMOS sensor (*SP*) are parallel, the flux of photons reaching the lens ($\Phi$) from an infinitesimal area dA can be calculated as [168]:

$$\Phi = \pi \cdot L \cdot u^2 \cdot dA \tag{2.2}$$

where L is the luminosity of the source. The fraction of the solid angle covered by the lens, $\Omega_f$, can be calculated by dividing the flux of photons by the emission solid angle. Since the emission solid angle is $4\pi$ steradians, we have:

$$\Omega_f = \frac{\Phi}{LdA} \cdot \frac{1}{4\pi} = \pi u^2 \frac{1}{4\pi} = \frac{\left(\frac{D}{2}\right)^2}{4s^2} \tag{2.3}$$

Following Figure 2.10, if $s \gg s_{ep}$, the relation between *s* and *s′* and the image magnification, *I*, can be considered identical to linear optics. This approximation holds

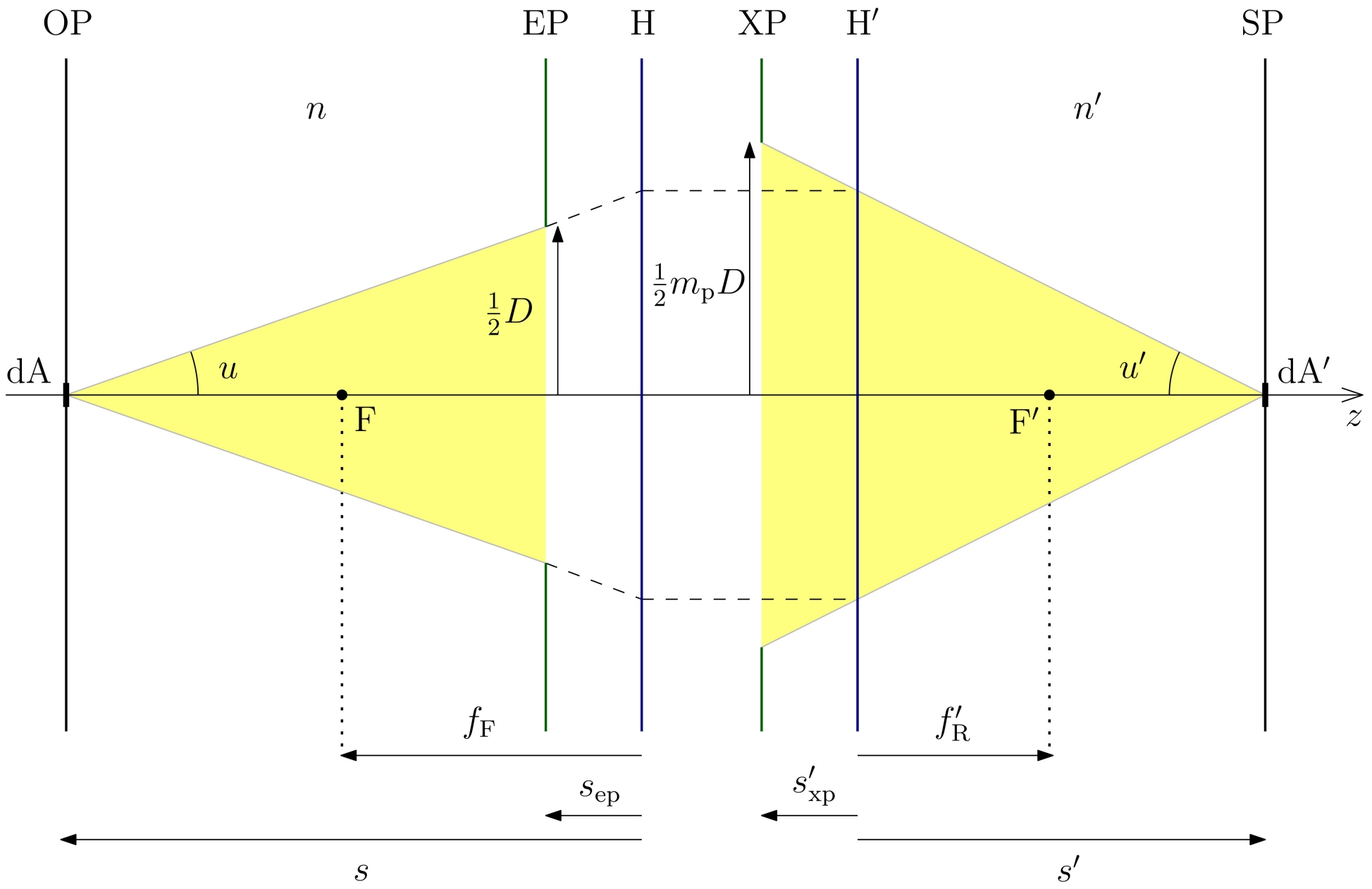

**Figure 2.10:** Schematic of a lens optical system identical to the one used in CYGNO detectors. The different components of the system are explained in the text. Image retrieved from [168].

for CYGNO, and thus:

$$\frac{1}{f} = \frac{1}{s} + \frac{1}{s'} \tag{2.4}$$

$$I = \frac{y'}{y} = \frac{s'}{s} \tag{2.5}$$

where $y'$ and $y$ are the dimension of the image and object, respectively.

Combining the previous equations, the solid angle covered by the lens can be derived, resulting in:

$$\Omega_f = \frac{1}{\left(4N\left(\frac{1}{I} + 1\right)\right)^2} \tag{2.6}$$

with $N$ being the f-number or aperture ratio.

The CYGNO/INITIUM project employs the Schneider Onyx Fast lens [169] in its detectors. This lens features a focal length of $f = 25.6$ mm and a maximum aperture ratio of $N = F/\# = 0.95$. The focal length defines the distance from the lens to the point where incoming light rays converge to form a sharp image of a distant object. It also determines the lens's field of view (FOV): shorter focal lengths result in a wider

FOV, while longer focal lengths yield a narrower, more zoomed-in view.

The f-number, also known as the aperture ratio or f-stop, is defined as the ratio of the lens's focal length ($f$) to its diameter ($D$), and it determines the amount of light that enters the lens. A lower f-number corresponds to a larger aperture, allowing more light to pass through. On the other hand, a higher f-number indicates a smaller aperture and reduced light intake. In CYGNO, the f-number is minimized to maximize the amount of light reaching the sensor, as the available light in particle detector applications is significantly lower than in conventional photography.

The granularity and solid angle of an optical readout in a TPC are constrained by the area to be imaged – the amplification region, in CYGNO – and the optical system. Taking CYGNO's current largest detector, LIME (see Section 2.2), as an example: in this setup, the camera is focused on an imaging area of $35.7 \times 35.7\ cm^2$, slightly larger than the GEM region where the light is produced, which measures $33 \times 33\ cm^2$. The camera used in LIME is the ORCA Fusion (see Section 2.1.3.2), featuring 2304 pixels along each axis and a total sensor size of $14.976 \times 14.976\ mm^2$. This configuration results in an effective spatial granularity of $155 \times 155\ \mu m^2$, meaning that each $6.5 \times 6.5\ \mu m^2$ pixel of the camera images this corresponding area on the GEM plane.

Under these conditions, the solid angle – or geometrical acceptance – of the optical system was calculated to be $\Omega = 1.2 \times 10^{-4}$. This value represents the fraction of photons detected by the camera relative to the total number emitted. Such a significant reduction highlights the need for high amplification gains to achieve energy thresholds on the order of $\mathcal{O}(keV)$, further justifying the use of a triple GEM system (see Section 2.1.2). To further improve light collection, the CYGNO collaboration is testing new lenses, such as the EHD-F0.85 series [170], which offer an even lower f-number of 0.85 ($N$, in Equation 2.6). This allows for increased light intake, although the practical implementation of this lens is still under evaluation.

#### Additional lens effects

The discussion above assumes ideal conditions and point-like sources emitting along the optical axis. In practice, however, additional effects can influence the images acquired through these lenses. Two of the most significant effects in CYGNO detectors are the *vignetting*, which causes a reduction in the measured light yield toward the edges of the image, and the *barreling*, which introduces geometrical distortions that can affect the reconstructed shape of ionization tracks. In this section, these two effects are briefly explained.

**Vignetting:** The flux calculated in Equation 2.2, based on the schematic shown in Figure 2.10, assumes ideal conditions where the source is point-like and located along the optical axis ($z$ in Figure 2.10), which crosses the centers of the source, lens, and sensor. If the source is positioned off-axis, the illuminance on the sensor plane decreases compared to the on-axis case – i.e., relative to a source located at the axial position on the object plane. This reduction occurs because, as the source moves farther from the optical axis, the angle between the emitted light and the lens plane increases, reducing the effective area of the lens that captures the light and thus lowering the photon flux.

For an off-axis source, the reduction in photon flux depends on the angle $\theta$, defined as the angle between the optical axis and the line connecting the off-axis source to the center of the lens entrance (located in the *EP* plane). In this case, the photon flux reaching the lens is given by [171]:

$$\Phi = \pi \cdot L \cdot dA \cdot u^2 \cdot \cos^4(\theta) \tag{2.7}$$

This effect is commonly referred to as *vignetting* and plays a significant role in the analysis of images acquired by the CMOS sensor in CYGNO detectors, as it introduces a non-uniformity in the light intensity observed across the image. To correct for this light distortion caused by the lens, a *vignetting-correction map* is applied to all images acquired.

To create this map, a white paper is uniformly illuminated, and several images of it are captured using the camera-lens system in question. The intensity value of each pixel is averaged across the images and normalized to the center pixel (i.e., the pixel at the *axial position*), where the flux is highest. Figure 2.11 illustrates the vignetting map obtained for the Schneider Onyx Fast lens, demonstrating how, under uniform illumination, the intensity at the corners of the image drops to about 20% of the center value. This vignetting map is then used to re-normalize any images taken during normal data acquisition by dividing each pixel intensity by the corresponding value from the vignetting map, thereby compensating for the loss of light towards the peripheral regions of the image.

**Barreling:** The *barreling distortion* is another effect that can negatively impact the images acquired in CYGNO/INITIUM detectors. Barrel distortion is a common optical aberration where straight lines near the edges of an image appear to curve outward, resembling the sides of a barrel. This distortion is primarily caused by the lens design and is particularly noticeable in wide-angle lenses. In these, the center of the lens bends light rays more strongly than the edges, leading to a decrease in magnification as you move away from the center (or optical axis). The resulting effect is that the image appears to be mapped around a sphere (or a barrel) [172].

In CYGNO detectors, where the light sources (i.e., the ionization tracks) are often extended, barrel distortion causes the edges of the tracks to bow outward, leading to an artificial bending. This effect is most noticeable in straight tracks (similar to lines), as illustrated in Figure 2.12a. For very short tracks, which can be considered almost point-like, the distortion instead shifts their perceived position inward in the image, making them appear closer to the center than they actually are. This phenomenon is shown in Figure 2.12b, where the original positions of the points (represented by red circles) are

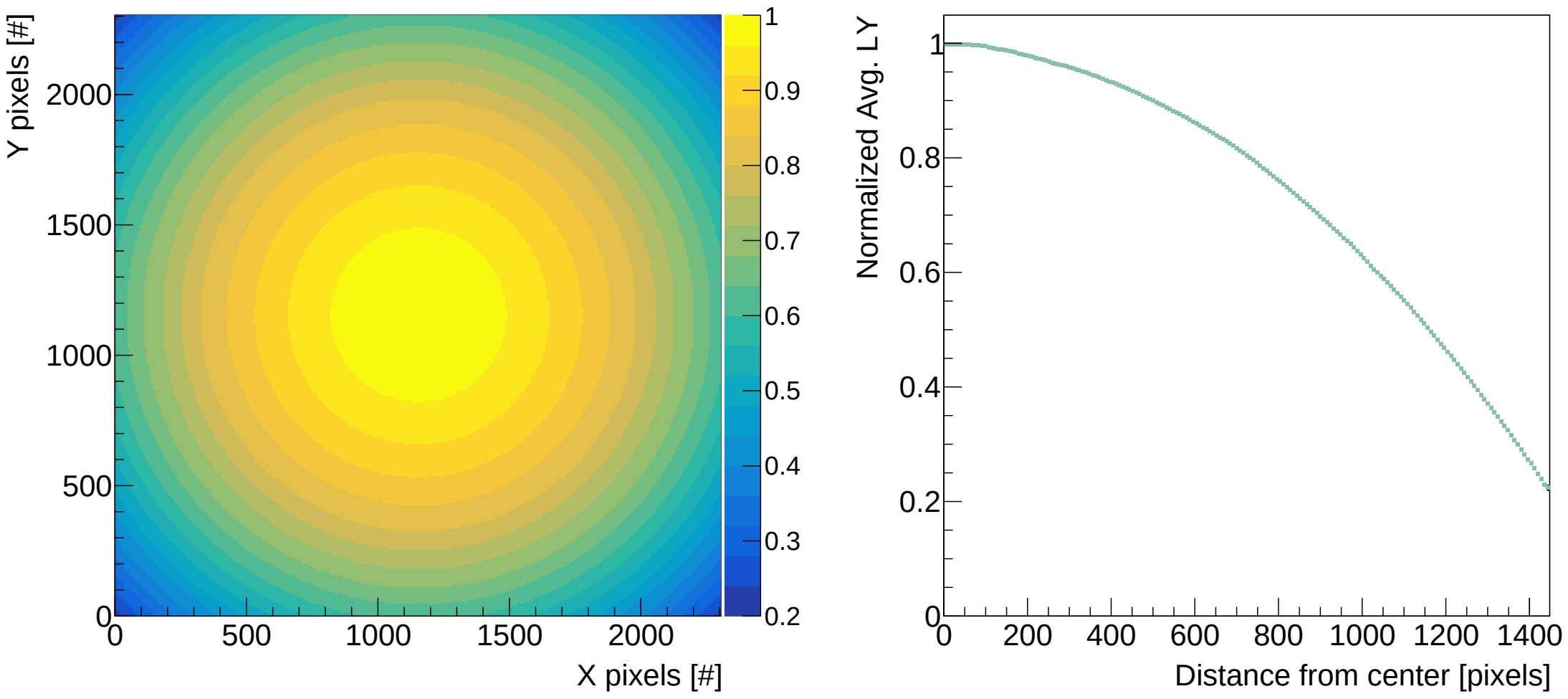

**(a)** **(b)**

**Figure 2.11:** (a) Vignetting map obtained for the Schneider Onyx Fast lens, and (b) respective values used to correct the pixel light intensity in each picture.

compared to their perceived or measured positions under the influence of the barreling effect (black circles).

This effect can impact the precision in determining the spatial localization and/or shape of events inside the CYGNO detectors as observed by the CMOS sensor. For example, extended tracks – such as those from alpha particles – may exhibit small distortions in their reconstructed XY length due to the induced bending. The correction of this effect is currently under study by the collaboration and is expected to lead to a slight improvement in the reconstruction of the XY properties of ionization tracks. As explained later in Section 3.3, this correction can also enhance the comparison and matching of information from the CMOS and PMT sensors, thanks to a more accurate interpretation of the localization of ionization events in the CMOS images. The planned correction will likely involve generating a correction map – similar to the one shown in Figure 2.12b – which will then be used to adjust the geometric positions of ionization events in the images, following approaches described in works such as [173]."

## 2.2 The CYGNO/INITIUM timeline

The CYGNO/INITIUM project was developed through a staged approach aimed at optimizing the apparatus and progressively improving its performance, with the ultimate goal of constructing a large-scale gaseous TPC for DM searches. Since 2015, the collaboration has developed several prototypes, as illustrated in the project timeline shown in Figure 2.13. The project is structured in three main phases [131]: *Phase_0* corresponds to the R&D stage, during which several small-scale prototypes were constructed to test and refine the CYGNO detector concept and analysis techniques. *Phase_1* involves the deployment of a demonstrator with an active volume of order $\mathcal{O}(1)$ $\mathrm{m}^3$, with the main focus being testing the scalability of the CYGNO approach and the potentialities of a large PHASE_2 detector to reach the expected physics goals. *Phase_2* foresees the construction of a large-scale detector, with a target active volume in the range of $\mathcal{O}(30-100)$ $\mathrm{m}^3$, fully dedicated to performing directional DM searches and solar neutrino spectroscopy [132].

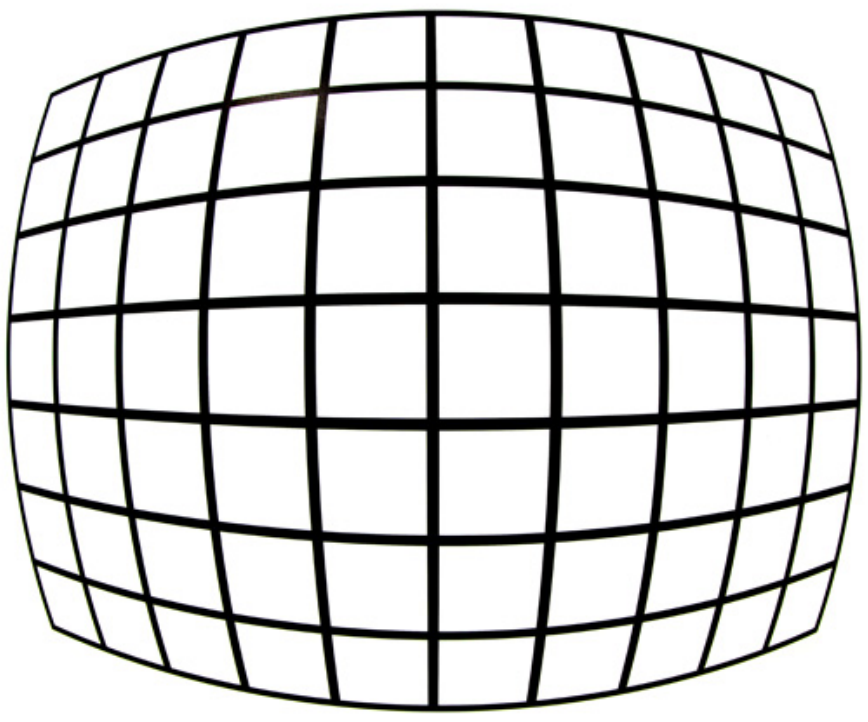

(a)

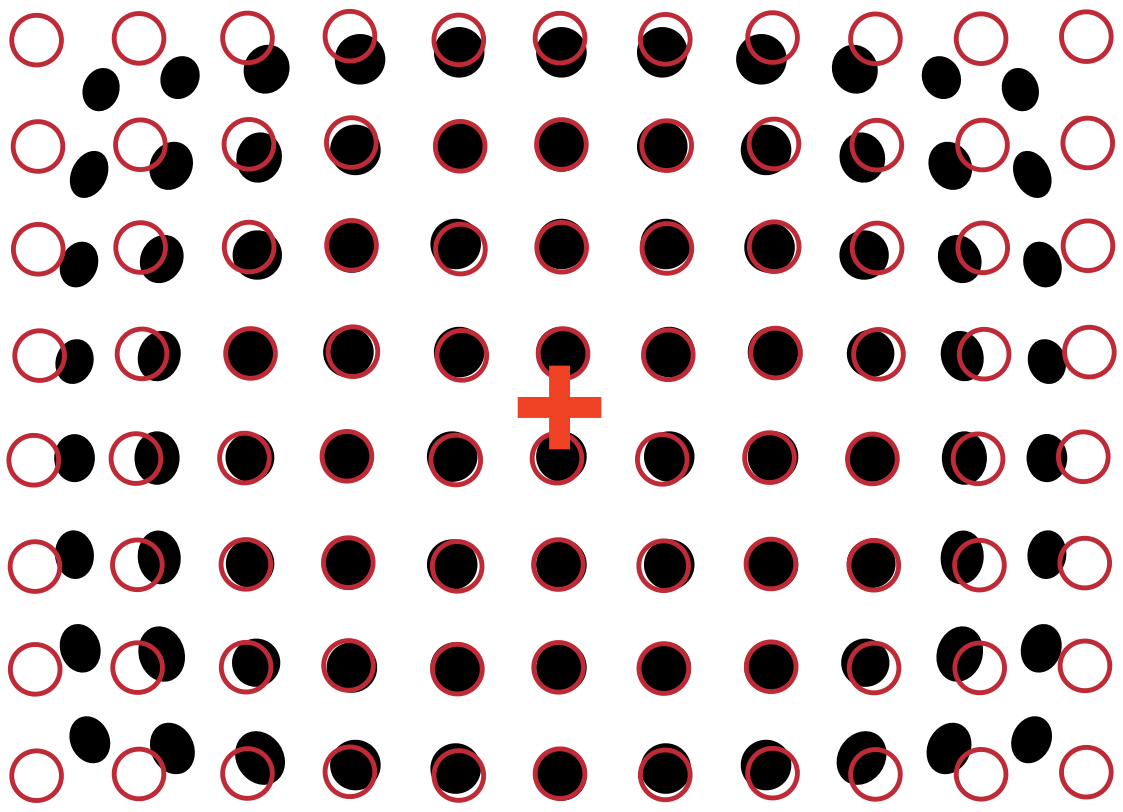

(b)

**Figure 2.12:** Example of the barreling effect showing (a) the outward bowing of straight lines in an image, and (b) the inward shift of the perceived position of points. In the latter, the red circles represent the original equally distributed points, while the black circles represent the perceived positions of these points under the influence of the barreling effect. Retrieved from [174].

The first prototypes developed during *Phase_0* were ORANGE and LEMOn, which served as initial benchmarks for the CYGNO detector concept. Specifically, ORANGE, with a sensitive volume of 100 $\mathrm{cm}^3$, was used for a preliminary assessment of track reconstruction capabilities and light yield performance [134, 163]. LEMOn, in contrast, featured a significantly larger drift length of 20 $\mathrm{cm}$ and was used to validate the technique at an intermediate scale. It focused on evaluating energy resolution and operational stability [139], provide the first identification of low energy nuclear recoils [175], and demonstrated nearly 100% detection efficiency for 5.9 $\mathrm{keV}$ electron recoils at drift distances up to 20 $\mathrm{cm}$ [176].

The two next prototypes in the timeline are MANGO (Section 2.2.1) and LIME (Section 2.2.2). These are the detectors used in the experimental studies presented in this thesis and, for that reason, a particular attention is dedicated to them in this section. MANGO, featuring a sensitive volume of $\mathcal{O}(1)$ $\mathrm{l}$, was initially used in studies focused on testing GEMs with various thicknesses and investigating the production of additional light by introducing an induction field after the final GEM in the amplification stage. The results of these studies are reported in [155, 177]. As part of the INITIUM project (see Section 2.3), MANGO was also used to explore the negative ion drift operation mode. This approach offers reduced diffusion during drift, as the charge carriers are now negative ions rather than electrons. These studies and the contribution provided

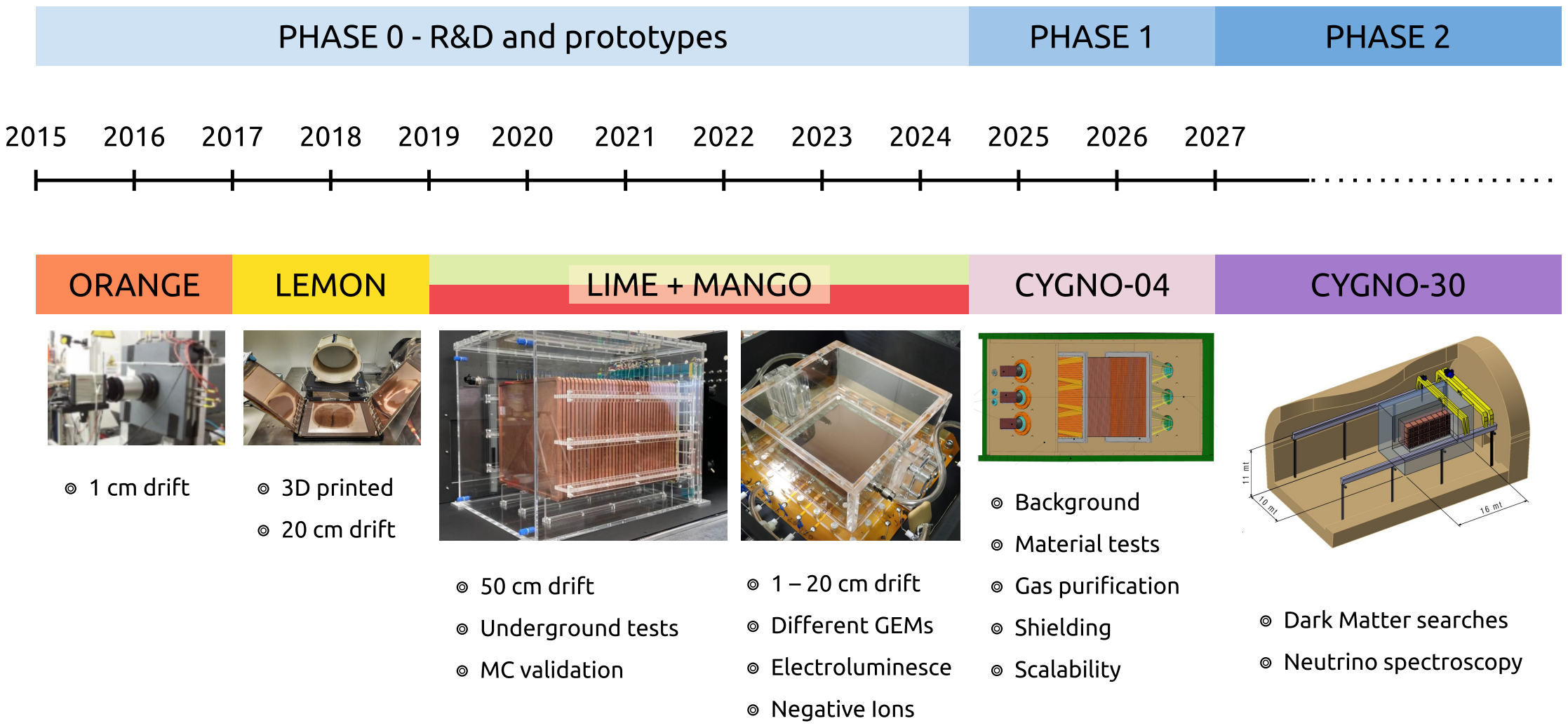

**Figure 2.13:** Schematic of the CYGNO roadmap, showing the different detectors built, their basic objectives, and operation timescales. Based on an image by Melba d'Astolfo.

by the author to these are discussed in more detail in Chapter 7.

LIME is the largest detector manufactured so far by the collaboration, featuring a sensitive volume of approximately 50 liters and a drift length of 50 cm. It was initially commissioned at the overground Laboratori Nazionali di Frascati (LNF), and subsequently placed and operated at the underground facilities of LNGS. The primary goals of LIME are to demonstrate the CYGNO technique capabilities in a low-background environment and to validate the Monte Carlo simulations against experimental data. Within the scope of this thesis, multiple studies were performed using LIME, with particular focus on data acquired with the PMTs. These studies are introduced in Chapter 4, which presents the results of overground measurements with the PMTs, including a detailed analysis of the cosmic muon flux recorded with this detector. The main conclusions from the background characterization performed with LIME underground by the CYGNO collaboration are discussed later in Section 6.1, as the context and motivation for the development of the 3D reconstruction algorithm presented in Chapter 5, as well as for the alpha background studies described in Chapter 6. Given the central role of LIME in this work, a comprehensive description of the detector, its data acquisition system, and its commissioning phases is provided in Section 2.2.2.

Finally, this section also includes a brief overview of the upcoming detectors, CYGNO-04 (Section 2.2.3) and CYGNO-30 (Section 2.2.4), to contextualize the future goals and scientific ambitions of the CYGNO project in the field of dark matter searches.

### 2.2.1 MANGO

The **M**ultipurpose **A**pparatus for **N**egative ions studies with **G**EM **O**ptically readout (MANGO) is, as the name suggests, a versatile detector designed with an "extendable" drift field cage, having been used with 1, 5, and 15 cm drift. Within the several R&D studies carried out with MANGO, this has been utilized to investigate different types and configurations of GEM amplification stages, including both 50 μm and 125 μm thick configurations [104]; and the potential use of an additional electric field after the final GEM to induce further light scintillation by the means of molecules excitation, without degrading the energy resolution since no multiplication is involved [155]. In the context of this thesis, MANGO was employed to perform the first ever operation of negative ion drift [178,179] at atmospheric pressure with an optical readout TPC (see Section 2.3) and, more specifically, to perform the first ever analysis of NID PMT signals. This topic is discussed in detail in Chapter 7.

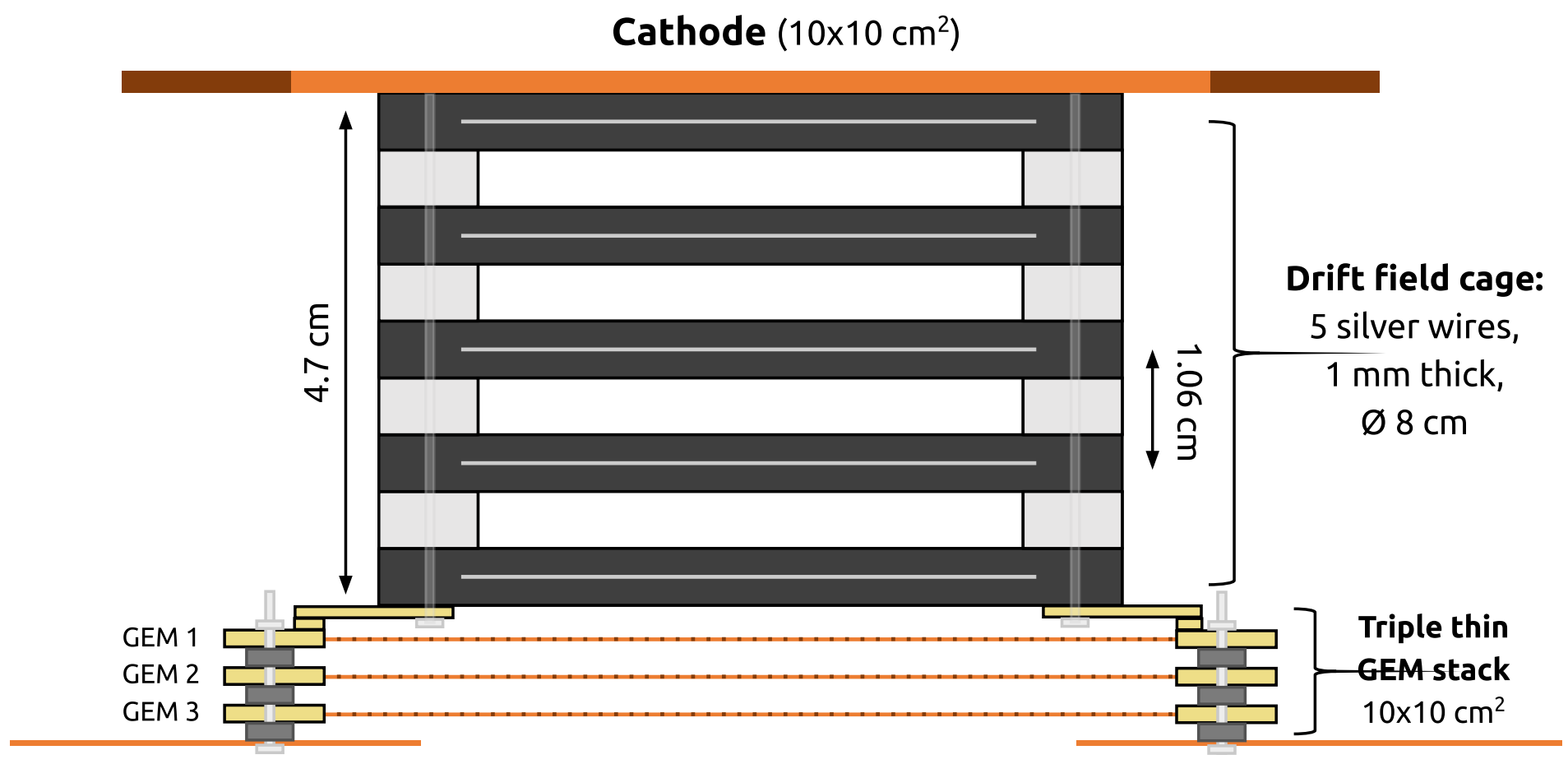

**Figure 2.14:** Schematic of the MANGO detector, showing the different components that constitute the amplification and drift regions.

MANGO implements the CYGNO technique using a stack of three GEMs for amplification, each with an area of $10 \times 10$ cm$^2$, a thickness of 50 μm, and spaced 2 mm apart with a transfer field of 2.5 kV/cm between them. Transfer field is the common name given to the electric field between the GEMs. The GEMs are powered by a custom-made power supply, the HVGEM [180], which allows for independent biasing of each GEM and current measurements down to approximately 1 nA, useful to monitor the correct functioning of the amplification stage and detect eventual sparks. The gas volume is contained within an acrylic vessel and further enclosed by a black box to prevent external light from entering. A schematic of MANGO is shown in figure Figure 2.14.

The optical system and components used in MANGO follow the description provided in Section 2.1.3, with a few modifications: a single Hamamatsu ORCA-Fusion camera is used and positioned at $20.5 \pm 0.3$ cm, resulting in a granularity of approximately $49 \times 49$ μm$^2$ when imaging an area of $11.3 \times 11.3$ cm$^2$. In contrast to LIME, MANGO is equipped with only one PMT (Hamamatsu R1635 [181]), located at the left upper corner of the camera.

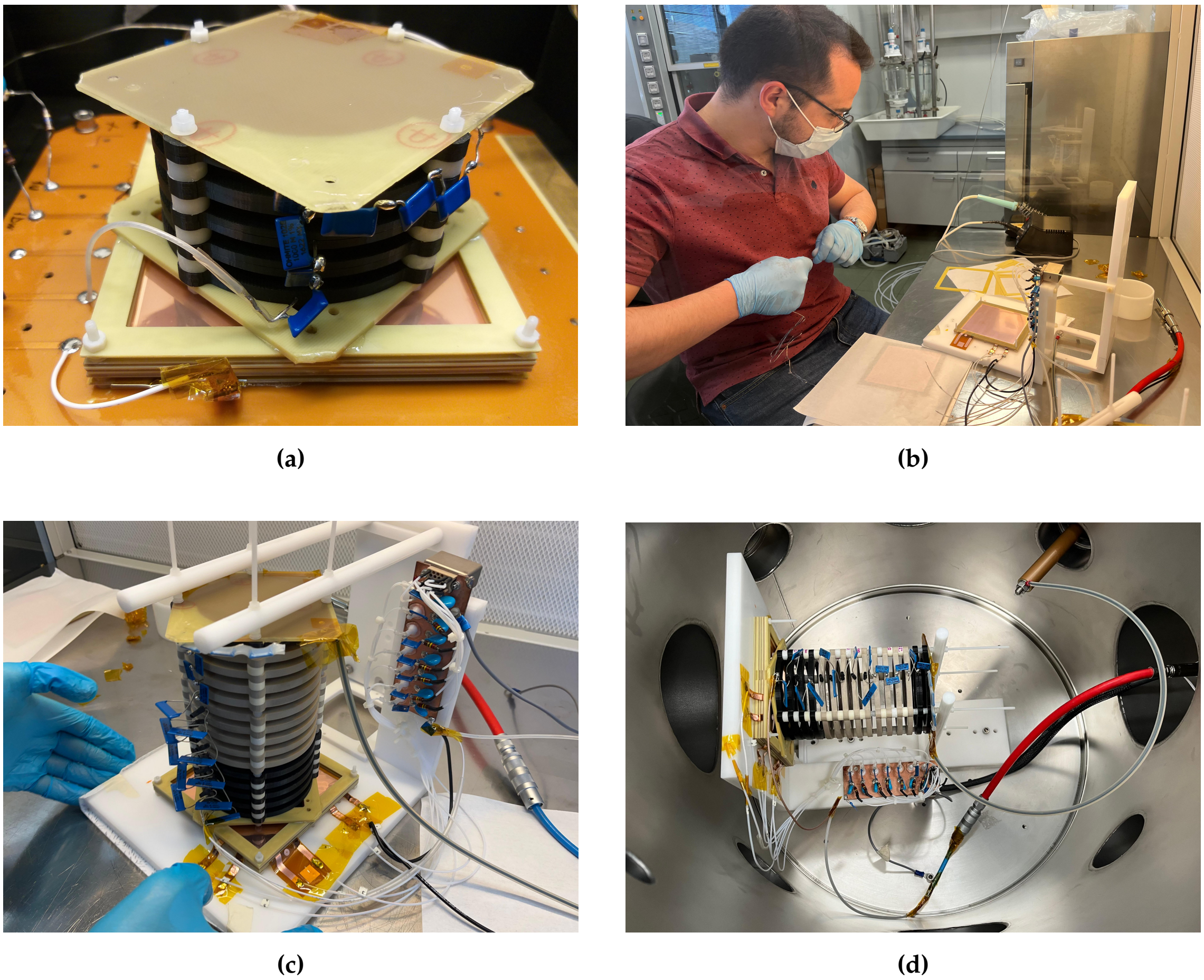

(a) (b)

(c) (d)

**Figure 2.15:** Several steps of the augmentation of the field cage of MANGO detector: (a) initial configuration of MANGO with 5 cm drift, as shown in Figure 2.14; (b) process of swapping the GEMs of MANGO in a clean environment to a larger support; (c) MANGO longer field cage of 15 cm assembled into new support structure; (d) final housing of MANGO inside the *keg*.

For more advanced studies in the context of negative ion drift, the length of the drift region of MANGO was increased from the initial 5 cm to approximately 15 cm. This was achieved by adding additional field rings to the existing field cage. These field rings are made of 1 mm thick silver and are enclosed by 3D printed rings that hold them in place. The rings are spaced 1.06 cm apart and have a diameter of 8 cm, creating a cylindrical sensitive region for ionization and drift. This new version of MANGO was placed inside a large gas-tight metal vessel, referred to as *keg*, which ensures more stable gas conditions, particularly in terms of pressure, temperature, and humidity. Pictures of the different assembly steps and the final result of the MANGO field cage augmentation can be seen in Figure 2.15.

### 2.2.2 LIME

LIME, the **L**ong **I**maging **M**odul**e**, is the largest prototype developed by CYGNO to date. It is expected to be the final prototype in CYGNO's main development, or *Phase_0* in the CYGNO/INITIUM timeline in Figure 2.13, preceding the CYGNO-04 demonstrator for full-scale DM searches (Section 2.2.3).

This section presented a thorough description of the LIME detector (Section 2.2.2.1) and its data acquisition system (Section 2.2.2.2). A special emphasis was placed on the latter, as a deep understanding of LIME's DAQ was required for developing the analysis tools used to extract information from the PMT waveforms, as detailed in the following chapter, Section 3.2. LIME was constructed at LNF, where a series of operational tests were carried out to study its stability, response linearity, and other performance aspects (Section 2.2.2.3). In February 2022, the detector was relocated to its current site at the underground LNGS laboratory. After an initial commissioning phase, several data-taking runs were conducted to assess LIME's performance in a low-background environment (Section 2.2.2.4). These tests were aimed at identifying both the strengths and the challenges of the CYGNO approach to full-scale dark matter searches.

#### 2.2.2.1 Detector description

LIME, represented in Figure 2.16a, is a square prism TPC with amplification provided by a stack of three GEMs, each with an area of $33 \times 33$ $\mathrm{cm}^2$ and a drift length of 50 cm, giving an active volume of approximately 50 l. The GEMs are 50 µm thick, with 50 µm holes and a 140 µm pitch, with a total spacing between then of 2 mm. The active volume is enclosed by 34 copper electric field rings, each 10 mm wide, with a 14 mm pitch (see Figure 2.16c). These rings are electrically interconnected using 100 MΩ resistors, ensuring a uniform electric field throughout the sensitive region. At the end of the drift region, a 0.5 mm thin copper sheet forms the TPC cathode, closing the electric field lines (see Figure 2.16b). Typical LIME operative conditions are a drift field of 800 V/cm and a fixed voltage of 440 V applied to each GEM.

Encasing the field cage is a transparent acrylic (PMMA) vessel with a capacity of approximately $\sim$ 100 l, which serves as an insulator to ensure the detector's electrical stability and structural integrity. At the top of the vessel, a 150 µm thin ETFE window (see Figure 2.16a) is installed to allow low-energy photons (a few keV) from external radioactive sources to pass through for detector calibration.

The optical readout of LIME consists of a single Hamamatsu ORCA Fusion CMOS camera [182] and 4 Hamamatsu R7378 PMTs [159]. The location and field of view of the camera and each PMT relative to the GEM plane (where light is produced) is shown in Figure 2.17, from both the side and the front views. The camera is positioned 62 cm away from and centered on the GEM plane. The ORCA Fusion has a resolution of $2304 \times 2304$ pixels and images an area of $35.7 \times 35.7$ $\mathrm{cm}^2$, resulting in a pixel granularity of $155 \times 155$ $\mu\mathrm{m}^2$. A Schneider lens with a focal length of 25.6 mm and an aperture of 0.95 is mounted on the camera. In this configuration, the lens (and thus the

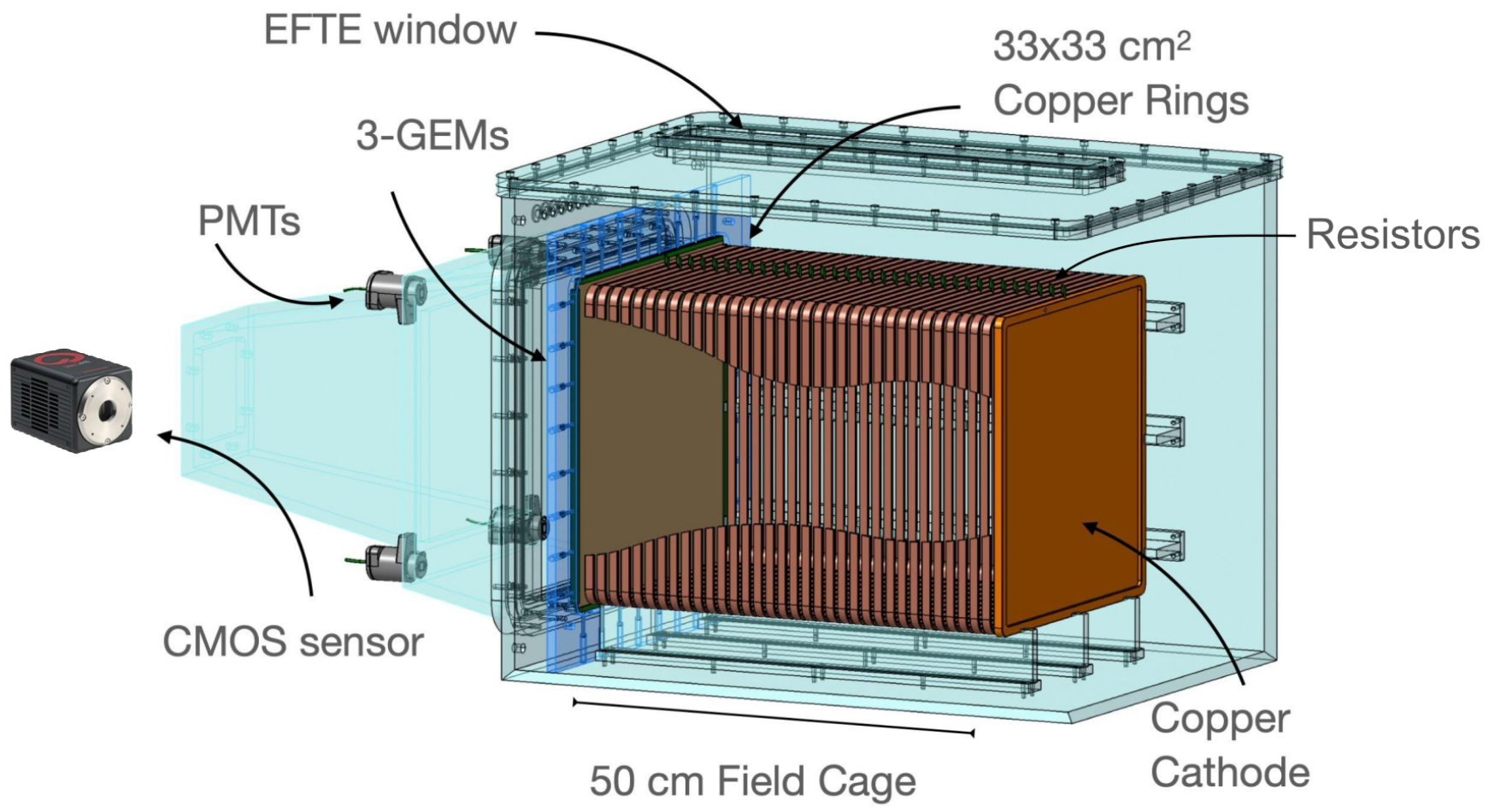

**(a)**

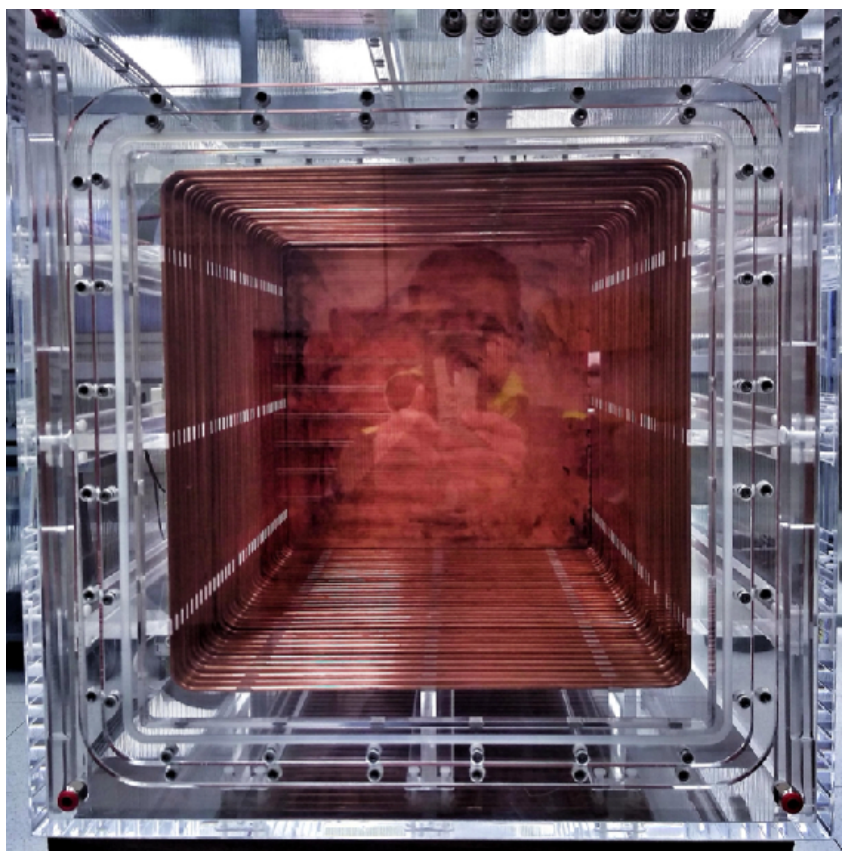

**(b)**

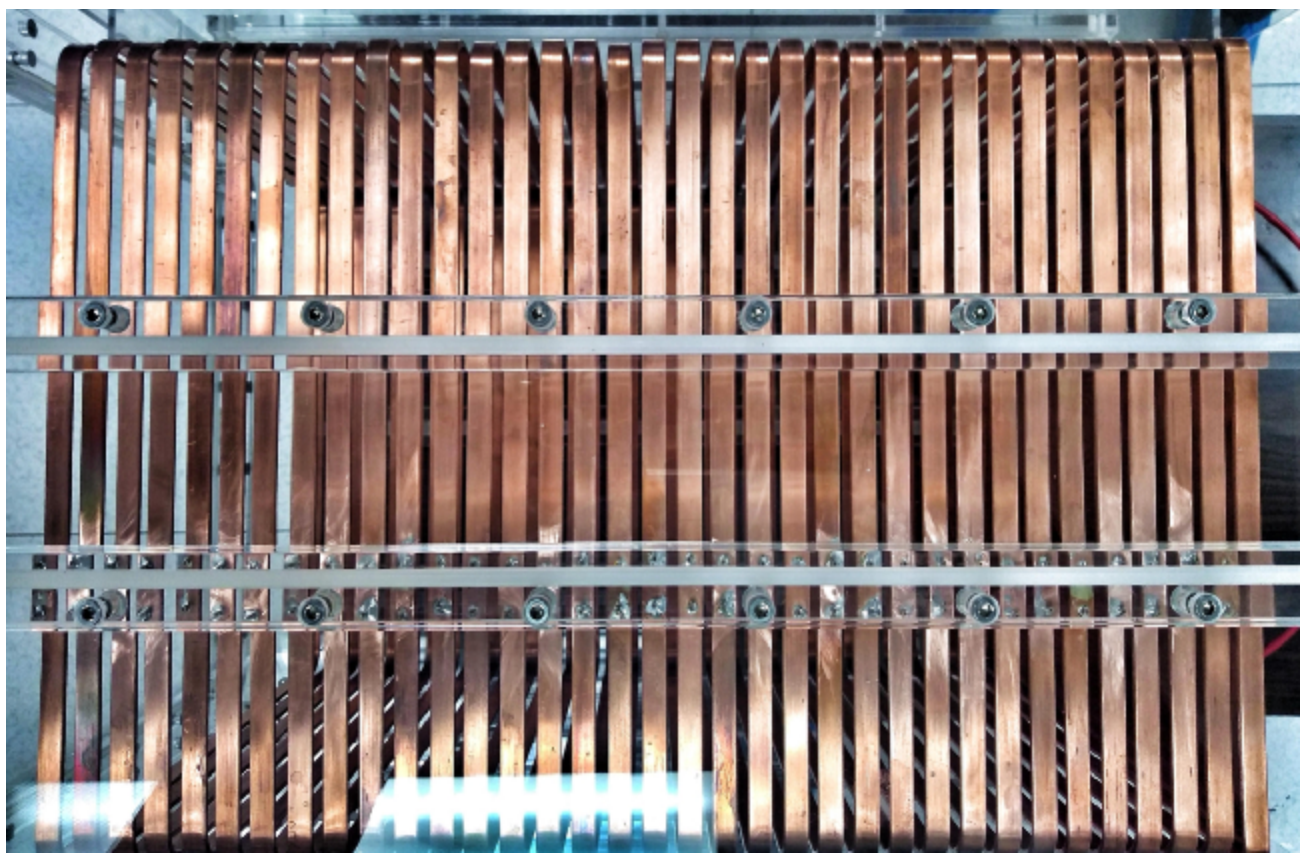

**(c)**

**Figure 2.16:** Illustrations of LIME from different perspectives: (a) a schematic highlighting the main components of LIME; and images showing the (b) front view and (c) side view of LIME. The images were retrieved from [131].

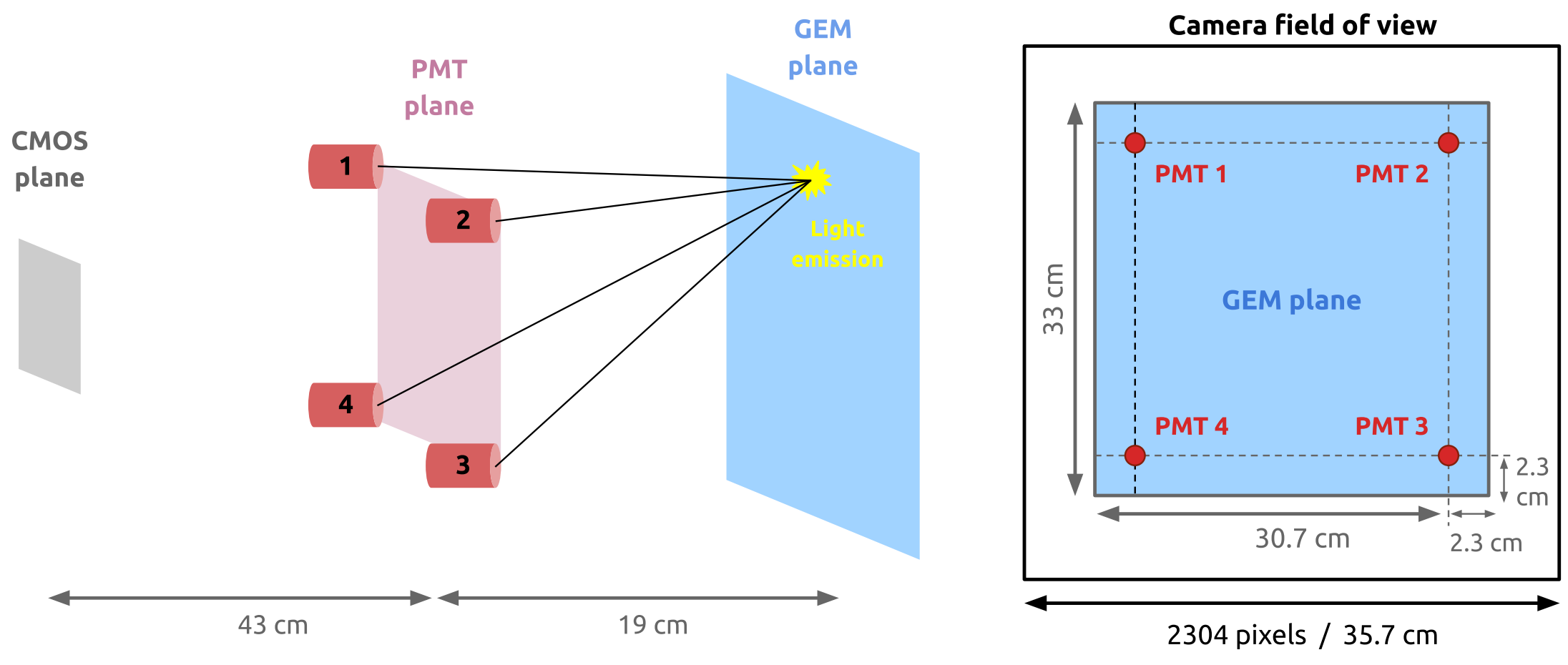

**(a)** Side view **(b)** Front view

**Figure 2.17:** Representation of the fields of view of the CMOS and PMT sensors, seen from (a) the side, and (b) the front.

CMOS sensor) has a geometrical acceptance of $\Omega = 1.2 \cdot 10^{-4}$ (see Section 2.1.3.3). The PMTs, each with a diameter of 22 mm, are positioned 19 cm away from the GEM plane, surrounding the camera, following the schematic in Figure 2.17b. These references are crucial for the later studies involving the matching of PMT and CMOS data, as an accurate conversion or transformation of coordinates between sensors is required.

#### 2.2.2.2 Data Acquisition system

The Data Acquisition (DAQ) system is a collection of instruments, comprising both hardware and software components, responsible for handling, acquiring, and storing data from the detector.

The LIME detector's DAQ system includes one CMOS camera, which serves as both the sensor and digitizer, and four PMTs, each connected to a separate channel of two digitizers. These digitizers convert the analog signal from the PMTs into a digital signal for storage and subsequent analysis. The GEMs in LIME are also connected to these digitizers via a capacitor-resistor (CR) circuit [183], allowing the recording of the current induced by electrons and ions during amplification. Since the CYGNO experiment primarily focuses on optical readout, the GEM signals are (at the moment) considered secondary information and were not used in any analysis or results presented in this thesis.

The two digitizers connected to the PMTs are the CAEN VME V1742 [184] and V1720 [185]. Both have 12-bit resolution, allowing for $2^{12} = 4096$ samples within a voltage range of 1 V for the V1742 and 2 V for the V1720. The V1742 is configured with a sampling rate of 750 MHz over 1024 samples, resulting in a sample size of (4/3) ns per sample, covering a total of 1365.3(3) ns. The V1720, on the other hand, has a fixed sampling rate of 250 MHz for 4000 samples, resulting in a sample length of 4 ns and covering a total of 16 μs. Due to the difference in sampling size, the V1742 is commonly referred to as the *fast digitizer*, while the V1720 is referred to as the *slow digitizer*.

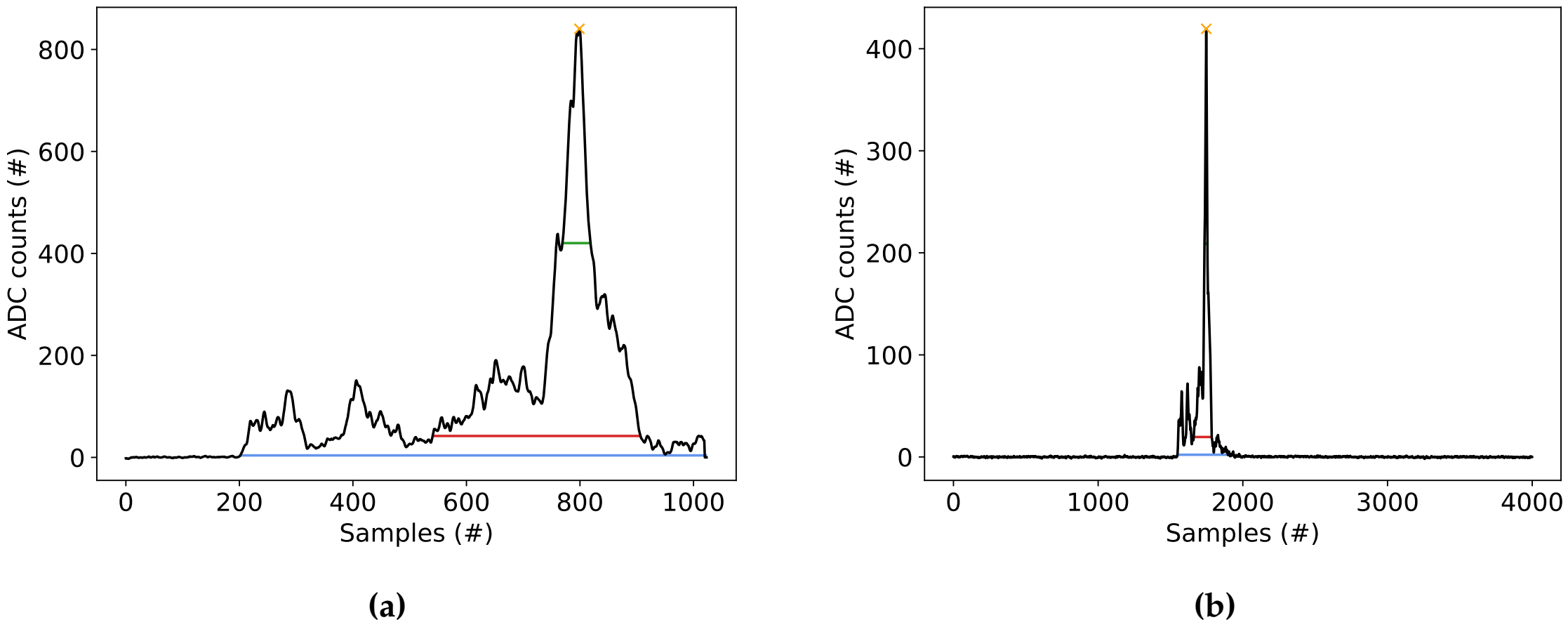

**Figure 2.18:** Example of an ionization signal observed by PMT #4 and recorded by the two LIME DAQ digitizers: (a) fast digitizer and (b) slow digitizer. The lines drawn on the waveforms represent variables defined during the CYGNO reconstruction process (Section 3.2). Notably, the two waveforms have different numbers of samples (1024 vs. 4000) and sampling frequencies (750 MHz vs. 250 MHz), which is why (b) appears as a compressed version of (a). The Y-axis is also halved in (b) since the slow digitizer has twice the dynamic range of the fast digitizer.

The sampling size (4000 samples) of the slow digitizer is optimized to cover the longest possible event in LIME, which would correspond to an event traversing the full 50 cm length of LIME. At a electron drift speed of ~ 5 cm/μs, this translates to a signal duration of about 10 μs. Such signals are most commonly associated with high-energy cosmics or ERs above 100 keV (MIP-like particles in LIME). Performing the reverse calculation, a time window of 1365.3(3) ns should capture about 5 cm of extension in the drift direction for the fast digitizer. This is sufficient to sample ER signals below 20 keV or alpha particles below 10 MeV. The use of both digitizers ensures nearly complete coverage of all signal types in LIME. However, for the physics objectives of LIME, which focus on low-energy NRs, the fast digitizer is likely the most suitable for analysis. An example of the same signal viewed by both digitizers (connected to the same PMT) is shown in Figure 2.18.

The scientific CMOS camera used in LIME has an exposure time for each image, in the lowest noise mode, ranging from 300 to 500 ms, depending on the type of study and the expected image occupancy. The exposure time of the camera is approximately $10^5$ times longer than that of the fast digitizer, which creates a synchronization challenge. Given these timings, it is possible to have multiple events captured in a single CMOS image, leading to multiple quartets of PMT waveforms (one per ionization event above threshold). As will be shown, due to the camera's operational mode, it is also possible for an event to be observed only by the PMTs.

**Digitizers amplitude to voltage conversion**

The small electrical current generated by the PMTs when receiving light is converted into a voltage signal and recorded by the digitizers as an amplitude in ADC units (or Analog-to-Digital Units, ADU). To convert the digitizer signal into an electrical voltage, the amplitude $A_i$ (in ADU) is multiplied by the digitizer's dynamic range ($DGTZ$) and divided by its resolution. This conversion can be expressed mathematically as:

$$A_i[V] \;=\; A_i[ADU] \cdot \frac{\text{DGTZ dynamic range [V]}}{\text{DGTZ resolution [bits]}} \;=\; A_i[ADU] \cdot \frac{1}{2^{12}} \tag{2.8}$$

The total charge collected by each PMT in Coulombs, Q [nC], is determined by integrating the signal's amplitude $A(t)$ (in Volts) over a given time window $\Delta t$, and dividing the result by the impedance $R$ of the fan-in/fan-out module to which the PMT is connected:

$$Q \;=\; \frac{1}{R}\int_0^{\Delta t} A(t)\; dt \tag{2.9}$$

To convert the time of integration into seconds, the number of samples in the waveform ($N_{\text{samples}}$) is multiplied by the digitizer's sampling frequency $f_s$ (in Hz):

$$\Delta t \;=\; N_{\text{samples}} \cdot \frac{1}{f_s} \tag{2.10}$$

Thus, the final expression for the total charge collected by each PMT becomes:

$$Q \;=\; \frac{1}{R} \cdot \sum_{i=1}^{N_{\text{samples}}} A_i \cdot \frac{1}{f_s} \tag{2.11}$$

where $A_i$ is the amplitude at the i-th sample, and $N_{samples}$ is the total number of samples in the waveform (or time window considered). $R$ is impedance of the fan-in/fan-out module (typically 50 Ω). The integrated charge of a waveform is a generally useful information when working with multiple digitizers with different dynamic ranges and/or resolutions.

#### 2.2.2.2.1 Trigger system & Event definition

The trigger system is designed to optimize the acquisition of signals within the detector and accommodate the significant differences in exposure and operation time between the sensors.

Figure 2.19 shows a schematic summarizing the operation of the LIME trigger system. The CMOS operates in continuous mode, capturing images repeatedly. The exposure of the sensor pixels is controlled electronically, with each row of pixels being sequentially (de-)activated from top to bottom. Each row takes 80 $\mu s$ to activate, so with the CMOS having 2300 rows of pixels, the full sensor takes 184 $ms$ to be completely activated and deactivated. During this time, if the PMTs detect a trigger – meaning that at least two PMTs record a signal below $-5$ $mV$ – the image is saved along with the corresponding waveforms from the triggering event, for all four PMTs and the two digitizers. Any subsequent event occurring before the CMOS sensor is fully deactivated (i.e., before the image is ”complete”) will also have its waveforms saved. The final result is a data bank containing a CMOS image and at least one (but potentially several) sets of PMT waveforms corresponding to different events. In summary, an event in LIME can be defined as:

- 1 CMOS picture of 300 or 500 ms
- X PMT waveforms, where $X = N_{triggers} * N_{PMTs} * N_{digitizers}$

where $N_{triggers}$ is the number of ionizing events occurring inside LIME that produce enough light to trigger the PMTs and occur at any point while the CMOS sensor was fully or partially exposed.

An example of all the data present in a LIME event is shown, for instance, in Figure 3.1. The methodologies used to reconstruct and analyze this data are presented later in Chapter 3. In addition to the CMOS and PMT data, the CYGNO DAQ system's data banks also contain other environmental and technical information, such as the temperature and humidity of the LIME room, as well as hardware details of the various components, including the type and sampling rate of the digitizers.

The main limitations of this trigger setup are twofold. First, as shown in Figure 2.19, it is possible for a signal to be detected by the PMTs but not by the CMOS sensor, and still trigger an acquisition. This occurs when the event takes place in the "shadow region" of the CMOS sensor, i.e., when the sensor is partially closed/opened. This particular limitation is expected to be resolved in CYGNO-04, where the (under development) DAQ scheme uses the CMOS in *continuous* mode, eliminating the existence of shadow regions (see Section 2.2.3). The second limitation arises when multiple events occur in one image since they are not inherently synchronized between sensors: while PMT triggers can be timed relative to each other, the image is saved as a whole, with no timing information available for the active pixels. This lack of synchronization poses a significant challenge for data reconstruction in the CYGNO experiment, particularly in associating CMOS and PMT signals to reconstruct a single event and define its three-dimensional properties. Several approaches are currently being tested to automatically associate CMOS and PMT signals, with the most promising method involving the use of a Bayesian fit to determine the position of a PMT signal within the camera's field of view, thus allowing the association of the corresponding "events". This approach is detailed in Section 3.3 and is subsequently applied in Chapter 5 for the 3D reconstruction of alpha particles.

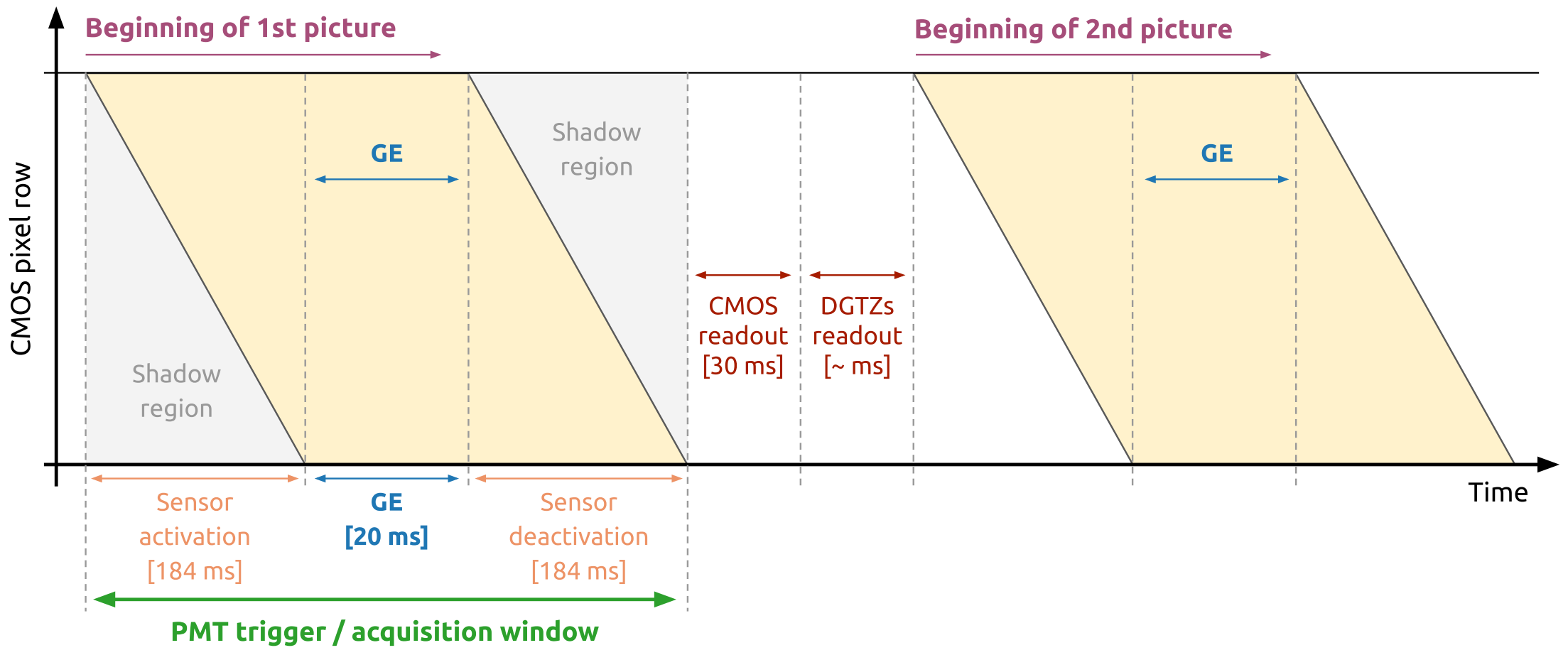

**Figure 2.19:** Schematic of the LIME trigger system. The exposure of the CMOS pixel rows occurs sequentially from top to bottom. While each part of the sensor is exposed, the PMTs can trigger the acquisition of images and waveforms if the light signal produced exceeds the $-5$ mV threshold on at least two PMTs. The "GE" region corresponds to the *Global Exposure*, where all the pixels of the sensor are exposed and activated. The *shadow region* represents the pixel/time region where the PMTs can trigger (thus saving the image), but no visible event is present in the image. The *CMOS/DGTZs readout* is the time required for the DAQ to save the information of the camera and the digitizers (PMTs).

Although the choice of this trigger system methodology might not appear optimal, it was chosen to minimize the system's deadtime. As shown in Figure 2.19, if only events occurring within the global exposure region – where all pixels are exposed and we would be sure to have a visible event in both sensors – were recorded, the deadtime would approach 90%. The actual deadtime of the LIME DAQ is a complex issue, as it depends on the readout, compression, and saving processes of the DAQ and the computer. For example, while the CMOS readout has a constant duration of 30 ms, a signal with a higher number of PMT triggers (i.e., waveforms) will increase the time required to digitize all waveforms (designated "DGTZ readout" in Figure 2.19), thereby increasing the deadtime.

#### 2.2.2.3 Overground commissioning

The first tests with LIME were conducted overground at LNF and provided a comprehensive characterization of its operational conditions and performance. During these tests, LIME was operated with a continuous inflow of fresh gas at a rate of 200 cc/min and maintained at an overpressure of 3 mbar. This setup ensured optimal and stable light yield by minimizing the presence of impurities. The drift field was set to 800 V/cm, and a bias voltage of 440 V was applied to each GEM. The results of these measurements created the standards of data-taking and operation parameters for subsequent runs.

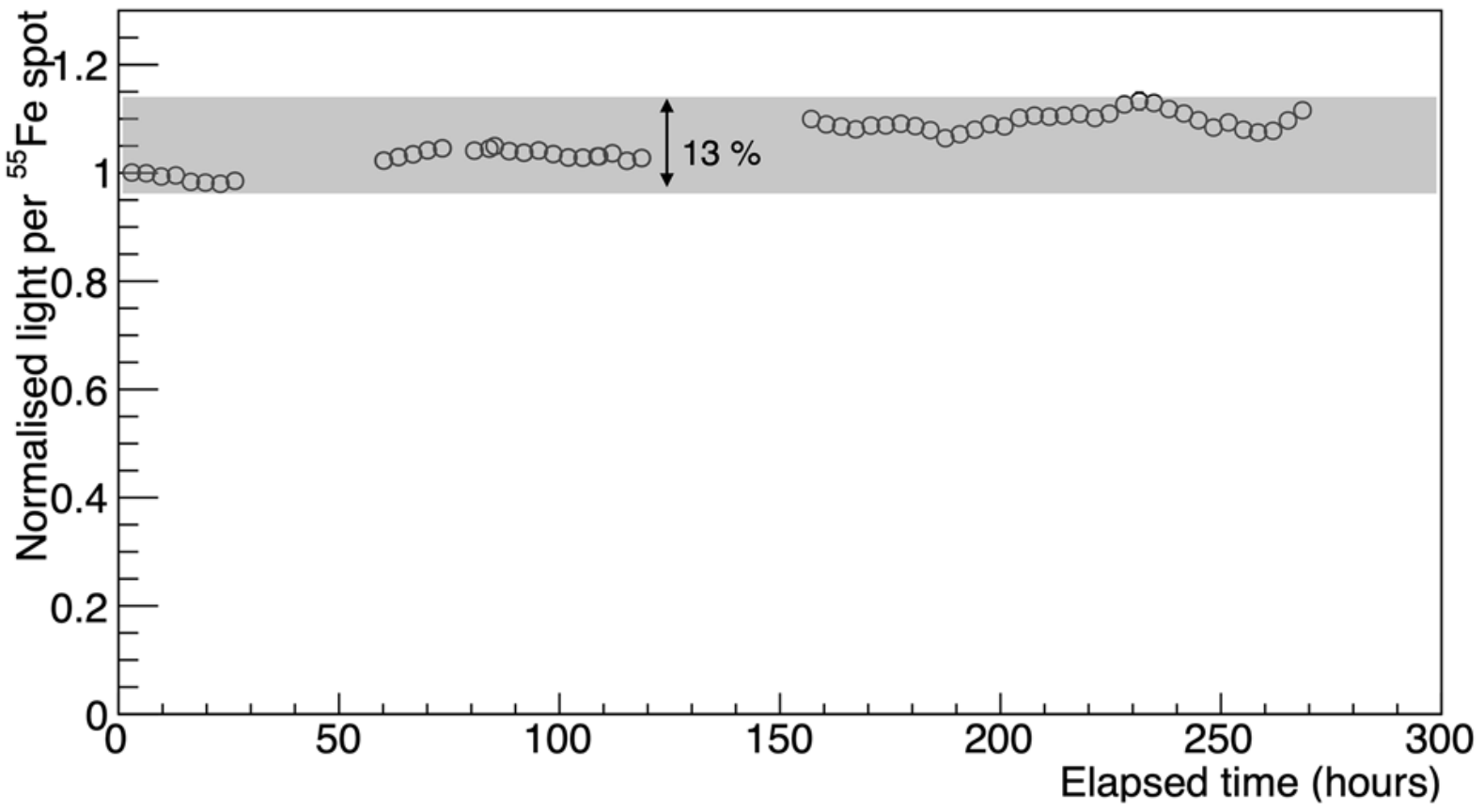

**Figure 2.20:** Normalized light yield of the $^{55}$Fe spots as a function of time during a long-term stability test. Adapted from [186].

In this section, a brief summary of the studies primarily conducted using the CMOS camera is presented. A more comprehensive description of these studies can be found in [186]. Later, in Chapter 4, a detailed characterization of the work performed with the PMTs in the overground LNF is provided, as those studies are directly related to the development of this thesis and the PMT analysis techniques described in Section 3.2.

**Light yield stability:** The energy calibration of the detector is performed using a $^{55}$Fe source, which emits 5.9 keV X-rays that produce spot-like tracks in the gas via photoelectric effect. These tracks are analyzed through a dedicated reconstruction procedure, described in Section 3.1. The pixel-integrated counts within each reconstructed spot are fitted with a Gaussian distribution, where the mean provides the light yield (LY) calibration factor. The LY can fluctuate over time due to environmental factors such as pressure, temperature, or the presence of impurities. Long-term (i.e., about 10 days) stability tests, during which the detector operated without human intervention, revealed a 13% variation in the LY of $^{55}$Fe spots, as shown in Figure 2.20 [186]. These variations were found to be primarily caused by changes in pressure.

**Energy response linearity:** The response linearity of LIME was tested using various low-energy X-rays. These X-rays are produced using the radioactive source described in Section 4.1.3, along with a custom setup for generating energies below 8 keV. For each measurement, the source was placed 25 cm from the GEM plane, i.e., at the center of LIME's drift region. The light integral of each ER track in the CMOS images was reconstructed using the CYGNO-reconstruction code, as in the previous study. The distribution of reconstructed energies for each element of the source was fitted with two Crujiff functions [187] – representing the $K_\alpha$ and $K_\beta$ lines (see Section 4.1.3 for more de-tails) of each element (Ca, Ti, Fe, Cu, Rb, Mo, Ag, Ba) – and an exponential component to account for background contributions [186]. Figure 2.21 shows the fitted means of the two Crujiff functions as a function of the expected energy, with a dotted line indicating a perfectly linear response. As demonstrated, at a fixed Z position, LIME exhibits good response linearity to low-energy ERs, in the range from 3.5 up to 37 keV.

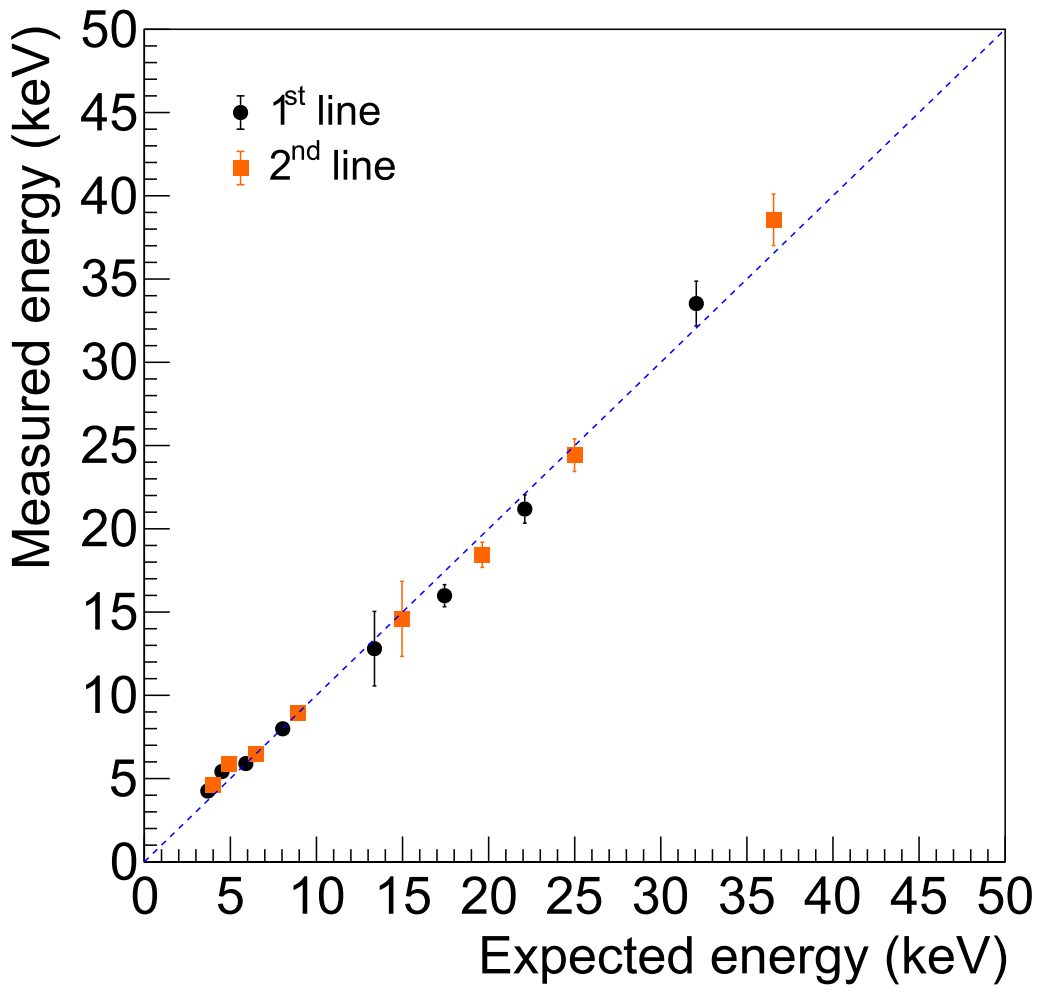

**Figure 2.21:** Estimated energy response compared to the expected energy from the $K_\alpha$ (black dots) and $K_\beta$ (orange dots) lines for each of the radioactive source elements used. The dotted line represents a linear response. Adapted from [186].

**Absolute Z:** The shape of the primary electron ionization cloud reaching the GEM plane depends on the drift distance covered or, in other words, the absolute position of the event along the drift direction (Z). In LIME, this was investigated by analyzing ionization tracks induced by $^{55}Fe$ X-rays. It was found that the variable $\xi$ – defined as the product of the track width (parameterized by the standard deviation of its Gaussian transverse profile) and the standard deviation of the number of counts per pixel within the same track – is the most sensitive indicator of the absolute Z position [186]. This dependence was verified by placing the $^{55}Fe$ source at different positions along the drift direction of LIME and measuring the variation of $\xi$. The results, shown in Figure 2.22, reveal a clear dependence on Z, with the uncertainty increasing from 4 cm to 8 cm for larger drift distances. This preliminary study demonstrated the feasibility of estimating the absolute Z position of an event in LIME with moderate accuracy. A similar study was performed for the writing of this thesis using alpha particles instead of $^{55}Fe$-induced ERs (Section 5.2.3). The results from that analysis provided important information for inferring the location of alpha-emitting materials within the detector.

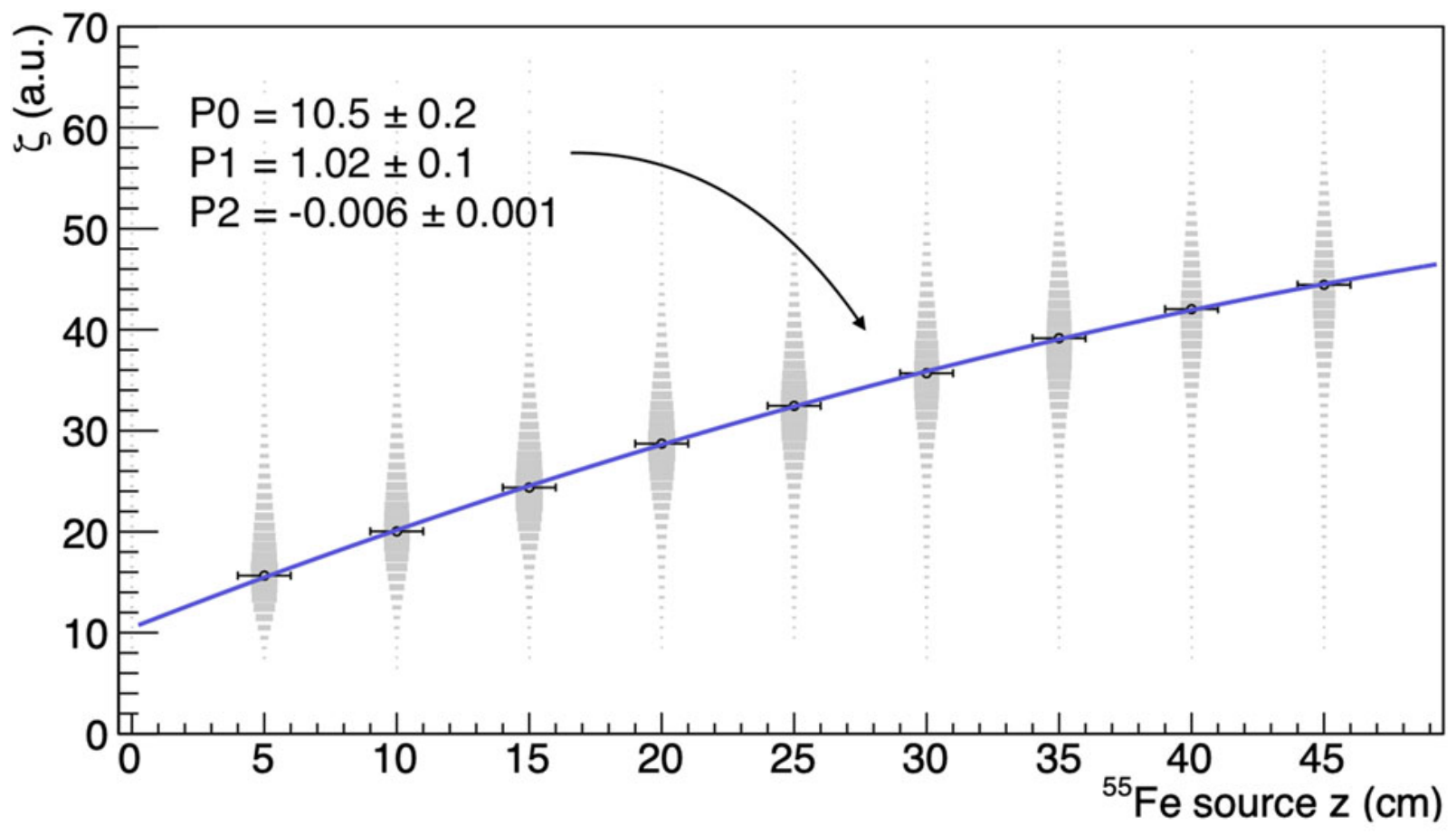

**Figure 2.22:** Distribution of the ξ variable for $^{55}$Fe spots at different drift distances, highlighting its dependence on the event's absolute Z position.

#### 2.2.2.4 Underground commissioning

One of the main objectives of LIME, within the CYGNO timeline (Figure 2.13), was to evaluate its performance in a low-background environment and provide an initial assessment of the effective background expected in CYGNO detectors. Such measurements are essential for understanding the overall sensitivity of CYGNO detectors to DM searches, and to optimize the data collection, reconstruction, and analysis processes. Additionally, the data obtained from LIME in this environment can be used to validate CYGNO's Monte Carlo simulations, while refining the current background models and predictions for future operations, runs, and detectors.

This section details LIME's underground commissioning, operation, and data-taking procedures. The findings concerning LIME's internal and external background, as well as its performance in a low-background environment, were crucial for the overall advancement of the CYGNO project. Additionally, these results, presented in more detail in Section 6.1, provided the basis for several studies conducted in this thesis, particularly the alpha 3D reconstruction and background analyses described in Sections 5 and 6, respectively.

Since 2022, LIME has been installed underground in a two-story container to fulfill its objective of operating in a low-background environment. The upper floor houses the electronic components, including the DAQ system, high-voltage modules, and other hardware required for the operation of LIME and its sensors. The lower floor contains the LIME detector itself and its shielding, and environmental factors in this room – such as temperature, humidity, and light – are carefully monitored. In this context, the deployment of LIME underground also provided an opportunity to test the CYGNO DAQ, slow control, and gas system during extended periods of data-taking.

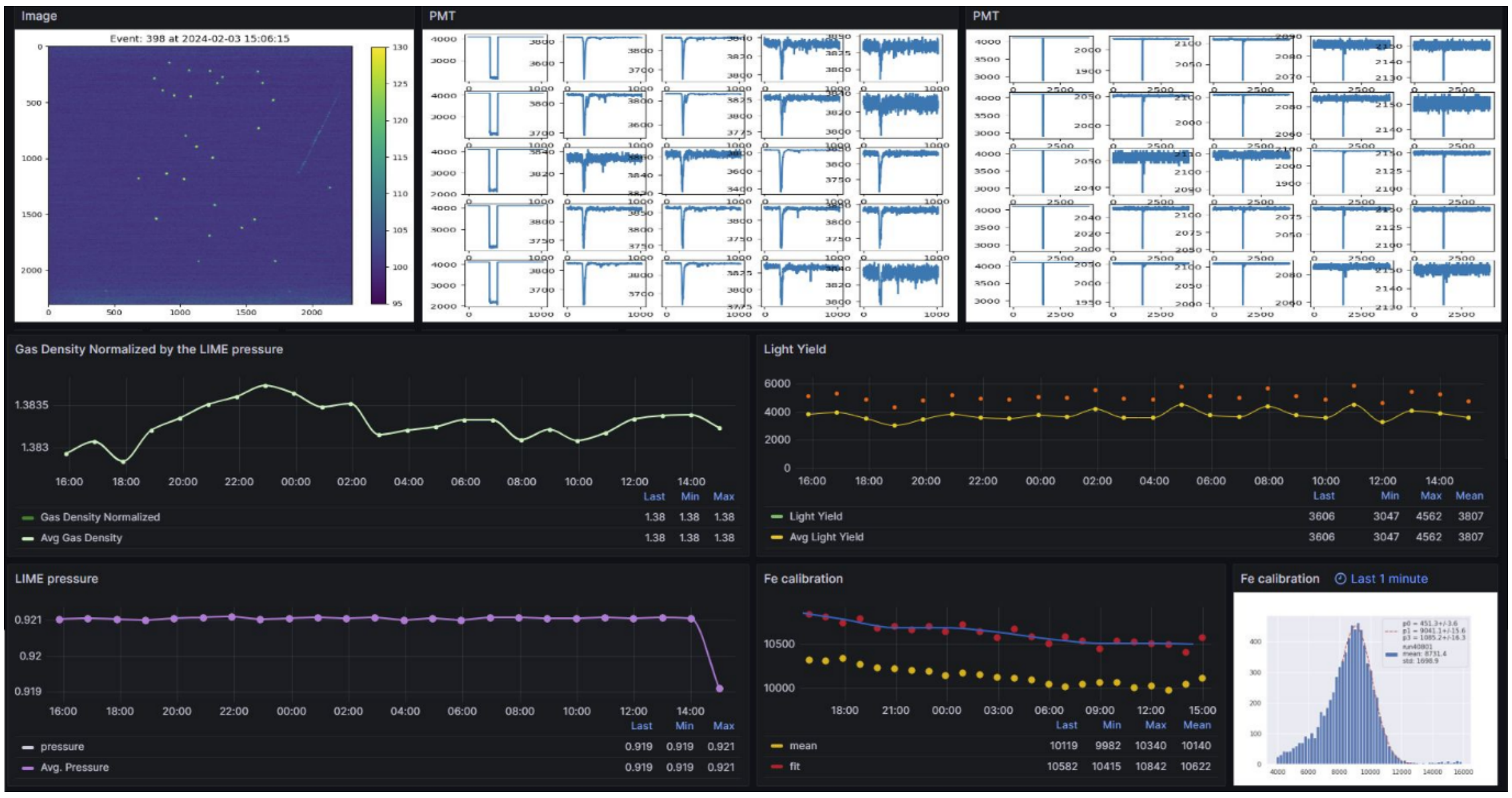

**Figure 2.23:** Example of the CYGNO underground live monitor panels. At the top, a CMOS image is displayed, showing multiple spot-like events, likely from an ongoing calibration run with an $^{55}Fe$ source. To the right, the multiple PMT waveforms from different triggers and digitizers are visible. Below, various real-time monitoring properties of LIME's operation are shown, including gas density, pressure, and the $^{55}Fe$ light yield, a typical parameter used to evaluate gas quality.

#### 2.2.2.4.1 Slow control

LIME's operation is managed by an open-source C++ slow control software built on the MIDAS framework [188]. This software interfaces with all the sensors and electronics of LIME, allowing for the configuration of detector operating conditions, monitoring of sensor performance, and ensuring stable functioning during data collection.

In the context of CYGNO, the software controls the high-voltage supplies for the PMTs, cathode, and GEMs, tracking their currents to detect instabilities. In the event of issues or system malfunctions, it triggers recovery procedures to minimize downtime. The system also monitors temperature and pressure sensors located in the LIME room on the ground floor, ensuring stable operational conditions and detecting any potential gas leaks. Additionally, it manages gas flow and recirculation protocols, while handling error messages that are forwarded to the experiments's DAQ experts for resolution.

The information retrieved by the slow control system can be monitored using Grafana, an open-source software for data analytics [189], as shown in Figure 2.23. Preliminary analysis of PMT waveforms and CMOS images are also conducted in real time to assess data quality. Key indicators, such as the number of image clusters and photon count, are used to monitor light yield, gas purity, and electronic stability.

#### 2.2.2.4.2 Gas system

An external gas handling system is required for the operation of LIME. This system performs several functions, including maintaining the appropriate 60:40 $He{:}CF_4$ gas mixture, chemical purification through filters, recirculation, and gas recovery. The system components are mounted on a cart connected to two large gas bottles for supply ($He$ and $CF_4$) and two smaller ones for gas recovery. The recovery system is designed to collect and store all used/wasted gas, ensuring environmentally responsible disposal. To minimize the entry of external impurities, LIME is operated slightly above atmospheric pressure (4 $mbar$). Since atmospheric pressure varies with altitude, the nominal pressure of LIME underground is between 910 and 920 $mbar$ at LNGS, which is at a slightly higher altitude than LNF (approximately 1000 $m$ versus 200 $m$), where the overground tests were conducted.

Initially, LIME was operated with continuous pure new gas flux, where the gas was flushed directly from the new gas bottles through LIME and into the recovery bottles. However, after July 2023, a recirculation system was introduced. This system routes the outgoing gas from LIME through a purification system and then returns it back to LIME. Approximately 80% of the total gas used in LIME is now recirculated, with the remaining 20% consisting of fresh gas. The introduction of fresh gas is essential to maintain a minimum level of purity and high light yield in LIME. This rate of new gas flux has been reduced over the months due to improvements in the sealing of the gas system components, as well as the addition of new filters to prevent the introduction of impurities into the system. As will be discussed in Section 6.1, the modification of the gas system's operating mode and the introduction of specific filters has significantly impacted the measured background activity of alpha particles and the overall light yield. In future phases of the CYGNO experiment, low-radioactivity molecular sieves, currently being developed by collaborators from the University of Sheffield [190,191], will be employed and tested to further enhance the purification process.

One of the key improvements in the LIME gas system during Run 4 (see Section 2.2.2.4.3) was the introduction of two gas filters: the Supelpure-O Oxygen/Moisture Trap [192] and the Molecular Sieve 5A Moisture Trap [193]. These filters are commonly referred to as the *red* and *blue* filters within the collaboration, based on their physical colors, as shown in Figure 2.24. The characteristics and effectiveness of these filters can be summarized as follows:

- **Supelpure-O Oxygen/Moisture Trap:** This filter combines materials designed to remove both oxygen and moisture from gas streams. It typically contains oxygen scavengers, such as reduced metals or catalysts, which chemically react with oxygen to eliminate it. Additionally, it includes molecular sieve materials that capture moisture. In the case of LIME, this filter is primarily used to remove oxygen, while a separate moisture filter is used for moisture removal. Oxygen can cause oxida-

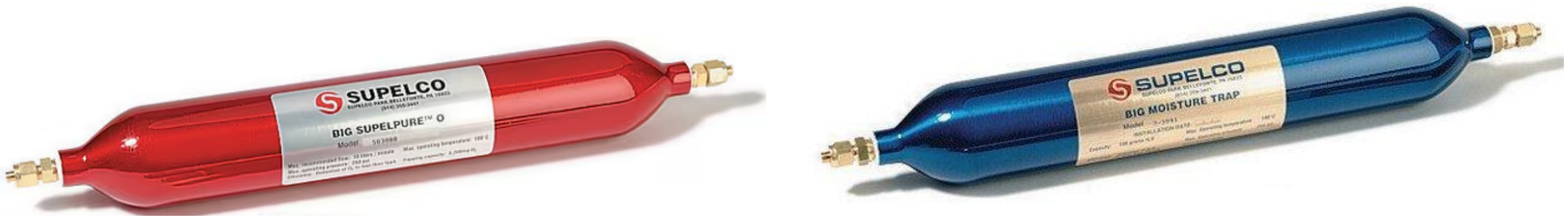

**(a)** Oxygen (red) filter

**(b)** Moisture (blue) filter

**Figure 2.24:** Filters implemented in the underground LIME gas system in order to remove the presence of (a) oxygen, and (b) moisture in LIME.

tion, equipment degradation, and other undesirable chemical reactions. This type of filter is commonly employed in systems that require a pure, inert atmosphere, such as in TPC-like systems.

- **Molecular Sieve 5A Moisture Trap:** This moisture trap uses synthetic zeolite with a uniform pore size of 5 angstroms. The molecular sieve operates through physical adsorption, where water molecules are captured and held within the porous structure. Its pores selectively adsorb molecules based on size, allowing it to trap water molecules while permitting other gases to pass through. This filter is essential in removing moisture from gas streams and is particularly important in systems where ultra-dry gas is required. In the context of gas TPCs, such as LIME, maintaining a moisture-free environment is crucial for stable operating conditions. In LIME, it has also been observed that light yield decreases as humidity increases, making this filter a significant improvement in the detector's overall performance.

One downside of the red filter, due to its chemical nature, is that it has a finite lifetime, and its regeneration process can be complex. Regeneration involves "cleaning" the filter as it becomes chemically saturated. In contrast, the blue filter, dedicated solely to moisture removal, has a much longer lifespan and can be regenerated through a simpler vacuum-pumping process. Additionally, it was observed that the amount of radon inside the detector decreased after the blue filter was installed (see Chapter 6 for details). This was further confirmed by introducing a third filter in series with the blue filter to reduce radon concentration, which resulted in a negligible further reduction of Rn. This led to the conclusion that the majority of radon removal was occurring through the blue filter. The hypothesis is that radon and its progeny attach to the water vapor that sips inside LIME [194], and then decay inside the detector. The reduction in radon levels can be tracked by monitoring the rate of alpha particles detected in LIME. The study of the rate, origin, and other 3D properties of alpha particles is one of the key topics of discussion of this thesis, and will be presented in Chapter 6.

#### 2.2.2.4.3 Runs & Objectives

The data acquisition strategy for LIME was designed to perform a comprehensive comparison between measured background and MC predictions, under different conditions, and to test the effectiveness of the external shielding strategy. In particular, the various science runs were conducted with progressively increasing shielding thicknesses, as illustrated in the picture of the setup shown in Figure 2.25. Combined with parallel tests performed using other radioactive sources ($^{55}\mathrm{Fe}$, $^{137}\mathrm{Ba}$, $^{152}\mathrm{Eu}$, and $^{241}\mathrm{Am}^{9}\mathrm{Be}$), this approach provided important conclusions regarding not only the LIME background and the validity of the MC models, but also the performance of LIME and the CYGNO technique under different operating conditions and for various types of particles and interactions. The different stages of data collection are summarized in Table 2.2, and can be outlined as follows:

- **Run 1:** Following the commissioning, Run 1 focused on the study and characterization of external backgrounds. For this run, no shielding was used, and LIME was kept in its original configuration: a gas region enclosed in a plexiglass box, all housed within a Faraday metal cage.
- **Run 2:** In Run 2, a 4 $\mathrm{cm}$ thick copper shielding was installed around LIME. This led to a noticeable reduction in image occupancy and event rate. At this stage, the background caused by external gamma radiation was estimated to have been reduced by a factor of approximately 40 times.
- **Run 3:** For Run 3, an additional layer of copper bars was installed around LIME, increasing the total shielding thickness to 10 $\mathrm{cm}$. With this enhanced shielding, the external gamma radiation background became comparable to the internal background. This phase marked the first extended data-taking period, lasting several months without interruption. The gas system recirculation parameters and filtering were also optimized during Run3, causing significant changes in the light yield and alpha background rate during the Run, as discussed in detail in Section 6.1.3. While detrimental to the stability of the detector response and the quality of the data, these tests were of the utmost importance, as they allowed for the development of an optimized and stable gas system configuration by the end of the Run. This configuration was subsequently employed successfully in Run4 and Run5, ensuring high-quality data for these two physics Runs. During Run 3, measurements were also carried out using an AmBe neutron source. The objective of this test was to optimize signal identification and background rejection using a clean sample of NRs. Unfortunately, a failure in the gas system reduced the AmBe data taking to less than 2 days of data, enough to provide a clear indication of NR detection but too small to be systematically analyzed.
- **Run 4:** Several water tanks were installed for Run 4, creating a 40 $\mathrm{cm}$ thick barrier. Water serves as an effective moderator for environmental neutrons, significantly

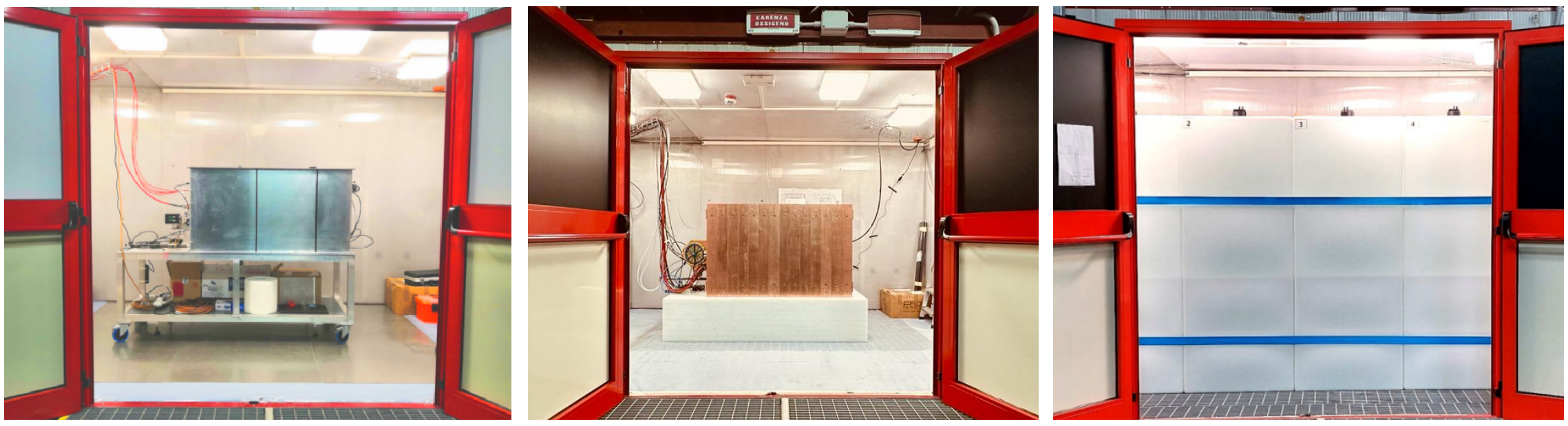

**(a)** Run 1 **(b)** Run 2 **(c)** Run 4

**Figure 2.25:** LIME in the underground LNGS cabinet during different runs. In Run 1 (a), no shielding was used; in Run 2 (b), a 4 cm copper layer was placed around LIME, which was later increased to 10 cm in Run 3. In Run 4 (c), 40 cm of water shielding were added around LIME. In Run 5, the layout reverted to 10 cm of copper shielding, as in Run 3.

reducing the external neutron-induced background. This setup reduces the external neutron flux contribution to about 2 NR/year, allowing a more accurate characterization of the internal neutron recoil background. The remaining tests conducted during Run 4 focused on the introduction of gas filters to minimize the impurities within the gas. As discussed in Section 2.2.2.4.2, this had the added benefit of reducing the alpha background by lowering the amount of radon inside LIME.

- **Run 5:** The main objective of Run 5, achievable only without the water shielding, is to perform a spectral and directional measurement of the underground LNGS neutron flux, as part of the project "Zero Radioactivity for Future Experiments" (PRIN Prot. 2017T54J9J). Based on current CYGNO background predictions, a quasi-background-free dataset is expected at energies above 20 keV, owing to the achieved ER-NR discrimination capabilities [195]. Preliminary simulations suggest an expected number of external neutron recoil events of about 250 over 6 months, which is significantly larger than the expected 19 internal background originated events in one year (after data cuts) [196]. This provides an opportunity to improve the current understanding of underground neutron flux at LNGS [86, 197, 198], particularly at low energies ($<$ 1 MeV), and to conduct an in-depth study of CYGNO's capabilities in 3D-reconstruction, directionality, and energy resolution for NR events, the expected signal from WIMPS in CYGNO detectors.

The LIME detector has been operational for over two years, during which a large amount of data has been collected. During the analysis of the data from Runs 2 and 3, we observed discrepancies between observed data and MC simulations which lead to a more in-depth study of the gas quality and related backgrounds, as discussed further in detail in Chapter 6. The analysis presented in this thesis primarily uses data from Run 4, which provided high statistics and stable gas and light yield conditions, resulting in a particularly clean dataset. These improvements were made possible by upgrades to the gas system – including improved piping and the introduction of two new filters, as described in Section 2.2.2.4.2 – as well as improved real-time monitoring of gas properties and light yield.

**Table 2.2:** Summary of the LIME underground data-taking campaign, including the dates, number of runs, and types of shielding used. Each recorded run consists of 400 pictures, each containing at least one triggered signal observed by the PMTs (for more details on the trigger, see Section 2.2.2.2).

| Phase | Dates | # runs recorded | Shielding |
|---|---|---|---|
| Run 1 | 08/01/2022 – 06/12/2022 | 3744 | None |
| Run 2 | 15/02/2023 – 09/03/2023 | 3487 | 4 cm Cu |
| Run 3 | 05/05/2023 – 16/11/2023 | 22276 | 10 cm Cu |
| Run 4 | 30/11/2023 – 22/04/2024 | 16099 | 10 cm Cu + 40 cm $H_20$ |
| Run 5 | 17/05/2024 – 31/12/2024 | >25000 | 10 cm Cu |

### 2.2.3 CYGNO-04

CYGNO-04 represents *Phase_1* of the CYGNO project, as shown in Figure 2.13, and involves the deployment of a 0.4 $m^3$ detector. This detector will serve as the first demonstrator for the large-scale experiment outlined in Section 2.2.4.

The CYGNO-04 design comprises two back-to-back TPCs with a central cathode. This detector layout is commonly seen in large detectors as it creates two independent TPCs that share a cathode, effectively minimizing the amount of detector material. With this strategy, the total detector active volume is also optimized with respect to the drift distance, as this is now halved in each of the twin TPCs. A smaller drift distance is preferred since: 1) it reduces the voltage required to be applied to the cathode to achieve the desired electric field; and 2) the effects of the diffusion suffered by the electrons while drifting are reduced, which in turn improves the 3D tracking and directionality capabilities of the detector.

The current design of CYGNO-04 is shown in an exploded view in Figure 2.26a. The internal PMMA vessel containing the gas is shown in light blue. Following the preliminary studies reported [196], validated and extensively characterized in this thesis in Chapter 6, a double sealed scheme is envisioned for CYGNO-04, where the ultra-pure copper is employed as a second gas vessel to minimize leakage and the diffusion of ambient radon in the active volume of the detector. This copper layer – which also works for background suppression – is depicted in Figure 2.26c as green and brown panels, which stand for regular and ultra-pure copper, respectively. The detector features a shared cathode constructed from a 50 μm Kapton foil clad with 5 μm of copper on both sides, resembling the structure of a GEM (see Section 2.1.2). As shown in Figure 2.26d, each of the twin TPCs will be read out by three ORCA-Quest cameras (see Table 2.1), evenly distributed across the imaging area, namely the $50 \times 80\ \mathrm{cm}^2$ GEMs used for amplification and light production. Compared with their predecessor, these camera models offer lower noise and higher resolution, featuring smaller pixels and larger sensors. Additionally, each side will be instrumented with eight PMTs positioned at the corners around the cameras and facing the GEMs, following a similar layout as LIME. The number and layout of sensors were optimized based on the required spatial coverage and granularity, while also considering cost-efficiency and spatial constraints. With the final dimensions of CYGNO-04, the expected granularity per image pixel is $125 \times 125\ \mu\mathrm{m}^2$, representing an approximate 25% improvement over the resolution achieved in LIME.

The strategies for reducing radioactive background in CYGNO-04 are based on the background simulations performed with LIME, as detailed in di Giambattista's thesis [196]. These studies identified the primary sources of internal background as the copper used in the drift field rings and cathode, ceramic resistors, GEMs, and the camera lenses. The measured contributions from these components are presented in detail in Table 6.1. Among all background sources, those coming from the decay chain of $^{238}\mathrm{U}$ were found to be the most prominent, as this isotope is present in relatively high concentrations in nearly all of the mentioned detector materials. This topic is further revisited in Chapter 6, in the context of the studies conducted on the alpha background observed in the LIME underground campaign.

As part of the effort to minimize potential backgrounds, several design improvements over LIME have been implemented in the development of CYGNO-04. One significant upgrade is the use of a customized lens made from ultra-pure silica (SUPRASIL), developed in collaboration with external partners, aiming to reduce the background contribution from the lens by a factor of $10^4$. The GEMs will be cleaned using high-pressure deionized water baths, a technique known to reduce surface radioactive contamination and already employed in other experiments, such as the T-REX collaboration [199]. The drift field cage will be made from a thin Kapton foil with copper strips interconnected by SMD[3] resistors, surrounding the active volume, as shown in the detector cross-section in Figure 2.26b. This design reduces the amount of copper and ceramic resistors in contact with the gas, thereby reducing internal backgrounds. In addition, all remaining components of CYGNO-04 will be constructed using radio-pure materials, with their activities individually measured and verified. The planned external shielding for this phase consists of 10 cm of copper, surrounded by 100 cm of water, as illustrated in Figure 2.26e, and is expected to reduce the contribution of external backgrounds by a factor of 20 relative to the internal one.

CYGNO-04 will test the scalability of the CYGNO technique on a realistic scale in several fronts, also including mechanical and ancillary systems, where ongoing tests are being carried out addressing aspects such as material deformation, internal power distribution, and GEM stretching. The data acquisition system for CYGNO-04 will also be upgraded. In the scheme currently under development, the deadtime caused by the sensor's exposure time will be eliminated. This is made possible by operating the CMOS camera in *continuous acquisition* mode, where each sensor row is readout immediately after its exposure. This method, illustrated in Figure 2.27, allows for uninterrupted image capture. In this scheme, if a track is generated outside the global exposure window but crosses a row that is being readout, it will appear within two consecutive frames, and the combination of the frames enables the full reconstruction of the track. However, as the system scales to include a larger number of cameras and PMTs, precise synchronization between the various readout systems will become essential. To address this, the collaboration is currently developing a prototype for the upgraded DAQ system, which will be implemented in CYGNO-04.

[3] Surface mount device.

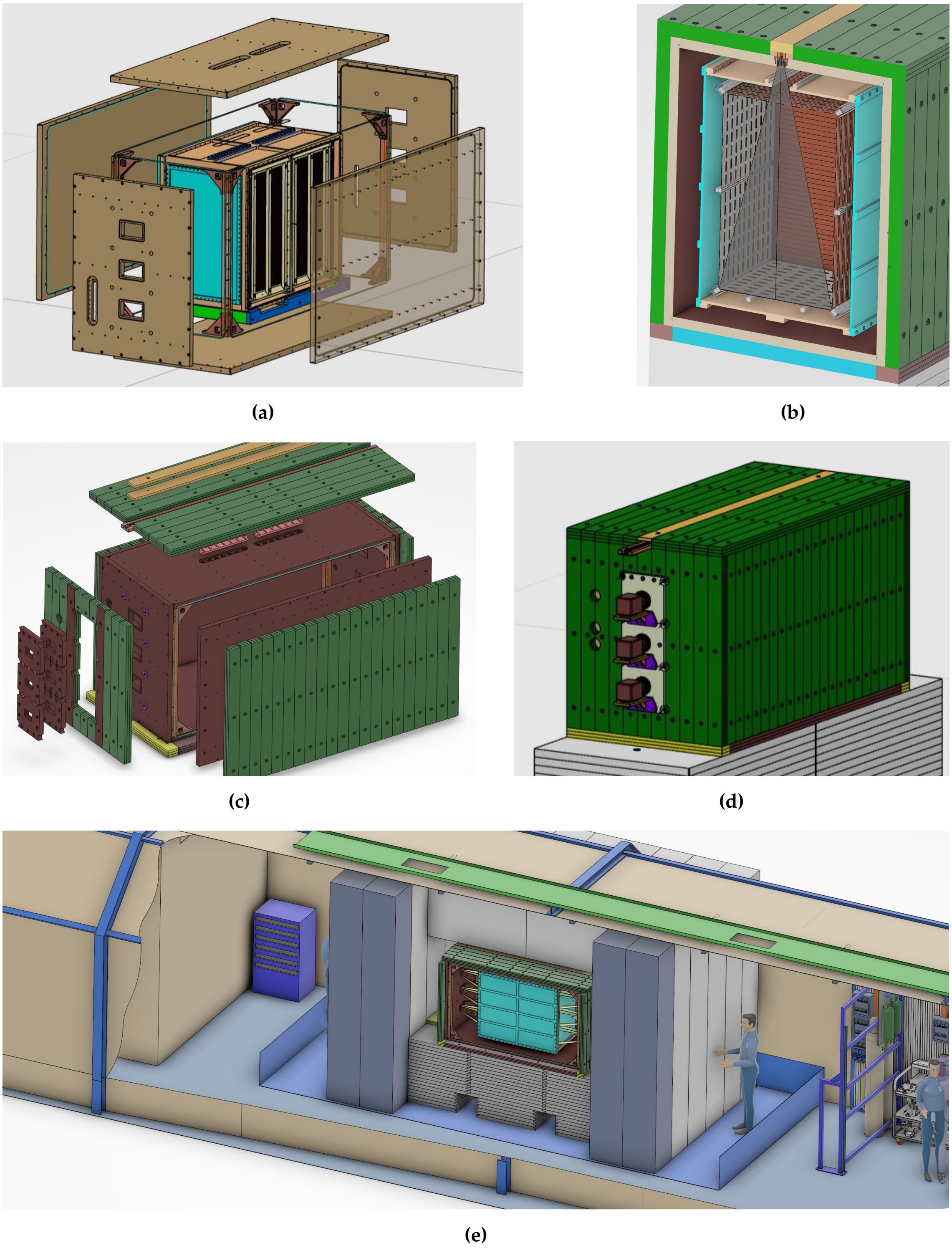

**Figure 2.26:** Technical designs of the CYGNO-04 detector: (a) internal exploded view showing the PMMA gas vessel in blue and the copper shielding in brown; (b) cross-section of the detector highlighting the internal field cage; (c) exploded view of the external background shielding, with green indicating standard copper and brown indicating ultra-pure copper; (d) closed external view, showing the configuration of the 3 CMOS cameras and 8 PMTs on each side of the vessel; (e) location at the underground LNGS facility where CYGNO-04 will be installed, including the surrounding water columns that form part of the external shielding.

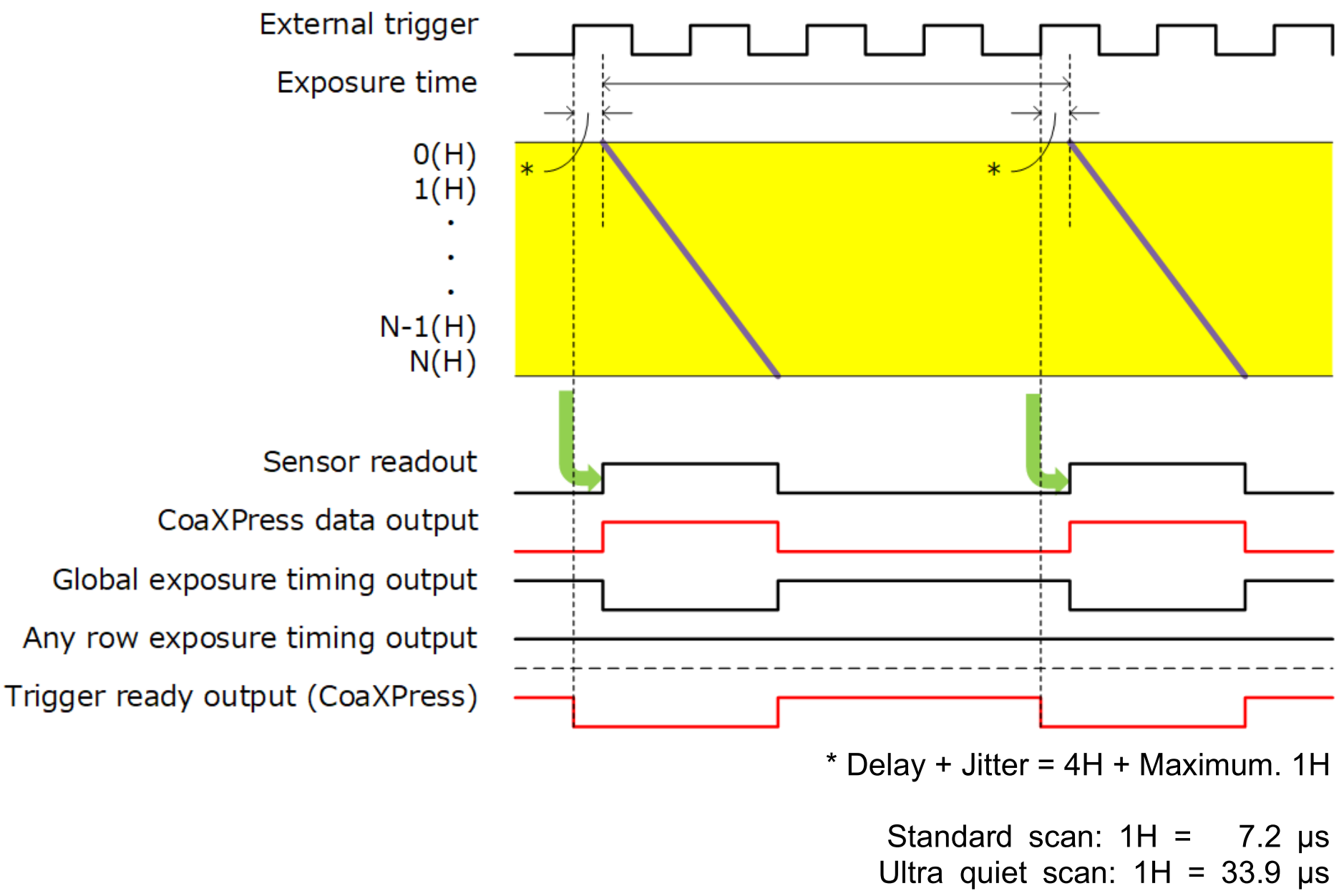

**Figure 2.27:** CMOS sensor readout schematic in *continuous acquisition* mode. Retrieved from [200].

Overall, CYGNO-04 is expected to demonstrate that the CYGNO/INITIUM project has the necessary expertise to construct a large, $\mathcal{O}(30-100)$ m$^3$, optically read out TPC with high granularity, with competitive sensitivity for DM searches. The use of low-radioactivity materials will provide important information into the actual sensitivity and physics reach of CYGNO detectors for WIMP interactions. The CYGNO-04 underground infrastructure in LNGS - Hall F has been completed in March 2025 and detector installation is foreseen to be finalized by end of 2025, with commissioning in early 2026.

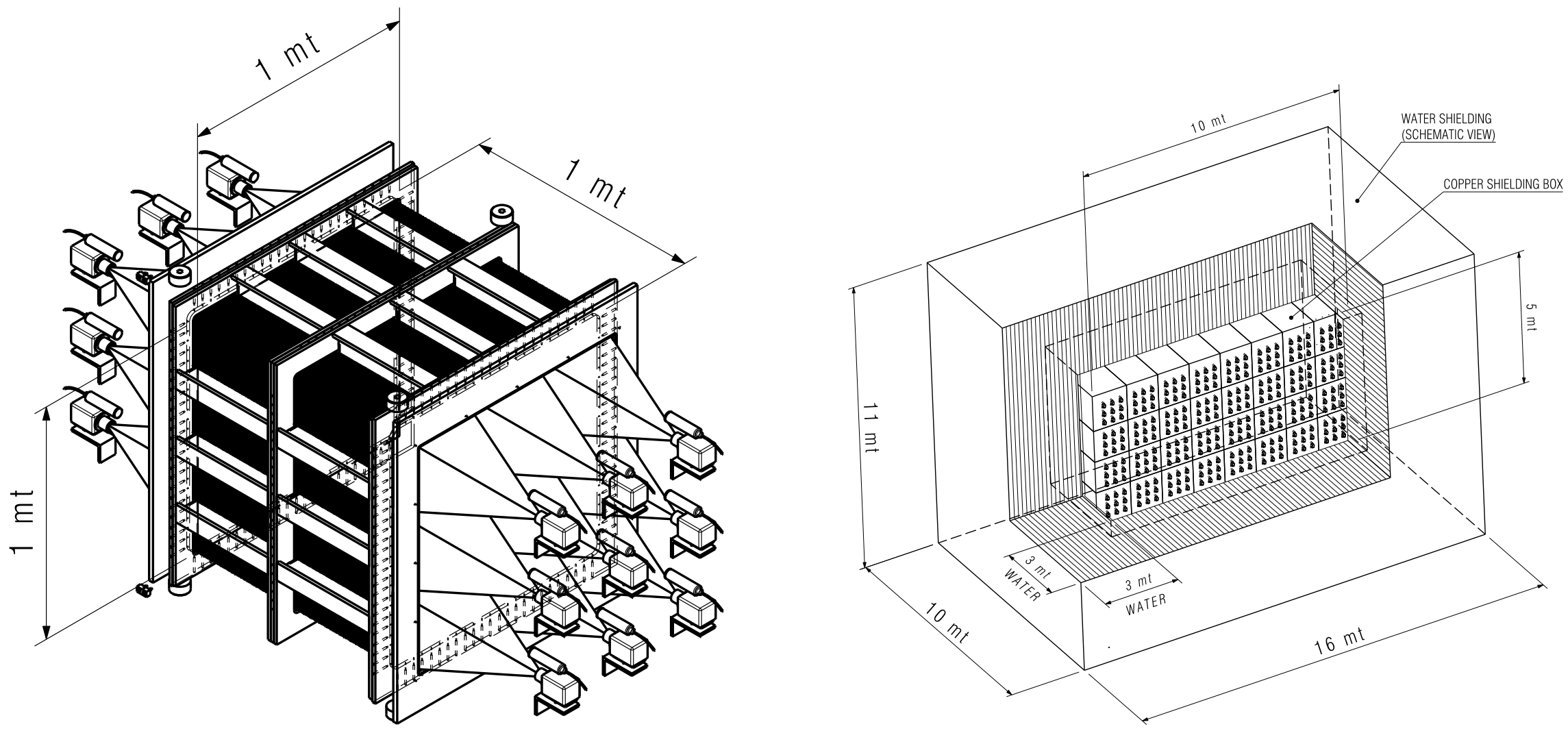

(a) (b)

**Figure 2.28:** Preliminary technical design of a possible CYGNO *Phase_2* detector, illustrating (a) the modular 1 $m^3$ units that could form (b) a large, wall-like detector configuration.

### 2.2.4 CYGNO-30

The goal of CYGNO-04 is to demonstrate the feasibility and physics potential of the CYGNO technique at a large and realistic scale. If the results are compelling and funding is secured, the next phase of the experiment would involve constructing a significantly larger detector, currently referred to as CYGNO-30, with a total volume on the order of tens of cubic meters. While the final design and geometry of CYGNO-30 are still under development, the current concept involves using a modular approach based on detectors with volumes of approximately $\mathcal{O}(1)$ $m^3$, such as CYGNO-04 or an larger variant, as shown in Figure 2.28a. These modules would be arranged side-by-side in a wall-like configuration. An example of such a layout is presented in Figure 2.28b, where an array of 8 $\times$ 4 modules, each with a volume of 1 $m^3$, achieves a total active volume close to 30 $m^3$ (thus the name CYGNO-30). This approach would simplify scaling, as it essentially involves reproducing and reading out the same detector multiple times.

CYGNO-30 has the potential to make a substantial impact in the search for dark matter WIMPs, particularly for masses below 10 GeV/$c^2$, in both SI and SD interaction channels. To estimate the projected exclusion limits for a 30 $m^3$ CYGNO-like detector, a Bayesian inference approach is employed to determine the upper bound on the number of DM events, under the assumption that the observed data are consistent with the expected background. A Monte Carlo method is used to generate pseudo-experiments across a range of background scenarios, with the number of background events drawn from a Poisson distribution with varying mean values. Each simulated event is assigned a direction: background events are assumed to be isotropic, as any intrinsic anisotropy would be smeared out in galactic coordinates due to Earth's rotation, while WIMP-induced events follow the anisotropic angular distribution predicted by the Standard Halo Model (see Section 1.3.2.3). These angular distributions are then smeared to account for the detector's finite angular resolution.

For each scenario, a binned likelihood function is constructed and fitted, profiling the angular distribution of the sample – assumed to contain both background and potential DM events. A 90% credible interval (C.I.) on the number of DM events is then derived within the Bayesian framework. The expected number of DM interactions is related to the SI and SD WIMP-nucleon cross sections through the nuclear recoil rate, which depends on detector-specific parameters such as gas composition, total active mass, exposure time, and energy threshold. Astrophysical assumptions, such as the local DM density and velocity distribution (Section 1.3.1.3), are also taken into account. By equating the 90% C.I. upper limit on the number of DM events to the predicted number for a given WIMP mass and cross section, it is possible to derive an exclusion curve in the WIMP mass – cross section parameter space [104].

The computed exclusion limits for SI and SD WIMP-nucleon couplings for a 30 $m^3$ detector with three years of exposure and assuming a 1 $keV_{ee}$ energy threshold are shown in Figure 2.29 [131]. This analysis assumes an angular resolution of 30 $deg^2$, full 3D directional reconstruction, and 100% head-tail recognition efficiency (see Section 1.4.2.1). The results are presented for varying numbers of background events ($10^2$ to $10^4$), and for two different energy thresholds (0.5 and 1 $keV_{ee}$).

For both SI and SD couplings, CYGNO-30 would be capable of probing regions of parameter space that remain unexplored. In the SI case, the main competitors are CRESST [201] and CDMSLite [202], which exhibit comparable projected discovery limits. At an energy threshold of 0.5 keV (bottom plot of Figure 2.29a), CYGNO can surpass these experiments, reaching sensitivities for WIMP masses as low as $\sim 0.7$ $GeV/c^2$. For SD interactions, CYGNO-30 would be competitive with PICO [203], currently the most sensitive experiment in this category. With an energy threshold of 0.5 keV, CYGNO-30 could already surpass PICO in sensitivity, as shown in the bottom plot of Figure 2.29b. A key advantage of a directional detector like CYGNO over conventional detectors comes from its ability to confirm the galactic origin of a dark matter signal through its directional signature, allowing for a decisive and unambiguous identification of a dark matter signal. The lower panels of the plots in Figure 2.29 also show the exclusion limits for a CYGNO-like detector using a gas mixture that includes isobutane ($iC_4H_{10}$) [204]. The inclusion of hydrogen significantly enhances sensitivity to low-mass WIMPs due to favorable kinematic matching, thus extending CYGNO-30's reach even more in this low-mass WIMP region.

Beyond dark matter searches, CYGNO-30 could also provide the first directional measurement of solar neutrinos from the pp-chain by exploring electron-neutrino scattering processes, extending and improving upon the results obtained by Borexino at lower energies [205]. Simulations suggest that a 20 keV detection threshold and directional capability correspond to an 80 keV neutrino energy, allowing CYGNO to probe

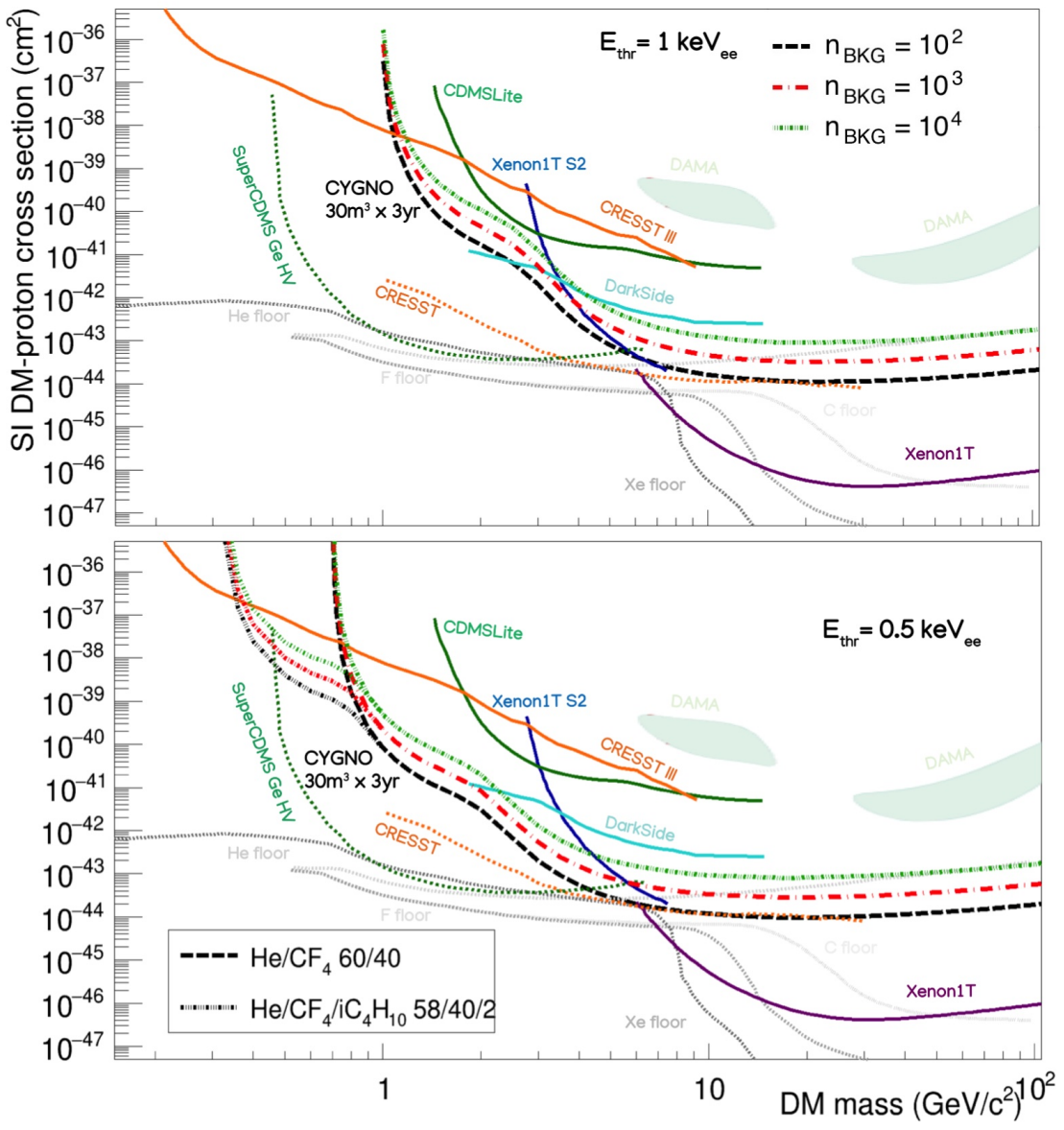

(a)

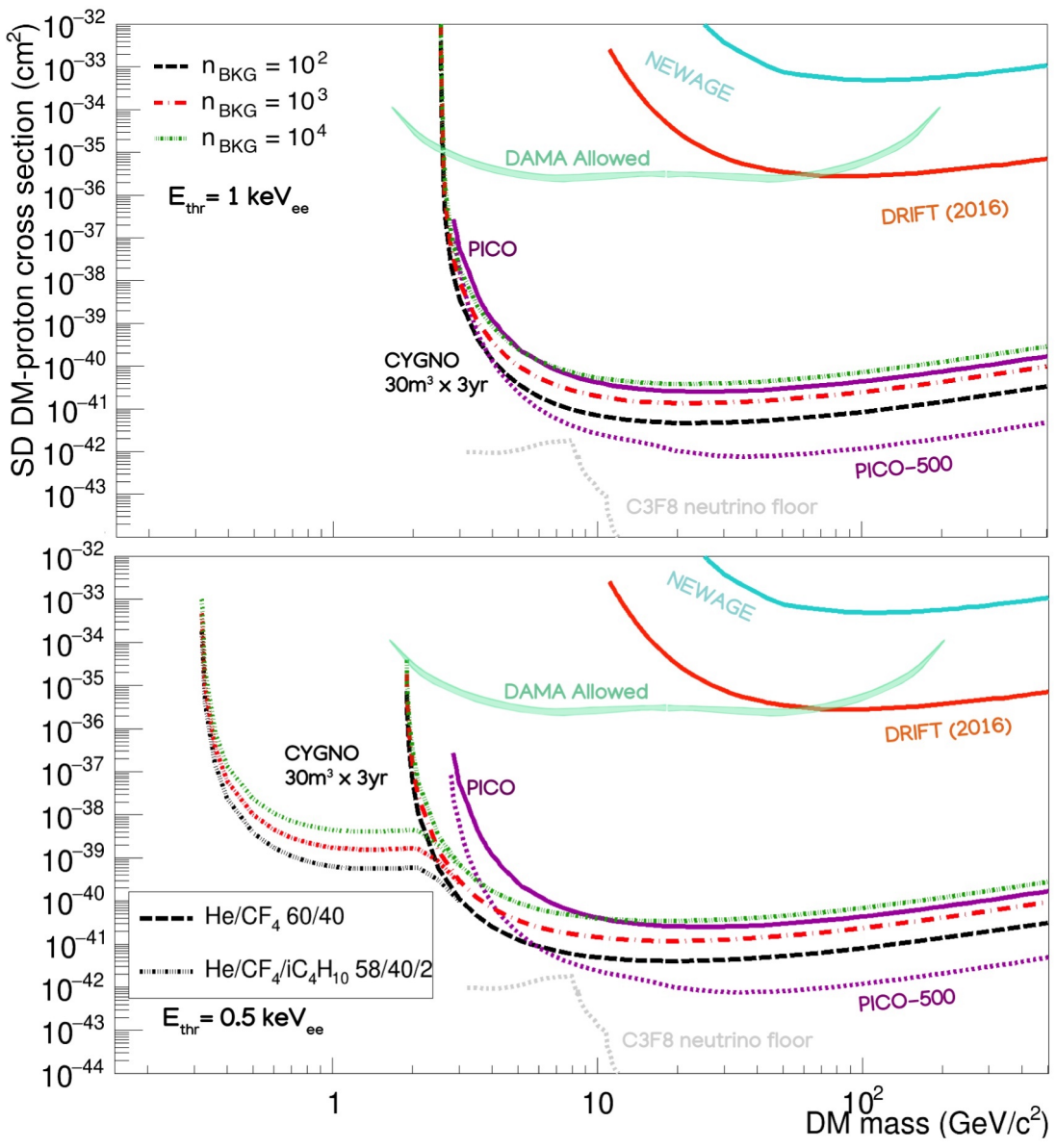

(b)

**Figure 2.29:** Spin-independent (a) and spin-dependent (b) 90% credible interval exclusion limits for CYGNO-30, assuming 3 years of exposure under different background scenarios and operational thresholds of 1 $keV_{ee}$ (top plots) and 0.5 $keV_{ee}$ (bottom plots). Dashed curves correspond to a He:$CF_4$ (60:40) gas mixture with 100 (black), 1000 (red), and 10.000 (dark green) background events. Dotted curves represent the sensitivity for a He:$CF_4$:isobutane (58:40:2) mixture. The remaining solid lines represent the currently published bounds from other experiments, while the dotted lines indicate the projected future limits. Figure adapted from [131] and references therein.

below the 160 keV neutrino energy threshold achieved by Borexino.

The feasibility of detecting solar neutrinos with a 30 $m^3$ CYGNO detector is explored in detail in Torelli's thesis [132]. In this study, a Bayesian analysis framework is adopted, incorporating both the energy and angular distributions of signal and background events, as estimated from LIME data and LIME simulation, respectively. The expected interaction rate in the CYGNO gas mixture for pp neutrinos is approximately 1 event/$m^3$/year, corresponding to roughly 30 signal events annually in a CYGNO-30 detector. A detailed GEANT4 simulation was used to characterize the internal background. Signal and background templates were then constructed based on the detector's angular and energy resolution, and employed in a Bayesian analysis of toy Monte Carlo datasets across various exposure durations and background levels.

The discovery potential was evaluated using the Bayes factor. Results indicate that a 3$\sigma$ detection of pp neutrinos down to 55 keV in neutrino energy (to be compared to the 300 keV Borexino threshold [205]) could be achieved with a 50% probability after 5.5 years of data-taking, assuming a Bayes factor exceeding 20 and internal background levels kept below 1760 events per year. Furthermore, the analysis shows that CYGNO-30 could tolerate background rates up to 60 times higher than the expected signal, corresponding to a noise-to-signal tolerance roughly 23 times greater than that of Borexino. These results underscore the advantages of directional detection and demonstrate the strong potential of CYGNO-30 for competitive solar neutrino measurements.

Overall, the studies presented in [132] provide compelling evidence that a large-scale CYGNO detector could uniquely combine sensitivity to both dark matter and low-energy solar neutrinos, underscoring the versatility of gas-based TPCs with optical readout and directional detection capabilities.

## 2.3 INITIUM

The **I**nnovative **N**egative **I**on **TI**me projection chamber for **U**nderground dark **M**atter searches (INITIUM) is an ERC[4] Consolidator Grant, aimed at developing and optimizing Negative Ion Drift (NID) operation for TPCs, within the CYGNO 3D optical TPC approach.

Negative ion drift operation is a modification of conventional TPC approach that involves introducing a highly electronegative dopant into the gas [178]. In this setup, primary electrons generated by an ionizing particle are rapidly captured by electronegative molecules (typically within distances of 10 – 100 µm), forming negative ions. These anions then drift towards the anode, where their extra electron is removed, leading to a standard electron avalanche. Unlike the traditional approach, negative ions have much higher masses than electrons, which significantly reduces their diffusion during drift, thus better preserving the original shape of the ionization cluster. In addition, minority charge carriers can be produced alongside the primary ions during the initial ionization.

[4]The ERC is the premier European funding organization for excellent frontier research (https://erc.europa.eu/homepage). ERC grant agreement No 818744.

These secondary ions, having different masses, exhibit different mobilities. By measuring the difference in arrival times of the various anions, it is possible to determine the event position along the drift direction [122]. This capability is a powerful tool for TPCs, as it allows for precise fiducialization of the detector, which is particularly important in rare-event searches.

The goal of INITIUM is to develop a $\mathcal{O}(1)$ m$^3$ gas TPC filled with a scintillating He:$CF_4$:$SF_6$-based mixture at atmospheric pressure for NID with optical readout, effectively creating a Negative Ion TPC (NITPC). $SF_6$ is particularly attractive due to its ability to form multiple minority charge carriers – $SF_X^-$ with $X \in \{5, 4, 3, ...\}$ [206] – as well as its non-toxicity and ease of handling, making it well-suited for integration with existing gas installations. If NID can be successfully implemented within the optical approach, the reduced diffusion and potential fiducialization effects, combined with the high-resolution readout of CYGNO, would significantly enhance the experiment's 3D tracking and directional capabilities.

In this thesis, the topics discussed within the INITIUM framework focus on the first ever operation in NID configuration of an optical readout TPC and on the analysis of the relative PMT signals, which, to the best of the author's knowledge, have not been observed or published by other groups. The methods developed for analyzing the peculiar NID PMT waveforms are presented in Chapter 7, along with preliminary results derived from these analyses.

Chapter 3

# Data Reconstruction & Analysis

The CYGNO experimental approach consists of a TPC in which electrons from primary ionization clusters are drifted towards a GEM plane, where they are then amplified, producing light in the process. This light is recorded by a CMOS camera and PMTs, producing two very distinct types of data – a 2D image and a 1D waveform, respectively – that require separate reconstruction approaches. The feasibility of a combined sensor readout for 3D tracking was initially demonstrated in [134]. In initial tests with small datasets, the measurement of the arrival time of the main clusters of a high-energy electron allowed an independent reconstruction of their absolute position in the drift direction, with an evaluated resolution on the reconstructed Z coordinate of about 100 $\mu m$.

Differently from the initial R&D prototypes of CYGNO (Section 2.2), which featured only a single PMT, the full realization of the CYGNO experiment (CYGNO-04 and its demonstrator, LIME) features several PMTs placed around the CMOS sensor, all looking at the same plane of light production. This allows the use of the light barycenter, obtained from the relative amplitudes between PMTs, to infer additional information about the track. Therefore, it was thanks to the availability of a long and thorough overground and underground LIME commissioning in the last couple of years that a systematic development of PMT signal reconstruction and its integration with CMOS analysis was possible. This became feasible only during the time of this PhD, motivating the large focus of this thesis on the analysis of multiple and combined PMT signals, and their combination with the CMOS sensor for the 3D reconstruction of ionization tracks.

To highlight the features of the signals recorded in a CYGNO detector, a typical event obtained with the LIME detector is shown in Figure 3.1, where both a CMOS image and the corresponding PMT waveforms are visible. An event in LIME consists of a combination of a camera image and a set of PMT signals. In LIME, the DAQ operates with a 300 $ms$ triggerless camera exposure while continuously monitoring the PMTs. A trigger is issued when at least two PMTs simultaneously cross a predefined threshold, resulting in the storage of the sCMOS image along with the corresponding PMT waveforms. Multiple PMT triggers (waveforms) within a single image exposure are linked to the same image (see Section 2.2.2.2.1).

This example signal shows the typical three types of events observed: an alpha particle (highly ionizing, dense track) at the leftmost part of the image (Figure 3.1a), which

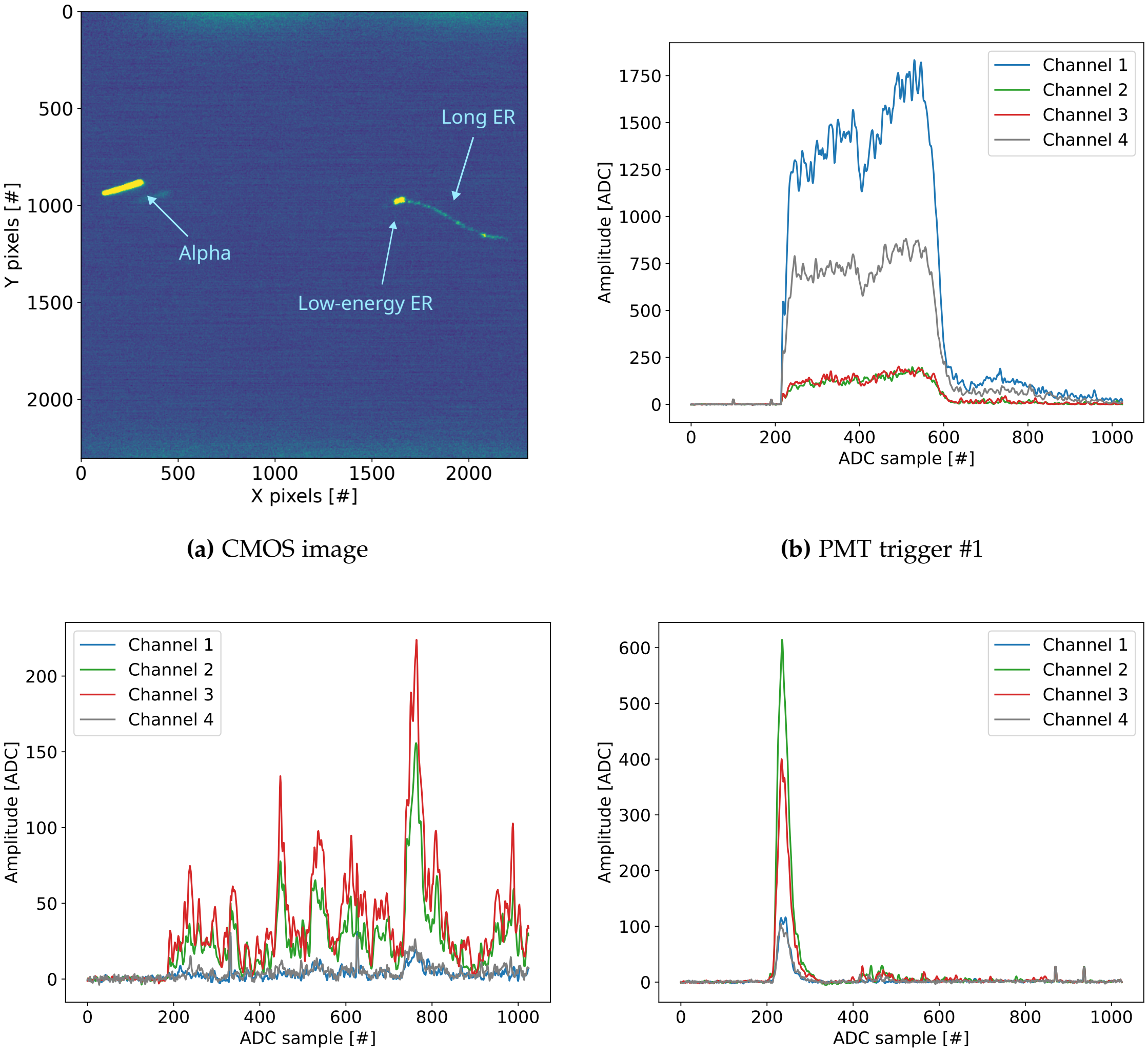

**(a)** CMOS image

**(b)** PMT trigger #1

**(c)** PMT trigger #2

**(d)** PMT trigger #3

**Figure 3.1:** Example of the CMOS and PMT information contained in an event data bank acquired with LIME: (a) CMOS image showing, presumably, three ionization events; (b), (c), and (d) show the waveforms acquired by the four PMTs for each of the triggered events in the CMOS image. Visually, an alpha particle is observed in the left part of the image, corresponding to trigger (b); a long ER/MIP-like event is visible on the right, corresponding to trigger (c); and *possibly* a low-energy ER is located around $(x, y) = (1500, 1000)$, corresponding to trigger (d).

corresponds to PMT trigger #1 (Figure 3.1b), assumed given its high-amplitude, sustained signal; a long ER at the rightmost part of the image, showing a low ionization profile, associated with trigger #2 (Figure 3.1c), given its extended, multi-peak waveform; and a spot-like event (likely a low-energy ER) visible at the end of the longer ER track, associated with the simple, single-peak waveform of trigger #3, which reflects a short (spot-like) signal. The three different PMT signals also highlight how these allow to distinguish events that could occupy the same position on the CMOS camera image, as the extended and low energy ERs on the right.

To analyze these ionization tracks, the information must first be extracted, or *reconstructed*. Starting with the CMOS, the reconstruction algorithm was developed to extract various characteristics of the interactions, such as energy, length, and stopping power ($dE/dx$), among others. In contrast, the PMTs provide the temporal profiles of the tracks, representing their path in the drift direction. In this chapter, a descriptive summary of the current status of the *CMOS-Reco* developed by the collaboration will be provided for context (Section 3.1). This section summarizes the procedure behind the reconstruction of information from CMOS images and the key variables retrieved, which are then used in subsequent analysis in this thesis. The PMT-Reco, instead, will be thoroughly discussed (Section 3.2), as it represents a significant portion of the work developed for this thesis. The discussion will focus on the methods and routines developed by the author to reconstruct important information not only from single PMT waveforms, but also from the combined analysis of multiple PMT signals. Finally, a discussion on the current methods used for matching the two sensors' information is presented (Section 3.3). To our knowledge, this is the first time a full 3D reconstruction has been developed by merging CMOS and PMT signals in an optical TPC.

## 3.1 CMOS

Every picture from the CMOS is effectively a matrix of pixels, each with an associated integer representing the ADC counts recorded by each pixel. The intensity and spatial arrangement of the pixels linked to an ionizing event provide information about the type of particle that crossed or interacted in the detector, as well as its energy. Tracks produced in the active gas volume of CYGNO detectors create different ionization patterns, which are recorded by the sCMOS camera. These patterns vary in both shape and light intensity. Minimum Ionizing Particles (MIPs) typically leave long, faint trails. In contrast, low-energy electrons (tens of keV) produce curly tracks that become denser toward their endpoints. Nuclear recoils, on the other hand, due to their high stopping power, generate short, straight, and dense tracks. As a result, the CMOS reconstruction needs to be flexible enough to handle these different particle types in order to effectively select interesting events and reject the rest. To achieve this, the CMOS-Reco process begins by addressing the sensor's intrinsic noise (Section 3.1.1), followed by the identification and grouping of the pixels that form a track, guided by directional criteria using a modified version of the IDBSCAN algorithm (Section 3.1.2). The output of the CMOS-Reco is a collection of pixels (called "clusters") associated with each detected track. From these clusters, information about the track topology, energy deposition pattern, and integral can be estimated and used for further processing and analysis. Some of the variables obtained from this process and relevant to this thesis are discussed in Section 3.1.3. A schematic of the CYGNO reconstruction algorithm is shown in Figure 3.2, and each step is explained in this section.

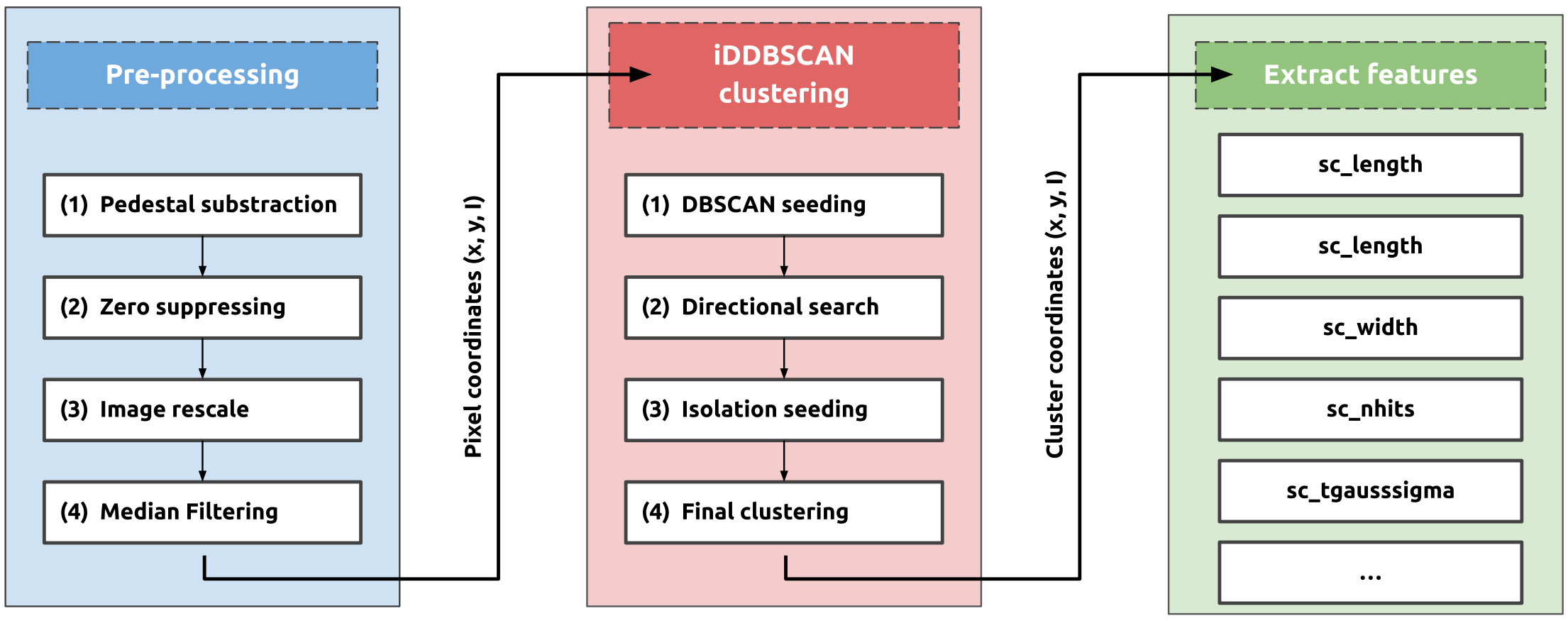

**Figure 3.2:** Schematic of the CYGNO-reconstruction algorithm, highlighting the major steps – pre-processing, iDDBSCAN, and feature extraction – and different sub-processes.

### 3.1.1 Preprocessing

Among the several CMOS camera models discussed in Section 2.1.3.2, all the data used in this thesis was acquired with the Orca Fusion. Each picture from this CMOS model is effectively a matrix of 2304 $\times$ 2304 pixels. Given the large number of pixels in each image, the data undergoes a pre-processing stage that extracts only the relevant information from each pixel, while accounting for single-pixel noise. Single-pixel noise refers to the noise associated with each individual CMOS sensor pixel, which is primarily determined by thermal energy, the manufacturing process, and the electrical connections during the readout of the charge collected in the sensor. This noise varies from pixel to pixel and must therefore be properly accounted for. To that end, a set of noise-only images is acquired – referred to as *pedestal runs*. These are taken with the GEMs turned off, resulting in images that capture only the intrinsic noise of each pixel. Pedestal runs are taken regularly (every couple hours) during physics runs so that they are closely related in date and time to those data, to account for possible variations in noise levels over time. The preprocessing procedure can be summarized in four steps [207]:

1. **Pedestal subtraction:** A pedestal image is generated by averaging the single-pixel intensities from 100 images acquired during a pedestal run. From this pedestal image, the mean value of each pixel, $\mu_i$, and its respective noise RMS, $\sigma_i$, are calculated. The pixel baseline $\mu_i$ is then subtracted from the corresponding pixels in the "normal/physics" runs, producing a new set of intensity values denoted as $I_i$.

2. **Zero suppression:** Both upper and lower thresholds are applied to the adjusted intensities, $I_i$. The upper threshold is intended to eliminate pixels with abnormally high intensities, often caused by leakage currents entering the sensor wells (referred to as *hot pixels*). The lower threshold can be adjusted and optimized depending on the data-taking conditions, in order to balance noise suppression and detection efficiency. For the data and camera used in this thesis, the lower threshold is set to $1.3 \times \sigma_i$. Pixels with intensities outside these thresholds are set to zero.

3. **Image Rescale:** The image is then down-sampled to a $512 \times 512$ pixel resolution for computational efficiency. Each $4 \times 4$ block of pixels, referred to as a "macro-pixel", is assigned an intensity value equal to the average intensity of the 16 pixels within that block.

4. **Median Filtering:** A $4 \times 4$ median filter is applied to the rescaled image, replacing the intensity of each macro-pixel with the median intensity of its neighboring macro-pixels ($w$). This filtering process helps suppress noise while maintaining computational efficiency [208]. As shown in Equation 3.1, where $f(x, y)$ represents the intensity of the macro-pixel at coordinates $(x, y)$, the new intensity $g(x, y)$ is determined by the median value of all surrounding macro-pixels.

$$g(x, y) = median\,\{f(x, y), (x, y) \in w\} \tag{3.1}$$

In CYGNO detectors, this filter has proven effective in reducing noise, showing a decrease in noise pixels by a factor of approximately $3.07 \pm 0.02$ [207].

After this procedure, the pixel matrix is further corrected for the vignetting effect, a natural reduction in light intensity toward the edges of an image produced by a lens, as described in Section 2.1.3.3. For this correction, a map is generated by averaging images of a uniformly lit white surface and normalizing the pixel values to that of the central pixel. The correction is applied by dividing each pixel's intensity in the data acquired images by the corresponding value in this map, compensating for peripheral light loss. The $(x, y)$ coordinates and vignetting-corrected intensities of the non-zero pixels are then passed to the clustering algorithm, iDDBSCAN.

### 3.1.2 iDDBSCAN

After preprocessing, the image is restored to full resolution. From this "cleaned" picture, a clustering algorithm is applied to identify and group pixels belonging to the same ionization track. One of the most widely used clustering algorithms for this task is Density-Based Spatial Clustering of Applications with Noise (DBSCAN) [209], which identifies clusters as regions of high data point density. In this context, a modified version of DBSCAN was developed – called intensity-based DBSCAN (iDBSCAN) [207] – which uses pixel intensities as weights to better identify high-density regions, enhancing the detection of relevant signals while rejecting background noise. Building on iDBSCAN, a further improved algorithm, directional intensity-based DBSCAN (iDDBSCAN) [210], was developed to better reconstruct particle tracks with defined directionality. These two steps of the clustering process can be summarized as follows:

1. **iDBSCAN:** Pixels located within regions of non-zero intensity in the final image – identified during the pre-processing phase – are selected and passed to the iDBSCAN clustering algorithm. This algorithm relies on two main parameters: a search radius $\epsilon$ and a minimum intensity threshold $N_{min}$. For each pixel, if the total intensity within a radius $\epsilon$ exceeds $N_{min}$, the central pixel and its neighboring pixels are grouped together to initiate a cluster. Unlike standard DBSCAN, which counts neighboring pixels, the iDBSCAN implementation in CYGNO uses the sum of pixel intensities $I_i$ as the clustering criterion. The parameters $\epsilon$ and $N_{min}$ are carefully tuned: too small a radius may cause noise to be misidentified as real clusters, while too large a radius can lead to noise merging into valid events. Similarly, a low $N_{min}$ increases sensitivity to faint noise fluctuations, while a high threshold may cause real low-energy events to be missed. To ensure optimal performance, $\epsilon$ and $N_{min}$ are calibrated at the beginning of each data-taking run, depending on the noise conditions and the typical track density in the images. Common values range from approximately 1 to 10 for $\epsilon$, and from 5 to 30 for $N_{min}$.

2. **iDDBSCAN:** After the initial clustering with iDBSCAN, the additional directional iDBSCAN (iDDBSCAN) algorithm [210] is applied to address cases where long tracks, such as those from muons or high-energy electrons, may have been fragmented into separate clusters. To identify such directional structures, a linear RANSAC procedure [211] is repeatedly applied to the clusters detected by iDBSCAN, searching for extended, straight, low-density tracks. If no such directional features are detected, the original iDBSCAN clusters are retained. Otherwise, the algorithm proceeds with a directional reconstruction using an enhanced version of RANSAC that incorporates a third-order polynomial fit: as illustrated in Figure 3.3, for each identified directional cluster, any pixel located within a fixed width $\omega$ around the fit line and within a radius $\epsilon_{dir}$ of a given point is added to the cluster. Remaining pixels that fall within a certain distance of a directional cluster are considered noise and excluded from further processing, effectively preventing the formation of additional clusters near cosmic-ray tracks. Finally, the set of selected pixels undergoes a second application of iDBSCAN to produce the final cluster configuration. This refined directional approach significantly enhances the reconstruction of long tracks, which would otherwise be fragmented by iDBSCAN, while maintaining the algorithm's efficiency for short tracks. The effectiveness of iDDBSCAN is demonstrated in Figure 3.4, where a before and after iDDBSCAN

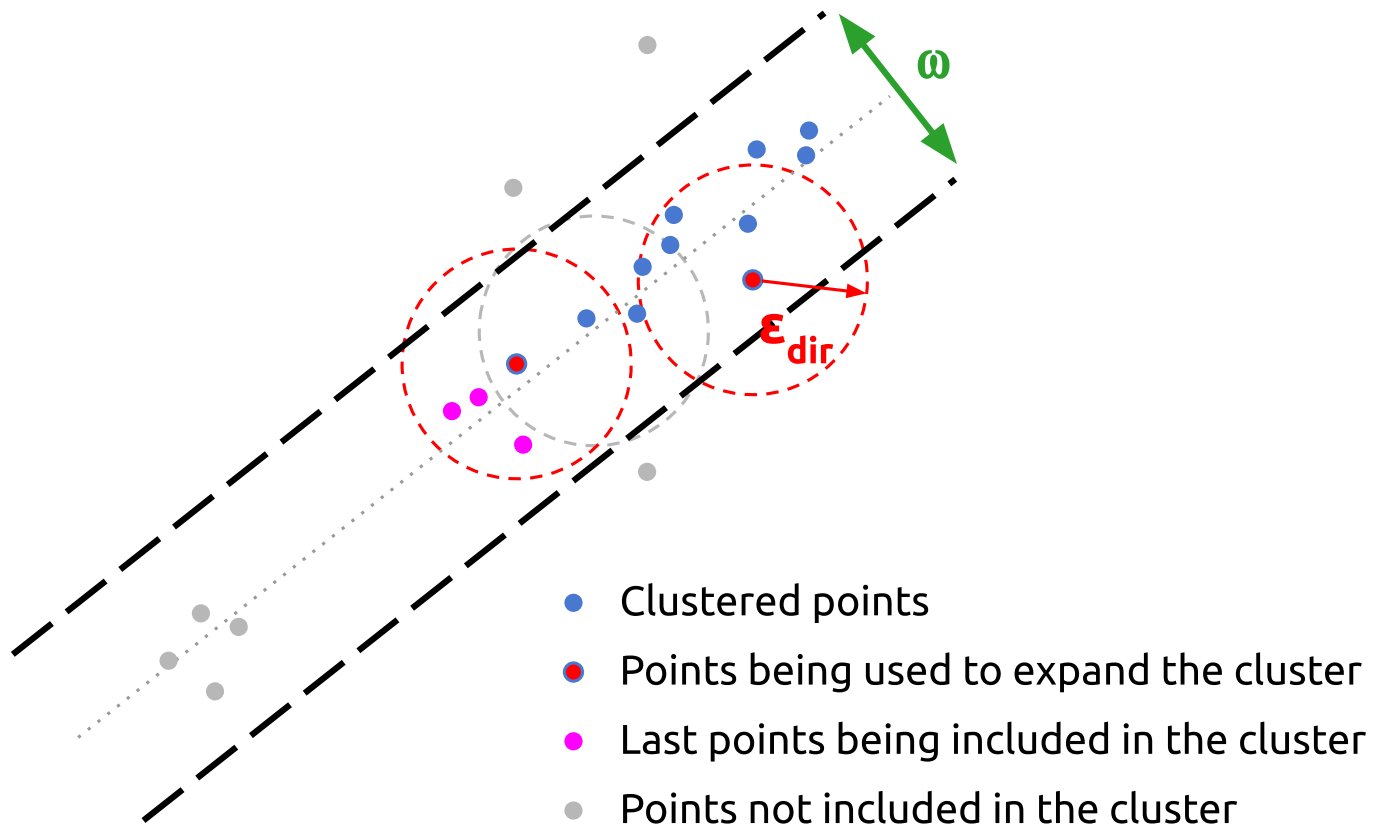

**Figure 3.3:** Illustration of the directional search performed by iDDBSCAN, highlighting the $\omega$ and $\epsilon_{\mathtt{dir}}$ parameters.

version of an image is shown.

### 3.1.3 Variables of interest

After the data reconstruction is finalized, a list of pixel clusters – each corresponding to an ionization track – is created. From the pixels associated with each track, a set of track features is computed, characterizing the physical properties of the track, which can then be used for subsequent analysis. Among these, the key track features retrieved and used in this thesis for initial event classification and localization through cuts are:

- **sc_integral:** The sum of the ADC counts in all pixels constituting a track. This value is proportional to the light emitted and can later be calibrated to correspond to the total energy deposited in the detector.
- **sc_nhits:** The number of pixels constituting an ionization track after the CMOS reconstruction processes described above. This variable is related to the spatial extension of the track inside the detector, convoluting the original track with the smearing caused by electron diffusion.
  - **Delta(δ):** Track/ionization density, which can be calculated by dividing the total light integral by the number of activated pixels in a given track. This variable provides a measurement of the track's average $dE/dx$, and for this reason, it can be used for general particle identification. $\delta$ is given by:

$$\mathtt{delta}(\delta) = \frac{\mathtt{sc_integral}}{\mathtt{sc_nhits}} \tag{3.2}$$

- **sc_length & sc_width:** Length in pixels of the major (*length*) and minor (*width*) axes

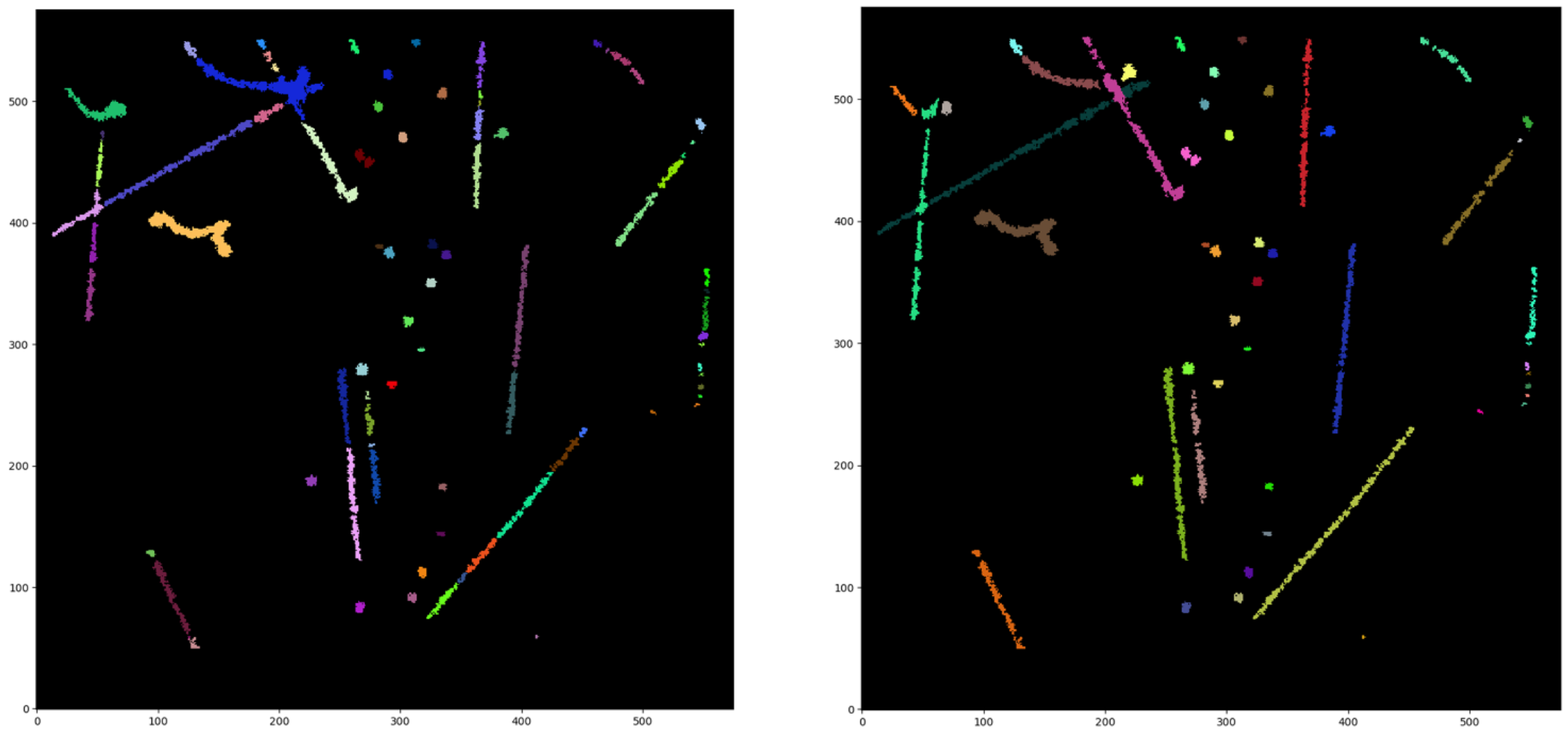

(a) (b)

**Figure 3.4:** Example of an image acquired with a CYGNO detector, and reconstructed with (a) iDBSCAN and (b) iDDBSCAN algorithms. Each cluster of pixels identified is shown in a different color. The improved code (b) shows better reconstructed tracks, without fragmentation.

of the track, determined using a principal component analysis (PCA) [212] applied to the track. This provides a general idea of the track's shape in the 2D (XY) plane.

- **Slimness ($\xi$):** With these two parameters, the track *slimness* can be calculated, defined as the ratio between $\mathtt{sc_width}$ and $\mathtt{sc_length}$. This metric is useful for quantifying how round or straight a track is, as given by the following equation:

$$\mathtt{slimness}(\xi) = \frac{\mathtt{sc_width}}{\mathtt{sc_length}} \quad (3.3)$$

- **sc_tgausssigma:** The sigma of the Gaussian fit applied to the transverse profile of the ionization track, obtained by projecting the pixel intensity distribution of the track's minor axis. This value can be used to estimate the absolute Z position of the track by correlating it with the electron diffusion.

- **sc_xmean & sc_ymean:** Average ("mean") position of the track, in pixel coordinates, based on the track's light barycenter.

These track shape variables are among the most relevant for CYGNO analysis and the ones used in this thesis for the remaining studies presented.

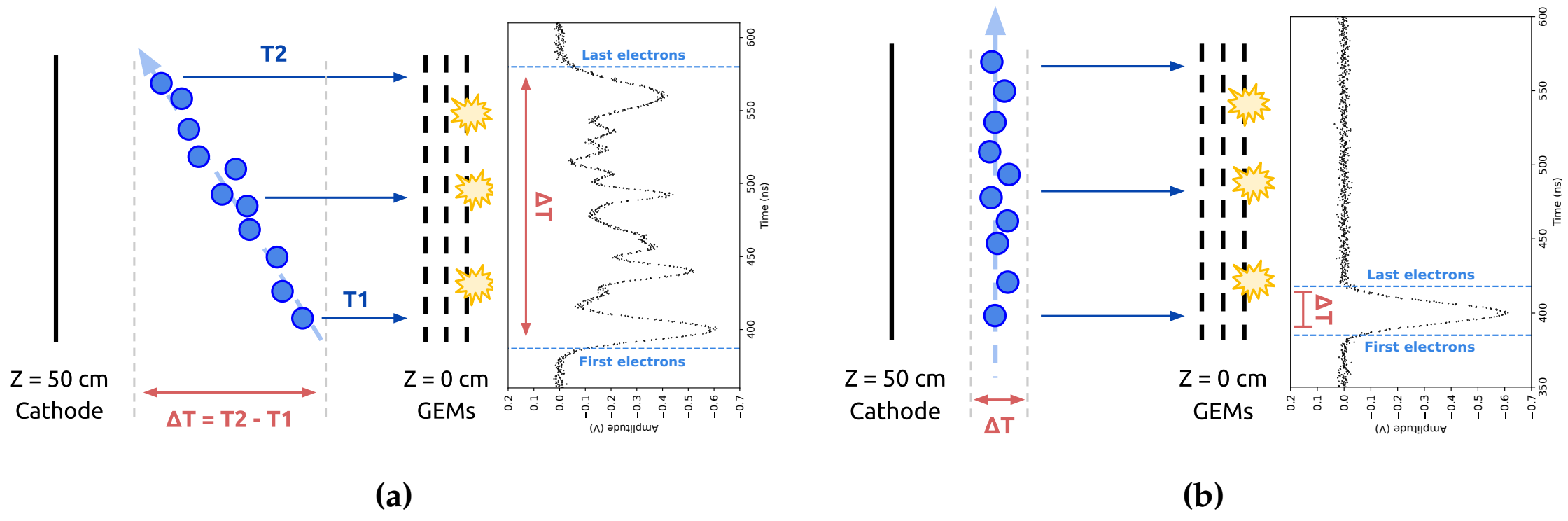

**Figure 3.5:** Schematic of the PMT signal generation in a TPC, highlighting how the waveform reflects the times of arrival of the different electrons of the ionization track, for two distinct case: a (a) tilted and (b) parallel track to the GEM plane.

## 3.2 PMT

The PMT signals represent the temporal profile of the collected light and encode several characteristics of the ionizing event, particularly related to the arrival time of the ionization cloud electrons at the GEM plane. This timing reflects the path traveled by the particle along the drift direction. A schematic illustrating the generation of a PMT signal in a TPC is shown in Figure 3.5. When a photon strikes the PMT photocathode, it is converted into an electron, which is then multiplied through a cascade process, resulting in a small electrical pulse (see Section 2.1.3.1). As more photons are detected, the signal accumulates, increasing in both amplitude and duration, until photon production ceases and the signal returns to baseline. In this way, the PMT waveform encodes the arrival times of the electrons originating from the ionization track. As the particle travels through the gas, it produces electrons at various Z positions (Figure 3.5a), each requiring a different drift time to reach the GEMs and be amplified. This produces multiple light peaks in the PMT signal. Conversely, if all electrons reach the GEMs simultaneously (Figure 3.5b), only a single peak is observed. From this rationale, by analyzing the shape of the PMT signal, it becomes possible to extract information about the particle's trajectory along the drift direction as well as the spatial distribution and density of the ionization cloud.

The analysis of PMT data from ionization tracks in CYGNO detectors was a major focus of this thesis, and the corresponding results are presented in this section. It begins by introducing the single-PMT waveform analysis tools developed to reconstruct the basic properties of the PMT signals acquired in CYGNO detectors, hereafter referred to as *PMT-reco* (Section 3.2.1). Next, the discussion moves to how the signals from multiple PMTs can be combined and how the waveform's relative intensities can be exploited to retrieve information about the XY position of the track within the CMOS field of view (Section 3.2.3). Finally, in the following section, the variables and tools defined here are used to match the information obtained from the CMOS sensor with the one from the PMTs (Section 3.3).

### 3.2.1 Waveform analysis – *PMT-Reco*

The waveform analysis that forms the basis of the *PMT-Reco* is a crucial component of the 3D analysis of CYGNO. The retrieval of important information from the PMT waveforms and their subsequent integration with the CMOS data represents one of the innovative and original aspects of this thesis. In this section, a detailed description of the waveform analysis pipeline is provided, starting with the DAQ-related pre-processing applied, followed by the methods and variables developed to extract information from each individual PMT waveform.

#### 3.2.1.1 Digitizer corrections

The signal produced by the PMTs is converted into a digital signal by two digitizers with different sampling frequencies, as described in Section 2.2.2.2. The DRS4 chips used in the digitizers for LIME have inherent differences in their construction, which require corrections to be applied to the acquired data [184]. Although the fast-sampling digitizer is typically used in most analyses, the corrections are applied to both digitizer signals and are usually handled at the software level. These corrections can be classified into three main types:

1. **Cell Index Offset Correction:** Compensates for signal offsets caused by differences in cell amplitudes.
2. **Sample Index Offset Correction:** Adjusts for signal offsets due to noise in the final 30 samples.
3. **Time Correction:** Accounts for variations in the delay line timing of the chips.

These routines have default correction tables stored in the board's flash memory, which the board uses to automatically correct the data. However, these tables need to be periodically updated or when the digitizer chips experience malfunctions. To address this, measurements were conducted to obtain the updated digitizer *correction tables*, first for the overground setup and later for the underground commissioning. These updated tables were then integrated into the general CYGNO-reconstruction analysis framework, enabling the proper correction and subsequent analysis of PMT waveforms obtained with CYGNO detectors. An example of these corrections, specifically the cell index correction (the most significant), is shown in Figure 3.6, as directly obtained from the official digitizer manual [184].

#### 3.2.1.2 Initial filtering

Once the waveforms are properly corrected, each undergoes an initial *inversion and centering* routine, which converts the signal to positive polarity (since PMT signals are

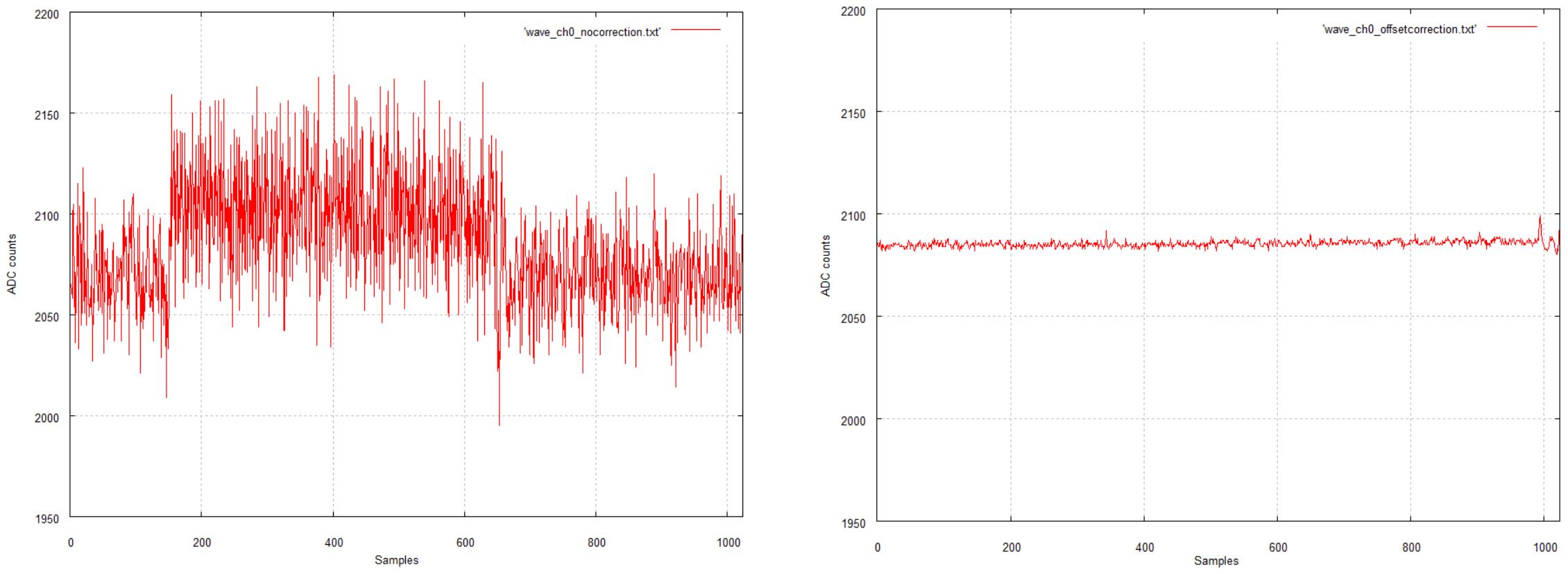

**(a)** Before correction

**(b)** After correction

**Figure 3.6:** Example of the effects of the cell index correction in a PMT waveform digitized with the CAEN V1742. Retrieved directly from the manual [184].

traditionally negative) and removes the offset baseline. Afterwards, a *moving average filter* is applied to remove high-frequency noise. The moving average filter is effective for reducing random noise while preserving the sharpness of the signal, making it ideal for time-domain encoded signals, like those from the PMTs [213]. The filter operates by averaging the signal's point-by-point amplitude over a predefined sample window. The larger this window, the smoother the signal becomes. In the PMT-Reco, this window is left as a free parameter that can be fine-tuned by the analyst, depending on the analysis being carried out. Initially, this value was optimized to 5 samples, as it effectively reduced high-frequency noise without removing the sharp peaks of the signals.

An example of these processes is shown in Figure 3.7, where the original waveforms retrieved from the 4 digitizers (each connected to a PMT), in Figure 3.7a, are compared with their respective filtered versions, in Figure 3.7b. As demonstrated, in the filtered version, all the channels have the same baseline and positive polarity. Additionally, the key characteristics of the waveform are still clearly visible, such as the main peaks and the sharp rise and fall, while some of the high-frequency noise is eliminated. This processing step simplifies the subsequent analysis by reducing the waveform to its most important features.

#### 3.2.1.3 Peak finder

The filtered PMT signals are then subsequently processed to identify peaks in the waveforms. In order to do this, the *find_peaks* routine from the Python SciPy library [214] is applied to the waveforms. This routine is often used given its efficiency and flexibility in identifying significant local maxima in 1D signals [215]. This routine takes a 1D array of values that constitute the waveform and identifies all local maxima (*peaks*) by performing comparisons with the neighboring values. The properties of the peaks can be specified using a set of tunable parameters on which the *find_peaks* routine depends.

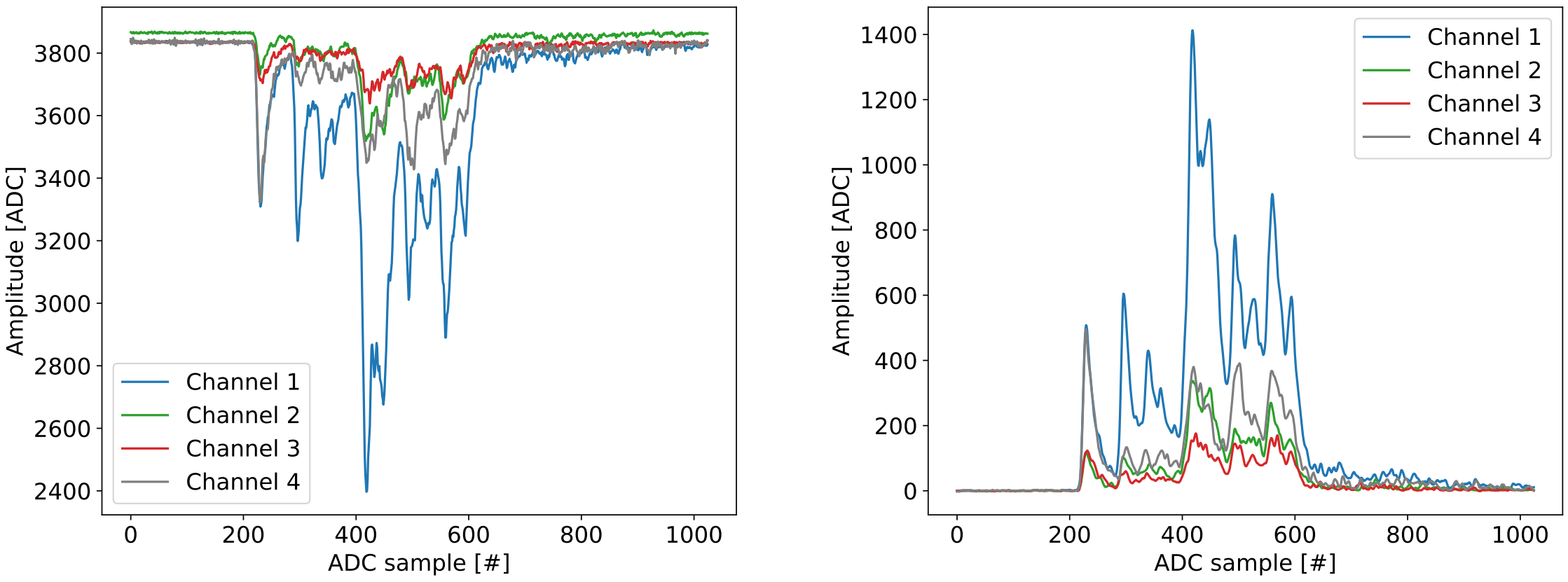

**(a)** Raw **(b)** Filtered

**Figure 3.7:** Example of a waveform as it is (a) directly saved by the digitizer, and (b) after applying the initial filtering procedures which include *inversion*, *centering*, and applying a *moving average filter*.

These parameters are related to peak features such as height, width, prominence, and distance between them.

The choice and optimization of these parameters depend on the specifics of the analysis being conducted and the type of particle or interaction to be selected. Therefore, while no definitive parameters are universally defined, the default ones are generally sufficient to select the "main" peaks of the waveform – that is, those corresponding to individual ionizations within the track, which exhibit high prominence and broad widths. In particular, peaks with prominences greater than one-quarter of the waveform's maximum and widths greater than five samples are identified. An example illustrating the working principle of this basic routine is shown in Figure 3.8. In this figure, two waveforms are shown with different variables identified by the peak-finder routine: yellow crosses indicate the peaks found – typically associated with the mentioned discrete energy depositions occurring as the particle traverses the detector – while green and red lines mark the respective widths at half and full amplitude maxima. These are additional useful variables employed throughout this thesis.

#### 3.2.1.4 Time-over-Threshold (*ToT*)

The PMT signal reflects the time of arrival of the electrons from the ionization track to the amplification plane. Therefore, if a particle track is monotonic in the drift direction, the length of the PMT signal becomes directly proportional to the path traveled by the particle in the longitudinal (Z) direction, or in other words, to $\Delta Z$. The time extension of the PMT signal can thus be used to estimate $\Delta Z$ for particles like alpha particles, nuclear recoils (NRs), or MIP-like tracks, which are expected to produce mostly straight tracks, with minimal deviation from their initial direction due to scattering. Low-energy ERs are expected to have curly trajectories in all directions, making their true path along Z

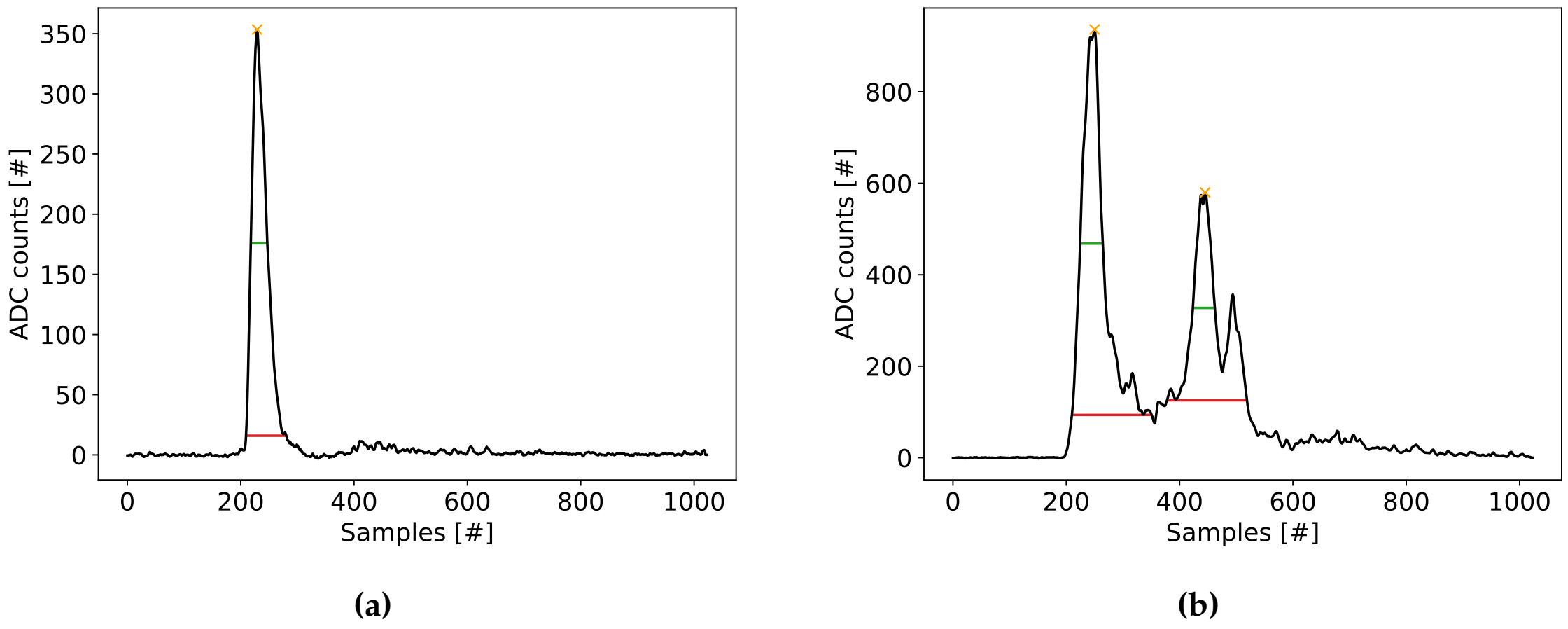

**Figure 3.8:** Example of two waveform to showcase the operating principle of the peak finder routine applied during the PMT-Reco pipeline. Yellow crosses represent the peaks found, and green and red lines represent the widths at half and full maxima of each peak.

more difficult to infer, although work is ongoing in this regard.

The simplistic full amplitude variable retrieved from the *find_peaks* routine is often insufficient (and not designed) to accurately capture the full extent of a PMT signal when multiple ionizations peaks are present in the waveform, as illustrated in Figure 3.8b. To address this, the *Time-over-Threshold (ToT)* variable was introduced. This method defines the start and end of a signal by identifying the fifth consecutive point where the signal exceeds or falls below a tunable threshold, initially optimized to be five times the waveform's RMS. An example of this variable is shown in a waveform in Figure 3.9 as a blue line, alongside the other variables extracted by the peak finder routine. As shown, this approach more accurately represents the total temporal width of the signal.

Assuming the particle maintains the initial direction while crossing the gas, this information can then be used for the determination of the ΔZ component of ionization tracks by correlating the ToT (in number of samples) with the digitizer's sampling frequency ($f_s$) and the electron drift velocity ($v_e^-$) in the gas (see Section 2.1.1), through the following expression:

$$\Delta Z \ [\mathrm{cm}] \ = \ \mathrm{ToT} \ [\mathrm{samples}] \cdot \frac{1}{f_s \ [\mathrm{MHz}]} \cdot v_e^- \ [\mathrm{cm}/\mu\mathrm{s}] \tag{3.4}$$

### 3.2.2 PMT-reco integration in the CYGNO-analysis framework

In addition to the variables mentioned in the previous section, other simpler variables are also extracted from the PMT information – such as the waveform's *ID* (containing its run, event, trigger, and channel number), its *RMS*, the position in time of the peaks, among others. A full list of all the variables retrieved through this waveform analysis is shown in Table 3.1. If requested, the waveforms can also be saved in an image format (e.g., *.png*) for visual inspection or scientific outreach. Users can also choose to save the raw waveform data for specialized analysis. If later deemed useful, any new routine developed can be easily integrated into the main reconstruction algorithm.

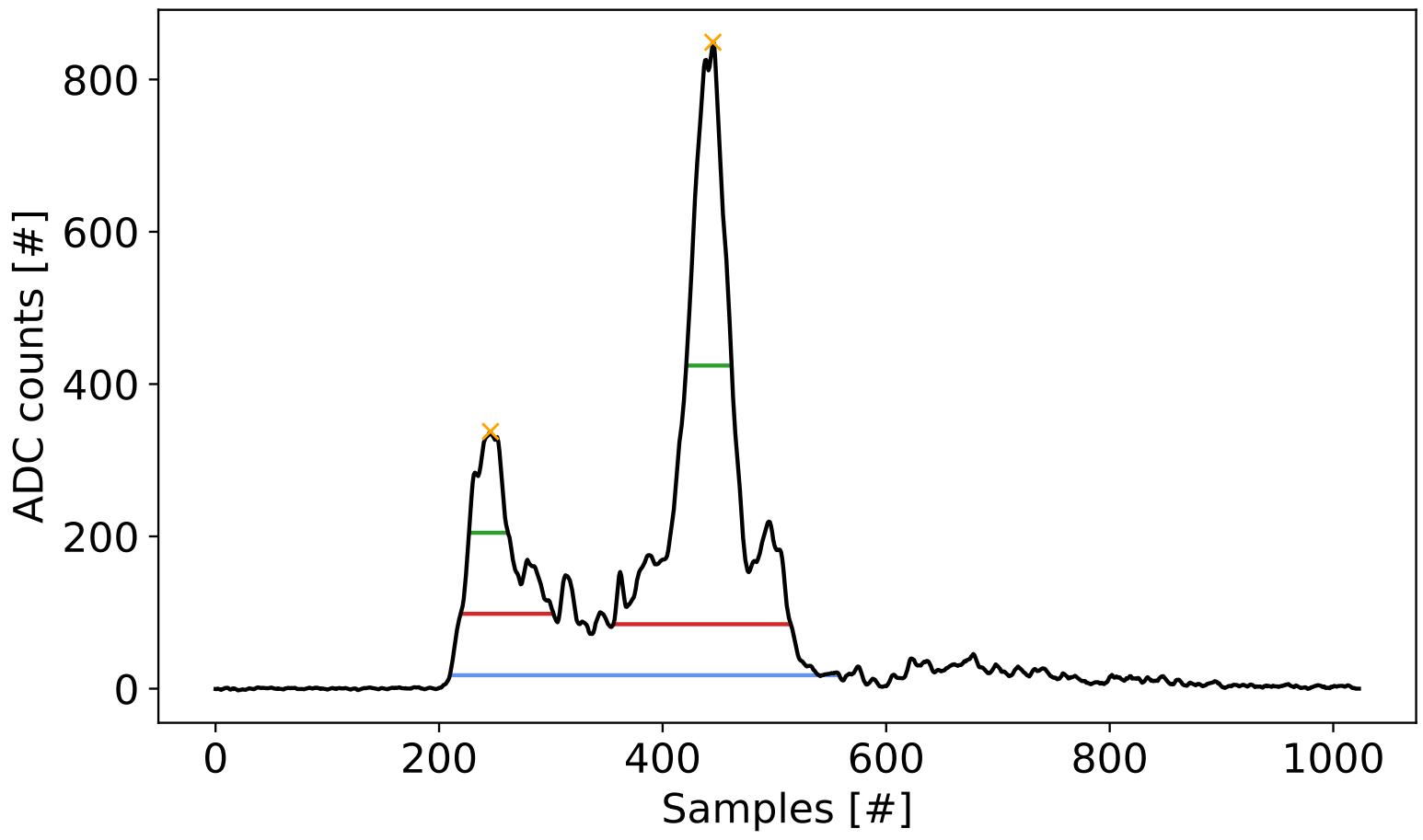

**Figure 3.9:** Example of a waveform showcasing the *Time-over-Threshold (ToT)* variable as a blue line. The remaining features are created through the peak finder routine.

The PMT-Reco was developed to be fully integrated within the general CYGNO reconstruction framework. For that, all the information retrieved is stored in a ROOT TTree [216] for further post-analysis. A ROOT TTree is a data structure within the ROOT framework [217] designed to efficiently store large amounts of structured data. It organizes the data into branches and leaves, allowing fast access and analysis of complex datasets. As a result, CYGNO researchers can now easily and efficiently use both the CMOS and PMT data individually for their analyses. For example, this information has been used by other researchers to study the longitudinal diffusion of electrons in LIME by examining the relationship between signal duration and drift distance [218].

The integration of the PMT waveform data into the general CYGNO reconstruction framework was also designed to be backward-compatible with older data and adaptable to other CYGNO prototype detectors. For example, the MANGO detector (Section 2.2.1), which has a different number of PMTs (one instead of four), digitizers (one type instead of two), and other subtle DAQ differences compared to LIME, is also supported by the PMT-Reco. To further enhance and optimize the PMT-Reco, another CYGNO subgroup began developing a simulation of PMT waveforms in CYGNO detectors. This is expected to significantly improve the currently implemented routines.

### 3.2.3 Multiple PMT combined analysis

As discussed so far, the single PMT waveform provides the information regarding the particle's $\Delta Z$ traveled pattern. When combining multiple PMTs signals, it is also possible to retrieve information about the XY position of the detected event. Indeed, the CYGNO experimental approach foresees the use of multiple PMTs looking at the same amplification plane in order to exploit the light barycenter as means to better infer the event position on that plane. With this information, it is possible to match the information of both PMT and CMOS sensor.

**Table 3.1:** List of all the variables retrieved by the PMT-Reco and then introduced in the CYGNO-reconstruction framework.

| ROOT::TTree variable | Definition |
|---|---|
| pmt_wf_[ID/run/event] | Basic waveform info regarding the event. |
| pmt_wf_[trigger/ch/samp] | Basic waveform info regarding the PMT and digitizer. |
| pmt_wf_insideGE | Check if PMT trigger is inside the camera's *Global exposure*. |
| pmt_wf_TTT | Trigger-time-tag. For identification of CMOS-cut events. |
| pmt_[baseline/RMS] | Waveform's baseline and RMS. For data quality monitor. |
| pmt_tot_[integral/charge] | Integral and charge of waveform |
| pmt_max_ampl | Waveform's maximum amplitude. For PID. |
| pmt_nPeaks | Waveform's number of peaks from *find_peaks*. For PID. |
| pmt_peak_Number | Peaks' ID. For peaks' coincidence check among PMTs. |
| pmt_peak_[position/height] | *find_peaks* variables. For PID and general analysis. |
| pmt_peak_[Half/Full_Width] | *find_peaks* variables. For PID and general analysis. |
| pmt_TOT_[time/area] | Time-over-threshold. For determination of $\Delta Z$. |

Using the LIME detector realization as an example, four PMTs are positioned near the corners of the CMOS field of view, as shown in Figure 2.17, and are used to record the light emitted at the GEMs. Each PMT detects a different signal intensity depending on its distance from the point where the light is emitted on the GEM plane.

The relationship between the total light $L_j$ emitted in the position $(x_j, y_j)$ of the GEMs and the amount received by each i-th PMT, $L_{ij}$, can be modeled using the general case of illumination from a perfectly diffuse (Lambertian) source [219]. A Lambertian surface appears equally bright from any viewing angle, as it reflects light uniformly in all directions, with the intensity following the cosine of the angle with respect to the surface normal. Using Figure 3.10 as a reference, the radiant power ($\Phi$) received by each PMT can be evaluated as:

$$\Phi = \frac{L_j \; A_j \; A_i \; \cos^2 \theta_{ij}}{R_{ij}^2}$$

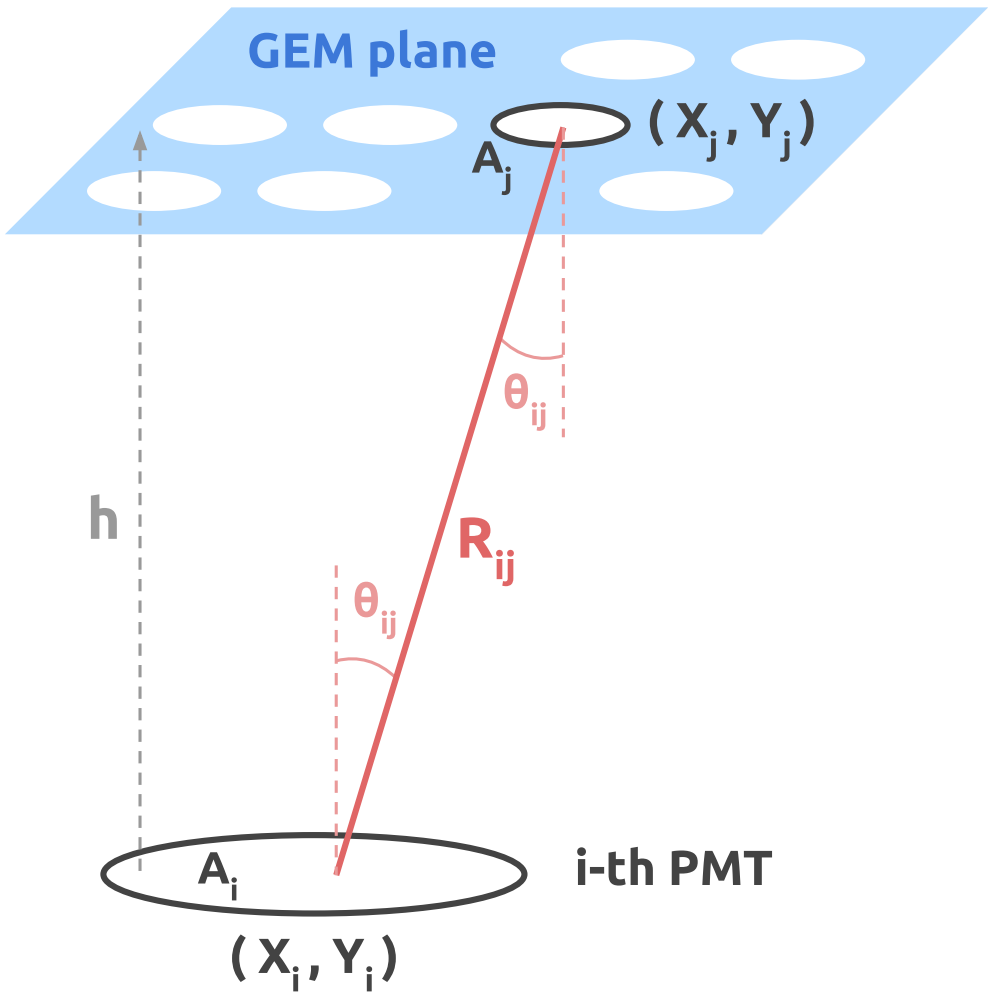

**Figure 3.10:** Schematic representation of the illumination of the $i$-th PMT of area $A_i$ with coordinates $(x_i, y_i)$ by the radiating source of area $A_j$ with coordinates $(x_j, y_j)$ on the GEM plane. The distance between the centers of the two surfaces is denoted by $R_{ij}$, and the angle as $\theta_{ij}$.

where $A_j$ is the area of the source (i.e., the light emitted at the GEMs), and $A_i$ is the PMT surface. Recognizing that $\cos(\theta_{ij}) = h/R_{ij}$, where $R_{ij}$ is the distance between the light emission point and each $i$-th PMT, and $h$ is the projection of that distance along the direction orthogonal to the two surfaces (which are parallel in this case), it is possible to express the light intensity received by each PMT ($L_{ij}$), over its surface $A_i$, relative to the original light emitted at the GEM plane ($L_j$), as:

$$L_{ij} \propto \frac{L_j}{R_{ij}^{\alpha}}, \text{ with } \alpha = 4 \tag{3.5}$$

This relation is also known as the *cosine fourth power* law of illumination [220]. This *approximate* law accounts for the inverse square law of light ($L \propto 1/R^2$), which describes how the solid angle subtended by the observer decreases with the square of the distance between the light source and the observer; and it also incorporates the reduction in flux as the source position on the object plane moves away from the optical axis [171]. In the scenario considered, this cumulative effect leads to a fourth power relationship between the emitted and received light. The angular response of the PMTs is not considered, as it remains approximately constant for the range of angles involved in LIME.

As this law is an approximation [219], and other non-physical factors may influence the amount of light reaching the PMTs (such as the transparency of the media between the light emission and the PMTs), dedicated measurements validated this approximated scaling law factor $\alpha$ in LIME [218]. In summary, this involved measuring the integrated charge recorded by each PMT for signals of equal intensity ($L_j$, in the formulas above) at different distances $R_{ij}$ from each $i$-th PMT. This was done using an $^{55}Fe$ calibration source, which emits photons with 5.9 keV energy, providing a stable light source. The event selection consisted in identifying events with a single cluster in the CMOS images and a single PMT trigger. This ensured an unambiguous matching between the CMOS and PMT data. The distances $R_{ij}$ were determined by identifying the positions of the $^{55}Fe$ clusters in the CMOS images and computing the corresponding distances to each PMT, using the geometry shown in Figure 2.17. By applying Equation 3.5, the relationship between the total light emitted and the amount received by each PMT can be calculated, and the power $\alpha$ fitted. The combined result of this procedure for the four PMTs is shown in Figure 3.11, along with the residuals from the best-fit line for each PMT. A final best-fit value of $\alpha = 4.01 \pm 0.06$ was obtained. This result validates the theoretical approximation of the cosine fourth-power law and is therefore adopted for all subsequent analyses.

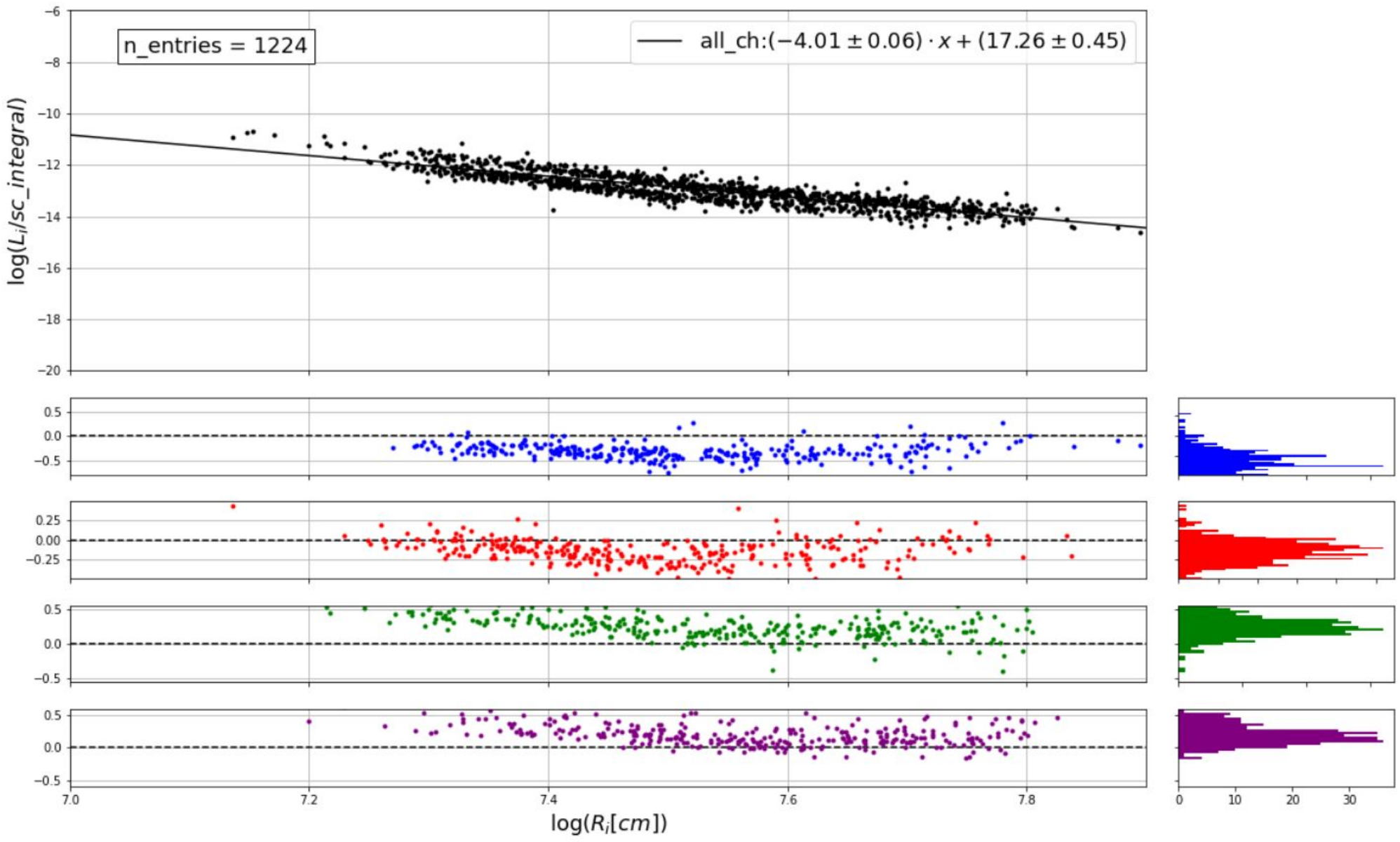

**Figure 3.11:** Best fit and relative residuals of the alpha power law in double log scale for the four PMTs combined. Retrieved from [218].

To better illustrate the geometrical dependence of the PMT signal amplitude on the distance between the event and each PMT, two examples are subsequently presented using the LIME detector configuration. These are intended to provide a clearer understanding of this phenomenon using real-world data. Specifically, examples of *short-* and *extended*-like signals are shown.

### Short events

Figure 3.12 shows an example of an event in which two particles are visible in the CMOS image (Figure 3.12a), and two PMT triggers were present (in Figures 3.12b and 3.12c).

The CMOS image is divided into four quadrants, organized clockwise starting from the upper-left corner, each representing the position of one PMT (following Figure 2.17). In the image, two relatively short (with respect to the whole image) pixel clusters are visible: a spot-like event in quadrant 1 and an alpha particle in quadrant 2. Each set of waveforms shows the light signal measured by each PMT, positioned in the corresponding quadrant. In Figure 3.12b, the PMT with the highest integrated charge is PMT #1 (channel 1, blue line), indicating that the closest PMT to the CMOS cluster corresponding to these four waveforms is PMT #1. Therefore, it can be concluded that the waveforms of trigger #1 correspond to the event occurring in quadrant 1 (the spot-like event).

Similarly, for the second triggered event (Figure 3.12c), the highest charge is recorded by PMT #2 (green line), indicating that this event is associated with the cluster in quadrant 2, the alpha particle. The remaining channels in this trigger also support this interpretation, with the second-highest charge coming from channel 1 (blue line), which is expected due to its relative proximity. Channels #3 and #4 show similar integrated charges, as these PMTs are roughly equidistant from the light emission.

This basic association between PMT triggers and CMOS clusters allows a quick matching between the two sensors' data. However, this approach is applicable mostly for short (few centimeters) signals that do not cross multiple quadrants, and with a resolution dependent on the number of "quadrants" (or macro pixels) defined in the image. Nonetheless, this relationship between the amplitudes of each PMT and the position of the event in the XY plane is a crucial property of the PMT signal and is the basis of the more robust version of this approach that is presented in Section 3.3.1. In this approach, a precise (centimeter resolution) definition of the XY position of the track as seen by the PMTs is achieved, allowing an accurate match between CMOS and PMT data.

#### Extended events

For long events which span several centimeters and cross multiple quadrants in the CMOS image, such as cosmic rays or high-energy ERs, the analysis of the waveform and subsequent association with CMOS clusters becomes more challenging. Particles induce the emission of photons continuously along their tracks, each at a different distance from the PMTs. As the particle gets closer to or farther away from a PMT, the amplitude (light) seen by the PMT grows or reduces, proportionally to this distance. For long tracks, this results in more complex waveforms, with multiple peaks, spread across a large time window. This prevents the use of simple total charge integration to estimate the event's XY position, as done for spot-like events, and requires a more in-depth analysis of the shape of the waveform.

Figure 3.13 shows an ionization track in the upper-right corner of the CMOS image, likely caused by a cosmic ray or high-energy ER, based on its straight path and uniform energy deposition (indicative of a MIP-like event). The positions of the four PMTs are marked on the CMOS image, following Figure 2.17, with the colors corresponding to

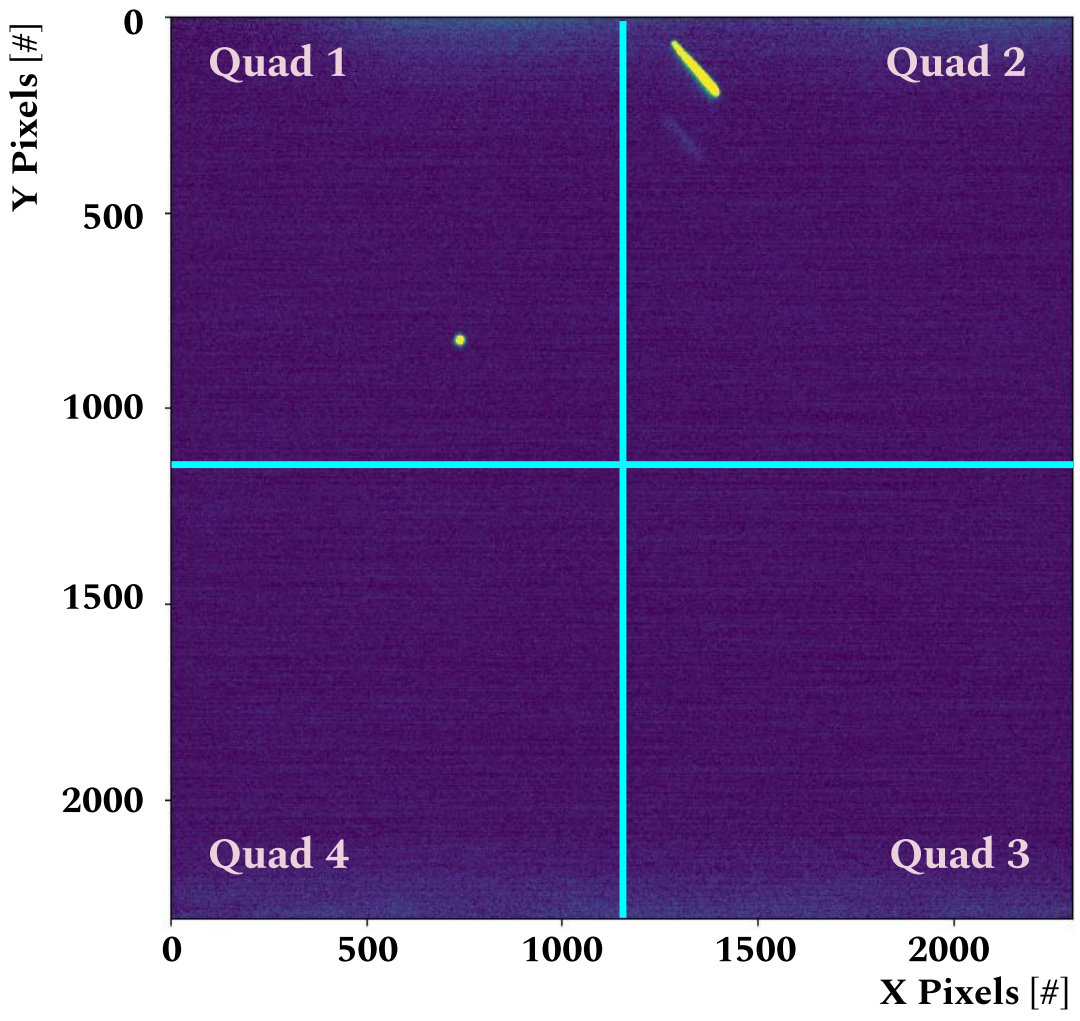

**(a)** CMOS image

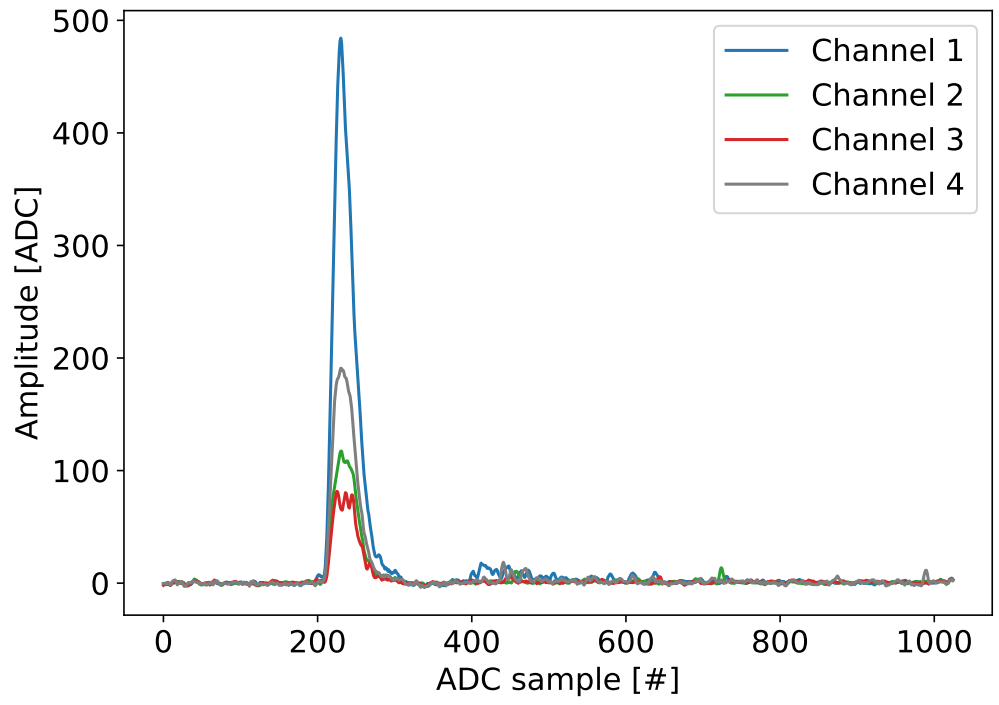

**(b)** Trigger #1

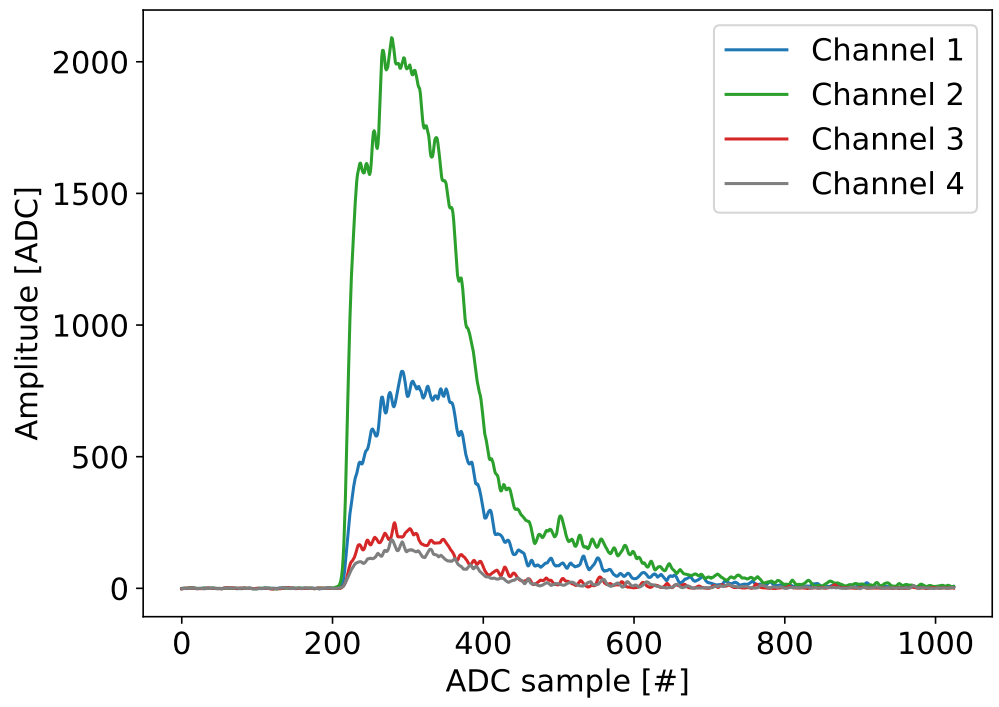

**(c)** Trigger #2

**Figure 3.12:** Example of an event where two pixel clusters are visible and two PMT triggers are present in the data bank. The image (a) is subdivided into four quadrants, corresponding to the general position of each PMT. The waveforms are associated with each cluster by identifying the channel with the highest integrated charge in each trigger: in (b), channel 1 has the highest integrated charge, associating it with the cluster in Quadrant/PMT #1, while in (c), channel 2 shows the highest charge, thus associating it with the cluster in Quadrant/PMT #2.

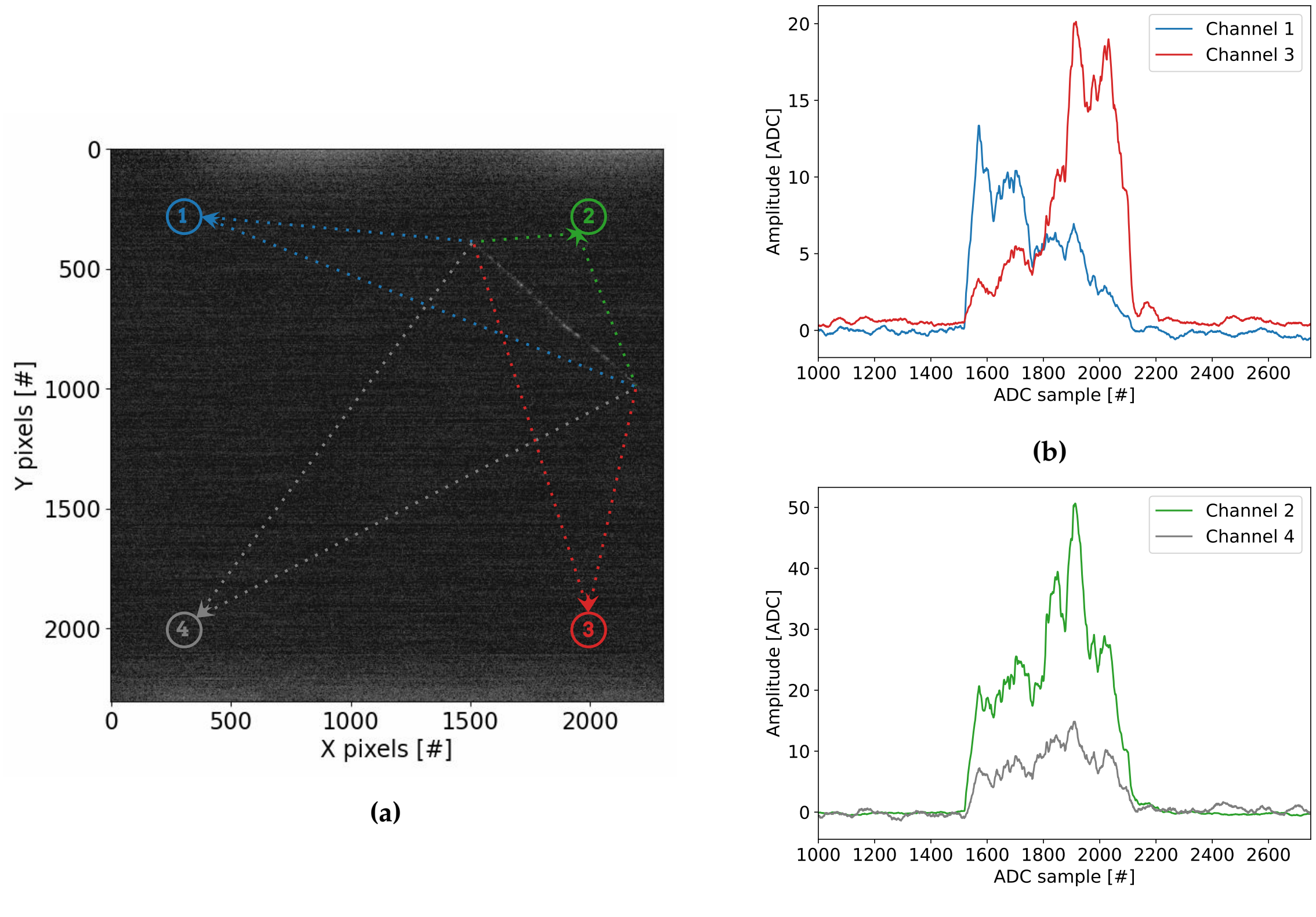

(c)

**Figure 3.13:** Example of a cosmic or high-energy electron recoil visible in the top right corner of the CMOS image, in (a). The position of the four PMTs of LIME is also shown in the CMOS image, with numbers and colors representing the respective channels and waveforms associated with each in (b) and (c). The shape of the PMT signals reflect the path traveled by the track in XY, as it gets closer or father way from each PMT.

their respective waveforms shown in the right panels, Figures 3.13b and 3.13c. The waveforms are grouped in pairs for better visualization. In this specific case, the *slow* digitizer is used due to the signal's extended duration. Unlike the *fast* digitizer, in which each ADC time sample corresponds to 1.3(3) ns, the slow one has each sample corresponding to 4 ns, as discussed in Section 2.2.2.2.

Looking at the waveforms, PMTs #1 and #3 (channels 1 and 3 in Figure 3.13b) show similar signal amplitudes but with opposite trends. This reflects the movement of the particle within the detector: as the particle moves farther from PMT #1, it approaches PMT #3, causing a decrease in light detected by PMT #1 and an increase in light detected by PMT #3. However, this pattern alone does not reveal whether the particle is moving left or right, as the PMTs are sensitive only to the tilt of the track relative to the longitudinal axis. Assuming the track is mostly straight in Z and has uniform energy deposition, it can instead be inferred that the first electrons to reach the GEM plane originated near PMT #1, as indicated by the initial peak in its waveform, although it is unclear whether the particle was moving towards the GEMs or towards the cathode.

For PMTs #2 and #4, the waveforms are similar in shape but differ in absolute amplitudes (note that the waveform of PMT #4 has been magnified by a factor of 3× for improved clarity). This occurs because the track is nearly perpendicular to the line connecting these two PMTs (the perpendicular bisector), making it so that the photons generated at different points in the track follow the same distancing pattern from PMTs #2 and #4, resulting in similar-shaped waveforms for both, differing only in the absolute amplitude, due to the different absolute distances of each PMT from the track. The shape of the waveforms shows a peak in the middle, which reflects the changing distance between each point along the track (i.e., the emitted photons) and the individual PMTs, which is smaller in the center of the track (thus more light) and greater at the borders.

These geometric observations assume that the track is mostly straight and has nearly uniform energy deposition along its path (MIP-like events). For lower-energy ERs ($<$ 100 keV), this assumption no longer holds, as electrons begin to scatter in various directions and "curl" in the longitudinal direction, increasing the complexity of the analysis of the waveform. For alpha particles and nuclear recoils, while being mostly straight in Z, this process becomes convoluted with the track's asymmetry in energy deposition, namely the Bragg peak for alphas and the head-tail effect for NRs, as will be illustrated in more detail in Section 5.1. For these types of events, both the geometric and deposition asymmetry dependencies of the PMT signals are correlated and must be considered together to accurately describe the 3D geometry of the event, as further discussed in Section 5.2.

These remarks on the relationship between the position of the light emission and PMT signals highlight the challenges in generalizing the 3D analysis across different particle types. Studies on long (higher-energy) ERs and cosmic rays have greatly benefitted from the developments discussed here, and ongoing work within the collaboration continues to improve the understanding and description of these events in three dimensions, although this is beyond the scope of this thesis.

#### 3.2.3.1 Average waveform

Given the dependence of the signal amplitude on the amount of emitted light, the efficiency in determining certain waveform variables depends on the signal-to-noise ratio in that specific waveform. In particular, the time-over-threshold can be significantly affected when analyzing very long signals, as will be shown. In such cases, combining the signals from multiple PMTs can be useful to better highlight and determine the full extension of the light signal.

With this in mind, a weighted average waveform is constructed using each waveform's signal-to-noise ratio (SNR), defined here as the ratio of the waveform's maximum amplitude to its RMS. Averaging the waveforms in this way reduces the influence of noise and mitigates the geometric dependence of the signal. This approach is particularly useful for particles that traverse large portions of the detector, where the signal amplitude can decrease rapidly as the particle moves farther from a given PMT. Using a weighted average rather than a simple average ensures that more weight is given to PMTs with stronger signals, which is especially important in cases where the event occurs near a single PMT, resulting in low-amplitude signals in the others.

While different properties of the weighted average waveform from multiple PMTs can be studied, here an example is illustrated focused on the time-over-threshold (ToT) calculation. Specifically, Figure 3.14 shows the individual signals recorded by each PMT and highlights: yellow crosses representing the peaks automatically identified by the *find_peaks* routine; red lines marking the full widths of each peak; and a blue line indicating the calculated ToT. As shown, the ToT determination fails for most individual waveforms due to their overall low signal amplitudes and the high number of peaks. However, in the weighted average waveform (Figure 3.14e), the ToT accurately captures the full time extension of the track. This effect is particularly noticeable in long tracks that traverse large portions of the detector, as the weighted average helps balance the contribution of each PMT and emphasizes the higher amplitude segments of the waveforms, thus improving the overall determination of the ToT, which can then be used to determine the traveled path of the particle in the longitudinal direction, through Equation 3.4. This technique is especially useful in Chapter 4, where the time extension of overground muons is analyzed to determine their incidence zenith angle relative to the Earth's surface.

## 3.3 PMT waveforms to CMOS images matching

With the integration of PMT-Reco into the general CYGNO analysis framework, both CMOS and PMT data can be reconstructed and used for post-analysis. The correlation between the information from these two sensors offers a two-fold benefit. First, it helps address cases where multiple ionization tracks appear in a single CMOS image. In such scenarios, associating CMOS clusters with PMT triggers is not straightforward, as there is no direct timing link between the two (see Section 2.2.2.2.1). While the analysis could still proceed using a single sensor (either CMOS-only or PMT-only), combining the information from both sensors significantly improves the reliability and accuracy of the reconstruction.

Secondly, correlating the information from the two sensors enables the reconstruction of the particle's path in three dimensions, by combining the 2D imaging from the CMOS (XY) with the time-resolved light signal along the drift direction ($\Delta Z$) from the PMTs. Furthermore, by analyzing additional shape variables from both the PMT and CMOS data, it becomes possible to infer the particle's initial direction and sense in both the transverse and longitudinal directions, resulting in full 3D vectorial tracking, as discussed later in Chapter 5. Since CYGNO is a directional dark matter search experiment, this capability is essential for identifying WIMP-like signals based on their incoming direction, while also enhancing background rejection, as discussed in Section 1.4.

An example that underscores the importance of integrating the two sensors' infor-

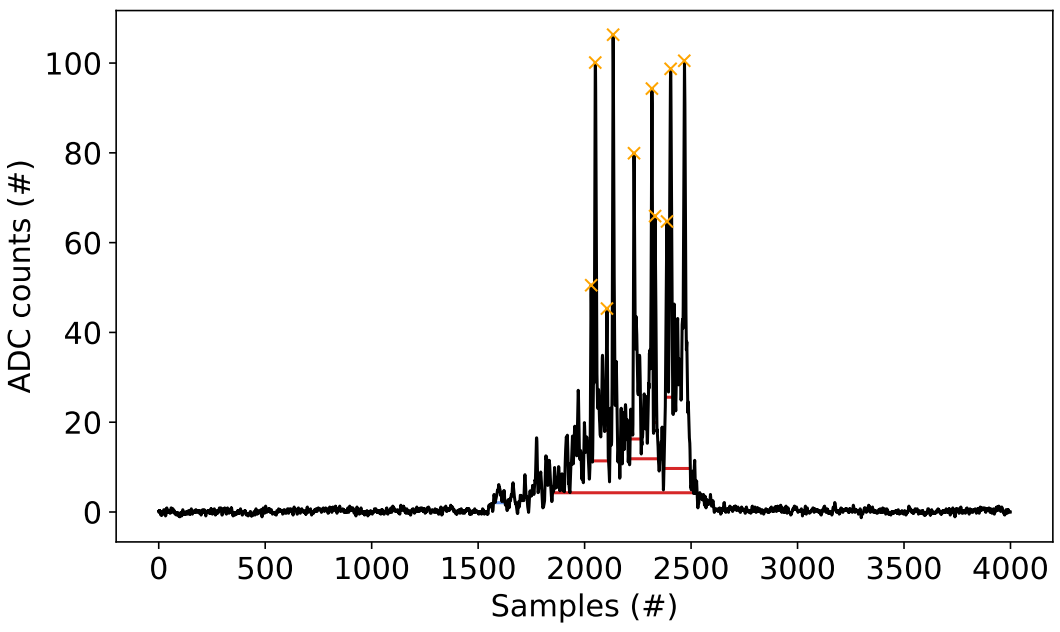

**(a)** PMT #1

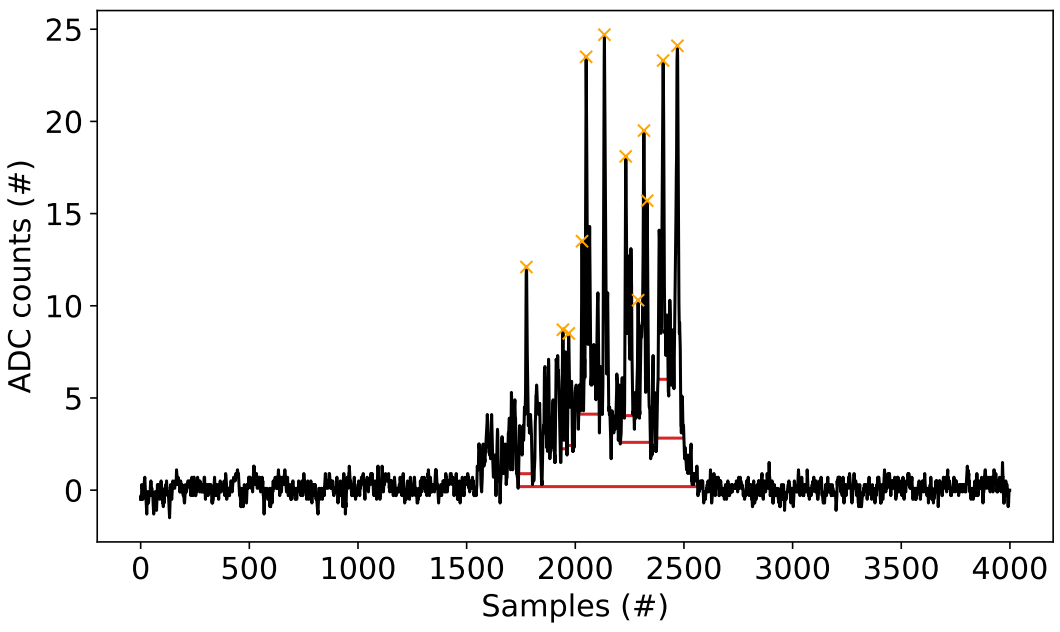

**(b)** PMT #2

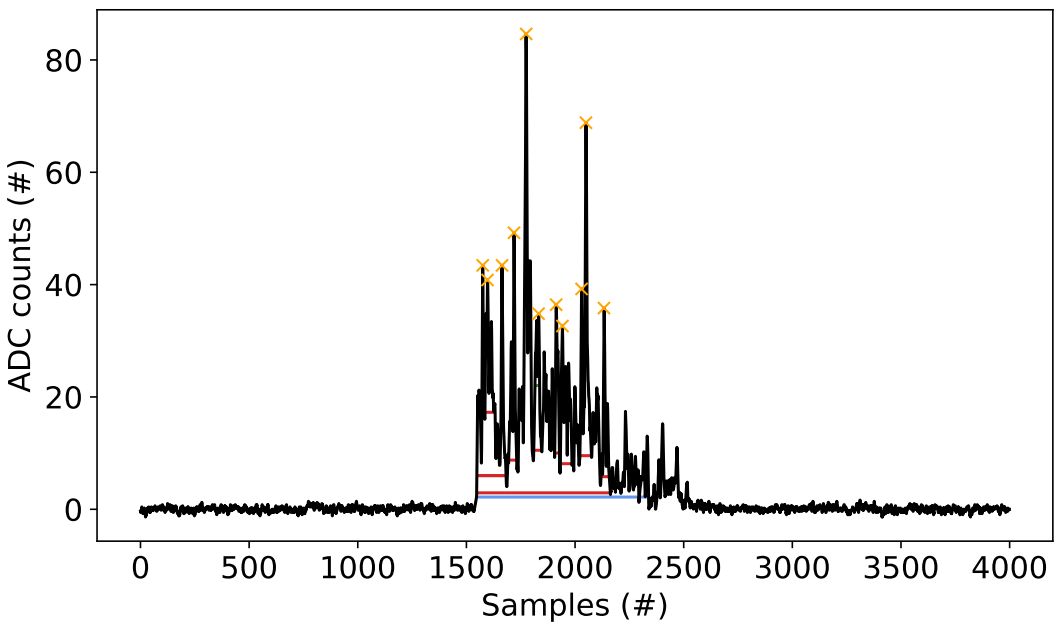

**(c)** PMT #4

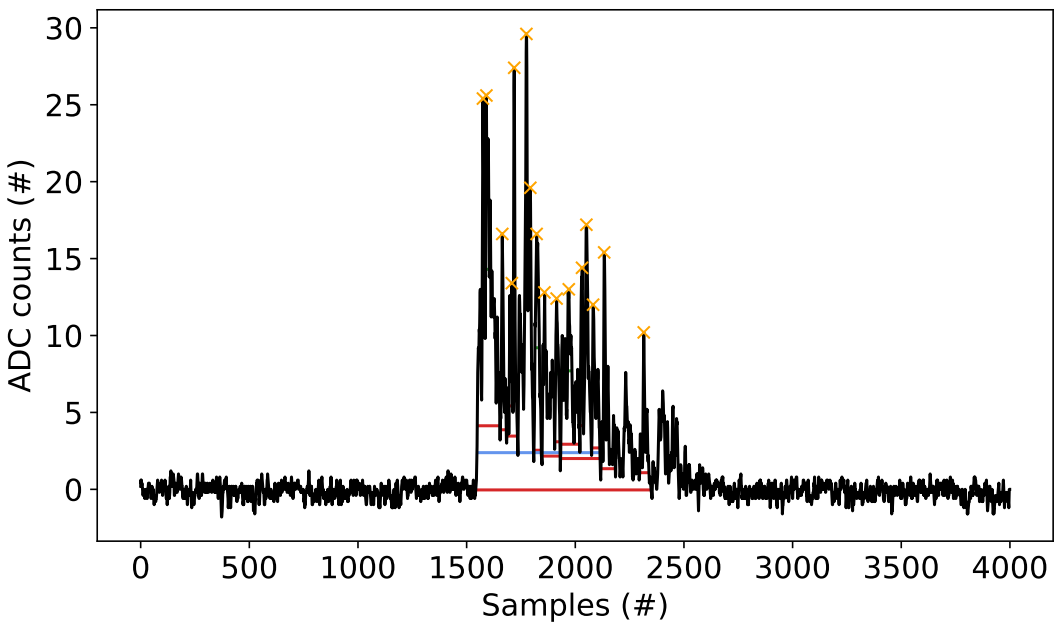

**(d)** PMT #3

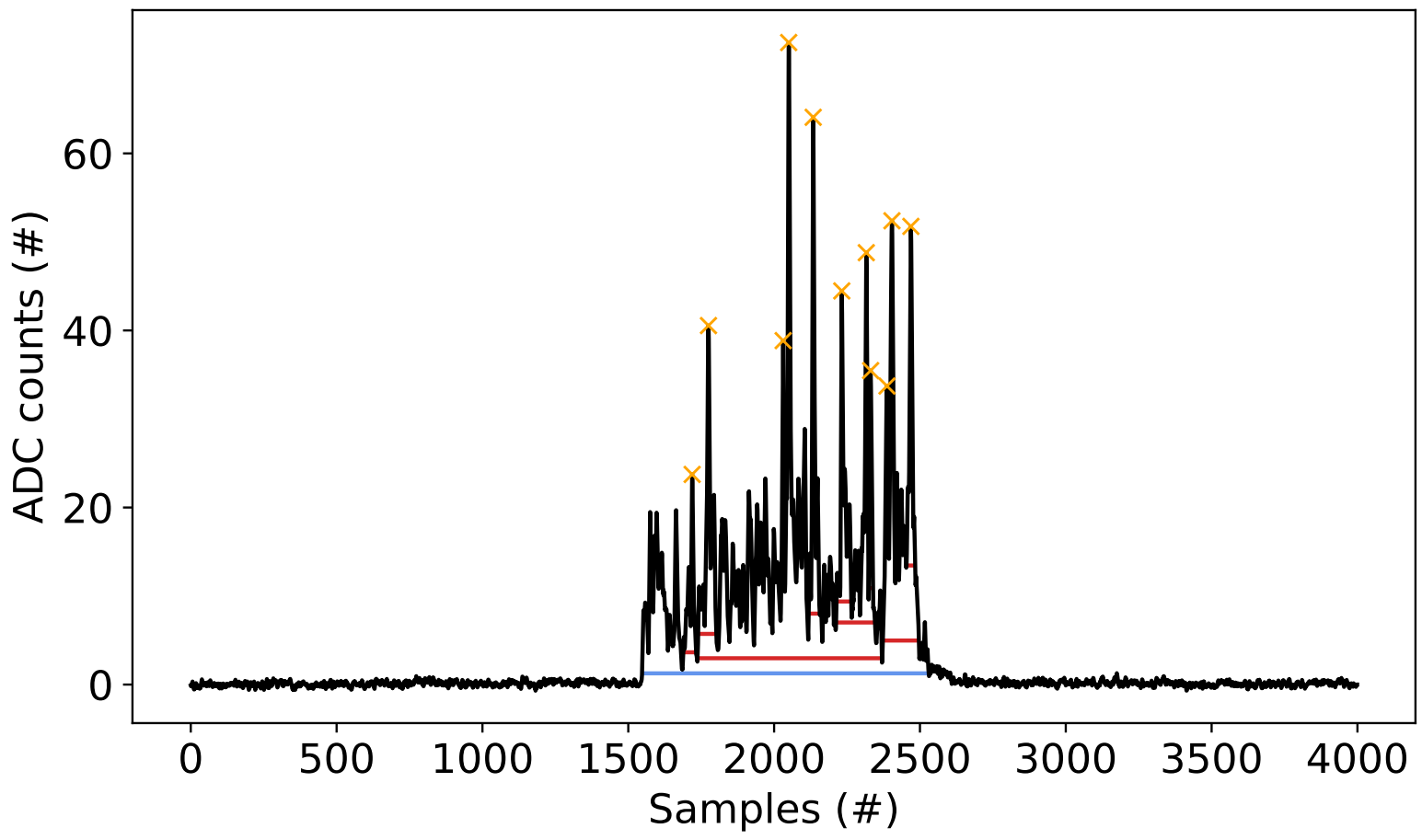

**(e)** Weighted average

**Figure 3.14:** Example of an ionization event observed through the four LIME PMTs ((a) - (d)) and the corresponding weighted average waveform, in (e). This large event is only fully contained by the *slow* digitizer, in which each time sample is 4 ns. In all waveforms, yellow crosses represent the identified peaks detected using the *find_peaks* routine, with their respective full widths indicated by red lines. The ToT variable is shown in blue in all waveforms. The weighted average is calculated using the normalized SNR of each waveform, defined as the maximum amplitude of the waveform divided by its RMS.

mation is shown in Figure 3.15. In this, two different CMOS images are presented, each showing an alpha particle track with very similar shapes in terms of length and position. However, when the PMT signals associated with these events are analyzed (right panels), it becomes clear that the two alpha events are quite distinct: one event has a short time extension (Figure 3.15b), indicating a small $\Delta Z$, while the other exhibits a much larger time extension (Figure 3.15d). This difference reveals that the second event traveled a longer 3D path throughout the gas volume and has a higher angle relative to the GEM plane when compared to the first event. Such information is only possible to retrieve once the two sensors' information is matched and considered together.

The matching of the two sensors' information for subsequent 3D reconstruction of ionization tracks saw steady improvements during the writing of this thesis and, with support from the author, a methodology was recently established and published in [221]. This method makes use of Bayesian Networks [222,223] to infer the XY shape and position of ionizing tracks on the GEM plane from PMT signals. This allows the comparison of event positions as seen by the two sensors, and enables the association of PMT and CMOS data, resulting in the possible combining the $\Delta Z$ light profile from the PMTs with the high-resolution XY information from the CMOS. This section presents the current developments of the BAT-fit approach (Section 3.3.1), along with the results obtained for the reconstruction of short (Section 3.3.2) and extended tracks (Section 3.3.3). It concludes with a discussion of the implications of this work for the future of 3D analysis in CYGNO (Section 3.3.4). This has been a collaborative effort between the author of this thesis and other students of the CYGNO group.

### 3.3.1 Bayesian Networks for PMT-based track reconstruction

To match the information between the CMOS and PMT sensors, a dedicated *one-to-one matching* algorithm was developed by F. Borra in his MSc thesis [224] with contribution from the author of this thesis, resulting in a published work [221]. The algorithm is based on Bayesian Networks and built on the the proportionality between the light recorded by each PMT and its distance to the emitted light (see Section 3.2.3). For this thesis, the algorithm has been additionally adapted and optimized by the author specifically for the analysis of alpha particles, in Chapter 5. Although the algorithm itself is not new, its implementation and testing within the 3D analysis framework for different particle types (alphas) is a novel contribution to the CYGNO collaboration.

In a scenario as LIME, for any light event occurring at the GEM plane, the set of parameters ($\theta$) used in the Bayesian inference is as follows:

$$\theta = (L, x, y) \tag{3.6}$$

where $(x, y)$ is the position of the event and $L$ is the total light emitted at the last GEM, as seen by the PMTs.

In Bayesian inference, Bayes' theorem is used to update the model parameters, $\theta$, as

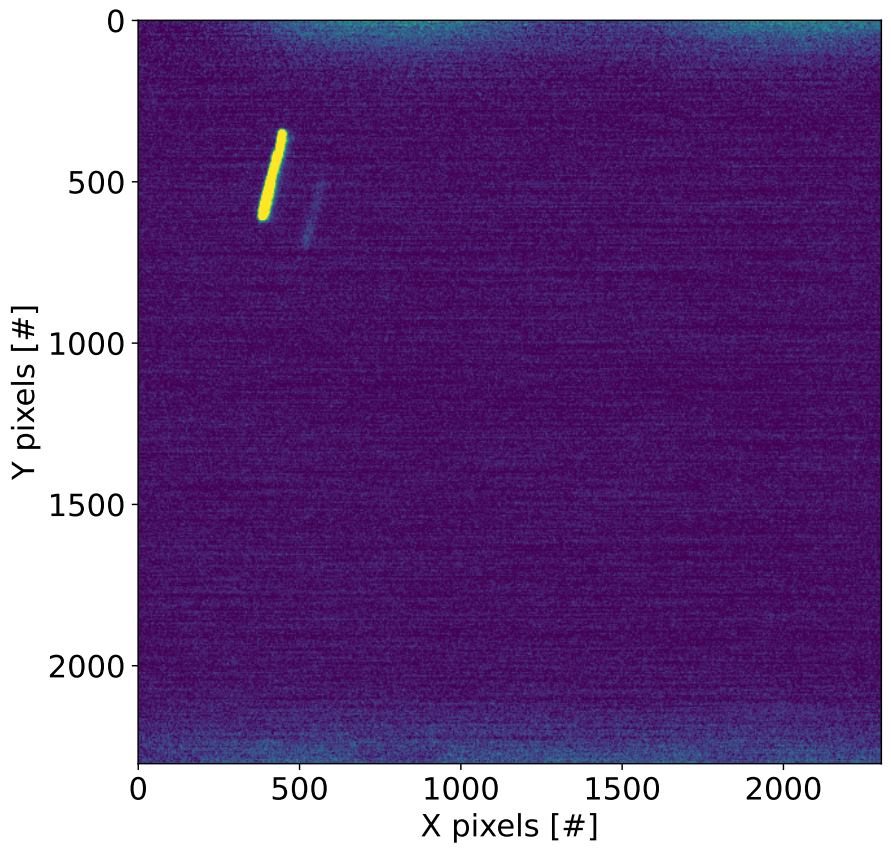

**(a)** Alpha #1 – CMOS

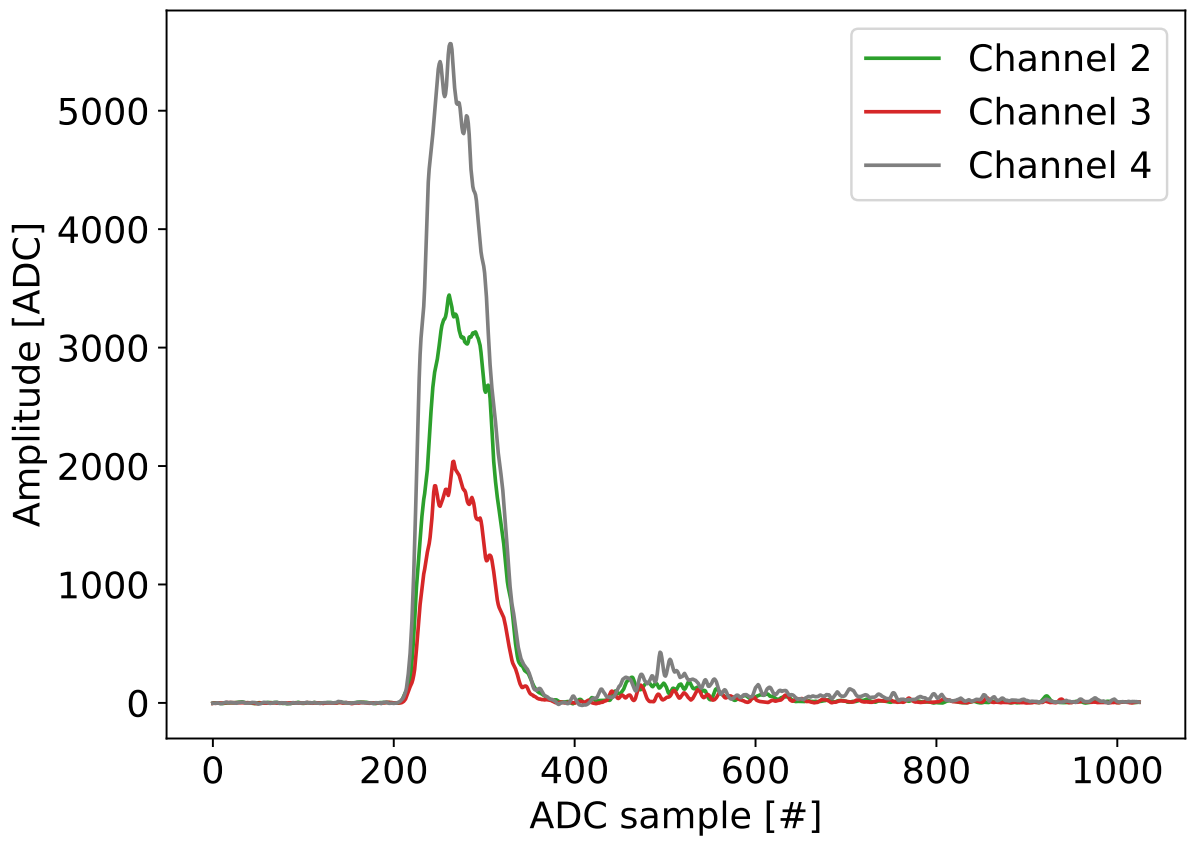

**(b)** Alpha #1 – PMTs

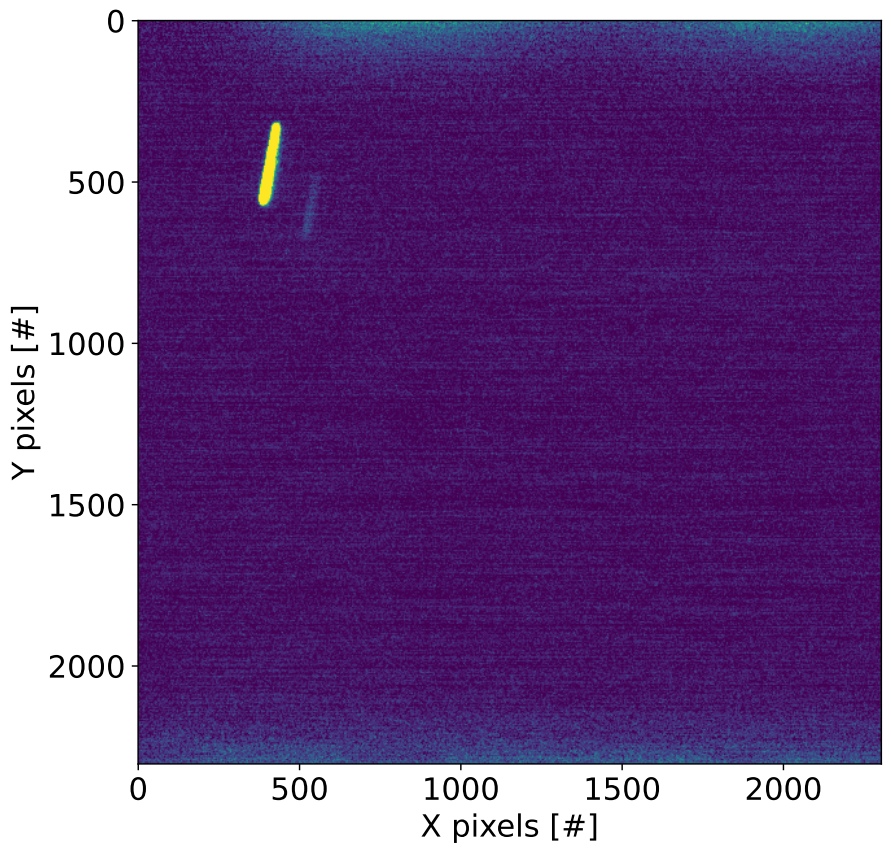

**(c)** Alpha #2 – CMOS

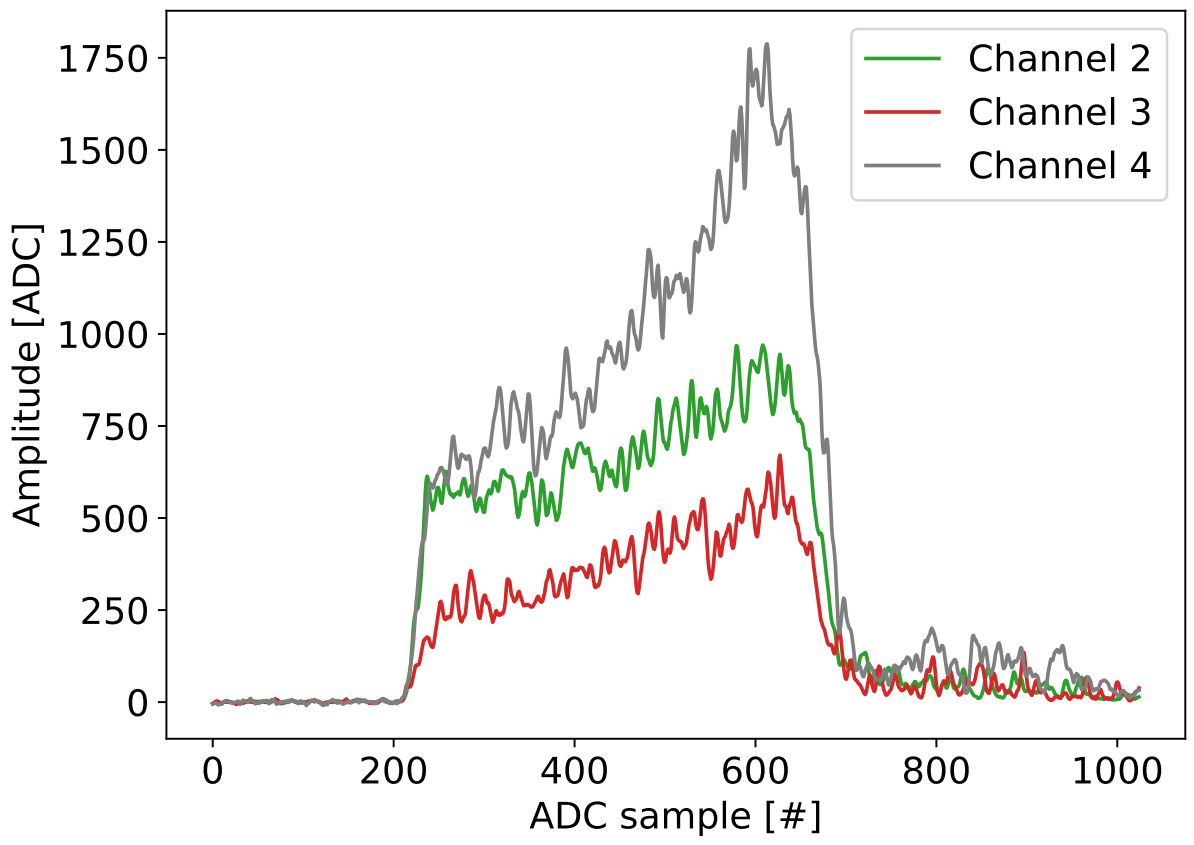

**(d)** Alpha #2 – PMTs

**Figure 3.15:** Example of two different alpha events that appear similar in the CMOS images (a and c) but show significant differences in the corresponding PMT signals (b and d). The signal from PMT #1 has been omitted for clarity.

new information becomes available in the form of experimental observations, following the expression:

$$p(\theta|\{x\}) = \frac{p(\{x\}|\theta) \cdot \pi(\theta)}{p(\{x\})} \tag{3.7}$$

where $p(\theta|\{x\})$ is the posterior probability density function (pdf) for the parameters $\theta$, given the observed data $\{x\}$. The prior probability is given by $\pi(\theta)$, and $p(\{x\}|\theta)$ is the likelihood, representing the probability of observing the data given $\theta$.

The priors for this fit restrict the X and Y parameters to the dimensions of the GEM plane, which are $33 \times 33\ \mathrm{cm}^2$. The only constraint on L is that it must be positive, with an additional loose upper limit applied to speed up the fitting process. The PMT response is assumed to follow a normal distribution with a standard deviation of $\sigma_{ij} = 10\%$. In this case, the likelihood function used in the inference process of Equation 3.7 becomes:

$$p(\{x\}|\theta) = \prod_{j=1}^{N} \prod_{i=1}^{4} \frac{1}{\sqrt{2\pi}\sigma_{ij}} \cdot \exp\left( - \frac{(Q_{ij} - L'_{ij}(\theta))^2}{2\sigma_{ij}^2} \right) \tag{3.8}$$

where $j$ runs over all light-emitting sources, $i$ is the PMT index, $Q_{ij}$ represents the integrated charge collected by the $i$-th PMT for the $j$-th emission point, and $L'_{ij}(\theta)$ is the expected light ("true value") seen by each PMT, based on the total emitted light ($L_j$) and its emission position $(x_j, y_j)$, following:

$$L'_{ij}(\theta) \propto L_j / R_{ij}^{\alpha} \tag{3.9}$$

where the value of the parameter $\alpha$ is fixed to 4 from the discussion in Section 3.2.3. $R_{ij}$ is the distance between the light emission and each PMT position $(x_i, y_i, h)$, based on Figure 3.10, and therefore given by:

$$R_{ij} = \sqrt{(x_j - x_i)^2 + (y_j - y_i)^2 + h^2}. \tag{3.10}$$

In this equation, $h = 19\ \mathrm{cm}$ is the longitudinal distance between the PMTs and the position of the last GEM, where the light is produced, following Figure 2.17.

The posterior distribution is computed using numerical methods, specifically the Metropolis-Hastings MCMC algorithm, implemented through the Bayesian Analysis Toolkit (BAT) [225]. This package has been widely used in the HEP field by numerous collaborations, including UTfit [226] and HEPfit [227].

Since each PMT may have a slightly different response to the same amount of light due to differences in intrinsic gain, additional calibration parameters ($c_i$) were intro-duced. This modifies $L'_{ij}(\theta)$ to:

$$L'_{ij}(\theta) = c_i \cdot \frac{L_j}{R^4_{ij}} \tag{3.11}$$

The $c_i$ parameters have been calibrated using a dataset where the original light intensity of the signals was known, leaving the fit to adjust only the event positions and calibration constants [221,224].

### 3.3.2 Results for spot-like interactions

The BAT-fit was initially tested by reconstructing the positions of the $^{55}Fe$ spot-like signals from calibration runs. $^{55}Fe$ spots are short-range electron recoils from 5.9 $keV$ photons that appear as round clusters in CMOS images and are associated with single-peak PMT signals, making them an ideal first candidate to evaluate the BAT-fit capabilities. For each set of four PMT signals, the parameters are inferred using Equation 3.8 with $N = 1$ source. The priors on $x$ and $y$ are flat and constrained to the GEM plane, while the prior on the light intensity is flat for positive values below a defined maximum, as mentioned above. The integrated charge of the PMT signal is evaluated over a time window of about 60 $ns$ around the main (and only) peak of the waveform, as shown in Figure 3.16a. This value translates to a resolution in Z of roughly $\sim 3.3$ $mm$. The inferred $(x, y)$ parameters of each interaction can then be overlaid on the CMOS image to visually compare the fitted positions with the CMOS ones. An example of this procedure is shown in Figure 3.16b, where the $^{55}Fe$ spots are visible in the picture (yellow dots on a blue background), and the BAT-fitted positions are marked as red crosses, with the size of the crosses representing the uncertainty.

The performance of the BAT-fit was assessed by measuring the distance between the $^{55}Fe$ clusters in the CMOS image and the $^{55}Fe$ PMT-reconstructed positions in the XY plane. This was performed with events featuring only a single visible spot in the image and a single PMT trigger associated with it, to avoid mismatching of signals, and thus allowing for a precise evaluation of the accuracy of the reconstruction process.

Figure 3.17 presents the distribution of the residuals ($\Delta x$ and $\Delta y$) between the PMT- and camera-based reconstructed positions. The mean and standard deviations of the two distributions are [221]:

$$\begin{aligned} \Delta x &= (-0.07 \pm 0.85)\ cm \\ \Delta y &= (0.2 \pm 1.6)\ cm \end{aligned} \tag{3.12}$$

The centimeter-order resolution on a $33 \times 33$ $cm^2$ plane highlights the efficacy of this method for identifying the position of the event through the PMT waveforms. The reconstruction of $^{55}Fe$ spots with the BAT-fit was also tested in images with multiple events.

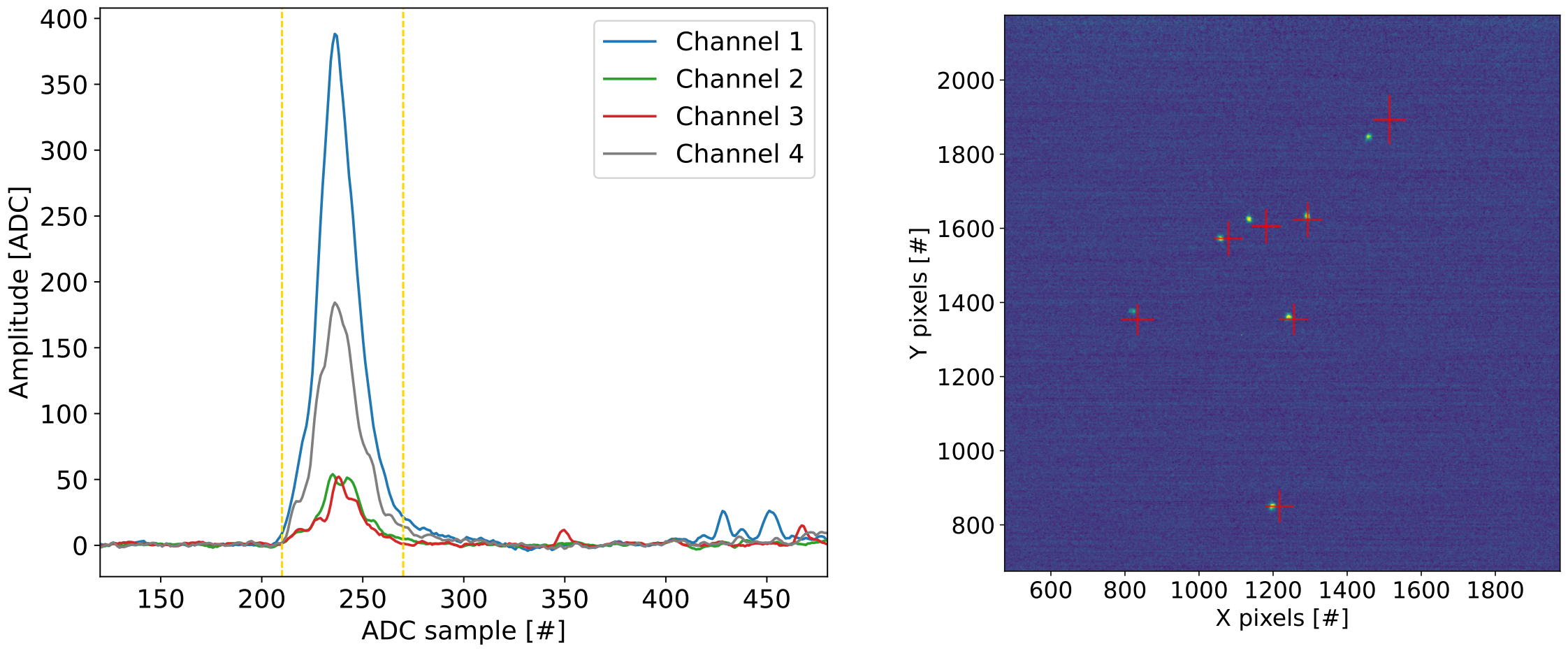

(a) (b)

**Figure 3.16:** Demonstration of the application of the BAT-fit in spot-like interaction induced by 5.9 keV emitted by an $^{55}$Fe source, showing (a) 4 waveforms corresponding to one of these events and the charge integration window; and (b) an example of the BAT-fitted $(x, y)$ positions of these events, in red crosses, overlaid on the CMOS image. The length of the crosses represent the uncertainty on the fitting procedure.

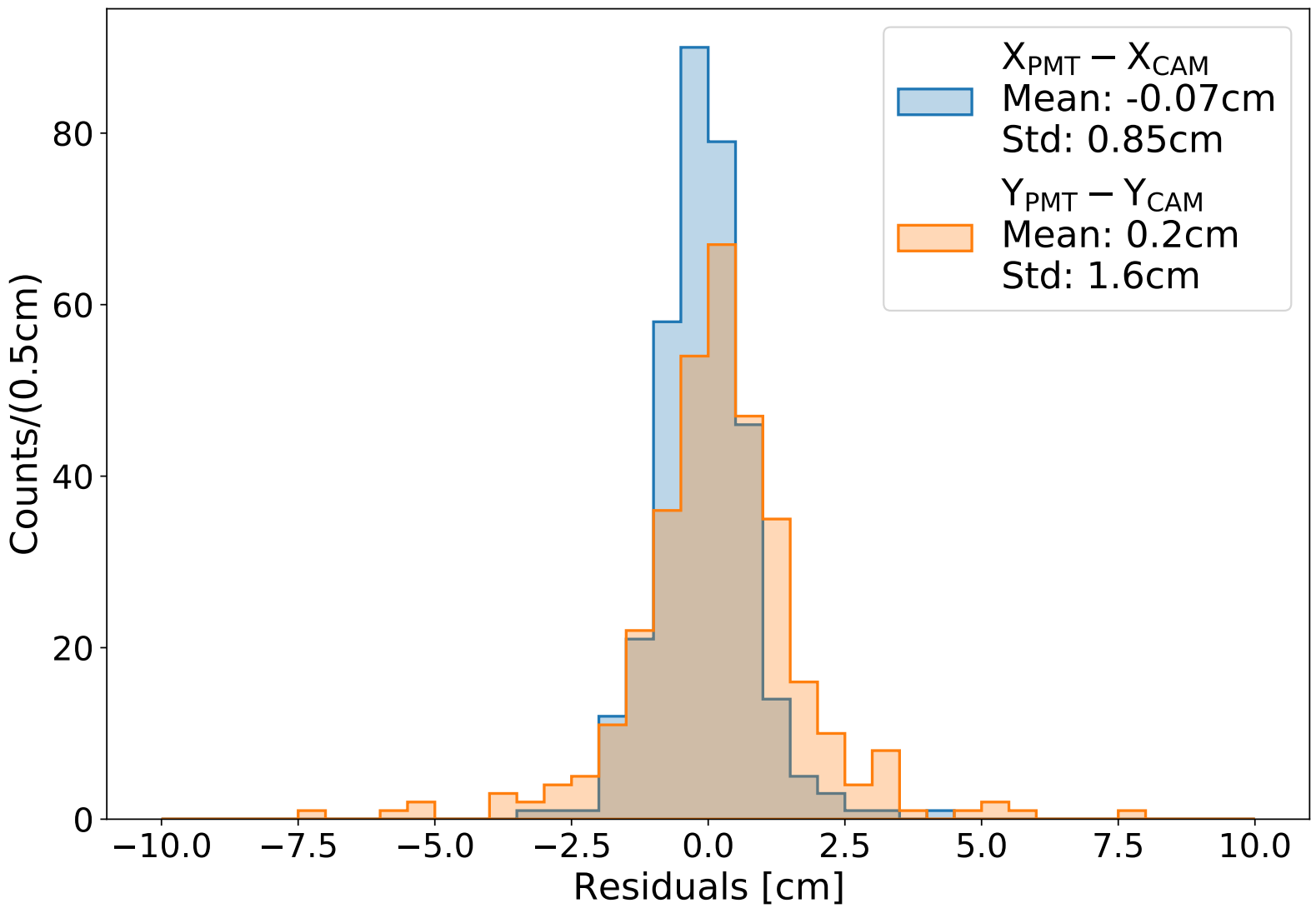

**Figure 3.17:** Distribution of the differences in $x$ and $y$ coordinates between CMOS and PMT reconstruction positions, for spot-like interaction induced by an $^{55}$Fe radioactive source. The dataset consists exclusively of LIME events with a single visible $^{55}$Fe spot in the image and one trigger (waveform) in the PMT data. Retrieved from [221].

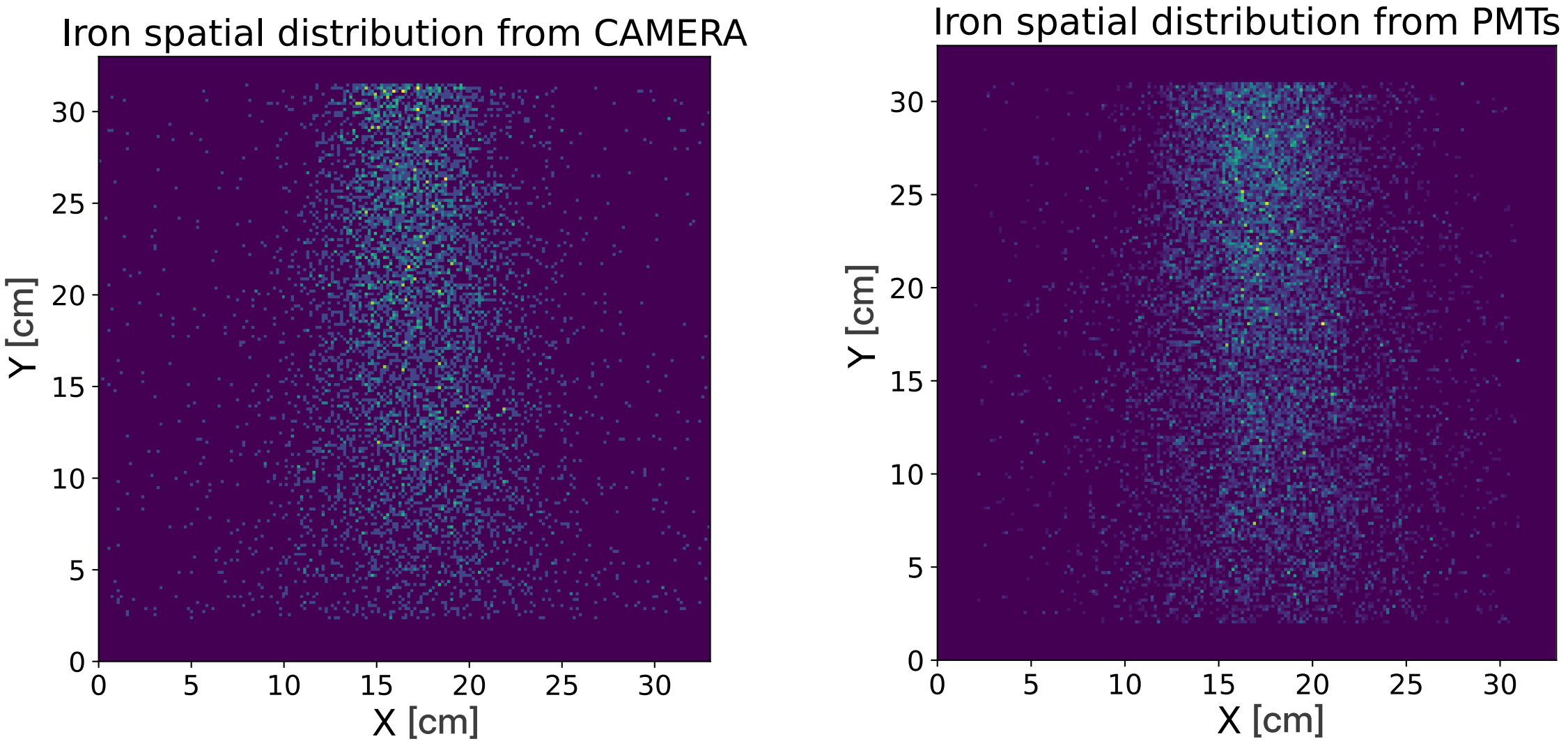

(a) (b)

**Figure 3.18:** Planar distribution of the $^{55}$Fe ionization tracks as reconstructed by (a) the CMOS and (b) the PMTs. Both images shows a similar shower-like shape, reflecting the positioning of the source at the top center of the GEM XY plane. Retrieved from [221].

Figure 3.18 shows two plots of the spatial distribution of $^{55}$Fe events reconstructed by the PMTs with the BAT-fit, and seen directly from the CMOS camera. The plots are confined to the CMOS field of view and, for better visualization, several events (CMOS images + PMT triggers) were overlapped. In this example, while a 1-to-1 comparison is not performed, it is possible to observe the agreement between the two reconstructed datasets: both distributions show a shower-like shape starting from the top, reflecting the position and emission of photons from the $^{55}$Fe source, and with a density of events decreasing with Y, which is expected from the mean free path of electrons in the gas, on the order of $\mathcal{O}(20)$ cm. The reconstructed events are also mostly concentrated within a narrow band in X due to the presence of a slit collimator in front of the source. As a result, the spot density is higher along X which leads to an improved spatial resolution in this direction when compared to Y, as seen in Figure 3.17.

### 3.3.3 Preliminary reconstruction of extended tracks

The optimization of the BAT-fit procedure for highly extended, ER-like tracks is still on-going. Here, a cherry-picked example is presented to illustrate the approach. When the PMT signal exhibits multiple peaks, as in Figure 3.19a, the peak-finding algorithm described in Section 3.2.1.3 is applied to identify the dominant peaks. From the peaks identified in all four waveforms from the LIME PMTs, only those that are observed within the same time window by at least two PMTs, exceed a defined threshold, and are separated by a minimum of 60 ns are retained. Each selected peak is then treated as an independent localized event and fitted using the BAT-fit as described in Section 3.3.2. The resulting (X, Y) coordinates from this fit can then be used to match or confirm the identity of the track in the CMOS image, as shown in Figure 3.19b, where the fitted peak positions are shown as gray crosses, closely overlapping the track in the CMOS image. Finally, the $\Delta Z$ component of the track is obtained by measuring the time difference between consecutive peaks and converting it into spatial distance using the electron drift velocity in LIME. Combining these two, a three-dimensional representation of the ionization track can be constructed, as shown in Figure 3.19c.

In the case of alpha particles, where the PMT signal appears as a continuous, step-like waveform without prominent peaks, a different strategy was developed by the author for the scope of this thesis, as illustrated in detail in Section 5.3. The signal is divided into short time windows, each approximately 60 ns wide, corresponding to the typical duration of a localized interaction. The integrated charge in each time slice is computed and fitted using the BAT algorithm, resulting in a set of (X, Y) points. This information is then used to match the PMT signals with the corresponding pixel cluster in the CMOS image. Since individual peaks are absent in this case, the $\Delta Z$ coordinate is extracted by measuring the full time-over-threshold width of the signal. This enables the reconstruction of the 3D shape of the alpha track, using the XY information from the CMOS and the $\Delta Z$ from the PMT. This procedure is the basis of the CMOS-PMT matching tools used in Chapter 5 to perform the 3D analysis of alpha particles.

### 3.3.4 Remarks PMT-CMOS matching

The development and integration of the PMT-Reco into the CYGNO general analysis allowed the first dual-sensor analysis of ionizing events in CYGNO detectors, particularly LIME. The analysis of CMOS images had already been extensively developed by the CYGNO collaboration, as documented in various publications [131, 186], which led to a more concise discussion of the topic at the beginning of the chapter. In contrast, the analysis of PMT waveforms was still in a primitive stage before the work carried out in this thesis, which motivated a more detailed and comprehensive description of the reconstruction techniques developed in this regard.

The PMT-reco has been successfully integrated into the general CYGNO analysis framework, and the new organization of sensor-mixed data has already been utilized by other members of the collaboration. The reconstruction techniques and variables developed have also been applied in this thesis, particularly in the study of cosmic muon flux (Chapter 4) and the alpha particle spectrum from the LIME underground scientific runs (Chapter 6). For the specific analysis of alpha particles, new waveform analysis methods were introduced to infer the particles' initial 3D vector direction (Chapter 5). This represents a significant step forward in the development of 3D directional reconstruction techniques for the CYGNO/INITIUM approach, which are essential for directional DM searches, hence producing a compelling result for the future development of the project.

The implementation of the Bayesian framework for analyzing PMT waveforms has

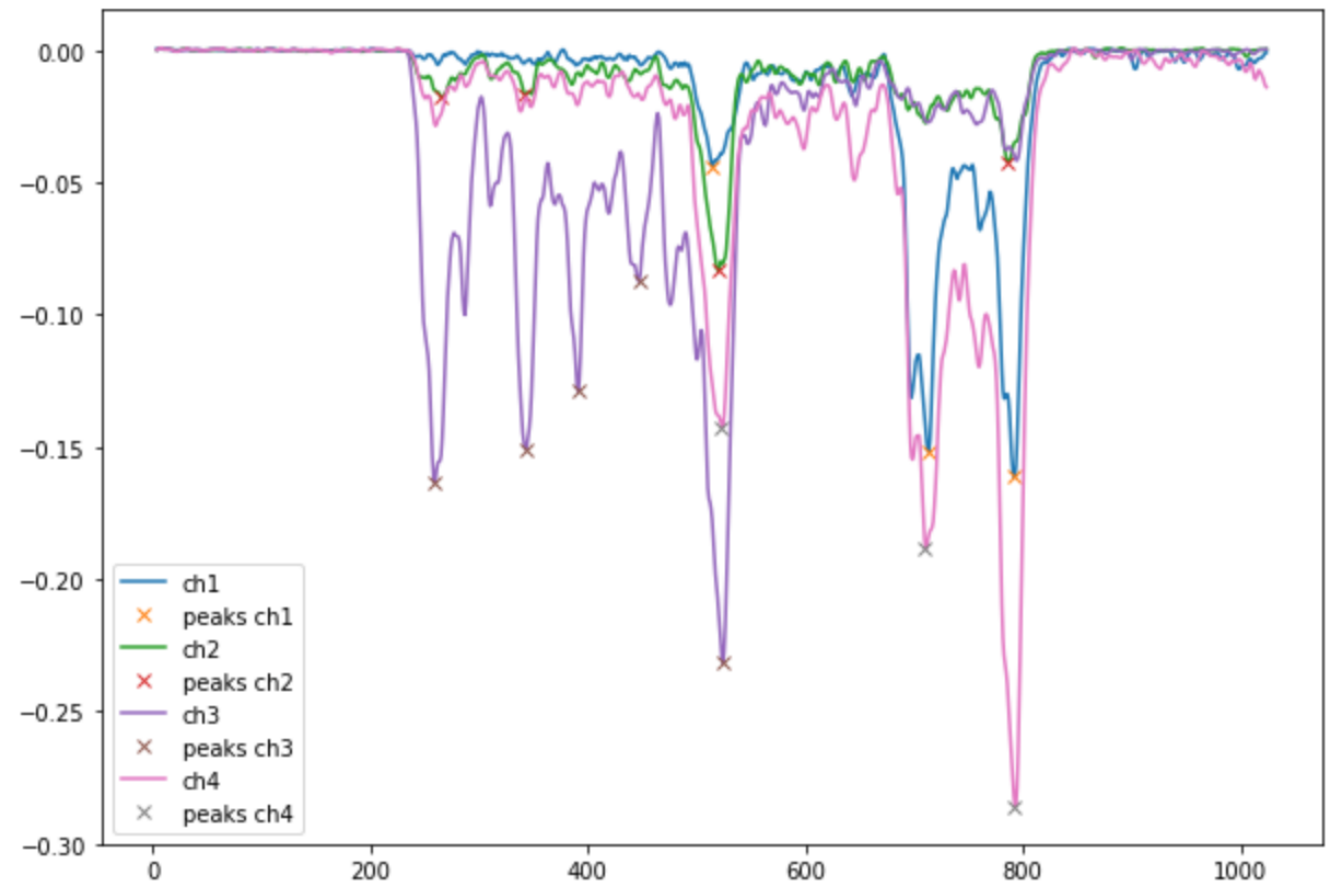

**(a)**

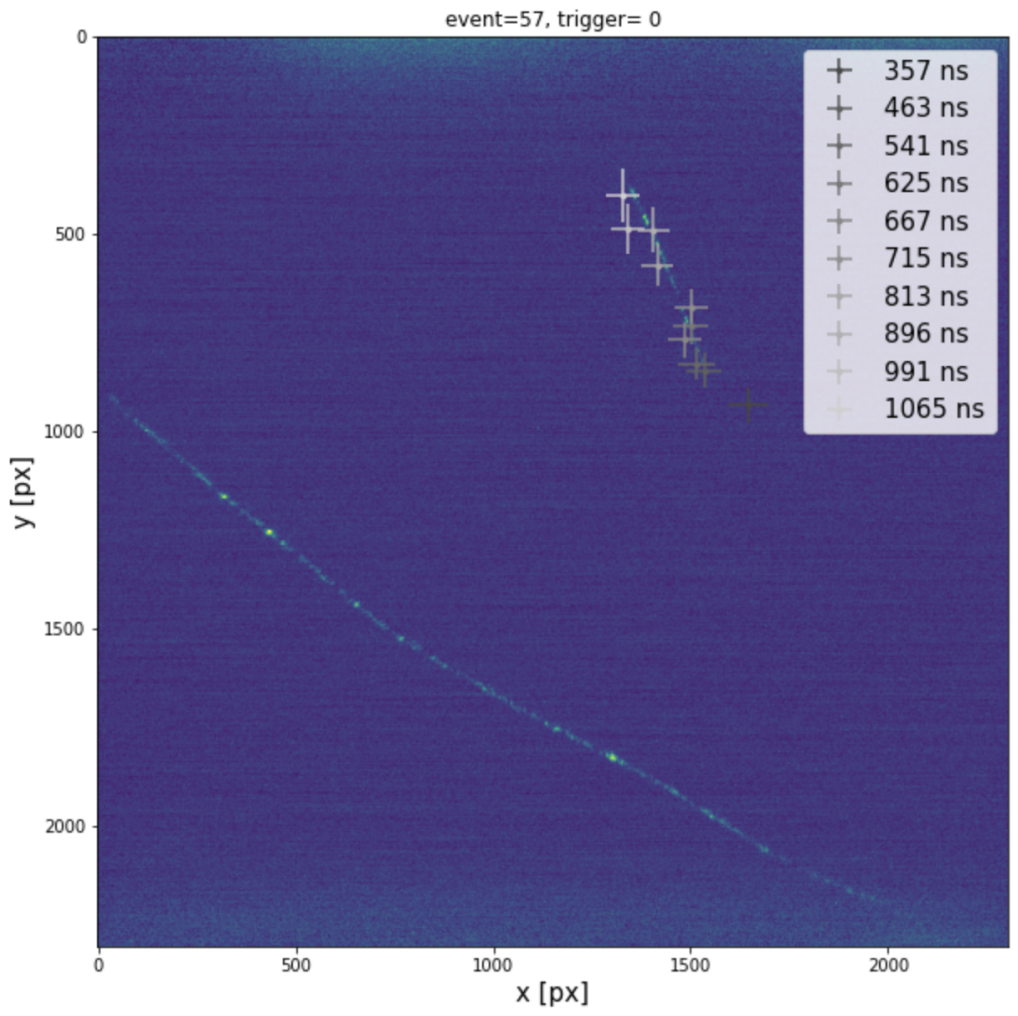

**(b)**

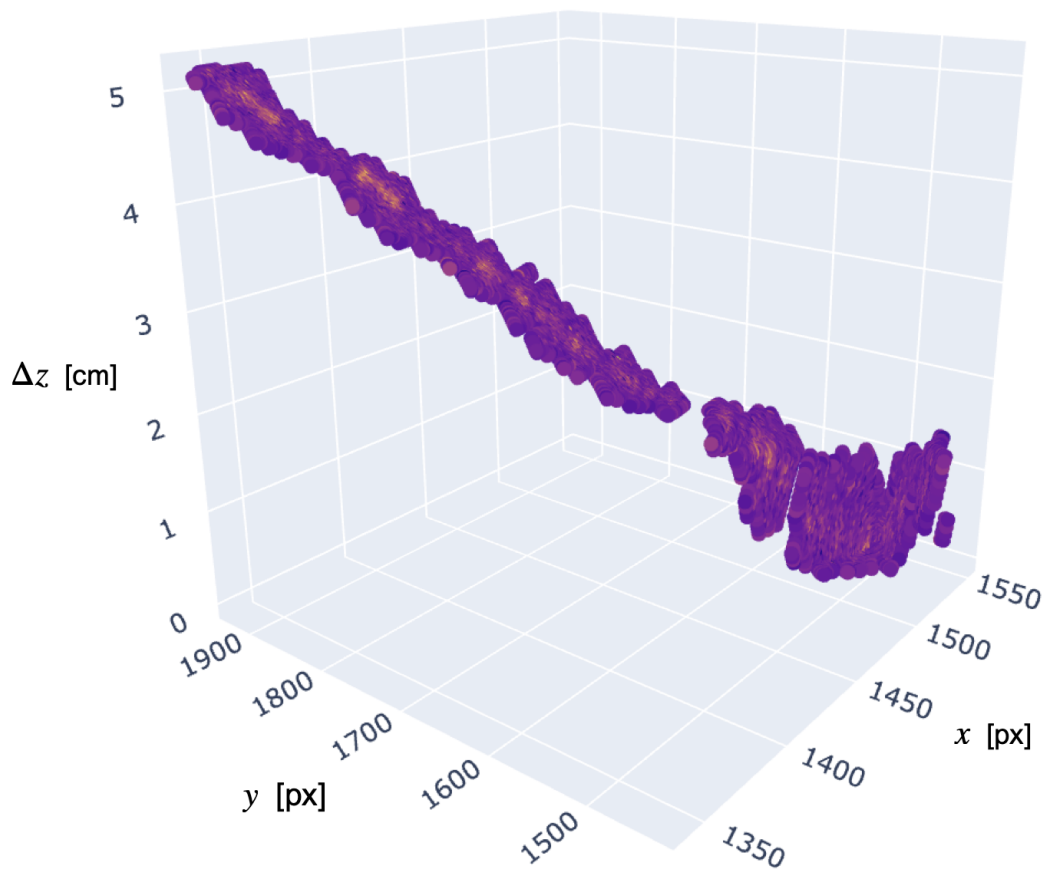

**(c)**

**Figure 3.19:** Example of the 3D reconstruction of an extended track: (a) PMT signals with highlighted peak identified; (b) overlay of the BAT-reconstructed positions from the peak identified in the waveform (gray crosses) on the corresponding CMOS image; (c) final 3D representation of the track combining PMT and sCMOS information.

demonstrated reliable event localization for spot-like events [221]. This enables the correct association of the temporal information retrieved from the PMTs with the corresponding high-resolution CMOS pixel tracks. Initial tests show promising results for the 3D reconstruction of extended tracks that are of interest for DM searches, (i.e., nuclear recoils). This work, effectively developed for alpha particles as a starting point, will be show in more detail in Chapter 5.

Chapter 4

# LIME overground PMT studies

Before the installation of LIME in the underground Laboratori Nazionali del Gran Sasso (LNGS), an extensive overground commissioning campaign was performed. This involved a range of studies focused on evaluating LIME's performance and the overall potential of the CYGNO technique. These studies were centered on aspects such as the efficiency of extracting relevant information from the CMOS and PMT sensors, and how each component integrates into the LIME DAQ system. Additional tests were conducted on ancillary systems – such as high voltage and gas supply – addressing the stability of LIME over extended periods, as well as the optimization of the trigger strategy to be used in the detector. A brief overview of the studies involving the CMOS sensor, carried out by other members of the collaboration, was previously presented in Section 2.2.2.3, and more details can also be found in [186].

Analyzing overground data proved more complex than underground data due to the higher occupancy of the CMOS images. Figure 4.1 illustrates the contrast between overground and underground image occupancy. Because of this high occupancy, the development of the PMT-CMOS information matching techniques discussed in Section 3.3 was not feasible with overground data. These techniques were instead developed starting from underground data, where the event rate is lower. As a result, the overground commissioning of the CMOS and PMT sensors was carried out separately. This chapter reports the commissioning studies conducted by the author concerning the development and optimization of the trigger for data acquisition based on PMT information, and the analysis of PMT-only data with LIME in the overground phase.

The chapter is divided into two main sections. The first describes the development of the PMT trigger strategy (Section 4.1), which was particularly important for the underground commissioning of LIME, as it helped optimize the data-taking procedure and avoid acquiring empty signals. The second section presents a more complex study of the cosmic muon flux performed with LIME (Section 4.2), which served as a testing ground for the PMT-Reco algorithm and some of the newly developed variables shown in Section 3.2.1.

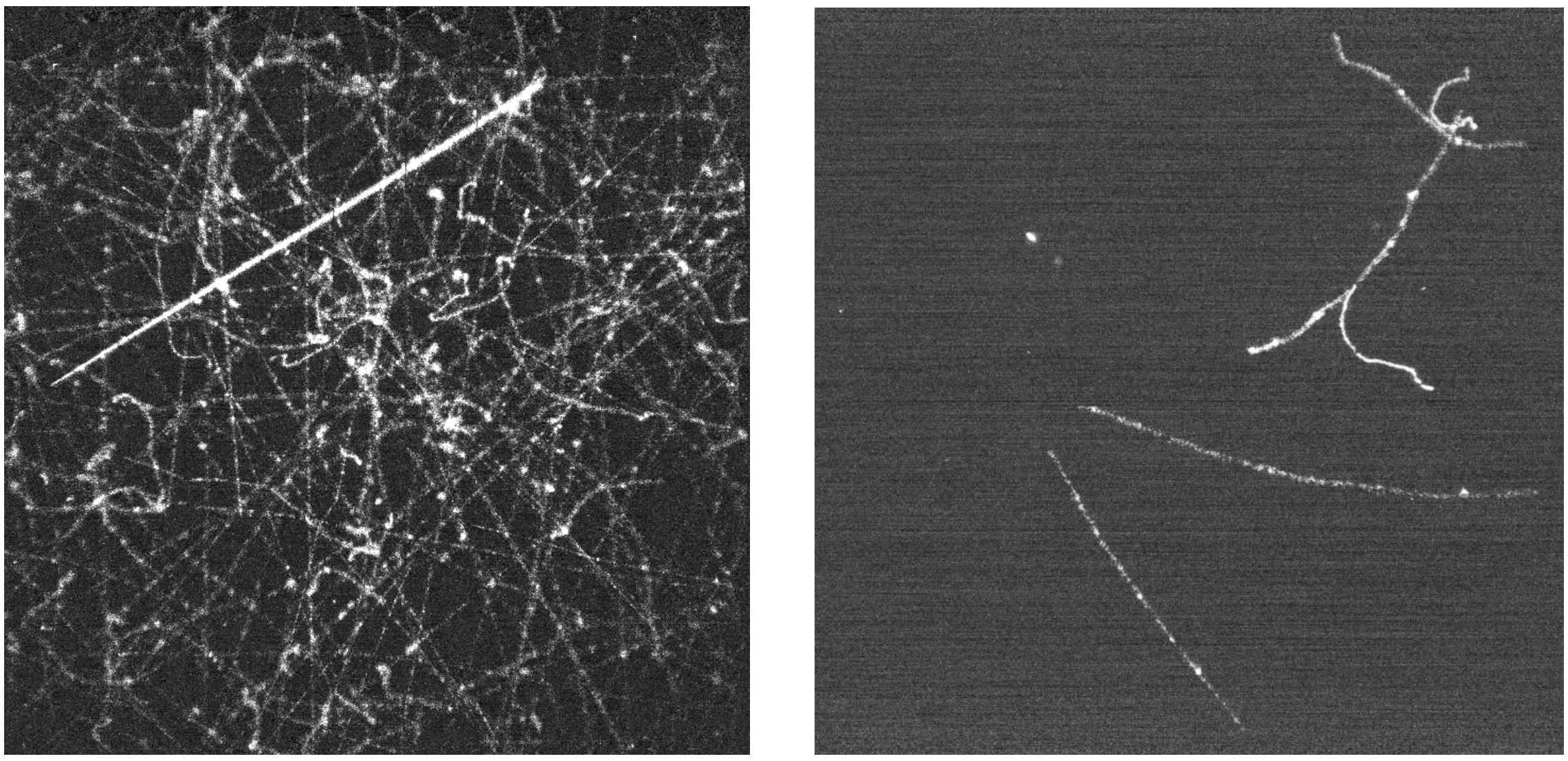

(a) (b)

**Figure 4.1:** Example highlighting the difference in the occupancy (number of events) observed (a) overground at the Frascati laboratories, and (b) underground at the LNGS.

## 4.1 PMT trigger strategy development

In the CYGNO timeline (Section 2.2), LIME represents the first prototype realization with the experiment's full sensor configuration, i.e., a CMOS camera and multiple PMTs. As such, it was only with LIME that it became possible to effectively develop a trigger strategy for the CYGNO experiment. This section presents the tests carried out for this purpose, in which the author was directly involved. It begins with the optimization of the trigger configuration using PMT signals (Section 4.1.1). With the optimized approach, the trigger rate was then evaluated in terms of stability over time (Section 4.1.2) and as a function of event energy and distance from the GEMs (Section 4.1.3).

### 4.1.1 PMT trigger configuration optimization

The trigger system of LIME is based on PMT signals since these offer a faster response to light detection compared to the CMOS sensor, as detailed in Section 2.1.3. As previously mentioned, it was only with LIME that a multi-PMT trigger system could be developed and implemented.

During the overground commissioning of LIME, the PMT trigger strategy and configuration were studied and optimized with the goal of achieving the highest efficiency for $\mathcal{O}(\mathrm{keV})$ energy deposits, while minimizing the occurrence of false triggers. To this end, an $^{55}\mathrm{Fe}$ source was positioned on top of the detector, aligned with the EFTE calibration window (see Section 2.2.2.1), to allow the 5.9 $\mathrm{keV}$ photons to enter LIME's active volume. A threshold of 5 $\mathrm{mV}$ was set on the PMT signal, and the event rate was studied as a function of the number of PMTs triggered and the applied PMT bias voltage. The results of this study are shown in Figure 4.2, where the background rate (i.e., without the $^{55}Fe$ source) was subtracted, and the error bars represent $\sqrt{N}$, with N being the number of triggered events.

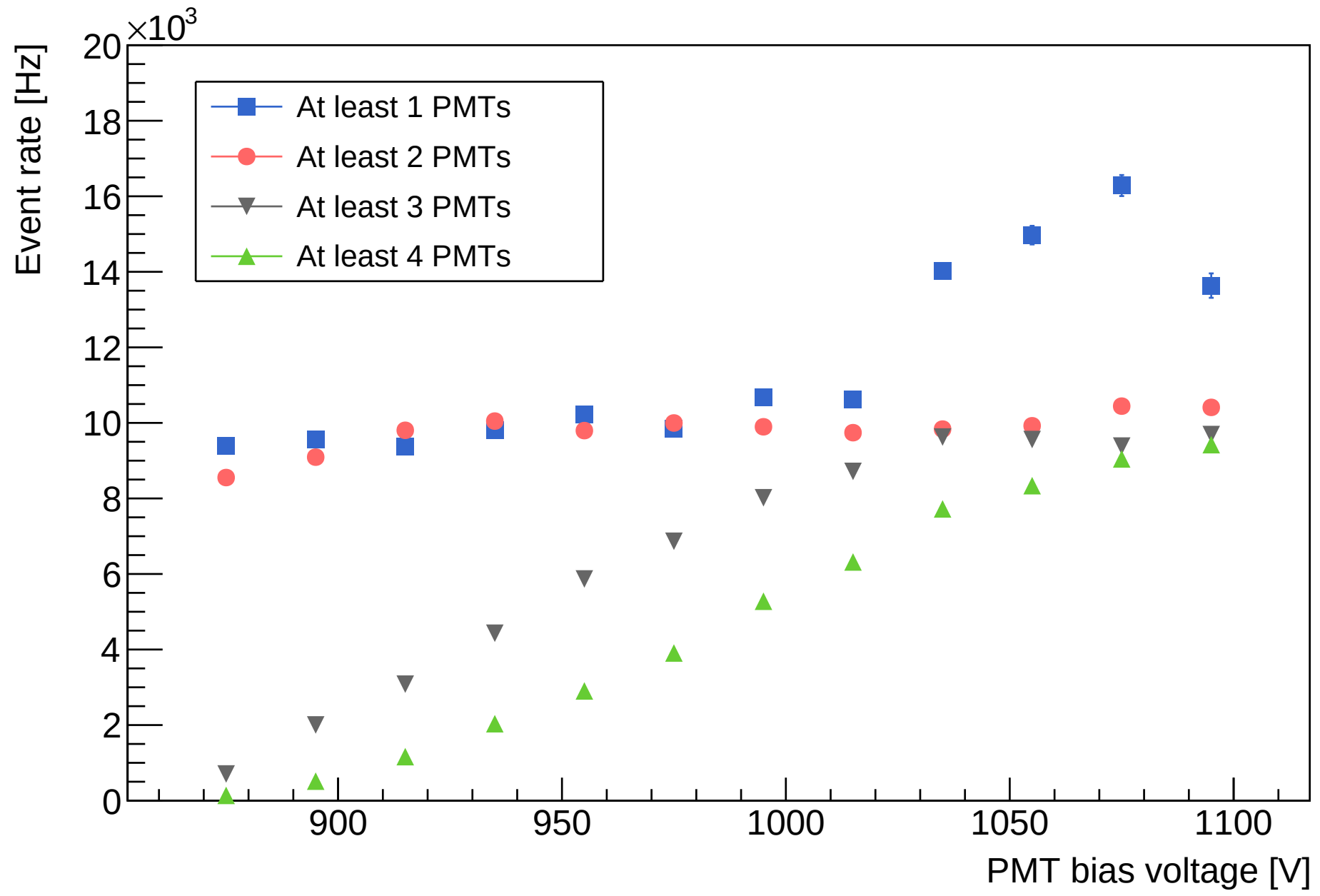

**Figure 4.2:** Rate of events acquired in LIME using a $^{55}Fe$ source as a function of the PMT bias voltage and trigger configuration.

Looking at the results, the requirement of *at least only 1 PMT* triggering becomes unstable above 1000 V, where the trigger rate significantly increases, likely due to false triggers caused by electrons naturally released inside the PMT due to the high gain, initiating an avalanche and leading to a trigger. While in principle these PMTs could also be operated at a bias voltage lower than 1000 V, possible degradation over time of the PMT response could require raising this, consequently increasing the fake rate. These false triggers do not activate the remaining PMTs, making the *at least 2 PMTs* configuration appear more stable. The requirement of *at least 3/4 PMTs* proves inefficient in terms of the low energy threshold, since the light produced by low-energy events may not be sufficient to trigger all the PMTs, as visible from the reduced event rate for these two configurations. As a result of this preliminary study, the *at least 2 PMTs* trigger configuration was found to be the most stable and efficient across all bias voltages tested, leading to the choice of this trigger scheme for the commissioning of LIME underground. These findings were subsequently published in a conference paper [228].

### 4.1.2 Trigger rate stability

Under similar experimental conditions, the stability of the trigger rate over time was also studied for each PMT. This test is important to ensure that the PMTs are working properly and can be used over long periods of time. The results obtained are visible in Figure 4.3, which shows the normalized trigger rate of each PMT, in the presence of the $^{55}\mathrm{Fe}$ source, over a period of 30 $\mathrm{min}$. All PMTs show a flat rate of triggers with time, with variations smaller than 10% with time. With no significant time dependency, this performances were deemed adequate for the upcoming studies.

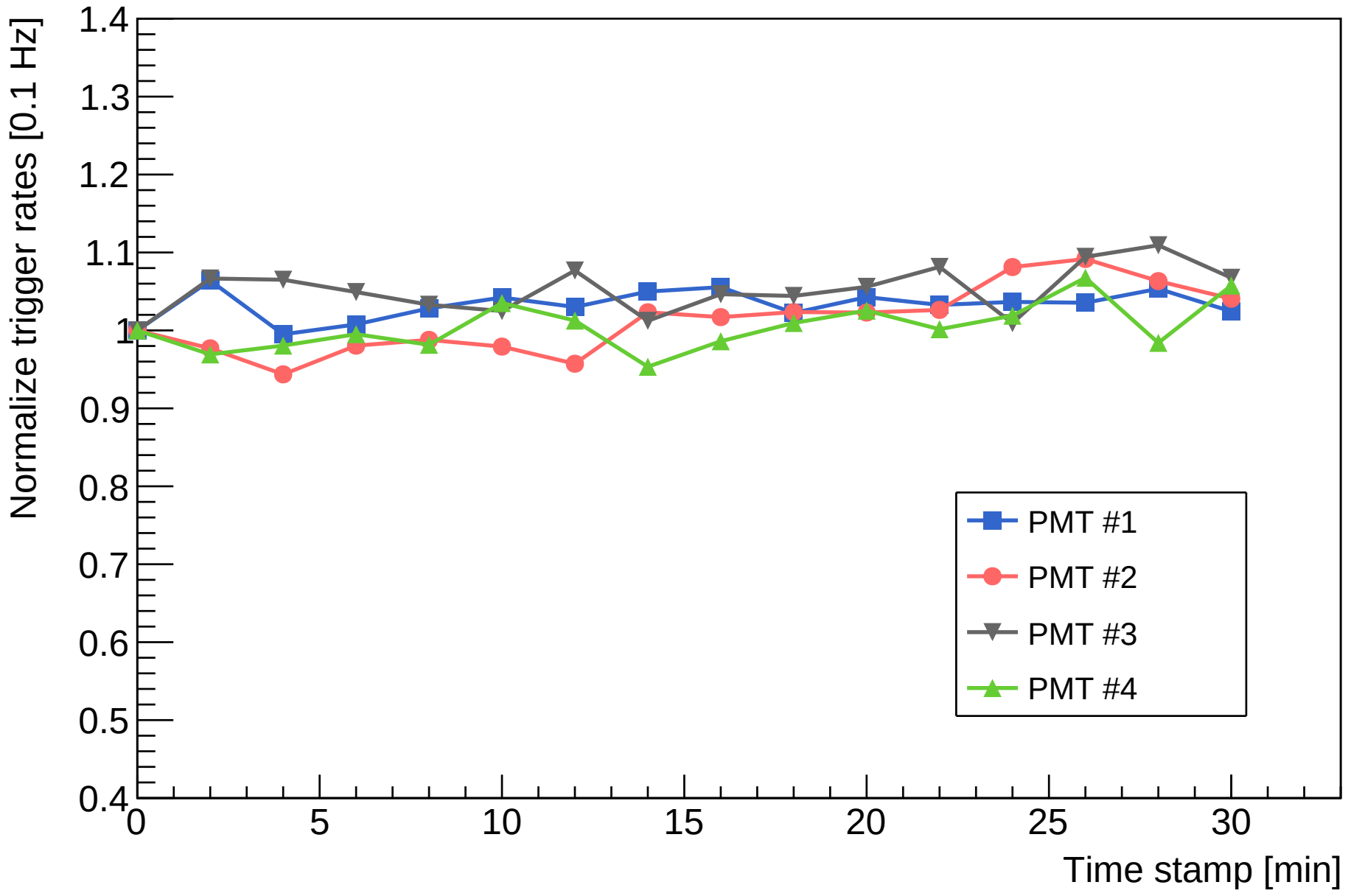

**Figure 4.3:** Rate of events acquired in LIME using a $^{55}\mathrm{Fe}$ source as a function of time, for each of the 4 PMTs.

### 4.1.3 Multi-source PMT trigger rate

The trigger strategy performance and stability was furthermore validated as a function of the event energy and distance from the GEMs. This test was performed using the Amersham AMC.2084 multi-energy X-ray radioactive source. This consists of a 370 $\mathrm{MBq}$ $^{241}\mathrm{Am}$ source that emits $\sim$ 5 $\mathrm{MeV}$ alphas into a target that is placed in a rotating support. This support contains multiple targets (different elements) which can be selected and placed under the flux of alphas. When these targets are hit, electrons are ejected from the inner shell of the atom, the $k$-shell. The outer shells ($l$ and $m$) rapidly fill the vacancy, leading to the emission of monochromatic X-rays, respectively named $K_\alpha$ and $K_\beta$. Figure 4.4 shows a schematic of this radioactive source, as well as the energies of the $K_\alpha$ and $K_\beta$ X-rays emitted by each target. The emitted gammas interact inside LIME mainly by the photoelectric effect, originating electron recoils in the active volume of the detector.

The measured event rate for the *at least 2 PMTs* trigger strategy resulted from the study in Section 4.1.1 as a function of energy at 3 distances (10, 25, and 45 $\mathrm{cm}$) from the

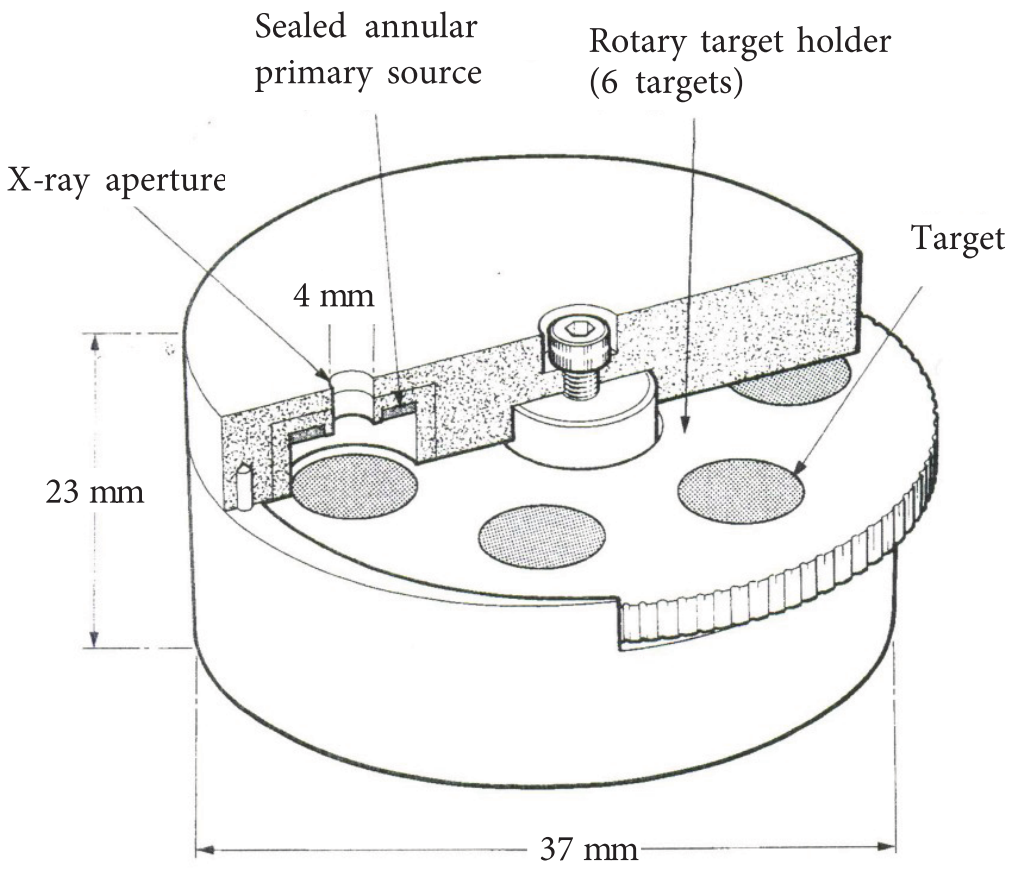

**(a)**

| Target material | $K_\alpha$ [keV] | $K_\beta$ [keV] | Photon yield [$s^{-1}sr^{-1}$] |
|---|---|---|---|
| Cu | 8.04 | 8.91 | 2500 |
| Rb | 13.37 | 14.97 | 8800 |
| Mo | 17.44 | 19.63 | 24000 |
| Ag | 22.10 | 24.99 | 38000 |
| Ba | 32.06 | 36.55 | 46000 |
| Tb | 44.23 | 50.65 | 76000 |

**(b)**

**Figure 4.4:** Amersham AMC.2084 multi-energy X-ray radioactive source used for the studies of the trigger rate for different ER energies and for different drifting distances, showing (a) a schematic of the source, and (b) the possible energies emitted by this source for each of its targets. Retrieved from [229].

GEMs is shown in Figure 4.5. The stability shown further validates the chosen trigger strategy, confirming that this scheme can be reliably used for LIME operations once deployed underground at LNGS. These results are also discussed in [228].

A final test was performed to assess the overall performance of the entire PMT system – from data acquisition to the analysis results obtained with the newly developed PMT-reco – by measuring the overground cosmic muon flux at LNF, as reported in Section 4.2.

## 4.2 Cosmic muons flux measurement

With the introduction of the PMT waveforms and PMT-Reco into the general CYGNO-reconstruction framework, further discussed in Section 3.2.1, it became important to test its potential and efficiency through targeted analyses. This process allowed for the identification of limitations or algorithm errors, which were then addressed, ensuring that the detector and overall analysis were ready for upcoming underground operation.

To test these capabilities, a measurement of the angular dependence of the atmospheric muon flux at LNF (near sea level) was conducted. Since cosmic muons follow a well-known $\cos^2(\psi)$ distribution of the zenith angle, measuring the relative rate at different angles with respect to the detector axis can provide a validation of the PMT DAQ and analysis described so far. To achieve this, scintillator bars were placed at different positions around the detector to serve as triggers, selecting events arriving from specific angles, as illustrated in Section 4.2.2. The theoretical context of cosmic rays arriving at Earth is discussed in Section 4.2.1, and the data analysis procedure – where the PMT waveform time-over-threshold is used to calculate the muon incidence angle – is presented in Section 4.2.3. The final results, shown in Section 4.2.4, effectively validated the CYGNO DAQ and PMT reconstruction.

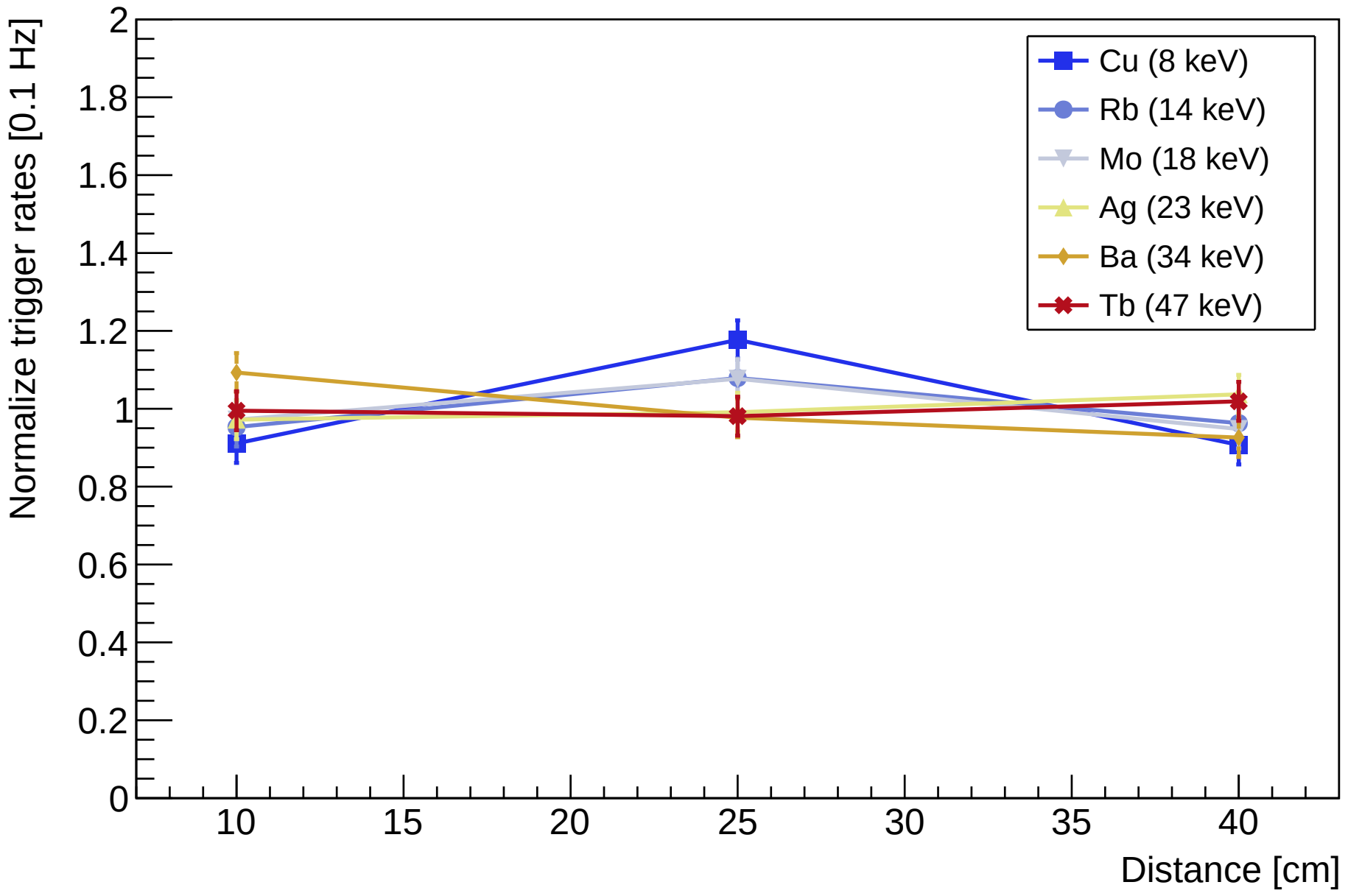

**Figure 4.5:** Rate of events acquired in LIME using the *at least 2 PMTs* trigger configuration for different electron recoil event energies (average energy shown, see Figure 4.4b) and different drifting distances.

### 4.2.1 Theoretical contextualization

Cosmic rays (CRs), primarily composed of high-energy protons, helium nuclei, and a small fraction of heavier nuclei, are originated from various energetic astrophysical processes. As they travel through the interstellar medium, they eventually reach Earth, where particles with energies above a few GeV interact with the magnetic field and collide with atmospheric molecules, initiating hadronic and electromagnetic cascades. During these cascades, pions and kaons are produced and rapidly decay into lighter leptons and gamma rays [230]. The study of these events is important for determining the nature, energy, and direction of the primary cosmic ray responsible for the cascade.

Atmospheric muons are the most abundant charged particles originating from these cascades reaching Earth surface, with an average flux of approximately 1 $\mathrm{cm^{-2} min^{-1}}$ and typical energies around 4 GeV [231]. These muons primarily lose energy through ionization, at a nearly constant rate of about 2 $\mathrm{MeV\ g^{-1}\ cm^{2}}$ [232], which classifies them as minimum ionizing particles (MIPs).

Measuring the distributions of secondary cosmic rays, particularly muons, at different altitudes and latitudes provides important insights into the physics of cosmic rays, and this research remains highly relevant today. The study of muons has led to applications in various scientific fields, such as the muon tomography technique, where muons create a projection of the volume they traverse. This technique has been employed in areas like civil engineering, for example, to construct two-dimensional maps of nuclear reactors [233], and in geophysics and archaeology, to study seismic activity recorded in ancient burial mounds [234]. Additionally, understanding the flux and characteristics of muons reaching the Earth's surface is crucial for assessing their impact on radiation-induced errors in microelectronic circuits [235]. In the realm of physics, muons are not only used to study the acceleration processes and origins of cosmic rays, but also frequently serve as calibration tools in experiments such as OPERA [236] and ANTARES [237]. This is due to their well-known energy spectra, deposition topology, and angular distribution.

The integral flux, $\Phi_I(P_\mu^m, \psi)$, of atmospheric muons with momentum greater than a minimum value of $P_\mu^m$ and direction $\psi$ – defined as the angle between the muon's direction and the normal to the Earth's surface – is given by [238]:

$$\Phi_I(P_\mu^m, \psi) = \frac{N_\mu}{\epsilon\ S\ \Delta\Omega\ \Delta t} \qquad \left[m^{-2}\ sr^{-1}\ s^{-1}\right] \tag{4.1}$$

where:

- $\mathbf{N_\mu}$: Number of detected muons with momentum $P_\mu > P_\mu^m$ arriving at the surface area S in the solid angle $\Delta\Omega$ during the time interval $\Delta t$.
- $\boldsymbol{\epsilon}$: Intrinsic detector efficiency (typically $> 99\%$, likewise assumed for LIME from the studies carried out in Section 4.1).
- **S**: The surface area of the detector.
- $\boldsymbol{\Delta\Omega}$: Solid angle of detection, adjusted to account for the system geometry, scattering effects, and secondary particles.
- $\boldsymbol{\Delta t}$: Time interval over which the muons are detected.

The distribution of muons at sea level is largely uniform with respect to the azimuthal angle $\phi$, as the effects of geomagnetic fields and solar modulation are minimal and have a negligible influence. However, the muon flux exhibits a significant dependence on the zenith angle $\psi$, due to the increasing atmospheric thickness with $\psi$. To describe the angular distribution of atmospheric muons at (near) sea level, the general consensus is that, for not too large zenith angles, the flux at a given angle, $\Phi_I(\psi)$, can be expressed as [239]:

$$\Phi_I(\psi) = \Phi_I(0^\circ) \cdot \cos^n(\psi) \tag{4.2}$$

where $\Phi_I(0^\circ)$ is the vertical flux of muons at the Earth's surface (with which the detector is typically aligned), and $n = 2$ is the commonly used value for quick estimates of the zenith dependence of the muon flux [240].

In this study, the parameters $\Phi_I(0^\circ)$ and $n$ were measured from LIME overground PMT data and compared with existing literature. The energy measurements and corresponding spectra of the muons, which give information regarding their momentum, were not studied due to high image occupancy in LIME overground, which didn't allow a direct link between PMT waveforms and CMOS signals, from which the energy is typically measured. Therefore, the analysis is focused on the dependence of $\Phi_I$ on $\psi$, integrating over all muon energies reaching LIME, and thus corresponding to a muon momentum cutoff of 1 $GeV$ in Equation 4.1 [240].

Since the goal of this study was not to perform a precise measurement of the cosmic muon flux, but rather to validate the full chain of PMT data acquisition and analysis – and given the time constraints due to the scheduled underground installation – some simplifications in the analysis were made, such as: the detection efficiency was only briefly tested, resulting in an estimated efficiency of approximately 99%; the time-of-flight of muons, which could be used to measure their velocity, was not directly measured; and the muon energy losses in the building and the LIME system were considered negligible, especially since LIME operates at a gas density similar to that of air. Nonetheless, this study resulted in a positive outcome, as discussed later in Section 4.2.4.

### 4.2.2 Experimental setup

The experimental setup for measuring the cosmic muon flux involved placing two plastic scintillators[1], each connected to a PMT, above and below LIME at varying distances between them. The two PMTs were connected to the LIME DAQ, and the system trigger was defined as a coincident signal (within 50 $ns$) in both scintillator bars. This configuration allowed the acquisition of PMT signals (those of the optical readout of LIME) for tracks with predefined and known inclination with respect to the GEM amplification plane, which could then be used to produce a measurement of the muon flux as a function of the zenith angle.

The placement of the scintillating bars and the distances between them were constrained by the mechanical limitations of the setup. While the active volume of LIME is $z \times l \times h = 50 \times 33 \times 33\ cm^3$, this volume is enclosed within a larger plexiglass box (see Section 2.2.2.1), which is itself surrounded by an even larger metal box that functions as a Faraday cage. This setup limits the proximity of the scintillators to the sensitive volume of LIME. A schematic of the assembled setup is shown in Figure 4.6. The scintillator bars measured $z \times l \times h = 6 \times 60 \times 2\ cm^3$, with a measurement accuracy of 3 $mm$, and were each connected to a THORN EMI 9814B series PMT. From the length of the bars (60 cm), it is evident that they extended beyond the width of LIME (33 cm), which – as discussed later in Section 4.2.3.1 – can lead to appearance clipped muons in the data, i.e., muons that do not traverse the full height of LIME (33 cm), thus leading to an incorrect determination of the muon's incidence angle. The positions of the top and bottom scintillators were measured from the edge of the Faraday cage enclosing LIME. The scintillators were oriented with their longer sides crossing the width of the detector. For the first two measurements, the top scintillator was fixed at $107 \pm 1$ cm from the Faraday cage, while the bottom scintillator was moved along the drift direction of LIME to the positions of $54 \pm 1$ cm and $31 \pm 1$ cm, effectively creating two different possible muon incidence angles, $\psi$. In this work, a simpler coordinate system is employed compared to the one used in Chapter 5. In particular, a two-dimensional reference frame is used, with the only relevant variable being the angle $\psi$, defined as the angle between the muon's direction and the normal to LIME's bottom plane (and the Earth's surface).

[1]Plastic scintillator bars are made from a plastic material doped with chemicals that scintillate when struck by ionizing radiation, making them a typical instrument for detecting the passage of charged particles.

**Table 4.1:** Summary of the three setups used to measure the atmospheric muon flux, highlighting the positions of each scintillator bar with respect to the faraday cage.

| Setup angle [#] | Top scintillator distance to Faraday cage [cm] | Bottom scintillator distance to Faraday cage [cm] |
|---|---|---|
| 1 | $107 \pm 1$ | $54 \pm 1$ |
| 2 | $107 \pm 1$ | $31 \pm 1$ |
| 3 | $80 \pm 1$ | $80 \pm 1$ |

For the third configuration, the scintillators were placed parallel to each other near the center of LIME's sensitive volume, at a distance of $80 \pm 1$ cm from the Faraday cage. In all configurations, the vertical distance between the two bars was $h_{scint} = 130 \pm 1$ cm. An uncertainty of 1 cm was assigned to all these measurements due to the difficulty of precisely aligning the scintillators in the experimental setup. The three geometrical configurations are summarized in Table 4.1. The combination of available space, bar dimensions, and geometrical constraints resulted in a very low event rate of about $\mathcal{O}(10^{-2})$ Hz, requiring approximately one week of data taking per angle to accumulate sufficient statistics.

### 4.2.3 Data analysis

The data analysis for this study involved evaluating the variables required to measure the integral flux above 1 GeV (i.e., Equation 4.1) as a function of the track zenith angle, from either data or MC simulation. Specifically, the zenith angles were measured using the ToT variable (Section 4.2.3.1) and later validated with Monte Carlo simulations (Section 4.2.5). The event rate was determined by constructing the distribution of events divided by the time length of each run (Section 4.2.3.2), and the solid angle for each configuration was evaluated by means of a GEANT4 MC simulation (Section 4.2.3.3).

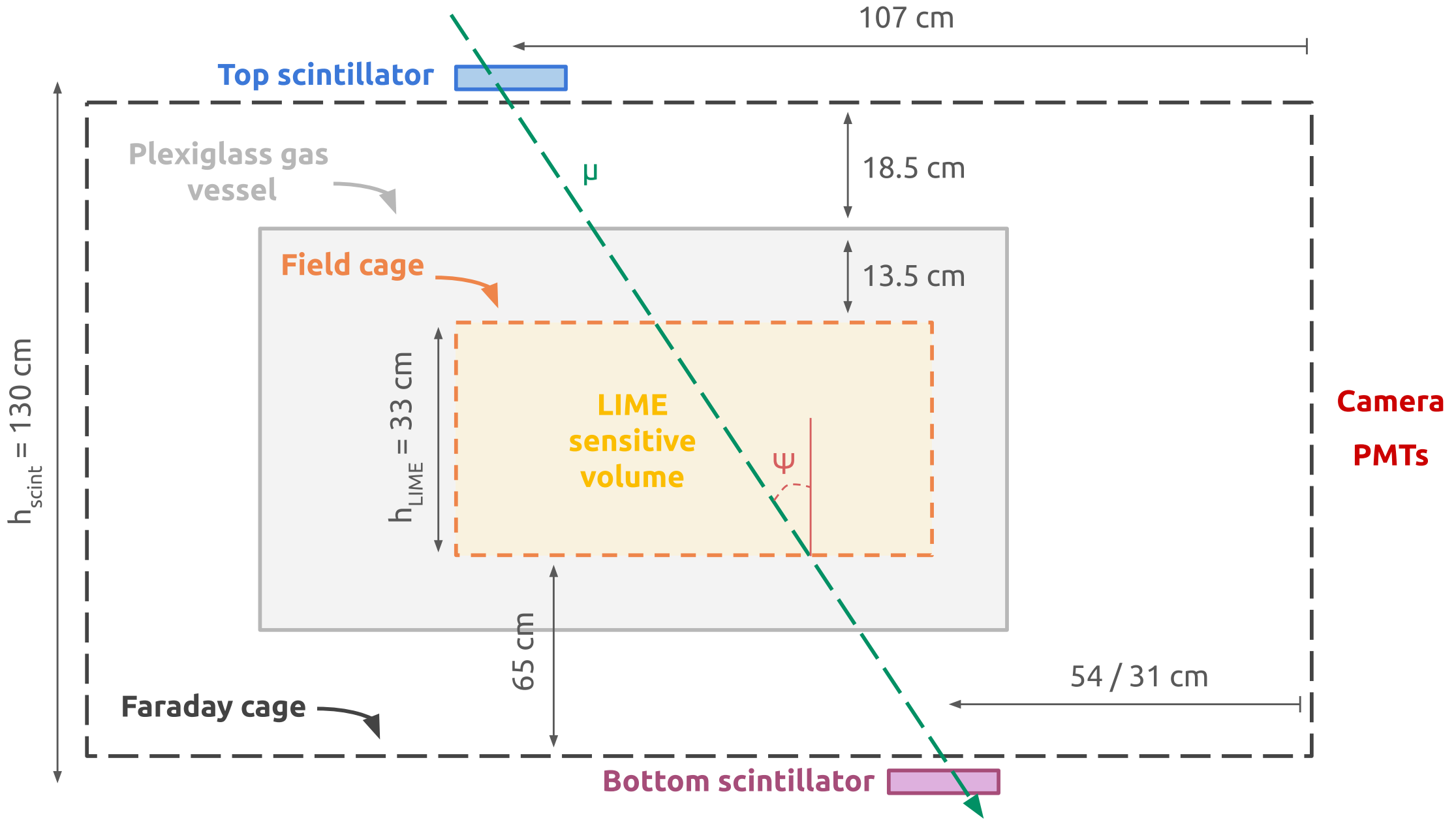

**Figure 4.6:** Schematic (not to scale) of the setup used to measure the atmospheric muon flux, as seen from the side. The various components are: the *LIME sensitive volume*, where particles interact; the *plexiglass gas vessel*, which encloses the gas; the *Faraday cage*, which shields the system from electromagnetic noise and visible light; and the *top/bottom scintillators*, which trigger the LIME DAQ system and define the crossing angle $\psi$ of the muons ($\mu$).

#### 4.2.3.1 Measurement of the zenith angle from PMT signals

The muon incidence zenith angles ($\psi$) in Equation 4.1 were evaluated for each configuration setup of Table 4.1 using the time-over-threshold variable of the weighted average waveform, introduced in Section 3.2.3.1. This variable, in the case of high-energy muons – where the tracks are strictly straight – allows a direct calculation of the traveled $\Delta Z$ of the muon, and consequently, the determination of its incidence angle. Following the schematic shown in Figure 4.6, the zenith ($\psi$) angle can be calculated using the following geometric argument:

$$\psi \;=\; \tan^{-1}\left(\frac{\Delta Z}{h_{LIME}}\right) \tag{4.3}$$

In this equation, $h_{LIME}$ corresponds to the height of LIME, specifically the height of the sensitive region of the GEMs (33 $cm$). This consideration is important because light is only produced in the sensitive area, which constrains the size of the muon ionization track seen by the PMTs. The remaining term, $\Delta Z$, can be calculated using Equation 3.4 with the sampling frequency of the slow digitizer (250 $MHz$). The slow digitizer was exceptionally used in this case due to the large spatial extension of the muon signals, thus ensuring they would be fully contained within the waveforms. The $ToT$ variable in Equation 3.4 is calculated over the weighted average waveform since, as discussed in Section 3.2.3.1, this better represents the overall time extension of signals that cross large portions of the detector, such as muons. An example of the $ToT$ variable, evaluated on the average waveforms obtained in this study, is shown in Figure 4.7, where the blue line represents the $ToT$, and the red lines and yellow crosses correspond to the automatic results from the peak-finder subprocess (Section 3.2.1.3) of the PMT-reco.

For the three datasets corresponding to the three geometrical angles used in this measurement, the zenith angles were calculated using the methodology described. The distributions are shown in Figure 4.8. These distributions were fitted with a Gaussian curve constrained to the region corresponding to the main visible peaks, leaving all the fitting parameters free. The long tails observed in the distributions are attributed to false-positive triggers, which populate the entire angle spectrum in a complex manner that is difficult to model, thus justifying the constraining of the fit to the central peak. False-positive triggers can occur due to ”clipped muons”, i.e., events in which muons cross the top scintillator and then enter (or exit) LIME from the side face, while still crossing the bottom scintillator. This is possible because the scintillators are longer than the width of LIME (see Section 4.2.2). In such cases, the measured muon's $\Delta Z$ is smaller than expected, resulting in a random reconstructed angle and contributing to the flat background and tails of the distributions. Additionally, fortuitous coincidences arising from natural radioactivity or secondary particles are also possible. These events could

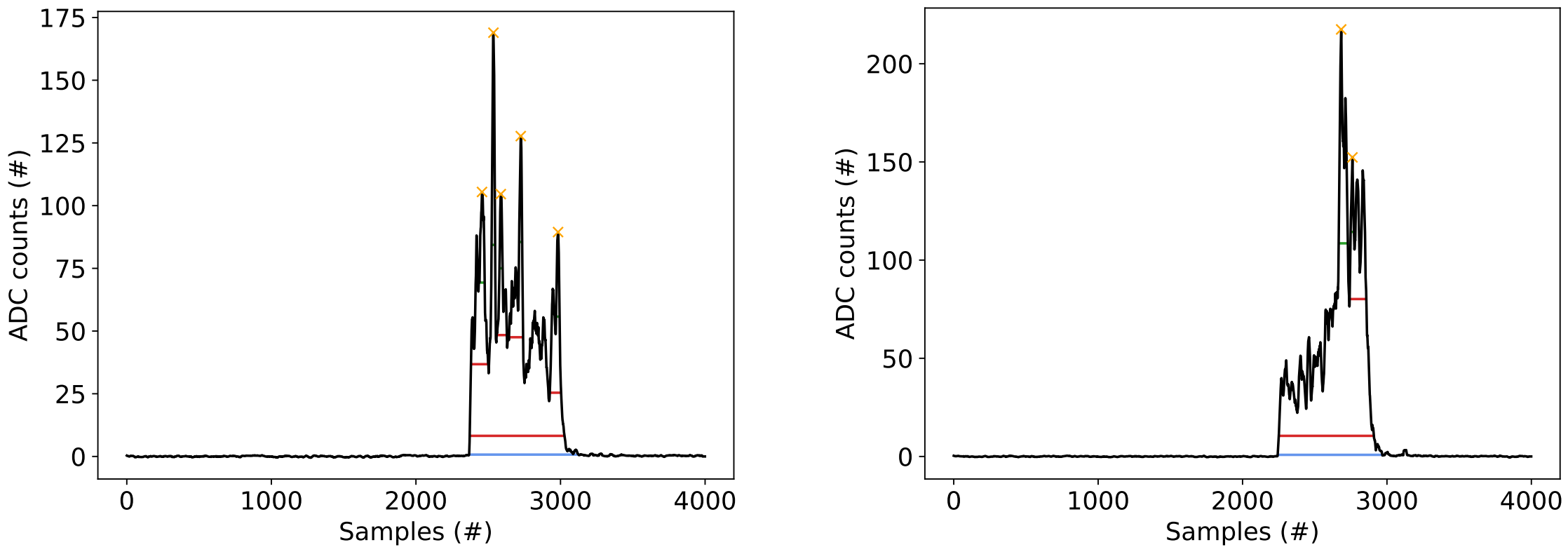

**(a)** **(b)**

**Figure 4.7:** Two weighted averaged waveforms obtained in this study with several identified variables discussed in Section 3.2.1, namely the ToT, as a blue line, and other peak-related variables, as red lines and yellow peaks.

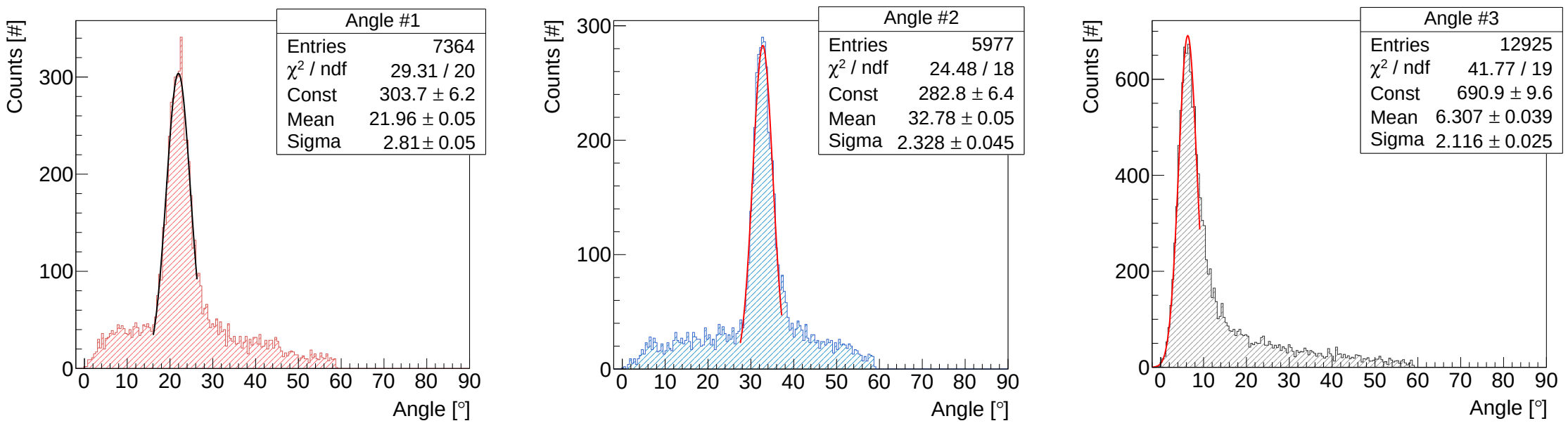

**(a)** Angle #1 **(b)** Angle #2 **(c)** Angle #3

**Figure 4.8:** Distributions of zenith angles obtained from the ToT variables, using Equations 4.3 and 3.4, for the three setups shown in Table 4.1. The distributions were fitted with a Gaussian curve, and the fitted results are overlaid on the data.

not be removed from the dataset, as no particle identification (PID) was applied. This is due to the difficulty of performing PID using *only* the PMT signals, and because the focus of this work was primarily on validating the PMT-DAQ interplay and basic waveform reconstruction techniques.

The Gaussian means obtained from the distributions of zenith angles for each setup of Table 4.1 were: $21.96 \pm 0.05°$, $32.78 \pm 0.05°$, and $6.31 \pm 0.04°$ for angles #1, #2, and #3, respectively. These values were later cross-checked and validated with a comparison with the geometrically allowed range of angles and the ones returned by the GEANT4 simulation developed, in Section 4.2.5.

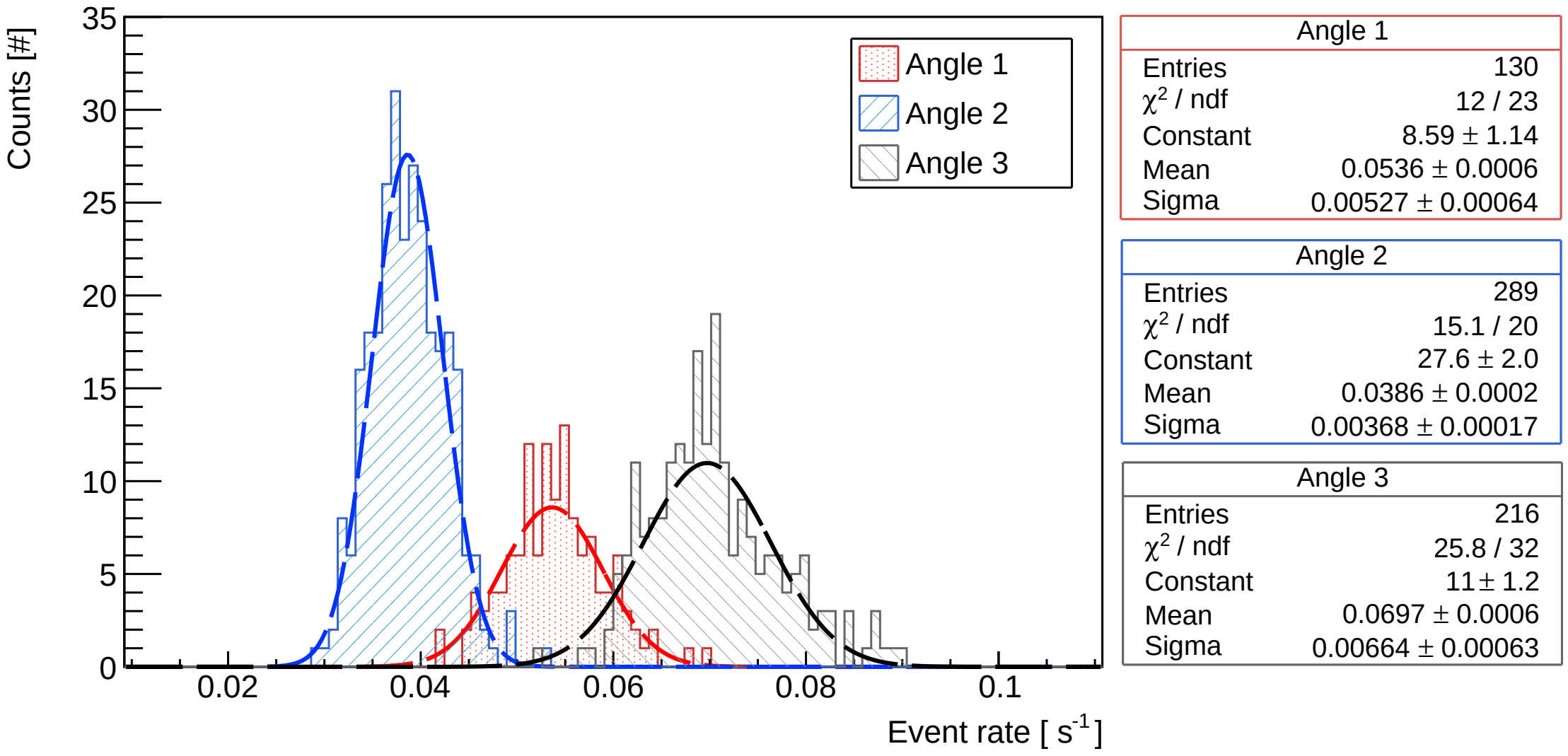

**Figure 4.9:** Distributions of event rates for the three angle setups used in this measurement: angle #1 (red), angle #2 (blue), and angle #3 (grey). Each distribution was fitted with a Gaussian curve, and the fitting results are displayed on the side of the figure.

#### 4.2.3.2 Calculation of the event rate

The event rate for each dataset is calculated by dividing the number of events in each run – predefined to contain 100 events – by its duration. The duration of a run is retrieved from the DAQ system, which records the timestamps of the beginning and end of the run, defined respectively by the recording of the 1st and 100th picture. The event rate distribution for each dataset is shown in Figure 4.9. Since these distributions represent event counts (normalized to the run length), a binomial distribution with a spread of $\sqrt{N}$ is expected. With runs fixed to 100 events, the expected spread is 10%, which is consistent with the observed spread ($\sigma$) from the Gaussian fits used to describe the event rate for each angle setup.

The difference in the mean of each event rate distribution highlights the variations in flux and geometrical acceptance between the datasets. Given the very low event rate observed (1 event in every tens of seconds), it can also be concluded that the loss in detection efficiency due to deadtime or signal overlap is negligible. This is because the time required by the DAQ to process events – of the order of tens to hundreds of milliseconds – is much shorter than the time between muon events. If the event frequency were on the order of $\mathcal{O}(10^0)$ Hz, similar to the timings associated with the DAQ deadtime and the opening and closing of the CMOS sensor, then the detection efficiency loss would need to be considered for such a measurement.

#### 4.2.3.3 Simulation of geometrical acceptance

Another important parameter needed to evaluate the atmospheric muon flux is the geometrical acceptance of the detector. Geometrical acceptance refers to the fraction of particles that the detector (or setup) can detect based on its geometry, position, and orientation relative to the incoming particle direction. In this case, the setup includes the combined condition of a coincident event in the two scintillator bars and LIME detector. Essentially, it describes the portion of the solid angle within which the detector can register particles. Mathematically, the geometrical acceptance, $GA$, can be expressed as:

$$GA = \Delta\Omega \cdot S \tag{4.4}$$

where $S$ is the area of the detector and $\Delta\Omega$ is the solid angle subtended by the detector, as also described in Equation 4.1.

Geometrical acceptances are commonly simulated using Monte Carlo softwares such as GEANT4 [241]. The advantage this software in particular, compared to analytical calculations, is that it not only models the acceptance derived from the primary particles (muons in this case) but also includes additional effects such as particle scattering and the emission of secondary particles from interactions of the primaries with the detector. Analytical calculations for two-detector systems commonly neglect factors like detector thickness, region-dependent detection efficiencies, or edge effects, all of which can influence the overall acceptance of the detector [242]. For a more detailed discussion on the analytical calculation of geometrical acceptance in detectors, see references [242–245]. Considering these complexities, a simplified model of LIME was created using GEANT4 to measure its geometrical acceptance in each setup.

The GEANT4 simulation was designed with a simple 3-body schematic representing the two scintillator bars and the LIME sensitive gas volume, following Figure 4.6. For simplicity, other detector components, such as the field rings, plexiglass vessel, and metal enclosure, were excluded from the simulation. This simplification is expected to have a negligible impact on the detector efficiency, given the low material thicknesses and the high penetration / low interaction rate of atmospheric muons in materials.

For each geometry, muons with an energy of 4 $GeV$ were randomly emitted from the upper scintillator in a downward direction, creating a hemisphere of emission angles. The events were generated to produce an isotropic flux of muons in 3D within the solid angle as, for this calculation, the only interest is the geometrical acceptance, since the shape of the muon flux is the variable to be measured [246]. It should be noted that if the goal were to compare the measured angular response of the detector to muons with simulation, or to estimate the expected number of *counts* of muons, the detector's angular acceptance ($GA(\psi)$) would have to be simulated using the actual flux shape, where $\psi$ follows a $\cos^2(\psi)$ distribution [246,247].

For each angle, a total of $N_{total} = 4 \cdot 10^7$ muons were simulated as being emitted from the upper scintillator. The geometrical acceptance at each angle was then calculated using the formula from [242, 246]:

$$GA = \left(\frac{N_{hit}}{N_{total}}\right) * \pi * S \tag{4.5}$$

where $N_{hit}$ is the number of muons that triggered the system, $S$ is the scintillator area ($(6 \times 60) \cdot 10^{-4}$ m$^2$), and $\pi$ represents the solid angle of the "source" of muon emission, i.e., the top scintillator. For each configuration (angles #1, #2, and #3), the following values were found: $N_{hit}/N_{total} = (3.295 \pm 0.009) \times 10^{-3}$; $(2.785 \pm 0.008) \times 10^{-3}$; $(3.780 \pm 0.010) \times 10^{-3}$. These values will serve as a normalization factor to the rates of events calculated in order to obtain the real flux of muons.

### 4.2.4 Results & Discussion

Using the collected data, the flux of atmospheric muons was calculated for the three tested angles using Equation 4.1. Figure 4.10 displays the calculated flux for each angle, determined using the ToT variable. The data was fitted to Equation 4.2 using the MINUIT package [248] in the ROOT framework [217], with $\Phi_I(0°)$ and $n$ as free parameters. This approach follows similar methods used in previous studies [238, 249]. The final values obtained were:

$$\Phi_I(0°) = 83.01 \pm 3.99 \text{ m}^{-2}\text{sr}^{-1}\text{s}^{-1} \qquad n = 1.698 \pm 0.438$$

with the errors accounting for both statistical fluctuations in the muon rate measurement and systematic uncertainties related to the geometry of the setup.

The results obtained for $\Phi_I(0°)$ and $n$ are globally in good agreement with values reported in the literature. For example, Table 3 in [238] summarizes several works on this topic, including those by [246, 247, 251–253]. These studies report vertical muon flux values ($\Phi_I(0°)$) ranging from 56.66 to 83.00 m$^{-2}$sr$^{-1}$s$^{-1}$, with some additional ones, like [254], showing values as high as 91.3 m$^{-2}$sr$^{-1}$s$^{-1}$, putting the result from this study well within this range. Similarly, the parameter $n$, which describes the $\psi$ dependence of the atmospheric muon flux, typically ranges from $n = 1.72$ to $n = 2.15$, depending on factors such as altitude and muon momentum cutoff. The $n$ value obtained in this study aligns with these values.

The large uncertainty in $n$ is primarily due to systematic errors, including imprecision in measuring the dimensions, distances, and potential misalignments of the components in the setup. Additionally, the limited number of angular points measured with this setup and uncertainties in the actual geometrical acceptance GA contribute to the error.

Overall, given the available data and the precision of the apparatus, the results obtained for the integral flux of atmospheric muons near sea level resulted satisfactory to validate the full PMT DAQ and reconstruction chain. The time-over-threshold variable, introduced with the PMT-reco, has proven to be a reliable tool for estimating the longitudinal distance traveled by particles ($\Delta Z$) and, consequently, the incident angle of the muons. This is especially true when using the weighted average waveform, which effectively reduces the contribution of noisy signals and improves the ToT calculation.

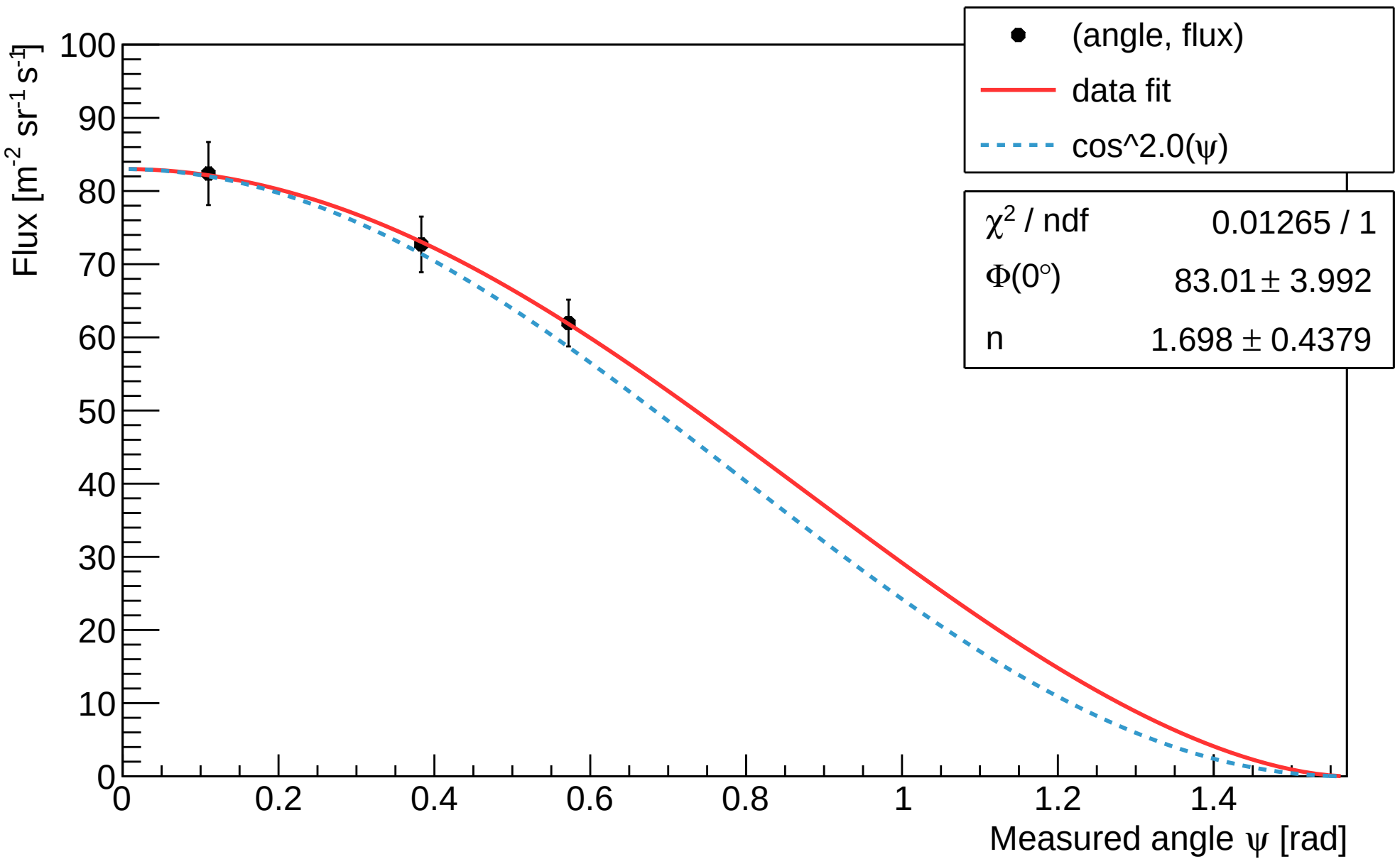

**Figure 4.10:** Zenith angle distribution of the atmospheric muon flux for the three scintillator configurations. The solid red line shows the data fitted using Equation 4.2. The dashed blue line represents the same curve with $n = 2.0$ fixed, illustrating the commonly assumed value for $n$ and emphasizing the model's sensitivity to this parameter. This approach was based on [250].

This study also serves as a reference point, being the first complex analysis to incorporate the recent updates to the CYGNO-reconstruction algorithm, specifically the integration of PMT analysis. For future work, since muons behave as minimum ionizing particles in CYGNO detectors, they could be used to investigate the transverse and longitudinal diffusion characteristics of the gas and amplification system for low-ionizing events at various drift distances. To this end, it is suggested to repeat these measurements by placing the scintillator bars perfectly parallel to each other, with the detector positioned in between, and at varying distances from the GEM plane. This setup would produce muon tracks with zenith inclinations close to 0° (i.e., $\Delta Z \sim 0$), generating PMT waveforms that reflect only the intrinsic spread of the ionization track (negligible for MIPs) and the diffusion experienced during drift. Additionally, reducing the size of the scintillator bars is worth considering, as it would narrow the geometrically allowed angular spread and thus reduce uncertainties related to $\Delta Z$. Such measurements would complement the ongoing parallel research in this area being carried out with low-energy ERs and alpha particles.

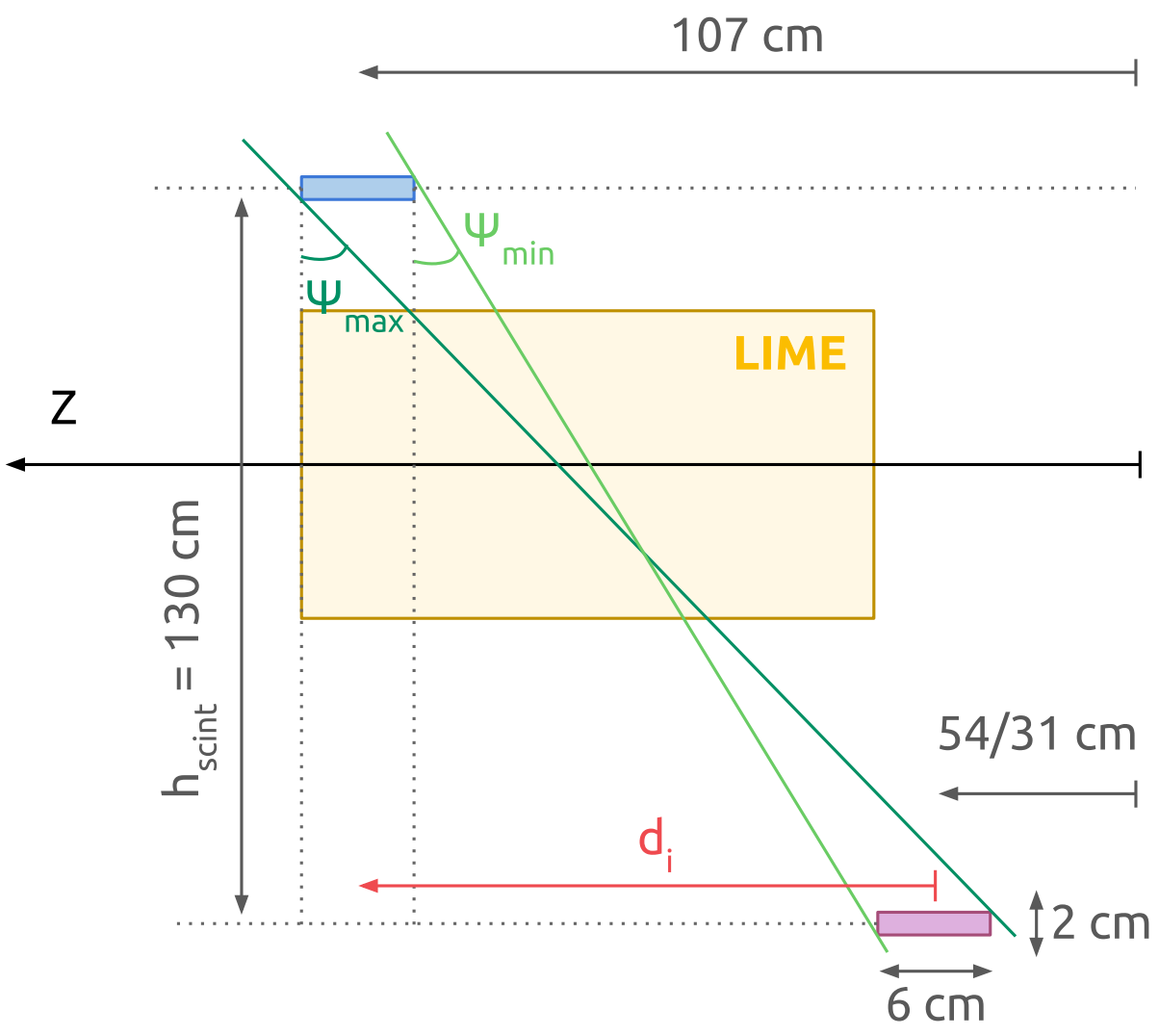

**Figure 4.11:** Schematic (not to scale) of the setup used to measure the atmospheric muon flux with LIME, highlighting the maximum ($\psi_{max}$) and minimum ($\psi_{min}$) allowed angles for muons to cross and trigger the detector, based on the variable distance between scintillators, denoted as $d_i$.

### 4.2.5 Analysis cross-check

As a cross-check of the analysis carried out in this chapter, the ranges of angles geometrically allowed by the experimental setup were determined. Given the configuration of the system, three geometrically distinct angles are possible for the crossing of muons. Figure 4.11 shows a simplified schematic of the measurement geometry. Two angles, $\psi_{max}$ and $\psi_{min}$, are depicted, corresponding to the maximum and minimum zenith angles that muons can have to cross LIME and trigger the detector's DAQ for each configuration in Table 4.1.

Following the geometry of the system and dimensions of the bars ($z \times l \times h = 6 \times 60 \times 2$ cm$^3$), Equation 4.6 show how $\psi_{max}$ and $\psi_{min}$ can be calculated for each case:

$$\psi_{min} = \tan^{-1}\left(\frac{d_i - z/2 * 2}{h_{scint}}\right) \qquad \psi_{max} = \tan^{-1}\left(\frac{d_i + z/2 * 2}{h_{scint}}\right) \tag{4.6}$$

where $d_i$ is the distance between the two scintillator bars. The results of the geometrically accepted range of angles for each configuration are summarized in Table 4.2. For the minimum and maximum angles within each range, the uncertainty is calculated through error propagation. Angle #3 is limited to positive values since $\Delta Z$ is always positive and no assumption is made regarding the track direction (head-tail). Therefore, Equation 4.3 returns only the absolute value of $\psi$. For this reason, the error on the lower bound of the range is also considered zero.

**Table 4.2:** Summary of the geometrically allowed range of angles for the three setups used in this measurement (Table 4.1). The absolute minimum and maximum incidence angles were calculated using the expressions in Equation 4.6.

| Setup angle [#] | $d_i$ [cm] | $[\psi_{min}, \psi_{max}]$ [°] |
|---|---|---|
| 1 | $53 \pm \sqrt{2}$ | $[19.88, 24.41]^{+0.55}_{-0.58}$ |
| 2 | $76 \pm \sqrt{2}$ | $[28.30, 32.24]^{+0.50}_{-0.53}$ |
| 3 | $0 \pm \sqrt{2}$ | $[0.00, 2.64]^{+0.64}_{-0.00}$ |

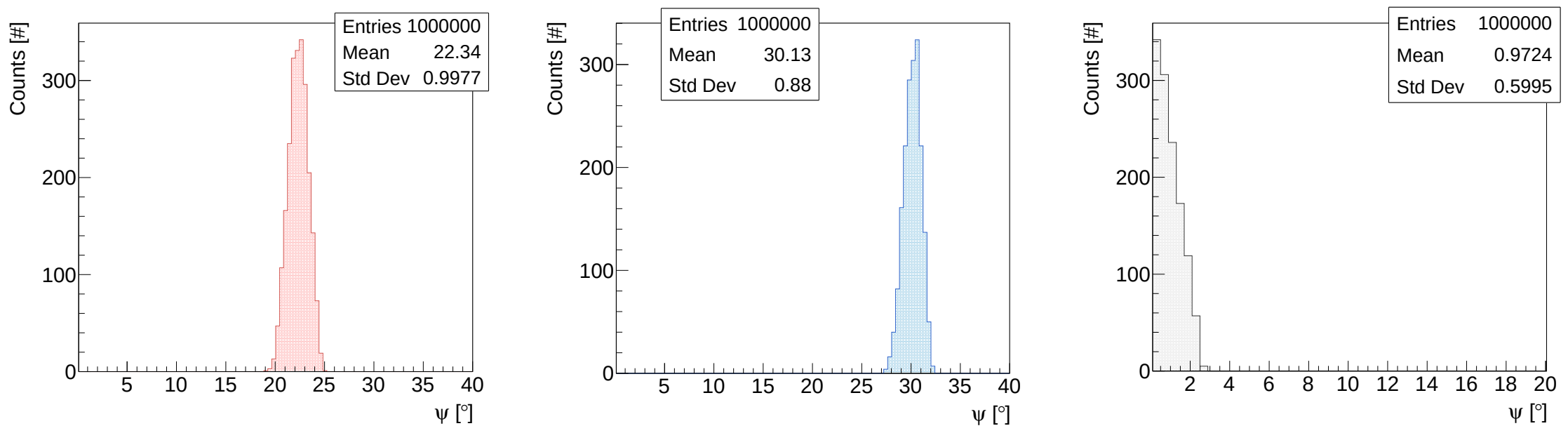

**(a)** Angle #1 **(b)** Angle #2 **(c)** Angle #3

**Figure 4.12:** Distributions of angles for the simulated muons obtained through the GEANT4 simulation for the calculation of the geometrical acceptance of each detector setup.

For angles #1 and #2, the geometry-defined angles are consistent with the measured values in Section 4.2.3.1 – 21.96° and 32.78°, respectively. However, for angle #3, the measured value of 6.31° deviates from the expected range. This is likely due to the intrinsic limitation encountered when measuring the length of a PMT signal that is expected to resemble a delta function in the PMT response (for a perfectly vertical track), but appears broadened due to electron diffusion during drift and amplification, as discussed in more detail in Section 6.3.1.2, resulting in a small angular offset. Concerning the width of the angle peaks measured, in all cases, the angles subtended by two times the sigma of the fits to the distributions in Figure 4.8 are generally consistent with the geometrically allowed range of angles. This cross-check demonstrates that the ToT variable can indeed be used to accurately assess the tilt angle of incoming muons in LIME.

Additionally, the correct implementation of the GEANT4 simulation used to determine the geometrical acceptance of each angle configuration was also verified. In Figure 4.12, the angles at which the simulated muons crossed the 3-body setup visible in Figure 4.11 are shown. These distributions are in good agreement with the analytically predicted range of values summarized in Table 4.2, demonstrating that the simulation was correctly implemented and, therefore, also the geometrical acceptances for each setup configuration.

Chapter 5

# Alpha particles 3D tracking

As illustrated in Section 1.4, the expected anisotropy in the Dark Matter wind resulting from the Earth's motion through the galactic halo provides a powerful signature to distinguish DM from background events. While inherently hard to achieve experimentally given the very low energies at play, the capability to measure the recoil direction can hence effectively provide a unique opportunity for establishing a positive claim of observation of a DM signal. As further elaborated in Section 1.4, while a $30 \times 30\ deg^2$ angular resolution results sufficient enough to distinguish the strong anisotropy expected from such signal in Galactic coordinate from an isotropically distributed background, the capability of properly infer the sense, or *head-tail*, of the direction of the recoils is one of the feature with the largest impact in terms of ability to claim discovery with the least possible number of events.

As illustrated in Chapter 3, the CYGNO collaboration initially focused on establishing the analysis of CMOS images due to the novelty of this light sensor, which required an original and tailored data analysis. One of the focus of this thesis, as discussed in Section 3.2, has been to complement this with the PMT waveform analysis, combination of multiple PMTs signals, and PMT-to-CMOS reconstructed track matching to establish a first systematic and standardized procedure for 3D tracking within the CYGNO/INITIUM experimental approach. In Chapter 4 the newly developed PMT-only analysis, including the merging on multiple PMT signal, was tested and validated by measuring an angular distribution of cosmic rays with the LIME detector at overground Laboratori Nazionali di Frascati consistent with expectations.

In this Chapter, all the ingredients discussed so far are effectively finally combined to obtain the first ever CYGNO/INITIUM full 3D reconstructed tracks sample, including sense (i.e. head-tail) and XYZ absolute position. This was developed on underground LIME data in its full shielding configuration (i.e., 10 $cm$ of copper + 40 $cm$ of water) in order to minimize the occupancy of the images to allow a full 3D event-by-event reconstruction. Alpha particles track candidates were chosen to establish the first version of the full 3D tracking procedure, due to the characteristics of their energy deposit (see Section 5.1), which guarantees a mostly monotonic track development along the drift direction (hence less complex $\Delta Z$ extension measurement from PMT TOT signal) and a clear signature of the sense of the direction through the Bragg peak. Additionally, the high alphas stopping power provides easy means of discrimination of these from other particle interactions from the track information retrieved from the CMOS images analysis. Last but not least, nuclear recoils resulting from interactions of WIMPs and neutral particles display a stopping power similar to alphas (especially the Helium ones), even though with opposite sense direction with respect to a Bragg peak-like structure (see Section 5.1), making this first development on alpha tracks a critical step propaedeutic to a fully established CYGNO/INITIUM directional Dark Matter search technique. While developed on the LIME detector, this work effectively establishes the basis towards full 3D tracking of any kind of particle interaction and energy within the CYGNO/INITIUM approach, providing a paramount contribution also to the development of CYGNO-04 demonstrator analysis (Section 2.2.3).

The choice of performing a 3D analysis of alphas in LIME was furthermore very strongly supported by the expectations that this could shed light on the unexpected background component observed in the comparison of LIME data to MC simulation (see Section 2.2.2.4), which we deemed associated to the presence of Radon and its progeny decays. This resulted much more effective than we would have ever predicted and allowed us to accurately characterize the alpha backgrounds material and isotope origin in LIME, providing compelling information for the optimization of the design and construction choices for CYGNO-04.

This Chapter starts with a brief introduction to particle interactions with matter to establish the peculiarity of the energy loss profile of alphas, nuclear recoils and electron recoils (Section 5.1). The different spatial and directional information that can be separately extracted from sCMOS images analysis (Section 5.2.1 - Section 5.2.3) and PMT signal (Section 5.2.4 - Section 5.2.5) are then discussed, before illustrating how they can be effectively matched with the BAT-fit presented in Section 3.3.1 to return the full 3D track extension and orientation with respect to the detector reference frame (Section 5.2). A final example of the integrated working principle of the alpha 3D reconstruction is presented in Section 5.4.

## 5.1 Particle interaction with matter

As already illustrated in Section 1.3.5, in the context of direct WIMP-like DM searches, despite the deployment of detectors in deep underground laboratories to shield against cosmic rays, a variety of backgrounds persist, originating from both environmental and intrinsic detector material radioactivity. Crucially, for the purpose of interpreting detector data, all particle interactions within a (gas-based) DM detector are ultimately observed as one of two categories of charged particle interactions: electron recoils (ER) or nuclear recoils (NR).

Electron recoils are induced by particles that interact primarily via electromagnetic processes. This category includes beta particles (electrons), which by definition are already electrons and hence directly result in ER when they deposit energy in the gas. Gamma rays, though electrically neutral, also contribute to ER backgrounds. Their energy is deposited in the detector through secondary interactions, predominantly the photoelectric effect and Compton scattering, both of which result in the production of electrons. Due to the relatively low energy of photons relevant in DM searches (typically below a few MeV), pair production is negligible and does not significantly contribute to ER in most detectors.

Nuclear recoils, on the other hand, arise from interactions that transfer momentum directly to atomic nuclei. This class includes neutrons and neutrinos – both neutral particles capable of inducing recoils through elastic or quasi-elastic scattering on nuclei – as well as WIMPs, which are expected to produce single-scatter nuclear recoils in target gases. Alpha particles also fall into the NR category, since they are essentially helium nuclei (i.e., doubly ionized helium ions).

Both NR and ER, when interacting with a medium, lose energy predominantly via two mechanisms:

1. **Inelastic collisions with atomic electrons**, leading to ionization and excitation of atoms in the material.

2. **Elastic scattering with atomic nuclei**, resulting in the transfer of kinetic energy to recoiling nuclei, which in turn are also ionizing.

Other energy loss mechanisms — such as Cherenkov radiation, nuclear reactions, and bremsstrahlung – become relevant only at much higher energies than those typically encountered in Dark Matter searches, and will therefore not be discussed in this context.

The well-known Bethe-Block formula [255] describe the stopping power per unit length $dE/dx$ in terms of medium characteristics (atomic mass and number, density, mean ionization potential) and particle nature (i.e. ER or NR), charge and velocity normalized to the speed of light (i.e. $\beta$ factor). Given points (1) and (2) from above, the effect, interplay, and relative importance of these processes in the overall energy loss mechanism strongly depend on the mass of the particle depositing energy. In particular, the behavior differs significantly depending on whether the particle mass is comparable (i.e. ER) or much larger than (i.e. NR) that of atomic orbital electrons.

Electrons and positrons lose a significant fraction of their energy in each individual collision. As a result, they undergo frequent and substantial angular deflections, producing tortuous, curly tracks with inhomogeneous energy loss. The majority of their energy is deposited near the end of their path, forming a characteristic end-point blob.

On the other hand, nuclei (or alphas) typically follow relatively straight tracks with minimal angular deflection, depositing energy in a dense and approximately uniform fashion. For $\beta \gtrsim 0.1$, as the particle slow down the energy loss increases rapidly according to the $1/\beta^2$ leading to a pronounced *Bragg peak* at the end of the range. The localized energy release at the Bragg peak makes heavy charged particles particularly useful in applications such as radiation therapy, where targeted energy delivery to tumors and minimal damage to the surrounding healthy tissue are required [256,257]. In the context of directionality determination, this feature can be exploited to determine the sense of the direction from the position of the Bragg peak. An example of this peculiar feature is displayed in Figure 5.1 for a 5.9 MeV alpha particle.

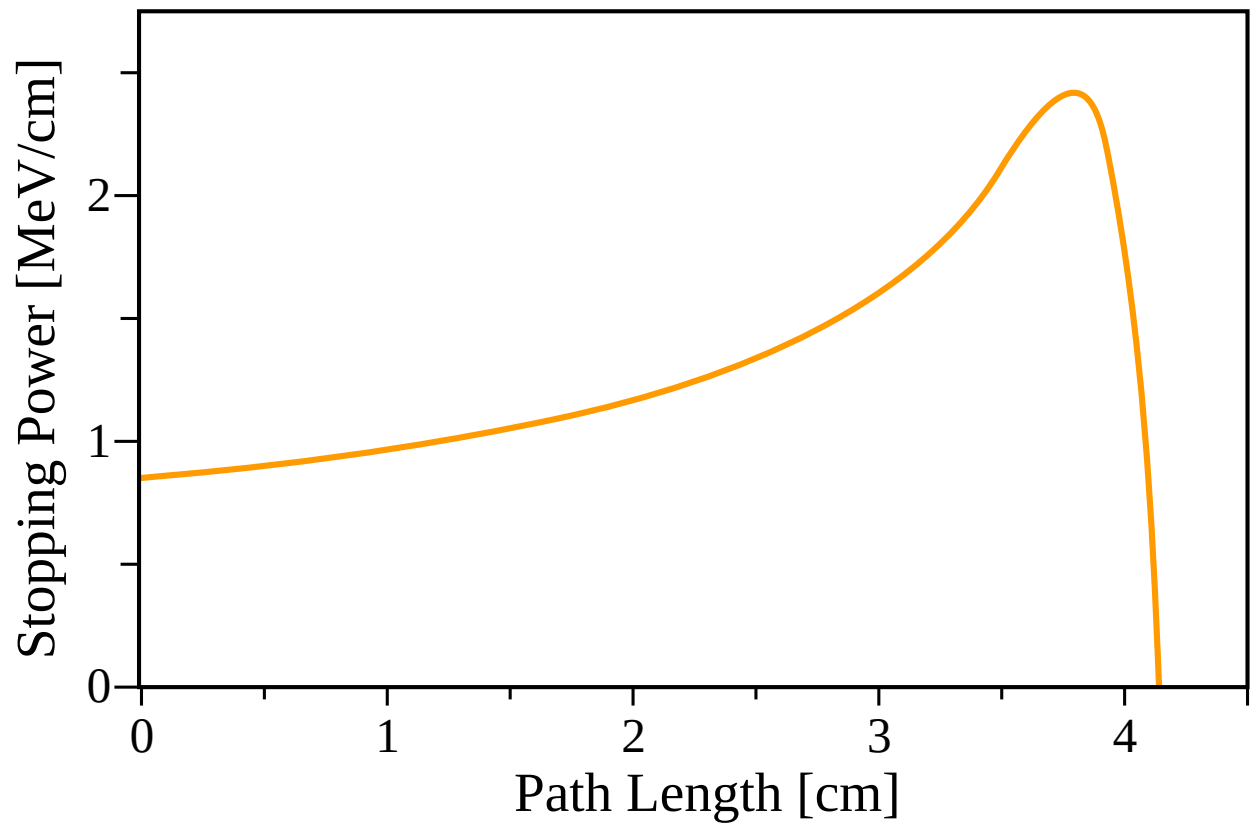

**Figure 5.1:** Stopping power of 5.9 MeV alpha particles in air as a function of the path length traveled. As it travels, the alpha increases its energy deposit until reaching its maximum – the *Bragg peak* – where it loses all the remaining energy.

A critical distinction has to be made between alpha particles and low energy NR from WIMPs, neutrons or neutrinos in the DM search context, since the latter have low kinetic energies (from a few to tens of keV), corresponding to $\beta \lesssim 0.01$. In this regime, the Bethe-Bloch formalism can no longer be considered valid and the energy loss is typically described by Lindhard theory [258], which models the partition of recoil energy between ionization and nuclear stopping. This theory, validated through measurements and simulations (e.g., SRIM software), predicts that NRs deposit more energy at the beginning of the track, and less as they slow down. This effectively produces an inverted Bragg peak compared to that of alpha particles. This inversion is of fundamental importance: it implies that, although NRs and alphas differ in absolute energy and particle type, they share a common directional feature – a varying $dE/dx$ along the track that can, in principle, be used for head-tail determination. The methods developed in this work for analyzing alpha tracks thus constitute a solid basis for extending head-tail reconstruction to signal-like nuclear recoils – once the approach is properly optimized and validated for the relevant energy regime.

In LIME, whenever a particle interaction occurs that results in gas ionization, the light produced at the amplification stage is recorded by a CMOS and PMT sensors. Figure 5.2 shows example of the three kind of tracks observed with the LIME detector for which we can reasonably infer the nature of the particle that induced them: from left to right these are a 5.9 keV ER produced in the absorption of $^{55}$Fe X-rays observed in calibration runs, a low energy ER and an high energy alpha. Using this figure as a reference, these interactions can be described and categorized as follows:

- **Spot-like ER:** In Figure 5.2, the left panels illustrate a spot-like event generated

by a 5.9 keV photon emitted by an $^{55}Fe$ source. The photon has been absorbed in the gas through the photoelectric effect, resulting in the emission/recoil of an electron. Interactions of this kind at this energy result in the so-called "spot-like" events, due to their characteristic rounded shape in the CMOS images. This shape occurs because the diffusion of the ionized electrons is comparable to or larger than the intrinsic range (size) of the ionization track. In LIME, electron recoils maintain this spot-like appearance up to approximately 8 keV, after which a slight elongation becomes noticeable to the naked eye. These $^{55}$Fe-induced interactions are typically used as low-energy event benchmarks in the CYGNO experiment, for studies related to the detector's energy thresholds, directionality at lower energies, and overall detector performance. The waveforms associated with these events typically feature a single peak with no additional notable characteristics, as visible in Figure 5.2d.

- **ERs:** The middle panels in Figure 5.2 display an ER event with energy around 50 keV. In LIME, ER events are generally associated with electron recoils with energies above 8 keV, which already exhibit a non-rounded shape. ERs form a significant part of LIME's background, and a considerable effort is dedicated to distinguishing them from NRs, the expected signal from DM scattering. At higher energies, ERs may also result from particles that undergo Compton scattering, receiving only a fraction of the initial particle's energy. The waveforms for ERs show multiple peaks, each corresponding to individual ionization clusters along the particle's path. The amplitudes of these peaks are similar to those of spot-like events. These features are visible in Figure 5.2e.

- **Alphas particles:** The right panels of Figure 5.2 show an alpha particle with an energy of approximately 5 MeV. Alpha particles exhibit very high ionization density clouds, releasing all their energy over a short distance. In LIME, alphas travel in nearly straight paths for several centimeters, and present energy ranges of the order $\mathcal{O}(1-10)$ MeV. Due to their short ranges, alphas are more likely to be found near the detector materials (cathode, field cage, etc.), resulting from the decay of primordial radioisotopes present in these materials, although the presence of Rn in the detector could lead to the presence of alphas also in the bulk of the sensitive volume, as discussed in Section 1.3.5.

  The waveforms of alpha events are characterized by a high-intensity signal, which reflects the large number of electrons arriving closely together at the amplification plane, and the consequent emission of photons. In the time domain, this results in a high-amplitude signal that persists for an extended period, somewhat similar to a step-function, with a significant amount of integrated charge in each PMT waveform, compared to ERs. Given the high-density deposition of energy, these waveforms don't show prominent peaks. These features are visible in Figure 5.2f.

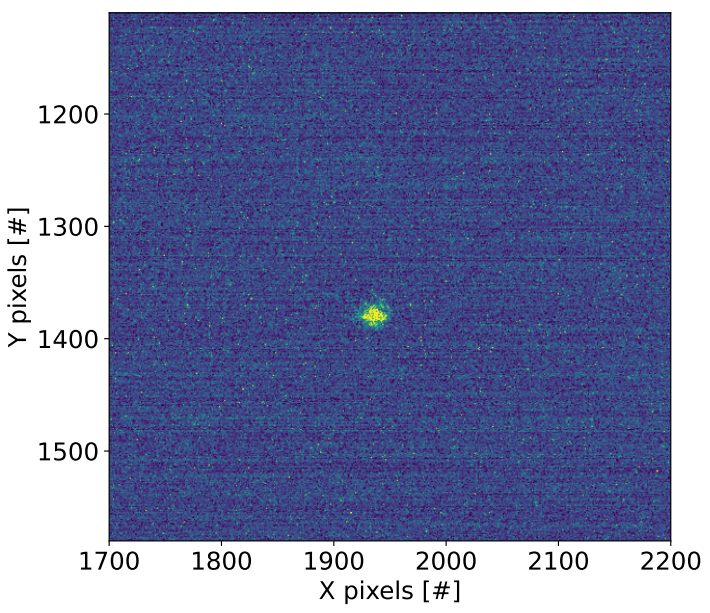

**(a)** Spot-like event – CMOS

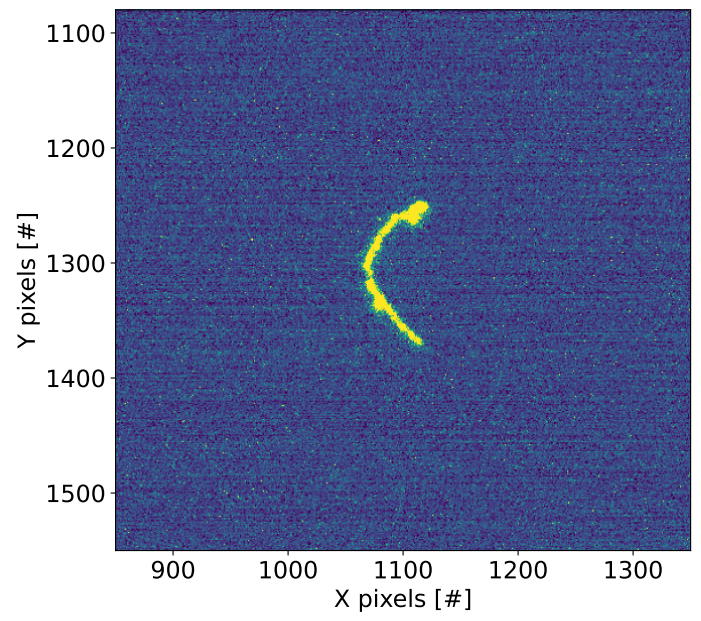

**(b)** ER event – CMOS

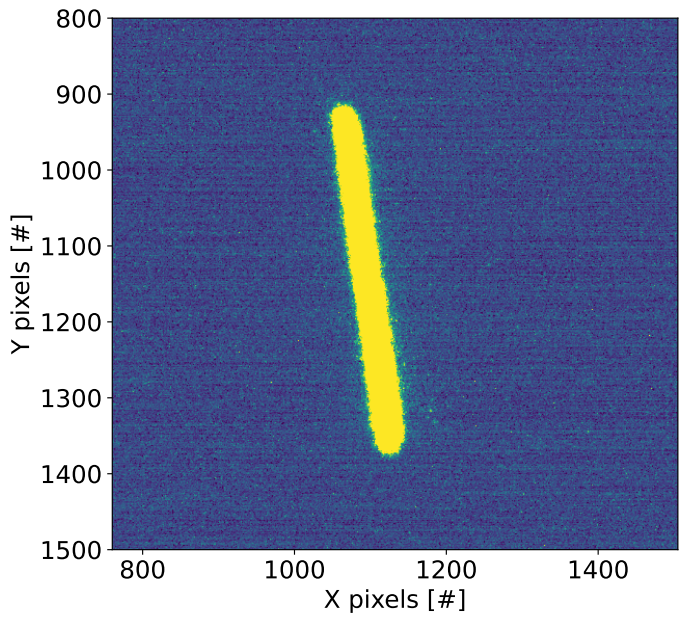

**(c)** Alpha event – CMOS

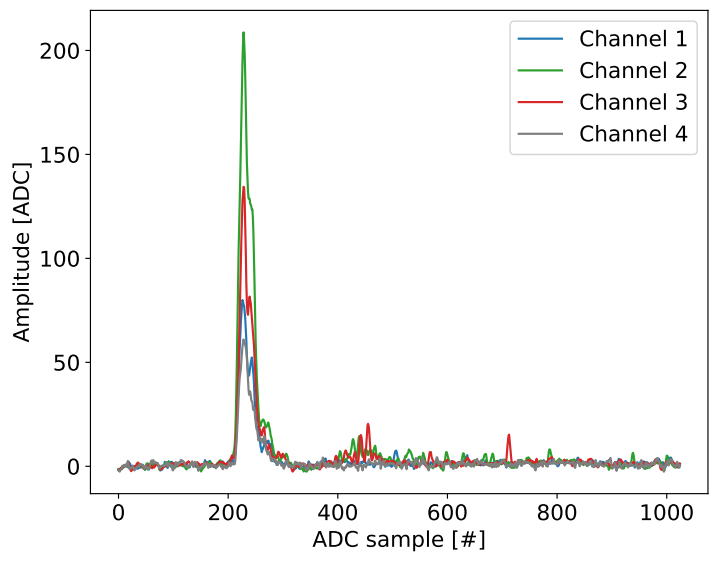

**(d)** Spot-like event – PMTs

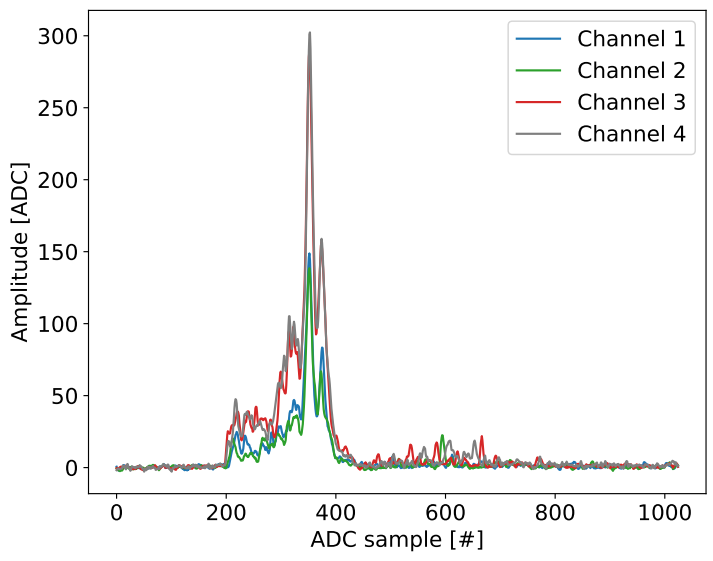

**(e)** ER event – PMTs

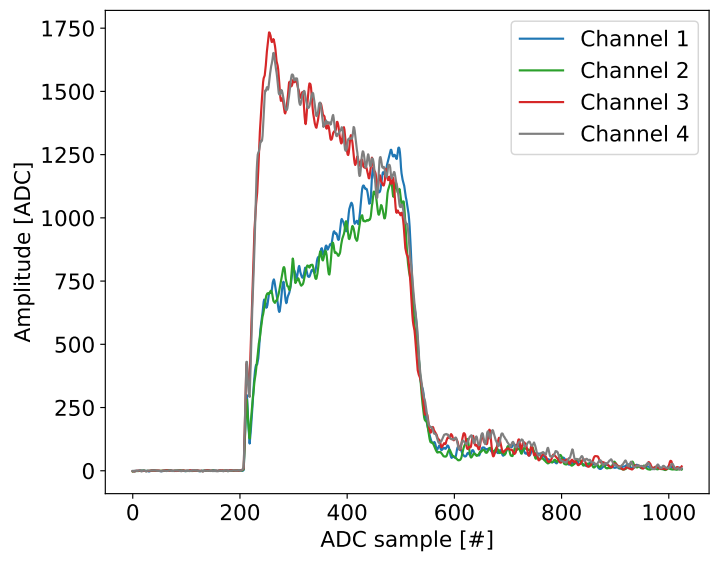

**(f)** Alpha event – PMTs

**Figure 5.2:** Examples of three different types of interactions in LIME. The top panels show the CMOS images, while the bottom panels display the corresponding waveforms from the four PMTs associated with each image. From left to right: a typical spot-like event induced by a $^{55}$Fe source, a higher-energy electron recoil ($\sim 50$ keV), and an alpha particle.

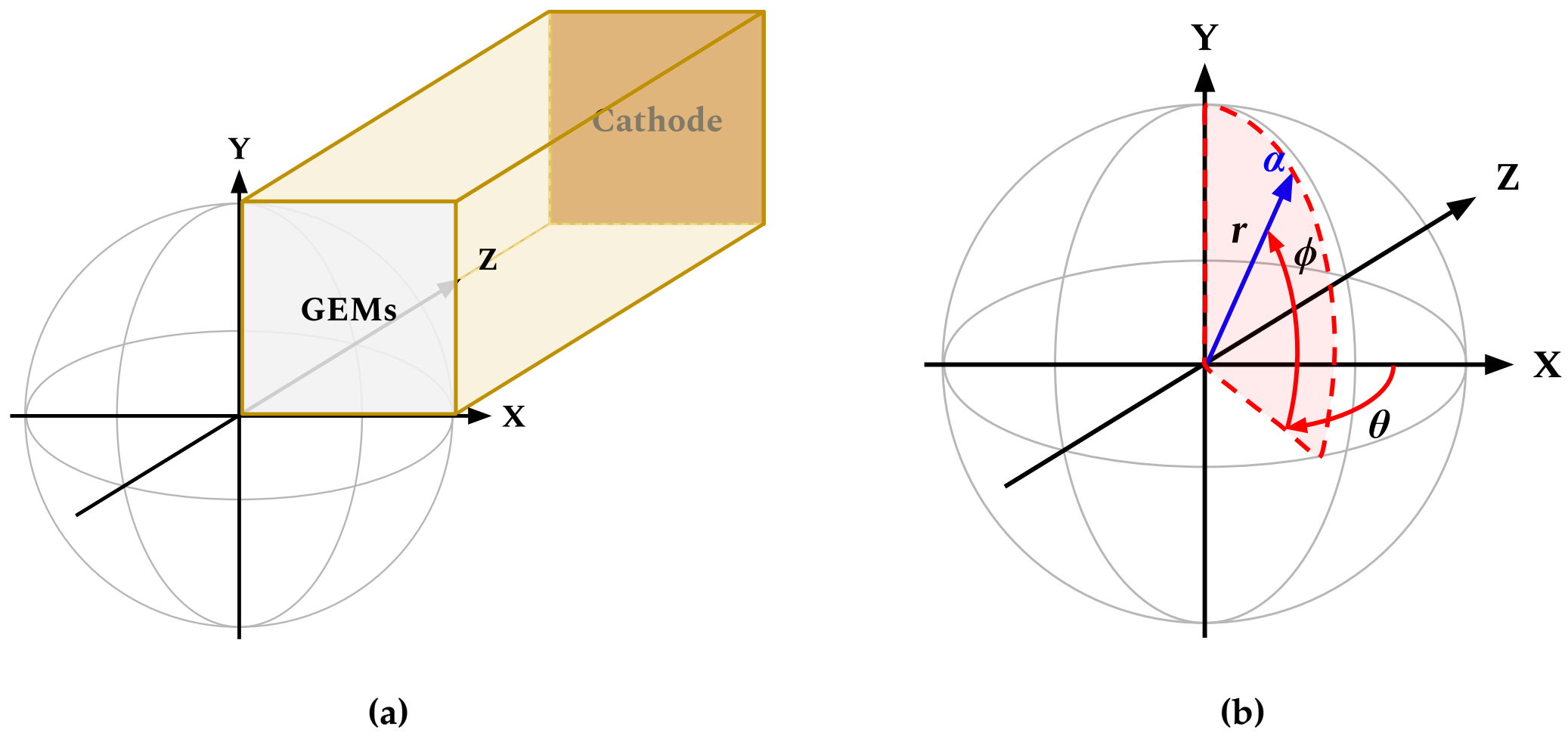

**Figure 5.3:** Coordinate system used to define (a) the absolute XYZ position and (b) the $\phi$ and $\theta$ angles of alpha particles in LIME.

### 5.1.1 3D reference frame

For all subsequent analyses in this thesis, the same coordinate system is used as described here. The CYGNO detector geometry and readout concept naturally define the coordinate system, with the XY plane corresponding to the projection of the light emitted at the GEMs, as imaged by the CMOS sensor, and the Z axis being perpendicular to it, along the drift direction. The Z coordinate represents the time of arrival of electrons at the GEM plane, as projected onto the PMT waveform. The origin of the XYZ reference system is located at the bottom left corner of the GEM plane, as illustrated in Figure 5.3a. The angles $\phi$ and $\theta$, which describe the trajectory of the particle in three dimensions, are defined with respect to the track's origin, as illustrated in Figure 5.3b, with a blue line representing an (alpha) particle track. The $\phi$ angle is defined as the angle between the track and the Y-axis in the 2D projection of the track in the CMOS image, being positive if the particle is moving upwards, and negative if moving downwards. The $\theta$ angle describes the inclination of the particle's trajectory with respect to the XY plane, being defined positive when the particle moves toward the GEMs and negative when moving toward the cathode, as illustrated. In this reference frame, the two angles, $\theta$ and $\phi$, are constrained within the following ranges:

$$\phi \in [-\pi, \pi) \qquad \theta \in \left[-\frac{\pi}{2}, \frac{\pi}{2}\right) \tag{5.1}$$

## 5.2 3D track reconstruction strategy

A track in 3D is fully defined by an oriented (i.e. with sense) 3D vector positioned in space. Given the specifics of CYGNO experimental approach, with the sCMOS data providing information on the XY projection of the track on the amplification plane, and the PMT signal being sensitive to the track development along the drift direction (i.e. $\Delta Z$), the preliminary strategy developed to achieve 3D track reconstruction illustrated along this chapter proceeds as follows:

- The XY absolute position and intensity of each pixel associated to a track are returned by the IDDBSCAN algorithm illustrated in Section 3.1;
- Exploit the associated pixel intensity pattern along the track to determine the $\phi$ angle relative to the Y-axis in the XY plane (Section 5.2.2);
- Use the transverse track profile pixel intensity of the CMOS image XY projection to infer the absolute Z position from a fit to track diffusion (Section 5.2.3);
- Evaluate the track extension along the drift direction (i.e. $\Delta Z$) from PMT time-over-threshold(Section 5.2.4);
- Extract the relative inclination with respect to the Z axis (i.e. $\theta$ angle *sign*) from the Bragg peak in PMT signal (Section 5.2.5);
- Match each alpha in the sCMOS image with the corresponding PMT signal with BAT fit (Section 5.3);
- Merge all info to obtain a 3D vector positioned in space (Section 5.4).

Given that a signal in both sensors is required for full 3D tracking and since the PMT has a lower intrinsic energy threshold with respect to the camera, a selection is applied first to the sCMOS images sample (Section 5.2.1) to remove fake events and ER-like tracks, and the PMT analysis is performed only on the events resulting from such requirements. A subsequent selection is applied to these PMT signals to further clean the sample from the remaining ER. The strategy focus is to maximize the purity of the sample rather than the efficiency of the selection. The analysis presented in this and the next Chapter is based on Run 4 underground LIME data (see Section 2.2.2.4.3 since, as illustrated in Section 6.2, this represents the first Run of the underground LIME campaign where the detector conditions and data quality were constant during the entire run.

### 5.2.1 CMOS image selection

As illustrated in Section 3.1, the CMOS image analysis returns as output the tracks found by the algorithm and the XY positions and intensity of the pixels associated with each of them. These are used to further select and analyze the events of interest by exploiting the characteristic features of the energy deposit along the track (see Section 5.1), geometrical correlations, and the known properties of primary ionization electron diffusion during drift in the CYGNO gas mixture. In order to do this, some minimal cuts to remove fake events and ER-like tracks are applied to the sCMOS image events and discussed in this Section.

#### 5.2.1.1 Removal of mis-reconstructed events

During the two years of LIME underground operation, an increase in the intrinsic pixel noise was observed in the top and bottom rows of the sCMOS sensor. Being such an effect of a stochastic nature over time, the application of standard noise-reduction filters and pedestal subtraction techniques proved inefficient in addressing this issue. Consequently, the CYGNO reconstruction procedure illustrated in Section 3.1 resulted in clustering together noisy pixels not belonging to any real energy deposit in the gas, which were falsely reconstructed as tracks. It was hence decided by the collaboration to completely remove the noisy upper and lower rows before applying the iDDBScan algorithm discussed in Section 3.1.2. As a result, tracks developing their ionization pattern in both the removed and the active regions resulted in a "truncated" CMOS image reconstruction, which could therefore not be used for 3D reconstruction. An example of one of these events is shown in Figure 5.4, where the full field of view of the CMOS camera is shown in the left panel, with the *noise-band-cut* marked in grey. A long electron recoil crossing the entire detector and an alpha particle emitted near the bottom edge of the detector are also visible. The middle and right panels illustrate, respectively, the original alpha track and the pixel cluster retrieved by the CYGNO reconstruction algorithm. In this example, it is evident that the true 2D dimensions of the alpha track have been incorrectly truncated by the *noise-band-cut* during the CYGNO reconstruction process.

To remove such events, any reconstructed tracks with the edgemost pixels coinciding with the border of the noise band were removed from the subsequent analysis. The edgemost pixels constitute the minimum and maximum X and Y pixels of an ionization track and are retrieved through the directionality algorithm applied to calculate the track's $\phi$ angle, in Section 5.2.2. This occurrence is not expected to happen with the sCMOS Orca Quest models purchased for the manufacturing of CYGNO-04 (see Section 2.2.3), and therefore we do not foresee having to deal with this issue in future demonstrator realizations.

Additionally, as illustrated in Section 2.2.2.2.1, the peculiarity of the sensor exposure performed along pixel rows can result in a track being "split" between two different CMOS images. Since all rows are sequentially opened and closed, this effect can potentially occur anywhere within the image. In principle, this issue could be addressed with a more sophisticated CMOS image analysis capable of merging tracks recorded in two different images. Such functionality is anticipated in the improved DAQ approach

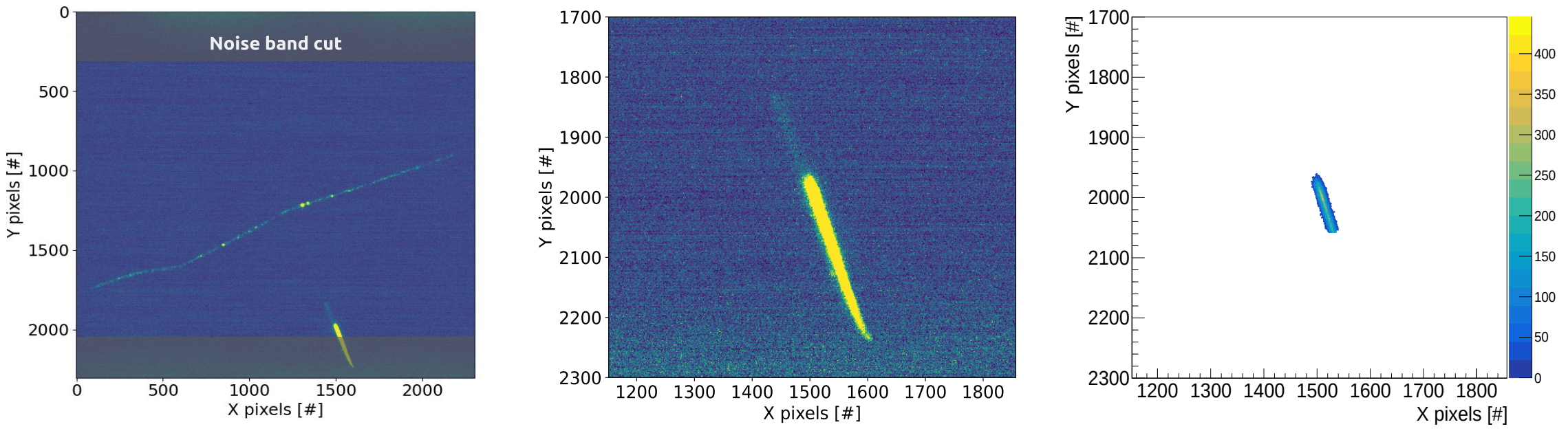

**(a)** CMOS field of view. **(b)** CMOS zoom-in. **(c)** Reconstructed alpha particle.

**Figure 5.4:** Example of an alpha particle emitted near the border of the detector and falling within the *noise-band-cut* region. (a) shows the full CMOS field of view, including the *noise-band-cut* indicated by the grey band. (b) provides a zoomed-in view of the alpha event, and (c) shows the same event after the CYGNO-reconstruction algorithm.

to be implemented in CYGNO-04 (see Section 2.2.3), where this situation is inherently avoided. However, due to the novelty, complexity, and DAQ requirements of this strategy, it was decided – within the scope of this thesis – to simply remove this type of event.

To identify (and reject) these events, the "Trigger-Time-Tag" (TTT) of the PMT waveforms can be used. The TTT is a time-stamp that connects each waveform to the beginning of the opening of the CMOS sensor. Each waveform falls into one of three possible time windows, depending on the number of activated pixel rows in the image. These windows correspond to the three regions visible in Figure 2.19: the sensor activation/opening, the sensor fully open (GE), and the sensor deactivation/closing. In time, these three time regions can be described as follows, where 184.4 $ns$ is the time required to expose all the sensor rows, 300 $ns$ is the time at which the sensor starts closing, 2304 is the total number of pixel rows, and $MR$ (moving row) is the row of pixels currently opening or closing:

$$\begin{aligned}
&1) \quad TTT < 184.4\ ns : sensor\ opening \Rightarrow MR = 2304 - 2304 * \frac{TTT}{184.4} \\
&2) \quad 184.4 < TTT < 300\ ns : sensor\ fully\ open \Rightarrow GE \\
&3) \quad TTT > 300\ ns : sensor\ closing \Rightarrow MR = 2304 - 2304 * \frac{TTT - 300}{184.4}
\end{aligned} \tag{5.2}$$

Similarly to the previous cut, the edgemost pixels of the track are compared with the current moving pixel row, and if they coincide, this indicates that the track was effectively cut by the opening or closing process of the CMOS sensor. These alpha

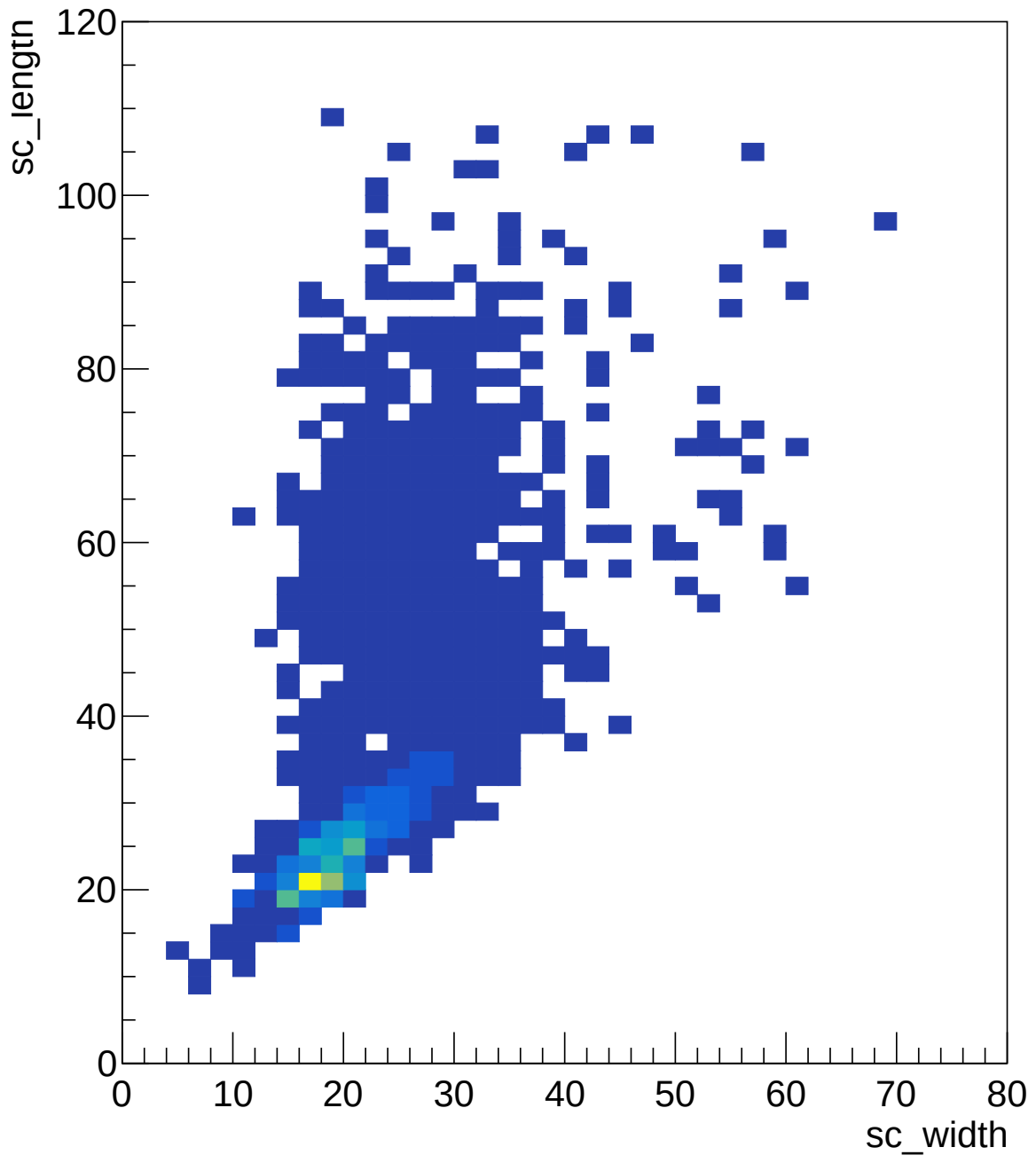

**Figure 5.5:** Distribution of the *sc_length* and *sc_width* variables from pedestal runs in LIME underground Run 4.

tracks are then tagged and excluded from further analysis.

#### 5.2.1.2 Fake cluster removal

Even after removing the high noisy band on top and bottom of the CMOS image, hot pixels and noise still remain present even in the center of it. Depending on the intensity and proximity of these, sometimes the IDDBSCAN code ends up reconstructing from these fake events as real tracks. In order to remove such fake occurrences, an analysis of the pedestal runs in Run 4 (see Section 3.1.1 for pedestal definition) was done within the CYGNO collaboration to identify some discriminating variables among the ones given by the output of the reconstruction code. Figure 5.5 display the reconstructed track length (i.e. sc_lenght) versus width (i.e. sc_width) as illustrated in Section 3.1.3 (in number of pixels) from pedestal runs, where it can be clearly observed how these concentrate at low length and width values. Stemming from this, a cut to Run 4 dataset was applied to select events with length $>$ 100 pixels and width $>$ 50 pixels in order to remove fake cluster presence in the data set.

#### 5.2.1.3 Alpha tracks selection

Thanks to their peculiar high density and mostly straight $\mathcal{O}(\mathrm{cm})$ long energy deposition pattern, alpha tracks can be easily selected from the events remaining after applying the quality cuts illustrated Sections 5.2.1.1 and 5.2.1.2 by exploiting the delta variable, i.e., the track light integral divided by the number of pixels associated to the track passing zero suppression, as illustrated in Section 3.1.3. Figure 5.6 display the distribution of events delta versus sc_integral (see Section 3.1.3 for definition) after these cuts. As already elaborated in [196], the band at low uniform delta extending up to high light yield integral is expected to be associated to high energy ERs, with the upper high delta population linear with energy represent high density dE/dx deposits typical of alpha particles and nuclear recoil. From Figure 5.6 a cut selecting events with $\delta > 25$, identified in the figure as a red line, appears to easily and clearly remove the ER candidates from the dataset and provide a clean alpha sample and it was hence selected for the work discussed in this thesis. Table 5.1 summaries all the cuts applied to the CMOS images and used in the rest of the chapter to identify alpha tracks, as well as in Chapter 6.

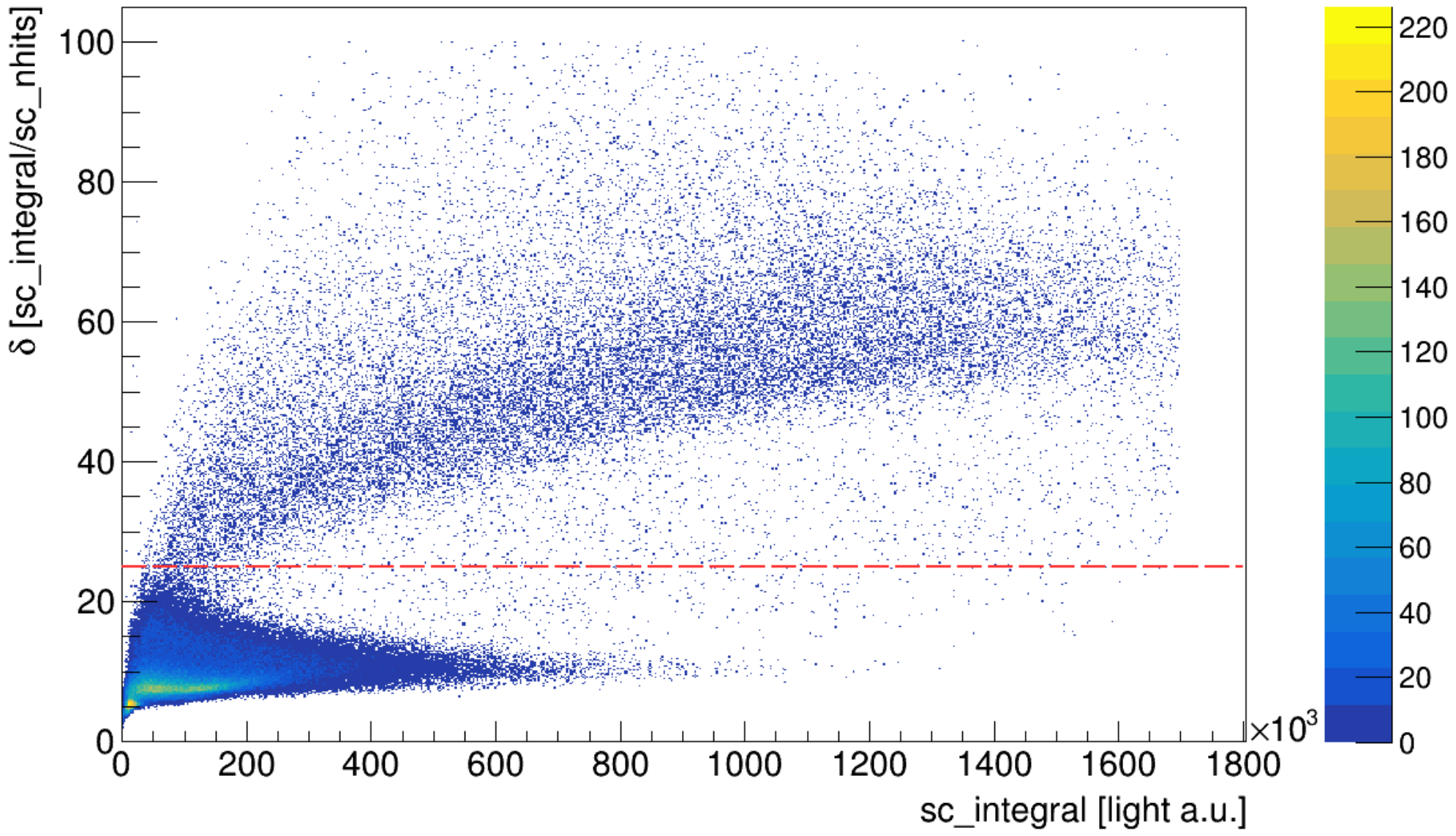

**Figure 5.6:** Distribution of the *delta* variable associated to each of the pixel clusters identified in the dataset used in the analysis. The data has been pre-selected using the quality cuts in Sections 5.2.1.1 and 5.2.1.2. The cut used is shown as a dashed red line at $\delta = 25$, highlighting the separation between high and low deposition density events.

### 5.2.2 Phi ($\phi$) angle determination

In order to determine the angle relative to the Y-axis in the XY plane (i.e. $\phi$ in Figure 5.3b) for the events passing the quality requirements illustrated in Section 5.2.1, an algorithm originally developed to infer low energy ERs direction [132] adapted from X-ray polarimetry methods [259] was employed, after having properly adapted its free parameters to suitably perform on alpha tracks.

**Table 5.1:** Summary of the cuts applied in order to select alpha particles from the CMOS images dataset.

| Particle | $\delta$ = sc_integral / sc_nhits [light a.u. / pixel] | sc_length [pixel] | sc_width [pixel] |
|---|---|---|---|
| Alpha | $> 25$ | $> 100$ | $> 50$ |

The algorithm's working principle can be summarized as follows:

1. **Track Major Axis determination:** The algorithm calculates the barycenter of the track (the center of mass of the light distribution) using the amount of light recorded in each pixel of the reconstructed cluster. It then identifies the major axis of the track by maximizing the second moment ($M2$) of the pixel intensity distribution. This is achieved by rotating a reference frame around the barycenter and calculating $M2$ for various angles. The maximization of $M2$ provides an approximation of the track's direction by identifying the axis along which the particle deposited most of its energy.

2. **Interaction Point (IP) determination:** The interaction point is determined by analyzing the energy loss behavior characteristic of different particles. Alpha particles deposit most of their energy near the end of their tracks due to the Bragg peak. To identify the "tail" of the track – opposite to the Bragg peak – the skewness of the light distribution is evaluated. A circumference is drawn around the barycenter of the track, and pixels located on the side opposite to the Bragg peak and at a specific distance from the barycenter are selected. The radius of this circle is defined such that a fixed number of points, $N_{pt}$, lie outside it. This free parameter, $N_{pt}$, is later optimized for alpha particles. The interaction point is then calculated as the weighted average of the selected pixels within this radius.

3. **Track Direction Evaluation:** The algorithm determines the track's direction starting from the IP and analyzing the light intensity along the track. For this purpose, the intensity of each pixel ($W$) is reweighted exponentially with the distance from the interaction point ($d_{ip}$), following the formula:

$$W(d_{ip}) = exp(-d_{ip}/w) \tag{5.3}$$

   where $w$ is free normalization factor, also optimized later for alphas particles. This method emphasizes the contribution of pixels closer to the interaction point, as they better preserve the initial direction of the recoil due to consecutive scattering. The track's direction is then calculated by determining the main axis using the re-weighted pixels, ensuring that the axis passes through the interaction point.

4. **Track Sense:** To determine the sense of the track along its major axis, the algorithm uses a method based on two barycenters: one at the IP, and another, denoted as *IPpr*, calculated further along the track by selecting more pixels during the iterative impact point reconstruction process. Specifically, IPpr is computed at the stage where the number of selected pixels reaches $2 \times N_{pt}$, effectively capturing the progression of light along the track. The algorithm then defines two lines perpendicular to the track's main axis – assumed to be $y = ax + b$ with $a = \tan(\phi)$ — each passing through one of the two barycenters. These lines are given by:

$$\begin{aligned} \text{IP}: \quad & y = m(x - x_{IP}) + y_{IP} = mx + q_{IP} \\ \text{IPpr}: \quad & y = m(x - x_{IPpr}) + y_{IPpr} = mx + q_{IPpr} \end{aligned}$$

with $m = -1/a$. By analyzing the relative positions of these lines – specifically, comparing $q_{IP}$ and $q_{IPpr}$ – and considering the sign of $a$, the algorithm determines whether the light pattern aligns or opposes the initial direction estimate. The final orientation angle $\phi$ is then taken as $\arctan(a)$ or corrected by $\pm 180°$, accounting for the fact that arctan is limited to the range $(-90°, +90°)$ and does not cover the full angular space.

The final result of the directionality algorithm is the determination of the initial direction of the ionization track in the XY (CMOS) plane, or the $\phi$ *angle*, following the reference frame shown in Figure 5.3. Additionally, the algorithm performs a "cleaning" step to remove isolated pixels from the track: each bin with a signal greater than zero is examined, and if none of its immediate neighbors contain significant light counts, it is considered noise and removed. From the cleaned track, the *edgemost pixels* are identified – defined as the minimum and maximum X and Y pixels of the ionization track in the CMOS plane. These are then used by the 3D analysis to determine the track's *XY length*, i.e., its spatial extension in the XY plane.

As demonstrated in [132], this directionality algorithm encounters limitations based on the angle $\theta$ of the event (see Figure 5.3). When the track is tilted more than 45°, the resolution in identifying the original direction decreases by about 30%. This occurs because the algorithm operates on a 2D projection of the track, which becomes less effective for tracks that are highly inclined. In the limiting case, when the track is nearly perpendicular to the GEM plane ($\theta \sim 90°$), the 2D projection appears as a spot, making it very hard to determine the track's (initial) direction.

#### 5.2.2.1 Optimization for alpha particles

When adapting this algorithm for alpha particles, several features remained similar. As mentioned in Section 5.1, both ERs and alphas exhibit their maximum energy deposit density at the end of their paths, allowing the algorithm's evaluation of the track di-rection and sense to be similarly applied to both particle types. Since alpha particles in LIME generally produce straight tracks, the limitations arising from hard curvature seen in ERs tracks are negligible, simplifying the assessment of the *IP* and initial direction. Still, due to the large differences in the electron cloud shape between ERs and alphas, modifications to the two free parameters of the algorithm, $N_{pt}$ and $w$, were required.

Since a recursive optimization of the $N_{pt}$ and $w$ was not possible for the scope of this thesis given the too demanding CPU usage and time required to fully simulate alpha tracks response in the CYGNO experimental approach, as also further elaborated in Section 6.3.1, a more simplistic approach based on the physic at play in this context was used to adapt these parameters to properly work on alpha tracks. It is important to stress how the very straight nature of these energy deposits compared to the curliness of ER tracks makes the directionality algorithm performances much less dependent on these parameters than in the ER case and therefore this elementary adjustment resulted suited for the scope of this thesis.

- $\mathbf{N_{pt}}$: This parameter represents the number of points (pixels) along the track used to define the *IP*. For alpha particles, which have a much larger electron cloud than ERs, $N_{pt}$ had to be increased by approximately 30 times to ensure enough initial pixels were considered for the accurate estimation of the *IP*. As shown in Figure 5.7b, the points $N_{pt}$ used for the determination of interaction point from the original track (Figure 5.7a) cover, indeed, only a small portion of the beginning of the track, as intended.

- The normalization factor $w$ adjusts the re-scaling of pixel intensity to determine the initial direction of the track, following Equation 5.3. The $w$ factor needs to be carefully tuned to give the correct weight to each pixel starting from the *IP*. Since multiple scattering (and straggling) is inversely proportional to the energy deposit, in [132] it was observed how larger $w$ values should be applied with increasing ER energy. The optimal $w$ factor estimated on MC simulation in [132] is in fact observed to increase from 1.5 to 2.5 from 16 keV to 70 keV ER energy. Given that, from back-of-the-envelope arguments and calculations, we can expect ER at about 150–200 keV to resemble alphas in terms of "straightness". Therefore, a $w$ factor equal to 3.5 was used for the application of the directionality algorithm to alpha particles. The reweighted track pixels are shown in Figure 5.7c, clearly indicating the sense of track, with the final determination of the track's $\phi$ angle shown in Figure 5.7d.

### 5.2.3 Absolute Z evaluation

The precise measurement of the ionization event's position inside the detector along the longitudinal plane – defined as *absolute* Z using the system of coordinates in Figure 5.3 – is crucial for any TPC-like rare event search experiment due to the inherent lack of an interaction time-stamp, $t_0$, as discussed in Section 1.4.3. By determining the absolute Z

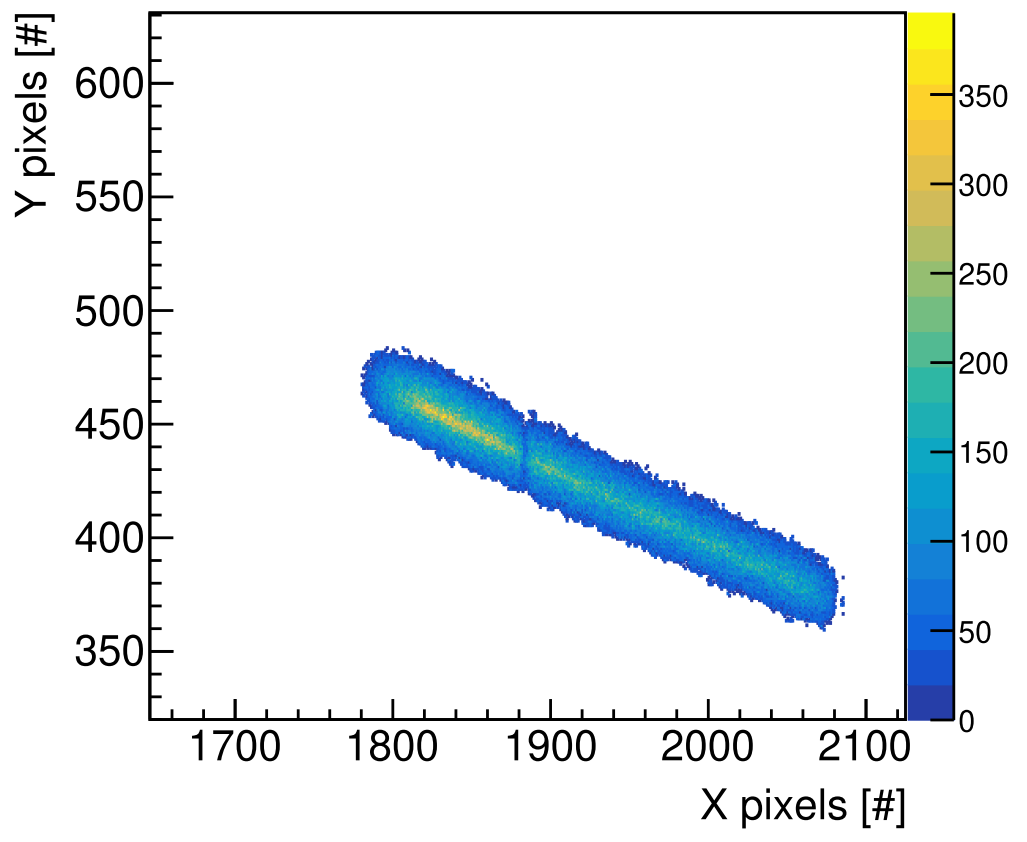

**(a)** Alpha – Original

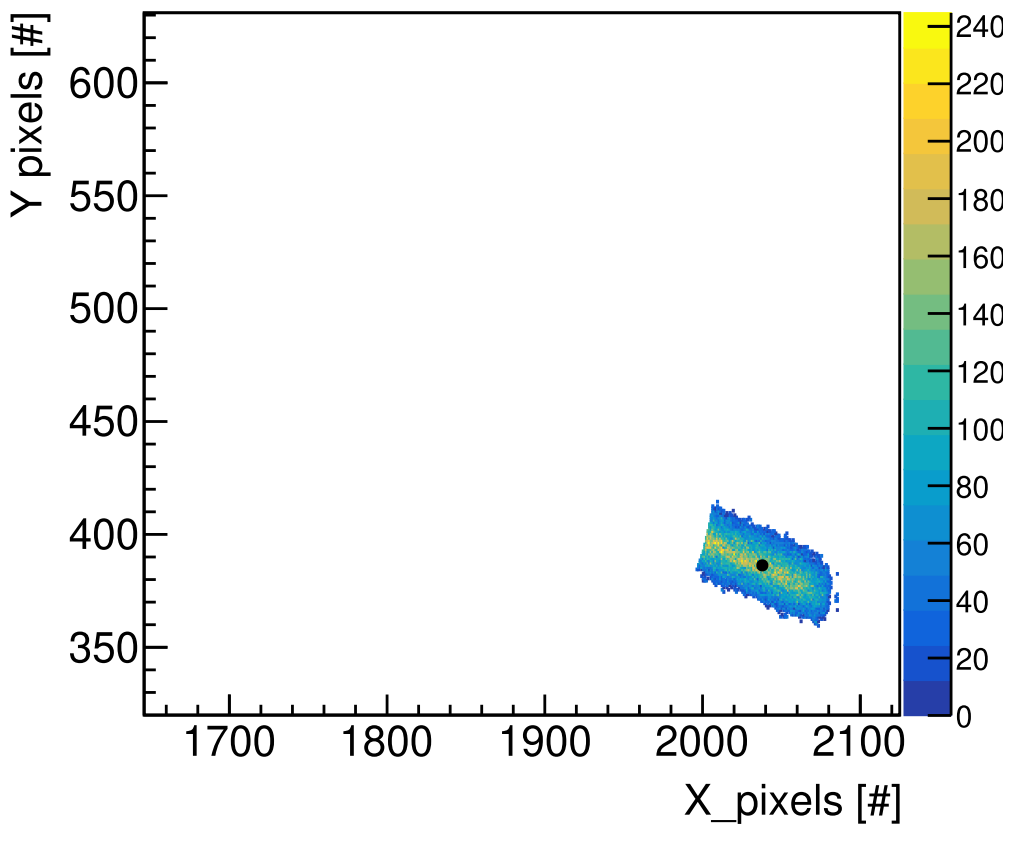

**(b)** Alpha – *IP*

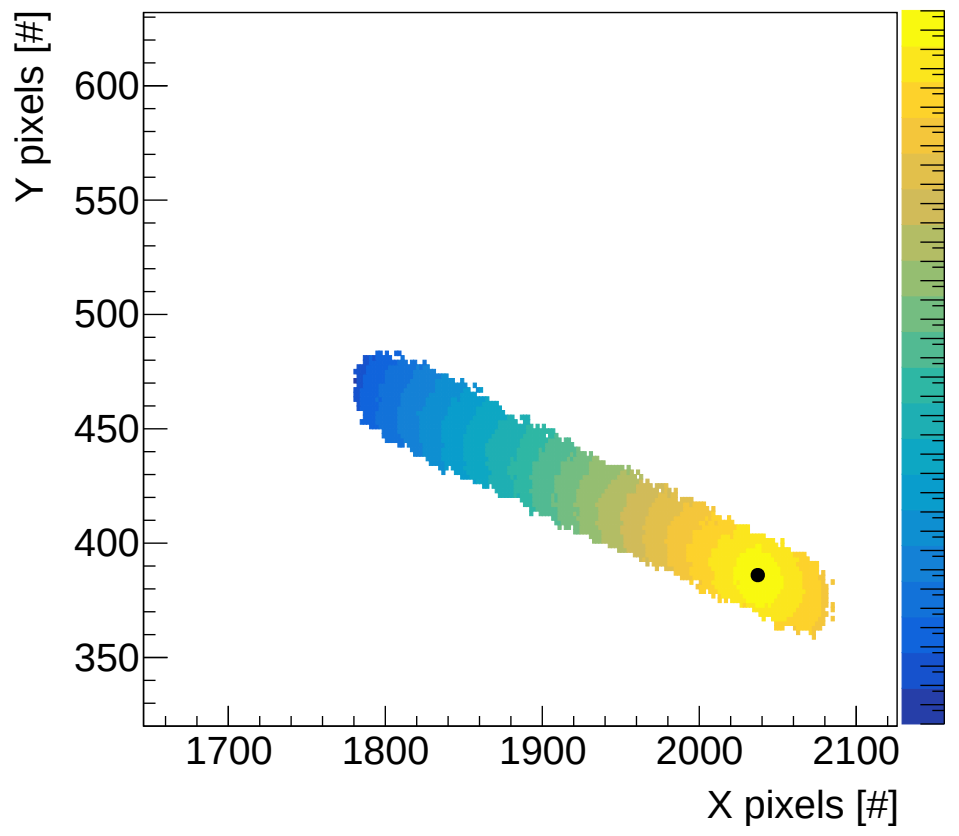

**(c)** Alpha - *IP* + re-weighted pixels

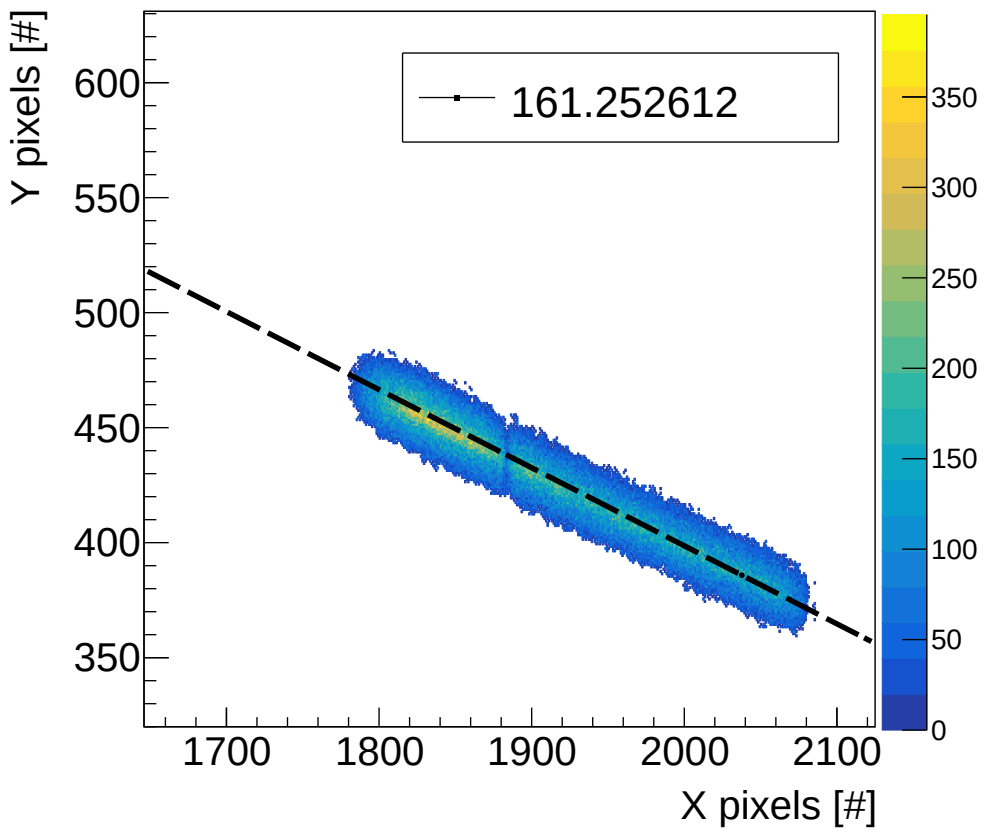

**(d)** Alpha – Initial direction angle

**Figure 5.7:** Examples of the directionality algorithm used to illustrate the adaptations of the $N_{pt}$ and $w$ parameters: (a) the original reconstructed track CMOS images; (b) the track region with $N_{pt}$ points used to find the *Impact Point*; (c) track with reweighted pixel intensities; and (d) calculated initial direction angle overlaid on the original track.

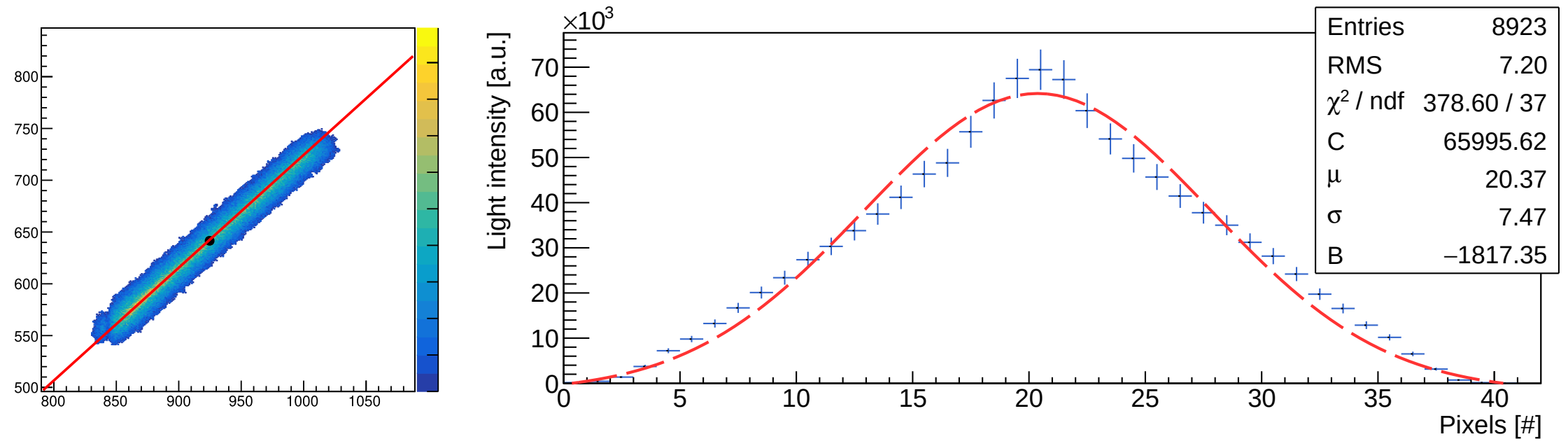

**(a)** **(b)**

**Figure 5.8:** Alpha track in (a) shown after passing through the CYGNO-reconstruction algorithm, with the main axis overlaid as a red line. In (b), the transverse profile of the ionization cloud, constructed perpendicular to the track's main axis at its center, is displayed, with a Gaussian fit overlaid as a dashed line. This fit clearly demonstrates how the use of the RMS appears more consistent in this context given the observed non-Gaussianity of the distribution.

of an event, a fiducial volume can be defined within the sensitive region, allowing for the rejection of background events originating near the detector's boundaries, particularly from the cathode and amplification region [260,261].

The position of an ionizing event in a gas detector can be obtained using various approaches, one of which involves analyzing the shape of the ionization cloud and performing a "fit to diffusion" [262]. The method developed in this thesis to infer the absolute Z for alpha tracks consists in measuring and fitting the width of the transverse light profile of the track, which reflects the amount of diffusion experienced by the ionization cloud's electrons as they drift between their point of creation and the anode. Since diffusion scales with $\sqrt{Z}$, this relationship allows for the calculation of the absolute Z. In practice, for the same alpha energy or range in the gas, events closer to the GEM plane experience less diffusion and appear "thinner", while events near the cathode appear wider. The absolute Z position can be derived by rearranging the standard diffusion formula, as originally used in [121] and later adapted by the CYGNO collaboration in [154]. This gives:

$$Z = \frac{\sigma^2 - {\sigma_0}^2}{{\sigma_T}^2} \tag{5.4}$$

where $\sigma_T$ is the transverse diffusion coefficient, $\sigma_0$ represents the minimum diffusion caused by the multiplication processes at the GEM region, and $\sigma$ is the measured diffusion for each event, obtained by fitting the transverse light profile of the alpha particles.

An example of the determination of the absolute Z coordinate in an alpha track is shown in Figure 5.8. This figure presents an alpha track and its transverse profile, obtained by first identifying the main axis of the track (red line in Figure 5.8a) and then plotting the orthogonal light intensity distribution at its center (Figure 5.8b). The transverse profile obtained does not seem to be well represented by a simple Gaussian curve. Two possible reasons could explain this: first, there is an intrinsic limitation in the measurement, as it relies on the 2D projection of the track. If the track is not parallel to the GEM plane, the ionization cloud starts collapsing onto itself in the 2D image, resulting in a profile that may not strictly follow a Gaussian distribution. The second possible reason, according to very recent studies within the CYGNO group, is that the transverse profile of these tracks, contrary to the theoretical expectation, does not follow a simple Gaussian curve derived solely from the diffusion process, but rather a convolution of two or more Gaussian curves, leading to a more triangular shape. This phenomenon could arise from the contribution of different processes impacting the electron cloud, including not only the charges' drift but also their multiplication. It has also been suggested that the high amount of charges released by the passage of the alpha particle could lead to space-charge effects [263] already at the second GEM stage during the multiplication, leading to an additional and likely non-Gaussian contribution to the shape of the light profile.

Due to this, for the application of Equation 5.4, it was decided to use the RMS of the transverse profile distribution as a measure of diffusion instead of the sigma ($\sigma$) from the Gaussian fit. The RMS can be a more appropriate descriptor for the spread of a distribution because, unlike $\sigma$, it does not assume any specific distribution shape. Since it measures the dispersion of all data points, including outliers and non-Gaussian features (asymmetries, long tails, etc.), the RMS can provide a more robust characterization of irregular distributions, such as this one. The coefficient $\sigma_T$ for the CYGNO gas mixture was instead calculated using Garfield++ and is shown in Figure 2.3a. Finally, $\sigma_0$ was evaluated from data obtained using the MANGO prototype. This was done by placing an alpha source very close to the GEM plane and shooting parallel to it, under conditions similar to those of LIME. The "minimum" diffusion is determined by analyzing the RMS of the transverse light profile of the alphas. Combining both simulation and measurements, the final values used in Equation 5.4 are: $\sigma_T = 129.7 \pm 3.1\ \mu m/\sqrt{cm}$ and $\sigma_0 = 780 \pm 30\ \mu m/\sqrt{cm}$. The distribution of absolute Z values, calculated using this equation with the RMS approach for all alpha particles identified in the datasets used for the 3D analysis, is shown in Figure 5.9.

The calculated absolute Z distribution of alphas in LIME shows a clear peak at a drift distance of 50 cm, perfectly coinciding with the longitudinal length of LIME, with a FWHM of about 15 cm. This correspond to a sigma of about 6.4 cm, which results consistent with previous work developed along the same line of fitting the diffusion on $^{55}$Fe events [264]. This is explained by the position of the expected sources of alpha background in the detector: a significantly higher number of alphas appears to be emitted at $Z = 50$ cm, likely originating from the cathode. At lower Z values, a nearly flat distribution is observed, suggesting that these alphas are being emitted uniformly distributed throughout the detector. In Chapter 6, a full characterization of the spatial distribution and orientation of alphas will be shown, further characterizing this observation.

Since diffusion depends on $\sqrt{Z}$, and the $\sigma_0$ term in Equation 5.4 results much larger than the Z-dependent $\sigma_T$ term, the accuracy of absolute Z determination is expected to deteriorate with decreasing Z and therefore to perform poorly on low Z tracks. The extension of the inferred absolute Z distribution to negative Z values is hence attributed and motivated by this feature. Although the distribution does not perfectly reflect the full length of LIME, the preliminary absolute Z evaluation method illustrated in this section still resulted suitable for the scope of characterizing the spatial distribution of alpha tracks in LIME, as will be illustrated in Chapter 6.

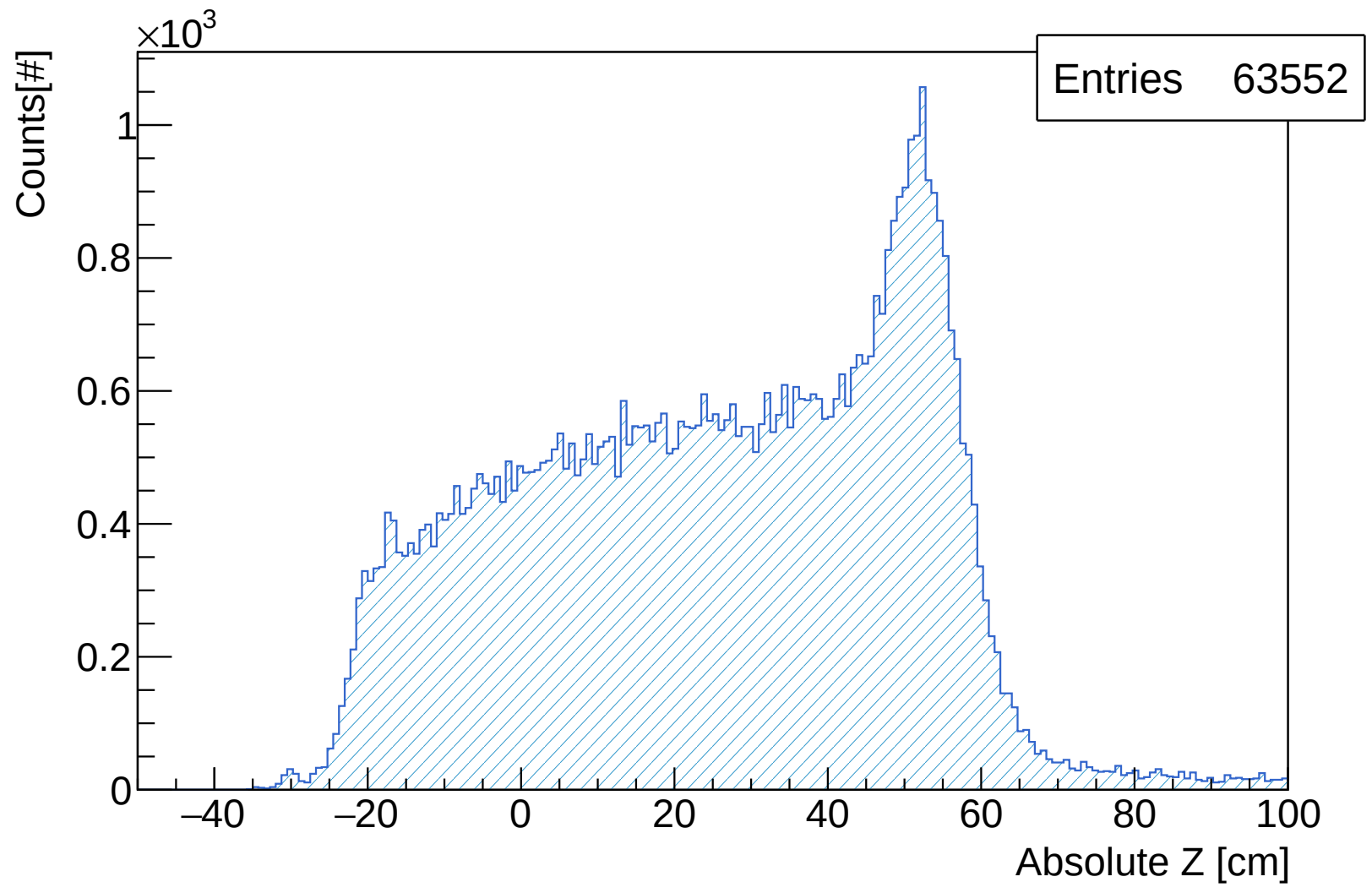

**Figure 5.9:** Distribution of absolute Z's for all the alpha particles analyzed, calculated using Equation 5.4 and the RMS of the transverse profile distribution as the spread of the ionization cloud ($\sigma$ in the formula).

### 5.2.4 PMT signal selection and improved TOT definition

The analysis of PMT waveforms presented in this section is an extension of the work developed and presented in Section 3.2. As illustrated in Section 5.2, it is performed only on the events passing the selection applied to the CMOS images discussed in Section 5.2.1. Since, given the trigger strategy illustrated in Section 2.2.2.2.1, even when selecting events in which only alphas are present in the CMOS image, other signals from different particles could be present in the associated PMT waveforms, additional variables have been developed and are discussed in this section. These were developed to further clean the PMT sample from ER-like events and optimize the alpha PMT analysis.

Differently from typical LIME PMT signals of low-energy ERs, it was observed that in many of the ones associated with alpha particles, signal tails appeared after the sharp fall times of these waveforms, characteristic of alpha particle signals. This is suspected to be associated with PMT electronic noise which, due to the fact that LIME was a prototype, resulted from the use of PMTs that were not fully optimized in terms of performance and quality. This issue is, therefore, not expected to be present in future detectors, such as CYGNO-04 (Section 2.2.3), where the selection of the PMTs used will be more thorough, stemming also from the conclusions presented here. The use of the standard strategy illustrated in Section 3.2.1.4, based on the noise RMS to evaluate the alpha tracks' TOT, would hence result in a biased estimation of this quantity towards larger values. For this reason, a preliminary study was developed based on the observation that alpha energy deposits have very sharp waveform rise and fall times resembling a step function, due to their characteristic energy deposition features shown in Figure 5.1. From this, various waveform TOT levels were tested to systematically evaluate and characterize this feature in all waveforms. Specifically, the variables *TOT 5RMS* (the standard one), *TOT10*, *TOT20*, and *TOT30* were tested. These variables represent the time over threshold at amplitudes 5RMS, and 10%, 20%, or 30% of the waveform's maximum amplitude. An example of these is shown in Figure 5.10. As shown by the TOT 5RMS and TOT10 lines, the full extension of the waveform can be overestimated when such low values are used due to the tails of the signal being included, while both TOT20 and TOT30 are already within the bulk (step-like) part of the signal.

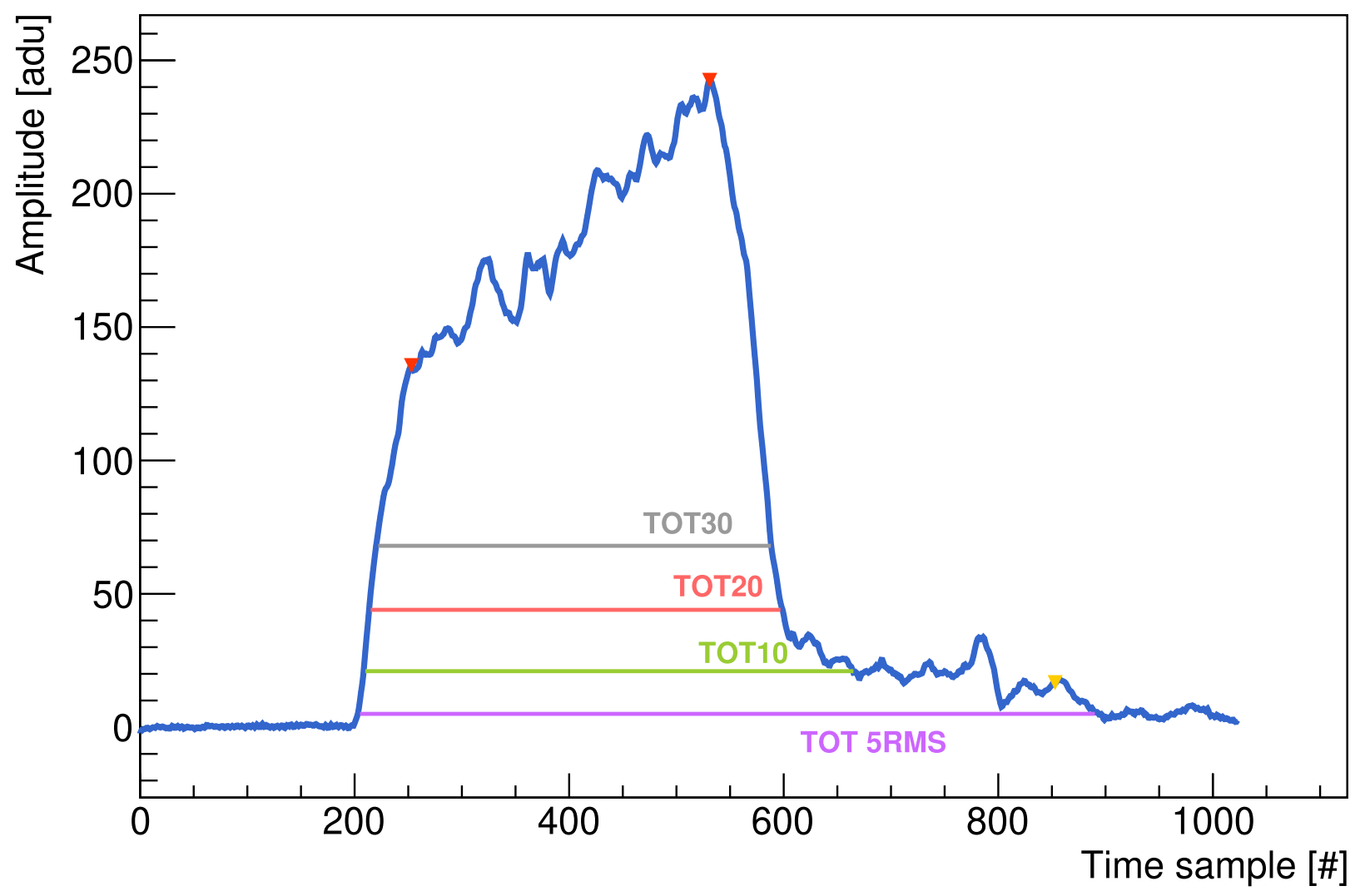

**Figure 5.10:** Example of an alpha particle waveform signal, illustrating the different definitions of the variables TOT 5RMS, TOT10, TOT20 and TOT30. These thresholds are marked in the waveform with, respectively, purple, green, red and grey full lines.

Since the CYGNO group does not yet have a fully validated PMT signal simulation, these four levels were visually inspected in hundreds of tracks. From this, the TOT20 was found to be the lowest threshold that consistently appeared within the sharp rise and fall of the waveforms, and was therefore chosen as the proxy to calculate the full time extension of the signal. Given that alphas travel in a mostly straight path, their extension in Z can be derived directly from their waveform's time extension, using

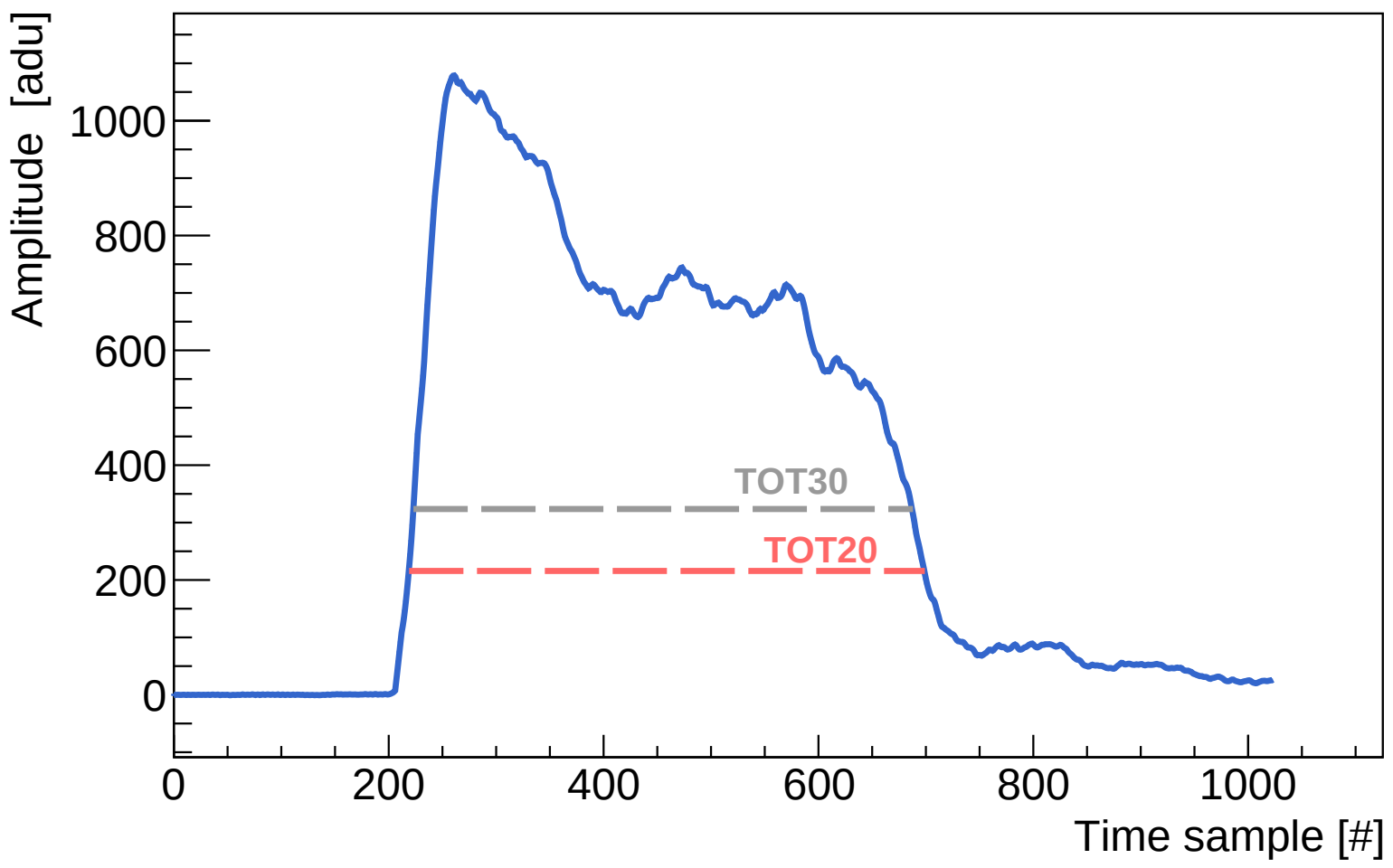

**Figure 5.11:** Example of an alpha particle waveform signal as detected by the PMT, illustrating the definitions of the variables TOT20 and TOT30. These thresholds are marked in the waveform with red and grey dashed lines, respectively.

Equation 3.4, taking the sampling frequency of the fast digitizer (750 MHz). As in the majority of the studies carried out with LIME, the fast digitizer was used for this measurement given its higher resolution and the length of the tracks being studied ($< 10$ cm), which allows them to be fully contained within the waveform. As a side note, the slow digitizer (250 MHz) is only exceptionally used when very long tracks are being analyzed, such as those of MIPs discussed in Chapter 4. For alphas, the PMT time extension is therefore evaluated through the *TOT20* variable. The TOT20 calculation is performed for all four PMT waveforms, and the results are then averaged to retrieve the final $\Delta Z$ value. The $\Delta Z$ values obtained from each PMT waveform are typically very similar (within a couple of time samples), and averaging reduces the bias toward a specific PMT and/or small variations induced by sporadic noise. This ensures a robust and reliable estimation of the extension in Z traveled by the alpha particle.

These newly developed variables also proved considerably effective in further removing ER-like events from the sample – specifically, by requiring that the ratio TOT20/ TOT30 is between 1 and 2. This cross-check ensures that the waveform exhibits a similar time over threshold at 20% and 30% of its maximum, which reflects both sharp rise and decay times, as well as an overall sustained high amplitude – characteristics typical of alpha signals, as illustrated in Figure 5.11. Electron recoil waveforms, on the other hand, are characterized by multiple energy deposition peaks and generally lack this well-defined shape. Furthermore, due to their sparse energy deposition, even if one PMT waveform passes the test, it is highly unlikely that all the remaining waveforms from the other PMTs in that event will also pass.

The number of peaks returned by the algorithm illustrated in Section 3.2.1.3 can be

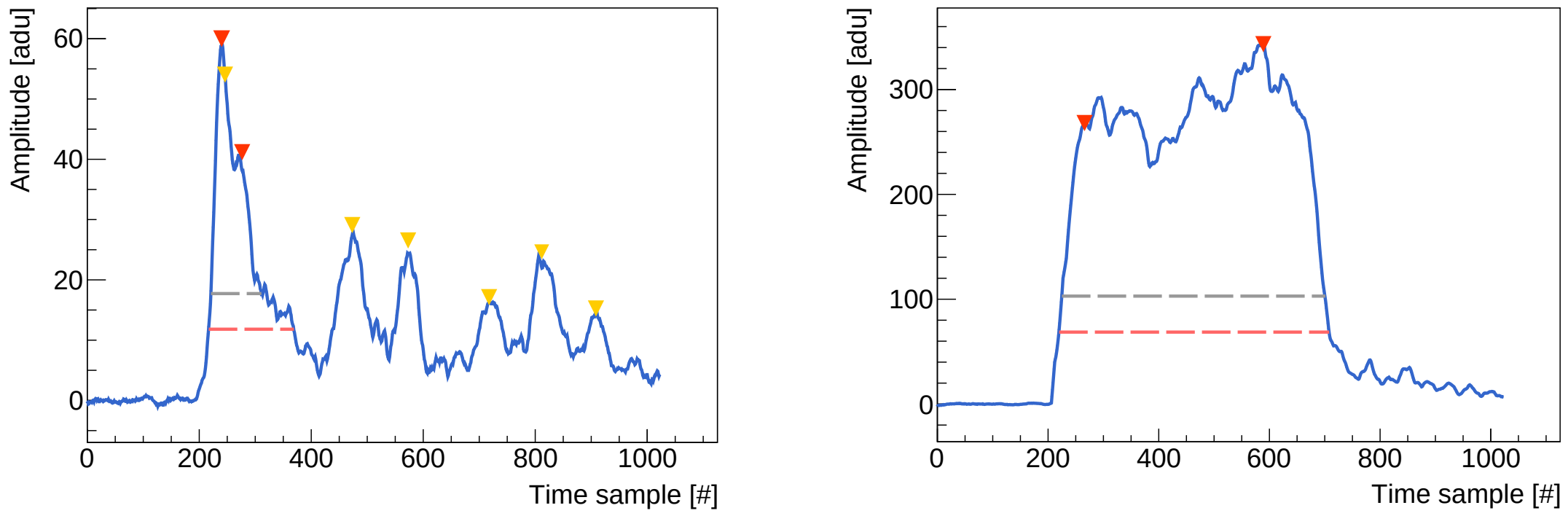

**(a)** ER event

**(b)** Alpha event

**Figure 5.12:** Example of two waveforms highlighting the result of the peak finder routine (yellow triangles) applied to the waveforms. In (a), an ER event is shown, with six small energy depositions identified, while in (b), no small depositions are found. The red triangles represent the points used to calculate the waveform crown skewness, as will be illustrated in Section 5.2.5, and the dashed grey and red lines correspond to the TOT30 and TOT20 variables, respectively.

exploited to further remove ER-like events from the sample. ER waveforms generally display peaks associated with singular energy depositions, clearly separated in time and exhibiting high prominence. In contrast, for alphas, the ionization cloud is more condensed, resulting in a more continuous signal featuring fewer prominent peaks, typically arising from small variations in the alpha's energy deposition and fluctuations in the electronic signal. In Figure 5.12, an example of this peak finder algorithm is given, where two waveforms are shown: one from an electron recoil and the other from an alpha event, with the yellow triangles representing the peaks identified. As shown, six peaks are identified in the ER waveform, while no peaks are found in the alpha waveform. This simple example demonstrates the effectiveness of the routine and how the number of identified peaks can help distinguish alpha particles from other types of events, especially ER. With this information, by requiring the number of found peaks to be strictly less than 3, an additional alpha selection is performed on the PMT signal.

Both the requirements illustrated so far effectively remove ER-like events above about 10 keV. Lower energy ERs, conversely, typically present PMT waveforms with a single peak (associated with a spot-like CMOS image), as shown in Figure 5.2 and discussed in Section 5.1. Since alphas are expected to have a longer minimal extension in the relative $\Delta Z$, these lower energy ERs can be effectively removed by requiring that the TOT20 extends over more than 70 digitizer samples. The requirement was optimized by studying $^{55}$Fe calibration PMT waveforms. Given the use of the fast digitizer for the acquisition of this data, corresponding to a sampling frequency of 750 MHz, and the electron drift velocity of 5.5 cm/μs as simulated by Garfield++ for LIME operation, this selection is equivalent to constraining the track $\Delta Z$ extension to be larger than about 5 mm. The resulting three selection cuts from this discussion are summarized in

**Table 5.2:** Summary of the PID checks applied to all the waveforms for identifying alpha events.

| PID check | Rule |
|---|---|
| 1 | $1 < \text{TOT20} / \text{TOT30} < 2$ |
| 2 | $\bar{N}_{peaks} < 3$ |
| 3 | TOT20 $>$ 70 samples |

Table 5.2.

### 5.2.5 Theta (θ) angle *sign* determination

Together with the geometric dependence on the distance to the light emission (see Section 3.2.3), the PMT signal also reflects the density of the ionization cloud: if electrons arrive at the GEMs more densely and in higher quantities, the light emitted will be greater, leading to a larger (in time and amplitude) PMT-generated signal, as is the case for alphas. On the other hand, if the electron cloud is less dense over the same time period, the PMT signal will be smaller, as is the case for ERs.

This property is especially useful when studying particles with asymmetric energy deposition, such as alpha particles. The position of the Bragg peak in the PMT waveforms can reveal the direction in which the alpha particle is moving in the longitudinal plane, specifically whether it is moving towards the GEMs or towards the cathode, effectively defining the *sign* of the θ angle of the track, following the system of coordinates in Figure 5.3b. This is because the Bragg peak occurs at the end of the alpha particle's path and therefore a higher amplitude region in the waveforms is expected, reflecting this phenomenon. Then, since PMTs detect this light, if the alpha particle is moving towards the GEMs, the first electrons to reach them (and thus produce photons) are those associated with the end of the particle's path, i.e., the Bragg peak. This causes the Bragg peak to appear *early* in time in the waveform, visible as a higher amplitude region followed by a decreasing slope, as shown in the first example in Figure 5.13a. The decreasing slope reflects the decrease in the ionization cloud's density over time (of arrival). In contrast, if the alpha particle is moving towards the cathode, the electrons associated with the Bragg peak will be the last to produce light, resulting in a waveform that shows an increasing amount of light with time, with the Bragg peak appearing *later* in time, as illustrated in the second example in Figure 5.13b. The example given here specifically analyzes the Bragg peak in alpha particles, although a similar argument can be made for NRs, with the key difference being that the highest deposition density occurs at the beginning of the track instead of at the end, as discussed in Section 5.1.

While the assessment of the position of the Bragg peak in the waveforms might seem straightforward, the interplay between charge deposition asymmetry and the geometrical dependencies of the PMT signal with the distance to the light emission makes deter-mining the sign of the θ angle more complex. For this reason, a basic method such as the one developed in [107], based on the difference of signal integral between the first and the second half of the waveform, would result ineffective. To address this, the skewness of the "crown" of the waveform is analyzed. The crown refers to the high-amplitude region of the waveform, confined within its sharp rise and fall, and, by definition, within the *TOT20* region. By analyzing the skewness of the four waveforms, a final score is assigned to evaluate the likelihood that the angle sign is calculated correctly. This method has been specifically developed by the author given the peculiarity of the combination of the PMT signals in LIME. It is important to note that this represents only the first attempt to evaluate the signals' skewnesses and the actual inclination with respect to the Z-axis from these, and that alternative strategies are under study within the collaboration. Still, the developed approach proved suitable for the purpose of characterizing LIME alpha backgrounds in terms of orientation with respect to detector regions and materials, as will be illustrated in Chapter 6.

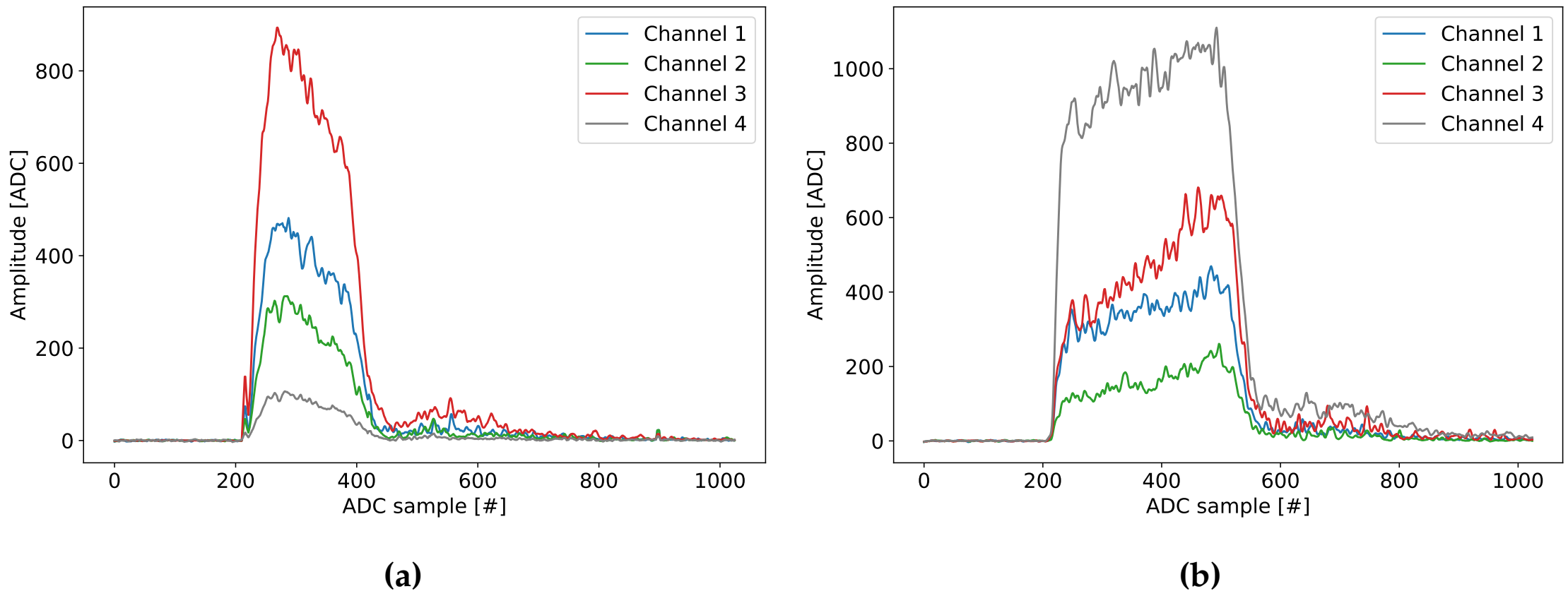

**Figure 5.13:** Example of two PMT signals from alpha events in LIME. In (a), an alpha particle is moving towards the GEMs, while in (b), the alpha is moving towards the cathode. This difference is reflected in the position of the Bragg peak within the waveforms.

To calculate the waveform skewness, the first step is to identify the highest and lowest amplitude points of the waveform crown, defined as the region enclosed between the fast rise and fall times of the waveform – i.e., the high-amplitude, step-like region of the signal. The highest point is naturally found as the highest amplitude waveform sample within the TOT20 region. Then, a second, lower point is searched for in the opposite half of the waveform (crown), relative to the maximum position. This second point is defined as the first sample where the derivative of the waveform changes, indicating the transition point from the fast rise (or fall) time of the waveform to the crown region. An example of this procedure is shown in Figure 5.14: in these waveforms, the two selected points used to calculate the crown skewness are represented as red inverted triangles. The red arrow visually illustrates the skewness, becoming more tilted as the skewness increases.

The crown / Bragg peak skewness is evaluated by calculating the ratio between the amplitudes of the two peaks identified. A sign is attributed to this ratio based on the position of the maximum amplitude: it is considered positive (+) if the maximum appears in the first half of the waveform, and negative (–) if it occurs in the second half, based on the explanation given above. After calculating the skewness ratios for all four waveforms, each ratio is normalized to the maximum among the four, and the resulting values are summed. The outcome value represents the *score* given to each set of 4 waveforms and reflects the confidence in the determination of the sign of the $\theta$ angle, as described below.

In the example shown in Figure 5.14, all four waveforms show the Bragg peak at the end of the waveform, indicating the Bragg peak electrons were the last to arrive at the amplification plane, and thus strongly hinting that the alpha particle was moving towards the cathode. The confidence of this hypothesis is calculated as described, resulting in a final score of -2.17, which signifies close to complete confidence that the particle is moving towards the cathode. Therefore, the sign attributed to the $\theta$ angle is negative (–). There are three possible score categories for the evaluation of the sign of $\theta$, which are described below and summarized in Table 5.3:

**A.** $|\mathbf{score}| > \mathbf{1}$ **:** The cases where the score's absolute value is greater than 1 occur when most of the PMT signals (at least 3 out of 4) show a clear Bragg peak in the same half of the corresponding waveform. In these events, one can be reasonably sure (close to 100%) of the alpha direction. As defined earlier, a positive score indicates the alpha is moving towards the GEMs, while a negative score indicates the alpha is moving towards the cathode. Figure 5.14 shows an example of this scenario.

**B.** $\mathbf{0.5} > |\mathbf{score}| > \mathbf{1}$ **:** In these cases, the crown topologies from the four PMTs exhibit a more complex shape, making the direction of the alpha particle less clear. However, a directional tendency is still generally present, which is why these events are still considered in subsequent analyses. An absolute score between 0.5 and 1 can occur in two situations:

- The waveforms present an opposing position of the Bragg peak due to its convolution with geometrical effects of the PMT signal. Figure 5.15a demonstrates this case, where two PMTs (channels #3 and #4) show a signal consistent with the track moving towards the GEMs, while the other two show the Bragg peak in the opposite position. The direction score for this event was -0.52, indicating lower certainty in the alpha's direction, although still suggesting a tendency toward the cathode.
- All waveforms show very flat crowns, with skewnesses close to 1.

**C.** $|\mathbf{score}| < \mathbf{0.5}$ **:** This scenario represents an extension of the previous one, where the waveforms show completely opposite results or very flat crowns, with no hint on

**Table 5.3:** Summary of the three possible cases for determining the sign of the $\theta$ angle of a given alpha event, based on a score calculated from the skewnesses of the crowns of the four waveforms corresponding to that event.

<table>
<tr><th>Scenario</th><th>Score (s)</th><th>Direction</th></tr>
<tr><td rowspan="2">A</td><td>$s < -1$</td><td rowspan="4">+ : Towards GEMs<br>- : Towards Cathode</td></tr>
<tr><td>$s > +1$</td></tr>
<tr><td rowspan="2">B</td><td>$-1.0 < s < -0.5$</td></tr>
<tr><td>$0.5 < s < 1.0$</td></tr>
<tr><td>C</td><td>$-0.5 < s < 0.5$</td><td>Ambiguous,<br>randomly assigned<br>sign (+ or -)</td></tr>
</table>

the actual direction of the alpha. An example of the latter is shown in Figure 5.15b, where the waveforms are mostly flat without a clear Bragg peak. The skewness calculation resulted in a score of +0.15. In cases where the absolute value of the score is smaller than 0.5, given the low certainty and following a more conservative approach, the confidence on the alpha's direction is assumed to be zero, and a random sign is assigned to $\theta$.

The flat-like waveforms in Figure 5.15b further illustrate why the binary (1 or -1) approach used in [107] is not ideal: dividing the waveform into two and integrating the charge in each half to determine the Bragg peak position would result in incoherent outcomes for flat waveforms, since small variations in amplitude could skew the results. This emphasizes the need for a number (*score*) that quantifies the "tilt" in the waveform's crown and the consequent confidence on the calculation of the $\theta$ sign.

In cases like these, it is important to note that although the particle's direction along Z is ambiguous, the value of the angle itself remains unaffected, as it is derived solely from geometrical arguments relating the particle's path in XY and Z, as further explained in Section 5.4. Therefore, these events do not need to be necessarily removed from the analysis dataset, as they still contain valuable information about the alphas. This motivates the assignment of a random sign for $\theta$ in these events. Later on, while analyzing the data, quality cuts are applied to deal with these cases when necessary: if the analysis focuses on the $\theta$ direction of the tracks, these events may be excluded, while if the focus is, for instance, on the 3D length distribution of alphas, these events can be kept, as the sign of the angle does not impact the traveled distance.

The score approach used for the determination of the $\theta$ sign was primarily designed

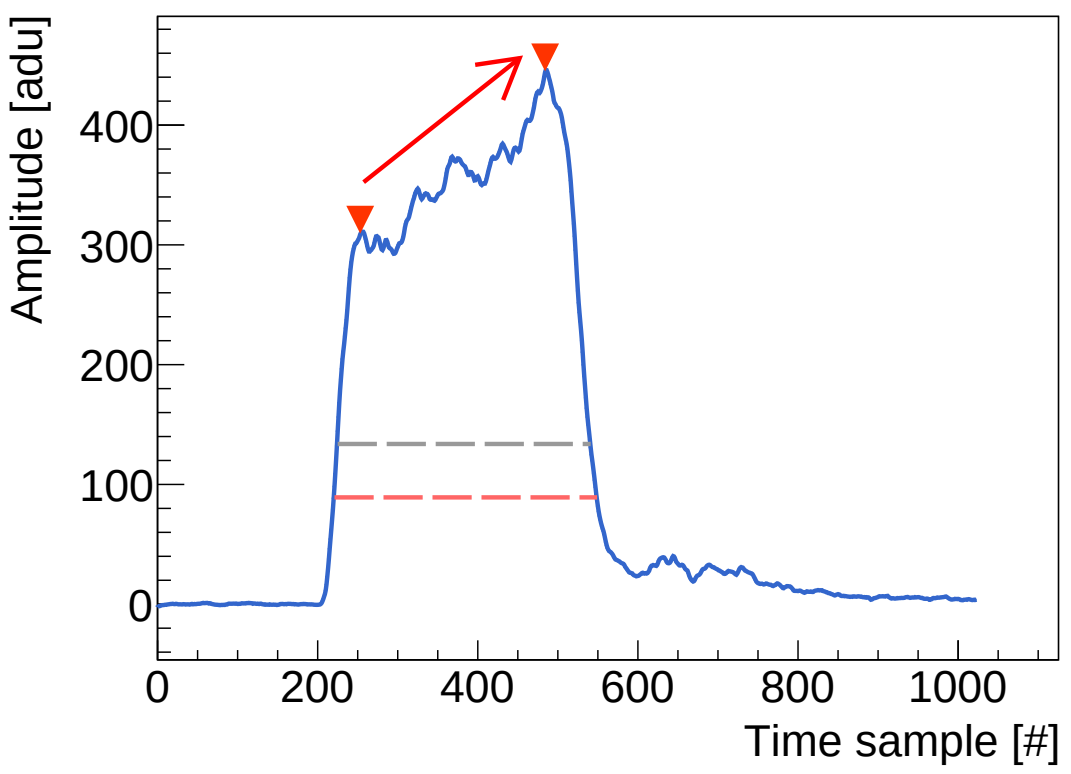

**(a)** PMT 1

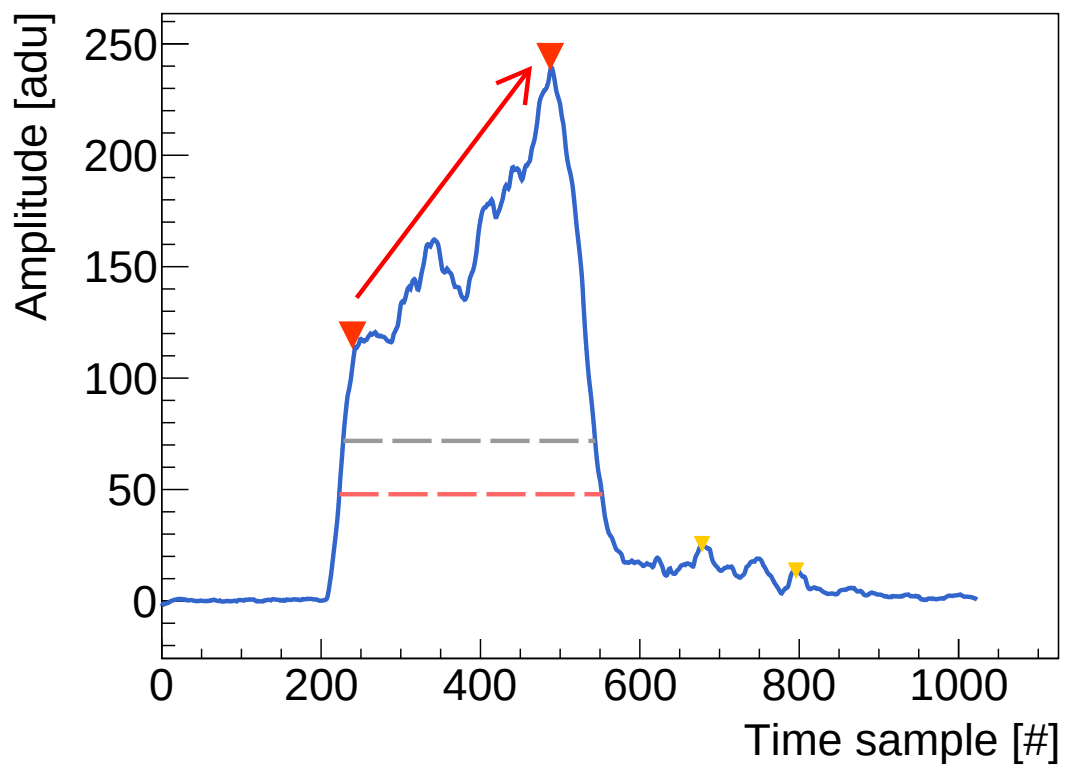

**(b)** PMT 2

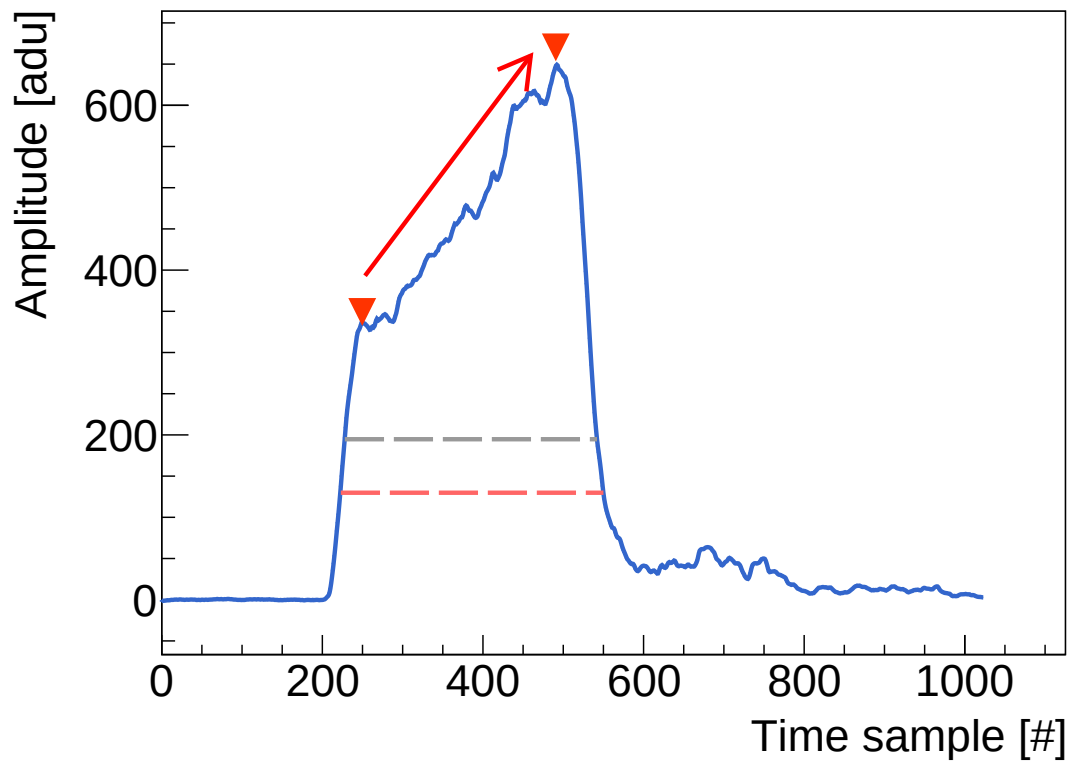

**(c)** PMT 3

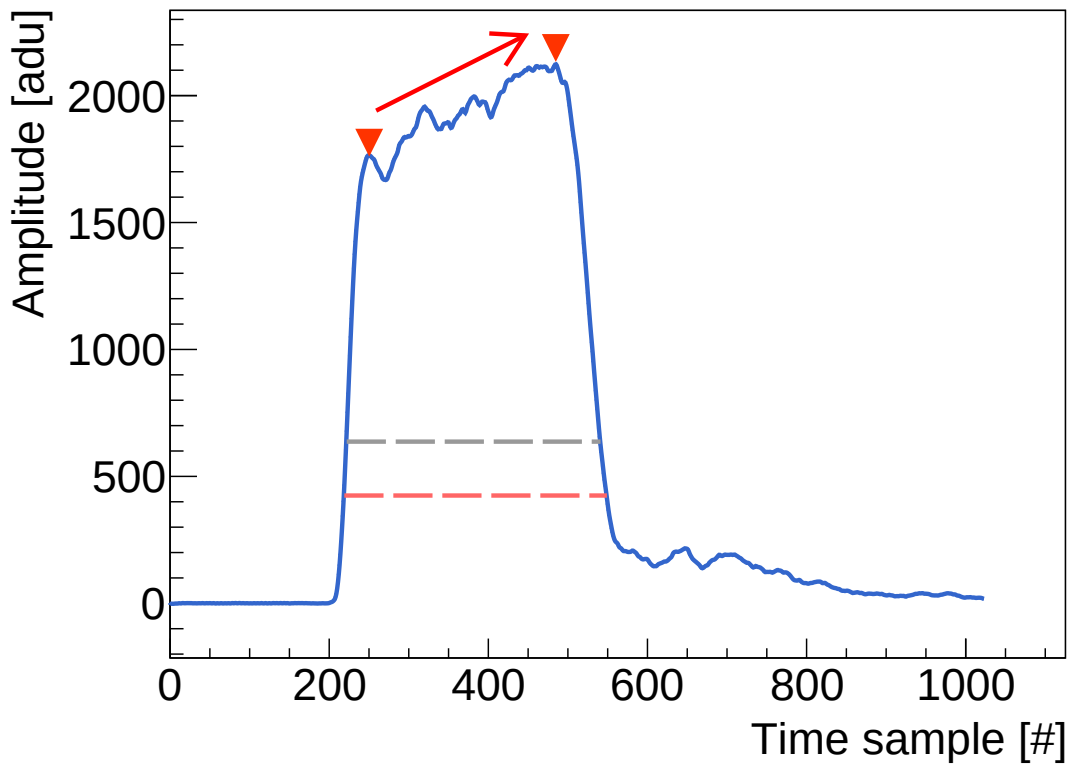

**(d)** PMT 4

**Figure 5.14:** Four waveforms of the same alpha particle event, showing the two peaks (red inverted triangles) used to calculate the Bragg peak / crown skewness. The red arrow represents the skewness of the waveform. In this example, the calculated score for the skewness is -2.17, indicating a close to full confidence that the alpha particle is moving towards the cathode. This is considered an "easy case", as all four waveforms show the Bragg peak in the same position – at the end of the waveform – illustrating that the last electrons to arrive (and produce light) are those associated with the Bragg peak.

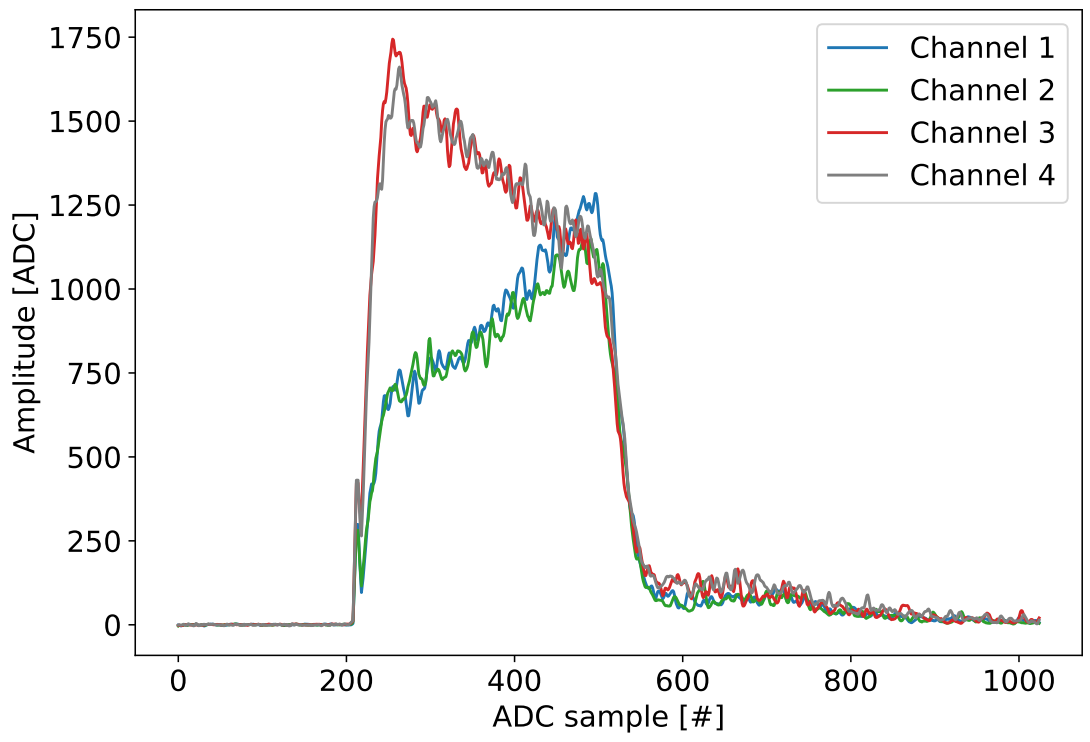

**(a)**

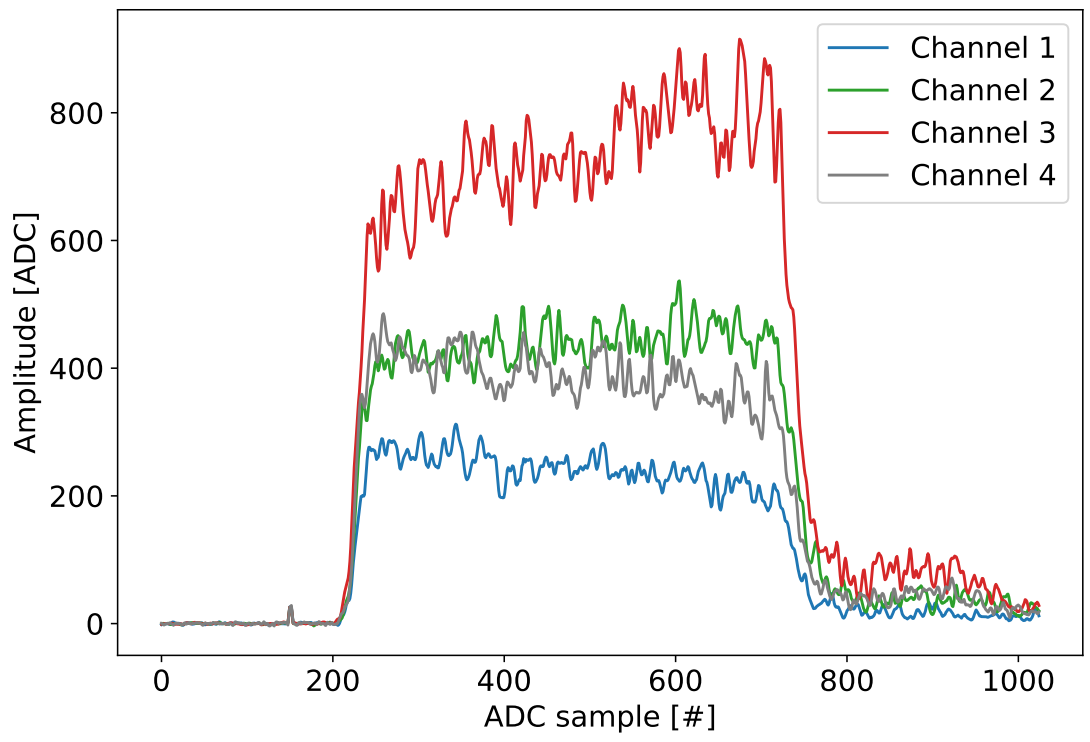

**(b)**

**Figure 5.15:** Two examples of events where the determination of the alpha direction using the skewness of the waveform crown is more complex. In (a), the Bragg peak skewness score was -0.52, indicating low confidence in the alpha's direction, although it still suggests it was moving toward the cathode. In (b), the score was +0.15, meaning no definitive conclusion about the direction of the track in the longitudinal plane can be drawn and, therefore, a random direction is assigned.

to address the geometrical dependencies of the PMT signals. In addition to these, there are other non-physical or detector-related factors that can complicate or limit the calculation of the waveform crown skewness, and thus the evaluation of the sign of $\theta$. For example, when an event occurs directly on top of a PMT (i.e., in the 2D projection, as shown in Figure 2.17), the waveform can become saturated in amplitude due to the high intensity of the signal produced. This, in turn, causes the crown skewness to become either undetectable or immeasurable. Another detector feature of LIME that affects this measurement is the GEM segmentation. The GEMs in LIME are equipped with vertical lines that electrically divide them into segments, intended to prevent the destruction of the entire GEM in the event of a strong spark. However, these electrically isolating lines create regions without amplification, resulting in artificial dips in the light profile of the alpha track, both transversely and longitudinally. For the PMT analysis, this can negatively impact the assessment of the position and skewness of the Bragg peak, as it alters the amount of charge and amplitude throughout the waveform.

Taking these factors into account, a score-based approach was found to be particularly well-suited for the 3D analysis presented in this thesis, as it provides a confidence level for assessing the sign of $\theta$. In general, a conservative strategy was adopted for calculating these scores, as the primary goal of these studies was to achieve a precise selection of alpha particles rather than a highly efficient one. Nonetheless, scenarios A and B in Table 5.3 – where the alpha direction is either very clear or shows a consistent directional tendency – account for approximately 75% of all identified alphas. This score can also be used as a tool for evaluating the performance of the 3D analysis under different detector operating conditions, with various datasets, or after introducing or updating new routines in the algorithm. Moreover, this approach represents the first strategy developed within CYGNO for head-tail determination of ionization tracks, and more advanced methods that exploit additional information and variables could be considered by the collaboration in the future.

## 5.3 PMT signal to CMOS track matching

At this stage of the algorithm, for a given event (composed of one CMOS image and one or more PMT triggers, each with 4 waveforms), two populations of selected alpha particles exist: one from the CMOS sensor and one from the PMTs. To complete the 3D description of the alpha ionization track, including its initial direction and sense, it is necessary to merge these two populations. As discussed in Section 2.2.2.2, the number of alphas detected by the CMOS sensor may not always match the number detected by the PMTs due to differences in how the camera sensor operates. A single CMOS image may also contain multiple alphas, making it essential to correctly associate the CMOS and PMT information referring to the same alpha. It is also possible that an alpha particle failed to pass the PID checks (false negative), or that spurious events incorrectly passed the selection criteria (false positive), either in one of the sensors or simultaneously in both. In all these cases, the association between PMT triggers and CMOS clusters is needed and is carried out using a routine based on the BAT-fit method discussed in Section 3.3.1.

In the initial BAT-fit studies (Section 3.3.2), spot-like interactions originating from a $^{55}$Fe source were used. In these studies, the charge of each waveform was integrated and input into the BAT-fit, which then determined the event's $(L, x, y)$ position. By comparing the event's position in the CMOS image with that seen by the PMTs (as determined by the BAT-fit), the different pixel clusters in the image could be correctly associated with the corresponding group of PMT waveforms [221]. The accuracy of this procedure is shown in Figure 3.17.

The interactions caused by the 5.9 $keV$ photons emitted by the $^{55}$Fe source are often referred to as spot-like events, with lengths below 5 $mm$, and are thus generally well represented by a single $(x, y)$ position (see Figure 5.2). In contrast, alpha particles are extended in space, traveling up to several centimeters within the gas, so the distance between the emitted light and each PMT varies along the ionization track. As a result, alphas cannot be represented by a single $(x, y)$ position, unlike spot-like events. To address this, a new approach – *waveform slice-and-fit* – was implemented by the author for the scope of this thesis to more accurately represent the position of ionization tracks induced by alphas in the XY plane as observed by the PMTs, allowing for a proper match with the CMOS pixel cluster counterpart.

Each waveform is divided into five equal temporal slices within the TOT20 region, as illustrated in Figure 5.16a. The integrated charge in each slice is calculated and then input into the BAT-fit. The BAT-fit treats each slice as a spot-like interaction (similar to an $^{55}$Fe-induced event) and retrieves the $(L, x, y)$ information for each slice. The fitted $(x, y)$ positions are then plotted on a 2D plane, representing the position of the alpha ionization track as seen by the PMTs, as shown in Figure 5.16b. This procedure is applied to each quartet of waveforms (triggers) identified as alphas in a given event.

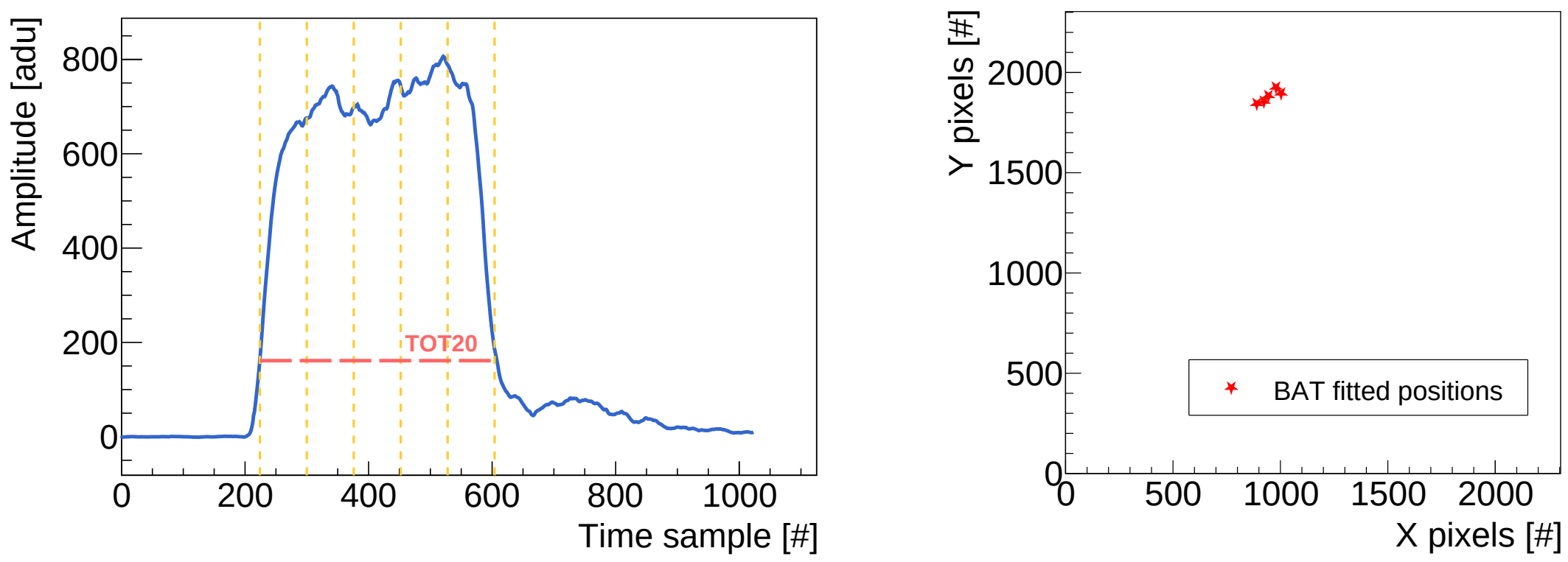

(a)

(b)

**Figure 5.16:** Example of the *waveform slice-and-fit* routine: in (a), the waveform is divided into 5 equal slices within the defined TOT20 region. This process is applied to the remaining 3 waveforms as well. In (b), the fitted positions (red stars) of each slice are shown in the CMOS field of view.

In a given event, for all possible combinations of alpha track matches between the CMOS and PMT data, the distances between the cluster positions and the trigger-fitted positions are calculated. To do this, the alpha track in the CMOS is subdivided into 5 points (corresponding to the number of time slices). These 5 points are equally spaced in 2D between the "edges" of the ionization track, i.e., the edgemost pixels along the track's major axis, as defined in Section 5.2.2.

The $(x, y)$ points from both sensors are then ordered by X, and the point-by-point distance is computed through the following expression:

$$d_i = \sqrt{(x_{PMT_i} - x_{CAM_i})^2 + (y_{PMT_i} - y_{CAM_i})^2} \quad , i \in \{0..4\} \tag{5.5}$$

The final distance value (D), used to characterize the distance between each combination of alphas in the CMOS and PMT datasets, is defined as the average of the point-to-point distances calculated. Mathematically, this is expressed as:

$$D = \sum_{i=0}^{N=4} d_i/N \tag{5.6}$$

The matching between camera clusters and PMT triggers is finalized by associating the pairs with the smallest distance (D, as defined in Equation 5.6) between them.

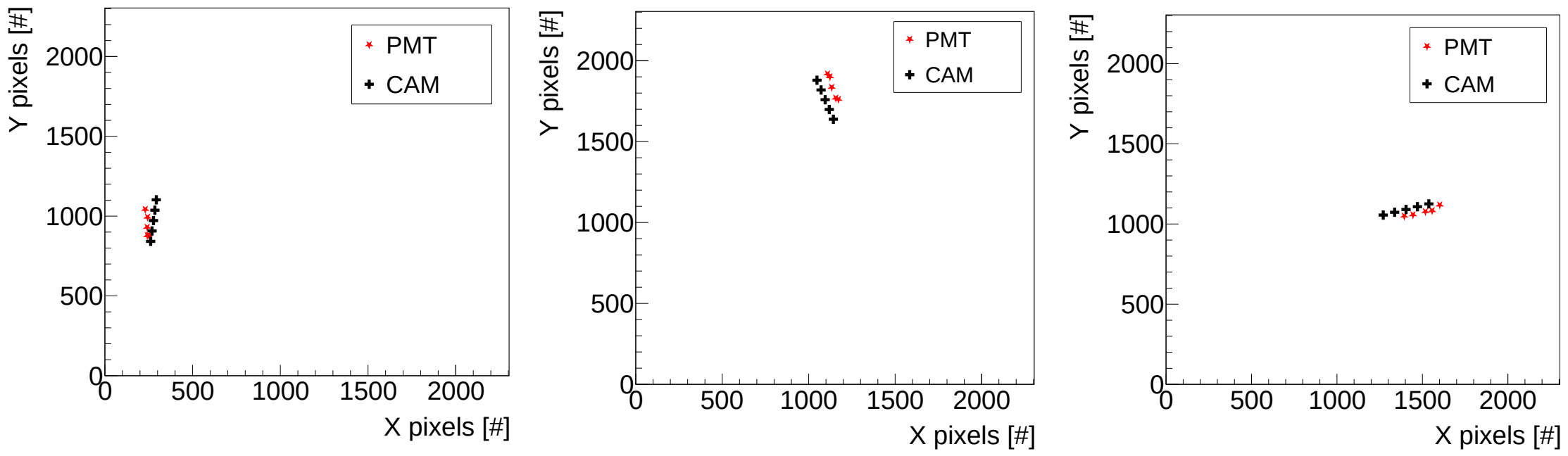

**(a)** **(b)** **(c)**

**Figure 5.17:** Three examples of the PMT-CAM matching using the BAT-fit, demonstrating the effectiveness of this approach across different track positions and orientations. The track position as seen by the PMTs through the BAT-fit is shown in red stars, while the position as seen by the CMOS is shown in black crosses. The axis represent the CMOS image.

### 5.3.1 Alphas matching accuracy

To illustrate the accuracy of the *slice-and-fit* approach, three examples of events containing a single alpha are shown in Figure 5.17. In these examples, the CMOS ("CAM") reconstructed alpha track position in the CMOS XY plane is plotted alongside the PMT-fitted track. As shown, the proximity of the two tracks is evident in different regions of the image, with some points even overlapping.

A second example is shown in Figure 5.18, where two alpha tracks were identified in the same event in both CMOS and PMT data. The original image is shown in Figure 5.18a, while Figure 5.18b illustrates the positions of these alphas as fitted by the PMTs using the procedure described above, alongside the positions of these alphas as seen by the CMOS. While the positions are not necessarily overlapping, it is clear which signals (represented by the red stars for PMT and black crosses for CMOS) should be associated with each other.

The point-to-point distances measured for all the *matched* alpha particles in the datasets analyzed are shown in Figure 5.19. These distances are subdivided into their X and Y components, rather than presenting the D variable, to allow for a more detailed analysis of the results.

The distribution of these distances, when averaged across both dimensions, has a standard deviation of 123.85 pixels, which corresponds to 1.92 cm. These distributions are not symmetrical, so a Gaussian fit was not attempted. Given the basic method used to determine the distances between the two views of the same event (point-by-point distance), this result is expected. The slicing of the waveform can introduce systematic effects, particularly in the first and last slices, where the waveform is less uniform. While more advanced methods or new variables could be introduced to better describe the distance between the "two alphas", increasing the complexity of the procedure would be time-consuming and may not be necessary. The resolution of a few centimeters

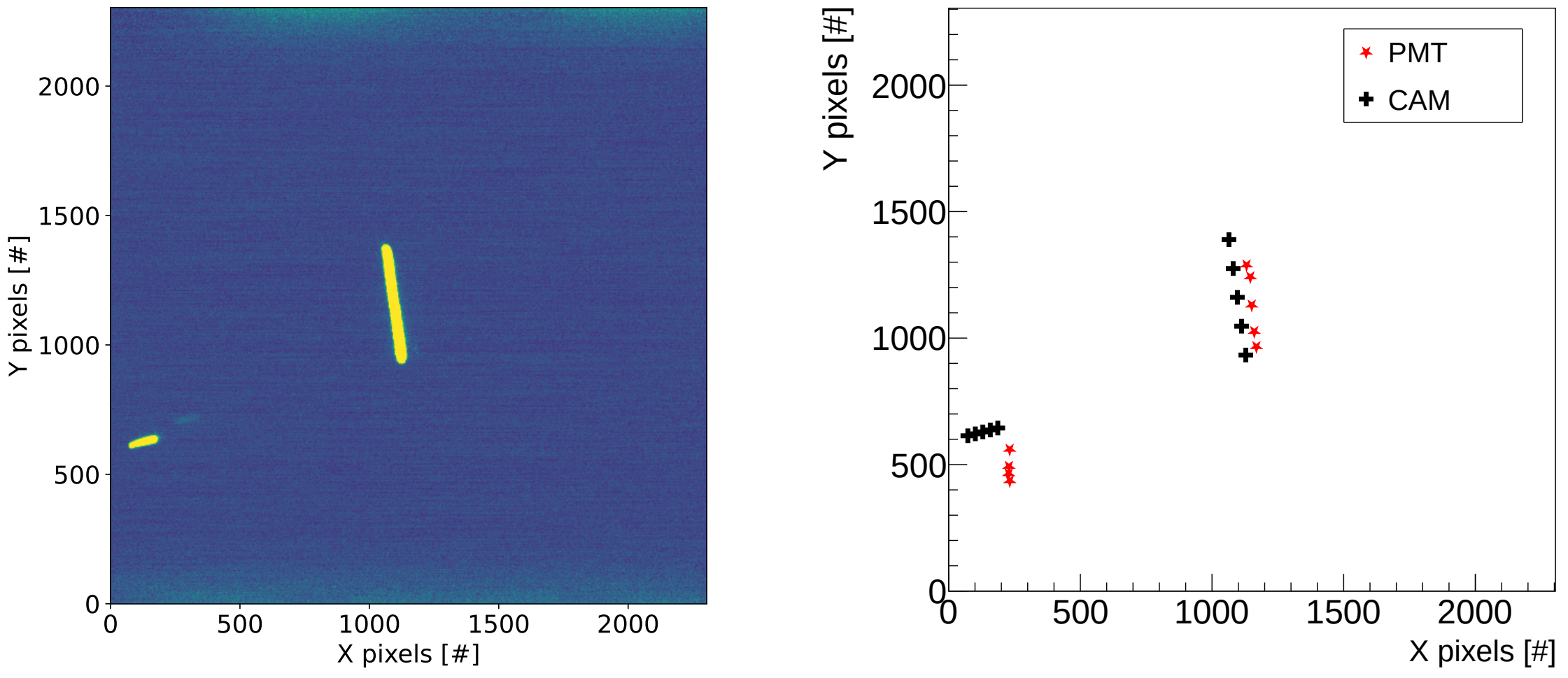

**(a)** **(b)**

**Figure 5.18:** Example of an event with two alpha particles as seen (a) in the original CMOS picture. In (b), the positions of the identified alphas are shown, as reconstructed with the BAT-fit (red stars) and seen by the CMOS sensor (black crosses), in the CMOS XY plane.

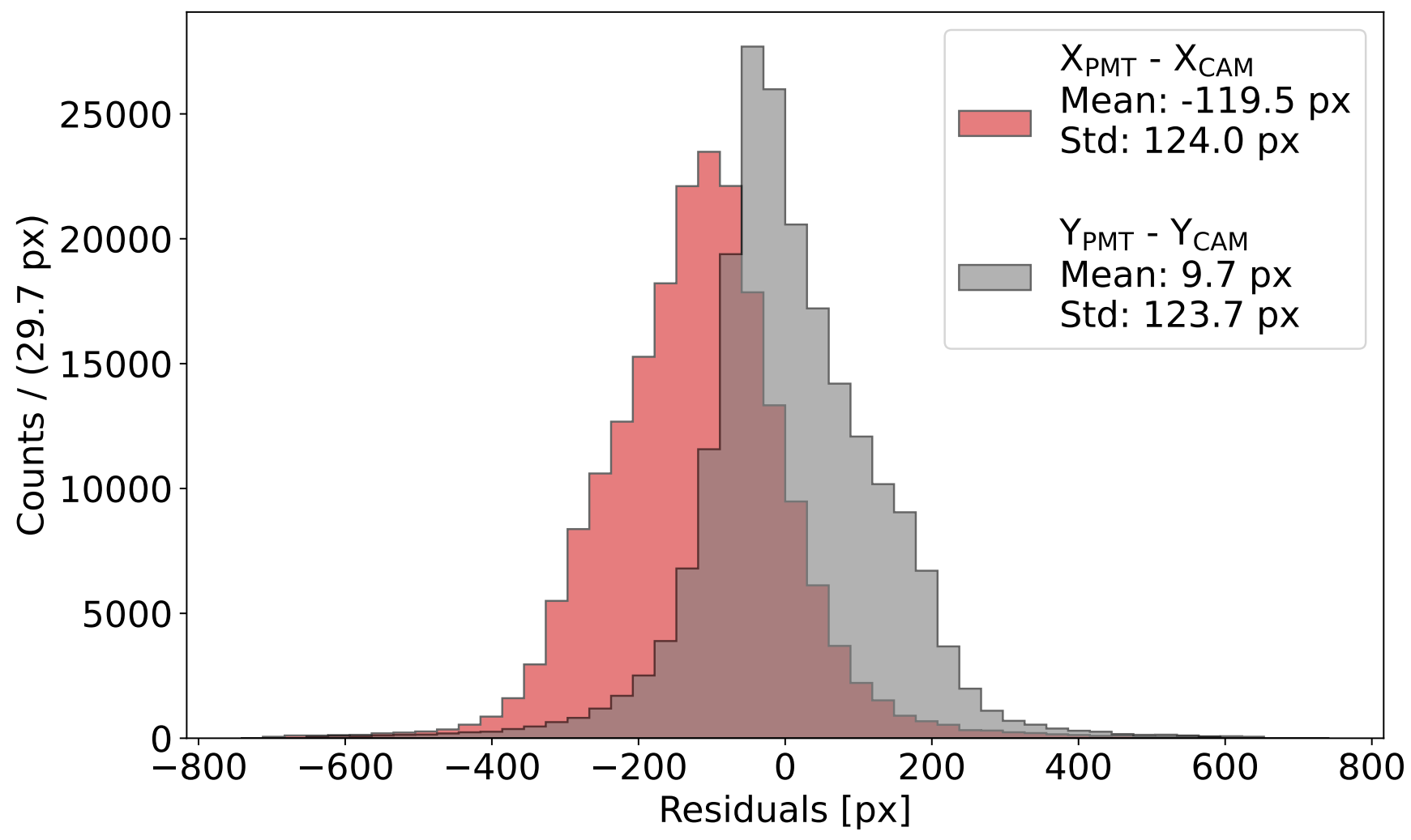

**Figure 5.19:** Distribution of the residuals $\Delta x$ and $\Delta y$ between the PMT-based and camera-based track reconstruction methods for extended events. The tracks in the camera images are resampled to match the number of points in the corresponding PMT waveform, and a point-by-point distance is computed. Found also in [221].

obtained is sufficient to distinguish between two events in the same image, which spans an area of 36 $\times$ 36 $cm^2$ in LIME. For the datasets analyzed (LIME underground), the total particle event rate was externally measured to be around 1 - 2 Hz, further reducing the likelihood of overlapping or excessive alphas in a single image. Finally, if two alpha events are located less than 2 cm apart, it is likely that the initial CMOS reconstruction itself becomes the limiting factor, by wrongfully merging two clusters of pixels from different events.

Another noticeable feature in Figure 5.19 is the offset in the $(x, y)$ coordinates of $-119.5$ and 9.7 pixels, corresponding to $-1.85$ cm and 0.15 cm, respectively. While the collaboration is still investigating the origin of this offset, it must be noted that the accuracy of this procedure matters only for the matching between PMT and CMOS information, and that this does not effectively affect the overall 3D tracking performance, since this is based on the information retrieved individually within each sensor's data. A possible explanation for the difference between the X and Y offsets could arise from the distortion and light reduction induced by the lens properties, as discussed in Section 2.1.3.3. Since the noise band cut on the top and bottom of the image, described in Section 5.2.1.1, effectively removes extreme Y values, the smaller Y offset could stem from this effect. Overall, the average standard deviation of 1.92 cm found for alpha particles using this process is quite satisfactory, considering the method used to calculate the distances between CMOS and PMT events. These findings demonstrate that, while the BAT-fit was originally optimized for $^{55}Fe$ signals, its integration with the slice-and-fit approach for extended tracks is not only feasible but also performs exceptionally well, despite the differences in size and charge density between $^{55}Fe$ and alpha signals.

## 5.4 Integrated CMOS-PMT 3D reconstruction

At this stage of the algorithm, all relevant variables calculated from both the CMOS and PMT analyses have been evaluated. From the CMOS side, this includes the XY length and $\phi$ angle, determined using the directionality algorithm, and the absolute Z position of the event, obtained through a fit to diffusion. From the PMT side, the traveled Z (or $\Delta Z$) has been calculated, as well as the track direction in the longitudinal plane, i.e., the sign of the $\theta$ angle. Once the corresponding information from the two sensors has been matched, it becomes possible to determine the actual value of the $\theta$ angle, correlating the XY and Z extensions of the track, following basic geometric arguments [265]:

$$\theta = \arctan\left(\frac{\Delta Z}{\Delta XY}\right) \tag{5.7}$$

Finally, the final 3D length of the ionization track is obtained by summing in quadrature each of its XYZ components. A summary of these variables and how they are calculated is shown in Table 5.4. It is important to note that other variables of interest, such as the cluster's integrated pixel counts and the waveform's number of peaks, are also saved from the 3D analysis. However, the variables mentioned here are the most relevant for the 3D and directional analysis of alpha particles.

To provide a comprehensive overview of the 3D analysis algorithm, a final example is presented in Figure 5.20, illustrating the complete process behind the reconstruction, selection, and analysis of alpha particles from LIME underground data. Moving from the top panels downward, Figure 5.20.(a) starts by showing the original CMOS image with two alpha particles, designated $\alpha$ #1 (in the upper half) and $\alpha$ #2 (in the lower half), as outlined in Table 5.4. The impact point and initial direction angle of these alphas are shown in the other two plots, determined using the *directionality algorithm*. In the next step, Figure 5.20.(b) presents the transverse light profile of the alphas, from which the absolute Z position is determined via a fit to diffusion, using the RMS of the distribution as described in Equation 5.4.

Following this, the PMT analysis in Figure 5.20.(c) and Figure 5.20.(d) identifies key features in the waveforms, such as the TOT20 and TOT30 variables, and the crown skewness. These features are used to determine the extension in Z traveled by the alphas and their direction in the longitudinal plane. The TOT20 variable is also used to subdivide the waveform into time slices, which are then integrated and converted into a charge signal to be fed into the BAT-fit (*slice-and-fit*). This fit retrieves the $(x, y)$ positions of the tracks as seen through the PMTs, which, as shown in Figure 5.20.(e), is then used to perform the *one-to-one matching* of the two sensors' information, leading to the final 3D reconstruction of the event. The final 3D variables calculated in this process for the two alpha particles – which include the 3D direction vector, sense, and absolute Z position of the track – are summarized in Table 5.4.

For the 3D *representation* of the alpha events in Figure 5.20.(e), a simplification is made: since it is not possible to track each individual electron and the corresponding photon created during amplification, random points are sampled in 3D space around the 3D direction vector of the alpha to represent the electron cloud and its characteristic ionization profile, including the Bragg peak at the end. These random points follow a Gaussian distribution with a width ($\sigma$) equal to the RMS of the alpha's light profile. This representation serves only to improve the visualization of the alpha track in 3D and does not influence the final analysis or results. The electron cloud is color-coded to indicate the direction of the alpha particle: darker blue symbolizes the start of the track, while brighter yellow represents the higher charge deposition near the Bragg peak at the end of the track.

## 5.5 Final remarks

The 3D analysis developed in this thesis represents a significant advancement in the tracking capabilities of CYGNO. By combining the high resolution of the CMOS and the fast response of the PMTs, it becomes possible to reconstruct the full spatial profile and direction of ionization tracks in CYGNO detectors. Such detailed information enables the identification of the origin of the event and enhances the particle identification techniques of CYGNO. The 3D algorithm builds upon the expertise and analysis routines developed by CYGNO collaborators for CMOS data and integrates them with the newly developed PMT analysis. Within this framework, existing routines were optimized and new variables were introduced to specifically address the analysis of alpha particles. This included:

1. Identifying alpha tracks truncated by the CMOS sensor, which can distort their reconstructed 3D length;

2. Adapting the directionality algorithm to accommodate the denser and longer ionization tracks of alphas compared to ERs, to finally determine the track's $\phi$ angle;

3. Determining the absolute Z position of alphas using the RMS of the distribution of their transverse light profile;

4. Developing new PMT variables to discriminate alphas from ERs down to $\mathcal{O}(\text{keV})$, establishing an innovative method for particle identification (PID) using the PMT signal;

5. Creating a novel head-tail determination strategy for PMT waveforms capable of handling the combined analysis of the four signals returned by the LIME prototype. With appropriate modifications to account for the inverted Bragg peak profile, this method could be optimized in the future for NRs;

6. Modifying the BAT-fit procedure to handle extended tracks, allowing the matching between PMT and CMOS data, and laying the groundwork for future adaptations to ER and NR extended tracks.

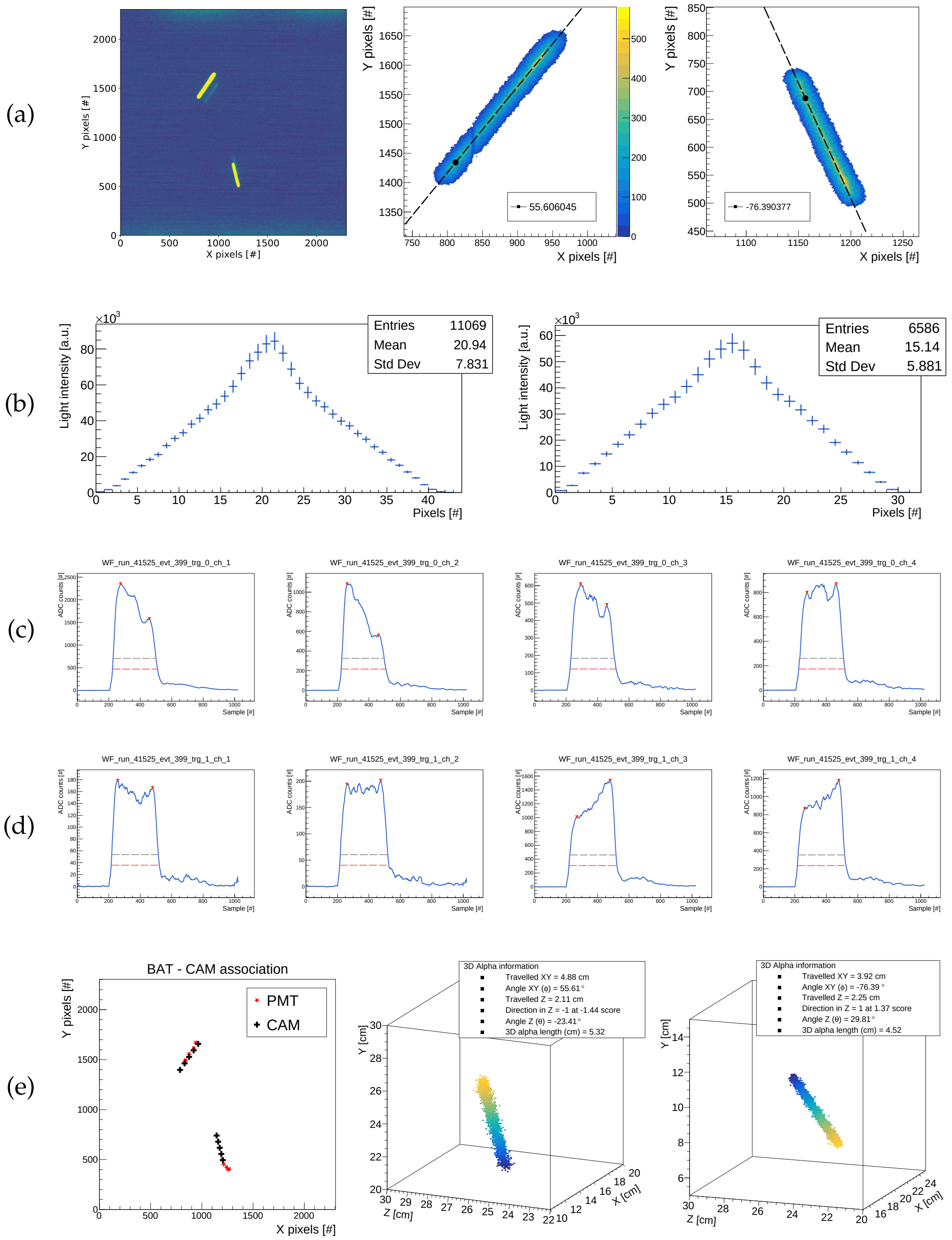

**Figure 5.20:** Complete example of the 3D analysis pipeline for alpha particles: (a) Original CMOS image and determination of the track's XY length and $\phi$ angle through the directionality algorithm; (b) Transverse light profile of the alpha tracks used to determine the absolute Z position of the event via a fit to diffusion; (c and d) Waveforms corresponding to each of the 2 triggers identified as alpha tracks, with relevant variables extracted; (e) One-to-one waveform-cluster association and final 3D representation of the two alpha events.

**Table 5.4:** Summary of the main calculated variables of alpha tracks using the 3D analysis.

| **Definition** | **Operation** | **$\alpha$ #1** | **$\alpha$ #2** |
|---|---|---|---|
| XY length [cm] | $\sqrt{(X_{end} - X_{start})^2 + (Y_{end} - Y_{start})^2}$ | 4.879 | 3.925 |
| Traveled Z [cm] | $1/4 \cdot \sum \Delta Z_i\ ,\ \text{with } i \in \{\text{PMT \#1 .. PMT \#4}\}$ | 2.112 | 2.249 |
| Absolute Z [cm] | *from* Fit to diffusion | 49.98 | 1.41 |
| $\phi$ or XY angle [°] | *from* Directionality algorithm | 55.61 | -76.39 |
| $\theta$ or Z angle [°] | $\tan^{-1}\left(\frac{\text{Traveled Z}}{\text{XY length}}\right) * \text{direction in Z}$ | 23.41 | -29.809 |
| Direction in Z | **(+) :** towards GEMs; **(–) :** towards cathode | + | - |
| Direction score | *from* Bragg-peak skewness | 1.44 | -1.37 |
| 3D length [cm] | $\sqrt{(\text{XY length})^2 + (\text{Traveled Z})^2}$ | 5.316 | 4.523 |

To optimize the various parameters involved in this analysis, the ideal approach would be to use simulated signals rather than real data. Simulations enable detailed fine-tuning while eliminating non-physical effects present in real data. However, at the time of this work, a simulation of PMT signals in CYGNO detectors was still at a very preliminary stage which did not allow its use. Once fully developed, such a simulation will improve the understanding of the interplay between Bragg peak asymmetry and geometrical dependencies in the waveforms, potentially allowing these effects to be disentangled. It will also support better tuning of the selection cuts for alpha particles on the PMT side.

Following this, all the choices of parameters and cuts employed in the 3D analysis were focused on maximizing accuracy in the selection of alphas, rather than prioritizing efficiency. By using a clean sample of alphas, it was possible to properly tune the parameters and evaluate the capabilities of the analysis. This revealed limitations in detection efficiency, such as the difficulty in identifying and analyzing alphas near certain angles ($\theta \sim 0°/ 90°$), leading to the subsequent decision to cut these events from the analysis. In a future upgrade of this analysis, such events could be categorized for CMOS-only or PMT-only analysis, which, although lacking data from the counterpart sensor, could still provide meaningful information about the ionizing events. That being said, this first version of the 3D analysis marks a milestone in CYGNO's 3D tracking capabilities, with the innovative results from this analysis presented in the following chapter.

Chapter 6

# LIME alpha background characterization

During the analysis of Runs 1, 2, and 3 of the LIME underground campaign, a notable discrepancy was observed between the measured and simulated event rates, raising questions about the accuracy of the initial background model. This issue is examined in detail in F. di Giambattista's thesis [196], where several potential sources of the discrepancy are discussed, including detector efficiency, unaccounted background contributions, and simulation parameters.

In these studies, preliminary results pointed to an unaccounted-for background component, likely due to the presence of $^{222}\mathrm{Rn}$ in the gas. The $^{222}\mathrm{Rn}$ decay chain comprises in fact both beta decays, typically accompanied by gamma emission, which end up populating the low energy (i.e., $< 100$ keV) region of the background spectrum, and $\mathcal{O}(\mathrm{MeV})$ alpha decays. These alphas, due to charge saturation effects (see Section 6.1.2) are measured with energies as low as a few hundred keV. In the same preliminary studies [196], the 2D lengths of these high-energy alphas were measured, revealing three possible length peaks (with limited resolution), which were associated with the expected presence of $^{222}\mathrm{Rn}$ and its three dominant alpha emissions at 5.49, 6.00, and 7.69 MeV. As also suggested by di Giambattista, incorporating PMT waveform analysis and adding the third dimension to the total measured length would significantly improve the reconstruction of the true range (and thus the energy) of high-energy alphas. This would allow for the confirmation (or exclusion) of the presence of $^{222}\mathrm{Rn}$ in LIME.

The introduction of PMT analysis, combined with the matching of CMOS information, allowed the first-ever 3D reconstruction of ionization tracks from alpha particles, as described in Chapter 5. The discrepancy between data and Monte Carlo in previous studies emerged as an opportunity to test the 3D reconstruction on real data and to validate the hypothesis and characterize the feature of $^{222}\mathrm{Rn}$ presence within LIME. With this in mind, a thorough study of the LIME alpha background is conducted using the newly developed 3D measured length variable, which is only accessible through the inclusion of PMT information in the overall analysis framework. In these studies, alongside the 3D measured length, other variables – such as the emission angles ($\phi$ and $\theta$) – are also considered to identify the detector components responsible for each alpha emission. These studies aim to provide a better understanding of the discrepancy between the Monte Carlo expectations and the acquired data, as initially reported, and a full characterization of the alpha background in LIME.

In this chapter, a brief summary of the findings presented in [196] is provided to offer context for the new results (Section 6.1). Then, the dataset used for these studies is introduced (Section 6.2), followed by the first-ever measurements of the 3D alpha track lengths in LIME (Section 6.3). Based on these results, and using geometric selection criteria, the contributions of different detector materials to the overall alpha background spectrum are identified (Section 6.4). Finally, a more in-depth study of the alphas from the $^{222}\mathrm{Rn}$ decay chain is presented, confirming the initial expectations regarding the presence of this gas in LIME (Section 6.5).

## 6.1 LIME background studies

As illustrated in Section 2.2.2.4, one of the main objective of LIME underground operation was to provide a first experimental assessment of internal and external backgrounds and to validate their MC simulation. Towards this goal, a comprehensive study was carried out as part of F. Di Giambattista PhD thesis [196]. Here the detector response in terms of CMOS images was applied to a GEANT4 simulation of the main backgrounds components in the different shielding configuration, and the resulting energy spectra were compared to the actual Run1, Run2 and Run3 data. A summary of these studies and a brief re-discussion of the results is provided here to offer context for the new findings presented in this thesis.

### 6.1.1 Monte-Carlo simulation

For the purpose of simulating the LIME underground setup backgrounds, the layout of the detector was implemented starting from its CAD design and the shielding configurations were added to this geometry as a series of concentric boxes. The radioactivity of the materials of the main components used for the construction of LIME (namely, the GEMs, the field cage rings and resistors, the cathode, the PMMA vessel, the CMOS camera and the lens) was measured underground by the LNGS Special Techniques Division from samples each of them with an High Purity Germanium detectors (HPGe) as reported in Table 6.1. The main contributors resulted to be radioisotopes from the natural decay chains of $^{232}\mathrm{Th}$, $^{232}\mathrm{U}$ and $^{235}\mathrm{U}$. For each considered component of the detector, radioactive decays of the identified contaminant isotopes were simulated with GEANT4, distributed homogeneously in the corresponding volume. LIME detector internal components inducing the largest contributions to the total background are the copper field rings and cathode, the resistors and the GEMs, which account for almost 95% of the total internal radioactivity-induced background. The camera body and the camera lens resulted to be subdominant with respect to these and correspond to only 1.9% of the Run3 expected background. It is important to note that LIME was constructed using standard materials, not specifically optimized for underground operation or low radioactivity. The background arising from external ambient gamma and neutron fluxes were simulated with an isotropic distribution from a large (i.e. 3 m) sphere with the LIME setup positioned in the center, starting from the relative spectra already measured at underground LNGS by other experiments [266–268].

**Table 6.1:** Measured radioactivity of the materials composing LIME and used for the LIME background simulation. The isotopes not reported belonging to the $^{238}$U, $^{232}$Th and $^{235}$U chains were assumed to be in equilibrium with their parent isotope. The effective mass of each material in LIME is also reported in the last row. Table adapted from [196].

| Isotope | Material Activity (mBq/kg) | | | | | |
|---|---|---|---|---|---|---|
| | Fields rings / Cathode | Resistors | GEMs | PMMA | Camera | Lens |
| $^{238}$U($^{234}$Th) | $<$ 210 | $(20 \pm 2) \cdot 10^3$ | 163 $\pm$ 65 | $<$ 3.5 | 3290 $\pm$ 940 | 4200 $\pm$ 1400 |
| $^{238}$U($^{226}$Ra) | $<$ 1.3 | 2160 $\pm$ 160 | 32 $\pm$ 16 | $<$ 3.5 | 850 $\pm$ 50 | 1920 $\pm$ 90 |
| $^{210}$Pb | $<$ 13 | $(592 \pm 70) \cdot 10^3$ | 32 $\pm$ 16 | $<$ 3.5 | 850 $\pm$ 50 | 1920 $\pm$ 90 |
| $^{232}$Th($^{228}$Ra) | $<$ 1.1 | 3500 $\pm$ 320 | $<$ 30.9 | $<$ 5 | 990 $\pm$ 90 | 360 $\pm$ 40 |
| $^{232}$Th($^{228}$Th) | $<$ 1.3 | 3360 $\pm$ 320 | $<$ 15.6 | $<$ 4.5 | 990 $\pm$ 50 | 365 $\pm$ 28 |
| $^{235}$U | $<$ 1.6 | 330 $\pm$ 60 | $<$ 15.8 | - | 190 $\pm$ 50 | 145 $\pm$ 37 |
| $^{40}$K | $<$ 6 | $<$ 1780 | $<$ 358 | $<$ 35 | 890 $\pm$ 140 | $(53 \pm 5) \cdot 10^3$ |
| $^{137}$Cs | $<$ 0.47 | $<$ 74 | $<$ 8.13 | - | 42 $\pm$ 14 | $<$ 27 |
| $^{60}$Co | $<$ 0.57 | $<$ 7.7 | $<$ 7.48 | - | $<$ 5.6 | $<$ 46 |
| **Material quantity (kg)** | 11.8 / 1.6 | 0.054 | 0.039 | 27.6 | 1.7 | 0.19 |

The result of the internal and external background simulation for the different LIME shielding configurations are summarized in Table 6.2. These results indicate that in the unshielded configuration (*Run1*), the external background dominates, leading to an event rate of approximately 30 Hz, integrated over energies above 1 keV. With the introduction of 4 cm copper shielding (*Run2*), the external background from gamma and beta sources begins to decrease, reaching a level comparable to the internal component at 10 cm of copper (*Run3*). The external contribution for the nuclear recoil component is significantly reduced only after the addition of a 40 cm water layer around LIME (*Run4*), as expected, since hydrogen in water is highly effective at moderating and capturing neutrons. In this maximal shielding configuration, the expected external background drops to approximately $5 \times 10^5$ ERs and 2 NR events per year, with the internal background becoming dominant.

### 6.1.2 Data/MC comparison

Figure 6.1 shows the measured and simulated energy spectra up to 100 keV for Runs 1-3, including both ER and NR events. The data/MC consistency of Run1 over its en-tire spectrum demonstrate that LIME response to the external gamma background is properly simulated. Conversely, the observed increasing discrepancies with increasing shielding thickness pin-points to an internal origin of the missing background components, which become relatively more significant with the reduction of the external component. This missing component results furthermore consistent between Run2 and Run3 despite nearly an order of magnitude difference of the overall rate, further validating the hypothesis of a common internal unforeseen radioactive contamination. An additional confirmation of this picture comes from the comparison of the simulated and measured spectra during a AmBe data taking campaign performed during Run3 and shown in Figure 6.2. In the AmBe data, we expect the AmBe gamma + neutrons to dominate over all the backgrounds below 20 keV, becoming comparable only beyond 50 keV. This can be observed even more clearly in Figure 6.2b, which separates the simulated background and AmBe-induced events. The AmBe data/MC consistency below 20 keV therefore further demonstrates the correct relative data/MC normalization and energy calibration, and confirms the presence of a non-simulated background component at higher energies.

**Table 6.2:** Summary of the simulated background, divided into internal and external contributions, for both ER and NR events. The values are provided for the different shielding configurations used in LIME. The final column ("Total") includes also the background contribution coming from the introduced shielding. This table has been adapted from [196].

| **Run #** | **Shielding configuration** | **External** | | **Internal** | | **Total** | |
|---|---|---|---|---|---|---|---|
| | | **ER/yr** | **NR/yr** | **ER/yr** | **NR/yr** | **ER/yr** | **NR/yr** |
| Run 1 | – | $1.14 \times 10^9$ | 1480 | $7.34 \times 10^6$ | $7.9 \times 10^4$ | $1.14 \times 10^9$ | $8.05 \times 10^4$ |
| Run 2 | 4 cm Cu | $2.66 \times 10^7$ | 870 | $7.87 \times 10^6$ | $7.9 \times 10^4$ | $3.45 \times 10^7$ | $7.99 \times 10^4$ |
| Run 3 | 10 cm Cu | $1.49 \times 10^6$ | 930 | $7.88 \times 10^6$ | $7.9 \times 10^4$ | $9.37 \times 10^6$ | $7.99 \times 10^4$ |
| Run 4 | 10 cm Cu + 40 cm $H_2O$ | $0.5 \times 10^6$ | 2 | $7.88 \times 10^6$ | $7.9 \times 10^4$ | $8.38 \times 10^6$ | $7.90 \times 10^4$ |

A data/MC comparison beyond 100 keV up to 1 MeV pointed to the presence of alpha particles in the high end of the energy spectrum, which were originally discarded in the MC simulation for computing reasons and since they do not directly represent background to DM searches at low energies. While alpha particles are expected in the $4-9$ MeV range, charge saturation effects in LIME may cause them to appear across the entire energy spectrum, even as low as hundreds of keV [196]. Charge saturation refers to a phenomenon in particle detectors where, when operating at very high gains, due to a high accumulation of charges in the amplification region (GEM holes in the case of CYGNO), space-charge buildup occurs, leading to an effective reduction in charge gain and, thus, a non-uniform energy response by the detector [269,270]. This phenomenon, particularly impactful for the large and dense energy deposits of alphas, is typical of any detector designed to detect low-energy ionization events, and even more in optical

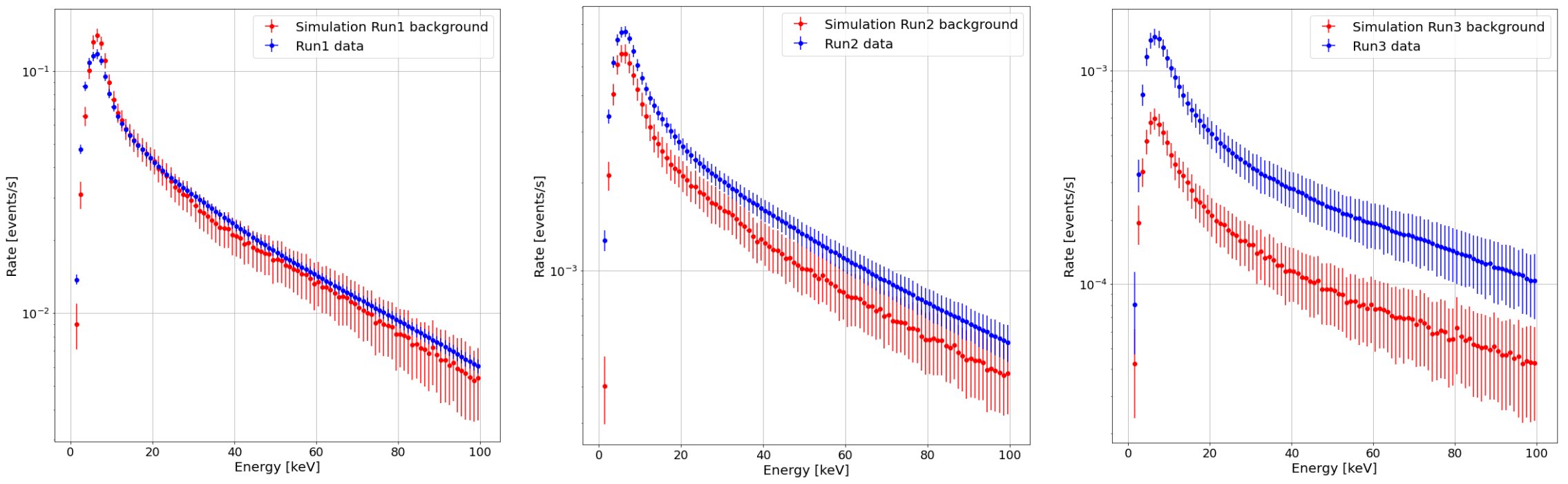

**(a)** Run 1. **(b)** Run 2. **(c)** Run 3.

**Figure 6.1:** Comparison between simulated (red) and measured (blue) energy spectra for different energies in Runs 1, 2 and 3 in LIME, up to 100 keV, with 1 keV bins. Retrieved from [196].

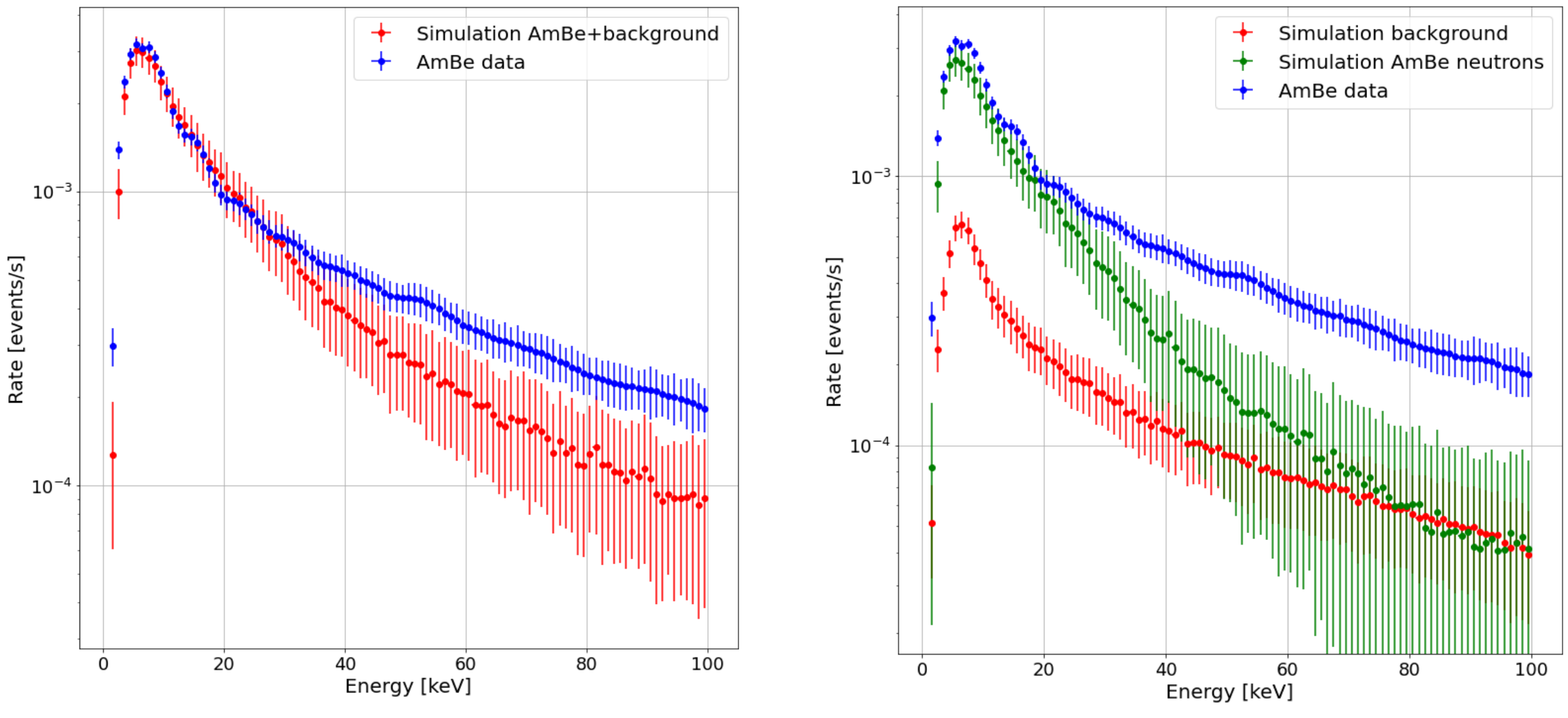

**(a)** **(b)**

**Figure 6.2:** Comparison between simulated (red) and measured (blue) energy spectra in AmBe run in LIME with AmBe and background components (a) combined, and (b) subdivided, with 1 keV bins. Retrieved from [196].

readout TPCs as CYGNO due to the very small solid angle covered by the lens + CMOS optical system.

### 6.1.3 Preliminary investigations of Data/MC discrepancy

To investigate the origin of the discrepancy between the Monte Carlo simulation and the data, alpha particles – selected based on their high energy deposition density, similarly to what was done in Section 5.2.1 – were studied as the potential source of the unaccounted-for component at high energies. Since saturation is expected to significantly affect the energy measurement of these events, track length was chosen as a more reliable metric for analysis. The track length is less influenced by saturation effects, making it preferable over energy for this study. However, the measurement of track length was limited by the use of CMOS-only data since the full 3D track reconstruction presented in this thesis was still under development. As one dimension (Z) is missing, the length distribution for a given energy shows a long tail, which is proportional to the inclination of the alpha track relative to the GEM plane.

These two effects are illustrated in Figure 6.3. On the left, the energy distribution of alpha-like events is shown, revealing a wide range of energies below 1500 keV, indicating significant charge saturation in these signals. The energy peaks are also visually misaligned which is due to the variation of light yield performances across the data-taking periods, induced by interventions on the gas system, as will be further elaborated below. On the right, the measured 2D track length of these events shows three potential peaks, which are generally aligned, as track length is not expect to significantly depend on saturation and light yield. As mentioned, these distributions exhibit large tails toward shorter track lengths, reflecting the track tilt, which is not considered in this analysis since it uses only CMOS (2D) information.

In these preliminary studies, the 2D lengths of the alpha particles revealed three potential length peaks, although with limited resolution. These 2D measured tracks length appeared consistent with SRIM simulations of the expected ranges in the CYGNO gas mixture for the three dominant alphas emission from the $^{222}\mathrm{Rn}$ decay chain. $^{222}\mathrm{Rn}$ is produced in the $^{238}\mathrm{U}$ decay chain and is a well-known contaminant in underground laboratories. At LNGS, $^{222}\mathrm{Rn}$ has been measured to reach $\mathcal{O}(10^2)$ $\mathrm{Bq/m^3}$ [271]. Due to the gas recirculation system in LIME, $^{222}\mathrm{Rn}$ can enter the gas volume through multiple pathways, including leaks in the piping, diffusion through plastic barriers, and traces of atmospheric air that infiltrate the gas system [88]. As will be further discussed, $^{222}\mathrm{Rn}$ progeny can also electrostatically attach to internal materials in the detector, leading to the long-term decay of the $^{218}\mathrm{Po}$ and $^{214}\mathrm{Po}$ daughters from both the gas bulk and the cathode and GEM surfaces.

The presence of $^{222}\mathrm{Rn}$ in the gas would further support the conclusion that its decay products populate the entire energy spectrum. This is due to the additional beta decays in the $^{222}\mathrm{Rn}$ decay chain, at energies of 1.2 MeV and 3.3 MeV, which generate high-energy electrons and gamma rays. These particles are unlikely to be fully contained

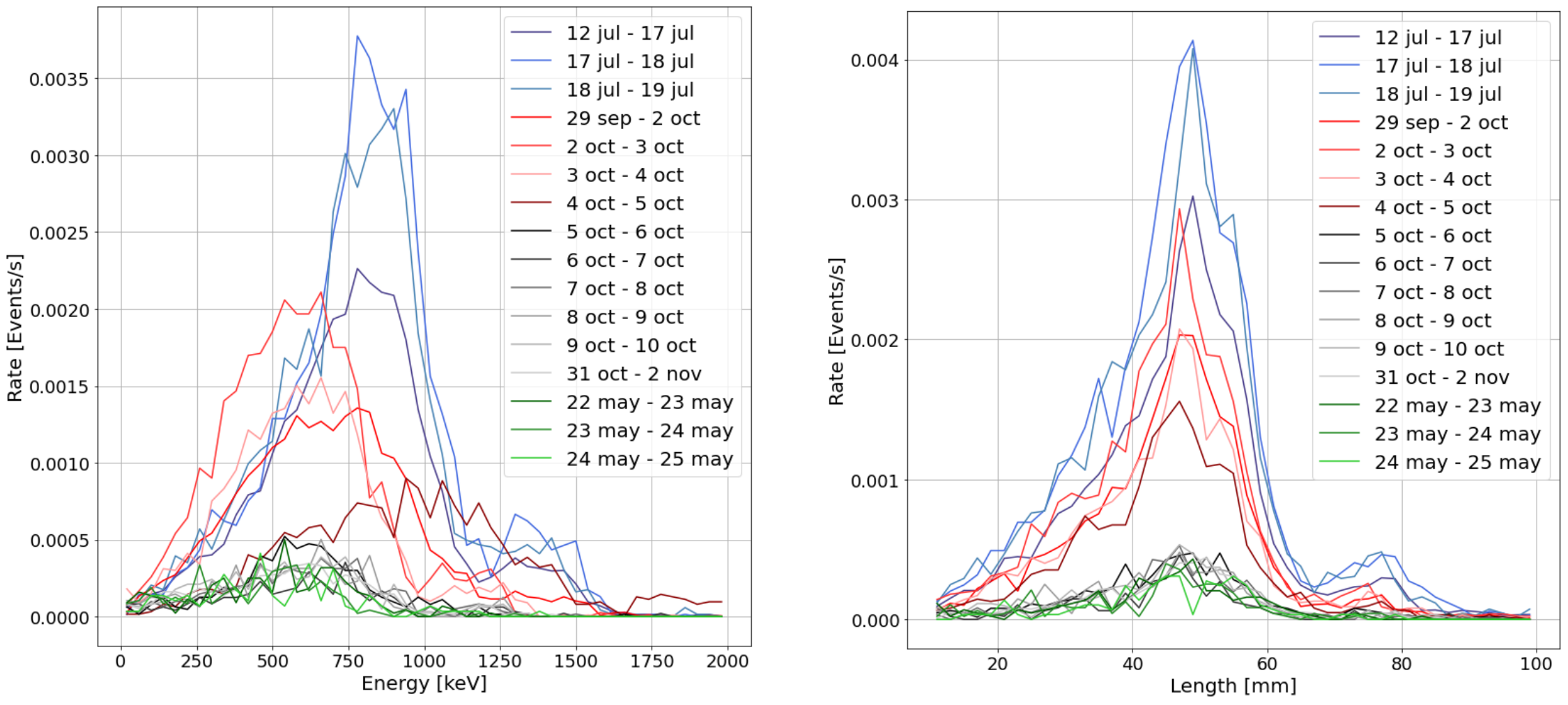

**(a)** **(b)**

**Figure 6.3:** Spectrum of alpha particles's (a) energy and (b) 2D lengths in Run 3. The different colors correspond to different data-taking periods.

within the LIME detector. As expected, beta particles and Compton-scattered ERs contribute to a broad range of visible energies, which explains the distribution of events throughout the lower region of the energy spectrum.

Further studies were conducted to explore the relationship between the observed rate of alpha particles and the operation of the LIME gas system. In these, several preexisting different type gas filters were tested as preliminary attempts to improve the gas quality, with the new and final ones being introduced only during Run 4. Figure 6.4 presents the rate of alpha events as a function of time (i.e. run number) in Run 3. In this plot, the size of the points corresponds to the light yield measured during calibration runs, to further highlight the overall effect of interventions of the gas system in the detector response. Five distinct regions can be identified:

- **Green points:** These represent the initial conditions, with no recirculation system, no filters, and a high flux of new gas. During this phase, the gas was thoroughly flushed in and out, ensuring its purity but leading to considerable gas waste. The light yield (LY) was high, and the alpha rate was low, indicating optimal gas conditions.

- **Red points:** In this region, a recirculation system was introduced to minimize the waste of new gas. An older preexisting gas filter was also added to remove impurities. Although the recirculation system recycled 80% of the gas through filters to maintain its quality, an increase in the alpha rate was observed. This suggests that the used filter had likely become saturated and already contaminated when it was implemented in the gas line.

- **Orange points:** After replacing the old filter with a different one, a decrease in the

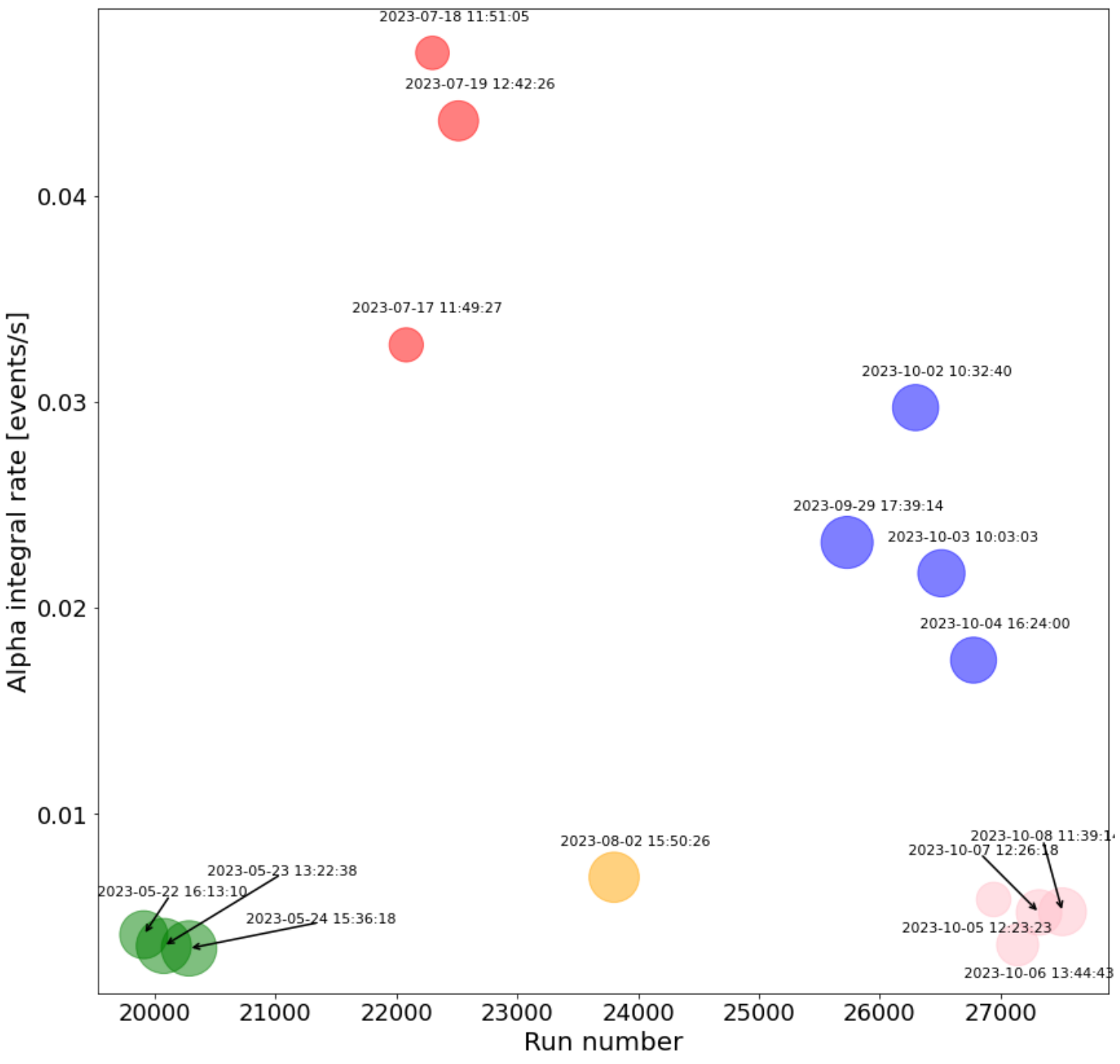

**Figure 6.4:** Measured rate of alpha particles in LIME during Run 3 The size of the points is proportional to the light yield (LY) measured from the $^{55}Fe$ peak in calibration runs. Different colors represent periods between modifications of the gas system: (green) high gas flux without recirculation nor filters; (red) recirculation system with the old, used filter; (orange) different filter replacing the old one; (blue) after a mechanical intervention of the gas system; (pink) after a further change of the previous gas filter.

alpha rate and a slight increase in the LY were observed, indicating improved gas quality and system performance.

- **Blue points:** A system malfunction required intervention in the gas system, leading to a slight increase in the alpha rate. This suggests a temporary issue with gas purity or flow.

- **Pink points:** Finally, in October 2023, the previous gas filter was changed again, resulting in a further reduction in the alpha rate while maintaining a generally high LY.

This information suggests that operations within the gas system are often accompanied by gas contamination, leading to both a decrease in light yield and an increase in alpha particle events. This effect is especially pronounced when the impurities are water-based, as humidity has been linked to a reduction in light yield in He:$CF_4$. Additionally, the presence of water can facilitate the attachment of radon progeny to water vapor molecules, further increasing the radon content in the system [194]. Furthermore, any radioactive contamination naturally present in the filters can be introduced in the gas active volume by such operations, further increasing the possibility of presence of unforeseen background component.

Overall, the studies conducted in di Giambattista's thesis [196] demonstrate a clear connection between the rate of alpha particles and the light yield with the "quality" of the gas. This quality is influenced by factors such as the filters in use, the flux of new gas being introduced, and any recent mechanical interventions. As a result of these findings, CYGNO has become more precise and attentive in maintaining and regulating the gas system, ultimately leading to the introduction of two filters – one for oxygen (red) and one for moisture (blue) – at the beginning of Run 4, as discussed in Section 2.2.2.4.2. These filters have proven to maintain stable light yield and alpha rate conditions over an extended period. For this reason, much of the analysis presented in this thesis focuses on data from Run 4, as it represents the period with the most stable gas conditions, allowing for a more accurate study of the origin of the alpha particles and potentially confirm the expectations for the discrepancies between Monte Carlo simulations and data.

## 6.2 Datasets

The datasets used for this analysis are from Run 4 (see Section 2.2.2.4.3), which aligns with the CYGNO timeline for studying the internal background by progressively increasing the detector shielding. Run 4 was selected due to its high statistics in stable gas conditions and low background resulting from the increased shielding, providing a very clean dataset for a detailed analysis of alpha particles.

From Run 4, the data selected corresponds to periods of data-taking that included long stretches of background data and daily calibrations, which were used to monitor the quality of the gas. The gain, drift field, and other operating parameters remained consistent across all data-taking, allowing for direct comparison and analysis of the results. As also mentioned in the introduction, following the conclusions regarding the detector's light yield conditions and the presence of impurities in the gas, an important turning point during Run 4 occurred with the introduction of the moisture filter (referred to as the blue filter, see Section 2.2.2.4.2) in the gas line to mitigate these issues and ensure more stable operation of LIME. This date in time (end of December, 2024) is used to separate the data into two groups, A and B, as shown in Table 6.3.

The last column of this table shows the alpha rate per run in these batches. The cuts applied to Run4 data to select the alpha sample are the same as those to developed the strategy to fully reconstruct alpha tracks in 3D as reported in Section 5.2. As highlighted in the table, the alpha rate decreased by a factor of approximately 5 from batch A to batch B. Since the detector operating parameters did not change between batches A and B except for the introduction of the filter, and given the observation of $^{222}$Rn contamination associated to the presence of water [194], it appears reasonable to assume that this difference it is mainly due to the reduction of the $^{222}$Rn component when the filter was in place. For this reason, batch B is first analyzed in Sections 6.3 and 6.4 to study alphas emission from all the expected radioisotopes present in LIME, while batch A is then used to better characterize only the presence of $^{222}$Rn and its progeny, in Section 6.5.

**Table 6.3:** Summary of the runs used for the analysis of the alpha background presented in this chapter. Each batch corresponds to an uninterrupted dataset, with the datasets selected to span different time periods to eliminate time-dependent systematics. The alpha rate is highlighted to illustrate the difference before and after the introduction of the moisture (blue) filter (see Section 2.2.2.4.2).

| Group | Dates | N runs considered | **Alpha rate** |
|---|---|---|---|
| A | 04/12/2023 - 14/12/2023 | 1820 | **29.01 ± 0.13** |
| B | 24/01/2023 - 05/03/2024 | 5676 | **5.82 ± 0.03** |

## 6.3 Alphas measured 3D length

As already preliminary demonstrated in F. Di Giambattista thesis exploiting only 2D sCMOS images [196], the length of alpha tracks results more effective in identifying alphas energy than their actual measured light yield, since the first is not significantly affected by charge saturation effects. Other experiments, such as DRIFT, also support and utilize this approach [92]. The introduction of the full 3D tracking illustrated in Chapter 5 can make this strategy significantly more powerful, since it removes the uncertainty on the track inclination with respect to the amplification plane and, thanks to this, provides a more precise estimation of the actual traveled path, as it will be illustrated in this Chapter. Additionally, the possibility of inferring the track absolute Z position and the angle relative to the Z axis (i.e. $\theta$) allows to associate each alpha component to a specific region of the detector and therefore to further characterize the different LIME internal background contributions, as will be elaborated in Section 6.4.

The measurement of the 3D lengths of alpha particles in the dataset of group B, as listed in Table 6.3, is shown in Figure 6.5. The distribution was fitted using six Gaussian curves, spanning from 3 to 9 cm. Similar to the approach described in Section 4.2.4, the fitting procedure was performed using MINUIT [248] within the ROOT data analysis framework [217], with all Gaussian parameters left free during the fit. The sum of the six Gaussians is represented by the thick black line, while the individual components are shown as differently colored dashed lines with the fitted parameters for each of them being reported in the legend. This fitting procedure was also applied to the other distributions presented in the rest of the chapter. As discussed in Section 6.2, the results presented here are only for group B in the datasets considered in order to highlight the various alpha components observed in the LIME alpha background spectrum.

To illustrate the effectiveness of the 3D reconstruction, the same distribution is built using the measured *2D length* instead, based on the CMOS-only reconstruction described in Section 3.1. The result is shown in Figure 6.6. As seen, the 2D lengths display a pronounced shoulder in the region below 4 cm, reflecting the absence of the $\Delta Z$ component from all alphas with longer 3D lengths. This occurs because the CMOS provides only a 2D projection of the ionization track, causing alpha particles from the same energy (thus length) population to appear with varying lengths and smearing any distinct peaks in the spectrum. While two peaks at approximately 4.4 cm and 5.1 cm are still visible, as in Figure 6.5, these are the only ones from which meaningful information could potentially be retrieved. The difference between the 2D and 3D distributions underscores the significant improvement in measuring ionization track lengths through the integration of PMT signals for the Z dimension, alongside with CMOS data for the XY plane.

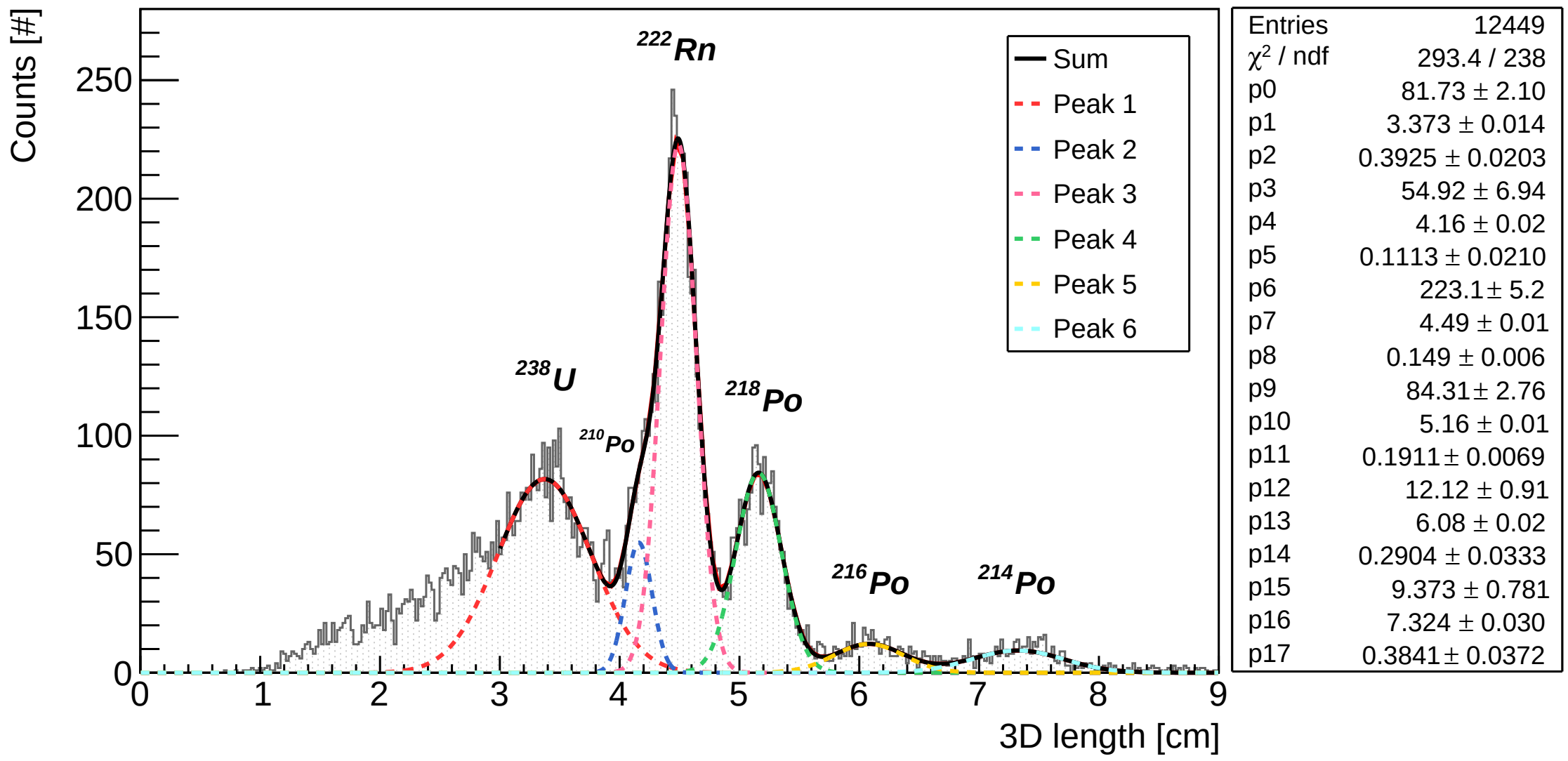

**Figure 6.5:** Distribution of 3D lengths of alphas measured in LIME, dataset B (see Table 6.3). The distribution was fitted with the sum of 6 Gaussian curves associated to each of the expected peaks. Each Gaussian fit is also individually plotted for better visualization. The fit parameters of each Gaussian is shown on the side. The isotopes associated with each peak represent the *candidate* decays discussed in the upcoming Section 6.3.2.

### 6.3.1 From measured 3D track length to particle range

In the context of particles traversing a medium and the studies presented here, three terms are commonly used: the *traveled path*, which refers to the total distance a particle covers along its actual trajectory, which can be curved or scattered due to interactions with the medium; the *range*, which is the straight-line distance between the particle's point of origin and the point where it comes to rest; and the *measured 3D length*, which is the track extension in 3D as returned from the reconstruction strategy illustrated in Chapter 5. For alpha particles, in a medium such as that in CYGNO detectors, the path traveled and range can be considered nearly identical due to the high dE/dx of these particles, which results in mostly straight ionization tracks. Conversely, the measured 3D length typically results greater than the actual range, since the original primary ionization cloud topology is degraded by diffusion during drift and amplification in the GEM and further modified by the detector response.

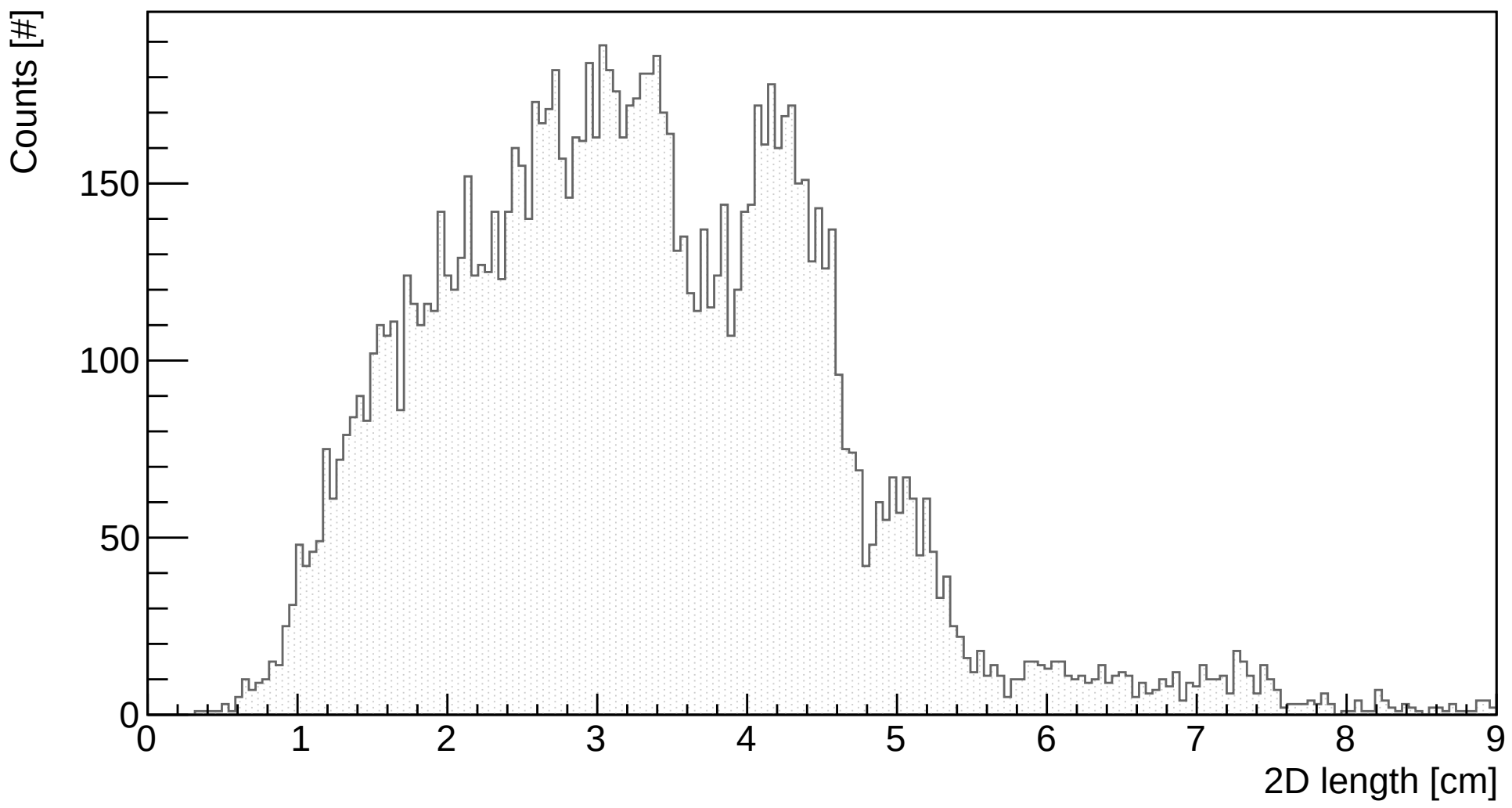

**Figure 6.6:** Distribution of *2D* lengths of alphas measured in LIME, dataset B (see Table 6.3).

The deconvolution of detector effect from the measured track length in order to obtain the actual traveled path (or range) of the original primary ionization cloud is a well know issue in imaging gaseous TPC, even the ones featuring high resolution pixels 3D readouts [272]. While the CYGNO collaboration is currently actively working on this topic testing various techniques including Machine Learning methods, no standard recipe has been established yet. For this thesis, two simplistic methods have been therefore developed to apply a correction to the length obtained from the CMOS images and PMT analyzes. These are then applied combined to the 3D length distributions rather than on an event-by-event and/or sensor-by-sensor basis given the primitiveness of the approach, and a significant systematic is assigned to take this into account. It is important to stress how this represents only a first attempt to a proper track length estimation, but that it still results well-suited to the scope of this work.

#### 6.3.1.1 CMOS XY length correction estimation

Given the extremely high CPU usage and time-computation required for the full simulation of CMOS images of alphas tracks in the standard CYGNO frameworks that includes all detector effects [196], a more basic strategy was chosen for this evaluation. This involved a Toy-MC simulation which included the major contribution of the detector effect in this context, represented by diffusion during drift and in the amplification process. In this Toy MC, the isotropic emission of 1000 alphas with 5.49 $\mathrm{MeV}$ energy in the center of LIME (i.e. $Z = 25$ cm) was simulated with GEANT4. The position of each primary electron of the ionization cloud was then smeared according to a Gaussian distribution with $\sigma = \sqrt{\sigma_0^2 + \sigma_T^2 \cdot Z}$ with $\sigma_0$ and $\sigma_T$ as reported in Section 5.2.3. The 2D length is then measured as the largest distance by any two smeared electrons to mimic the strategy illustrated in Section 5.2.2. The difference between this and the MC-truth 2D length from GEANT4 is then evaluated and the resulting distribution is shown in Figure 6.7. From this, the CMOS XY length correction is taken as $7.05 \pm 0.06$ mm. The error is calculated as $\mathtt{std}/\sqrt{N}$, with $\mathtt{std}$ being the standard deviation of the distribution and N being the number of simulated events.

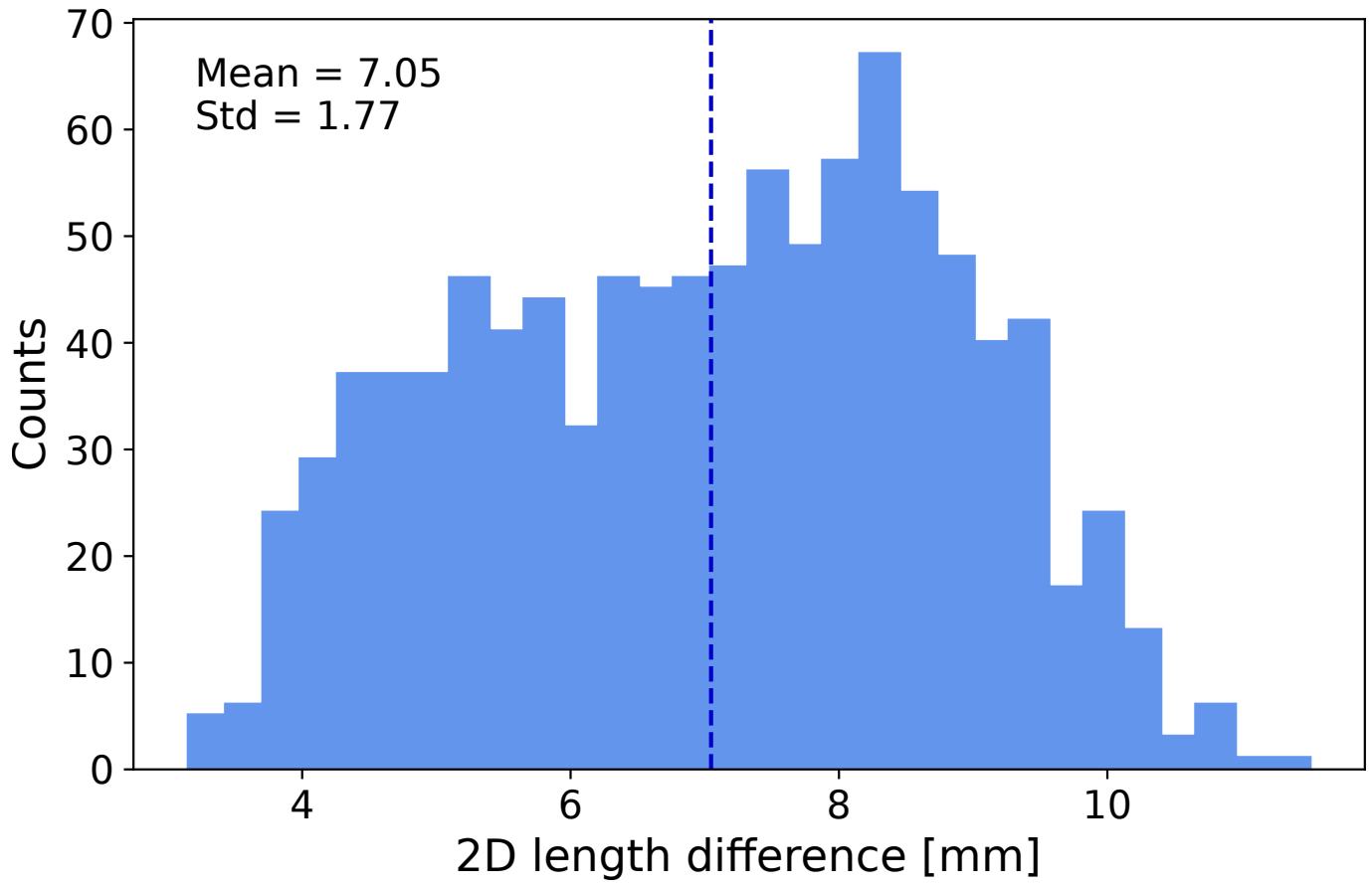

**Figure 6.7:** Distribution of the difference between Gaussian smeared tracks emulating the detector effects and MC-truth 2D lengths, on a 1000 alpha sample.

This result provides a first-order estimate of the detector effects when comparing the range and measured length of a alpha particles in LIME. The tests were also performed (with limited statistics) for alpha particles with different energies, showing consistent results. To improve the evaluation of this measurement systematic on the 3D length of alpha particles, a much more computationally efficient simulation is being developed and, at the time of writing of this thesis, close to finalized by the CYGNO group. This will allow the generation of a much larger alpha dataset, creating the opportunity to systematically study this effect.

#### 6.3.1.2 PMT Z length correction estimation

To assess the contribution of the PMT signal to the total 3D measured length, a more empirical approach was adopted due to the absence of a fully developed simulation of PMT signals in CYGNO detectors (which is currently under development).

The accuracy in the measurement of the coordinate in the drift direction in the experimental TPC approach is well known to possess an intrinsic limitation due to the diffusion during drift and amplification of the charges [273]. This means that even if an alpha particle (or other type of particle) were to cross the detector perfectly parallel to the GEM plane – resulting in a virtually zero extension in the Z direction, i.e., a delta function in terms of PMT signal – the PMT waveform would still appear as the Gaussian distribution of the diffused charges with $\sigma = \sqrt{\sigma_0^2 + \sigma_L^2 \cdot Z}$, convoluted with the intrinsic PMT response.

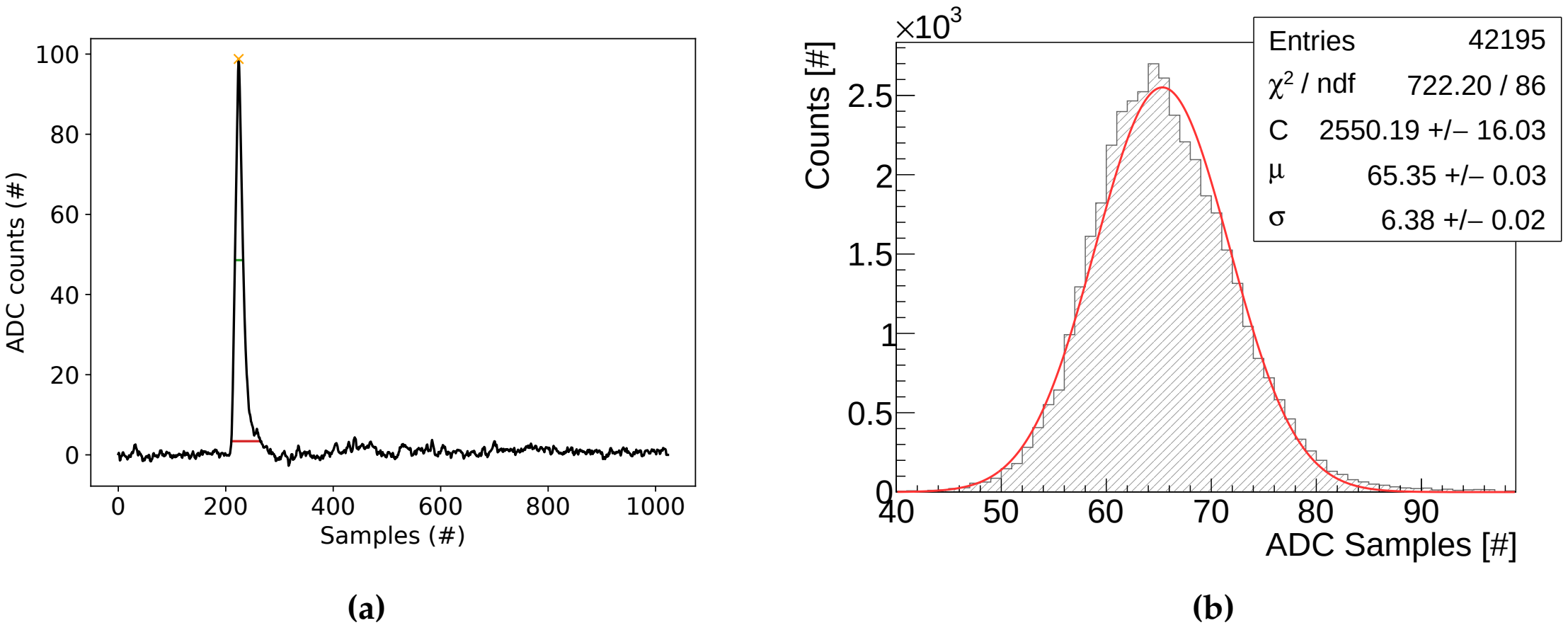

**Figure 6.8:** Assessment of the minimum length of a signal measured by the PMT sensor: (a) example of a PMT signal from a spot-like $^{55}$Fe interaction, and (b) distribution of the measured *peak_fullWidth* for a large dataset of these signals. The distribution is fitted with a Gaussian curve.

Given the lack of the possibility to know the exact original angle of alpha tracks with respect to the GEM plane, a reasonable approximation to assess the minimum extension of the PMT signal is to use the signals induced by an $^{55}$Fe radioactive source. $^{55}$Fe emits 5.9 keV photons and produces ERs with total ranges in 3D of about 0.4 mm in the CYGNO gas (see Figure 2.4), resulting in a very average small extension in the longitudinal direction. These signals can therefore be considered a "minimal signal" for the evaluation of the accuracy, i.e. minimal signal, achievable from the PMT analysis. Signals from an $^{55}$Fe-induced ER typically exhibit basic waveforms with a single peak and no additional features. To evaluate the time extension of these signals, the *peak_FullWidth* variable (see Section 3.2.1.3) was used. An example of this variable applied to an $^{55}$Fe signal is shown as a red line in Figure 6.8a. This measurement was performed during LIME calibration runs, where an $^{55}$Fe source was placed at various positions along the longitudinal axis. Since this correction is intended to be applied together with the one estimated for the CMOS images, the data acquired at the center of LIME ($Z = 25$ cm) was used for consistency.

The distribution of the full widths of the single peaks found in $^{55}$Fe PMT signals is shown in Figure 6.8b. The distribution was fitted with a Gaussian function, and the fitting parameters are displayed in the figure. The fitted value can be converted to cm following Equation 3.4. The final value for the Z range traveled by the $^{55}$Fe-induced ERs, and thus the "minimum" signal a particle produces in LIME, is $4.767 \pm 0.002$ mm.

#### 6.3.1.3 Measured 3D length correction

The values obtained from both the CMOS and PMT measurements can now be combined to estimate the correction factor for the measured alpha track lengths, allowing to retrieve the actual range of these particles in the gas. The statistical error stems from the quadratic combination of the singular error from the CMOS (Section 6.3.1.1) and PMT (Section 6.3.1.2) corrections. In addition to the statistical error, a systematic uncertainty of 10% is assigned to the final value to account for limitations in the approximation used in this assessment. For example, the varying effects of longitudinal and transverse diffusion on the electron cloud for different drift distances are not considered, as the correction is applied to the distribution rather than on a track-by-track basis. These and other systematics are currently being studied. Until a more refined and optimized simulation is available, this preliminary value is considered sufficient for evaluating the effect of the difference between the traveled range and measured length of alpha particles. The final value used to correct the measured 3D length of alpha particles is given by:

$$\text{3D length correction factor} = 8.51 \pm 0.06\ (\text{stat.}) \pm 0.85\ (\text{syst.})\ mm \tag{6.1}$$

From this correction, the particle range is obtained by subtracting this value to the fitted Gaussian mean of Figure 6.5, which is then converted to an estimated decay energy, as described in the following.

### 6.3.2 Track range to candidate decay association

In order to correlate the estimated track ranges to their deposited energy and identify from this the relative candidate decay process, the SRIM software [274] was used to obtain the tables of ranges of the various alphas energies in the CYGNO gas mixture. SRIM (Stopping and Range of Ions in Matter) is a software package that simulates the interactions of ions with matter, providing predictions on energy loss, range, and damage production as ions pass through various materials. SRIM takes as input the medium composition ($He$:$CF4$ 60:40 in CYGNO case) and density. LIME gas monitor measured a gas density varying between 1.46 and 1.50 $kg/m^3$ during Run4. Therefore, an average density value of $\rho = 1.48 kg/m^3$ was used in the SRIM simulation.

The SRIM software was run varying the energy deposited by alpha tracks in the CYGNO gas mixture until the various estimated ranges were obtained, and the results for each peak are displayed in Table 6.4. The error on the estimated energies was evaluated repeating the procedure for the minimum and maximum ranges obtained summing and subtracting the errors associated to each peak range. The obtained errors result asymmetric due to the non-linear scaling of the energy deposit with range. The errors on the measured 3D length is given by the $1\sigma$ deviation from the Gaussian fit applied to each peak in Figure 6.5. This intends to account for systematic uncertainties related to this first-ever implementation of the 3D reconstruction and measurement of the 3D length of alpha particles, which, due to various constraints, could not be systematically assessed. The error on the corrected 3D length corresponds to the combination in quadrature of the the error on the fit ($1\sigma$) with the sum in quadrature of the statistical and systematic error of the correction factor in Section 6.3.1.3.

**Table 6.4:** Summary of all identified peaks in the distribution of measured 3D length in Figure 6.5. For each peak, the corrected length (range) is shown (see Section 6.3.1), together with the estimated energy for that range obtained using the SRIM software. On the second half, the decay candidates for each peak are shown together with energy of the alphas emitted during that decay. The SRIM range for each candidate is also shown to facilitate the comparison.

| **Measured values** | | | | **Candidate decay** | | |
|---|---|---|---|---|---|---|
| Peak # | 3D length ($\pm 1\sigma$) [mm] | Corrected 3D length [mm] | Estimated energy [MeV] | Isotope | $\alpha$ energy [MeV] | SRIM range [mm] |
| 1 | 33.73 ± 3.93 | 25.22 ± 4.02 | $4.160^{+0.470}_{-0.480}$ | $^{238}$U | 4.198 | 25.34 |
| 2 | 41.60 ± 1.11 | 33.09 ± 1.40 | $5.055^{+0.110}_{-0.120}$ | $^{210}$Po | 5.304 | 35.45 |
| 3 | 44.90 ± 1.49 | 36.39 ± 1.72 | $5.400^{+0.175}_{-0.180}$ | $^{222}$Rn | 5.489 | 37.26 |
| 4 | 51.60 ± 1.91 | 43.09 ± 2.09 | $6.055^{+0.195}_{-0.205}$ | $^{218}$Po | 6.002 | 42.52 |
| 5 | 60.80 ± 2.90 | 52.29 ± 3.02 | $6.880^{+0.255}_{-0.265}$ | $^{216}$Po | 6.778 | 51.10 |
| 6 | 73.24 ± 3.84 | 64.73 ± 3.93 | $7.890^{+0.300}_{-0.305}$ | $^{214}$Po | 7.687 | 62.10 |

In order to associate these estimated energy to a candidate radioisotope alpha emission, following the preliminary findings reported in [196] and discussed in Section 6.1, the decay chain of $^{222}$Rn expected to contaminate LIME gas was considered, in addition to the ones relative to the radioactive contaminant of the main LIME detector component reported in Table 6.1. The $^{222}$Rn chain (see Figure 6.27) comprises four alphas emissions, namely from $^{222}$Rn to $^{218}$Po (5.49 MeV), from $^{218}$Po to $^{214}$Pb (6.00 MeV), from $^{214}$Po to $^{210}$Pb (7.69 MeV) and from $^{210}$Po to $^{206}$Pb (5.30 MeV). These energies correspond the energies of the alphas emitted in these decays. Given the minute-scale half-lives of these isotopes – as well as those of their intermediate daughters ($^{214}$Pb and $^{214}$Bi) – it is highly probable that all of these alpha decays will be observed in a detector which contains $^{222}$Rn.

Concerning the additional radioisotopes displayed in Table 6.1, only the $^{238}$U and $^{232}$Th decay chains were considered, since the $^{235}$U contamination result of about 1 order of magnitude lower than these two and the other contaminants do not produce alphas. The $^{222}$Rn, $^{238}$U and $^{232}$Th full decay chains are shown in Figures 6.27, 6.28, and 6.29, as taken from Laboratoire National Henri Becquerel database[1]. Within these three decay chains, following the typical alpha background observed in gaseous TPC for directional

[1] http://www.lnhb.fr/accueil/donnees-nucleaires/module-lara/

DM searches and characterized in the studies by DRIFT [92] and MIMAC [93], only the alphas produced in the decays with relative intensities (probabilities) greater than 75% in the *main* branching ratios of these chains were considered. Within these, the decays with the closest alpha energy to the estimated ones from the 3D length is taken as primary candidate. For each candidate, its resulting range according to SRIM is also obtained for a more direct comparison.

The large sigma of peak #1, compared to the other peaks, is not solely due to low statistics or limited resolution. In addition to the alpha decays already considered as candidates, the $^{238}\mathrm{U}$ decay chain includes several other alpha-emitting isotopes spanning a broad energy range. In particular, the decays of $^{234}\mathrm{U}$, $^{230}\mathrm{Th}$, and $^{226}\mathrm{Ra}$ produce alpha particles with energies between those of $^{238}\mathrm{U}$ (4.198 MeV) and $^{222}\mathrm{Rn}$ (5.489 MeV), placing them between these two peaks in the distribution. While these intermediate peaks are not individually distinguishable, they likely contribute to the broad sigma observed for the first peak, thereby helping to justify its width.

In the same distribution of Figure 6.5, from the $^{232}\mathrm{Th}$ chain, only the $^{216}\mathrm{Po}$ decay is clearly observed. The alphas with energies of approximately $\sim 5.4$ MeV and $\sim 6.3$ MeV, associated with the decays of $^{228}\mathrm{Th}$ and $^{220}\mathrm{Rn}$ in the thorium chain, respectively, are expected to be hidden behind the much larger contributing peaks of $^{222}\mathrm{Rn}$ and $^{218}\mathrm{Po}$, respectively, due to the proximity of their energies and the significantly lower contamination of $^{232}\mathrm{Th}$ compared to $^{238}\mathrm{U}$ – by approximately $\mathcal{O}(100)$ (see Table 6.1) – in the materials in close contact with the gas (cathode, GEMs, and field cage), thus making the thorium-chain decays subdominant. The $^{216}\mathrm{Po}$ peak stands out primarily because its energy is well separated from those of the other decays, allowing it to be more easily identified. Furthermore, as will be shown in Section 6.4.4, once the detector is segmented into different regions, additional peaks begin to emerge, such as the one likely corresponding to the decay of $^{224}\mathrm{Ra}$ from the thorium chain, demonstrating also how spatial segmentation enhances peak identification.

A final summary of all the identified peaks and their corresponding candidate decays is presented in Table 6.4. The table demonstrates a generally good agreement between the estimated energies derived from the measured lengths and the expected values from the candidate decays. The identified candidate decays results consistent with what presented in similar studies by the MIMAC [93] and DRIFT [92] collaborations, further supporting our findings. The results presented in this Section highlight the effectiveness of using 3D track length as a means for performing alpha spectroscopy in LIME, and representing a novel result for the CYGNO experiment.

## 6.4 Alphas localization

Thanks to the full 3D tracking illustrated in Chapter 5, the origin of each alpha decays can be studied in terms of spatial distribution and orientation to further characterize and identify the internal LIME backgrounds. As already discussed in Section 6.1 and highlighted by the measurement reported in Table 6.1, the largest radioactivity contribution to LIME background budget is expected to come from the field cage rings and resistors, the cathode and the GEMs. Since the CMOS camera and lens are located far from the gas and given that alphas ranges are all smaller than 10 cm, no contribution from these are expected to be observed in the active volume. Field cage resistors, while being closer to the gas, are located on top of the cage rings (see Figure 2.16). Therefore, considering such distance and the small 14 mm rings pitch, most of the alphas emitted from these are very likely to be absorbed in the rings themselves before entering the gas and therefore their contribution to the alpha spectrum is expected to be minimal. Starting from these considerations, LIME active gas volume was divided into four non-overlapping separate sectors, as illustrated in Section 6.4.1, and the alpha length distribution and orientation was studied separately for each of this (Section 6.4.2 – Section 6.4.5) as reported in the following.

### 6.4.1 LIME active volume sectors definition

In order to study the contributions of the main detector components presented in Section 6.4 to the LIME alphas background, the gas active volume was divided into three sectors stemming from their proximity to the field cage, GEMs and cathode materials, plus a fourth one used to represent the gas-only contribution in order to further validate and characterize the $^{222}$Rn presence in LIME. The different regions are defined using the measured alpha track absolute positions. In this, for the XY localization, the average X and Y position as returned by the IDDBSCAN algorithm were used (see Section 3.1.3), and the absolute Z was defined as illustrated in Section 5.2.3. By applying a selection to the *average* X and Y, the external XY region associated to the field cage is defined within 5 cm from the image border. Since the longest alpha tracks considered in this study travel about 8 cm in a straight line, as shown in Figure 6.5, this selection is considered sufficient to isolate the field cage from the other detector components. As already illustrated in Section 5.2.1.1, Run4 data featured a noise-band-cut in Y which is already automatically excluded in the reconstruction procedure. The resulting selection in the XY plane is displayed in Figure 6.9, which illustrates both the average XY position of the alphas considered in this study, and the final cutout region in the CMOS field-of-view.

Given the low accuracy of the absolute Z determination, a Gaussian fit was performed in the low (high) end region of the distribution of absolute Z positions of the alphas considered in this study, as displayed in Figure 6.10, and the GEMs (cathode) position along the longitudinal axis was defined within $\pm 1\sigma$ from the fitted means.

The four regions selected with these cuts are reported in Table 6.5 and shown in Figure 6.11, where is possible to observe how the field cage is removed not only from the gas but also from the GEMs and cathode region to prevent contamination from this.

### 6.4.2 Gas sector alphas characterization

The measured 3D lengths of alphas emitted within the region associated to the gas sector are shown in Figure 6.13. Two main distinct peaks are visible, consistent with the expected alpha lengths from the $^{222}$Rn and its daughter $^{218}$Po decays. These isotopes emit alpha particles with energies of 5.49 MeV and 6.00 MeV, respectively, as indicated in Table 6.4. No other contribution appear to be significantly present, except for a small bump around 6.1 cm measured 3D length, which results compatible with a small alphas peak emitted in the decay of $^{216}$Po. This isotope belongs to the $^{232}$Th decay chain, and in particular is produced by $^{220}$Rn with an half life of 55.6 s. Given the $^{232}$Th material activity reported in Table 6.1, this could be the result of a surface contamination where the $^{220}$Rn in this chain escape the material and diffuse in the gas subsequently decaying into $^{216}$Po. Nonetheless, given the small statistic of this contribution, no more in dept analysis could be performed to further support this hypothesis.

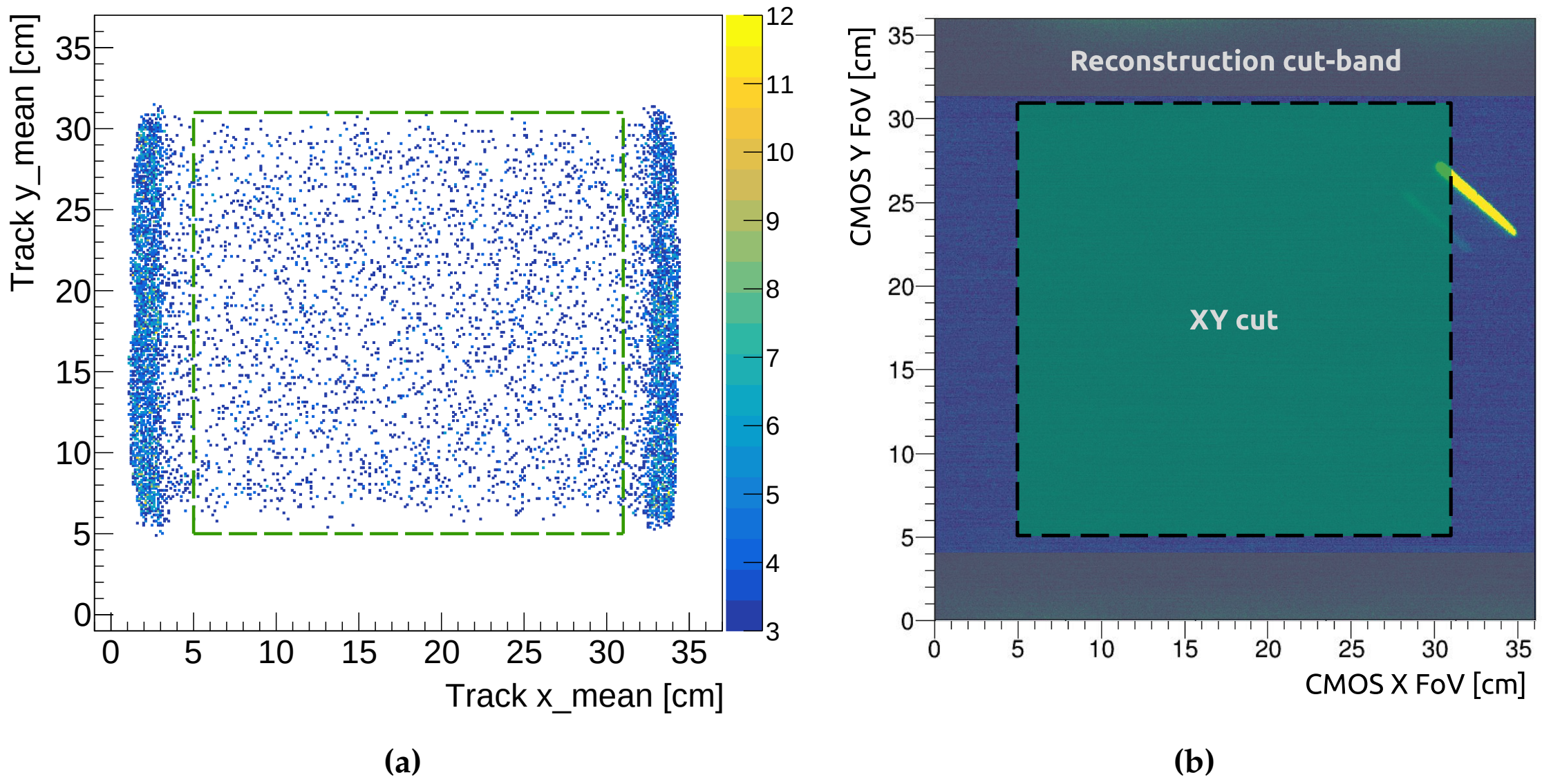

**Figure 6.9:** Geometrical cuts in XY illustrated in (a) as a green dashed line box overlaid on the distribution of the average XY positions of alpha particles in the LIME, and (b) as seen in the CMOS images, used to separate the alphas associated to the field cage from the remaining ones.

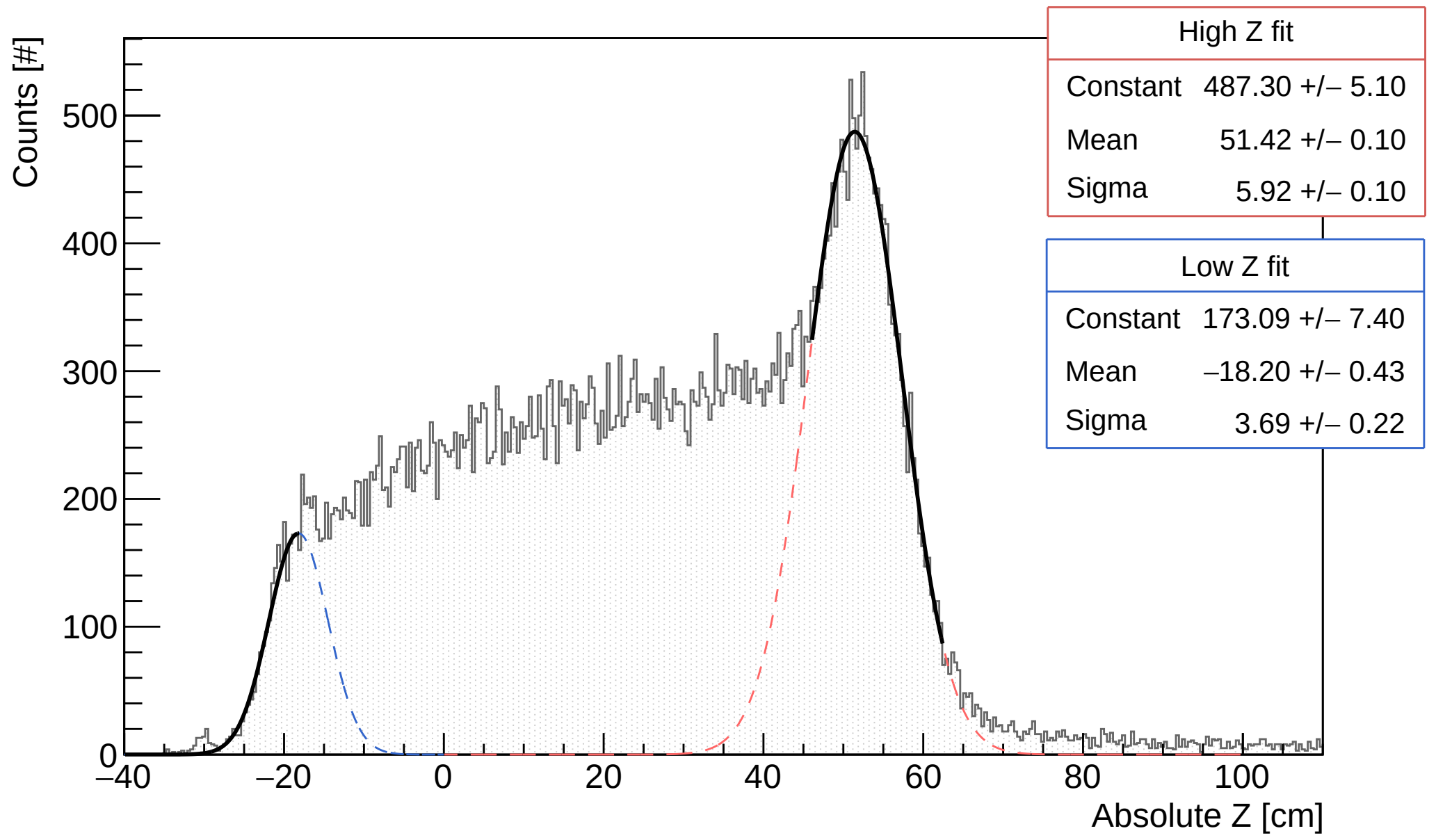

**Figure 6.10:** Geometrical cuts in applied in Z. Specifically, from the distribution of absolute Zs found for the alphas analyzed, two regions corresponding to a low and high Z are defined to isolate the alphas emitted by the GEMs and cathode, respectively.

**Table 6.5:** Summary of the geometrical cuts used to select different regions inside the LIME gas volume, visible in Figure 6.11.

| Detector region | X/Y cut | Z cut |
|---|---|---|
| (1) Gas volume | [5, 31] cm | [−14.51, 45.50] cm |
| (2) Cathode | [5, 31] cm | $> 45.50$ cm |
| (3) GEMs | [5, 31] cm | $< -14.51$ cm |
| (4) Field cage | $< 5$ cm $\vee > 31$ cm | [−14.51, 45.50] cm |

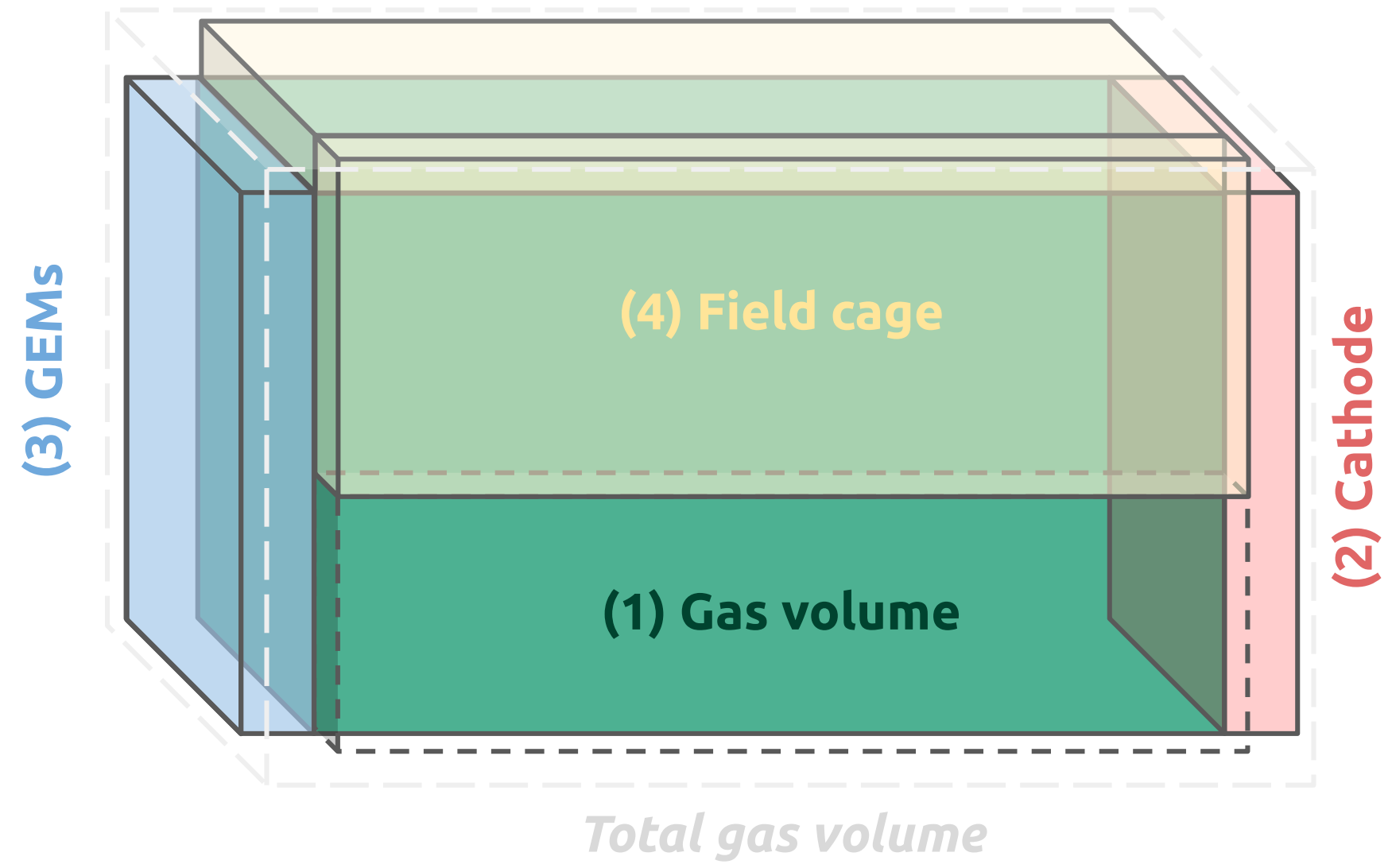

**Figure 6.11:** Schematic of the different regions defined inside the *total gas volume* of LIME to study the different contributions of each to the total alpha background spectra of LIME.

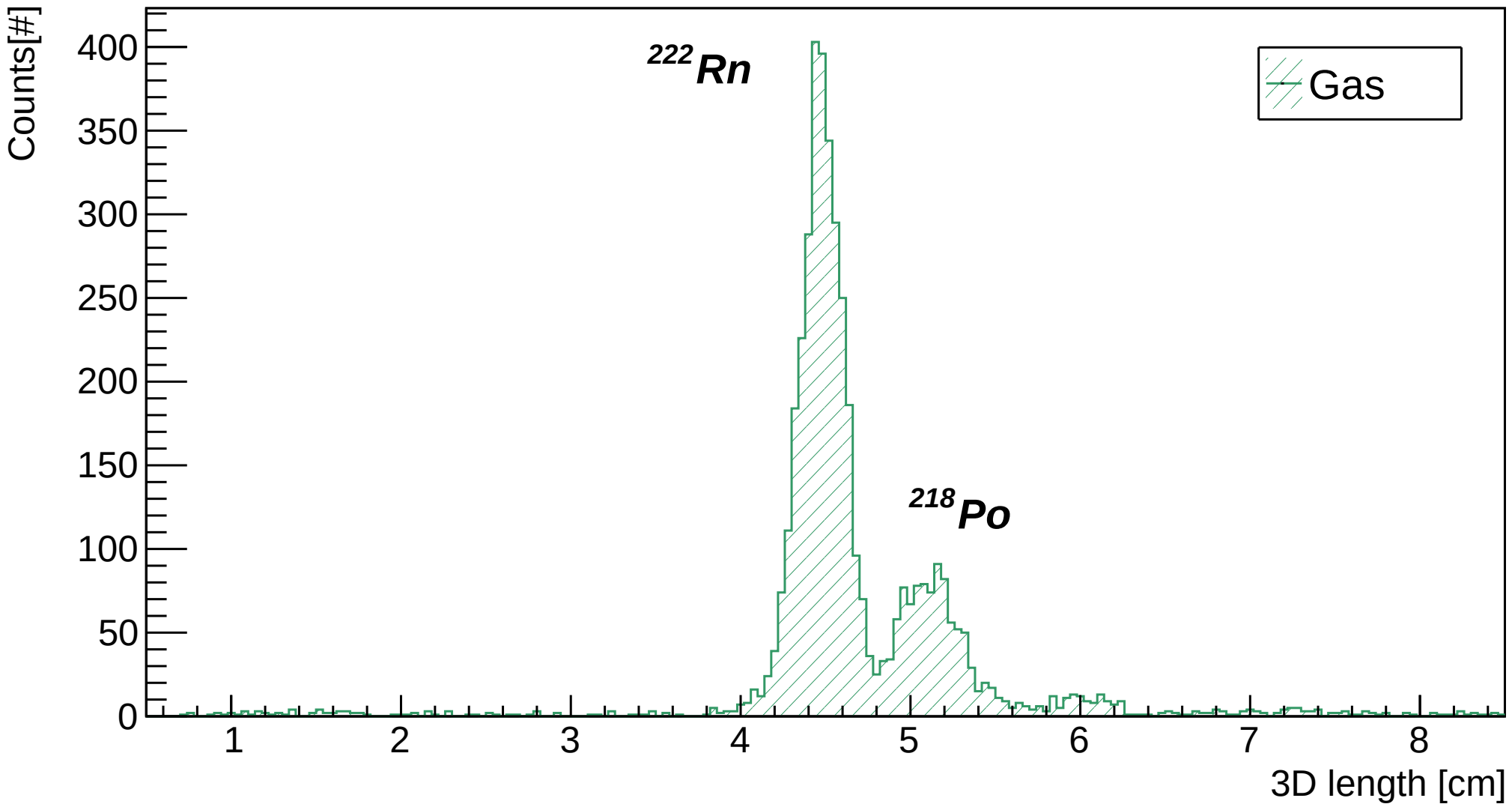

**Figure 6.12:** Measured 3D length of alpha particles in LIME from dataset group B of Table 6.3, using the selection cuts for alpha events in the gas volume region as defined in Table 6.5. The isotopes associated with each peak represent the *candidate* decays discussed in Section 6.3.2.

Since the cathode region is not considered in this volumetric selection, the significant difference in amplitude/area between the two main peaks, and the absence of the $^{214}$Po at around 8 cm 3D length, further support the hypothesis that most ($\sim$ 80%) of the $^{222}$Rn daughters are produced with a positive electric charge [275,276], as discussed in more detail in Section 1.3.5.2. These charged daughters drift towards the cathode once produced, and eventually decay there, provided their drift time is much shorter than their decay time. As will be further elaborated in Section 6.4.3, if this hypothesis is correct, the alpha decay associated with $^{214}$Po should be more visible in the cathode selection. Additionally, the presence of $^{218}$Po decays should increase with increasing Z, as these isotopes may also decay while drifting. This dependence is indeed observed, and a more detailed discussion is provided in Section 6.5, where the alphas from the Rn decay chain are treated in greater detail.

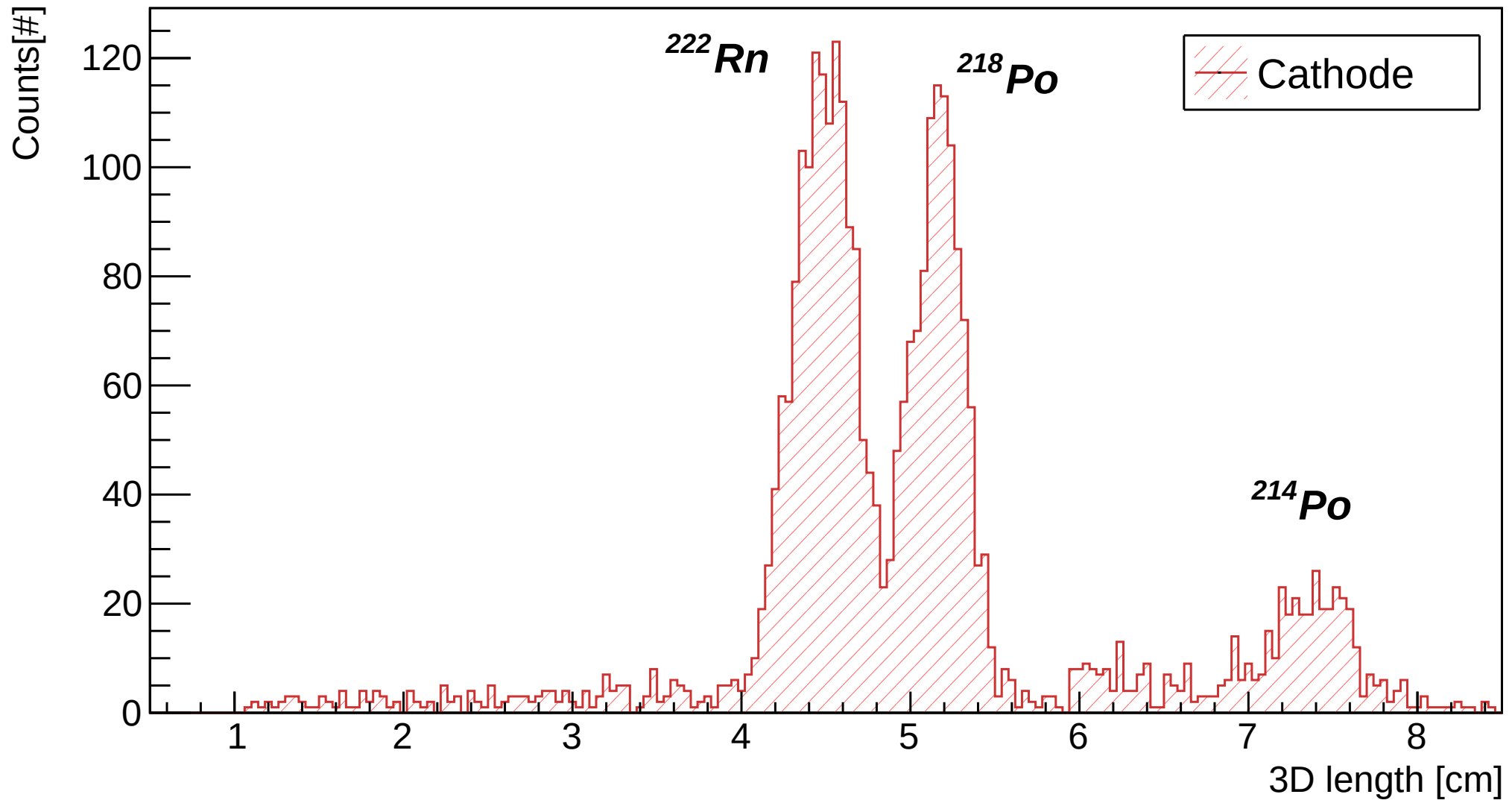

**Figure 6.13:** Measured 3D length of alpha particles in LIME from dataset group B of Table 6.3, using the selection cuts for alpha events in the cathode region as defined in Table 6.5. The isotopes associated with each peak represent the *candidate* decays discussed in Section 6.3.2.

### 6.4.3 Cathode sector alphas characterization

The cathode selection includes all events occurring within a distance of approximately 5 cm from the cathode, excluding the lateral regions where events from the field cage might originate. The distribution of the measured 3D lengths of alpha particles in the cathode region is shown in Figure 6.13. Three main peaks are visible at 3D lengths of approximately 4.5 cm, 5.1 cm, and 7.3 cm. These peaks, with corresponding energies of 5.40, 6.01, and 7.89 MeV, are identified in Table 6.4 as originating from respectively $^{222}$Rn, $^{218}$Po and $^{214}$Po. As discussed in Section 6.4.2, this distribution further supports the presence of $^{222}$Rn in the gas, as the final decay product ($^{214}$Po) is visible *only* in the measured 3D length distribution at the cathode.

The increased area of the $^{218}$Po peak relative to the main $^{222}$Rn peak with respect to the distribution observed in the gas sector additionally support the hypothesis of the $^{218}$Po daughter being produced typically charged and decaying mostly at the cathode. The distribution of $\theta$ emission angle for the cathode-identified alphas is shown in Figure 6.15b. In this, it is observed that most alphas are emitted in the direction of the GEMs, suggesting they originate from the cathode, further supporting the hypothesis. Some contamination from alphas traveling in the reverse direction is also present, likely due to imperfect selection cuts and given that in this region many $^{218}$Po nuclei may decay before reaching the cathode.

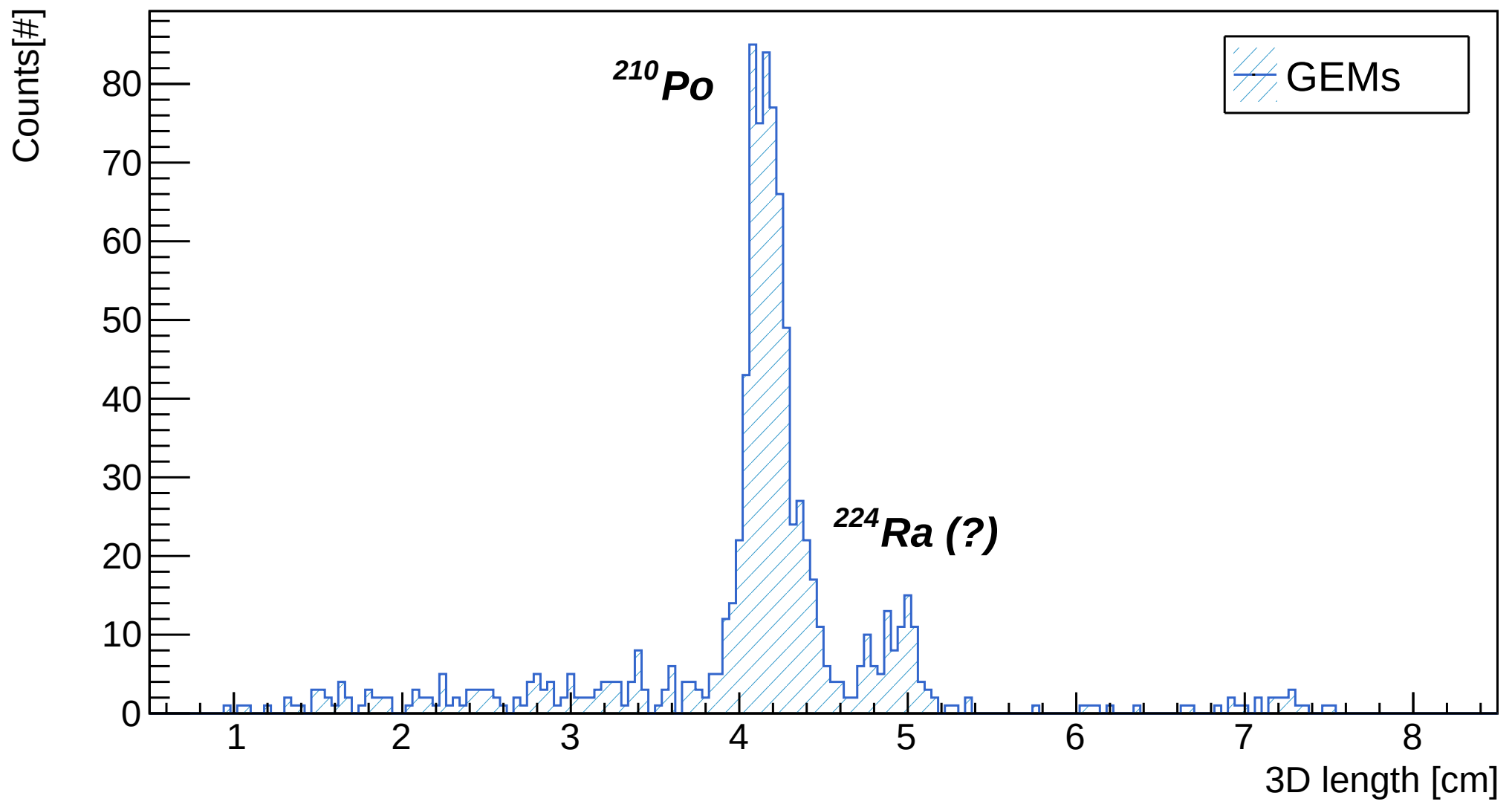

**Figure 6.14:** Measured 3D length of alpha particles in LIME from dataset group B of Table 6.3, using the selection cuts for alpha events in the GEMs region as defined in Table 6.5. The isotopes associated with each peak represent the *candidate* decays discussed in Section 6.3.2.

### 6.4.4 GEMs sector alphas characterization

The GEM selection is the opposite of the cathode selection in terms of absolute Z requirements and is designed to identify the alphas emitted by the GEMs, as these are also an expected source of alpha background. The 3D length distribution of alphas in this region is shown in Figure 6.14. In this distribution, one main peak is visible. Notably, this peak is *not* the same as the ones observed in Figures 6.12 or 6.13, as it is centered closer to 4.1 cm rather than 4.4 cm. In fact, this peak, which is barely noticeable in the total distribution shown in Figure 6.5, is expected to originate not from the decay of $^{222}$Rn, but rather from the decay of $^{210}$Po. This isotope appears in the $^{222}$Rn decay chain, but its source in the case of the GEMs is expected to be from the embedded $^{238}$U in these materials, which decays over time, since the charged daughters from $^{222}$Rn in the gas would not attach to the GEMs due to their positive charge. Additionally, the presence of the isotope $^{210}$Pb in the list of contaminants in the GEMs (Table 6.1) indicates that it is not in secular equilibrium with its parent, suggesting an additional source of $^{210}$Pb contamination in the materials. This further supports the clear and distinct appearance of the $^{210}$Po decay peak in the spectra, given its proximity to $^{210}$Pb in the decay chain (see Figure 6.27). The origin of these alphas can also be inferred from their $\theta$ angle distribution, as shown in Figure 6.15a. As seen, the vast majority of alphas selected with the GEM-region geometrical cuts exhibit a negative $\theta$, indicating they are emitted toward the cathode, thereby confirming their origin in the GEMs. The additional comparison between $\theta$ emitting angles for alphas identified in the GEMs with respect to cathode regions also validates the 3D analysis techniques.

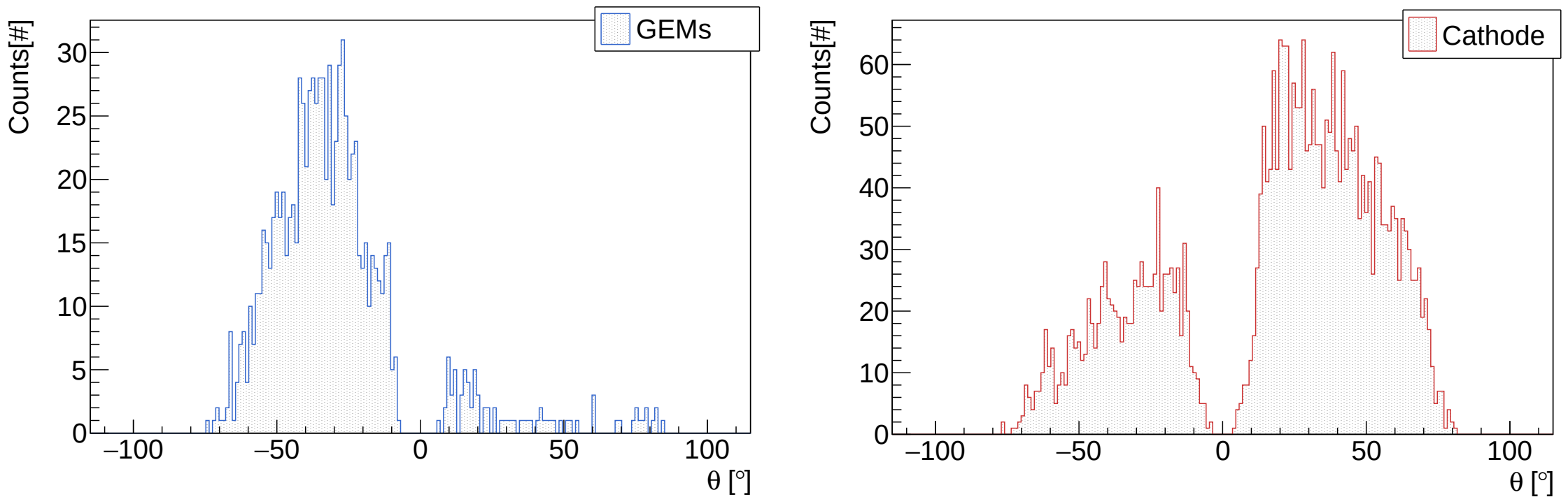

**(a)** GEMs region.

**(b)** Cathode region.

**Figure 6.15:** Distributions of the θ angle for the alpha selected through the geometrical defined defined for the (a) GEMs, and (b) cathode regions. A negative θ represents an alpha moving or emitted towards the cathode, and vice-versa.

In the spectrum shown in Figure 6.14, a secondary peak appears at a 3D track length of approximately 4.9 cm. Based on the radioactivity levels in the GEMs reported in Table 6.1, and considering the decay chains of $^{238}U$ and $^{232}Th$, the most probable origin of this peak is the isotope $^{224}Ra$ (from the thorium chain), which emits an alpha particle with an energy of 5.69 MeV, at a branching ratio of 95%. From the SRIM tables, this energy corresponds to an alpha range of 3.92 cm, which is consistent within 1.3 mm with the small peak observed, which has a corrected length of approximately 4.05 cm. Although other potentially competitive alpha decays exist in the $^{235}U$ chain, they occur with significantly lower branching probabilities. The observation of this peak further emphasizes the importance of the 3D tracking methods developed: thanks to the inherent geometrical cuts, it was possible to spatially subdivide LIME into sectors, which allowed the identification of this specific contamination that was not visible in the overall spectrum in Figure 6.5 where the decays from the $^{238}U$ chain are dominant.

### 6.4.5 Field cage sector alphas characterization

To select alphas originating from the field cage rings, geometrical cuts are applied to exclude the GEMs, cathode, and the bulk of the gas volume, leaving only the lateral bands of the detector. The distribution of the measured 3D lengths for alphas in this field cage selection is shown in Figure 6.16. In this distribution, a distinct new region emerges below 4 cm, peaking around 3.3 cm. This indicates that the significant number of alphas visible at low 3D lengths in Figure 6.5 originates almost entirely from the field rings. These alphas are primarily attributed to the $^{238}U$ decay chain, a primordial isotope commonly found in materials, including those used in the field cage rings, as detailed in Table 6.1. The strong contribution of these decays to the overall alpha background is justified by the large mass of the field cage materials in LIME and their direct proximity to the gas volume.

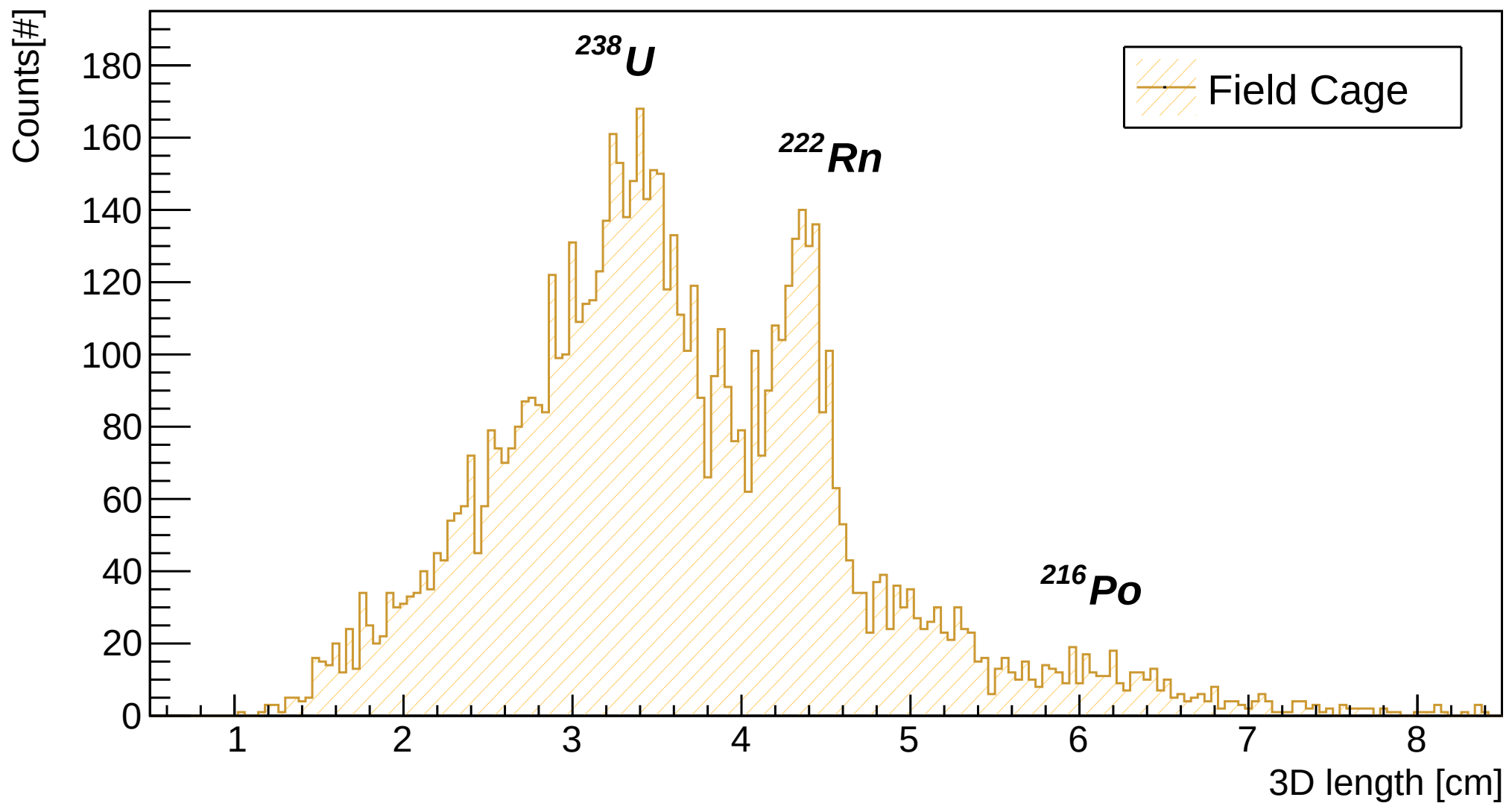

**Figure 6.16:** Measured 3D length of alpha particles in LIME from dataset group B of Table 6.3, using the selection cuts for alpha events in the field cage region as defined in Table 6.5. The isotopes associated with each peak represent the *candidate* decays discussed in Section 6.3.2.

As discussed in Section 6.3.2, the broad tail extending toward shorter 3D lengths is likely caused by alpha particles originating from other less pronounced decays in the $^{238}$U chain and/or that are emitted within the bulk of the materials, resulting in truncated tracks as they enter the gas. In this distribution, peaks corresponding to $^{222}$Rn and $^{218}$Po decays are also visible, although they are likely artifacts of the selection cuts, since these isotopes may also decay in proximity to the field rings. Lastly, a small bump at a 3D length of 6.0 cm, associated with the decay of $^{216}$Po (a product in the $^{232}$Th decay chain), suggests the presence of thorium contamination in the copper that constitutes the field rings.

Another informative variable when analyzing alphas from the field cage is the $\phi$ angle distribution. Figure 6.17b presents this distribution with an additional selection layer used to improve the visualization. In addition to the standard geometrical cuts defined in Table 6.5, a further requirement is applied based on the alpha particle's starting position in X and Y. The alpha starting position is determined using the XY edges of the track and its direction (including sense) determined with the $\phi$ angle in Section 5.2.2. With this information, the $\phi$ distribution is shown for two subsets: one where the alpha originates from an X position smaller than 2 cm, and another where it starts from a position larger than 34 cm, creating two sides bands as shown in Figure 6.17a. With this selection, the expected behavior is that alphas emitted from the left side of the detector ($X < 2$ cm) predominantly travel toward the right side, producing angles $\phi \in (-\pi/2, \pi/2)$, while those from the right side ($X > 34$ cm) should exhibit the opposite behavior. The measured distributions in Figure 6.17b confirming this expected trend. Another notable feature in this distribution is the dip observed around angles $\pm 90°$. This is naturally explained by geometric constraints: alphas emitted perfectly vertically from the side field rings are unlikely to enter the gas volume, as they would either remain inside the ring material or immediately impact the adjacent ones.

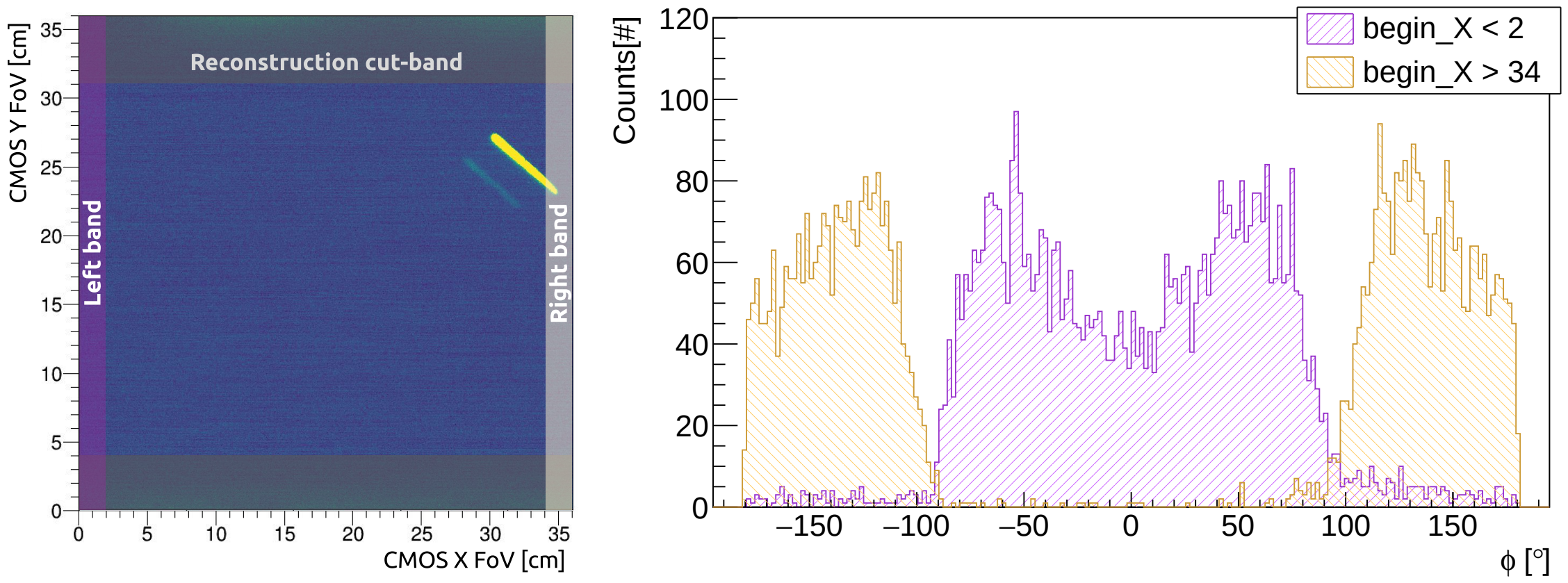

(a)

(b)

**Figure 6.17:** (a) CMOS field-of-view with the *Left* and *Right bands* used to select the lateral alphas, and (b) showing the distributions of $\phi$ angles retrieved in each of these bands, within the field cage geometrical cuts.

The results presented in Figure 6.17 also demonstrate the capability of the CMOS component in the 3D analysis to reliably reconstruct the direction of alpha particles. As discussed in Section 5.2.2.1, the algorithm was not originally optimized for alphas and required specific modifications to function correctly. Therefore, the successful reconstruction of this angular distribution not only validates the measurement itself but also reaffirms the performance and reliability of the 3D directional analysis framework.

A final confirmation that the new alpha population observed at lengths below 4 $\mathrm{cm}$ in Figure 6.16 originates from the detector's field cage rings is obtained by examining the Z distribution of alpha particles with 3D lengths smaller than 4 $\mathrm{cm}$, shown in Figure 6.18. In this distribution, it is evident that these alphas are emitted throughout the entire active volume (with the cathode and GEM regions excluded by the applied geometrical cuts). This uniform presence along the Z axis confirms their origin in the field rings, which, like the alphas, are also distributed uniformly along this direction. The denser horizontal band centered around a 3D length of approximately 3.5 $\mathrm{cm}$ corresponds to the prominent peak at this value seen in Figure 6.16.

## 6.5 Rn-222 decay chain

As discussed in Section 1.3.5.2, Radon Progeny Recoils (RPRs) pose a significant challenge for dark matter searches. Due to the tendency of charged $^{222}\mathrm{Rn}$ daughters to plate out onto the cathode, they can become a persistent background source capa-ble of mimicking WIMP-like interactions. The initial studies of LIME's alpha background presented so far, and in particular the alphas 3D length distribution observed in the gas (Section 6.4.2), cathode (Section 6.4.3) and GEM (Section 6.4.4) sectors, effectively validate the preliminary hypothesis present in [196] of unexpected $^{222}\mathrm{Rn}$ entering LIME from the small detector and gas system leakage and significantly affecting the overall background rate. This represents therefore a paramount result for the CYGNO/INITIUM collaboration in the aim of assessing and validating LIME background budget and a critical information for the future development of CYGNO-04.

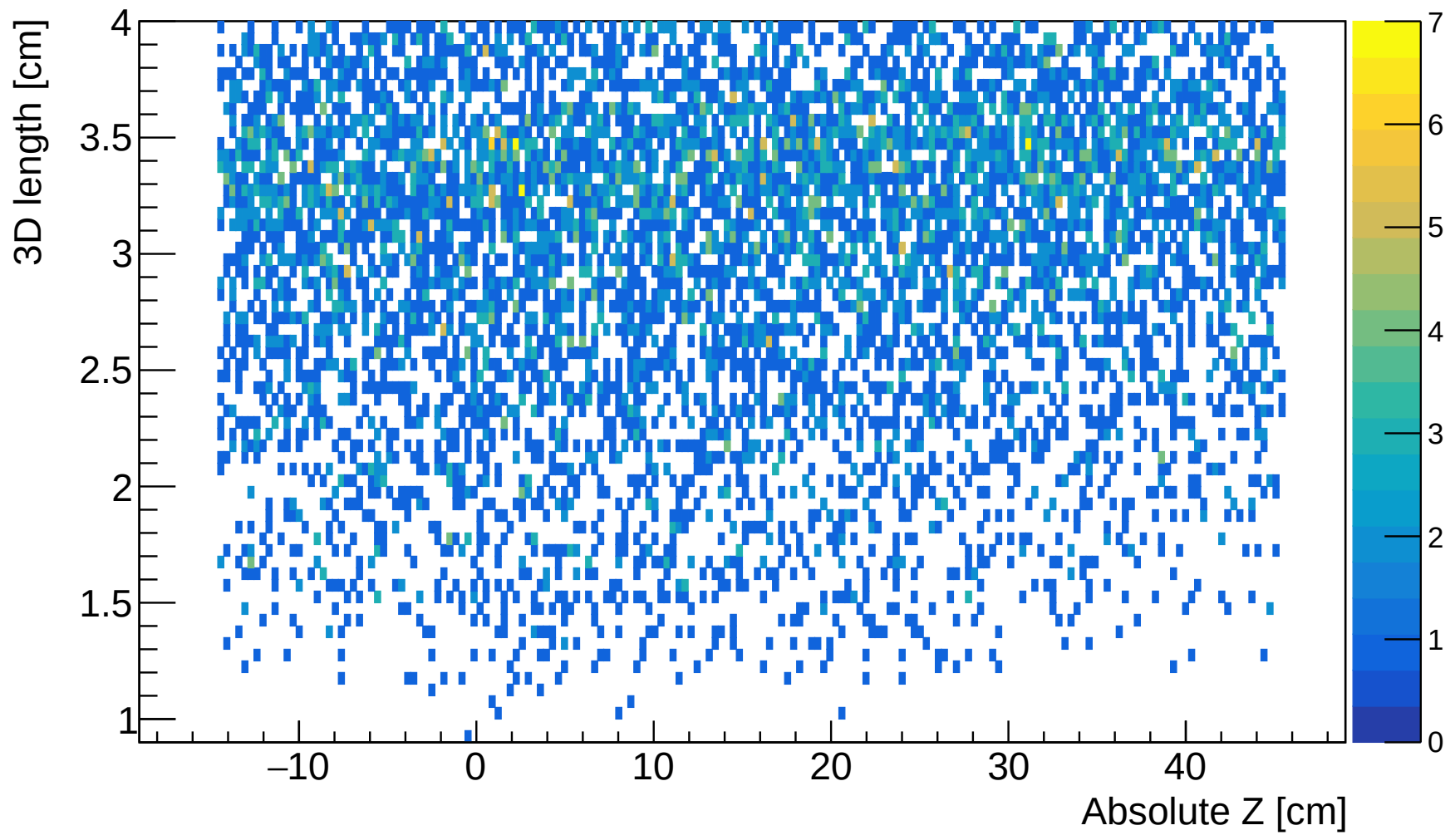

**Figure 6.18:** Relation between measured 3D lengths and absolute Z of the low energy alpha particles identified within the field cage geometrical cuts.

**Table 6.6:** Geometrical cuts used to select alphas from the $^{222}\mathrm{Rn}$ decay chain.

| Detector region | X/Y cut | Z cut |
|---|---|---|
| Rn chain | [5, 31] cm | $> -14.51$ |

In order to further characterize the RPR background only, a second analysis was performed using the batch A illustrated in Section 6.2, since here the $^{222}\mathrm{Rn}$ component is expected to significantly dominate over the other radioisotope contamination. In this context, the gas and cathode sectors previously discussed were merged with the selection shown in Table 6.6 in order to study the full $^{222}\mathrm{Rn}$ chain. The alpha length distribution along the drift direction (i.e. absolute Z) and alphas relative orientation with respect to the detector materials is inspected in order to characterize more extensively this contribution.

Using this selection, the distribution of measured 3D lengths was constructed and is shown in Figure 6.19. In this, only three prominent peaks are observed with their mean positions around 4.5, 5.1, and 7.4 cm, aligning well with the previously studied Rn alphas – namely the decays of the isotopes $^{222}Rn$, $^{218}Po$, and $^{214}Po$ (Figure 6.27) – and their expected lengths in the gas shown in Table 6.4.

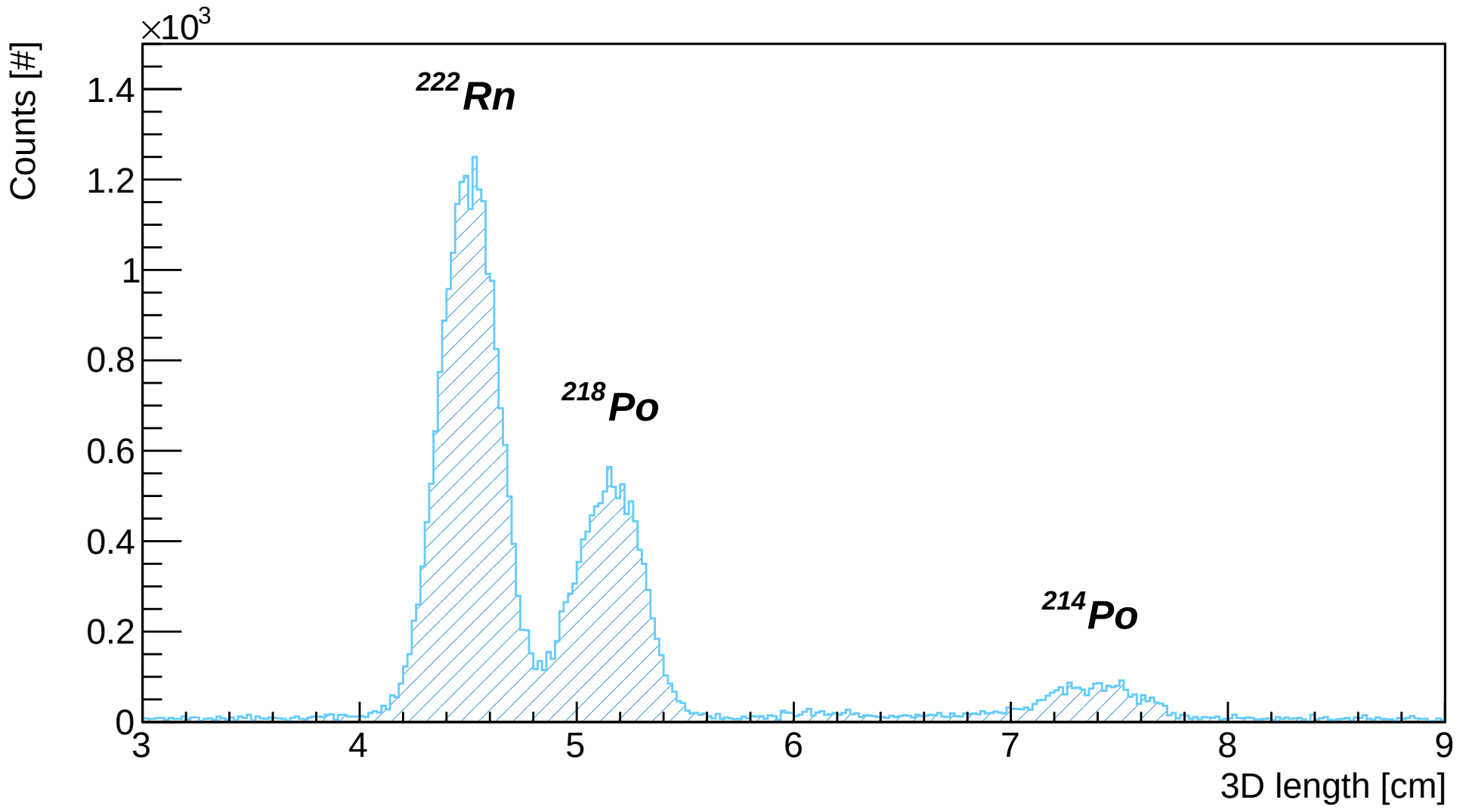

**Figure 6.19:** Measured 3D length of alpha particles in LIME in group $A$ of Table 6.3 using the cuts applied to select the $^{222}Rn$ decay chain shown in Table 6.6. The isotopes associated with each peak represent the *candidate* decays discussed in Section 6.3.2.

Figures 6.20 and 6.21 show the dependence of the alpha track length on the $\theta$ angle and the absolute $Z$ position, respectively. The cuts applied are those focused on the Rn chain, as shown in Table 6.6. Since the analysis concerns the $\theta$ angle (relative to the drift direction), alphas with a direction score of $< 0.5$ – indicating a randomly determined $\theta$ sign (see Section 5.2.5) – are removed from the sample. Additionally, a noticeable "hole" appears in the center of Figure 6.20. This is caused by very flat (parallel) tracks relative to the GEM plane, where the PMT's ability to positively identify alphas is significantly degraded. In this region ($\theta \sim 0°$), the waveforms lose their typical alpha shape and resemble more spot-like interactions. As a result, these alphas are excluded from the analysis, since their 3D shape cannot be reliably reconstructed. The angular cutoff is estimated to be around 10°, increasing for very short tracks ($< 3$ cm). A similar effect is occurs for angles greater than $\theta > 80°$ in the CMOS images.

Starting with Figure 6.20, three well-defined bands can be observed, corresponding to the three main peaks of the Rn decay chain, previously identified in Table 6.4 and visible in Figure 6.19. These bands are distributed differently in $\theta$: the 45 mm peak – corresponding to the $^{222}Rn$ decay – shows a higher population of alphas emitted with a negative $\theta$, i.e., directed towards the cathode. This could suggest that these alphas are emitted from the GEMs. However, as seen in Figure 6.21, the distribution of these alphas in Z is fairly uniform, indicating that they are likely from Rn decays occurring randomly and uniformly throughout LIME, rather than exclusively from the GEMs. The reason behind this non-uniform distribution in $\theta$ has not been yet explained within this study and additional work is being carried on by the collaboration to assess this. Among the possible explanations, the effect of the gas flow inside the detector was considered: in LIME, the gas enters near the GEMs and exits near the cathode, aligning with the drift direction and matching the sense of these additional alpha emissions – i.e., toward the cathode. This could potentially explain the observed non-uniform distribution, although no definitive conclusion has been reached so far.

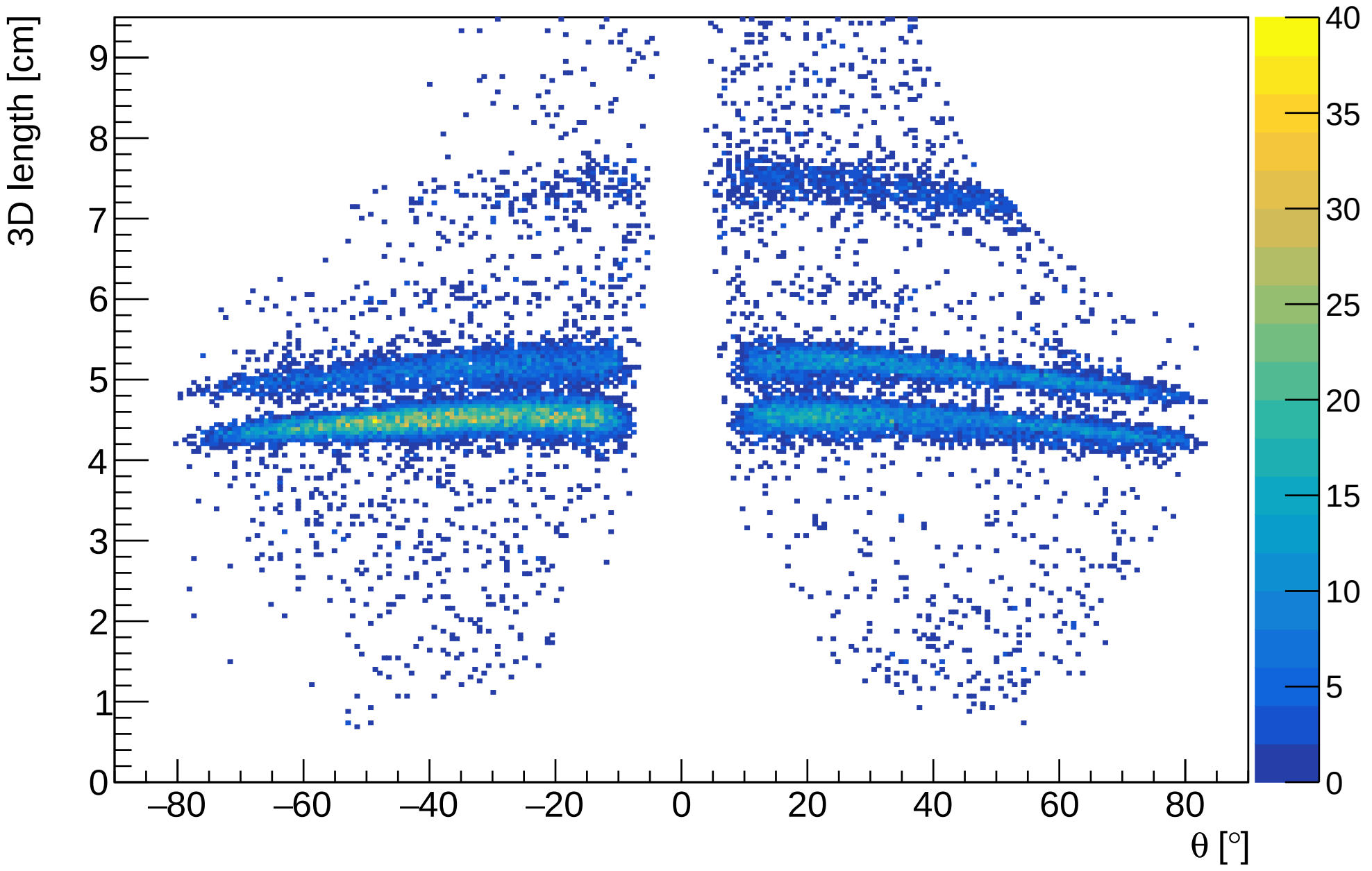

**Figure 6.20:** Relation between measured 3D length and θ emission angle of alpha particles in LIME within the cuts used to identify alphas from the $^{222}$Rn decay chain. Negative values of θ represent alpha moving towards the cathode, while positive represent alphas moving towards the GEMs.

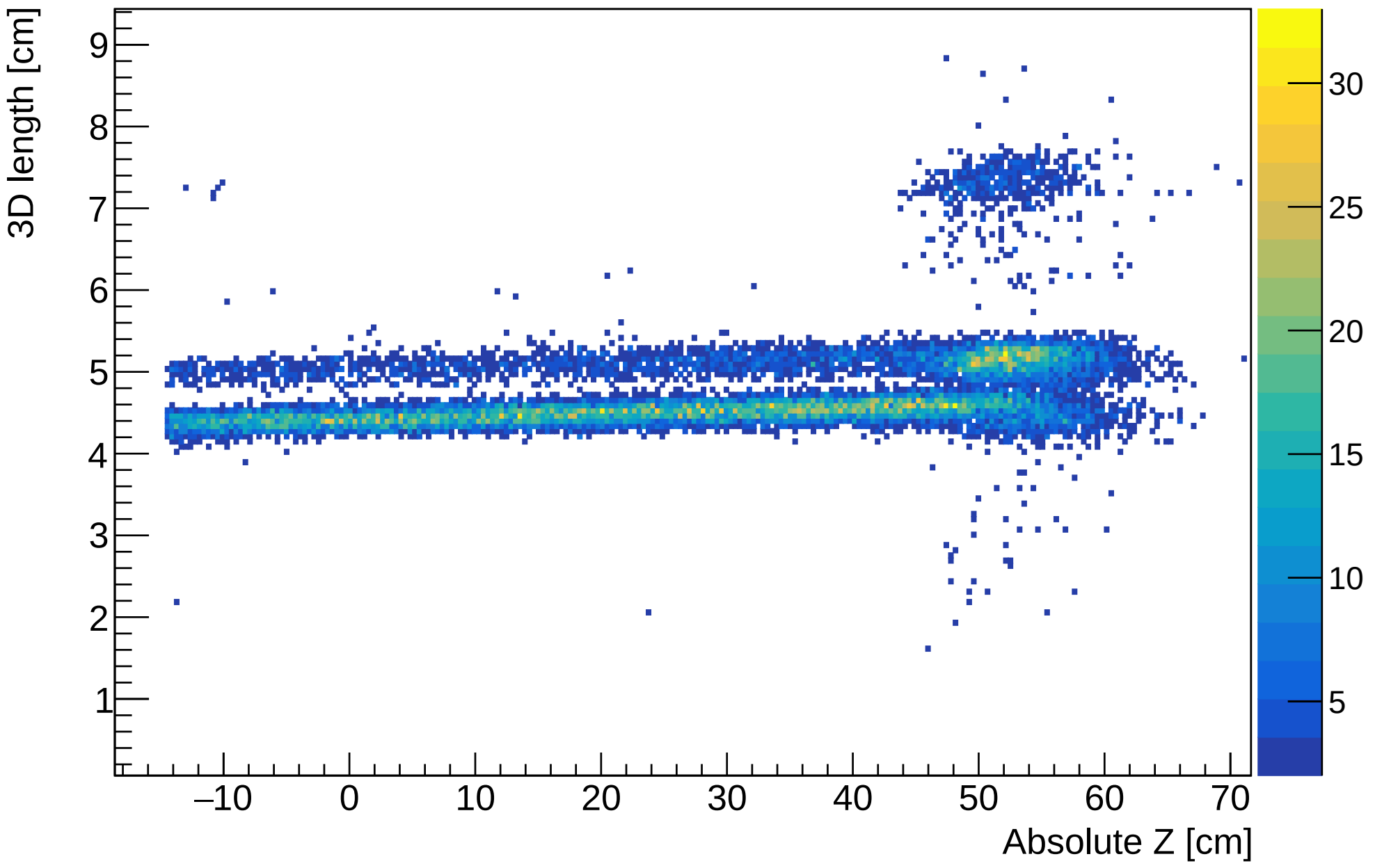

**Figure 6.21:** Relation between 3D length and absolute Z of alpha particles in LIME within the cuts used to identify alphas from the $^{222}$Rn decay chain.

The 51 $\mathrm{mm}$ peak – associated with the $^{218}\mathrm{Po}$ decay – shows a relatively uniform distribution in the emission angle $\theta$, with perhaps a slight increase in the positive $\theta$ emission. This could be explained by the fact that this alpha is the second in the $\mathrm{Rn}$ chain, and since we observed that the $\mathrm{Rn}$ decay occurs uniformly in Z, its daughter is also expected to be emitted uniformly in $\theta$, assuming no preferential decay direction. In contrast, looking at the distribution in Z, this peak shows an increasing preference for emission at higher Z, peaking at the cathode. This observation supports the argument made earlier in Section 6.4.2 about the expected polarization of the $\mathrm{Rn}$ nuclear daughters: if positively polarized, the nuclear daughters would drift towards the cathode, where they could attach and eventually decay. This hypothesis is supported by the increasing number of $^{218}\mathrm{Po}$ alphas being emitted as Z increases, compared to the parent $\mathrm{Rn}$ alphas. This also explains the slight increase in positive $\theta$ emission of these alphas, associated with the daughters that stick to the cathode before decaying.

To further support this hypothesis, the time needed for a charged $^{218}\mathrm{Po}$ to reach the cathode was estimated. Assuming a drift velocity three orders of magnitude slower than that of electrons (i.e., $\mathcal{O}(\mathrm{cm/ms})$), an ion would take approximately 50 milliseconds to travel 50 cm. Given that the half-life of $^{218}\mathrm{Po}$ is around 3 minutes, most of the daughters would likely reach the cathode before decaying. The presence of $^{218}\mathrm{Po}$ throughout the length of LIME can be explained by the exponential nature of its decay (time), which results in a non-zero probability of $^{218}\mathrm{Po}$ decaying before 50 $\mathrm{ms}$. Additionally, this also confirms the expectation that not *all* daughters are produced as charged, but rather only about 80–90%, as reported in other studies [92, 194, 275, 276].

The 73 $\mathrm{mm}$ length band appears predominantly at higher longitudinal positions in Figure 6.21, reinforcing the argument that the $^{218}\mathrm{Po}$ daughters ($^{218}\mathrm{Po} \to {}^{214}\mathrm{Pb} \to {}^{214}\mathrm{Bi} \to {}^{214}\mathrm{Po}$) are *also* produced with positive polarization, causing the final element in the chain ($^{214}\mathrm{Po}$) to decay directly at the cathode. Additionally, in Figure 6.20, this higher-energy band is more prominent in the positive $\theta$ region (emitted towards the GEMs), further supporting the idea that these alphas are primarily emitted from the cathode itself. In summary, the increasing rate of alphas at higher Z from each consecutive daughter in the $\mathrm{Rn}$ chain, as seen in Figure 6.21, shows that these daughters are created with positive polarization and accumulate at the cathode region.

In conclusion, the relationship between alpha lengths, their longitudinal emission angle, and the absolute Z position – namely the fact that $^{222}\mathrm{Rn}$ decays uniformly throughout LIME, $^{218}\mathrm{Po}$ decays at an increasing rate for higher Z, and $^{214}\mathrm{Po}$ is emitted entirely at the cathode level – supports the presence of $^{222}\mathrm{Rn}$ inside LIME and indicates that the

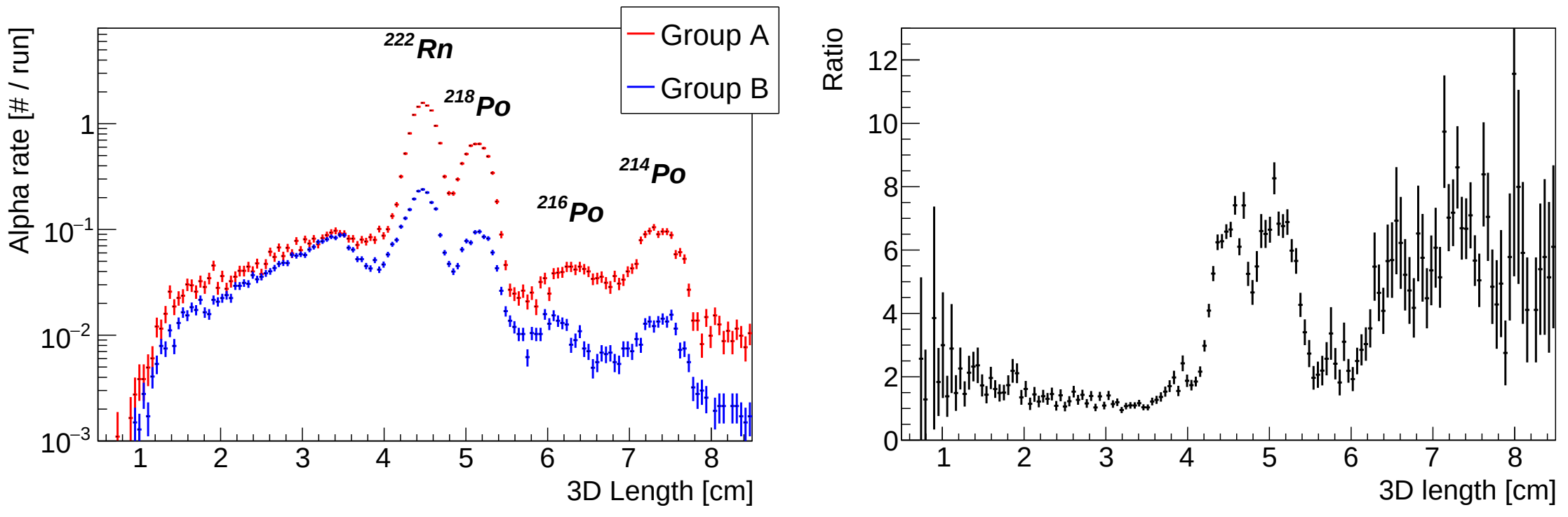

**(a)** **(b)**

**Figure 6.22:** Distribution of the 3D measured lengths of alpha particles subdivided in two datasets groups in Table 6.3: in (a) the number of alphas per run is presented, while (b) shows the ratio between the distributions for the two groups. The isotopes associated with each peak represent the *candidate* decays discussed in Section 6.3.2.

daughters of these isotopes are primarily created with a positive charge.

### 6.5.1 Filter effect

As illustrated in Section 6.2, during Run4 of the LIME underground data-taking campaign, a moisture filter ("blue", in Section 2.2.2.4.2) was introduced in the gas line at the end of 2023. This filter, initially thought to reduce the humidity inside LIME (which generally degrades the operating conditions of the detector), was found to also strongly reduce the rate of alpha background present in LIME, as illustrated more clearly in Table 6.3. Given the discussion of the origin of alpha particles in LIME so far, the two groups of data, A and B, were compared to pinpoint and identify the portion of the alpha background being mitigated by this filter.

For this comparison, since each detector component has already been discussed in detail, the entire gas volume is considered, and no additional significant cuts to the data are performed. As the measured 3D length is, so far, the most accurate variable to distinguish between different families of alphas, it was used to compare the two groups of data. Figure 6.22a presents the measured 3D length for alphas in the two groups mentioned, with Figure 6.22b showing the ratio between them. By analyzing these distributions, we can observe a clear difference in the rate of alpha particles per run in group B, especially in the peaks related to the Rn content (at lengths of 4.5, 5.1, and 7.3 cm). A difference at the peak expected from the decay of $^{216}$Po (at a length of 6.1 cm) is also visible, although, as mentioned earlier, little information can be retrieved from this peak due to the low statistics.

The striking difference in the rate of alphas originating from the Rn decay chain, approximately 5 times higher, demonstrates that the introduction of the moisture filter did indeed reduce the amount of Rn inside the detector. As expected, the alphas originating from the detector materials (mainly the field rings at 3D lengths $< 4$ cm) remain similar across the two groups, since these are unaffected by the presence of filters in the gas line. The hypothesis behind this reduction of Rn is that the gas enters LIME through the gas system, either directly by permeating the plastic gas tubes or by attaching to oxygen/water vapor molecules [194], which could explain the effectiveness of the moisture filter.

### 6.5.2 Calibration of LIME energy response at the MeV scale

As discussed at the end of Section 6.1.2, the charge saturation effect occurring in CYGNO detectors for alpha particles limits the accuracy of a direct energy measurement of these through the light detected by the CMOS or PMTs. The studies presented in this Chapter have shown how alphas 3D length results more much efficient in assessing their energy by matching this to the expected range. In this Section, an attempt is made to calibrate LIME response to $\mathcal{O}(MeV)$ energy deposits by relating the sc_integral (see Section 3.1.3) measured on CMOS images to the expected alphas energy inferred from their track length.

To minimize the effects of saturation on the measured spectra, specific cuts were applied to the data. Saturation is directly related to the spatial charge density of the ionization cloud at the amplification stage. As the charges diffuse while drifting, the longer the path traveled by the primary electrons, the more the initial alpha electron cloud spreads, which reduces the spatial electron density. As a result, saturation behavior is expected to be more pronounced for tracks emitted at a lower absolute $Z$ position. Following the same reasoning, alpha tracks with a high $\theta$, i.e., tracks close to perpendicular to the GEM plane, will experience more saturation due to the higher concentration of the electron cloud, as seen from the GEM plane. Based on this rationale, two cuts on the absolute $Z$ position and $\theta$ angle of the alphas were employed. In addition to these specific selections, the cuts used to select alphas from the $^{222}$Rn chain were applied (see Table 6.6), since these alphas can be clearly identified and their expected position within the detector is well known. The cuts on $Z$ and $\theta$ are as follows:

- **Absolute $Z$ position:** The dependence of the measured sc_integral by the CMOS on the alpha's absolute $Z$ position is shown in Figure 6.23a. If saturation were not present, three horizontal bands corresponding to the three main alphas from the $^{222}$Rn decay chain would be expected. Instead, a linearly increasing trend is observed, indicating that as $Z$ increases, the measured energy also increases. This illustrates how, due to higher diffusion of the electron cloud, charge saturation decreases. To eliminate this dependence on the absolute $Z$ position, alphas were selected within a narrow $Z$ range to ensure relatively uniform saturation. Since it is known that all $^{222}$Rn alphas appear at the cathode level to some extent (see Figure 6.21), a cut was applied to include only alphas with $Z$ within $1\sigma$ of the fit performed in Figure 6.10.

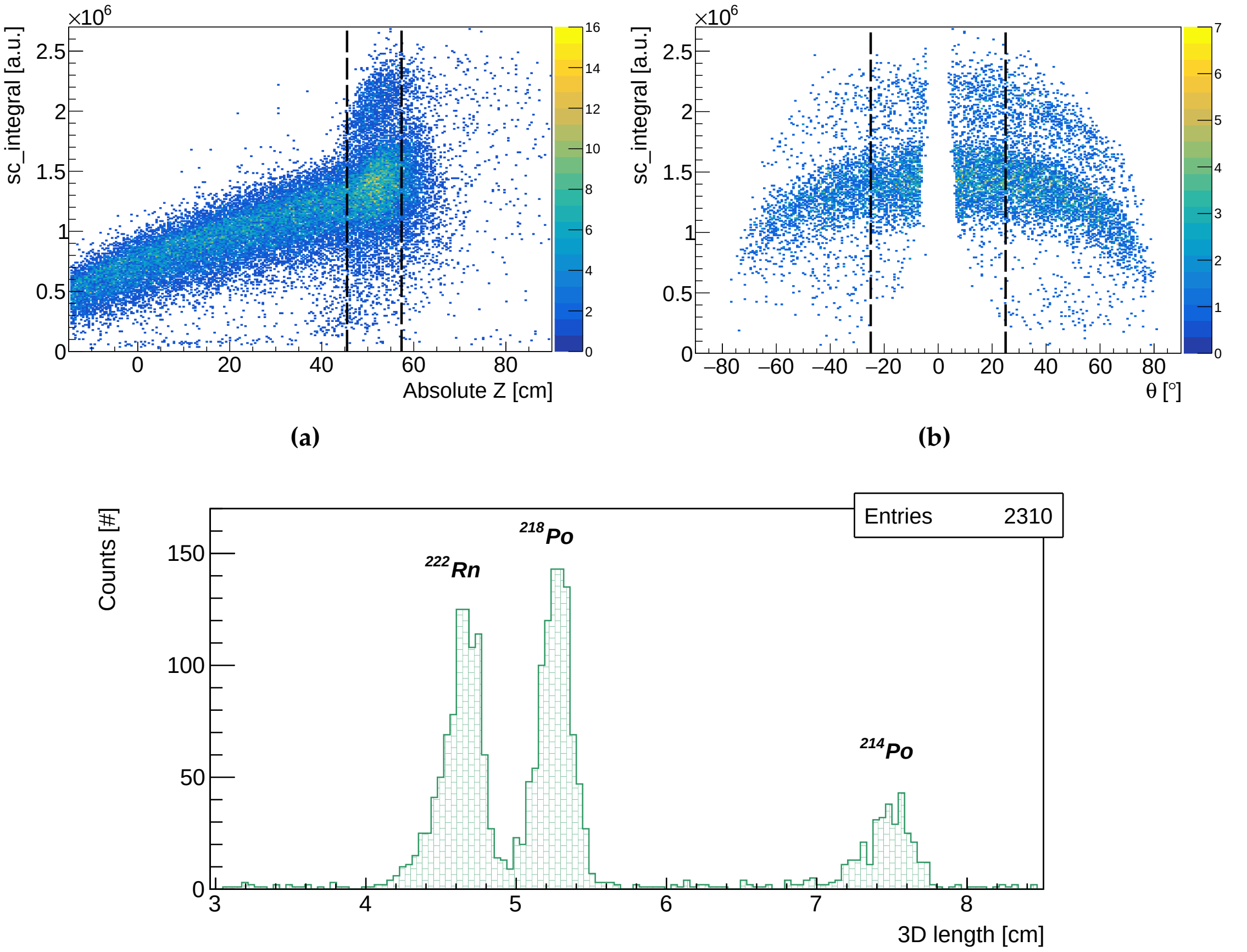

(a) (b) (c)

**Figure 6.23:** Relations between CMOS measured sc_integral and (a) absolute Z, and (b) θ angle, with the cuts applied in each relation overlaid as black lines. (c) Obtained distribution of alpha 3D lengths using the cuts showed in (a) and (b). The isotopes associated with each peak represent the *candidate* decays discussed in Section 6.3.2.

- **θ angle:** The dependence on the θ angle is shown in Figure 6.23b, where instead of the three horizontal alpha energy bands, a bending towards lower energies is observed at higher θ angles. This again illustrates how a higher θ leads to a higher spatial concentration of charges at the GEM plane, resulting in lower measured energy due to charge saturation. Note that the results in this figure are already filtered using the previous Z-position selection (see Figure 6.23a). To mitigate this dependency, a range of angles ($-25 < \theta < 25$), where the bending towards lower energies is less pronounced, was selected.

Given this alpha selection, the resulting distribution of measured 3D lengths is also plotted and shown in Figure 6.23c. The distribution of CMOS energies after applying this event selection is shown in Figure 6.24. Even when selecting high-diffusion alpha tracks, the saturation still significantly affects the measured light integral, introducing pronounced tails in the distribution. To address this, the distribution was fitted using the Crystal Ball function[2], which provides a better representation of the data given the strong charge saturation. The choice to fit three components is motivated by Figure 6.23c, which shows the same already identified three $^{222}$Rn alpha peaks of Figure 6.19.

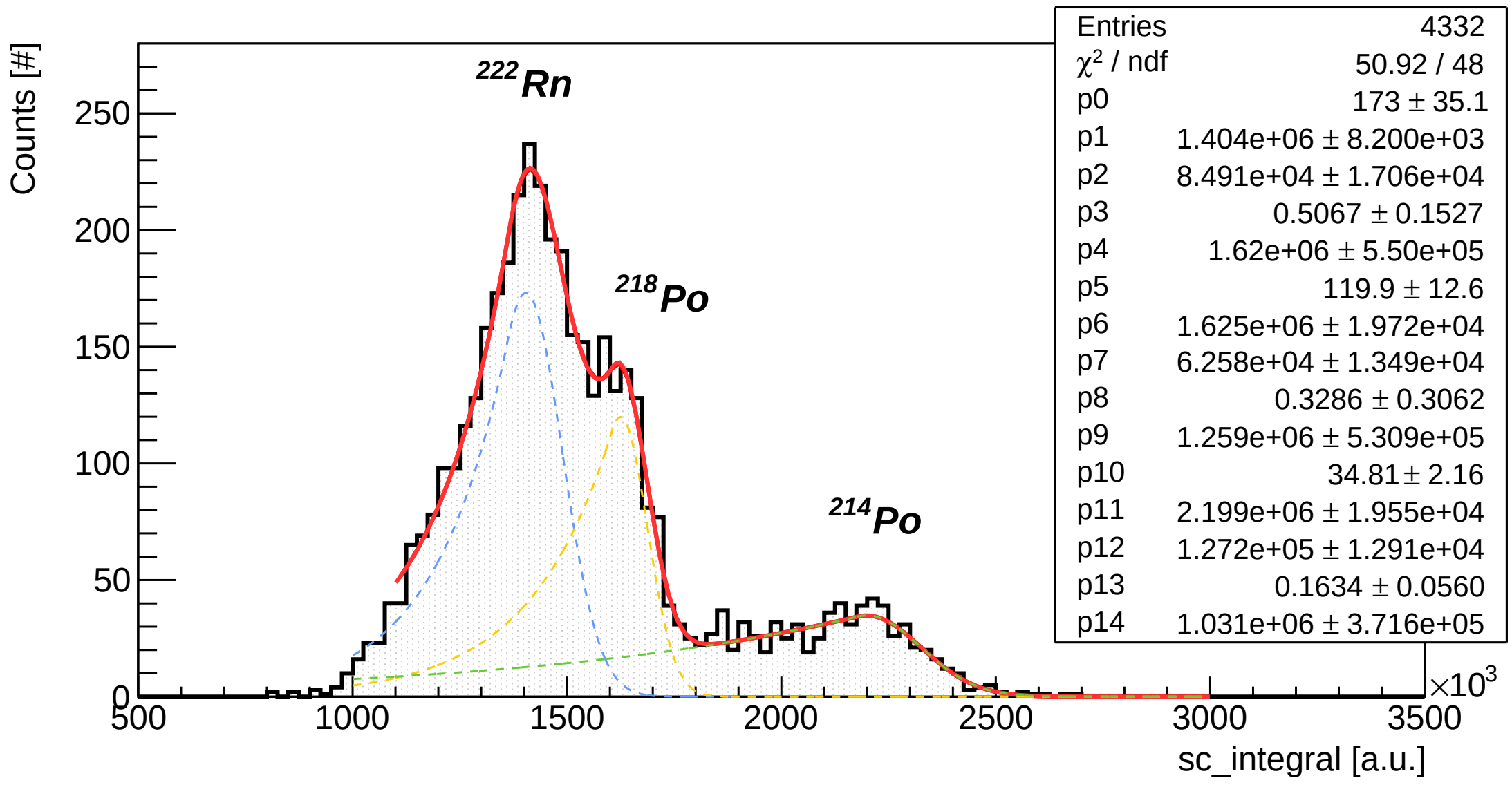

**Figure 6.24:** Distribution of CMOS measured sc_integral using the data cuts mentioned in Figure 6.23. The data was fitted with 3 Crystalball functions. Each of these fits is also showed in colored dashed lines, and the fitted parameters are reported in the legend, where the order is from lower to high energy. The isotopes associated with each peak represent the *candidate* decays discussed in this section and Section 6.3.2.

In order to validate the fit performed in Figure 6.24, each alpha energy peak was subsequently inspected by explicitly selecting on the 3D length shown in Figure 6.23c in order to isolate the three contributions. To this aim, the three energy peaks are selected by requiring the 3D length to be in the range [4, 5] cm, in the range [5, 6] cm and in the range [6, 9] cm, respectively. The result of this selection is displayed in Figure 6.25, with the Crystal Ball distribution resulting from the fit to each peak in Figure 6.24 overlaid (not fitted). The agreement between the two demonstrates that the global fit on the energy distribution reasonably identifies the three expected alpha peaks, validating the procedure illustrated in this Section.

By exploiting the low saturated alphas selected with this strategy, LIME energy response and energy resolution at the MeV scale was studied. Figure 6.26a shows the mean values of the three main peaks from the energy distribution shown in Figure 6.24 (fit parameters $p_1$, $p_6$, and $p_{11}$), plotted against their expected energies, assuming we are observing the three $^{222}$Rn alphas – a reasonable assumption at this stage. Although only three data points are available, the energy peaks appear to align well along a straight line when allowing the offset as a free parameter. The resolution of each energy peak was also calculated using $R = \sigma/\mu$, where $\sigma$ corresponds to the fit parameters $p_2$, $p_7$, and $p_{12}$ from the overall energy distribution fit. The resulting peak resolutions are shown in Figure 6.26b.

[2]The Crystal Ball function models asymmetric distributions with a Gaussian core and a power-law tail. It is commonly used in particle physics to account for detector effects, making it particularly suitable for fitting data with sharp peaks and skewed tails.

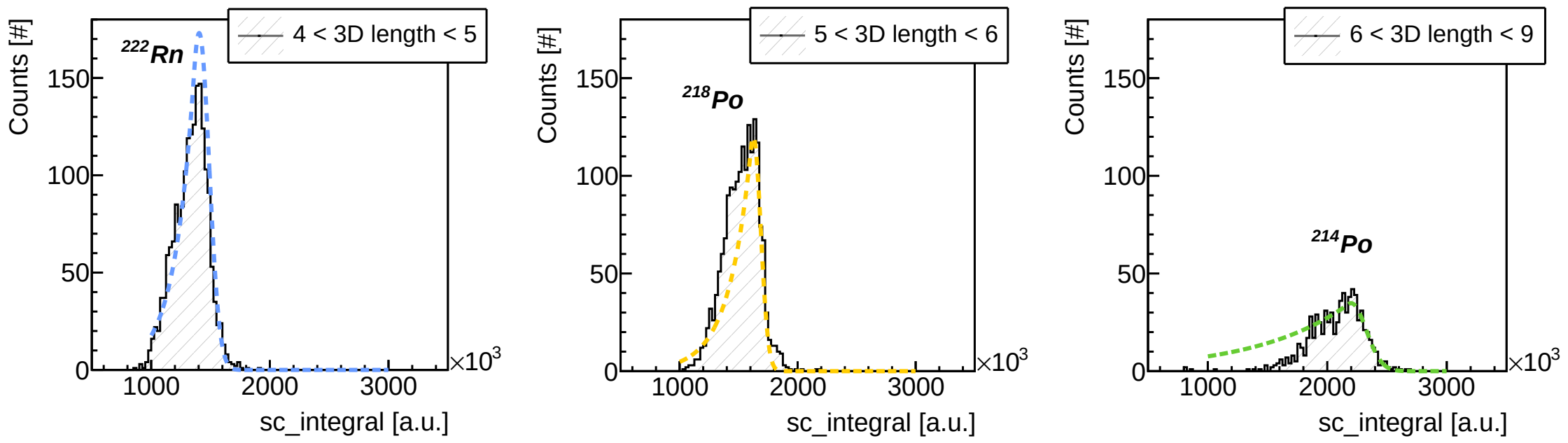

**Figure 6.25:** CMOS measured sc_integral distribution subdivided in three ranges using a cut on the 3D length of the alphas. Based on Figure 6.23c, from left to right and defining "$l$" as 3D length in cm, the cuts are: $4 < l < 5$; $5 < l < 6$; and $6 < l < 9$. In each new distribution, the individual Crystalball fit components of Figure 6.24 are *overlaid* on the corresponding expected peak. The isotopes associated with each peak represent the *candidate* decays discussed in this section and Section 6.3.2.

Compared to the results obtained during the extensive overground characterization of LIME [186] – where the energy resolution for $^{55}$Fe clusters (5.9 keV ERs) was found to be between 7% and 10% depending on the absolute Z – the results at the MeV scale are notably worse. At these higher energies, where a much larger number of primary charges are produced, significantly better resolutions would be expected in a gas detector. This preliminary result underscores the limitations introduced by saturation effects when measuring the energy of high-dE/dx particles. Conversely, it also demonstrates that variables retrieved from the 3D analysis – such as the 3D length or the $\theta$ angle – can be used to describe and potentially correct these saturation effects, thereby allowing a more comprehensive energy-based event analysis. In future studies, additional calibration points could be introduced to better evaluate the energy linearity in this high-energy and high-dE/dx particle regime.

## 6.6 Conclusions & Final remarks

The motivation for the studies presented in this chapter stemmed from discrepancies observed between experimental data and simulations in Runs 1–3 of LIME underground. These discrepancies were largely attributed to an unaccounted high-energy component in the internal background model. Preliminary analysis using CMOS-only data identified $^{222}$Rn as a potential contributor to these inconsistencies. The $^{222}$Rn decay chain

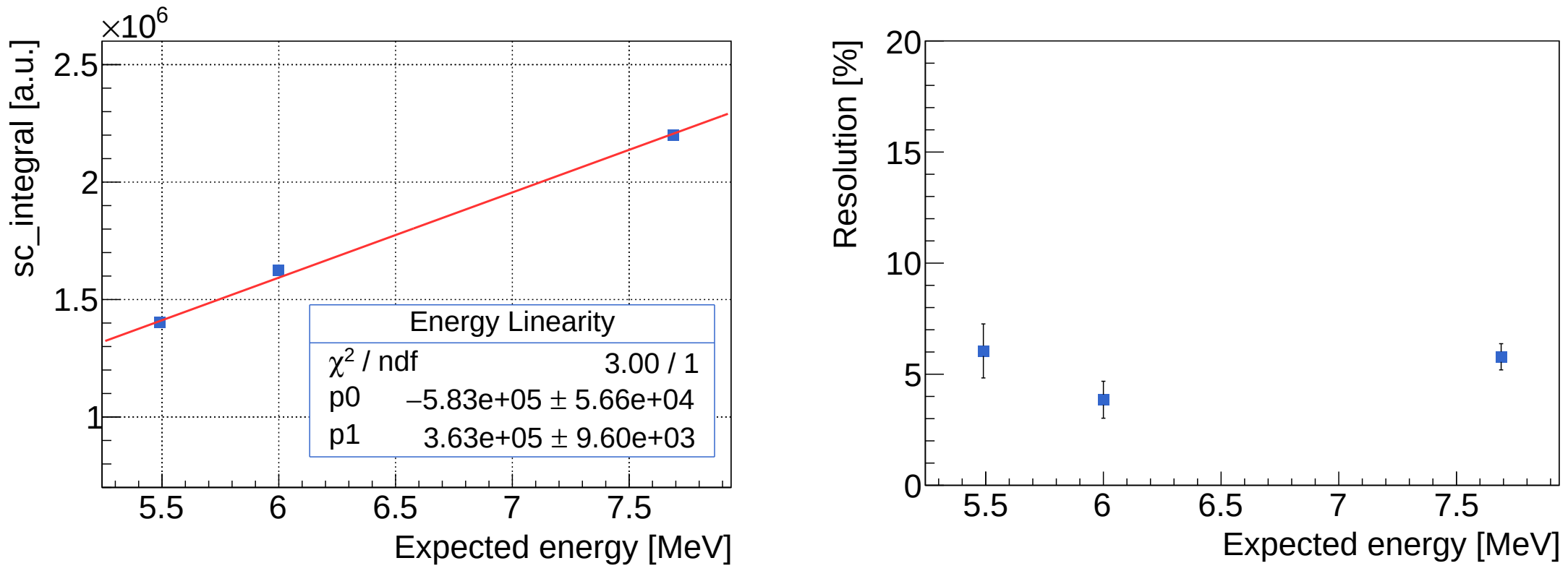

**(a)** Linearity

**(b)** Resolution

**Figure 6.26:** Energy (a) linearity and (b) resolution obtained for the $^{222}$Rn decay chain alphas, using the fit parameters showed in Figure 6.24.

emits both high-energy alpha particles and low-energy backgrounds through beta and gamma decays. Due to saturation effects, the combination of these two sources appears flat in the background spectra. To better characterize the origin of these backgrounds, I focused on studying its alpha component. In particular, the 3D analysis developed in this thesis was employed to reconstruct and identify alpha particles in Run 4 of the LIME underground campaign. The primary objective was to address these questions and evaluate the performance of the 3D analysis on real-world data, particularly its efficiency and capabilities.

The results for the measured 3D lengths of alphas in these datasets revealed several peaks in the alpha spectrum, namely attributed to decays of the isotopes appearing in the $^{222}$Rn decay chain – $^{222}$Rn, $^{218}$Po, and $^{214}$Po– as well as others such as $^{238}$U, $^{216}$Po, and $^{210}$Po. The identification of these isotopes was carried out by estimating the energy of the alphas through their 3D length using the SRIM software. The analysis also extended to studying the emission position – through fiducial cuts – and angles of the identified alphas. By confirming which alphas were emitted from different materials of LIME, such as the copper electric field rings, cathode, and GEMs, the origin of each alpha's isotope was further clarified and confirmed.

The $^{222}$Rn decay chain, with its three well-defined alpha emissions, was studied in more detail. The emission angle distribution confirmed the expected drift behavior of the $^{222}$Rn charged daughters, with the final alpha in the $^{222}$Rn decay chain showing a significantly higher emission angle towards the GEM plane. Cross-analysis of the Z position of these alphas further supported the presence of $^{222}$Rn, with the positions of the alphas following the expected distribution pattern: $^{222}$Rn decaying uniformly throughout the whole length of LIME; $^{218}$Po decaying at an increasing rate for higher Z, i.e., closer to the cathode; and $^{214}$Po being emitted entirely at the cathode level. This further confirmed the initial expectations of the presence of $^{222}$Rn in LIME.

A preliminary study was furthermore performed to define LIME energy response at the $MeV$ scale employing the measured $^{222}Rn$ and its progeny recoil alphas as calibrating sources. By applying a stringent selection based on charge saturation arguments, a linear relationship was found.

Overall, the "3D length spectroscopy" presented in this chapter has demonstrated the capabilities of the 3D analysis, both from a technical point of view and for the identification of backgrounds in LIME. These studies also demonstrated how LIME is highly sensitive to the geometrical properties of ionization tracks and the crucial role of 3D directionality and head-tail recognition in these types of detectors. The ability to identify the 3D shape and angle of particles in detectors enables the detection of problematic background isotopes in certain materials, which can then be tagged or removed during analysis. In this regard, the studies presented in this chapter are also expected to help prepare the next generation of CYGNO detectors by identifying the most relevant sources of background within the detector.

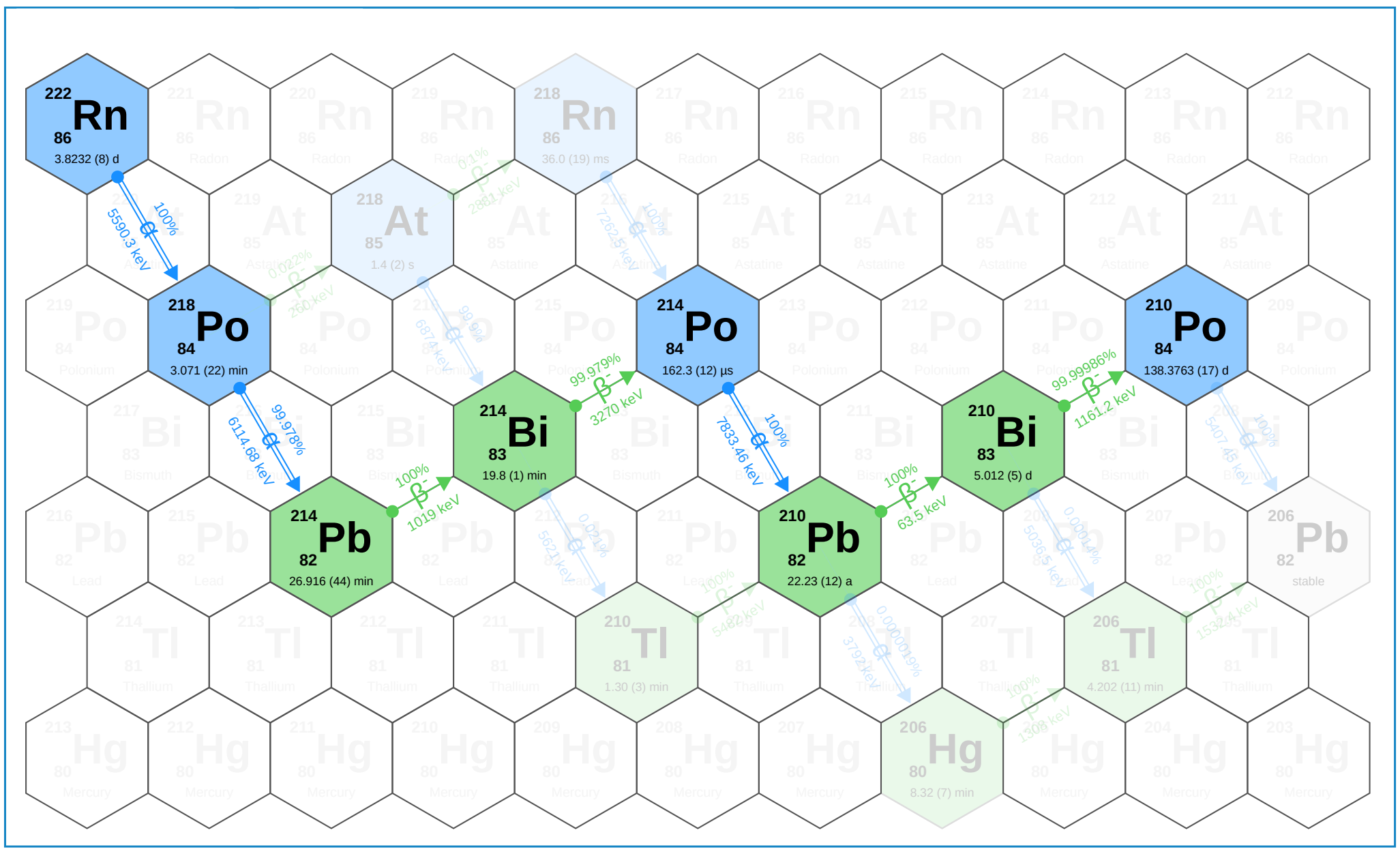

**Figure 6.27:** Decay chain of $^{222}$Rn. The probabilities and types of decay between isotopes are shown (blue for alpha and green for beta decays). The half-life of each element is also indicated. The energies correspond to the Q-values of the decays. Figure taken from [277].

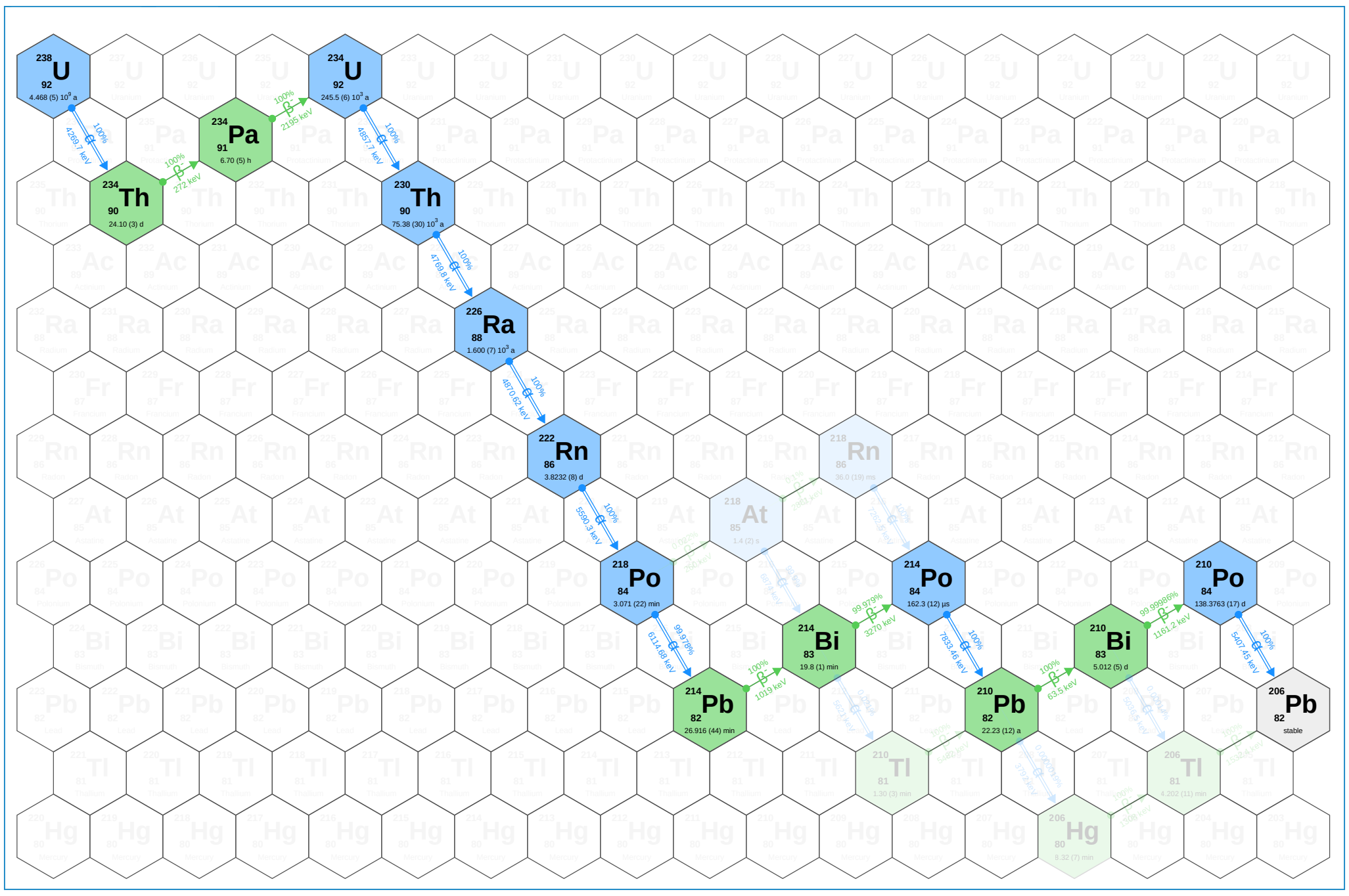

**Figure 6.28:** Decay chain of $^{238}$U. The probabilities and types of decay between isotopes are shown (blue for alpha and green for beta decays). The half-life of each element is also indicated. The energies correspond to the Q-values of the decays. Figure taken from [277]

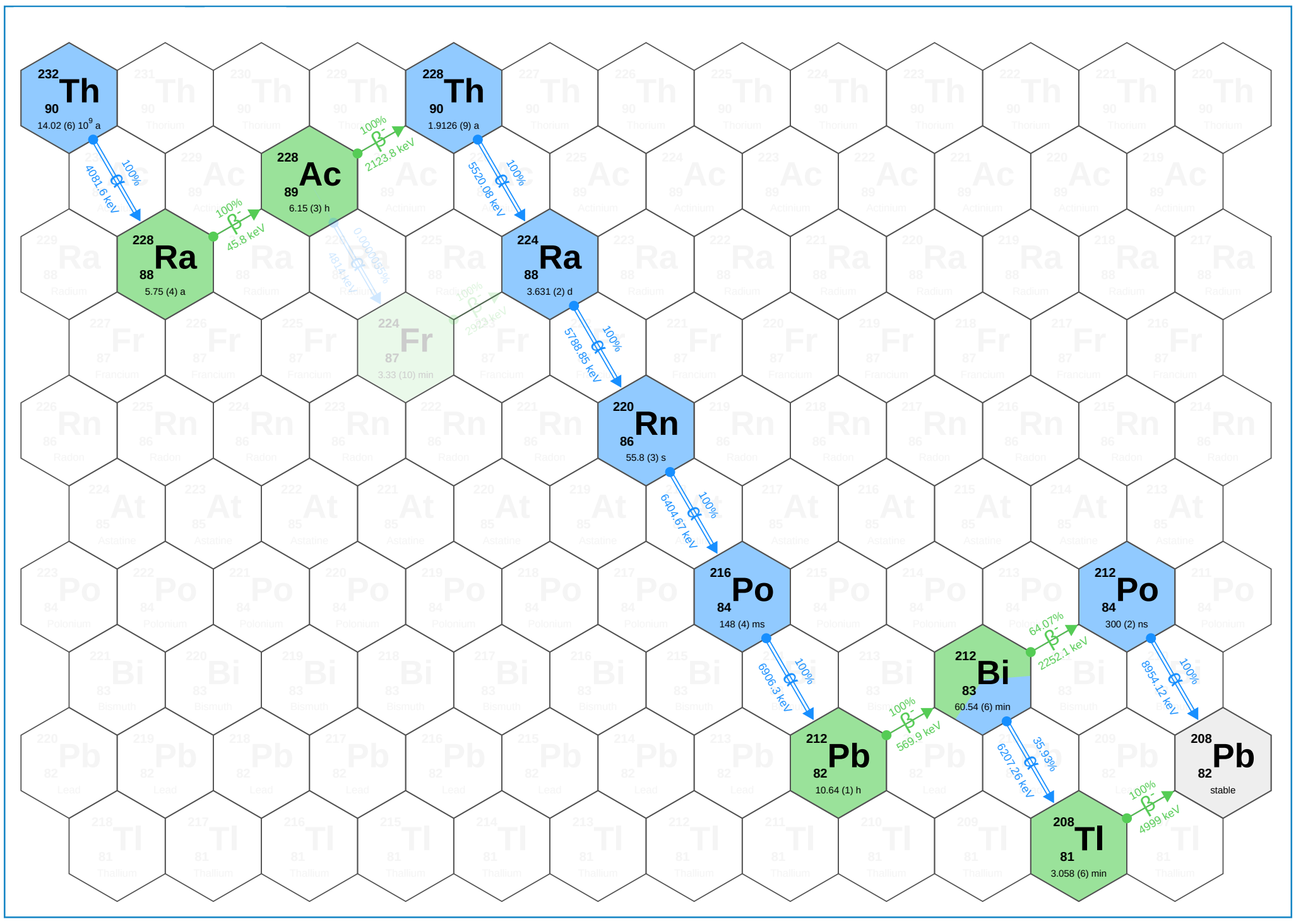

**Figure 6.29:** Decay chain of $^{232}$Th. The probabilities and types of decay between isotopes are shown (blue for alpha and green for beta decays). The half-life of each element is also indicated. The energies correspond to the Q-values of the decays. Figure taken from [277]

Chapter 7

# Negative Ion Drift operation with MANGO

This doctoral research was supported by the INITIUM project (see Section 2.3), funded through the European Union's Horizon 2020 research and innovation programme under the European Research Council (ERC) Grant Agreement No. 818744. Within this framework, part of the work presented in this thesis focused on implementing Negative Ion Drift (NID) operation with the MANGO prototype (see Section 2.2.1) at LNGS, with an emphasis on the analysis of PMT waveforms acquired in NID conditions.

In experiments with large gaseous detectors, the diffusion of charge carriers (typically electrons) during drift is a limiting factor in achieving high spatial resolution. This is because the longer the drift distance, the more spread will be the initial ionization cloud once it arrives and the amplification and/or readout planes, therefore degrading the ability to reconstruct its shape and profile. One approach to mitigate this issue is Negative Ion Drift, a variation of the typical TPC operation, which involves adding a highly electronegative dopant to the gas mixture [178,179].

The introduction of an electronegative gas causes electrons produced by the initial ionization to be captured at very short $\mathcal{O}(\mu m)$ distances by the electronegative molecules, forming negative ions. These ions drift toward the anode similarly to electrons but experience less diffusion, thus retaining better the spatial characteristics of the ionization cloud. Some electronegative gases also allow the formation of different ion species. From the drift time differences of these species, it is possible to determine the event position along the drift direction, thus offering the capability of detector fiducialization. These characteristics make NID particularly well suited for detector applications that require high-resolution track reconstruction over large volumes, such as rare low-energy event searches.

NID operation is typically employed at low pressures, below 100 $mbar$, as discussed in Section 7.1.2. In this chapter, for the first time, NID operation at atmospheric pressure with an optical readout is presented. This is achieved by adding a small percentage of $SF_6$ (1.6%) to the standard CYGNO $He$:$CF_4$ 60:40 gas mixture. The ability to operate in NID mode at atmospheric pressure within a large detector provides improved tracking capabilities – due to reduced diffusion – while also maintaining a large exposure for competitive sensitivities in the DM realm.

In this chapter, the preliminary results of the studies performed with MANGO

within the NID framework are presented. The chapter begins with a brief description of the advantages and the state-of-the-art of NID operation (Section 7.1), followed by the description of the first observation of NID operation at atmospheric pressure (Section 7.2). Then, the initial studies performed with a longer version of MANGO are discussed (Section 7.3), together with the original preliminary analysis developed for the purpose of this thesis to study the NID PMT signals. Finally, some conclusions and prospects from these studies are presented (Section 7.4).

## 7.1 NID operation

The addition of an electronegative species to the TPC gas mixture has the potential to modify its operation by changing the charge carriers from the ionization track from electrons to negative ions, giving rise to the term Negative Ion Time Projection Chamber (NITPC) [178]. In a NITPC, the primary electrons generated by ionization events along the particle path are quickly captured by the electronegative molecules present in the gas, leading to the formation of negative ions within a few micrometers (typically $< 10$ – $100\ \mu m$). As in a regular TPC, these anions then drift toward the anode under the influence of an applied electric field, serving as the primary carriers for the track image rather than the electrons themselves. When the anions reach the anode, where the typical charge amplification processes occur, the high electric fields present cause the release (or detachment) of the molecule's additional electron, triggering the standard electron avalanche process.

### 7.1.1 Advantages

This modification of the traditional working principle of a TPC, i.e., using anions instead of electrons to carry the charge information from the ionization track to the amplification and readout planes, brings two main advantages over regular electron drift (ED) TPCs: namely, a reduction in the diffusion felt by the drifting charges, and the ability to fiducialize the detector. These two advantages are described in this section.

#### Reduced diffusion

For any typical TPC operation, the diffusion of charge carriers is expected to be proportional to the square root of the average energy of the drifting particles [273]:

$$\sigma_D = \sqrt{\frac{4\epsilon L}{3eE}} \tag{7.1}$$

where $\sigma_D$ represents the 3D RMS diffusion spread of carriers with average energy $\epsilon$ and charge $e$, after drifting a distance $L$ in an electric field $E$. Electrons are several orders of magnitude lighter than the components of the gas, which results in inefficient momentum transfer during collisions. As a result, when electrons collide with the gas molecules, their direction is randomized, but they retain most of the energy acquired from the electric field between collisions. This component of their energy is the dominant contributor to their total energy, far surpassing the thermal energy. Anions, on the other hand, have much larger masses, comparable to those of the neutral gas molecules, leading to a much more efficient energy transfer with the surrounding medium. As a result, the effective value of $\epsilon$ is significantly reduced in each collision, allowing the diffusion ($\sigma$) experienced by the electron cloud to approach the thermal limit [273]. The thermal limit for diffusion in TPCs represents the minimum diffusion that results from thermal motion alone, where the particles are in thermal equilibrium with the surrounding medium. In this limit, the diffusion is unaffected by external fields, and it typically represents the best possible diffusion performance for a given gas under low-field or field-free conditions.

This reduction in the diffusion suffered by anions during drift represents a significant improvement over the classical electron drift behavior. Minimizing diffusion significantly improves the detector's directionality capabilities, as the topology of the ionization cloud is less distorted. Since diffusion also depends on the drift distance ($L$) in Equation 7.1, the use of NID also provides the opportunity to increase the maximum drift length in a TPC while maintaining low diffusion. Depending on the gas mixtures, diffusion in NID gases on the order of $70 - 80\ \mu m/\sqrt{cm}$ [178,179] can be achieved, representing a reduction of 3 to 5 times that of ED operation with cold gases (as $CF_4$, see Section 2.1.1) in similar setups [149].

### Fiducialization

Fiducialization refers to the ability to identify the 3D position where an ionization event occurred within a TPC. In rare event searches, this information is typically used to identify and reject background events occurring near the borders of the detector.

In NID operation, it has been shown that different species of negative ions can be created during the initial attachment process of the primary electrons [278]. Since there is typically a main ion responsible for transporting the majority of the charge, the secondary ions are often referred to as *minority charge carriers*. Following Figure 7.1, the drift velocity of an ion depends on its mass, so each of the anions created during the attachment process (Figure 7.1a) will take a different amount of time to arrive at the amplification and/or readout planes (Figure 7.1b). The difference in time of arrival between anions ($\Delta T$, Figure 7.1c) can be used to deduce the absolute position of the event along the drift direction ($Z$), through the following expression [279]:

$$Z = \frac{v_s \cdot v_p}{v_s - v_p} \cdot \Delta T \tag{7.2}$$

where $v_p$ and $v_s$ are, respectively, the drift velocities of the primary and secondary ions.

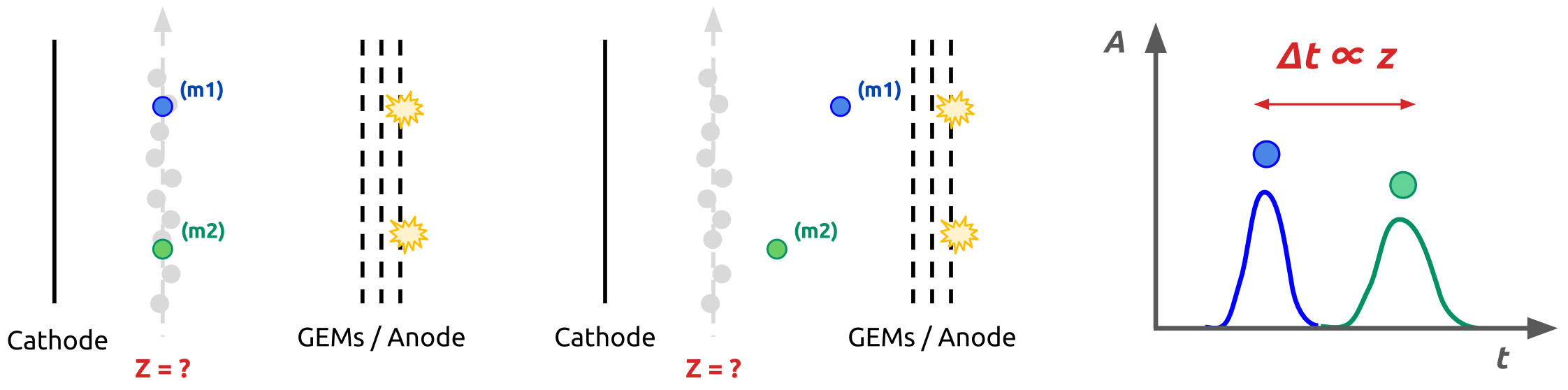

(a) (b) (c)

**Figure 7.1:** Schematic illustrating the potential of minority charge carriers for detector fiducialization in a NITPC: (a) a particle creates an ionizing track where different ions (with different masses) are created at an unknown Z; (b) the ions drift towards the amplification region at different velocities; (c) the PMT (or other fast time-resolving sensor) waveforms show two time-of-arrival peaks, each corresponding to a different ion. The time difference between the two is proportional to the absolute Z position of the event inside the detector.

This technique was demonstrated by D. P. Snowden-Ifft in [278] through a charge-sensitive amplification and readout system (a multi-wire proportional chamber) with $CS_2$ and $O_2$, at 40 Torr. By analyzing the time-of-arrival differences between the anions created by the passage of alpha particles, the author was able to reconstruct the absolute position of the event in the drift direction with an RMS $<$ 1 cm.

### 7.1.2 State-of-the-art

The concept of employing Negative Ion Drift to reduce diffusion in TPCs was first explored in 2000 by Martoff, Snowden, Ohnuki, and Spooner when they added $CS_2$, an electronegative gas, to xenon and argon gas mixtures in a MultiWire Proportional Chamber operating at low pressures of 40 $\mathrm{torr}$ [178]. In this pioneering work, drift velocities on the order of $\mathrm{cm/ms}$ were observed, consistent with the speeds of ions (around $10^3$ times slower than electrons). Additionally, extremely low diffusion coefficients (below 100 $\mu\mathrm{m}/\sqrt{\mathrm{cm}}$) were also reported for drift fields above 235 $\mathrm{V/cm}$, demonstrating reduced diffusion compared to traditional electron drift gas mixtures (see section Section 2.1.1).

In subsequent studies, the negative ion properties of $CS_2$ were investigated in combination with various gases such as $\mathrm{He}$, $CF_4$, and Ar at 40 Torr using MWPCs, further confirming the ionic nature of the charge carriers and their reduced diffusion [179,280,281]. NID operation was also successfully demonstrated at near-atmospheric pressures with mixtures like $\mathrm{He}$:$CS_2$ in MWPCs [282], and more recently, at 1030 mbar with a gas mixture of $\mathrm{Ar}$:$\mathrm{i}C_4H_{10}$:$CS_2$ coupled with GridPix charge readout [283].

NID operation with a gas mixture of $CS_2$:$CF_4$:$O_2$ (30:10:1 Torr) was successfully employed in the DRIFT experiment. Through the fiducialization capabilities offered by the measurement of the arrival times of minority charge carriers, the DRIFT experiment managed to reject all background nuclear recoils coming from the cathode and achieve a background-free DM limit [122].

Compared to $CS_2$, which has high vapor pressure and toxicity [284], $SF_6$ is safer to handle and has higher electronegativity. $O_2$, depending on the mixture, can become flammable and/or explosive [285]. $SF_6$, on the other hand, is a non-toxic and non-flammable gas [286], which makes it, at first glance, a better choice for the study of negative ion drift in Negative Ion TPCs. $SF_6$ is a well-known electronegative gas, mostly used to contain large discharges or streamer avalanches, namely in Resistive Plate Chambers. Due to its high electronegativity and large fluorine content (and thus sensitivity to spin-dependent WIMP-nucleon interactions, see Section 1.3.1.2), $SF_6$ has gained increasing interest in DM searches. NID operation in pure $SF_6$ was first demonstrated in 2016 at pressures between 20 and 40 Torr, using a 400 $\mu m$ thick GEM setup [287]. This demonstrated that operating this inert and non-toxic gas (compared with the highly toxic $CS_2$) was viable for NID studies, also proving the feasibility of fiducialization through minority carriers with this gas. The NEWAGE collaboration also performed studies with pure $SF_6$ at 20 Torr using GEMs, achieving 130 $\mu m$ resolution for alpha particle fiducialization [279]. More recently, NID operation in $SF_6$ was tested with triple thin GEMs and TimePix2 charge readout at pressures up to 610 Torr using a $He{:}CF_4{:}SF_6$ mixture (360:240:10). In this setup, despite the small amount of $SF_6$, NID operation was observed, and the mobility of the $SF_6^-$ ion was measured [288].

In the context of the CYGNO/INITIUM experiment, reducing diffusion greatly improves tracking and directionality capabilities. Since the charge gain and luminescence properties of the $He{:}CF_4$ mixture are crucial for detector operation (see Section 2.1.1), the studies carried out with CYGNO detectors and presented in this chapter involved introducing only a small percentage of 1.6% of an electronegative gas – $SF_6$ – into the CYGNO gas mixture to avoid altering too much these properties, following previous studies that demonstrated NID operation in $He{:}CF_4$ [288].

#### 7.1.2.1 SF6 properties

The pioneering work with $SF_6$ in 2017 by S. Phan et al. [287] demonstrated that $SF_6$ is a worthwhile candidate for negative ion drift, as it showed the possibility of signal amplification with high gains and fiducialization through the presence of minority $SF_5^-$ charge carriers. Given its ready availability, $SF_6$ has been the preferred choice for many recent works in the context of NID, including those performed within the INITIUM project and presented in this thesis. It should nonetheless be noted that $SF_6$ is one of the most harmful gases in terms of greenhouse effect potential [289], and for that reason, the scientific community encourages its replacement in large-scale experiments with less harmful alternatives. With this in mind, the use, handling, and disposal of $SF_6$ in the studies presented here have been carefully considered.

When an ionizing particle crosses the sensitive volume of a TPC, it leaves a trail of electron-ion pairs along its path. If a small quantity of $SF_6$ is present in the gas, the released electrons are quickly captured by the $SF_6$ molecules within a few micrometers of the initial ionization position [206, 290–292]. The process of electron capture – or *attachment* – in $SF_6$ leads to the creation of a metastable excited state of the ion $SF_6^{-*}$, following the process:

$$e^- + SF_6 \rightarrow SF_6^{-*} \qquad \text{(attachment)} \tag{7.3}$$

The cross-section for this process is shown in Figure 7.2a and, as visible, it peaks at approximately 0.1 eV, then rapidly decreases as the electron energy increases, with the cross-section dropping by a factor of $10^3$ at 0.18 eV. The $SF_6^{-*}$ ion then stabilizes either through radiative or collisional stabilization, or undergoes electron autodetachment, following:

$$SF_6^{-*} + SF_6 \longrightarrow SF_6^- + SF_6 \qquad \text{(collisional stabilization)} \tag{7.4}$$

$$SF_6^{-*} \longrightarrow SF_6^{(*)} + e^- \qquad \text{(auto-detachment)} \tag{7.5}$$

In the literature, the mean lifetime of the excited state $SF_6^{-*}$ is not consistent across studies, with reported values ranging from the µs to ms scale [290–292], although all indicate lifetimes exceeding 1 µs. In contrast, the average time between particle collisions in a gas is on the order of nanoseconds, which favors the process of collisional stabilization (Equation 7.4) over electron autodetachment (Equation 7.5), with the latter being counterproductive to the desired NID effect in a TPC.

As mentioned, the presence of minority charge carriers when using $SF_6$ has also been demonstrated. Specifically, the most probable minority ion species originating from $SF_6$ is $SF_5^-$, which is produced through dissociative electron attachment. In this process, an electron is captured by an $SF_6$ molecule, resulting in the breaking of chemical bonds and the ejection of a fluorine atom from the molecule, following [206]:

$$e^- + SF_6 \longrightarrow SF_5^- + F \qquad \text{(dissociative electron attachment)} \tag{7.6}$$

The cross-section for this process is shown in Figure 7.2b. As visible, a local maximum occurs at 0.38 eV, followed by a rapid decrease at higher electron energies. By analyzing the cross-sections for the production of $SF_6^-$ and $SF_5^-$, it becomes clear that these two processes are in competition, with $SF_6^-$ formation generally favored – particularly under the typical electric drift fields used in TPCs (on the order of hundreds of V/cm), where the energy gained by electrons before being absorbed promotes the formation of $SF_6^-$, especially at low collision energies.

The relative abundance of $SF_6^-$ and $SF_5^-$ ions generated during ionization thus depends on the energy acquired by electrons between collisions. This energy, in turn, depends on the gas density (which affects the mean free path) and the applied electric field (or more precisely, the reduced electric field, $E/p$), with higher energies favoring the formation of $SF_5^-$. At higher impact energies, dissociative attachment can also lead to the formation of other ion species such as $SF_4^-$, $SF_3^-$, or $F^-$, as discussed in studies such as [206,293]. In addition to these, other mechanisms – such as charge transfer processes or the formation of $SF_6$ molecular clusters around the main anion – can further complicate the charge transport process involving $SF_6^-$ ions [294]. However, these effects are generally considered to be of second-order importance in the overall signal formation. A summary of the processes discussed so far, along with other relevant mechanisms, is presented in Table 7.1.

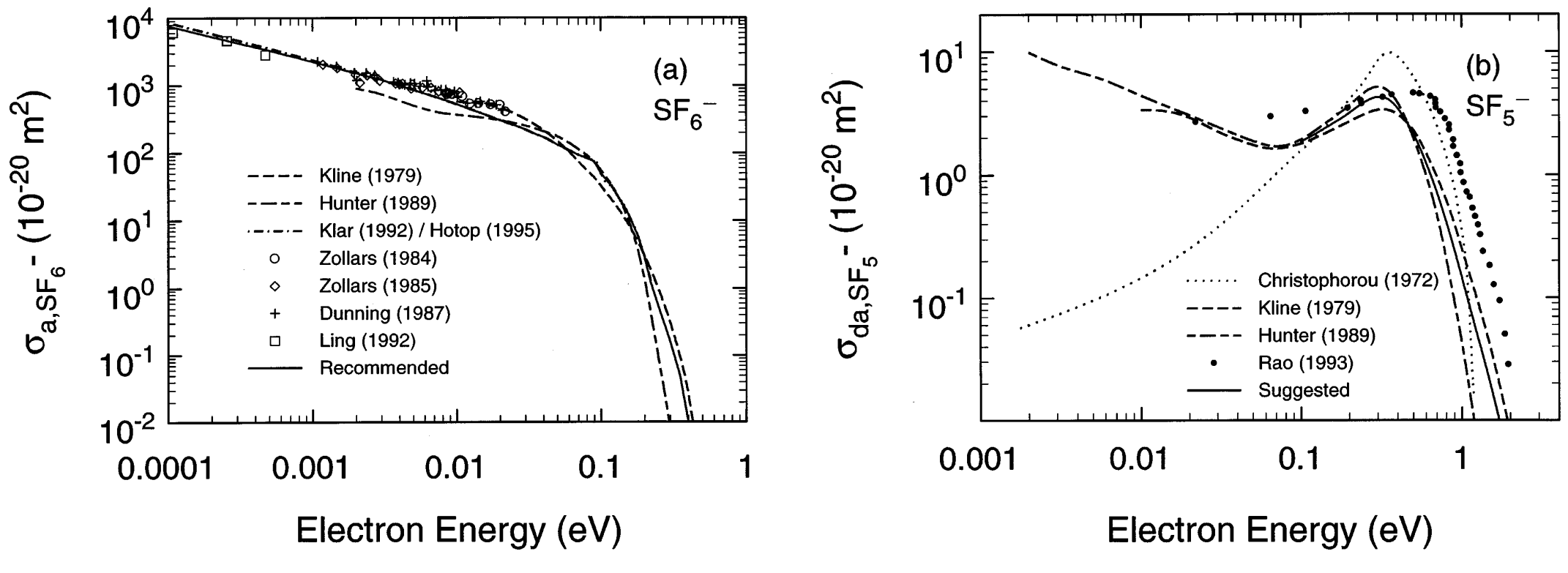

(a) (b)

**Figure 7.2:** Cross sections for the formation of (a) $SF_6^-$ through electron attachment, and (b) $SF_5^-$ through dissociative electron attachment, as a function of the electron energy. Retrieved from [206].

Once these negative ions are created, they drift toward the anode under the electric field applied in the TPC, just as electrons would. However, when the anions reach the amplification stage, it is necessary to strip the extra electron from the ion in order to initiate a standard electron avalanche. This requirement introduces an intrinsic reduction in charge gain for NID operation compared to normal ED, as it adds an additional step – electron detachment – before amplification can begin. The detachment processes have been studied in [206,295], which identified collision detachment, particularly processes VIII and IX in Table 7.1, as having the largest cross sections for removing the extra electron from $SF_6^-$. These processes typically require much higher energies – on the order of $\mathcal{O}(10)$ eV in the center-of-mass frame – to occur. Such energies can be reached in strong electric fields, such as those found inside GEM holes (on the order of $\mathcal{O}(40-60)$ kV/cm, see Section 2.1.2). Under these conditions, charge gains of $10^{3-4}$ have been achieved in NID operation [287,296].

All the interactions and cross-sections discussed up to this point are specifically for pure $SF_6$ gas. In a gas mixture such as the one used in CYGNO, which also includes a large proportion of He and $CF_4$, additional interactions and processes will affect the formation rate of each ion species. Variations in gas properties such as temperature, pressure, or component ratios will also influence these processes and their respective cross sections, potentially favoring some over others. Due to these factors, providing a complete description of all processes involved in NID operation for the specific case of the gas mixture used in these studies is challenging, especially since the cross sections for the interactions between $SF_6$ and He:$CF_4$ have not been studied or measured yet. Among the different possible processes are the identification of which ions are initially created during ionization, how they interact while drifting, and which mechanisms dominate electron detachment at the amplification stage. In this chapter, preliminary conclusions are drawn based on the current understanding of $SF_6$ behavior within the context of NITPCs.

**Table 7.1:** Summary of the reactions behind the creation of ion from the $SF_6$ molecules. Adapted from [206,293].

| ID | Reaction | Process | Energy |
|---|---|---|---|
| I | $e^- + SF_6 \rightarrow SF_6^{-*}$ | Electron attachment | $< 1$ meV |
| II | $SF_6^{-*} \rightarrow SF_6 + e^-$ | Autodetachment | |
| III | $SF_6^{-*} + SF_6 \rightarrow SF_6^- + SF_6$ | Collisional stabilization | |
| IV | $e^- + SF_6 \rightarrow SF_5^- + F$ | Dissociative electron attachment | 0 - 2 eV |
| V | $e^- + SF_6 \rightarrow SF_4^- + 2F$ | | 3 - 8 eV |
| VI | $e^- + SF_6 \rightarrow F^- + SF_5$ | | 1 - 14 eV |
| VII | $e^- + SF_6 \rightarrow F_2^- + SF_4$ | | 1 - 14 eV |
| VIII | $SF_{5/6}^- + SF_6 \rightarrow SF_{5/6} + SF_6 + e^-$ | Collisional detachment | ~ 90 eV |
| IX | $F^- + SF_6 \rightarrow F + SF_6 + e^-$ | | ~ 10 eV |
| X | $SF_6^- + SF_6 \rightarrow SF_6 + SF_6^-$ | Charge transfer | |
| XI | $SF_6^- + SF_6 \rightarrow SF_5^- + F + SF_6$ | Dissociative charge transfer | |

## 7.2 NID operation at atmospheric pressure

The detector used for the NID operation studies in a CYGNO-like setup is the MANGO prototype, described in Section 2.2.1. The drift gap was set to 5 cm – the maximum allowed by the gas-tight acrylic internal vessel. The field cage was made of 1 mm silver wires, embedded in 0.5 cm thick polycarbonate field rings with a 7.4 cm inner diameter, spaced at 1 cm intervals. One end of the field cage is enclosed by a cathode consisting of a thin copper layer deposited on a PCB, and the other by the GEM amplification stage. The silver wires are interconnected with 1G$\Omega$ resistors to evenly distribute the voltage supplied by an external CAEN N15701 power supply, ensuring a uniform drift field.

For the initial and first-ever studies of NID operation in MANGO, the detector was operated at the atmospheric pressure of LNGS (900 ± 7$mbar$), using a $He$:$CF_4$ 60:40 gas mixture, for electron drift (ED), and a $He$:$CF_4$:$SF_6$ 59:39.4:1.6 mixture, for negative ion drift (NID), based on the findings in [288]. The detector ran in continuous gas flow mode to ensure a high purity gas mixture and stable operating conditions. A $^{241}Am$ source, which emits 5.485 $MeV$ alpha particles was used for various tests. Specifically, is was placed between the field cage rings, facing the detector's active volume in order to generate ionization tracks inside the detector. The choice to use alpha particles for the studies conducted is connected to the lower gas gain observed in these initial tests. Lower gains are expected when operating with negative ions, as discussed in Section 7.1.2.1, due to the additional step required at the amplification stage – namely extracting the extra electron from the ion – before initiating the standard electron avalanche. Alpha particles also produce straight and relatively short tracks that can be fully contained within the MANGO sensitive volume, thus facilitating data analysis. The GEMs were operated at 310 $V$ each in the ED case, and at 550/545/540 $V$ for GEM1/GEM2/GEM3 for the NID one. The choice of the GEM voltages for the ED case was dictated by the aim to obtain comparable light yields in the two cases.

### 7.2.1 Data acquisition strategy

Given the novelty of these studies and the constraints related to the acquisition of the NID PMT signal (discussed below in Section 7.3.1), the information from the two CYGNO optical sensors was recorded and analyzed separately. Specifically, the PMT and the output of a CAEN A422A charge sensitive preamplifier connected to the bottom electrode of last GEM amplification plane were acquired through a highly performing Teledyne LeCroy oscilloscope. The Orca Fusion CMOS images were collected by the custom Hamamatsu software Hokawo 3.0 in free running mode and with 0.5 $s$ exposure.

For what concern the PMT signal, in the ED case they were acquired, through an Teledyne LeCroy oscilloscope with a negative edge threshold of 40 $mV$ on the PMT waveform amplitude and a sampling of 400 $ps/pt$. Figure 7.4b shows an example of an alpha track PMT waveform for ED, displaying a typical time extension of hundreds of $ns$, as expected from the $\mathcal{O}(cm/\mu s)$ ED drift velocity.

The data acquisition of NID PMT signal resulted significantly more challenging. The fast response of the PMT, combined with the slow drift velocity of NID anions, results in a very sparse signal that is significantly different from the ED signals. An example of these signals is shown in Figure 7.4c, which clearly demonstrates that these are indeed ions, given the $\mathcal{O}(ms)$ duration of the signals. Instead of a typical waveform, NID PMT signals consist of a series of thousands of small peaks, with amplitudes ranging from 5 to 30 $mV$ and time durations of a few nanoseconds, extending over several milliseconds (depending on the drift field applied). These peaks are spaced hundreds of nanoseconds or even microseconds apart. We hypothesize that each of these small peaks corresponds to the arrival of one or a few primary ionization clusters at the amplification plane. The time separation, due to the slow anion drift velocity, might allow for their individual identification. Although this topic is still under study, it is important to highlight how this feature could enable cluster counting, an analysis technique that could offer superior energy resolution and particle identification performance [297], particularly in He-based detectors [298, 299].

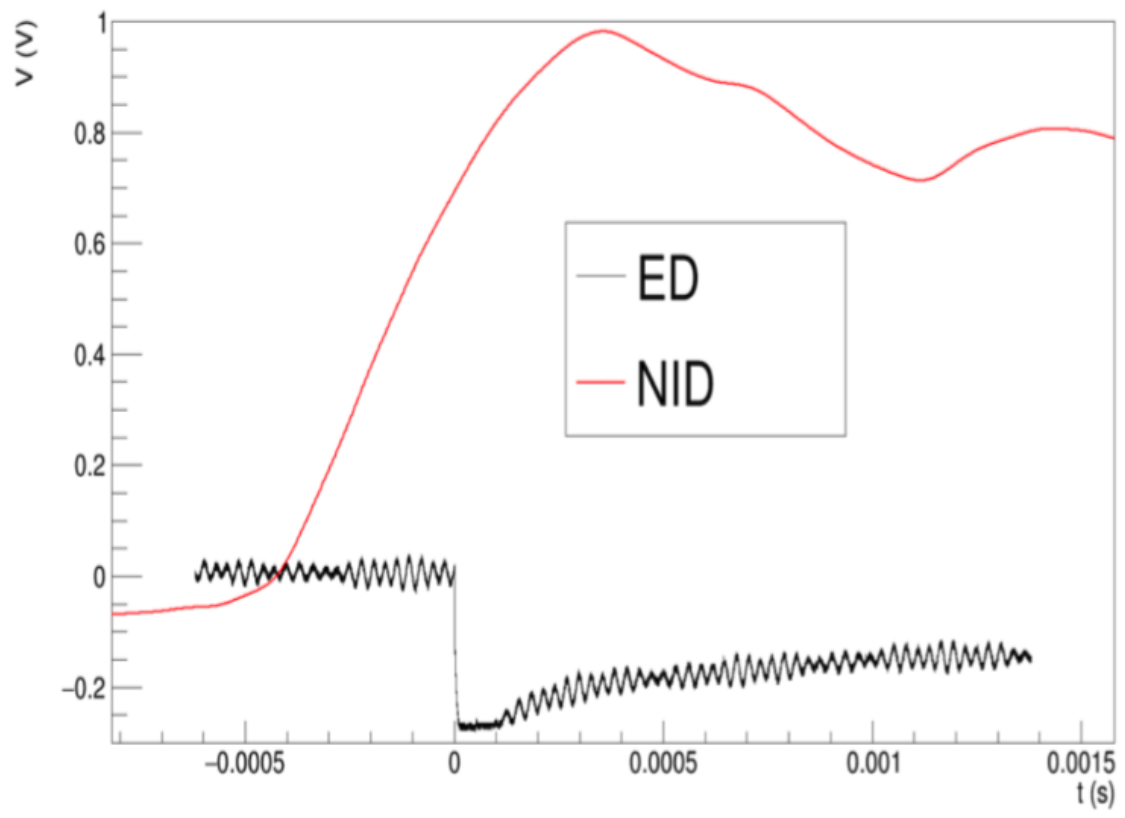

**Figure 7.3:** Overlay of signals acquired through the charge preamplifier connected to the last GEM in the MANGO amplification region, highlighting the significant temporal difference between ED and NID signals.

Since each of these small signal peaks closely resembles those caused by single photoelectron background, a simple threshold-based trigger applied directly to the PMT is ineffective for acquiring NID waveforms. In the first observations of NID signals, nothing was visible in the PMT waveforms, while clear signals appeared in the CMOS, undoubtedly due to negative ion drift. Several attempts were made to develop a suitable trigger mode for detecting these signals in the PMT, but all were unsuccessful due to the too high resemblance of each peak to noise. For this reason, a different trigger strategy was developed to acquire NID PMT waveforms based on the output of a CAEN A422A charge sensitive preamplifier (decay time of 300 μs) connected to the bottom electrode of the last GEM. The signal produced at the GEM reflects the movement of charges toward the amplification plane and can be seen as an integrated version of the PMT signal, thereby making it suitable for triggering. Consequently, in the studies conducted, NID events were recorded by applying a 250 mV positive-edge threshold to the output of the GEM preamplifier signal. In the case of an alpha track, this setup generates a long and continuous signal due to the extended decay time of the preamplifier, allowing a reliable triggering method. An example of the GEM preamplifier signal for both ED and NID is shown in Figure 7.3. This clearly illustrates the much lower drift velocity in NID operation compared to ED – a difference of nearly three orders of magnitude – which results in a significantly larger rise time of the preamplifier and an overall longer signal duration.

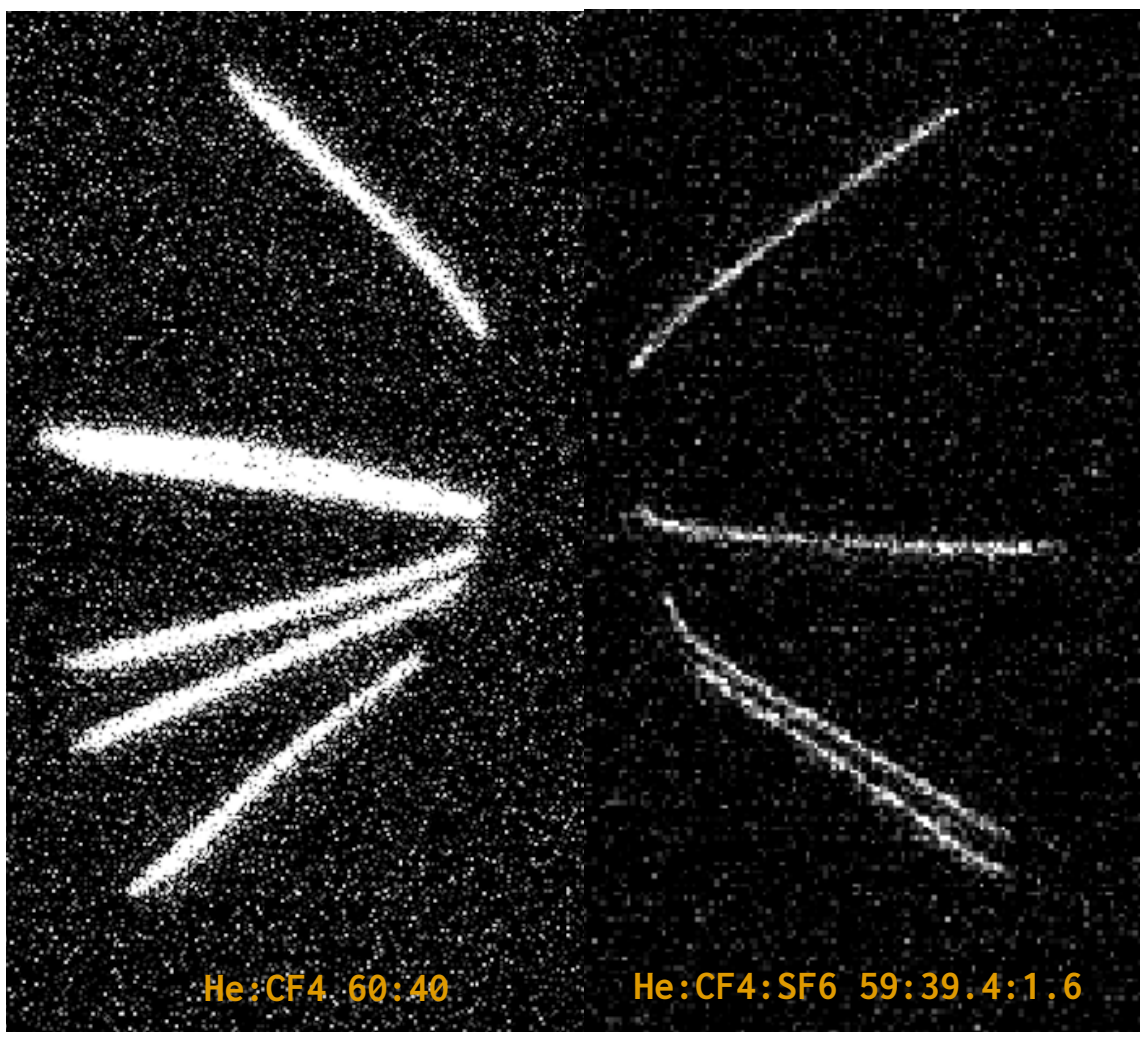

**(a)** CMOS image

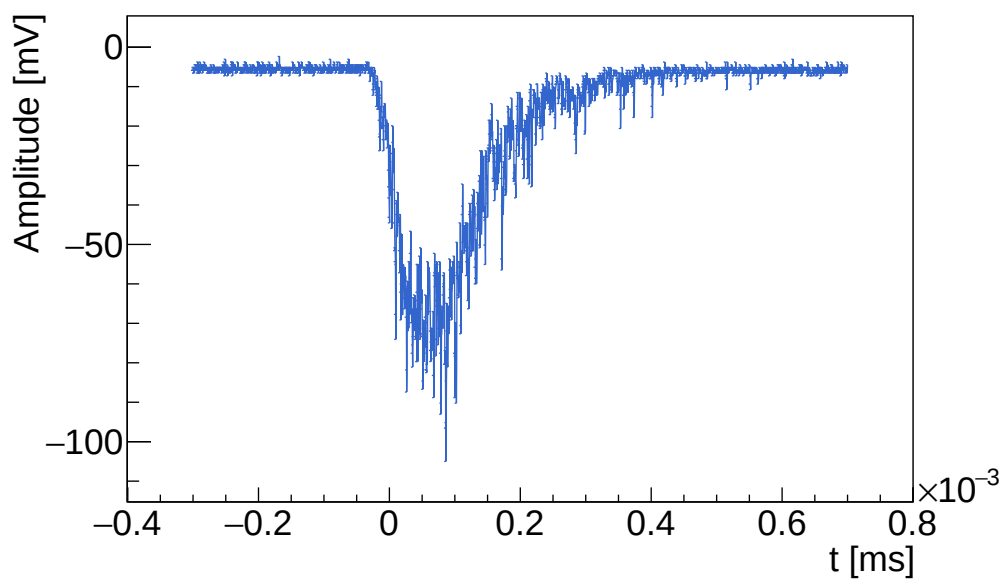

**(b)** ED PMT signal

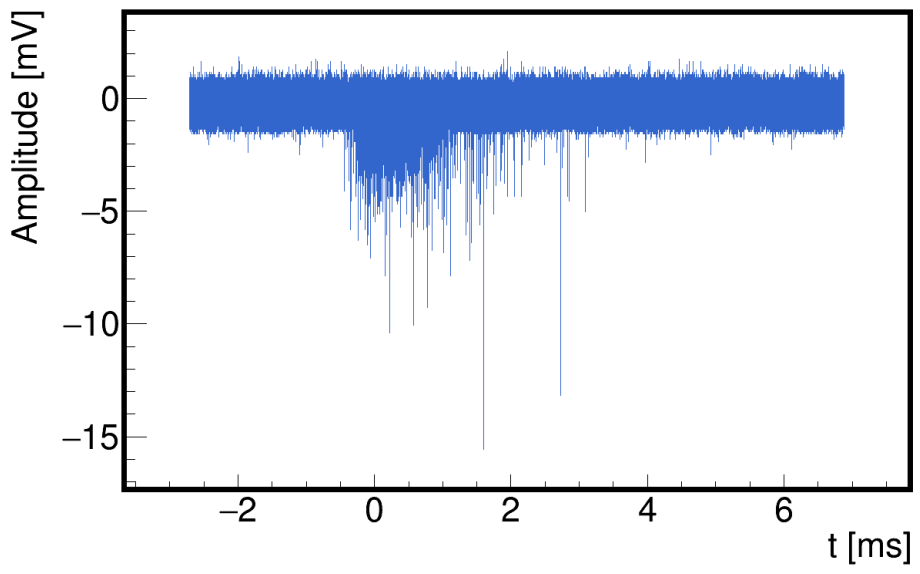

**(c)** NID PMT signal

**Figure 7.4:** Example of the difference between electron and negative ion drift operation, at LNGS atmospheric pressure (900 mbar), as seen by the (a) CMOS, and (b,c) PMTs.

### 7.2.2 Initial observations

As mentioned, the first indication of NID operation came from the CMOS images. Figure 7.4a shows an example of a 0.5 s exposure sCMOS image taken with the MANGO detector with a $^{241}Am$ source positioned 2.5 cm from the GEM plane, operated with $He{:}CF_4$ 60:40 (left) and $He{:}CF_4{:}SF_6$ 59:39.4:1.6 (right) at LNGS atmospheric pressure (900 mbar). In both images, the total light in each track was measured using the CYGNO CMOS analysis toolkit described in Section 3.1, and the light intensity for the alpha particles in both scenarios was found to be similar. However, the difference in diffusion experienced by the alpha tracks during drift in the two gas mixtures is clearly visible, marking the first strong indication of NID operation in MANGO. In parallel, the PMT signals were also recorded using the trigger methodology described in Section 7.2.1.

In order to further validate our hypothesis of NID operation with the $He{:}CF_4{:}SF_6$ mixture, PMT data were acquired varying the drift field between 200 and 600 V/cm. An example of PMT waveforms at such different fields are displayed in Figure 7.5, further supporting the interpretation of genuine NID signals. As evident in the figure, in fact, increasing the electric field results in a reduction of the temporal width of the PMT signals. This behavior indicates that the ionization charges arrive at the amplification region within a narrower time window, reflecting the higher fields strengths, while still remaining on the millisecond scale. These observations confirm both the presence of negative ions and the interpretation of the observed PMT signals, marking the first evidence ever obtained of NID operation at atmospheric pressure with an optical readout.

Additionally, to the author knowledge, this is the first time the drift of negative ions in a TPC is observed through a photomultiplier. This created both an opportunity and a challenge to develop an innovative and tailored PMT analysis to extract meaningful information from these signals, as will be illustrated in Section 7.3.1. While these first tests demonstrated undoubtedly the feasibility of NID operation at atmospheric pressure with an optical readout, realizing a decisive breakthrough for directional DM searches and more in general low diffusion TPC applications, the maximum 5 $\mathrm{cm}$ drift length attainable within the standard MANGO gas-tight acrylic internal vessel results too short to fully validate the preliminary NID PMT waveform algorithm also as a function of the drift distance. For this reason, an alternative setup featuring a longer field cage was realized and the PMT waveform analysis was developed and applied to the data acquired within this, as will be illustrated in Section 7.3.

## 7.3 NID operation inside a vacuum vessel

In order to extend the preliminary results presented in Section 7.2 to longer drift distances, the MANGO amplification stage was equipped with a 15 $\mathrm{cm}$ long field cage – built exactly as the short 5 $\mathrm{cm}$ version employed for the measurements in Section 7.2 – and the TPC structure was installed in an already existing 150 $\mathrm{l}$ stainless steel vacuum vessel (Figure 2.15d, incidentally, the same used for the measurement with charge pixel readout reported in [288]). In this new configuration, the setup was renamed MANGOk (MANGO in a keg). The optical readout system (CMOS + PMT) is mounted outside the vacuum vessel and optically coupled to the amplification plane through a quartz window with 90% transparency. Due to geometrical constraints given by the vacuum vessel dimensions, the ORCA-Fusion camera is positioned at a distance of 26.6 $\pm$ 0.4 $\mathrm{cm}$ from the last GEM, and is focused directly on it. In this configuration, the camera images a 14.1 $\times$ 14.1 $\mathrm{cm}^2$ area, resulting in an effective pixel size of 61 $\times$ 61 $\mu\mathrm{m}^2$. Due to the increased distance between the photo-sensors and the GEM amplification plane, the solid angle coverage is reduced by approximately a factor of three. To compensate for this reduction in light yield, the operating pressure is lowered to 650 $\pm$ 1 $\mathrm{mbar}$, which increases the gain at the amplification region.

In this setup, data was taken at five different drift distances: 2.5, 3.5, 4.5, 6.5, and 9.5 $\mathrm{cm}$ from the GEMs, corresponding to different "holes" in the drift field cage (i.e., the spacing between the rings) and, for each distance, measurements were performed at multiple drift fields: 300, 350, 400, 500, and 600 $\mathrm{V/cm}$. The $^{241}\mathrm{Am}$ source was positioned to generate alpha tracks as close as possible to being perpendicular to the drift direction, thereby minimizing angular projection effects. Additionally, the source was further collimated to reduce the spread in the emission direction. The data acquisition strategy employed in these measurements was the same as the one reported earlier in

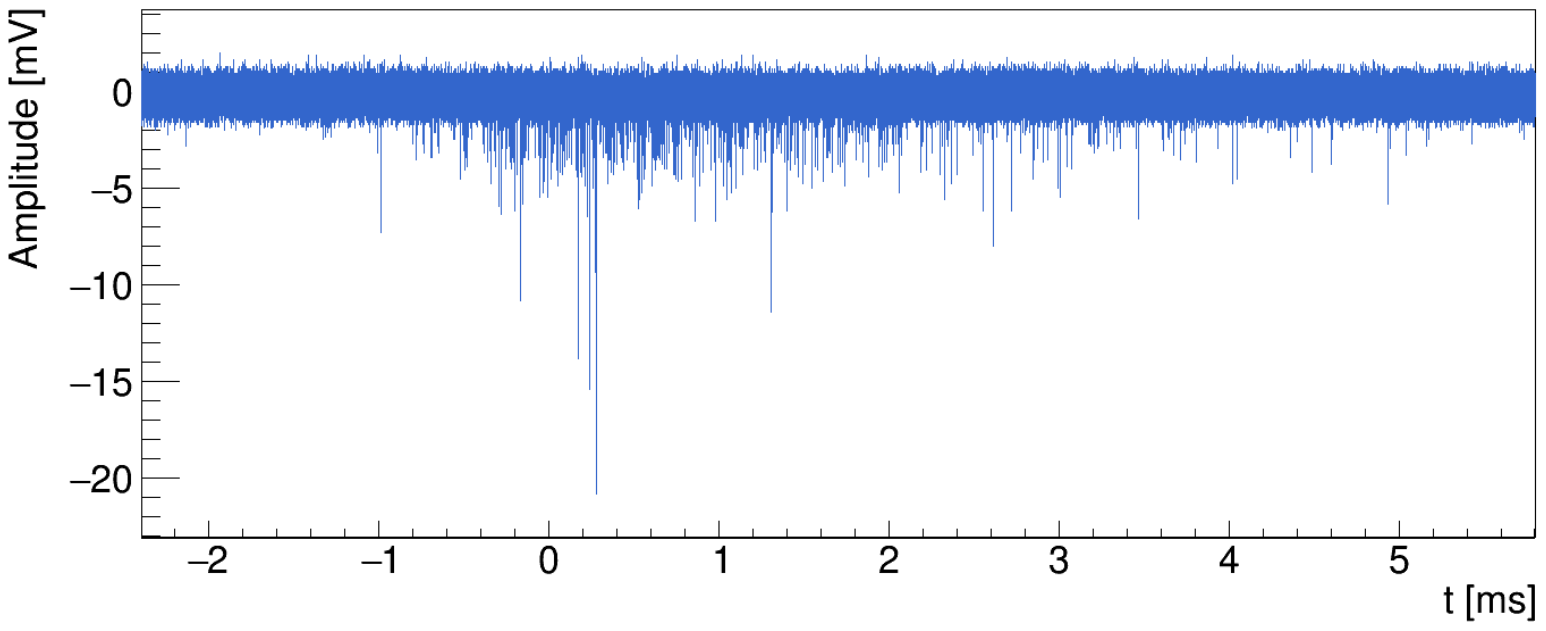

**(a)** 200 V/cm.

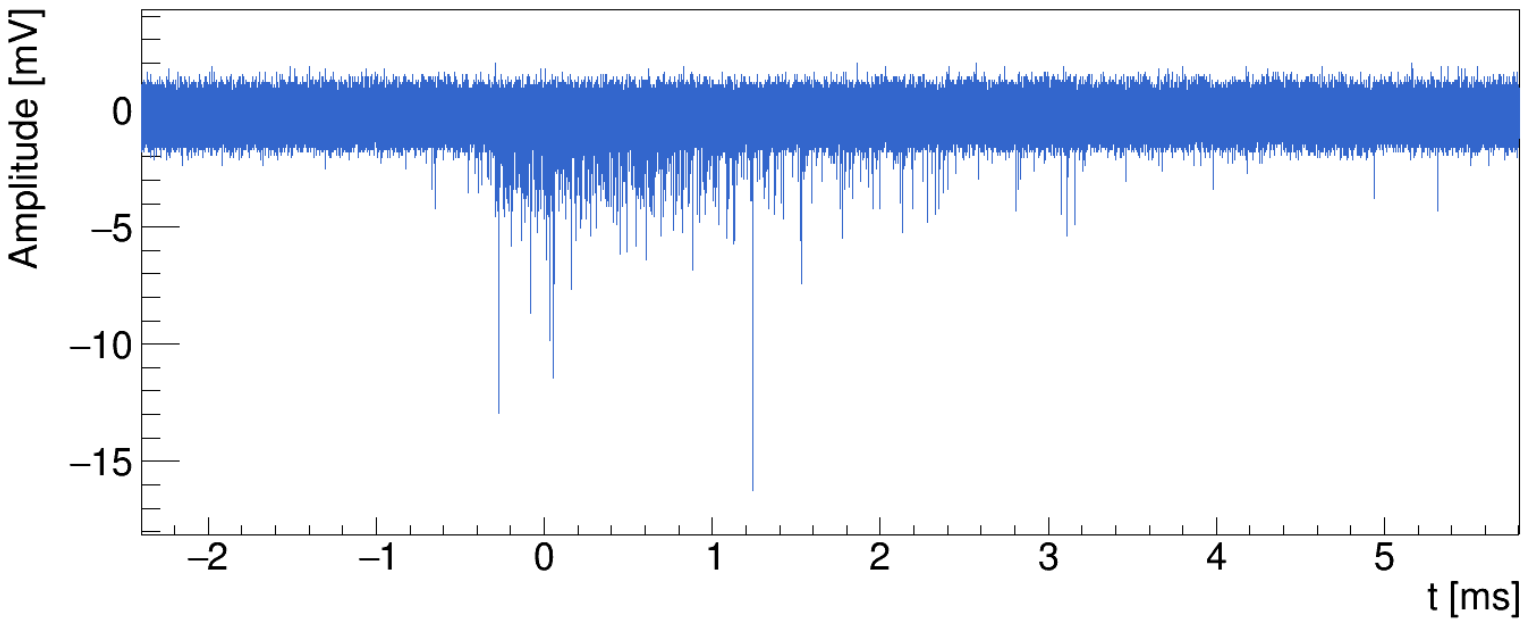

**(b)** 300 V/cm.

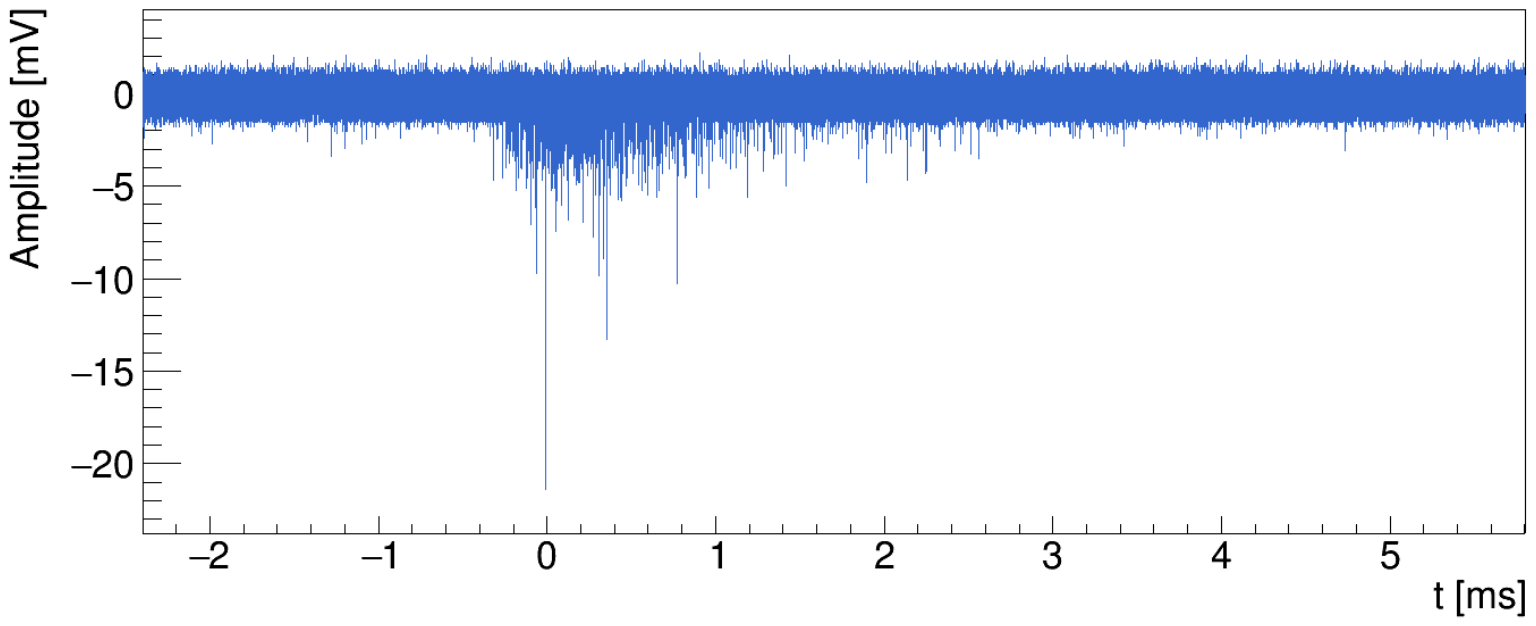

**(c)** 400 V/cm.

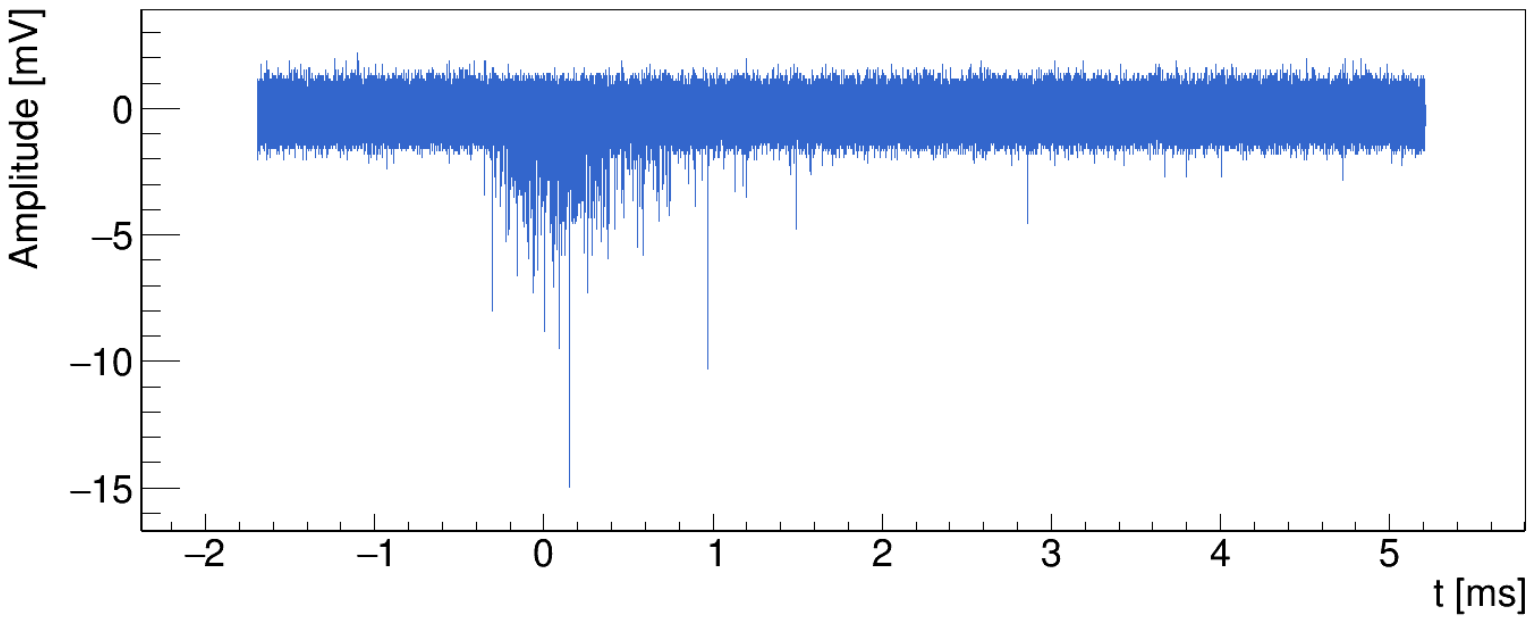

**(d)** 600 V/cm.

**Figure 7.5:** Four examples of NID PMT waveforms acquired at increasing electric field strengths, illustrating the shortening of the signal and the increasing density of peaks as the field increases.

Section 7.2.1 for both CMOS images and PMT waveforms.

The data acquired within this configuration were used to develop the original algorithm to analyze NID waveforms illustrated in Section 7.3.1. The analysis of the CMOS images is still under finalization within the collaboration and therefore will not be discussed further in this thesis.

### 7.3.1 NID PMT waveform time extension analysis

This section presents the original analysis procedure developed by the author to evaluate the time extension of the PMT waveforms acquired with MANGO in NID operation. As mentioned, these signals are unprecedented, or at least no literature has been found where they are addressed. Therefore, the analysis developed serves as an initial attempt to extract meaningful information from these waveforms.

Given the peculiar nature of these signals, as described in Section 7.2.1, determining the *start* and the *end* of the waveform signal in order to extract its time extension does not result straight forward. The large time span of these signals, combined with the small width of the individual peaks, requires recording at a high sampling rate on the order of GS/s with the oscilloscope. However, this high sampling rate introduces additional electronic noise, which interferes with the already low-amplitude peaks. To address these challenges, an original algorithm was developed to manage both the electronic noise introduced by the oscilloscope and the irregular, discrete nature of the NID signals. Furthermore, due to the high sampling rate and the long time span of the signals, the PMT waveform recordings result in large data file sizes, making the analysis time-consuming and, for that reason, also harder to systematically optimize.

To explain the algorithm procedure, two full examples are provided in Figures 7.6 and 7.7. These signals were acquired at the same drift distance of 3.5 cm, for two different electric fields (350 and 600 V/cm), illustrating two different scenarios with longer and shorter waveforms. Starting from the original raw waveform shown in Figures 7.6a/7.7a, the various steps of the developed algorithm are illustrated in the following:

1. From the original signal, a new one is created by retaining only the peaks that exceed six times the RMS of the waveform, which is approximately $\sim 2.5$ mV. The RMS is calculated over the first 500 points of the waveform, which are expected to be free of signal. The remaining points are set to zero, resulting in the signal shown in Figures 7.6b/7.7b. This step is crucial because, due to the high sampling rate, most of the points in the waveform correspond to the baseline electronic noise. Without this thresholding, a simple rebinning of the entire waveform would just blur the already few visible signal peaks.

2. The new peaks selected from this process are used to construct a histogram, creating a rebinned version of the waveform with a total of 200 bins over a 10 ms time span. This rebinned signal is shown in Figures 7.6c/7.7c.

3. In the rebinned histogram, the beginning (end) of the signal is defined when two consecutive bins exceed (fall below) 10 mV. These two time stamps are then used to calculate the time width of the signal, as shown in Figures 7.6d/7.7d.

4. As a cross-check of this procedure, the *begin* and *end* time stamps can then be plotted on the original signal, as shown in Figures 7.6e/7.7e.

As visible in these two examples, the algorithm selects mostly the *bulk* part of the signal, i.e., the region where the majority of the peaks are present or, in other words, where the peak density is highest and where we expect the main part of the signal to be present. The complete validation and optimization of this algorithm was particularly challenging due to the limitations imposed by the large data file size and the lack of literature on the topic and more generically of an in depth understanding of the exact physical and chemical processes at play in such complex, never tested before, ternary NID mixture. Nonetheless, we believe that this preliminary analysis strategy resulted satisfactory enough to be able to detect the main expected dependencies of the measured waveform time extension, as will be discussed in the following Section.

### 7.3.2 Preliminary results from NID PMT waveform analysis

The distribution of the PMT waveform time extension evaluated with the algorithm discussed in Section 7.3.1 were fitted with a Gaussian distribution, and the fitted means as a function of drift field and distance from the GEM are shown in Figure 7.8, where the full line represent the error on the mean as returned by the fit. As discussed in Section 7.3.1, we are aware that these estimates of the waveform time extension are affected by a systematic uncertainty from the algorithm, however, since no physics parameters are attempted to be extracted from this data (as will be discussed below), the systematics are not shown.

Drift velocity and mobility measurement (for both ED and NID mixtures) are typically performed by measuring the time difference between the creation of the primary ionization (i.e., $t_0$) and its arrival at the amplification plane [287] divided by the drifted length. The setup realized for the measurements presented in this Chapter lacked of the possibility to determine the $t_0$, since it was intended only to establish NID operation with 3D optical readout at (nearly) atmospheric pressure and develop the first reconstruction of NID PMT waveform, rather than to perform precise time measurements.

The algorithm developed and illustrated in Section 7.3.1 evaluates the waveform time extension as the time difference between the first and the last charge carrier that arrives at the amplification plane. While this effectively encodes the interesting physics parameters of the problem, the long alpha range $R$ of $3-4$ cm in the active gas volume introduces a complex convolution of diffusion and geometrical effects which prevents the measurement of mobility from the acquired data. Nonetheless, a general argument can be made to highlight how the observed dependencies on the drift fields and drift distances reported in Figure 7.8 follow the expected trends, further demonstrating the

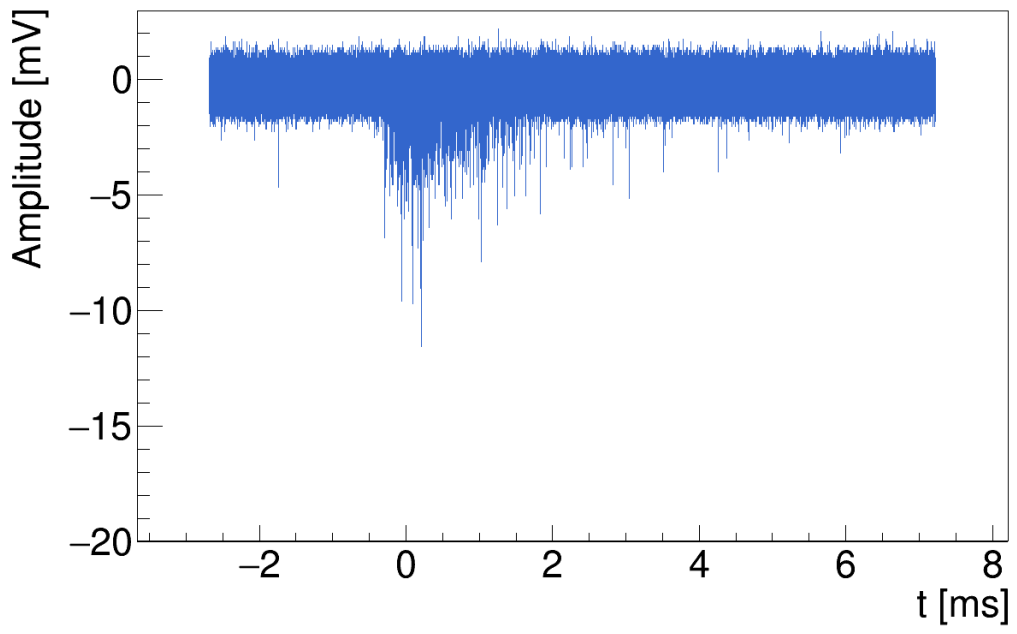

**(a)** Original signal

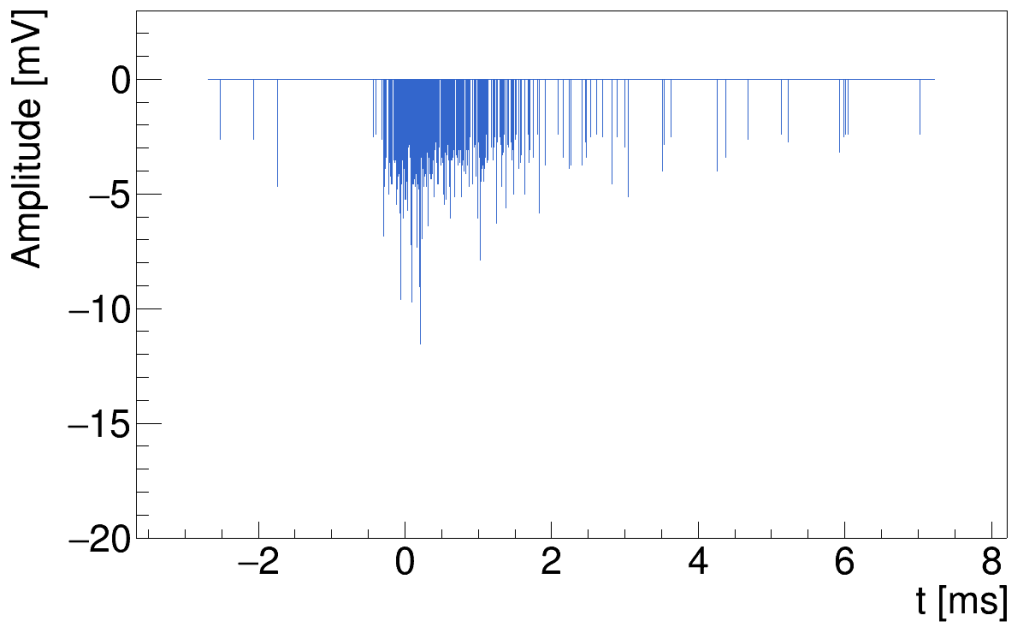

**(b)** Only peaks above 6 $*$ RMS

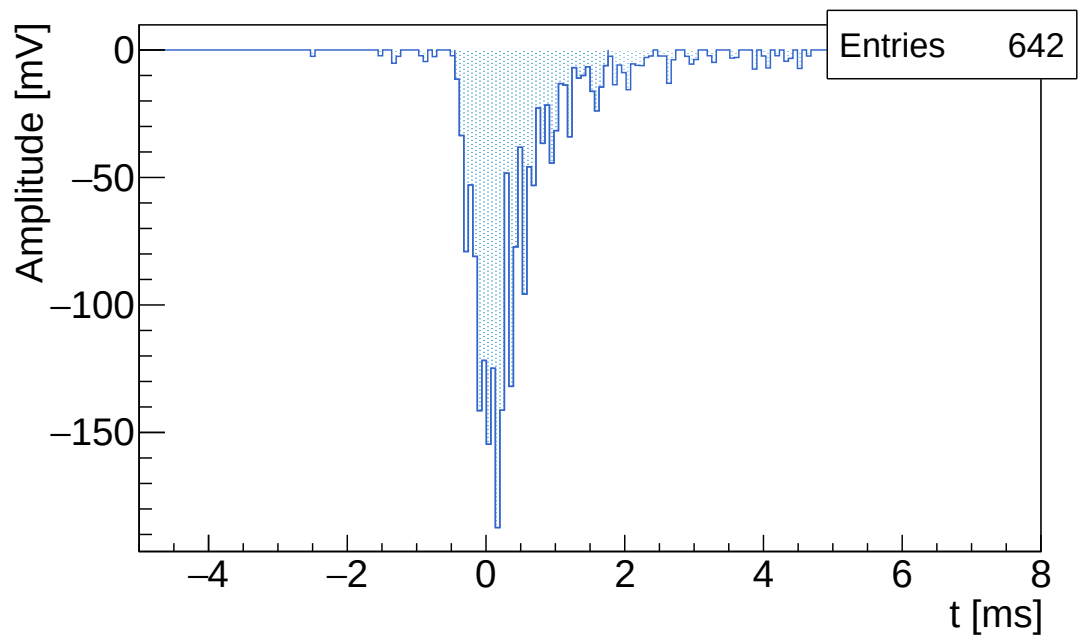

**(c)** Rebinned waveform

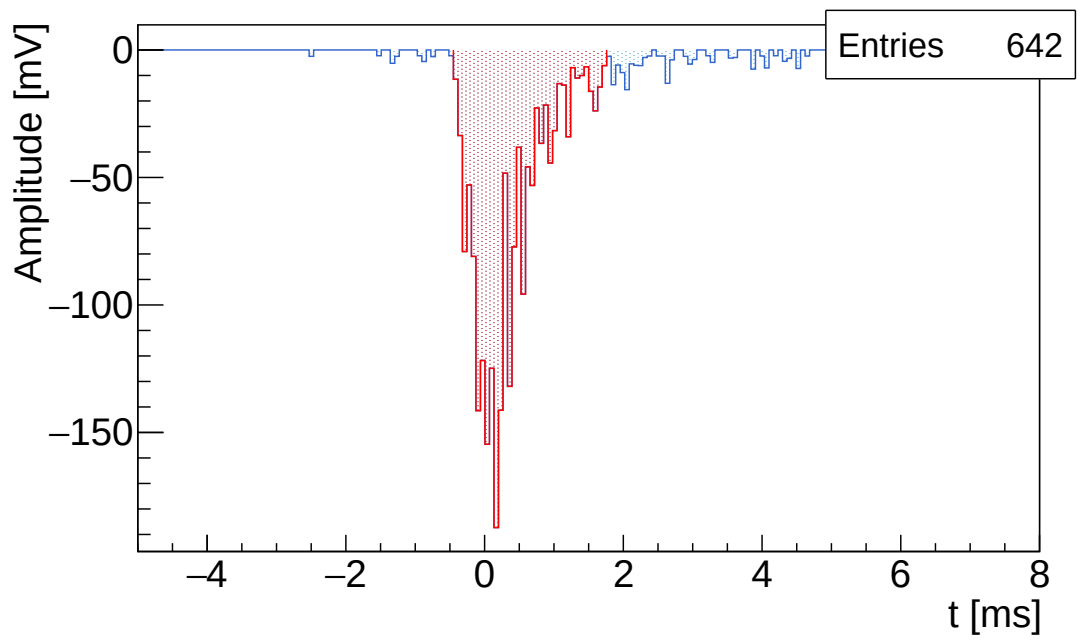

**(d)** Reconstructed signal in red

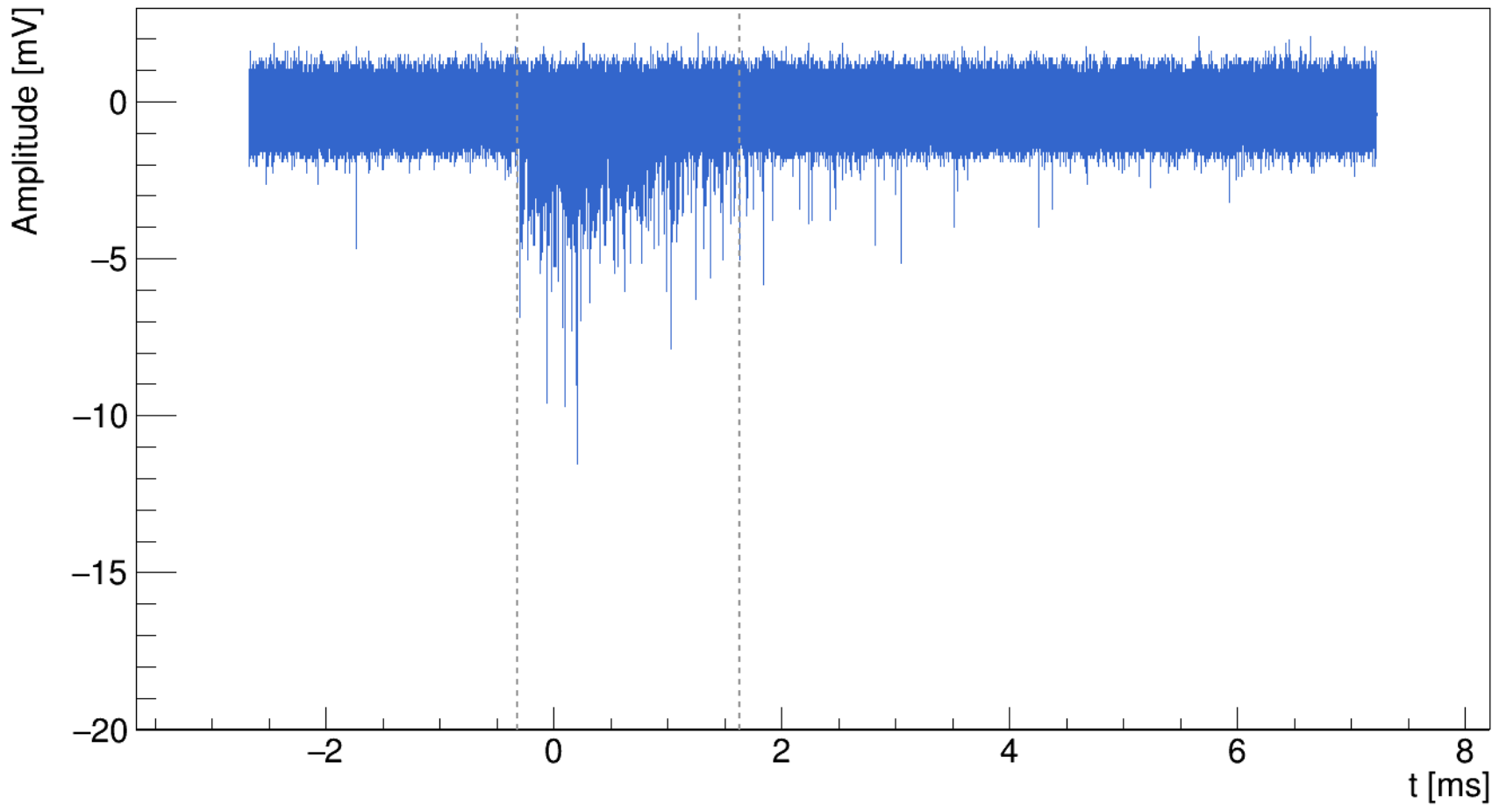

**(e)** Original signal with *start* and *end* timestamps overlaid as gray dashed lines.

**Figure 7.6:** Example of the analysis performed on NID PMT waveforms: (a) the original signal undergoes a (b) peak selection, where only those above 6 $*$ RMS of the noise level are retained. From this, (c) a rebinning procedure is applied, and (d) the *start* and *end* timestamps are determined, from which the signal duration is calculated. (e) These timestamps are then overlaid onto the original signal.

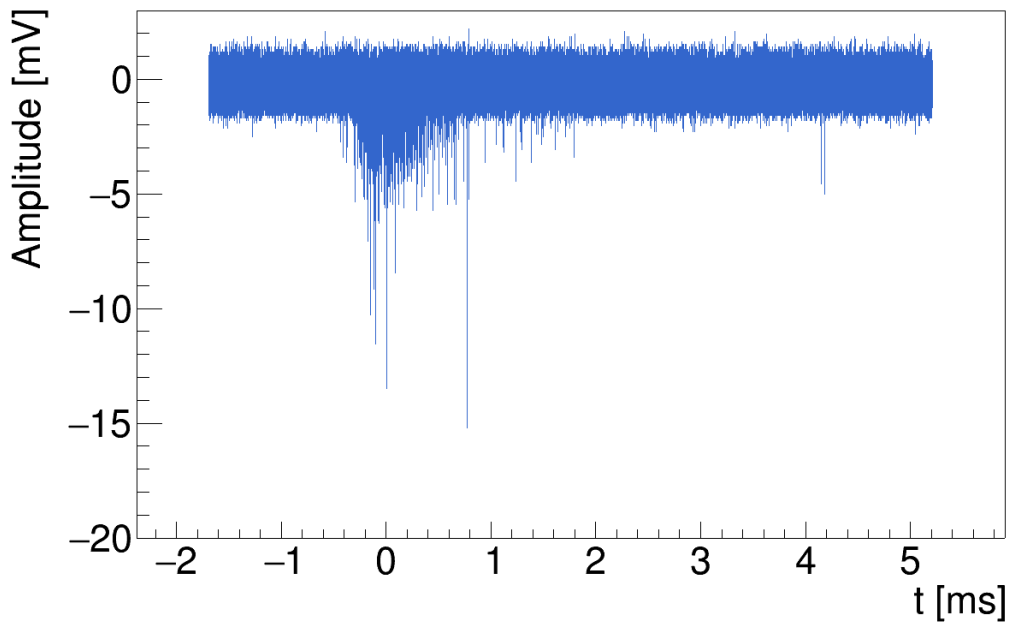

**(a)** Original signal

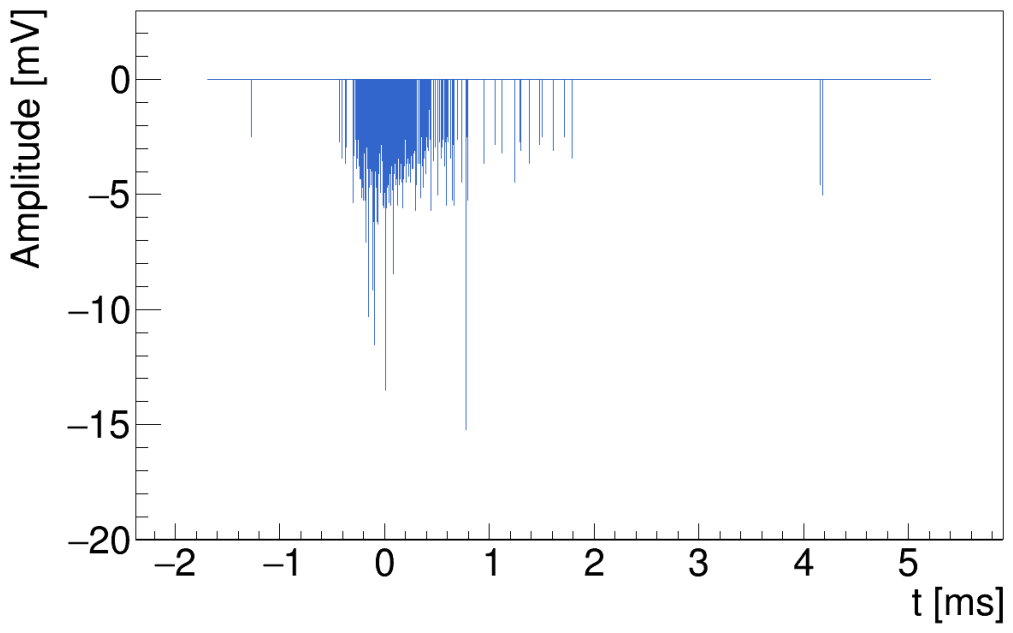

**(b)** Only peaks above 6 $*$ RMS

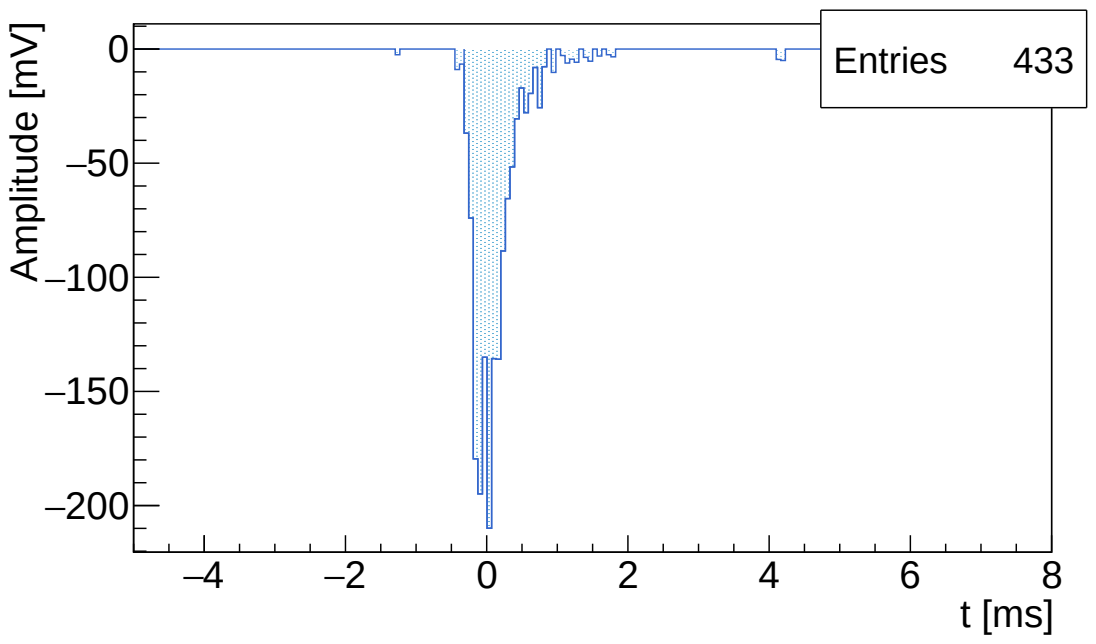

**(c)** Rebinned waveform

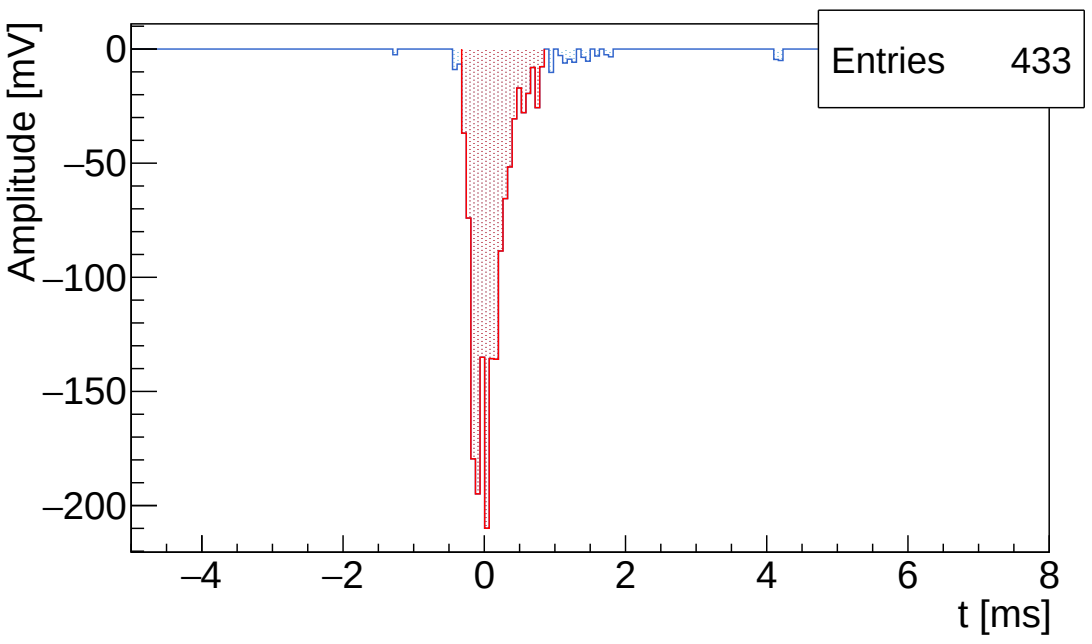

**(d)** Reconstructed signal in red

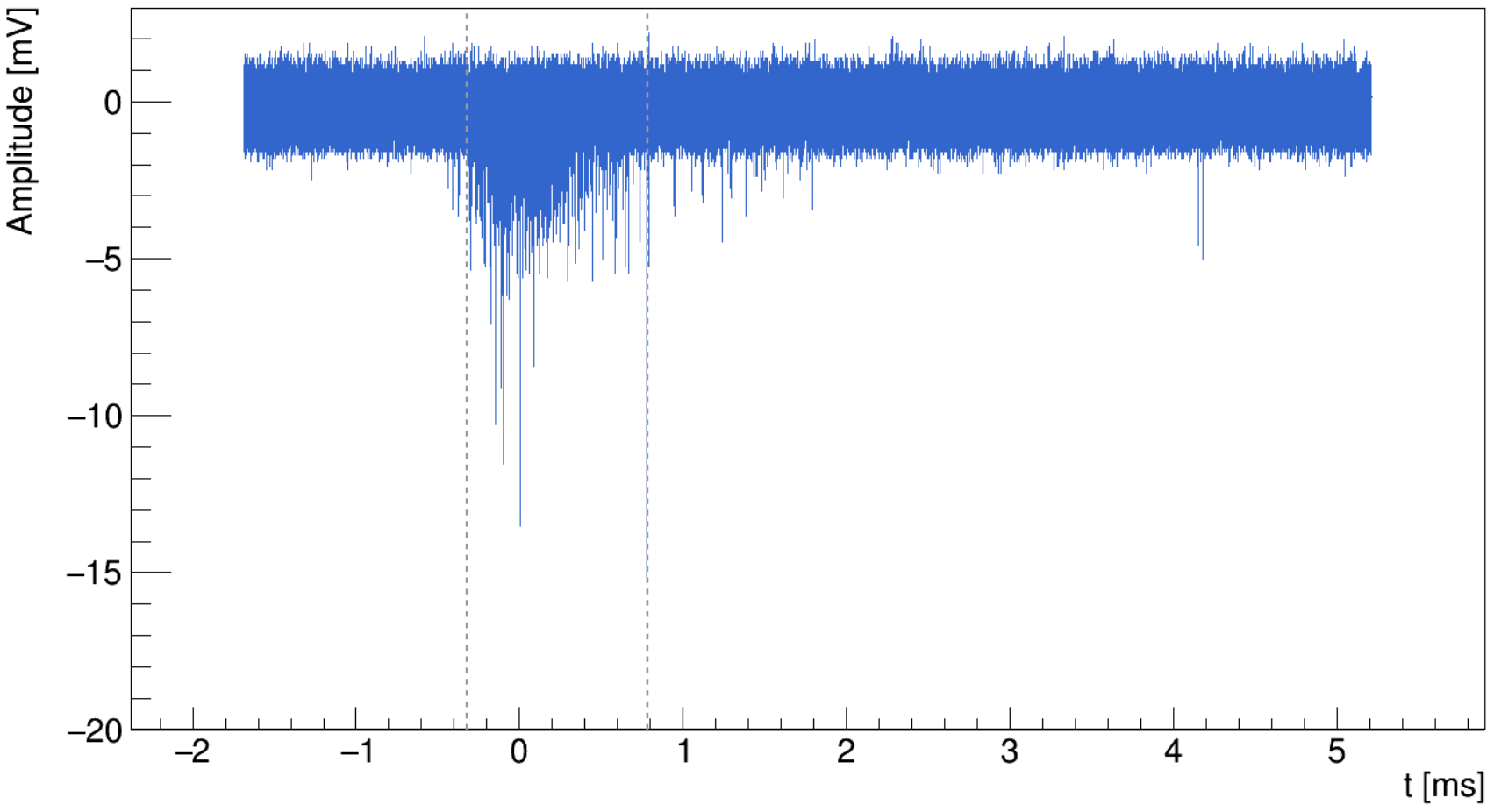

**(e)** Original signal with *start* and *end* timestamps overlaid as gray dashed lines.

**Figure 7.7:** Example of the analysis performed on NID PMT waveforms: (a) the original signal undergoes a (b) peak selection, where only those above 6 $*$ RMS of the noise level are retained. From this, (c) a rebinning procedure is applied, and (d) the *start* and *end* timestamps are determined, from which the signal duration is calculated. (e) These timestamps are then overlaid onto the original signal.

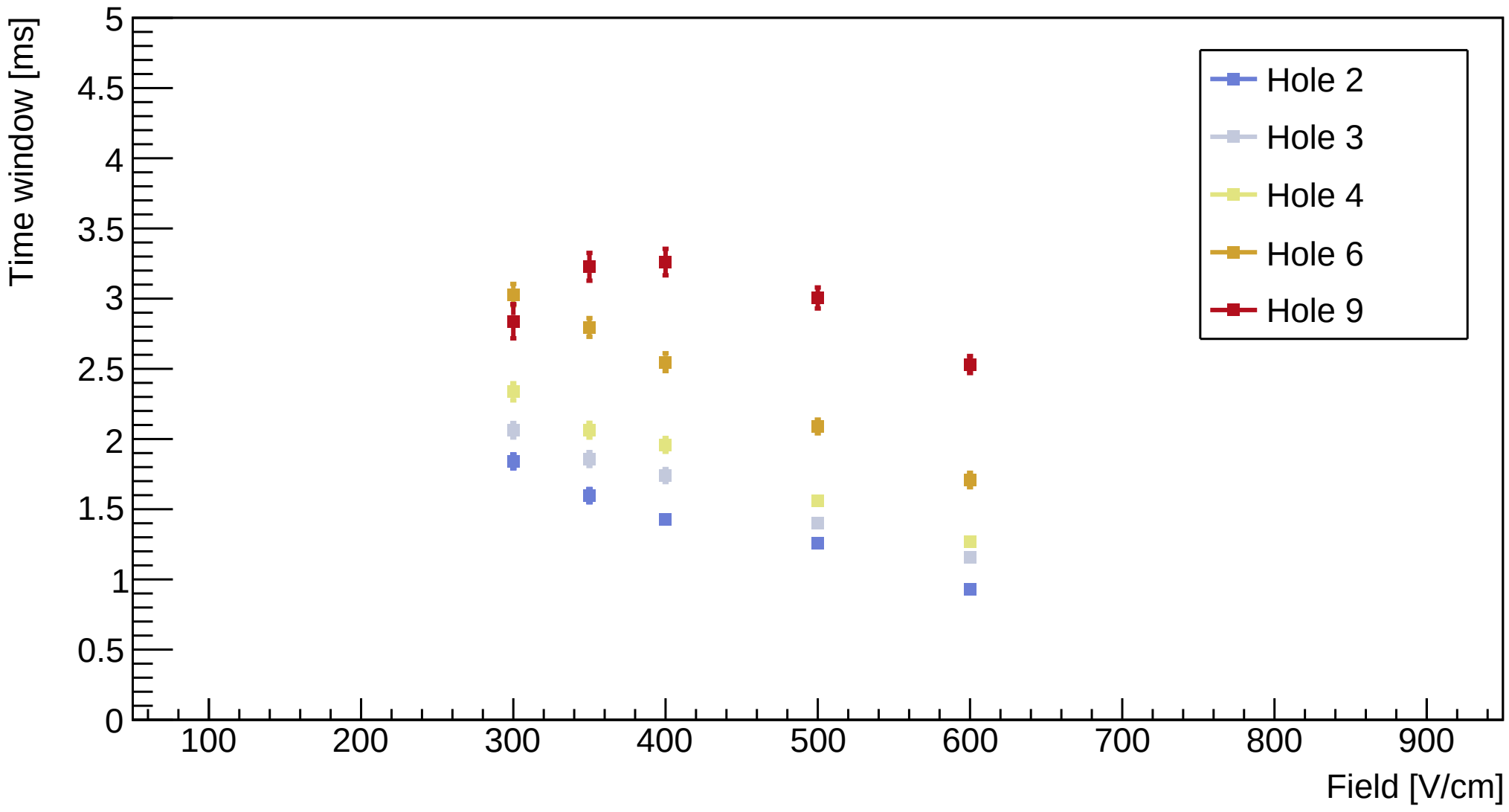

**Figure 7.8:** Time windows of NID PMT signals measured using the algorithm described in Section 7.3.1, for various drift field strengths and drift distances. Each "hole" corresponds to the spacing between the field rings of the electric field cage. Data were acquired at 650 mbar and room temperature.

validity of the preliminary NID PMT analysis developed and the NID nature of the observed events.

Generally speaking, and as already extensively elaborated from Section 3.2 through this thesis, the time extension of the PMT signal in the CYGNO experimental approach is determined by the spread of the arrival times of the primary ionization charges (i.e. electrons in ED operations and anions in the NID case) to the amplification plane. Physically, two main effects can contribute to the NID waveform time extension measurement in this context:

- **Diffusion of charges:** As charges drift, they undergo random thermal motion that broadens their spatial distribution along the drift direction Z following a diffusion coefficient $D_L$ that depends on the drifting charges energy. In addition to this, the amplification process within the GEM adds in quadrature a constant spread $\sigma_0$ independent of Z. Such broadening can be modeled as a Gaussian with a $\sigma_t$ of the form:

$$\sigma_t(Z, E) = \sqrt{\sigma_0^2 + \frac{\sigma_z^2(Z, E)}{v^2}} = \sqrt{\sigma_0^2 + \frac{2\, D_L\, Z}{\mu^2 E^2}} \tag{7.7}$$

where the drift velocity $v = \mu E$ dependence on the drift field is explicitly expressed in terms of mobility. For a perfectly collimated track perpendicular to the drift direction, that produces N primary ionization clusters uniformly spaced, the only contribution to the measured time extension of the waveform would come from the difference between the *earliest* and *latest* charge carrier arrival among those N due to their different diffusion. For $N \gg 1$ (actually of $\mathcal{O}(10^6)$ for an alpha particle in the NID gas), the extreme-value theory can be applied, yielding to an expected mean of the measured time extension of:

$$\langle \Delta t \rangle (Z, E) \approx 2 \cdot \sigma_t(Z, E) \cdot \sqrt{2 \cdot \ln N} \tag{7.8}$$

This estimation possesses anyway an intrinsic significant systematic due to the fact that the energy deposit along an alpha track is not constant (i.e. the N primary ionization clusters are not uniformly spaced), but is actually a stochastic process, which produces a Bragg peak creating more primary ionization towards the end rather than the start of the track (that will effectively differ event by event). Additionally, it is important to notice how, while we can reasonably expect $D_L$ to be thermal or nearly thermal, no physical arguments exist to assume that $\sigma_0$ should be the same as for alphas measured with ED operation, given the intrinsic differences of the NID amplification mechanism. For this reason, the *minimal signal* concept discussed and estimated in Section 6.3.1.2 cannot be employed here, nor exists NID $^{55}Fe$ or other *spot-like* data that can be used for an alternative evaluation of this contribution.

- **Track geometry:** In any realistic experimental setup (even with an highly collimated source) the measured tracks will not all be perfectly parallel to the amplification plane, but will be produced within a cone of maximal aperture $\theta_{max}$, and therefore each track might experience a tilt $\theta$ with respect to the GEMs. This will introduce a geometric path difference contribution $\Delta t_{geom}$ between the *first* and *last* charge to the time spread, depending on the inclination $\theta$, the effective range R of the ionizing track in the gas active volume, and the drift velocity, as:

$$\Delta t_{geom}(E, R, \theta) = \frac{R \cdot \sin\theta}{v} = \frac{R \cdot \sin\theta}{\mu E} \tag{7.9}$$

Practically speaking, each track in the measured sample with a $\theta$ small enough that the diffusion term significantly exceeds the geometrical path difference $\Delta t_{geom}$ induced by the tilt, will contribute to the measured waveform time extension only through the total diffusion spread $\sigma_t(E, Z)$ of the charges with Equation 7.8. In the opposite regime, whenever $R \sin\theta / v \gg \sigma_t$, the first and last arrivals can be reasonably assumed to come always from the track geometrical endpoints, therefore contributing only with the $\Delta t_{geom}$ term of Equation 7.9 to the measurement. The crossover between the two regimes happens in proximity of a critical angle $\theta_c$ where the two contributions are comparable,

that can be defined as:

$$\frac{R \cdot \sin\theta_c}{v} = \sqrt{\frac{2D_L}{v^2}(2Z + R \cdot \sin\theta_c) + 2\sigma_0^2} \tag{7.10}$$

Since $\theta_c$ depends on the drift field and the drift distance simultaneously, and since the unknown $\sigma_0$ can play a significant role in this, the relative contribution of the two regimes in the measurements shown in Figure 7.8 varies point to point. To properly model the data to extract the physics parameters of interest, especially the intermediate zone where the two contributions are comparable, a dedicated and comprehensive Monte Carlo simulation of the setup and the source collimation and inclination would be necessary.

While the lack of detailed simulation of alpha tracks already extensively discussed in this thesis prevents such elaborate analysis, qualitative arguments can still be applied. Both the diffusion and the geometrical contributions present a strong $1/E$ dependence (mainly through the drift velocity $v$), clearly visible in Figure 7.8, with measured $\Delta t$ decreasing at higher fields. In addition, the diffusion effects adds a $\sqrt{Z}$ term to the waveform time extension, which is consistent with the observed $\Delta t$ increase at longer drift distances. It is important to notice how the potential presence of minority carriers in the measured data would further complicate picture introducing an additional term $\Delta T_{v_{diff}}$ proportional to:

$$\Delta T_{v_{diff}} \propto \left(\frac{1}{v_{slow}} - \frac{1}{v_{fast}}\right) \cdot Z \tag{7.11}$$

emerging from the drift velocity difference between the fastest and the slowest negative ion species, which would further contribute to such increase. Given the uncertainties on the proportionality parameters, the unknowns of the problem and the relative low $Z$ values at which the measurements were performed, no further attempt can be made to extract the interesting physics quantity of the problem or interpret the data.

Nonetheless, we believe that the observed behaviors in terms of $Z$ and $E$ dependencies significantly confirms the overall picture of NID operation and effectively demonstrate the validity of the NID PMT algorithm developed in this thesis. The collaboration is currently working to repeat these measurement with a precise determination of the $t_0$ of the event (possibly from the primary scintillation of the alpha track in the gas detected with a dedicated PMT) in order to obtain data free from these uncertainties.

## 7.4 Conclusions & Prospects

The results presented in this chapter provide the first validation of NID operation with optical readout at atmospheric pressure, as well as the first-ever recording and analysis of NID PMT waveforms. This represents a significant milestone for directional rare event detection and low-diffusion TPC applications, with potential implications extending beyond optical readout technologies. Reducing diffusion during drift could allow for an increased drift length within the CYGNO framework while preserving current tracking performance. Such progress would substantially enhance tracking resolution and directionality for both electron and nuclear recoils.

The first-ever observation of PMT signals in NID operation mode presented several challenges due to the peculiar nature of these signals. The novelty of these studies is also reflected in the lack of literature available to guide the analysis or provide insights into the observed results. A preliminary analysis approach was developed to address this, showing promising initial results, although further optimization and validation within the collaboration and with other experts in the field are still required. The possibility that these PMT signals contain the imprint of each individual ion produced in the primary ionization process would represent a decisive breakthrough, significantly enhancing particle identification performance [297–299].

While still preliminary, it is important to emphasize that the waveform analysis of NID signals presented here is the first of its kind, serving as groundwork for future approaches by the community, and that the results reported represent only an initial step in exploring the full potential of NID mixtures for optical readout. Ongoing and future studies aim to further characterize the entire process. Upcoming work will focus on a systematic exploration of gain, diffusion, and directional sensitivity by varying the gas mixture composition and testing alternative amplification structures. The objective is to enhance the light yield and better understand the diffusion behavior of $He:CF_4:SF_6$ mixtures. Candidate amplification technologies under consideration include thicker GEM configurations, COBRA-GEMs [300], and ThickGEM-multiwire hybrids [301]. Additional efforts will also explore charge readout using Timepix3 [283] and the use of $He:SF_6$ mixtures with varying $SF_6$ concentrations. Regarding the PMTs, future tests might include the use of sensors capable of detecting the primary scintillation – such as SiPMs – to determine the $t_0$ and enable direct measurements of the drift velocity of $SF_6^-$ (and possibly other) ions.

The groundbreaking results presented in this chapter, along with recent advancements in the field of directional TPCs and the successful 3D reconstruction of alpha particles in LIME discussed in Chapter 5 and Chapter 6, demonstrate that a large-scale directional experiment with sensitivity to both Spin-Dependent and Spin-Independent couplings like CYGNO could, one day, represent a path to venture into the neutrino fog and ultimately determine the true nature of Dark Matter.

# Conclusions

Over the past decades, evidence from astrophysics, cosmology, and particle physics has established a consistent picture pointing to the existence of DM, which makes up most of the matter in the Universe. Among the leading candidates are WIMPs, which could produce nuclear recoils detectable in underground experiments. However, no conclusive detection has been achieved, and all current strategies face an intrinsic limitation from neutrino-induced recoils, known as the "neutrino fog". Directional detection offers a promising solution, making use of the expected anisotropy in WIMP-induced recoils due to the Earth's motion through the galaxy. This signature is unique and difficult to mimic with backgrounds, enabling not only improved background rejection but also the potential for an unambiguous DM discovery.

The CYGNO project is developing a large-volume gaseous TPC with a $He:CF_4$ mixture, using a hybrid optical readout of CMOS cameras and PMTs. This setup provides both directional sensitivity and ER/NR discrimination. In parallel, the INITIUM project – under which this thesis was supported – is implementing NID with gases like $SF_6$ to reduce diffusion and enhance tracking performance, further strengthening CYGNO's capabilities for directional DM detection. In the context of directional dark matter searches, the precise reconstruction of the 3D direction vector (including head-tail recognition) of ionizing events is crucial for identifying potential WIMP signals. Since the WIMP wind is expected to always originate from the same direction in galactic coordinates, this would cause an anisotropy in the angular distribution of NRs induced in the detector. This anisotropy would allow for the direct discrimination of background events that lack a directional dependence, as well as solar neutrinos, which arrive at Earth with a distinct, yet different, directional pattern compared to the WIMP wind. Ultimately, the directionality information is a key feature for positively identifying dark matter signals.

In this thesis, in parallel with the CMOS-based studies, a detailed analysis of the PMT signals was conducted to extract the fundamental information encoded in the waveforms, which describe the path traveled by the particles along the drift field direction. This included waveform integration and modeling, the development of pulse-shape discriminators, and the application of noise-filtering techniques. The PMTs proved essential not only for triggering but also for capturing subtle features of energy deposi-tion along the drift axis. Future developments in this area include improved modeling of the intrinsic PMT pulse shape – more accurately described by a Gaussian with an exponential tail – which could enhance the reconstruction of the temporal width of the signals.

Building on this, a complete 3D reconstruction method was developed by combining the high-resolution spatial imaging capabilities of the CMOS camera with the fast timing and light-collection efficiency of the PMTs. This hybrid approach enabled, for the first time, the full spatial reconstruction and localization in 3D (i.e. fiducialization) of ionization tracks in CYGNO detectors, including both the direction and the sense (head-tail) of the tracks. The 3D analysis begins with the application of the *directionality algorithm* – optimized for alpha particles in this thesis – which extracts the geometrical characteristics of the 2D ionization track observed by the CMOS sensor, namely the direction and sense in the XY plane. The transverse light profile of the track is also used to estimate its absolute Z position within the detector. The PMT analysis then contributes information about the longitudinal development of the track. New variables were introduced to extract the relative Z coordinate from the PMT signals using the Time-over-Threshold method, and to discriminate between electromagnetic interactions and NRs, including alphas. This information is combined with angular observables derived from Bragg-peak skewness and track topology, providing robust head-tail recognition for extended tracks.

The full 3D geometry – including parameters such as track length and angle relative to the GEM plane – is obtained by matching each PMT waveform (trigger) with the corresponding CMOS cluster. This sensor matching process is a key component of the CYGNO data reconstruction pipeline, as events recorded by the two sensors are not inherently synchronized. The work carried out on this topic was recently published in [221], with the author of this thesis as corresponding author. The additional successful adaptation of the BAT-fit to alpha particles carried out for this thesis – extending its applicability beyond $^{55}$Fe-like signals – represents a significant milestone in this regard. It demonstrates that the BAT-fit is effective not only for compact, point-like events but also for extended tracks spanning several centimeters, allowing for a clear association between signals observed in both sensors. Furthermore, the BAT-fit may enable a deconvolution of the geometrical dependence on track distance from the PMTs, potentially leading to a unified waveform that accurately reflects the longitudinal profile of the track, independent of PMT placement.

These advancements in 3D reconstruction lay the groundwork for the realization of a fully fiducialized and direction-sensitive optical TPC tailored for rare-event searches. Toward the end of Run 5 of LIME, an extended AmBe run was conducted. This dataset is expected to provide a clearer sample of nuclear recoils, offering further opportunities to refine the reconstruction of these type of events and their initial 3D direction – one of CYGNO's key signature measurements.

With the introduction of 3D reconstruction for alpha particles, it became possible to measure their full 3D length for the first time. This measurement is particularly valuable, as it provides a more reliable estimate of the alpha particle energy than the total light observed by the CMOS sensor. For nearly straight trajectories like those of alpha particles, the path length is directly proportional to their energy. In CYGNO detectors, where high gain settings are necessary to detect low-energy ionization events, alpha particles are prone to charge saturation, which compromises the linearity of energy estimation based on light yield. In contrast, the reconstruction of the track's 3D length is independent of the total collected charge and is instead limited only by the spatial granularity of the optical readout – on the order of $\mathcal{O}(10^2)$ $\mu m$ – making it a more robust proxy for particle energy under these conditions.

A preliminary analysis of the LIME underground data revealed discrepancies between experimental observations and Monte Carlo simulations, attributed largely to an unaccounted contribution in the background model. The main background sources considered in the simulations were radioisotopes from natural decay chains – namely $^{232}Th$, $^{238}U$, and $^{235}U$ – as well as common contaminants such as $^{40}K$ and $^{137}Cs$. While these were incorporated into the simulations, the contribution from gaseous $^{222}Rn$, introduced through infiltration into the gas system, could not be predicted and was therefore not considered. The decay of $^{222}Rn$ and its progeny produces multiple gamma and beta emissions – which appear as a flat background at low ($<$ 100 keV) energies – and $\mathcal{O}(MeV)$ alphas. The newly developed 3D tracking motivated a targeted study of the internal LIME material backgrounds through the analysis of the energy and spatial distribution of the associated alphas decays, with the goal of clarifying the presence of $^{222}Rn$ and of the other isotopes expected from material screening.

Using this framework, the MeV-scale alpha background in the LIME detector was studied in detail. For the first time, distinct peaks in the 3D track-length distribution were successfully associated with isotopes from the $^{222}Rn$ decay chain – such as $^{218}Po$ and $^{214}Po$ – as well as with $^{238}U$ and $^{210}Po$. These identifications were validated through energy conversion using SRIM software. Specifically, each measured 3D length was corrected using a length correction factor, then matched to its corresponding energy by identifying the alpha energy that returns the same range in SRIM. This allowed confident association of track-length peaks with specific decay products based on prior knowledge of LIME's background sources and comparison with existing literature.

The spatial distribution of alpha-emitting isotopes further confirmed these identifications. A significant concentration of events was observed near the electric field cage rings, consistent with their proximity to the sensitive volume, and the large $^{238}U$ and $^{232}Th$ contamination measured through material screening of these components. Moreover, the reconstructed Z-position and angular distributions aligned with physical expectations: $^{222}Rn$ decays appeared uniformly throughout the drift volume, further validating the hypothesis of its presence in the gas, while positively charged progeny were found to decay closer to the cathode and GEM planes, due to drift and plating effects. These daughter products are collectively known as Radon Progeny Recoils (RPRs). The presence of RPRs is a known challenge in rare-event search experiments, making this analysis and its results essential for assessing such backgrounds in current and future CYGNO detectors. It also enables the implementation of fiducial cuts, providing a valuable tool for background suppression in upcoming data-taking campaigns. A preliminary MeV-scale energy calibration was also established using the identified alpha peaks. While a linear response was observed, it was limited by the resolution due to sensor saturation at high energies. Nevertheless, these results solved prior inconsistencies between data and simulation and established a solid foundation for improved background modeling in CYGNO detectors.

The 3D reconstruction results unequivocally confirmed the presence of radon in the LIME gas system and directly influenced the design of the next-generation CYGNO-04 detector. During Run 4 of LIME, the use of specific gas filters – particularly those targeting moisture – was shown to significantly reduce the presence of $^{222}Rn$, likely due to the tendency of radon and its daughters to attach to water molecules. Following the introduction of such filters, the rate of detected alpha events dropped by approximately a factor of five, clearly demonstrating the effectiveness of this method. For the construction of CYGNO-04, these filters will be used in combination with a double-sealed design to further limit radon infiltration and reduce the related backgrounds. These findings reaffirm the central importance of 3D reconstruction in the CYGNO program, both for characterizing alpha backgrounds and, later, as a tool to identify DM signals over known backgrounds.

A major milestone was achieved with the first successful operation of a negative-ion drift (NID) TPC at atmospheric pressure using an optical readout. By adding a small fraction (1.6%) of $SF_6$ to the standard $He:CF_4$ (60:40) gas mixture, the MANGO prototype demonstrated stable NID operation at approximately 1 bar. This configuration significantly reduced transverse diffusion, therefore enhancing spatial resolution over long drift distances. Notably, PMT waveforms in NID mode were observed and analyzed for the first time – within the available literature – during this study. These waveforms exhibited complex, multi-peak structures likely corresponding to the arrival of individual ions at the amplification plane. A preliminary algorithm was developed to interpret these features, opening the doors for 3D optical reconstruction in NID operation for the first time. This breakthrough paves the way for scalable, low-diffusion gaseous detectors capable of maintaining high resolution over meter-scale drifts, a crucial step toward the construction of large-volume directional detectors.

In the near future, several promising developments are expected from the 3D analysis. First, once the CYGNO PMT simulation is fully developed, the entire analysis pipeline can be re-optimized, focusing on the physical characteristics of the PMT signals. Since the current analysis was originally optimized for alpha particles, more suitable parameters can be defined for studying low-energy nuclear recoils – CYGNO's primary candidate for WIMP-induced signals. In this context, the similarity between alphas and NRs provided a solid starting point for this refinement. Furthermore, the analysis of PMT waveforms has proven useful in distinguishing between electronic and nuclear recoil signals, a crucial capability for dark matter detectors. The use of the PMT for this purpose can complement the ongoing developments by CYGNO using the CMOS sensor, ultimately enabling a dual-sensor PID strategy. Lastly, although the NID-related studies presented in this thesis are still preliminary, a natural next step would be to integrate the current 3D reconstruction framework developed in LIME with the NID PMT analysis techniques. This would enable full 3D track reconstruction using negative ions. While this goal remains somewhat distant, achieving it would be a major advancement for tracking TPCs in general, and especially for those used in dark matter searches.

Overall, the results presented in this thesis demonstrate that high-resolution 3D reconstruction, precise background characterization, and low-diffusion drift can be concurrently achieved in optical TPCs. These capabilities form the core of next-generation rare-event detectors that can effectively discriminate signal from background through both fiducialization and directionality. The technologies and methodologies developed here position the CYGNO/INITIUM framework as a promising pathway toward directional dark matter detection below the neutrino fog, targeting both spin-dependent and spin-independent interactions. This work not only establishes a landmark in the analysis of ionization events within the CYGNO/INITIUM project, but also advances the state of the art in gaseous TPC readout and analysis, providing essential tools for future dark matter search experiments.